\documentclass[12pt,notitlepage,fleqn,openbib]{article}%
\usepackage{amssymb}
\usepackage{amsfonts}
\usepackage{amsmath}%
\usepackage{graphicx}
\newtheorem{theorem}{Theorem}

\newtheorem{notation}[theorem]{Notation}

\newenvironment{proof}[1][Proof]{\noindent\textbf{#1.} }{\ \rule{0.5em}{0.5em}}
\ifx\pdfoutput\relax\let\pdfoutput=\undefined\fi
\newcount\msipdfoutput
\ifx\pdfoutput\undefined\else
\ifcase\pdfoutput\else
\ifx\paperwidth\undefined\else
\ifdim\paperheight=0pt\relax\else\pdfpageheight\paperheight\fi
\ifdim\paperwidth=0pt\relax\else\pdfpagewidth\paperwidth\fi
\fi\fi\fi
\begin{document}

\title{Yang Mills model of interacting particles in the classical field theory}
\author{Jean Claude Dutailly\\Paris (France)}
\maketitle

\begin{abstract}
The purpose is to study systems of interacting particles, in the General
Relativity context, by the principle of least action, using purely classical
concepts. The particles are described by a state tensor, accounting for the
kinematic part (rotation) using a Clifford algebra, and the interaction part
(charges). The force fields, including gravitation and other forces, are
described by connections on principal bundles. A solution has been found to
account in the lagrangian for individual, pointlike, particles. The
constraints induced by the equivariance (gauge) and covariance are reviewed.
Modified Lagrange equations are written for general lagrangians. A model,
based on scalar products, Dirac operator and chirality is studied in more
details. Problems related to symmetries, including the Higgs mechanism, are introduced.

With a comprehensive coverage of topics, and the purpose of finding a physical
meaning to the mathematical tools used, it can be a useful pedagogical
study.\ But it opens also some new paths.

\end{abstract}

\begin{center}
\newpage

{\LARGE CONTENTS}
\end{center}%

\begin{tabular}
[c]{lll}%
PART 1 : FOUNDATIONS &  & \\
Geometry.................................... & ... & \pageref{Geometry}\\
Particles..................................... & ... & \pageref{Particles}\\
Force fields................................ & ... & \pageref{Force Fields}\\
Part 2 : LAGRANGIAN &  & \\
Principles................................... & ... &
\pageref{Lagrangian Principles}\\
Gauge equivariance................... & ... & \pageref{Guage equivariance}\\
Covariance................................. & ... & \pageref{Covariance}\\
Part 3 : LAGRANGE\ EQUATIONS &  & \\
Principles................................... & ... &
\pageref{Lagrange equations principles}\\
Equations.................................. & ... &
\pageref{Lagrange equations}\\
Noether currents....................... & ... & \pageref{Noether currents}\\
Energy-Momentum tensor......... & ... & \pageref{Energy momentum}\\
Part 4 : MODEL &  & \\
Scalar Products......................... & ... & \pageref{Scalar products}\\
Dirac Operator........................... & ... & \pageref{Dirac operator}\\
Chirality..................................... & ... & \pageref{Chirality}\\
Lagrangian................................ & ... & \pageref{Lagrangian Model}%
\\
Equations.................................. & ... &
\pageref{Model Lagrange equations}\\
Choosing a gauge..................... & ... & \pageref{Choosing a gauge}\\
PART 5 : SYMMETRIES &  & \\
CPT\ invariance.......................... & ... & \pageref{CPT}\\
Signature.................................. & ... & \pageref{Signature}\\
Spatial Symmetries................... & ... & \pageref{Spatial symmetries}\\
Physical Symmetries................. & ... &
\pageref{Contant physical characteristics}\\
Symmetry Breakdown............... & ... & \pageref{Symmetry breakdown}\\
PART\ 6 : APPLICATIONS &  & \\
General Relativity...................... & ... & \pageref{General relativity}%
\\
Electromagnetism.................... & ... & \pageref{Electromagnetism}\\
CONCLUSION......................... & ... & \pageref{Conclusion}%
\end{tabular}

BIBLIOGRAPHY :..........................................\pageref{Bibliography}

\newpage

The principle of least action has been the workhorse of theoretical physics
for decades. Both its high versatility and prodigious efficiency largely make
up for the weakness of its foundations. In its many implementations two
different paths can be broadly discerned. The classical approach (in newtonian
or relativistic geometry) encompasses mechanics and the theory of fields, and
provides sound basis to statistical mechanics and thermodynamics. There are
many ways to address the lagrangian specification (Morrison [19 ], Soper
[24]), but the key is to proceed quickly to the phase space, endowed with a
symplectic structure, where all the mathematical tools can be efficiently
deployed (Hofer [9]).\ The various Einstein-Vlasov equations are an example of
this approach (Choquet-Bruhat [3]). On the other hand quantum mechanics and
the quantum theory of fields make also an intensive use of the principle of
least action, as an hamiltonian or lagrangian is required as starting point.
The two main differences are that the distinction between particles (matter
fields) and "force fields" (bosonic fields) is blurred, and that the basic
axioms of quantum mechanics (such as summed up by Weinberg [30]) and the
Wigner theorem open the way to a more direct analysis of the equations. It is
the only theory that gives us some predictions for the physical
characteristics of the particles and how they change but, if there is no need
to aknowledge its power, we are still left with one the biggest enigma of
modern physics : "Where does the first quantization come from ?". As both the
classical and quantum approaches lead, through Poisson brackets and the likes,
to Banach algebras, one way to answer this question is to circumvent the
principle of least action and go straight to C*-algebra.\ It is roughly what
is attempted with the algebraic quantum field theory (Halvorson [8]). One
issue is that in the simplest of physical system (1 spinless particle) the set
of observables is not a C*-algebra...and anyway one is still far away from
understanding the axioms of quantum physics.

\ 

So, whatever one's personal philosophical belief about "realism in physics",
it seems useful to pursue further the classical approach with the principle of
least action.\ All the more so that decades of hard labour and progresses in
mathematics have brought to us new schemes, such as the Yang-Mills description
of the fields, and powerful tools such as fiber bundle or Clifford algebra. It
is the main objective of this paper : check if it is possible to build a
useful and comprehensive model of particles interacting with fields, in a
purely classical way, using the tools of to day (or at least yesterday),
hoping to get some hint at the meaning of quantum physics. Much work has
already been done about the mathematical foundations of a "modern" classical
theory of fields (Giachetta [5]) but I want to focus here on putting together
the ingredients to get the full picture of a physical system.

To be useful and comprehensive the model should :

- adopt the geometry of General Relativity, without any exotic feature (no
extra dimension)

- describe the kinematic (meaning here rotation and any other self centered
geometric movements), the dynamic (meaning here displacement in space-time)
and the physical characteristics (such as mass and charge) of the particles,
prior to any quantization

- include gravitation and "other" forces, seen separately (no "GUT" in stock)
and treated as gauge fields in a Yang-Mills formalism

- stay at the "laboratory level" (no cosmology).

As much as possible the different mathematical objects and hypothesis should
be clearly defined and related to physical or experimental procedures that
could be used to get numerical values of the variables.

Two difficulties arise :

- The "point particle" issue : the need to manage simultaneously force fields
and matter fields raises some mathematical difficulties which should not be
discarded lightly.\ I overcome them with an adaptated Green function in what
seems to be a new solution. However the self-radiation reaction issue (Poisson
[21], Quinn [22]), which is more about solving the equations, has been kept
out of the scope of the paper.

- The metric issue : the formalism of fiber bundle does not fit well with the
traditional treatment of gravitation based upon the metric. Furthermore, as we
will see, the connexion should not be torsion free. So I stay firmly in the
scheme of fiber bundle and connection, expressed in the tetrads formalism, and
the metric tensor is seen as a by-product of the orthonormal basis.

The first part gives the description of the geometric model (a gaussian normal
coordinates system), the kinematic model (a representation of a Clifford
algebra), the physical characteristics of the particles (through the
representation of a unidentifed compact group U) ,the associated tensor bundle
of their "state", and the covariant derivative.

The second part starts with a description of the configuration bundle and
gives a solution to the treatment of individually interacting particles. Then
it addresses the lagrangian issues : gauge invariance and general covariance,
and sets up the most general constraints on a lagrangian.\ 

The procedures to solve the variational problem are reviewed in the \ third
part, from the general fiber bundle and the functional derivative formalisms,
the equations are listed and the Noether currents for gravitation and the
other fields are evidenced with their super-potentials. The definitions and
properties of the energy momentum tensor are reviewed including a
"super-conservation law".

In the fourth part the scheme is implemented in a simple, but quite general,
model.\ It requires the definition of scalar products on the fiber bundles, of
the Dirac operator, and the introduction of chirality, which enables to
further specify the representation spaces in the vector bundles. A lagrangian
is specified. In this simplified, but still comprehensive, model the Noether
currents lead to moments (linear and angular momentum, charge and "magnetic
momentum") which characterize the particles. The gravitational connection has
an explicit and simple formulation from the structure coefficients and the
moments, showing that the connection will not usually be torsion free.

Symmetries are reviewed in the fifth part : the CPT problem has a simple
explanation. Spatial symmetries (spin particles) induce a strong dependance of
the state of the particles upon the 3 parameters defining a spatial rotation,
and its "paradoxical" properties are evidenced. Physical symmetries (involving
the group U) are studied first as defining families of particles, then through
the breakdown of symmetries in the Higgs mechanism.

In the final part the model is implemented to do the junction with general
relativity and electromagnetism.

Overall the paper shows most of the fundamental concepts of field theory in a
consistent and comprehensive scheme, and in a fully classical
picture.\ Implementing all the tools which are now common in theoretical
physics, it should be a useful pedagogical instrument. This paper requires a
common knowledge of the principles of fiber bundles, connections and Lie
groups.\ As often as possible the basic definitions are recalled. Some
calculations are a bit cumbersome, but I feel better to follow a simple if
lengthy path than to risk shortcuts which would require sophisticated
mathematical concepts.

As such the model can be a good starting point to investigate further

\newpage

\part{FOUNDATIONS}

Our aim is to build a model describing a system of N-particles interacting
with gravitation and other forces fields. The first key word is "system". In
physics it has great implications : it means that the particles can be
identified, their trajectories and properties measured and followed up for
some time, by observers who are networked in order to get a full picture of
what is happening. Perhaps some people would object that it is an impossible
task from the quantum point of view, but in classical physics, where we stand,
it is a sensible one, and any scientist should start by defining what it means
by the "system" that he (or she) is modelling. So the system is supposed to be
included in a not too big region of the universe (that excludes cosmology),
clearly defined in space and time (no infinities, but large enough so that it
can be considered as isolated from external interactions), inhabited by
particles (which, as usual, are physical objects without any internal
strucures involved - no scale is implied), gravitation and "other" forces
fields (electromagntic and the likes). A single, or a network, of observers
has defined a map and frames to measure all geometric quantities (such as
position, speed and angular momentum) and procedures to measure the physical
quantities such as charge and fields. It is clear that the second are deduced
from the first using some test particles or fields.

The first step is so to define the geometric part of the model.

\section{GEOMETRY}

\label{Geometry}

The geometry is that of general relativity : the space-time universe is a
smooth connected Hausdorff manifold M endowed with a metric g which has the
signature - +++ or + - - -.\ The signature is not a trivial issue as we shall
see.\ I will use the less conventional, but here more convenient (and more
natural), - + + + signature.

In general relativity it is traditional to take g as a starting point. With
the additional hypothesis that matter particles have a constant 4-velocity
equal to -1 (with the - + + + signature) and photons a null 4-velocity one can
build a causal structure of events over M which, with some generally accepted
assumptions (called hyperbolicity), leads for M to the structure of a trivial
fiber bundle SxR where S is a 3-dimensional space-like hypersurface (a Cauchy
surface) (Wald [29]).

Another way is to start from a bundle of orthonormal bases (the tetrads). The
key ingredient is a principal fiber bundle $O_{M}$ modelled on the connected
component of the identity $SO_{0}(3,1)$ of the Lorentz group SO(3,1). Local
trivializations charts are maps : $\varphi_{o}:M\times SO_{0}\left(
3,1\right)  \rightarrow O_{M}.$ From it one builds an associated vector bundle
$G_{M}=O_{M}\times_{SO_{0}(3,1)}R^{4}$ \ using the standard representation
$\left(  R^{4},\jmath\right)  $ of SO(3,1).\ The orthonormal bases
$\partial_{i}\left(  m\right)  $ are the images of the canonical base
$\varepsilon_{i}$ of R$^{4}$ : $\left(  \varphi_{o}\left(  m,1\right)
,\varepsilon_{i}\right)  \simeq\left(  \varphi_{o}\left(  m,h^{-1}\right)
,\jmath\left(  h\right)  _{i}^{j}\varepsilon_{j}\right)  $ but with a Lorentz
metric and are defined with respect to an holonomic frame by a matrix $\left[
O\right]  \in GL(4):\partial_{i}\left(  m\right)  =O_{i}^{\alpha}%
\partial_{\alpha}.$ The vector $\partial_{0}$ defines a time like distribution
T(O) and the 3 vectors $\partial_{1},\partial_{2},\partial_{3}$ a space-like
distribution S(O).\ Together they define a metric $g_{\alpha\beta}=\eta
^{ij}O_{i}^{\alpha}O_{j}^{\beta}$ with which the basis is orthonormal. If this
strucure has a physical meaning, the distributions are integrable over M and
define a foliation of M by the hypersurfaces orthogonal to T, and we get back
the previous topology with hyperbolicity. The necessary and sufficient
condition for that is that the 1-form $O_{0}^{\alpha}dm^{\alpha}$ is closed.
If M is simply connected then there is a scalar map $N(m):O_{0}^{\alpha
}dm^{\alpha}=\left(  \partial_{\alpha}N\right)  dm^{\alpha}$ \footnote{I will
usually use the Einstein indices summation convention}which gives a unique
time to a point in the universe.

This second approach seems a bit abstract. But actually it is closer to the
way an observer can see the structure of the universe. Indeed it is not
sufficient to define mathematical objects : we should also give some
procedure, however farfetched, to link them to physical measurements.

\subsection{Building a chart}

A point in the universe is situated by 4 components, and we need to build a
map : $R^{4}\rightarrow M.$ It is easier as we have limited the system in a
"not too big" region of M. This map is a classical "gaussian" chart. We recall
how it works.

\paragraph{1)}

The starting point is a connected space-like hypersurface S(0) : it represents
the "present" of an observer at the time t=0. The induced metric upon S(0) is
riemanian. Over each point x of S(0) there is a unique unitary future oriented
vector n(x) normal to S(0) and in a neighbourood of x there is a unique
geodesic tangent to n(x) . So we can define a family of geodesics
$\gamma\left(  x,t\right)  $ going through x and tangent to n(x) and a vector
field n(m) in each point m in the future of S(0). This vector field is the
infinitesimal generator of diffeomorphisms from $x\in S(0)$\ to $m=\exp
tn(x)$\ which for each $t\geq0$\ maps S(0) in a hypersurface S(t), the set of
points in M for which the time coordinate is t. Let us prove that the vector
field n(m) is orthonormal to S(t).

Let $\partial_{i}\left(  x,0\right)  ,i=1,2,3$\ be a set of othonormal bases
in S(0) and an arbitrary holonomic basis $\partial_{\alpha}\left(  m\right)
\in T_{m}M$. The derivative $\left(  \exp tn(x)\right)  ^{\prime}$ maps the
basis $\partial_{i}\left(  x,0\right)  $\ in a basis (exp is a diffeomorphism)
$\partial_{i}\left(  x,t\right)  =O_{i}^{\alpha}\partial_{\alpha}$\ of
$T_{m(t)}S(t)$\ in m(t)=exptn(x). Let $\phi\left(  t\right)  =g_{\alpha\beta
}(m(t))n^{\alpha}\left(  m(t)\right)  O_{i}^{\beta}\left(  x,t\right)  $\ be
the scalar product between n(m(t)) and\ $\partial_{i}\left(  x,t\right)
$\ with the metric g(m(t)) and $\nabla$\ the covariant derivative on M.

We have :

$\frac{d}{dt}\phi\left(  t\right)  =n^{\alpha}\left(  m\left(  t\right)
\right)  \nabla_{\alpha}\phi\left(  t\right)  =n^{\alpha}\nabla_{\alpha
}\left(  g_{\beta\gamma}n^{\beta}O_{i}^{\gamma}\right)  $

$=n^{\alpha}g_{\beta\gamma}n^{\beta}\nabla_{\alpha}O_{i}^{\gamma}+n^{\alpha
}g_{\beta\gamma}O_{i}^{\gamma}\nabla_{\alpha}n^{\beta}$

but :

$n^{\alpha}\nabla_{\alpha}n^{\beta}=0$ because n(m) is tangent to a geodesic

$n^{\alpha}\nabla_{\alpha}O_{i}^{\gamma}=O_{i}^{\alpha}\nabla_{\alpha
}n^{\gamma}$ because the vectors n et $\partial_{i}$ are linearly independants

$\frac{d}{dt}\phi\left(  t\right)  =\partial_{i}^{\alpha}g_{\beta\gamma
}n^{\beta}\nabla_{\alpha}n^{\gamma}$

$=O_{i}^{\alpha}\frac{1}{2}\left(  g_{\beta\gamma}n^{\beta}\nabla_{\alpha
}n^{\gamma}+g_{\gamma\beta}n^{\gamma}\nabla_{\alpha}n^{\beta}\right)  $

$=O_{i}^{\alpha}\frac{1}{2}\nabla_{\alpha}\left(  g_{\beta\gamma}n^{\beta
}n^{\gamma}\right)  =0$ because $g_{\beta\gamma}n^{\beta}n^{\gamma}=-1$

So the scalar product $g_{\alpha\beta}(m(t))n^{\alpha}\left(  m(t)\right)
O_{i}^{\beta}\left(  x,t\right)  $ is constant over m(t), it is null in t=0,
and the vectors $\partial_{i}\left(  x,t\right)  $ are orthogonal with n and n
is orthogonal with S(t).$\blacksquare$

\paragraph{2)}

We can run in two troubles in the process. The geodesics may not be complete:
they start or stop in some finite time. It is a singularity in the universe.
Or the geodesics may cross each other. It is another singularity (which always
exist : see Wald [29 ] chapt.9). But if we limit our system to a non-exotic
region (without black-hole) and a sensible time period $\left[  0,T\right]
$\ (no "big bang") we should not run into such troubles. So we can define our
system as enclosed in a region $\Omega\subset M$ generated by exptn(x) from an
open domain $\Omega\left(  0\right)  $ with compact closure in S(0). $\Omega$
is a connected 4-dimensional manifold, relatively compact and geodesically
complete. With the trivialization map $\varphi_{\Omega}:\Omega\left(
0\right)  \times\left[  0,T\right]  \rightarrow\Omega::m=\varphi_{\Omega
}\left(  y,t\right)  $ it has also the structure of a trivial fiber bundle
with base $%
\mathbb{R}
^{+}.$

\paragraph{3)}

How could we experimentally build such a chart ?\ \ A good example is given by
the GPS system (Ashby [1]). Spatial coordinates in S(0) can be measured by any
conventional method, such as electromagnetic signals if one assume that light
travels at constant speed. Notice that events occuring in S(0) cannot be
reported "live" so a co-ordinated network of observers is required. A (not too
big) "free body", which stays still without any other force than gravitation,
travels on a geodesic. To be sure to stay on a geodesic, or more generally to
know how much he deviates from a geodesic, an observer can follow the
movements of a free body in his local frame : it can be done with
accelerometers (as in the IPhone).

\paragraph{4)}

Notice that the "time" t is just a coordinate, wich is measured by coordinated
clocks by each observer.

\subsection{The system}

The initial conditions are defined by their values over S(0). Let us check
that the set of particles is well defined. They travel on their world-line m
with a proper time $\tau$\ specific to each of them, defined within an
additive constant. We assume that $\Omega$\ is large enough so that any
particle entering the system will stay within (or disappear). $\Omega$ is a
fiber bundle with base R so for each point $m\left(  \tau\right)  $ there is
one unique time $t_{M}=\pi\left(  m\left(  \tau\right)  \right)  .$ The
4-velocity u\ of the particle is a time-like, future oriented vector (it is a
matter particle) so it is projected over R by a positive scalar :
$\frac{dt_{M}}{d\tau}=\pi^{\prime}\left(  m\left(  \tau\right)  \right)
\frac{dq}{d\tau}=\pi^{\prime}\left(  m\left(  \tau\right)  \right)  u>0$ and
the map $t_{M}\left(  \tau\right)  $ is one-to-one. If a particle is observed
at some time t%
$>$%
0 it can be observed (short to disapear entirely) at any other time. This
seems obvious but has a strong consequence in the relativistic context.

We will assume that the system is closed, in that there is no interaction from outside.

\subsection{The principal bundle}

Any physical quantity is eventually measured through changes in tensorial
quantities in some vector bundle over M. So we need a procedure to define
local frames at each point of M.

\paragraph{1)}

It comes naturally from the chart : in each point $x\in S\left(  0\right)  $
the local observer chooses an orthonormal (euclidean) frame $\left(
\partial_{i}\right)  _{i=1,2,3}$\ . The fourth vector $\partial_{0}\left(
x\right)  =n(x)$ is parallel to the 4-velocity of the observer. The frame is
parallel transported along the geodesic (so it stays orthonormal), to get at
each $m\in\Omega$\ a standard frame $\partial_{i}=O\left(  m\right)
_{i}^{\alpha}\partial_{\alpha}$\ from which the local observer can deduce, in
a knowledgeable way, its own local frame. The parallel transport being
continuous the orientation is preserved : the bases have the same spatial and
time orientation which is defined as direct. Experimentally the parallel
transport is done by checking the movements of a test body (such as a
gyroscop) in a local transported frame.

\paragraph{2) The mathematical objects are :}

\bigskip

the principal fiber bundle $O_{M}$\ ,base $\Omega,$ modelled over
SO$_{0}\left(  3,1\right)  ,$ with trivialization charts : $\varphi_{o}%
:\Omega\times SO_{0}\left(  3,1\right)  \rightarrow O_{M}$

the associated vector bundle $G_{M}$ : $G_{M}=O_{v}\times_{SO_{0}(3,1)}R^{4} $
with the standard representation $\left(  R^{4},\jmath\right)  $ of SO(3,1)

the local orthonormal basis \ $\left(  \partial_{i}\right)  _{i=0,..3}%
:\partial_{i}\left(  m\right)  =O_{i}^{\alpha}\partial_{\alpha}$

\begin{notation}
:
\end{notation}

The greek letters will always refer to an holonomic basis $\left(
\partial_{\alpha}\right)  _{\alpha=0,..3}$ , with an arbitrary chart, unless
otherwise specified

The latin letters i,j,... will refer to the non holonomic orthonormal basis
$\partial_{i}=O\left(  m\right)  _{i}^{\alpha}\partial_{\alpha}.$ i=0,1,2,3

The latin letters a,b,..p,q will refer to bases of Lie algebras

Whenever necessary matrices are enclosed in backets : $\left[  O\right]
=\left[  O\right]  _{i}^{\alpha},\left[  O^{\prime}\right]  =\left[  O\right]
^{-1}=\left[  O^{\prime}\right]  _{\alpha}^{i}$

The dual holonomic basis is denoted : $dx^{\alpha}:dx^{\alpha}\left(
\partial_{\beta}\right)  =\delta_{\beta}^{\alpha}$

The dual non holonomic basis is denoted: $\partial^{i}:\partial^{i}\left(
\partial_{j}\right)  =\delta_{j}^{i}$

The fundamental form is : $\Theta=\sum_{j}\partial^{j}\otimes\partial_{j} $

The matrix (indexed over 0,1,2,3):

\bigskip

$\left[  \eta\right]  =%
\begin{bmatrix}
-1 & 0 & 0 & 0\\
0 & 1 & 0 & 0\\
0 & 0 & 1 & 0\\
0 & 0 & 0 & 1
\end{bmatrix}
$

\bigskip

The set of sections on a fiber bundle $G_{M}$ will be denoted $\Lambda
_{0}G_{M},$ the set of vector fields over a manifold M is $\Lambda_{0}TM$ and
the set of r-forms over M is $\Lambda_{r}TM^{\ast}.$

\subsection{Gauge equivariance}

The principle of relativity says that the laws of physics do not depend on the
observer : if two observers study the same system, using different sets of
frames (or gauge), their measurements can be deduced using the mathematical
relations transforming one set of frame into the other. This is the practical
meaning of the gauge equivariance. Let us see what can be transformed in the model.

\paragraph{1)}

The gaussian chart $\varphi_{\Omega}:\Omega\left(  0\right)  \times\left[
0,T\right]  \rightarrow\Omega$ ,the trivializations $\varphi_{o}:\Omega\times
SO_{0}(3,1)\rightarrow O_{M}$ and the frames $\partial_{i}\left(  m\right)
$\ are experimental constructions. But from them any observer can choose other
charts or frames, following procedures and rules which can be made known to
any other observer, so that they can compare their data.

\paragraph{2)}

In the theory of fields, quantum or classical, it is generally assumed that
the law of physics are local (if the entanglement of particles has questionned
this point, all gauge theories are local). It follows that all physical
quantities defined over M take the mathematical form of sections of a bundle
associated with $O_{M}.$ So all physical quantities which can be expressed as
tensor must belong to some vector bundle, modelled over a vector space which
is a representation of the group defining the geometry, here SO(3,1). This is
a useful but very general presciption, as there are infinitely many
representations of the same group and not all have a physical meaning.

\paragraph{3)}

The frames are parallel transported, so their transformations must be
continuous and the orientation preserved.\ But one can consider a
\textit{global} tranformation, involving the other connected components of
SO(3,1). They are defined with one of the 3 matrices :

\bigskip

$T=%
\begin{bmatrix}
-1 & 0 & 0 & 0\\
0 & 1 & 0 & 0\\
0 & 0 & 1 & 0\\
0 & 0 & 0 & 1
\end{bmatrix}
;P=%
\begin{bmatrix}
1 & 0 & 0 & 0\\
0 & -1 & 0 & 0\\
0 & 0 & -1 & 0\\
0 & 0 & 0 & -1
\end{bmatrix}
;PT=%
\begin{bmatrix}
-1 & 0 & 0 & 0\\
0 & -1 & 0 & 0\\
0 & 0 & -1 & 0\\
0 & 0 & 0 & -1
\end{bmatrix}
$

\bigskip

The frame is transformed according to : $\widetilde{\partial}_{i}\left(
m\right)  =J_{i}^{j}\partial_{i}\left(  m\right)  $ with $\left[  J\right]
=T,P,PT.$

P is the "space-inversion" matrix : it reverses the orientation of the space
frame (it can be done by using another orientation on S(0)). T is the
"time-reversal" matrix : it changes t in -t in the equations (the time is
measured from 0 to \ -$\infty$).\ Experiments show that they are not
physically admissible gauge transformations. In both cases the orientation of
the space-time is reversed : so it leads to the conclusion that the 4
dimensional physical universe is oriented.

The third matrix reverse space and time orientations, but preserves the
space-time orientation.\ Experiments show that this is an admissible
transformation if simultaneously particles are changed into antiparticles (the
"C" symmetry). We will come back to these matters in the fifth part.

\paragraph{4)}

Another open choice is the signature of the metric : at first there is no
physical justification for using - + + + or + - - -.The linear groups SO(1,3)
and SO(3,1) are identical, but manifolds equipped with one or the other metric
are not isometric, and the Clifford algebra Cl(3,1) and Cl(1,3) are not isomorphic.

\paragraph{5)}

The choice of the Cauchy hypersurface S(0) is crucial : another surface
defines another system, with possibly different particles and fields.\ 

\paragraph{6)}

The model can be easily transposed to Special Relativity and Galilean Geometry.

In Special Relativity the geodesics are straigth lines, S(t) are hyperplanes,
observers are inertial with constant 4-velocity u. The choice of the system is
the choice of an hyperplane plane S(0) and its unitary future oriented normal
u, and it fully determines the physical content of the system. The gauge group
is the restricted Poincar\'{e} group, the semi-product of the orthochronous
Lorentz group (the component of the identity of SO(3,1)) and the group of
translations in the Minkovski space. The time t is just a coordinate, without
specific meaning.

In Galilean geometry the universe can be viewed as a 4-dimensional affine
space.\ The hypersurfaces S(t) are parallel and there is only one
4-dimensional velocity. Time becomes a physical, independant variable,
identical for all the observers, defined within an affine similitude. The
gauges groups are on one hand the semi direct product of SO(3), the group of
rotations in 3-dimensional space, and of the group of translations in the
3-dimensional affine space, and on the other hand the group of similitudes in time.

\section{PARTICLES}

\label{Particles}

A particle is a point-like physical object moving along a world line with a
constant, future oriented, 4-dimensional velocity. Experience shows that,
besides any specific assumption about their "internal" structure, particles
come with "states" depending on:

- how they "rotate" around their center mass in the universe : we will call
these movements the kinematic part of the state

- how they behave when interacting with force fields : mass, charge,...We will
call these their "physical characteristics".

These characteristics are modelled independantly. As we are in the classical
picture they are not quantized, so there is no need to distinguish different
kinds of particles.

\subsection{Kinematic}

The most efficient way to model the kinematic part is by using a Clifford
algebra (Lasenby [17]) which gives a natural foundation for all kinds of spinors.

\subsubsection{ Clifford Algebra}

\paragraph{1)}

Let F be a real vector space endowed with a symetric bilinear non degenerated
function $\left\langle {}\right\rangle $ valued in the field K. The Clifford
algebra Cl(F,$\left\langle {}\right\rangle )$ and the canonical map
$\imath:F\rightarrow Cl(F,\left\langle {}\right\rangle )$ are defined by the
following universal property : for any associative algebra A with unit 1 and
linear map $f:F\rightarrow A$ such that :

$\forall v\in F:f\left(  v\right)  \times f\left(  v\right)  =\left\langle
v,v\right\rangle \times1$

$\Leftrightarrow f\left(  v\right)  \times f\left(  w\right)  +f\left(
w\right)  \times f\left(  v\right)  =2\left\langle v,w\right\rangle )\times1$

there exists a unique algebra morphism : $\varphi:Cl(F,\left\langle
{}\right\rangle )\rightarrow A$ such that $f=\varphi\circ\imath$

It always exists a Clifford algebra, isomorphic, as algebra, to the exterior
algebra $\Lambda F.$ Its internal product, noted by a dot $\cdot$\ is such
that :

$\forall v,w\in F:v\cdot w+w\cdot v=2\left\langle v,w\right\rangle $

The Clifford algebra includes the scalar K and the vectors F. As vector space
its bases can be taken as ordered products of vectors of an orthonormal basis
(with $\left\langle {}\right\rangle $\ \ ) of F and it has the dimension
$2^{\dim F}.$

\paragraph{2)}

Clifford algebras built over the same vector space F are isomorphic if and
only if the bilinear functions have the same signature. So Cl(3,1), Cl(1,3)
over $%
\mathbb{R}
^{4},Cl(4,%
\mathbb{C}
)$ over $%
\mathbb{C}
^{4},$are not isomorphic.\ $Cl(4,%
\mathbb{C}
)$ is the complexified algebra of both Cl(3,1) and Cl(1,3).

Let $\left(  \varepsilon_{i}\right)  _{i=0,..3}$ the canonical basis of $%
\mathbb{R}
^{4}$\ (and $%
\mathbb{C}
^{4}$\ with complex components).\ In this basis an element of the Clifford
algebra is written :

$w=s+x_{1}\varepsilon_{1}+x_{2}\varepsilon_{2}+x_{3}\varepsilon_{3}%
+x_{0}\varepsilon_{0}+y_{3}\varepsilon_{1}\cdot\varepsilon_{2}+y_{2}%
\varepsilon_{1}\cdot\varepsilon_{3}+y_{1}\varepsilon_{2}\cdot\varepsilon
_{3}+z_{1}\varepsilon_{1}\cdot\varepsilon_{0}+z_{2}\varepsilon_{2}%
\cdot\varepsilon_{0}$

$+z_{3}\varepsilon_{3}\cdot\varepsilon_{0}+t_{0}\varepsilon_{1}\cdot
\varepsilon_{2}\cdot\varepsilon_{3}+t_{3}\varepsilon_{1}\cdot\varepsilon
_{2}.\varepsilon_{0}+t_{2}\varepsilon_{1}\cdot\varepsilon_{3}\cdot
\varepsilon_{0}+t_{1}\varepsilon_{2}\cdot\varepsilon_{3}\cdot\varepsilon
_{0}+u\varepsilon_{1}\cdot\varepsilon_{2}\cdot\varepsilon_{3}\cdot
\varepsilon_{0}$

$\left(  s,x_{j},y_{j},t_{j},u\right)  \in K\left(  =%
\mathbb{R}
,%
\mathbb{C}
\right)  $

but the internal product follows the fundamental relations which differ
according to the scalar product :

Cl(4,C) : $\varepsilon_{i}\cdot\varepsilon_{j}+\varepsilon_{j}\cdot
\varepsilon_{i}=2\delta_{ij}$

Cl(3,1) : $\varepsilon_{i}\cdot\varepsilon_{j}+\varepsilon_{j}\cdot
\varepsilon_{i}=2\eta_{ij}$

Cl(1,3) : $\varepsilon_{i}\cdot\varepsilon_{j}+\varepsilon_{j}\cdot
\varepsilon_{i}=-2\eta_{ij}$

\paragraph{3)}

Let N be $%
\mathbb{R}
^{4}$ endowed with the bilinear function of signature - + + + and
$\widetilde{\Upsilon}:N\rightarrow CL(4,C)$ the linear map defined by :

$j=1,2,3:\widetilde{\Upsilon}\left(  \varepsilon_{j}\right)  =\varepsilon
_{j};\widetilde{\Upsilon}\left(  \varepsilon_{0}\right)  =i\varepsilon_{0}$

The vectors v,v' in N read :

$v=x_{1}\varepsilon_{1}+x_{2}\varepsilon_{2}+x_{3}\varepsilon_{3}%
+x_{0}\varepsilon_{0},v^{\prime}=x_{1}^{\prime}\varepsilon_{1}+x_{2}^{\prime
}\varepsilon_{2}+x_{3}^{\prime}\varepsilon_{3}+x_{0}^{\prime}\varepsilon_{0}$

and :

$\widetilde{\Upsilon}\left(  v\right)  =x_{1}\varepsilon_{1}+x_{2}%
\varepsilon_{2}+x_{3}\varepsilon_{3}+ix_{0}\varepsilon_{0}$

$\widetilde{\Upsilon}\left(  v^{\prime}\right)  =x_{1}^{\prime}\varepsilon
_{1}+x_{2}^{\prime}\varepsilon_{2}+x_{3}^{\prime}\varepsilon_{3}%
+ix_{0}^{\prime}\varepsilon_{0}$

It is easy to check that :

$\widetilde{\Upsilon}\left(  v\right)  \cdot\widetilde{\Upsilon}\left(
v^{\prime}\right)  +\widetilde{\Upsilon}\left(  v^{\prime}\right)
\cdot\widetilde{\Upsilon}\left(  v\right)  =2\left(  x_{1}x_{1}^{\prime}%
+x_{2}x_{2}^{\prime}+x_{3}x_{3}^{\prime}-x_{0}x_{0}^{\prime}\right)
=2\left\langle v,v^{\prime}\right\rangle _{N}$

So by the universal property of Clifford algebras there is a unique morphism
$\Upsilon:Cl(3,1)\rightarrow Cl(4,C)$ such that $\widetilde{\Upsilon}%
=\Upsilon\circ\imath$ where \i\ is the canonical map $\imath:N\rightarrow
Cl(3,1).$ The function $\Upsilon$ is not on-to : the image $\Upsilon\left(
Cl(3,1)\right)  $ is a sub-algebra Cl$_{c}$(3,1) in Cl(4,C) with elements :

$w=s+x_{1}\varepsilon_{1}+x_{2}\varepsilon_{2}+x_{3}\varepsilon_{3}%
+ix_{0}\varepsilon_{0}+y_{3}\varepsilon_{1}.\varepsilon_{2}+y_{2}%
\varepsilon_{1}.\varepsilon_{3}+y_{1}\varepsilon_{2}.\varepsilon_{3}%
+iz_{1}\varepsilon_{1}.\varepsilon_{0}+iz_{2}\varepsilon_{2}.\varepsilon
_{0}+iz_{3}\varepsilon_{3}.\varepsilon_{0}$

$+t_{4}\varepsilon_{1}.\varepsilon_{2}.\varepsilon_{3}+it_{3}\varepsilon
_{1}.\varepsilon_{2}.\varepsilon_{0}+it_{2}\varepsilon_{1}.\varepsilon
_{3}.\varepsilon_{0}+it_{1}\varepsilon_{2}.\varepsilon_{3}.\varepsilon
_{0}+iu\varepsilon_{1}.\varepsilon_{2}.\varepsilon_{3}.\varepsilon_{0}$

where the components s,u,... are real.

We will identify $\imath\left(  N\right)  \subset Cl(3,1)$ with N and
$\widetilde{\Upsilon}=\Upsilon\circ\imath$ with $\Upsilon.$

There is a similar result with N', $%
\mathbb{R}
^{4}$ endowed with the bilinear function of signature + - - - and the map
$\widetilde{\Upsilon}^{\prime}\left(  \varepsilon_{j}\right)  =i\varepsilon
_{j};\widetilde{\Upsilon}^{\prime}\left(  \varepsilon_{0}\right)
=\varepsilon_{0}.$ The subalgebra $\Upsilon^{\prime}\left(  Cl(1,3)\right)
=Cl_{c}(1,3)\neq Cl_{c}\left(  3,1\right)  $ but the two algebra have common
elements such as $i\varepsilon_{3}\cdot\varepsilon_{0}.$ One goes, inside
Cl(4,C), from images of Cl(3,1) to images of Cl(1,3) by the rule :
$\Upsilon^{\prime}\left(  \varepsilon_{j}\right)  =-i\eta^{jj}\Upsilon\left(
\varepsilon_{j}\right)  $ with $\eta^{00}=-1,j>0:\eta^{jj}=1$

\subsubsection{ Spin Group}

\paragraph{1)}

The involution $v\rightarrow-v$ on the vector space is extended in an
involution $\alpha$\ in the Clifford algebra.\ For homogeneous elements :
$\alpha\left(  v_{1.}\cdot v_{2}..\cdot v_{r}\right)  =\left(  -1\right)
^{r}\left(  v_{1}\cdot v_{2}..\cdot v_{r}\right)  .$ It follows that the set
of the elements of the Clifford algebra which are the sum of homogeneous
elements which are themselves product of an even number of vectors is a
subalgebra $Cl_{0}.$ The complement vector space denoted$\ Cl_{1}$\ is not a
subalgebra. There is another involution : the transposition which acts on
homogeneous elements by : $\left(  v_{1}\cdot v_{2}...\cdot v_{r}\right)
^{t}=\left(  v_{r}\cdot v_{r-1}...\cdot v_{1}\right)  .$

\paragraph{2)}

In Clifford algebra any element which is the product of non null norm vectors
has an inverse :

$\left(  w_{1}\cdot...\cdot w_{k}\right)  ^{-1}=\alpha\left(  \left(
w_{1}\cdot...\cdot w_{k}\right)  ^{t}\right)  /%
{\displaystyle\prod\limits_{r=1}^{k}}
\left\langle w_{r},w_{r}\right\rangle $

The Spin group of a Clifford algebra is the subset of $Cl_{0}$ of elements
which are the sum of an even product of vectors with norm +1. We will denote
Spin(3,1) the connected component of the identity of\ the spin group of
Cl(3,1) :

$Spin(3,1)=\left\{  \;\sum w_{1}\cdot.....\cdot w_{2k}:w_{l}\in N:\left\langle
w_{l},w_{l}\right\rangle _{N}=1\right\}  $

It is isomorphic to SL(2,C). Its identity is denoted $1.$

\paragraph{3)}

As a group Spin(3,1) has a linear representation $\left(  Cl(3,1),\mathbf{Ad}%
\right)  $ through the adjoint operator (denoted here in bold case to make the
difference with the function Ad which is introduced below) :

$\mathbf{Ad}:Spin(3,1)\rightarrow\hom(Cl(3,1);Cl(3,1))::\mathbf{Ad}\left(
s\right)  w=s\cdot w\cdot s^{-1}$

which has the properties that the image a vector is still a vector and that it
preserves the scalar product :

$v,w\in%
\mathbb{R}
^{4}\subset Cl(3,1):\mathbf{Ad}\left(  s\right)  v\in%
\mathbb{R}
^{4};\left\langle v,w\right\rangle =\left\langle \mathbf{Ad}\left(  s\right)
v,\mathbf{Ad}\left(  s\right)  w\right\rangle $

Spin(3,1) is the universal covering group of the connected component of
SO(3,1) with the double cover : $\mu:Spin(3,1)\rightarrow SO_{.0}\left(
3,1\right)  ::\mu\left(  \pm s\right)  =g$

So $\left(  Cl(3,1),\mathbf{Ad}\circ\mu^{-1}\right)  $\ is a linear
representation of SO$_{0}$(3,1) on Cl(3,1) (as a vector space) and $\left(
R^{4},\jmath\circ\mu\right)  $ is a linear representation of Spin(3,1) on
$R^{4}:$

$\forall v\in%
\mathbb{R}
^{4}:\mathbf{Ad}\left(  s\right)  v=s\cdot v\cdot s^{-1}=\left[
\text{\textexclamdown }\left(  \mu\left(  s\right)  \right)  \right]  v$

\paragraph{4)}

The operator $ad:A\rightarrow L(A;A)$ on a Lie algebra is defined by
$ad\left(  X\right)  \left(  Y\right)  =\left[  X,Y\right]  $ , and if A is
the Lie algebra of the Lie group G the operator $A:G\rightarrow L(A;A)$ is
defined by $Ad\left(  \exp X\right)  =\exp\left(  ad\left(  X\right)  \right)
$ (Knapp [12] p.80). One has the identity :

$g\in G,X\in A:g(\exp X)g^{-1}=\exp(Ad_{g}(X)).$ The couple (A,Ad) is the
adjoint representation of G.

A linear representation $\left(  F,\rho\right)  $ of a Lie group G induces a
linear representation $(F,\rho^{\prime}\left(  1\right)  )$\ of its Lie
algebra A :

$\forall X,Y\in A:\rho^{\prime}\left(  1\right)  \left(  \left[  X,Y\right]
_{A}\right)  =\left[  \rho^{\prime}\left(  1\right)  X,\rho^{\prime}\left(
1\right)  Y\right]  =\left(  \rho^{\prime}\left(  1\right)  X\right)
\circ\left(  \rho^{\prime}\left(  1\right)  Y\right)  -\left(  \rho^{\prime
}\left(  1\right)  Y\right)  \circ\left(  \rho^{\prime}\left(  1\right)
X\right)  $

and $\rho^{\prime}\left(  1\right)  X=\frac{d}{d\tau}\rho\left(  \exp_{G}\tau
X\right)  |_{\tau=0}$ where exp is defined over G (as a manifold) and A. If G
is a group of matrices then exp can be computed as the exponential of matrices.

Spin(3,1) being the covering group of SOl(3,1) they share the same Lie algebra
o(3,1) and :

$\forall X\in o(3,1),s\in Spin(3,1):Ad_{s}X=Ad_{\mu\left(  s\right)  }%
X=\exp\left(  ad\left(  X\right)  \right)  $

To sum up we have the following representations of Lie groups and Lie algebra :

$SO\left(  3,1\right)  :\left(  o(3,1),Ad\right)  \rightarrow o(3,1):\left(
o(3,1),ad\right)  $

$SO(3,1):\left(
\mathbb{R}
^{4},\jmath\right)  \rightarrow o(3,1):\left(
\mathbb{R}
^{4},\jmath^{\prime}(1)\right)  $

$Spin(3,1):\left(
\mathbb{R}
^{4},\jmath\circ\mu\right)  \rightarrow o(3,1):\left(
\mathbb{R}
^{4},\jmath^{\prime}\circ\mu^{\prime}(1)\right)  $

$Spin(3,1):\left(  Cl(3,1),\mathbf{Ad}\right)  \rightarrow o(3,1):\left(
Cl(3,1),\mathbf{Ad}^{\prime}\left(  1\right)  \right)  $

$SO_{0}\left(  3,1\right)  :\left(  Cl(3,1),\mathbf{Ad}\circ\mu^{-1}\right)  $

\paragraph{5)}

A great advantage of Clifford algebra is that vectors, groups and Lie algebra
can be expressed as sum of products of vectors. This is the case for the
algebra o(3,1), which is characterized by the brackets relations :

$\left[  \overrightarrow{\kappa}_{i},\overrightarrow{\kappa}_{j}\right]
=\epsilon\left(  i,j,k\right)  \overrightarrow{\kappa}_{k}$

$\left[  \overrightarrow{\kappa}_{i+3},\overrightarrow{\kappa}_{j+3}\right]
=-\epsilon\left(  i,j,k\right)  \overrightarrow{\kappa}_{k}$

$\left[  \overrightarrow{\kappa}_{i},\overrightarrow{\kappa}_{j+3}\right]
=\epsilon\left(  i,j,k\right)  \overrightarrow{\kappa}_{k+3}$

where the indexes i,j,k run over 1,2,3 and $\epsilon\left(  i,j,k\right)  =0$
is two indexes are equal, and equal to the signature of (i,j,k) if not. In its
standard representation the 6 following matrices are a basis for o(3,1):

\bigskip

$\left[  \widetilde{\kappa}_{1}\right]  =%
\begin{bmatrix}
0 & 0 & 0 & 0\\
0 & 0 & 0 & 0\\
0 & 0 & 0 & -1\\
0 & 0 & 1 & 0
\end{bmatrix}
;\left[  \widetilde{\kappa}_{2}\right]  =%
\begin{bmatrix}
0 & 0 & 0 & 0\\
0 & 0 & 0 & 1\\
0 & 0 & 0 & 0\\
0 & -1 & 0 & 0
\end{bmatrix}
;\left[  \widetilde{\kappa}_{3}\right]  =%
\begin{bmatrix}
0 & 0 & 0 & 0\\
0 & 0 & -1 & 0\\
0 & 1 & 0 & 0\\
0 & 0 & 0 & 0
\end{bmatrix}
$

$\left[  \widetilde{\kappa}_{4}\right]  =%
\begin{bmatrix}
0 & 1 & 0 & 0\\
1 & 0 & 0 & 0\\
0 & 0 & 0 & 0\\
0 & 0 & 0 & 0
\end{bmatrix}
;\left[  \widetilde{\kappa}_{5}\right]  =%
\begin{bmatrix}
0 & 0 & 1 & 0\\
0 & 0 & 0 & 0\\
1 & 0 & 0 & 0\\
0 & 0 & 0 & 0
\end{bmatrix}
;\left[  \widetilde{\kappa}_{6}\right]  =%
\begin{bmatrix}
0 & 0 & 0 & 1\\
0 & 0 & 0 & 0\\
0 & 0 & 0 & 0\\
1 & 0 & 0 & 0
\end{bmatrix}
$

\bigskip

Then their image in Cl(3,1) by $\widehat{j}:o(3,1)\rightarrow Cl(3,1)$ are the following:

\bigskip

$\widehat{\jmath}\left(  \overrightarrow{\kappa}_{1}\right)  =\frac{1}%
{2}\varepsilon_{3}\cdot\varepsilon_{2};\widehat{\jmath}\left(  \overrightarrow
{\kappa}_{2}\right)  =\frac{1}{2}\varepsilon_{1}\cdot\varepsilon_{3}%
;\widehat{\jmath}\left(  \overrightarrow{\kappa}_{3}\right)  =\frac{1}%
{2}\varepsilon_{2}\cdot\varepsilon_{1};$

$\widehat{\jmath}\left(  \overrightarrow{\kappa}_{4}\right)  =\frac{1}%
{2}\varepsilon_{0}\cdot\varepsilon_{1};\widehat{\jmath}\left(  \overrightarrow
{\kappa}_{5}\right)  =\frac{1}{2}\varepsilon_{0}\cdot\varepsilon_{2}%
;\widehat{\jmath}\left(  \overrightarrow{\kappa}_{6}\right)  =\frac{1}%
{2}\varepsilon_{0}\cdot\varepsilon_{3};$

\bigskip

One can write : $\widehat{\jmath}\left(  \overrightarrow{\kappa}_{a}\right)
=\frac{1}{2}\varepsilon_{p_{a}}\cdot\varepsilon_{q_{a}}$ with the following
table :

\bigskip

TABLE 1 :

$%
\begin{bmatrix}
a & 1 & 2 & 3 & 4 & 5 & 6\\
p_{a} & 3 & 1 & 2 & 0 & 0 & 0\\
q_{a} & 2 & 3 & 1 & 1 & 2 & 3
\end{bmatrix}
$

\bigskip

and it is easy to check that :
\begin{equation}
\left[  \widetilde{\kappa}_{a}\right]  _{j}^{i}\mathbf{=}\left(
\delta^{ip_{a}}\delta_{jq_{a}}-\delta^{iq_{a}}\eta_{jp_{a}}\right) \label{E1}%
\end{equation}

Spin(3,1) acts on the vectors (in Cl(3,1)) by $\mathbf{Ad}$ and the result is
such that :

$\forall v\in%
\mathbb{R}
^{4}:\mathbf{Ad}\left(  s\right)  v=s.v.s^{-1}=\jmath\circ\mu\left(  s\right)
v$

where $\jmath\circ\mu\left(  s\right)  $ is expressed as a matrix $\jmath
\circ\mu\left(  s\right)  $\ belonging to the standard representation of SO(3,1).

An element of Spin(3,1) is the sum of the products of an even number of
vectors with norm 1 :

$v_{k}=v_{1}\varepsilon_{1}+v_{2}\varepsilon_{2}+v_{3}\varepsilon_{3}%
+v_{0}\varepsilon_{0},v_{\alpha}\in R$

$\left\langle v_{k},v_{k}\right\rangle =v_{1}^{2}+v_{2}^{2}+v_{3}^{2}%
-v_{0}^{2}=1$

Direct computation gives an expression :

$S=T+Y_{3}\varepsilon_{1}\varepsilon_{2}+Y_{2}\varepsilon_{1}\varepsilon
_{3}+Y_{1}\varepsilon_{2}\varepsilon_{3}+Z_{1}\varepsilon_{0}\varepsilon
_{1}+Z_{2}\varepsilon_{0}\varepsilon_{2}+Z_{3}\varepsilon_{0}\varepsilon
_{3}+U\varepsilon_{0}\varepsilon_{1}\varepsilon_{2}\varepsilon_{3}$ with real
coefficients, which are not independant (there are only 6 degrees of freedom).

Let $\overrightarrow{\kappa}=\sum_{a}\kappa^{a}\overrightarrow{\kappa}_{a} $
be a fixed vector in o(3,1), with components $\kappa^{a}$ in a base
$\overrightarrow{\kappa}_{a},$ $\tau\in%
\mathbb{R}
$ and $s\left(  \tau\right)  =\exp\tau\overrightarrow{\kappa}$ . We have :

$\exp\tau\overrightarrow{\kappa}=T+Y_{3}\varepsilon_{1}\varepsilon_{2}%
+Y_{2}\varepsilon_{1}\varepsilon_{3}+Y_{1}\varepsilon_{2}\varepsilon_{3}%
+Z_{1}\varepsilon_{0}\varepsilon_{1}+Z_{2}\varepsilon_{0}\varepsilon_{2}%
+Z_{3}\varepsilon_{0}\varepsilon_{3}+U\varepsilon_{0}\varepsilon
_{1}\varepsilon_{2}\varepsilon_{3}$ with $T\left(  \tau\kappa\right)
,Y_{i}\left(  \tau\kappa\right)  ,Z_{i}\left(  \tau\kappa\right)  ,U\left(
\tau\kappa\right)  .$

differentiation in $\tau=0$ gives :

$\frac{d}{d\tau}\left(  \exp\tau\overrightarrow{\kappa}\right)  |_{\tau=0} $

$=\sum_{a}\kappa^{a}\frac{\partial}{\partial\kappa_{a}}T\left(  1\right)
+\kappa^{a}\frac{\partial}{\partial\kappa_{a}}Y_{3}\left(  1\right)
\varepsilon_{1}\varepsilon_{2}+\kappa^{a}\frac{\partial}{\partial\kappa_{a}%
}Y_{2}\left(  1\right)  \varepsilon_{1}\varepsilon_{3}$

$+\kappa^{a}\frac{\partial}{\partial\kappa_{a}}Y_{1}\left(  1\right)
\varepsilon_{2}\varepsilon_{3}+\kappa^{a}\frac{\partial}{\partial\kappa_{a}%
}Z_{1}\left(  1\right)  \varepsilon_{0}\varepsilon_{1}+\kappa^{a}%
\frac{\partial}{\partial\kappa_{a}}Z_{2}\left(  1\right)  \varepsilon
_{0}\varepsilon_{2}$

$+\kappa^{a}\frac{\partial}{\partial\kappa_{a}}Z_{3}\left(  1\right)
\varepsilon_{0}\varepsilon_{3}+\kappa^{a}\frac{\partial}{\partial\kappa_{a}%
}U\left(  1\right)  \varepsilon_{0}\varepsilon_{1}\varepsilon_{2}%
\varepsilon_{3}$

which can be written :

$\kappa=\widehat{j}\left(  \overrightarrow{\kappa}\right)  =t+y_{3}%
\varepsilon_{1}\varepsilon_{2}+y_{2}\varepsilon_{1}\varepsilon_{3}%
+y_{1}\varepsilon_{2}\varepsilon_{3}+z_{1}\varepsilon_{0}\varepsilon_{1}%
+z_{2}\varepsilon_{0}\varepsilon_{2}+z_{3}\varepsilon_{0}\varepsilon
_{3}+u\varepsilon_{0}\varepsilon_{1}\varepsilon_{2}\varepsilon_{3}$

where :

$t=\sum_{a}\kappa^{a}\frac{\partial}{\partial\kappa_{a}}T\left(  1\right)
,y_{j}=\sum_{a}\kappa^{a}\frac{\partial}{\partial\kappa_{a}}Y_{j}\left(
1\right)  ,z_{j}=\sum_{a}\kappa^{a}\frac{\partial}{\partial\kappa_{a}}%
Z_{j}\left(  1\right)  ,$

$u=\sum_{a}\kappa^{a}\frac{\partial}{\partial\kappa_{a}}U\left(  1\right)  $

are the components of an element of o(3,1) expressed in the base of Cl(3,1).

On the other hand the differentiation of :

$\mathbf{Ad}\left(  \exp\tau\overrightarrow{\kappa}\right)  v=\left(  \exp
\tau\overrightarrow{\kappa}\right)  \cdot v\cdot\left(  \exp\tau
\overrightarrow{\kappa}\right)  ^{-1}=\left[  \jmath\circ\mu\left(  \exp
\tau\overrightarrow{\kappa}\right)  \right]  v$

gives : $\left(  \mathbf{Ad}^{\prime}\left(  1\right)  \overrightarrow{\kappa
}\right)  v=\kappa\cdot v-v\cdot\kappa=\left[  \jmath^{\prime}(1)\mu^{\prime
}(1)\overrightarrow{\kappa}\right]  \left(  v\right)  =\left[  \widetilde
{\kappa}\right]  v$

where $\left[  \widetilde{\kappa}\right]  =\left[  \jmath^{\prime}%
(1)\mu^{\prime}(1)\overrightarrow{\kappa}\right]  $ is the matrix representing
$\overrightarrow{\kappa}$\ in the standard representation.

Expressing $\kappa$ as above and v as $v=v_{1}\varepsilon_{1}+v_{2}%
\varepsilon_{2}+v_{3}\varepsilon_{3}+v_{0}\varepsilon_{0}$\ we get :

$\kappa\cdot v-v\cdot\kappa=\sum_{kl}\left[  \widetilde{\kappa}\right]
_{l}^{k}v_{l}\varepsilon_{k}$

$=\left(  -2\right)  \{\left(  -v_{0}z_{1}-v_{3}y_{2}-v_{2}y_{3}\right)
\varepsilon_{1}+\left(  v_{1}y_{3}-v_{3}y_{1}-v_{0}z_{2}\right)
\varepsilon_{2}$

$+\left(  v_{1}y_{2}+v_{2}y_{1}-v_{0}z_{3}\right)  \varepsilon_{3}-\left(
v_{1}z_{1}+v_{2}z_{2}-v_{3}z_{3}\right)  \varepsilon_{0}\}$

$+\left(  2\right)  u\left(  \left(  v_{0}\right)  \varepsilon_{1}%
\varepsilon_{2}\varepsilon_{3}+\left(  v_{3}\right)  \varepsilon
_{0}\varepsilon_{1}\varepsilon_{2}-\left(  v_{2}\right)  \varepsilon
_{0}\varepsilon_{1}\varepsilon_{3}+\left(  v_{1}\right)  \varepsilon
_{0}\varepsilon_{2}\varepsilon_{3}\right)  $

$\Rightarrow u=0$

\bigskip

$\kappa\cdot v-v\cdot\kappa=\left(  -2\right)
\begin{bmatrix}
0 & -z_{1} & -z_{2} & -z_{3}\\
-z_{1} & 0 & -y_{3} & -y_{2}\\
-z_{2} & y_{3} & 0 & -y_{1}\\
-z_{3} & y_{2} & y_{1} & 0
\end{bmatrix}%
\begin{bmatrix}
v_{0}\\
v_{1}\\
v_{2}\\
v_{3}%
\end{bmatrix}
=\sum_{a}\kappa^{a}\left[  \widetilde{\kappa}_{a}\right]  _{l}^{k}%
v_{l}\varepsilon_{k}$

\bigskip

where $\left[  \widetilde{\kappa}_{a}\right]  =\left[  \jmath^{\prime}%
(1)\mu^{\prime}(1)\overrightarrow{\kappa}_{a}\right]  $

$\sum_{a}\kappa^{a}\left[  \widetilde{\kappa}_{a}\right]  =\left(  -2\right)
\left(  -z_{1}\left[  \widetilde{\kappa}_{4}\right]  -z_{2}\left[
\widetilde{\kappa}_{5}\right]  -z_{3}\left[  \widetilde{\kappa}_{6}\right]
+y_{1}\left[  \widetilde{\kappa}_{1}\right]  -y_{2}\left[  \widetilde{\kappa
}_{2}\right]  +y_{3}\left[  \widetilde{\kappa}_{3}\right]  \right)  $

$=\sum_{a}\kappa^{a}\left[  \jmath^{\prime}(1)\mu^{\prime}(1)\overrightarrow
{\kappa}_{a}\right]  $

By identifying with $\kappa=\sum_{a}\kappa^{a}\kappa_{a}=y_{3}\varepsilon
_{1}\varepsilon_{2}+y_{2}\varepsilon_{1}\varepsilon_{3}+y_{1}\varepsilon
_{2}\varepsilon_{3}+z_{1}\varepsilon_{0}\varepsilon_{1}+z_{2}\varepsilon
_{0}\varepsilon_{2}+z_{3}\varepsilon_{0}\varepsilon_{3}$ we get the expected
expression of a basis of a 6-dimensional space vector in Cl(3,1) which defines
the image of o(3,1).$\blacksquare$

\paragraph{6)}

These results can be extended to the Cl(4,C) Clifford algebra and its Cl$_{c}
$(3,1) subalgebra. The function $\Upsilon:Cl(3,1)\rightarrow Cl(4,C)$ is an
isomorphism, so Spin(3,1) has an image $Spin_{c}(3,1)$ which is a subgroup of
Spin(4,C) isomorphic to Spin(3,1). There is a similar construction with
Cl(1,3) with the function $\Upsilon^{\prime}:Cl(1,3)\rightarrow Cl(4,C)$ which
gives a group Spin(1,3) isomorphic to Spin(3,1), and we have $Spin_{c}%
(3,1)\neq Spin_{c}\left(  1,3\right)  $ but the two are also isomorphic.

\subsubsection{The spin bundle}

\paragraph{1)}

Most of the geometric transformations can be described by operations in
Clifford algebra.\ Reflections on a plane of normal vector u is just
$\mathbf{\alpha}\left(  u\right)  \cdot v\cdot u^{-1}.$ Rotations can be
expressed as the product of reflexions, and indeed this fact is at the
foundation of the Spin group through the map $\mathbf{Ad.}$ Clifford algebras
give a deep insight at the role of Spin groups to express rotations in
physics. Rotation of a body around its center mass can be expressed in the
local frame by the axis j and the angular speed $\omega.$ For an observer who
is at rest with the body the couples $\left(  j,\omega\right)  $ and $\left(
-j,-\omega\right)  $ represent the same rotation. But if we think to a network
of observers, who have to coordinate their data and the two representations
are equivalent but not identical : measuring the rotation implies moving the
local frame (posibly in "the time dimension") and the choice between $\left(
j,\omega\right)  $ and $\left(  -j,-\omega\right)  $ matters : the observer
must choose the orientation of the axis which will be parallel transported.
This issue is related to the fact that the group SO(3) is not simply connected
(as well as SO(3,1)) (see Penrose [20] 11.3 for a nice experiment on the
subject) . The full description of the spatial rotation of a body, in its
local environment, needs to be done in the Spin(3) group, which is the
universal covering group of SO(3) (as Spin(3,1) is to $SO_{0}(3,1)$). For each
$g\in SO_{0}\left(  3,1\right)  $there are 2 members of Spin(3,1) $\pm s$
which differ by their sign.

\paragraph{2)}

The construction of a Clifford bundle over a manifold endowed with a metric
occurs naturally. At each point m of the manifold where there is an
orthonormal basis $\partial_{i}\left(  m\right)  $ one defines the map :
$\varphi_{c}:M\times Cl(3,1)\rightarrow Cl(M)$ by identifying the generators :
$\partial_{i}\left(  m\right)  =\varphi_{c}\left(  m,\varepsilon_{i}\right)
.$ Cl(M) is an algebra bundle, and the associated vector bundle $G_{M}$ is
identified with the vector space part of Cl(M). Spin(3,1) acts upon Cl(M) by
\textbf{Ad}, which, for the vectors, results in a change of othonormal base :
$s\times\varphi_{c}\left(  m,\varepsilon_{i}\right)  =\varphi_{c}\left(
m,\mathbf{Ad}_{s}\varepsilon_{i}\right)  =\varphi_{c}\left(  m,\mu\left(
s\right)  \varepsilon_{i}\right)  =\left[  \jmath\circ\mu\left(  s\right)
\right]  _{i}^{j}\partial_{j}$

\paragraph{3)}

The construction of a principal fiber bundle $S_{M}$ modelled over Spin(3,1)
is more subtle. Starting from the principal fiber bundle $O_{M}$ it can can be
done if there is a spin structure : a map $\Xi:S_{M}\rightarrow O_{M}$ \ such
that \ $\Xi\left(  \varphi_{S}\left(  m,s\right)  \right)  =\varphi_{O}\left(
m,\mu\left(  s\right)  \right)  $\ such that $\Xi\left(  \varphi_{S}\left(
m,s\right)  .s^{\prime}\right)  =\Xi\left(  \varphi_{S}\left(  m,s\right)
\right)  \mu\left(  s^{\prime}\right)  $ meaning that one can choose one of
the two members +s or -s corresponding to an element of SO(3,1) in a
consistent and continuous manner over M.\ There are topological obstructions
to the existence and unicity of spin structures over a manifold (Giachetta [5]
7.2, Svetlichny [25]) but, as our system covers a limited, not too exotic,
region of M, we can assume that the procedure can be implemented to get the
principal fiber bundle $S_{M}=\left(  S_{M},\Omega,\pi_{S},Spin(3,1)\right)  $
based upon $\Omega$ with the projection $\pi_{S}:S_{M}\rightarrow\Omega$\ ,
local trivialization charts $\varphi_{S}:\Omega\times Spin(3,1)\rightarrow
S_{M}.$

The associated vector bundle $G_{M}$ can then be extended to an associated
vector bundle $S_{M}\times_{Spin(3,1)}R^{4}$ which is still denoted $G_{M}$
for simplicity. The sections $\partial_{i}\left(  m\right)  $ are associated
with the identity element $1_{S}$ of Spin(3,1).

\subsubsection{Representation of Clifford algebra}

\paragraph{1)}

There are two ways to deal with Clifford algebra.\ Either directly, using the
operations of the algebra.\ Or through a linear representation of the algebra,
using the fact that Clifford algebra are isomorphic to matrices algebra. For
the computation side the choice is mainly a matter of personal preference. But
we have here several issues.

i) The theory of fields, with connexions and variational calculus is, so far,
not well suited to algebra per se.

ii) As usual it can be useful to add more physical content to the model. It is
easier to do it by some specification of the vector space over which the
algebra is represented (we will see this with chirality).

iii) There are strong evidences which suggest the use of a complex structure,
besides the fact that most of the mathematical tools require it.

iv) Cl(1,3)\ and Cl(3,1) are not isomorphic : Cl(3,1) is isomorphic to the
algebra of 4x4 matrices over $%
\mathbb{R}
,$ Cl(1,3) to the 2x2 matrices over the quaternions, so if we stay to the pure
algebra we need to choose the signature.

For all these motives the choice done here is to use \textit{a representation
of the Cl(4,C) algebra}, the complexified of both Cl(1,3) and Cl(3,1), which
is isomorphic to the 4x4 matrices over $%
\mathbb{C}
.$ This is an irreducible representation of Cl(4,C).

\paragraph{2)}

So the kinematic state of a particle is assumed to be a vector $\phi$ of a 4
dimensional complex space F, so far unspecified, and $\left(  F,\rho\right)  $
is a representation of Cl(4,C), meaning that :

$\rho:Cl(4,C)\rightarrow L(F,F)::$

$\forall w,w^{\prime}\in Cl(4,C),\alpha,\beta\in%
\mathbb{C}
:\rho\left(  w\cdot w^{\prime}\right)  =\rho\left(  w\right)  \circ\rho\left(
w^{\prime}\right)  $

$\rho\left(  \alpha w+\beta w^{\prime}\right)  =\alpha\rho\left(  w\right)
+\beta\rho\left(  w^{\prime}\right)  .$

By restriction :

$\left(  F,\rho\right)  $ is a representation of the algebra $Cl_{c}\left(
3,1\right)  ,Cl_{c}\left(  1,3\right)  ,$ and of the groups
$Spin(4,C),Spin_{c}\left(  3,1\right)  ,Spin_{c}\left(  1,3\right)  ,$

$\left(  F,\rho\circ\Upsilon\right)  ,\left(  F,\rho\circ\Upsilon^{\prime
}\right)  $ are representations of the algebra Cl(3,1) and Cl(1,3)

$\left(  F,\rho\circ\Upsilon|_{Spin(3,1)}\right)  ,\left(  F,\rho\circ
\Upsilon^{\prime}|_{Spin(1,3)}\right)  $\ are complex representations of the
groups Spin(3,1), Spin(1,3)

$\rho\circ\Upsilon^{\prime}(s)|_{s=1}:o(3,1)\rightarrow\hom(F;F)$ is a complex
representation of the Lie algebra o(3,1)

\paragraph{3)}

F being 4 dimensional, in a basis denoted $\left(  e_{j}\right)  _{j=1}^{4}%
$\ $\rho\left(  w\right)  $ is a 4x4 matrice over $%
\mathbb{C}
$ fully determined by fixing the matrices $\gamma_{k}=\rho\left(
\varepsilon_{k}\right)  ,k=0,..3$ which shall meet the relations :%

\begin{equation}
\mathbf{\varepsilon}_{j}\mathbf{\in%
\mathbb{C}
}^{4}\mathbf{:\gamma}_{j}\mathbf{=\rho}\left(  \varepsilon_{j}\right)
\mathbf{;\gamma}_{i}\mathbf{\times\gamma}_{j}\mathbf{+\gamma}_{j}%
\mathbf{\times\gamma}_{i}\mathbf{=2\delta}_{ij}\mathbf{I}_{4}\label{E2}%
\end{equation}

These matrices are defined up to conjugation by a matrix. It is always
possible to choose hermitian matrices, so we assume that : $\gamma_{k}%
=\gamma_{k}^{\ast}.$ As $\gamma_{k}\gamma_{k}=I$ they are also unitary.

Remark : these $\gamma$\ matrices, which will be specified later, are not the
usual "Dirac matrices", which correspond to a representation of Cl(3,1) (or Cl(1,3)).

\paragraph{4)}

To the matrices $\widehat{\jmath}\left(  \overrightarrow{\kappa}_{a}\right)
=\frac{1}{2}\varepsilon_{p_{a}}\cdot\varepsilon_{q_{a}}$ representing elements
of o(3,1) in Cl(3,1) correspond the matrices :

$\left[  \kappa_{a}\right]  =\left(  \rho\circ\Upsilon\right)  ^{\prime
}(1)\left(  \widehat{\jmath}\left(  \overrightarrow{\kappa}_{a}\right)
\right)  =\left(  \rho\circ\Upsilon\right)  ^{\prime}(1)\left(  \frac{1}%
{2}\varepsilon_{p_{a}}\cdot\varepsilon_{q_{a}}\right)  $

$\Upsilon$ is a linear morphism $Cl(3,1)\rightarrow Cl(4,C)$ :

$\Upsilon^{\prime}\left(  1\right)  \left(  \frac{1}{2}\varepsilon_{p_{a}%
}\cdot\varepsilon_{q_{a}}\right)  =\frac{1}{2}\Upsilon\left(  \varepsilon
_{p_{a}}\right)  \cdot\Upsilon\left(  \varepsilon_{q_{a}}\right)  $

$\rho$ is a linear morphism $Cl(4,C)\rightarrow L\left(  F;F\right)  $ :

$\rho^{\prime}(1)\left(  \frac{1}{2}\Upsilon\left(  \varepsilon_{p_{a}%
}\right)  \cdot\Upsilon\left(  \varepsilon_{q_{a}}\right)  \right)
=\rho\left(  \frac{1}{2}\Upsilon\left(  \varepsilon_{p_{a}}\right)
\cdot\Upsilon\left(  \varepsilon_{q_{a}}\right)  \right)  =\frac{1}{2}%
\rho\left(  \Upsilon\left(  \varepsilon_{p_{a}}\right)  \right)  \rho\left(
\Upsilon\left(  \varepsilon_{q_{a}}\right)  \right)  $

with $\Upsilon\left(  \varepsilon_{0}\right)  =i\varepsilon_{0},\Upsilon
\left(  \varepsilon_{k}\right)  =\varepsilon_{k}:k=1,2,3$ we get :%

\begin{subequations}
\begin{align}
\left[  \kappa_{1}\right]   & \mathbf{=}\frac{1}{2}\mathbf{\gamma}%
_{3}\mathbf{\gamma}_{2}\mathbf{;}\left[  \kappa_{2}\right]  \mathbf{=}\frac
{1}{2}\mathbf{\gamma}_{1}\mathbf{\gamma}_{3}\mathbf{;}\left[  \kappa
_{3}\right]  \mathbf{=}\frac{1}{2}\mathbf{\gamma}_{2}\mathbf{\gamma}%
_{1}\mathbf{;}\label{E3}\\
\left[  \kappa_{4}\right]   & \mathbf{=}\frac{1}{2}\mathbf{i\gamma}%
_{0}\mathbf{\gamma}_{1}\mathbf{;}\left[  \kappa_{5}\right]  \mathbf{=}\frac
{1}{2}\mathbf{i\gamma}_{0}\mathbf{\gamma}_{2}\mathbf{;}\left[  \kappa
_{6}\right]  \mathbf{=}\frac{1}{2}\mathbf{i\gamma}_{0}\mathbf{\gamma}_{3}%
\end{align}

Notice that the matrices in the standard representation $\left(
\mathbb{R}
^{4},\jmath\right)  $ are denoted with a tilde : $\left[  \widetilde{\kappa
}_{a}\right]  $

\paragraph{5)}

An associated vector bundle modelled over F : $S_{M}\times_{Spin(3,1)}F$ can
be defined if there is a spin structure, and the kinematic state of a particle
is a represented by a vector of this vector bundle. Spin(3,1) acts over the
bundle by $\rho\circ\Upsilon.$

\paragraph{6)}

$\left(  F,\rho\right)  $ is a representation of Cl(4,C) so does not depend on
the signature which is used, and the $\gamma$\ matrices are the same. But all
the items computed through $\Upsilon$ have different value.\ The basic rule is
: $\Upsilon^{\prime}\left(  \varepsilon_{j}\right)  =-i\eta^{jj}%
\Upsilon\left(  \varepsilon_{j}\right)  $ so

$\left[  \kappa_{a}^{\prime}\right]  =\frac{1}{2}\rho\left(  \Upsilon^{\prime
}\left(  \varepsilon_{p_{a}}\right)  \right)  \rho\Upsilon^{\prime}\left(
\varepsilon_{q_{a}}\right)  =-\eta^{p_{a}p_{a}}\left[  \kappa_{a}\right]  $

$a<4:\left[  \kappa_{a}^{\prime}\right]  =-\left[  \kappa_{a}\right]
,a>3:\left[  \kappa_{a}^{\prime}\right]  =\left[  \kappa_{a}\right]  $

Kinematic states are supposed to be physical objects represented in F. If a
kinematic state is represented by a vector $\phi\in F$ with the signature (- +
+ +), it would be represented by $\phi^{\prime}=\sum_{j=1}^{4}-i\eta^{jj}%
\phi^{j}e_{j}$ with the signature (+ - - -).

\subsection{Physical characteristics}

\subsubsection{Definitions}

\paragraph{1)}

The interaction beween force fields and particles depend upon some features
specific to each kind of particle, such as electric charge. We know that these
features are quantized, but their quantization in the classical picture, and
indeed their exact value in any picture, should be the result of the model, so
I assume that the physical characteristics of a particle are part of its
state, and can take a priori a continuum of values, and therefore we do not
need to distinguish several kind of particles. We know that these
characteristics can be described in the linear representations of some groups.
I will not be too specific about the nature of the forces fields, and make the
following assumptions :

i) The physical characteristics of a particle can be modelled by a vector
$\sigma$ of a complex vector space W

ii) The couple $\left(  W,\chi\right)  $ is a linear representation of some
group U

iii) U is a connected compact Lie group

The last assumption is reasonable and it brings us nice properties : there is
a scalar product over W, such that the representation is unitary, and any
unitary representation is the sum of irreducible unitary finite dimensional
representations which are orthogonal. Such irreducible representations are
natural candidates to describe different families of particles. So I will
assume that iv)\ the representation $\left(  W,\chi\right)  $ is unitary and W
is m-dimensional, but not necessarily irreducible.
\end{subequations}
\begin{notation}
:
\end{notation}

The identity in U is denoted : $1_{U}$ or 1

The Lie algebra of U is denoted $T_{1}U$ . It is a real vector space with
basis $\left(  \overrightarrow{\theta}_{a}\right)  _{a=1}^{m}$ so
\ $T_{1}U=\left\{  \overrightarrow{\theta}=\sum_{a=1}^{m}\theta^{a}%
\overrightarrow{\theta}_{a},\theta^{a}\in%
\mathbb{R}
\right\}  .$ Its structure coefficients $C_{bc}^{a}$ are real numbers. U being
a compact Lie group the exponential map is on-to and $u\in U$ can be written
$u=\exp\left(  \sum_{a=1}^{m}\theta^{a}\overrightarrow{\theta}_{a}\right)  $
for some components $\theta^{a}.$

The hermitian scalar product in W is denoted $\left\langle \lambda\sigma
_{1},\sigma_{2}\right\rangle =\overline{\lambda}\left\langle \sigma_{1}%
,\sigma_{2}\right\rangle $ with an orthonormal basis $\left(  f_{i}\right)
_{i=1}^{m}$ and the matrices $\left[  \chi\left(  u\right)  \right]  $ are
unitary : $\left[  \chi\left(  u\right)  \right]  \left[  \chi\left(
u\right)  \right]  ^{\ast}=I_{m}$

$\left(  W,\chi^{\prime}\left(  1\right)  \right)  $ is a linear
representation of the Lie algebra $T_{1}U$\ and for each $\overrightarrow
{\theta}\in T_{1}U$\ the matrix $\left[  \theta\right]  =\chi^{\prime}\left(
1\right)  \overrightarrow{\theta}$ in the basis $\left(  f_{i}\right)
_{i=1}^{m}$ is antihermitian : $\left[  \theta\right]  ^{\ast}=-\left[
\theta\right]  .$ The matrices corresponding to the basis $\overrightarrow
{\theta}_{a}$ are denoted $\left[  \theta_{a}\right]  .$ There are the
relations :

$\forall\overrightarrow{\theta}\in T_{1}U,\tau\in%
\mathbb{R}
:\left[  \theta\right]  =\left[  \chi^{\prime}\left(  1\right)
\overrightarrow{\theta}\right]  =\frac{d}{d\tau}\left[  \chi\left(  \exp
\tau\overrightarrow{\theta}\right)  \right]  |_{\tau=0}$

and $\left[  \chi\left(  \exp\tau\overrightarrow{\theta}\right)  \right]
=\exp\left[  \tau\overrightarrow{\theta}\right]  $ can be computed as
exponential of matrices.

\paragraph{3)}

We will need the complexified $T_{1}^{c}U=T_{1}U\oplus iT_{1}U$\ of the Lie
algebra, with the same basis (but complex components) and the complex
extension of the bracket. The representation $\left(  W,\chi^{\prime}\left(
1\right)  \right)  $ can be extended to a representation $\left(  W,\chi
_{c}^{\prime}\left(  1\right)  \right)  $ of $T_{1}^{c}U$ by making
$\chi^{\prime}\left(  1\right)  $ a complex linear map : $\chi_{c}^{\prime
}\left(  1\right)  i\overrightarrow{\theta}=i\chi^{\prime}\left(  1\right)
\overrightarrow{\theta}.$ But the matrices $\left[  \theta\right]  $\ are
antihermitian only if $\overrightarrow{\theta}\in T_{1}U.$

U being a connected compact Lie group is isomorphic to a linear group (of
matrices) closed in GL(m) and there is a complex analytic group of matrices
$U_{c}$\ which admits $T_{1}^{c}U$\ as Lie algebra. $U_{c}$\ is not unique,
but all these groups are isomorphic (Knapp [11] 7.12). I will assume that
$\left(  W,\chi\right)  $\ can be extended to a linear representation (not
necessarily unitary) of $U_{c}.$ It can be done if U is simply connected or
admits a Cartan decomposition. This is the case for the usual groups SU(N),
which have su(N) as Lie algebra (which is a real algebra) and SL(N,C) and
sl(N,C) as complexified structures.

\subsubsection{The bundle of particles states}

The state of a particle is described in the tensor product $E=F\otimes W.$ It
is a tensor $\psi=\sum_{i=1}^{4}\sum_{j=1}^{m}\psi^{ij}e_{i}\otimes f_{j}.$

We need a geometric structure over M bearing this representation.

\paragraph{1)}

We have the principal fiber bundle $S_{M}$ . We assume that there is a
principal bundle $U_{M}$ over M modelled over the group U. These two fiber
bundles are manifolds, so we can, starting from one or the other, define the
fiber bundle $Q_{M},$ with base M, typical fiber Spin(3,1)xU and
trivializations :

$q=\varphi_{Q}\left(  m,\left(  s,u\right)  \right)  \times\left(  s^{\prime
},v\right)  \rightarrow\varphi_{Q}\left(  m,\left(  ss^{\prime},uv\right)
\right)  $

$\varphi_{Q}\left(  m,\left(  1_{Spin},u\right)  \right)  ,\varphi_{Q}\left(
m,\left(  s,1_{G}\right)  \right)  $ are trivializations of $U_{M},S_{M}$ and
the bundles $S_{M},U_{M}$ can be seen as sub-bundles of the principal bundle
$Q_{M}.$

We denote the section $q\left(  m\right)  =\varphi_{q}\left(  m,1_{Spin}%
\times1_{G}\right)  \in\Lambda_{0}Q_{M}$

\paragraph{2)}

We define the action of Cl(3,1)xU on FxW by :

$\left(  \phi,\sigma\right)  \times\left(  w,u\right)  \rightarrow\left(
\rho\circ\Upsilon\left(  w\right)  \phi,\chi\left(  u\right)  \sigma\right)  $

For any (w,u) this action is a bilinear map over FxW.\ So by the universal
property of the tensor product there is one unique linear map :

$\vartheta:Cl(3,1)\otimes U\rightarrow L\left(  F\otimes W;F\otimes W\right)
$

such that :

$\forall\left(  \sigma,\phi\right)  \in F\times W:\vartheta\left(  w,u\right)
\left(  \phi\otimes\sigma\right)  =\left(  \rho\circ\Upsilon\left(  w\right)
\phi\right)  \otimes\left(  \chi\left(  u\right)  \sigma\right)  $

The action of $\vartheta$\ restricted to Spin(3,1)xU is an action of the
direct product and $\left(  F\otimes W,\vartheta\right)  $ is a linear
representation of the group Spin(3,1)xU.

\paragraph{3)}

From there the \textbf{vector bundle }$E_{M}$\textbf{\ associated to }$Q_{M}%
$\ is defined through the action :

$\left(  q,\psi\right)  \times\left(  s,u\right)  \rightarrow\left(  q\left(
s,u\right)  ^{-1},\vartheta\left(  s,u\right)  \left(  \psi\right)  \right)  $

and the equivalence relation : $\left(  q,\psi\right)  \simeq\left(  q\left(
s,u\right)  ^{-1},\vartheta\left(  s,u\right)  \left(  \psi\right)  \right)  $

We denote the sections i=1..4;j=1..m $:$

$e_{i}\left(  m\right)  \otimes f_{j}\left(  m\right)  =\left(  \varphi
_{q}\left(  m,1_{Spin}\times1_{U}\right)  ,e_{i}\otimes f_{j}\right)  $

$\simeq\left(  q\left(  m\right)  \left(  s,u\right)  ^{-1},\left(  \rho
\circ\Upsilon\left(  s\right)  e_{i}\right)  \otimes\left(  \chi\left(
u\right)  f_{j}\right)  \right)  $

which define a local basis of the fiber $E_{M}\left(  m\right)  .$

A section of $E_{M}$ is a map :$\Omega\rightarrow E_{M}:\psi\left(  m\right)
=\psi^{ij}\left(  m\right)  e_{i}\left(  m\right)  \otimes f_{i}\left(
m\right)  .$

\paragraph{4)}

So far the choice of the vector spaces F and W is open, but we need some
procedure to measure the components $\psi^{ij},$ that is a way for an observer
to define the vectors $\left(  e_{i}\left(  m\right)  ,f_{j}\left(  m\right)
\right)  .$

The basis $e_{i}\left(  m\right)  $ transform according to the same rules as
$\partial_{i}\left(  m\right)  $\ and has clearly a geometric meaning.\ So we
assume there is some procedure to relate the two bases. The basis
$f_{j}\left(  m\right)  $ is related to the action of the force fields on the
particles, and should be defined form the trajectories of test particles.\ The
existence of the principal fiber bundle $U_{M}$ needs some procedure to
compare the measures done by observers in different locations. This issue will
be addressed below.

\subsubsection{Spinor and Clifford algebra}

It is useful to link the present model to the usual "spinors" used in quantum
physics to describe spinning particles. The situation of quantum physics is
indeed a bit complicated.

\paragraph{1)}

As was said previously, in a local field theory any physical quantity which is
expressed as a tensor must be a section of a vector bundle, associated with a
principal bundle of the world manifold and modelled over a vector space which
is a representation of the gauge group. There are well known, but fairly
technical, methods to find all linear representations of a group. Usually the
solution is a representation of the covering group, meaning a multi-valued
representation. The double cover of $SO_{0}\left(  3,1\right)  $ is Spin(3,1),
which is isomorphic to SL(2,C). Its representations are the direct product of
the usual "spin" representations of SO(3), and are indexed by 2 integers or
half integers (see Tung [28] and Knapp \ [11] ). So in the relativistic
picture (special or general) the physical vectors (whenever they are supposed
to represent a geometric quantity) belong to a vector space which is a
representation of SL(2,C), assuming that there is a "spin structure". The
"Weyl spinors" are $%
\mathbb{C}
^{2}$ vectors corresponding to one of the two non equivalent representations
(1/2,0), (0,1/2). The "Dirac spinors" are $%
\mathbb{C}
^{4}$\ vectors corresponding to the $(0,1/2)\oplus\left(  1/2,0\right)  $
representation, which is the 4 complex dimensional representation of the
Spin(3,1) group.

Notice that this prescription follows from the principles of locality and
relativity, and stands for classical as well as for quantum models.

\paragraph{2)}

But in quantum mechanics there is also the Wigner theorem, which states that,
whenever there is some gauge group, observables must be expressed in a
projective representation of this group. It is possible to get rid of the
phase factor, and go for a regular representation, if the Lie group is
semi-simple and simply connected (see Weinberg [30] I.2). This second
condition, infortunately, is not met by SO(3) or SO(3,1). There are some ways
around this issue, coming eventually to a representation of the covering
group, which is what one gets anyway, and impose a "super-selection" rule
between the 2 states. The problem is that the only unitary representations of
SL(2,C) (and SO(3,1)) are infinite dimensional (see Knapp [11]).

\paragraph{3)}

In Special Relativity the gauge group can be extended to the Poincar\'{e}
group.\ There is still no finite dimensional unitary representation but, if
one fixes one 4-vector $\overrightarrow{p}\neq0$ the irreducible unitary
representation of SO(3) (if $\left\Vert p\right\Vert \neq0$)\ of SO(2) (if
$\left\Vert p\right\Vert =0$\ ) are also irreducible unitary representation of
the subgoup of the Poincar\'{e} group leaving $\overrightarrow{p}$\ invariant.
These representations are labelled by $\overrightarrow{p}$ and the spin s (for
fermions) or helicity (for massless particles). They are infinite dimensional
unitary representations over an Hilbert space of functions of p (and labelled
by s). Their Fourier transform gives back functions of the coordinates, which,
for the massive particles, can be expressed as functions of space-time
coordinates valued in one of the finite-dimensional representation of
SL(2,C).These relativistic wave functions can be seen as plane-waves which
combine to give the actual particles, through a process of anhiliation and
creation. They are labelled by both the representation (s) and by other
quantum numbers which characterize the particle. In quantum theory of fields
observables are localized operators acting on these wave-functions.

\section{FORCE\ FIELDS}

\subsection{Principles}

\paragraph{1)}

Force fields interact with particles (remind that here particles are matter
particles) : they change their trajectories (and possibly their physical
characteristics) and conversely the particles change the strength of the
fields.\ Moreover the force fields are defined all over the universe, and
propagate without staying the same, even is there is no source : they interact
which each others. In a local field theory these interactions are purely local
: they are determined by the value of the fields and the states of the
particles which are present at the same location of the space-time.

\paragraph{2)}

The action of a force field on a particle depends on and changes the state
$\psi\left(  m\right)  $\ and the velocity\ of the particle. The simplest
assumption is that this action is linear, and can be modelled by some map over
$\Lambda_{0}E_{M}.$\ It depends also on the trajectory of the particle :
indeed particles always move on their world line, so the value of the field
that the particle meets is changing and by the same mechanism the presence of
the particle changes the value of the field. If we keep the assumption of
linearity the action of the field is reasonably modelled by a 1-form over M,
valued in $E_{M}:$ that is by a connection. Gauge equivariance implies
equivariance of the connection, which is therefore a connection associated to
a principal connection on the principal fiber bundle $Q\left(  m\right)  .$

\paragraph{\ \ 3)}

According to General Relativity inertial forces are equivalent to
gravitational forces and related to the curvature of space-time. As far as we
know they change the trajectories and the kinematic state of particles, but
not their physical characteristics. So the gravitational field will be
modelled as a principal connection \textbf{G} over $S_{M}$\ acting on the
kinematic part of $\psi$ (in F) and on the velocity of the particle. The
"other field forces" will be modelled as a principal connection \textbf{A}
over $U_{M}$\ acting on the other part (in W) of the state and on the velocity.

\subsection{Gravitation}

There are different approaches to the modellization of gravitation, related to
the two different pictures of the geometry of the universe.

\paragraph{1)\`{a}}

The traditional way stems from the description of M as a manifold endowed with
a metric g, and so g is the central piece. An affine connection (also called a
"world connection") can be seen as a linear connection on the tangent bundle,
which is no other than the vector bundle TM associated to the principal bundle
modelled on GL(4). It induces a covariant derivative $\widetilde{\nabla}%
$\ acting on the sections of TM (the vector fields) characterized by the
Christoffel coefficients $\Gamma_{\beta\gamma}^{\alpha}$\ in an holonomic
basis, and an exterior covariant derivative $\widetilde{\nabla}_{e}$ acting on
the forms over TM*, characterized by the Riemann tensor R and the torsion. So
far there is nothing which requires a metric. The connection is metric if it
preserves the scalar product, which is equivalent to the condition :
$\widetilde{\nabla}g=0\Leftrightarrow\Gamma_{\alpha\beta}^{\varepsilon
}g_{\varepsilon\gamma}+g_{\beta\varepsilon}\Gamma_{\alpha\gamma}^{\varepsilon
}=\partial_{\alpha}g_{\beta\gamma}.$

It is symmetric if the torsion is null, which is equivalent to :
$\Gamma_{\beta\gamma}^{\alpha}=\Gamma_{\gamma\beta}^{\alpha}$. There is a
unique affine connection which meets these two conditions : the
L\'{e}vy-Civita connection whose Christoffel coefficients are a function of
the first order partial derivatives of g.

From there if one takes g as the key variable, and imposes that $\Gamma$\ is
defined by some operator $g\rightarrow\Gamma$\ which cannot depend on the
choice of an holonomic basis (it is a "natural operator" in the categories
parlance) the unique first order solution is the L\'{e}vy-Civita connection
(Kolar [14] 52.3). A theorem by Utiyama says that if the lagrangian depends
only on the first derivatives of g it must factorize through the scalar
curvature R. Additional algebraic conditions then lead for the lagrangian to
the specification : $L_{G}\varpi_{4}=a\left(  R+\Lambda\right)  \sqrt
{\left\vert \det g\right\vert }\varpi_{0}$ (with a cosmological constant
\ $\Lambda$). So the problem is fairly delimited and we are in the usual
framework of General Relativity.

If no relation is imposed a priori between g and $\Gamma$ one has the
so-called Einstein-Cartan models (Trautman [27]). The compatibility\ between
the connection and the metric is an external constraint, and generally the
connection is not torsionfree.

In both cases the variational calculus can be done in an holonomic basis
(Soper [24]) or in an non-holonomic orthonormal basis. The latter method (by
"tetrads") has numerous variants (Wald [29]) but the use of an orthonormal
basis is mainly a way to simplify calculations which are always difficult.

\paragraph{2)}

In the alternate approach to the geometry let us assume that there is a
principal connection \textbf{G} on the principal bundle $S_{M}$ represented by
its connection 1-form $\widehat{G}:S_{M}\rightarrow\Lambda_{1}\left(
TM^{\ast};o(3,1)\right)  $\ and its potential : $G\left(  m\right)
=\widehat{G}\left(  \varphi_{S}\left(  m,1\right)  \right)  .$ Under a local
jauge transformation $\varphi_{S}\left(  m,1\right)  \rightarrow\varphi
_{S}\left(  m,h\left(  m\right)  ^{-1}\right)  $ (h varies with m)
$\widehat{G}$ changes as :

$\widehat{G}(\varphi_{S}\left(  m,h^{-1}\right)  )$

$=Ad_{h}\widehat{G}\left(  \varphi_{S}\left(  m,1\right)  \right)
+L_{h}^{\prime}\left(  h^{-1}\right)  \left(  h^{-1}\left(  m\right)  \right)
^{\prime}$

$=Ad_{h}\widehat{G}\left(  \varphi_{S}\left(  m,1\right)  \right)  -R^{\prime
}{}_{h^{-1}}\left(  h\right)  \left(  h\left(  m\right)  \right)  ^{\prime}$

This connection induces a covariant derivative $\widehat{\nabla}$ over the
associated vector bundle $G_{M}.$ The covariant derivative of a section
$V=\sum_{k}V^{k}\partial_{k}$ of $G_{M}$ is the 1-form :

$\widehat{\nabla}V=\left(  \partial_{\alpha}V^{i}+\left(  \jmath^{\prime
}(1)G\left(  m\right)  V\right)  ^{i}\right)  \partial_{i}\otimes dx^{\alpha}$

$=\left(  \partial_{\alpha}V^{i}+\left(  \sum_{a=1}^{6}G_{\alpha}^{a}\left[
\widetilde{\kappa}_{a}\right]  _{j}^{i}\right)  V^{j}\right)  \partial
_{i}\otimes dx^{\alpha}$

\subparagraph{a)}

There is a one-one correspondance between principal connections over $G_{M}%
$\ and affine connections over TM.

Indeed the section V is a vector field $V=\sum_{k}V^{k}\partial_{k}=$%
\ $\sum_{\alpha}v^{\alpha}\partial_{\alpha}$ with $v^{\alpha}=\sum_{k}%
V^{k}O_{k}^{\alpha}$ and equating both derivatives :

$\widehat{\nabla}V=\left(  \partial_{\alpha}V^{i}+\left(  \sum_{a=1}%
^{6}G_{\alpha}^{a}\left[  \widetilde{\kappa}_{a}\right]  _{j}^{i}\right)
V^{j}\right)  \partial_{i}\otimes dx^{\alpha}$

$\widetilde{\nabla}v=\left(  \partial_{\alpha}v^{\gamma}+\Gamma_{\alpha\beta
}^{\gamma}V^{j}\right)  \partial_{\gamma}\otimes dx^{\alpha}$

$\widehat{\nabla}V=\widetilde{\nabla}v$ \ gives : $\left[  \widetilde
{G}_{\alpha}\right]  _{i}^{j}=\sum_{a=1}^{6}G_{\alpha}^{a}\left[
\widetilde{\kappa}_{a}\right]  _{i}^{j}=O_{\beta}^{\prime j}\partial_{\alpha
}O_{i}^{\beta}+\Gamma_{\alpha\gamma}^{\beta}O_{\beta}^{\prime j}O_{i}^{\gamma
}$ which can be written in matrix notation with $\left[  \Gamma_{\alpha
}\right]  =\left[  \Gamma_{\alpha}\right]  _{\beta}^{\gamma} $%

\begin{equation}
\left[  \Gamma_{\alpha}\right]  =\left(  \left[  O\right]  \left[
\widetilde{G}_{\alpha}\right]  -\left[  \partial_{\alpha}O\right]  \right)
\left[  O^{\prime}\right]  \Leftrightarrow\left[  \widetilde{G}_{\alpha
}\right]  =\left[  O^{\prime}\right]  \left(  \left[  \Gamma_{\alpha}\right]
\left[  O\right]  +\left[  \partial_{\alpha}O\right]  \right) \label{E4}%
\end{equation}

It is easy to check that conversely an affine connection defines uniquely a
potential, and from there a principal connection : in the gauge transformation
: $\partial_{j}\rightarrow\widehat{\partial}_{j}=S_{j}^{k}\partial_{k},\left[
S\right]  \in SO\left(  3,1\right)  $ we have :

$\left[  O\right]  \rightarrow\left[  \widehat{O}\right]  =\left[  O\right]
\left[  S\right]  $

$\left[  \widetilde{G}_{\alpha}\right]  \rightarrow\widehat{\left[
\widetilde{G}_{\alpha}\right]  }$

$=\left[  \widehat{O}^{\prime}\right]  \left(  \left[  \Gamma_{\alpha}\right]
\left[  \widehat{O}\right]  +\left[  \partial_{\alpha}\widehat{O}\right]
\right)  $

$=\left[  S^{-1}\right]  \left[  \widetilde{G}_{\alpha}\right]  \left[
S\right]  -\left[  S^{-1}\right]  \left[  O^{\prime}\right]  \left[
\partial_{\alpha}O\right]  \left[  S\right]  +\left[  S^{-1}\right]  \left[
O^{\prime}\right]  \left[  \partial_{\alpha}O\right]  \left[  S\right]
+\left[  S^{-1}\right]  \left[  O^{\prime}\right]  \left[  O\right]  \left[
\partial_{\alpha}S\right]  $

$=Ad_{S^{-1}}\left[  \widetilde{G}_{\alpha}\right]  +\left[  S^{-1}\right]
\left[  \partial_{\alpha}S\right]  \blacksquare$

\subparagraph{b)}

A principal connection \textbf{G}\ is metric if the corresponding affine
connection is metric. The necessary and sufficient condition is that :
$g_{\varepsilon\gamma}\Gamma_{\alpha\beta}^{\varepsilon}+g_{\beta\varepsilon
}\Gamma_{\alpha\gamma}^{\varepsilon}=\partial_{\alpha}g_{\beta\gamma
}\Leftrightarrow\left(  \left[  g\right]  \left[  \Gamma_{\alpha}\right]
\right)  ^{t}+\left[  g\right]  \left[  \Gamma_{\alpha}\right]  =\partial
_{\alpha}\left[  g\right]  .$ Let us show that it is met if $\left[
\Gamma_{\alpha}\right]  =\left(  \left[  O\right]  \left[  \widetilde
{G}_{\alpha}\right]  -\left[  \partial_{\alpha}O\right]  \right)  \left[
O^{\prime}\right]  $ \ and $\left[  g\right]  =\left[  O^{\prime}\right]
^{t}\left[  \eta\right]  \left[  O^{\prime}\right]  $

$\left(  \left[  g\right]  \left[  \Gamma_{\alpha}\right]  \right)
^{t}+\left[  g\right]  \left[  \Gamma_{\alpha}\right]  $

$=\left(  \left(  \left[  O\right]  \left[  \widetilde{G}_{\alpha}\right]
-\left[  \partial_{\alpha}O\right]  \right)  \left[  O^{\prime}\right]
\right)  ^{t}\left(  \left[  O^{\prime}\right]  ^{t}\left[  \eta\right]
\left[  O^{\prime}\right]  \right)  ^{t}+\left[  O^{\prime}\right]
^{t}\left[  \eta\right]  \left[  O^{\prime}\right]  \left(  \left[  O\right]
\left[  \widetilde{G}_{\alpha}\right]  -\left[  \partial_{\alpha}O\right]
\right)  \left[  O^{\prime}\right]  $

$=\left[  O^{\prime}\right]  ^{t}\left[  \widetilde{G}_{\alpha}\right]
^{t}\left[  \eta\right]  \left[  O^{\prime}\right]  -\left[  O^{\prime
}\right]  ^{t}\left[  \partial_{\alpha}O\right]  ^{t}\left[  O^{\prime
}\right]  ^{t}\left[  \eta\right]  \left[  O^{\prime}\right]  +\left[
O^{\prime}\right]  ^{t}\left[  \eta\right]  \left[  \widetilde{G}_{\alpha
}\right]  \left[  O^{\prime}\right]  -\left[  O^{\prime}\right]  ^{t}\left[
\eta\right]  \left[  O^{\prime}\right]  \left[  \partial_{\alpha}O\right]
\left[  O^{\prime}\right]  $

$=\left[  O^{\prime}\right]  ^{t}\left(  \left[  \widetilde{G}_{\alpha
}\right]  ^{t}\left[  \eta\right]  -\left[  \partial_{\alpha}O\right]
^{t}\left[  O^{\prime}\right]  ^{t}\left[  \eta\right]  +\left[  \eta\right]
\left[  \widetilde{G}_{\alpha}\right]  -\left[  \eta\right]  \left[
O^{\prime}\right]  \left[  \partial_{\alpha}O\right]  \right)  \left[
O^{\prime}\right]  $

$=-\left[  O^{\prime}\right]  ^{t}\left(  \left[  \partial_{\alpha}O\right]
^{t}\left[  O^{\prime}\right]  ^{t}\left[  \eta\right]  +\left[  \eta\right]
\left[  O^{\prime}\right]  \left[  \partial_{\alpha}O\right]  \right)  \left[
O^{\prime}\right]  $

where we used the fact that $\left[  \widetilde{G}_{\alpha}\right]  \in
o(3,1)\Leftrightarrow\left[  \widetilde{G}_{\alpha}\right]  ^{t}\left[
\eta\right]  +\left[  \eta\right]  \left[  \widetilde{G}_{\alpha}\right]  =0$

On the other hand we have :

$\partial_{\alpha}\left[  g\right]  $

$=\left[  \partial_{\alpha}O^{\prime}\right]  ^{t}\left[  \eta\right]  \left[
O^{\prime}\right]  +\left[  O^{\prime}\right]  ^{t}\left[  \eta\right]
\left[  \partial_{\alpha}O^{\prime}\right]  $

$=-\left[  O^{\prime}\right]  ^{t}\left[  \partial_{\alpha}O\right]
^{t}\left[  O^{\prime}\right]  ^{t}\left[  \eta\right]  \left[  O^{\prime
}\right]  -\left[  O^{\prime}\right]  ^{t}\left[  \eta\right]  \left[
O^{\prime}\right]  \left[  \partial_{\alpha}O\right]  \left[  O^{\prime
}\right]  $

$=-\left[  O^{\prime}\right]  ^{t}\left(  \left[  \partial_{\alpha}O\right]
^{t}\left[  O^{\prime}\right]  ^{t}\left[  \eta\right]  +\left[  \eta\right]
\left[  O^{\prime}\right]  \left[  \partial_{\alpha}O\right]  \right)  \left[
O^{\prime}\right]  $ $\blacksquare$

\subparagraph{c)}

A principal connection \textbf{G}\ is symmetric is the corresponding affine
connection is symmetric. Which reads :

$\Gamma_{\alpha\beta}^{\gamma}=\Gamma_{\beta\alpha}^{\gamma}\Leftrightarrow
\left[  O\right]  _{i}^{\gamma}\left[  \widetilde{G}_{\alpha}\right]  _{j}%
^{i}\left[  O^{\prime}\right]  _{\beta}^{j}-\left[  \partial_{\alpha}O\right]
_{i}^{\gamma}\left[  O^{\prime}\right]  _{\beta}^{i}=\left[  O\right]
_{i}^{\gamma}\left[  \widetilde{G}_{\beta}\right]  _{j}^{i}\left[  O^{\prime
}\right]  _{\alpha}^{j}-\left[  \partial_{\beta}O\right]  _{i}^{\gamma}\left[
O^{\prime}\right]  _{\alpha}^{i}$

$\left[  O\right]  _{i}^{\gamma}\left(  \left[  \widetilde{G}_{\alpha}\right]
_{j}^{i}\left[  O^{\prime}\right]  _{\beta}^{j}-\left[  \widetilde{G}_{\beta
}\right]  _{j}^{i}\left[  O^{\prime}\right]  _{\alpha}^{j}\right)  =\left[
\partial_{\alpha}O\right]  _{i}^{\gamma}\left[  O^{\prime}\right]  _{\beta
}^{i}-\left[  \partial_{\beta}O\right]  _{i}^{\gamma}\left[  O^{\prime
}\right]  _{\alpha}^{i}$

$\left[  \widetilde{G}_{\alpha}\right]  _{j}^{k}\left[  O^{\prime}\right]
_{\beta}^{j}-\left[  \widetilde{G}_{\beta}\right]  _{j}^{k}\left[  O^{\prime
}\right]  _{\alpha}^{j}$

$=\sum_{\gamma}-\left[  O^{\prime}\right]  _{\gamma}^{k}\left[  O\right]
_{i}^{\gamma}\left[  \partial_{\alpha}O^{\prime}\right]  _{\beta}^{i}+\left[
O^{\prime}\right]  _{\alpha}^{i}\left[  O\right]  _{i}^{\gamma}\left[
\partial_{\beta}O^{\prime}\right]  _{\gamma}^{k}$

$=-\left[  \partial_{\alpha}O^{\prime}\right]  _{\beta}^{k}+\left[
\partial_{\beta}O^{\prime}\right]  _{\alpha}^{k}$%

\begin{equation}
\forall\alpha,\beta,k:\left[  \widetilde{G}_{\alpha}\right]  _{j}^{k}\left[
O^{\prime}\right]  _{\beta}^{j}-\left[  \widetilde{G}_{\beta}\right]  _{j}%
^{k}\left[  O^{\prime}\right]  _{\alpha}^{j}=\left[  \partial_{\beta}%
O^{\prime}\right]  _{\alpha}^{k}-\left[  \partial_{\alpha}O^{\prime}\right]
_{\beta}^{k}\label{E5}%
\end{equation}

\paragraph{3)}

In a consistent theory of fields the connection must be metric, to guarantee
that the scalar product is preserved along a geodesic. But the condition that
it is torsionfree is less obvious.\ There is no experimental evidence on this
issue (which could be a difficult one) and it seems better to keep the option
open. Moreover the alternate approach, starting from a principal fiber bundle,
and orthonormal basis, leads logically to put the connection itself as the key
variable, and to deduce the metric from the orthonormal frames. Indeed should
the metric be measured, it could be done through the relation :%

\begin{equation}
\mathbf{g}^{\alpha\beta}\mathbf{=}\sum_{jk}\mathbf{\eta}^{jk}\mathbf{O}%
_{j}^{\alpha}\mathbf{O}_{k}^{\beta}\mathbf{\Leftrightarrow g}_{\alpha\beta
}\mathbf{=}\sum_{jk}\mathbf{\eta}_{jk}\mathbf{O}_{\alpha}^{\prime j}%
\mathbf{O}_{\beta}^{\prime k}\label{E6}%
\end{equation}

So we will keep the connection \textbf{G}\ and the matrix $\left[  O^{\prime
}\right]  $ as key variables, the metric g being a byproduct given by the
relation above. It is clear that O' is determined within a matrix of SO(3,1) :
the number of degrees of freedom with g is 6 and 16 with O', which leaves 10
degrees of freedom to fix a \ gauge suiting the problem.\ It is one of the
main advantage of the tetrad method, and actually the chart which has been
built previously already pre-empted such a choice. With these assumptions the
connection is metric, but not necessarily symmetric.

\paragraph{4)}

I take the opportunity to introduce here some conventions and notations.

\subparagraph{a)}

There is the irritating issue of the conventions about exterior product and
antisymmetric tensor products. For a clear definition of the algebras of
symmetric and antisymmetric tensors see Knapp ([12] A).

\ Here I use the following :

- I denote by $\left(  \alpha_{1},..\alpha_{r}\right)  $ any set of r indexes
(taken in the pertinent set), and by $\left\{  \alpha_{1},..\alpha
_{r}\right\}  $ the set of r ordered indexes : $\alpha_{1}<\alpha_{2}%
..<\alpha_{r},$ by $\epsilon\left(  \alpha_{1},..\alpha_{r}\right)  $ the
quantity null if two of the indexes are equal, and equal to the signature of
$\left(  \alpha_{1},..\alpha_{r}\right)  $ if not,

- the exterior algebra $\Lambda F$ of a vector space F is the set of
anti-symmetric tensors :

$\lambda=\sum_{\left(  \alpha_{1},..\alpha_{r}\right)  }\lambda_{\alpha
_{1}...\alpha_{r}}e^{\alpha_{1}}\otimes...\otimes e^{\alpha_{r}}$ with
$\lambda_{\alpha_{1}...\alpha_{r}}=\epsilon\left(  \alpha_{1},..\alpha
_{r}\right)  \lambda_{s\left(  \alpha_{1}...\alpha_{r}\right)  }$ where
$\left(  e^{i}\right)  $ is a basis of F

$\Lambda F=\oplus_{r=0}^{n}\Lambda_{r}F$ notice that the field of scalars
belong to $\Lambda F$

- $\Lambda F$ becomes an algebra with the exterior product defined as :

$u,v\in F:u\wedge v=u\otimes v-v\otimes u$

$u_{k}\in F:u_{1}\wedge u_{2}...\wedge u_{r}=\sum_{\left(  \alpha_{1}%
,..\alpha_{r}\right)  }\epsilon\left(  \alpha_{1},..\alpha_{r}\right)
u_{\alpha_{1}}\otimes u_{\alpha_{2}}...\otimes u_{\alpha_{r}}$

$\left(  u^{_{1}}\wedge...\wedge u^{p}\right)  \wedge\left(  u^{p+1}%
\wedge...\wedge u^{p+q}\right)  $

$=\sum_{\left(  \alpha_{1},..\alpha_{p+q}\right)  }\epsilon\left(  \alpha
_{1},..\alpha_{p+q}\right)  u_{\alpha_{1}}\otimes u_{\alpha_{2}}...\otimes
u_{\alpha_{p+q}}=u^{_{1}}\wedge...\wedge u^{p}\wedge u^{p+1}\wedge...\wedge
u^{p+q}$

In the antisymmetrization process I \textit{do not use} the factor 1/r!.

- if $\left(  e_{i}\right)  _{i=1}^{n}$ is a basis of F, the set
$e^{\alpha_{1}}\wedge...\wedge e^{\alpha_{r}}$ of ordered products is a basis
for $\Lambda_{r}F$ and an antisymmetric tensor :

$\lambda=\sum_{\left(  \alpha_{1},..\alpha_{r}\right)  }\lambda_{\alpha
_{1}...\alpha_{r}}e^{\alpha_{1}}\otimes...\otimes e^{\alpha_{r}}$

$=\sum_{\left\{  \alpha_{1},..\alpha_{r}\right\}  }\lambda_{\left\{
\alpha_{1}...\alpha_{r}\right\}  }\left(  \sum_{s}\epsilon\left(  \alpha
_{1},..\alpha_{r}\right)  e^{\alpha_{1}}\otimes...\otimes e^{\alpha_{r}%
}\right)  $

$=\sum_{\left\{  \alpha_{1},..\alpha_{r}\right\}  }\lambda_{\left\{
\alpha_{1}...\alpha_{r}\right\}  }e^{\alpha_{1}}\wedge...\wedge e^{\alpha_{r}%
}$

and we have to pay heed to :

$\sum_{\left(  \alpha_{1},..\alpha_{r}\right)  }\lambda_{\alpha_{1}%
...\alpha_{r}}e^{\alpha_{1}}\wedge...\wedge e^{\alpha_{r}}=\left(  r!\right)
\sum_{\left\{  \alpha_{1},..\alpha_{r}\right\}  }\lambda_{\left\{  \alpha
_{1}...\alpha_{r}\right\}  }e^{\alpha_{1}}\wedge...\wedge e^{\alpha_{r}}$

- with these conventions the exterior product of the p antisymmetric tensor
$\lambda_{p}$ and the q antisymmetric tensor $\mu_{q}$ is :

$\lambda_{p}\wedge\mu_{q}=\sum\lambda_{\left\{  \alpha_{1}...\alpha
_{p}\right\}  }\mu_{\left\{  \beta_{1}...\beta_{q}\right\}  }e^{\alpha_{1}%
}\wedge...\wedge e^{\alpha_{p}}\wedge e^{\beta_{1}}\wedge...\wedge
e^{\beta_{q}}$

This product is associative and $\lambda_{p}\wedge\mu_{q}=\left(  -1\right)
^{pq}\mu_{q}\wedge\lambda_{p}$

\subparagraph{b)}

I will denote :

$\varpi_{0}$ the 4-form derived from a holonomic chart : $\varpi_{0}%
=dx^{0}\wedge dx^{1}\wedge dx^{2}\wedge dx^{3}$

so $\varpi_{0}=\sum_{\alpha_{1}\alpha_{2}\alpha_{3}\alpha_{4}}\epsilon\left(
\alpha_{1},\alpha_{2},\alpha_{3},\alpha_{4}\right)  dx^{\alpha_{1}}\otimes
dx^{\alpha_{2}}\otimes dx^{\alpha_{3}}\otimes dx^{\alpha_{4}} $

The volume form deduced from a metric g is the following :%

\begin{equation}
\mathbf{\varpi}_{4}\mathbf{=}\left(  \det O^{\prime}\right)  \mathbf{dx}%
^{0}\mathbf{\wedge dx}^{1}\mathbf{\wedge dx}^{2}\mathbf{\wedge dx}%
^{3}\mathbf{=\partial}^{1}\mathbf{\wedge\partial}^{2}\mathbf{\wedge\partial
}^{3}\mathbf{\wedge\partial}^{4}\label{E7}%
\end{equation}

The volume form on a manifold endowed with a metric g is defined as the 4-form
such that an orthonormal basis has volume 1:

$\varpi_{4}=\varpi_{4;0123}dx^{0}\wedge dx^{1}\wedge dx^{2}\wedge dx^{3}$

$\varpi_{4}\left(  \partial_{1},\partial_{2},\partial_{3},\partial_{4}\right)
$

$=\varpi_{4;0123}\left(  dx^{0}\wedge dx^{1}\wedge dx^{2}\wedge dx^{3}\right)
\left(  \partial_{1},\partial_{2},\partial_{3},\partial_{4}\right)  $

$=\sum_{\alpha_{1}\alpha_{2}\alpha_{3}\alpha_{4}}\epsilon\left(  \alpha
_{1},\alpha_{2},\alpha_{3},\alpha_{4}\right)  \varpi_{4;0123}\left(
dx^{\alpha_{1}}\otimes dx^{\alpha_{2}}\otimes dx^{\alpha_{3}}\otimes
dx^{\alpha_{4}}\right)  \left(  \partial_{1},\partial_{2},\partial
_{3},\partial_{4}\right)  $

$=\varpi_{4;0123}\sum_{\alpha_{1}\alpha_{2}\alpha_{3}\alpha_{4}}%
\epsilon\left(  \alpha_{1},\alpha_{2},\alpha_{3},\alpha_{4}\right)
O_{1}^{\alpha_{1}}O_{2}^{\alpha_{2}}O_{3}^{\alpha_{3}}O_{4}^{\alpha_{4}%
}=\varpi_{4;0123}\det O=1$

$\Rightarrow\varpi_{4}=\left(  \det O^{\prime}\right)  dx^{0}\wedge
dx^{1}\wedge dx^{2}\wedge dx^{3}$

Indeed we know that the volume form is $\varpi_{4}=$ $\sqrt{\left\vert \det
g\right\vert }dx^{0}\wedge dx^{1}\wedge dx^{2}\wedge dx^{3}$ and we have :

$\left[  \eta\right]  =\left[  O\right]  ^{t}\left[  g\right]  \left[
O\right]  \Rightarrow\left(  \det\left[  O\right]  \right)  ^{2}\det\left[
g\right]  =-1$

$\Rightarrow\det\left[  g\right]  =-\frac{1}{\left(  \det\left[  O\right]
\right)  ^{2}}=-\left(  \det O^{\prime}\right)  ^{2}$ $\Rightarrow
\sqrt{\left\vert \det g\right\vert }=\left\vert \det O^{\prime}\right\vert $

The orthonormal bases are direct, so $\left\vert \det O^{\prime}\right\vert
=\det O^{\prime}$

The volume form can be expressed in the orthonormal basis :

$\varpi_{4}=\widehat{\varpi}_{4}\partial^{0}\wedge\partial^{1}\wedge
\partial^{2}\wedge\partial^{3}=\sum_{i_{1}i_{2}i_{3}i_{4}}\epsilon\left(
i_{1},i_{2},i_{3},i_{4}\right)  \widehat{\varpi}_{4}\partial^{i_{1}}%
\otimes\partial^{i_{2}}\otimes\partial^{i_{3}}\otimes\partial^{i_{4}}$

and

$\varpi_{4}\left(  \partial_{0},\partial_{1},\partial_{2},\partial_{3}\right)
=\widehat{\varpi}_{4}\sum_{i_{1}i_{2}i_{3}i_{4}}\epsilon\left(  i_{1}%
,i_{2},i_{3},i_{4}\right)  \partial^{i_{1}}\otimes\partial^{i_{2}}%
\otimes\partial^{i_{3}}\otimes\partial^{i_{4}}\left(  \partial_{0}%
,\partial_{1},\partial_{2},\partial_{3}\right)  $

$=\widehat{\varpi}_{4}\sum_{i_{1}i_{2}i_{3}i_{4}}\epsilon\left(  i_{1}%
,i_{2},i_{3},i_{4}\right)  \delta_{0}^{i_{1}}\delta_{1}^{i_{2}}\delta
_{2}^{i_{3}}\delta_{3}^{i_{4}}=\widehat{\varpi}_{4}\det I_{4}=1\blacksquare$

\subparagraph{c)}

The partial derivative of the determinant is computed as follows :

$\partial_{\alpha}\sqrt{\left\vert \det g\right\vert }=\partial_{\alpha}\det
O^{\prime}=\sum_{\lambda,\mu}\left(  \frac{d\det\left[  O^{\prime}\right]
}{d\left[  O^{\prime}\right]  _{\mu}^{\lambda}}\right)  \partial_{\alpha
}\left[  O^{\prime}\right]  _{\mu}^{\lambda}$

For any invertible matrix M one has : $\frac{d\det M}{dM_{j}^{i}}=\left[
M^{-1}\right]  _{i}^{j}\det M$ (the order of the indexes matters)

$\frac{d\det\left[  O^{\prime}\right]  }{d\left[  O^{\prime}\right]  _{\mu
}^{\lambda}}=\left[  O\right]  _{\lambda}^{\mu}\left(  \det\left[  O^{\prime
}\right]  \right)  $

$\partial_{\alpha}\sqrt{\left\vert \det g\right\vert }=\partial_{\alpha}%
\det\left[  O^{\prime}\right]  =\sum_{\lambda,\mu}\left[  O\right]  _{\lambda
}^{\mu}\left(  \det\left[  O^{\prime}\right]  \right)  \partial_{\alpha
}\left[  O^{\prime}\right]  _{\mu}^{\lambda}=\left(  \det\left[  O^{\prime
}\right]  \right)  Tr\left(  \left[  O\right]  \left[  \partial_{\alpha
}O^{\prime}\right]  \right)  $%

\begin{equation}
\partial_{\alpha}\sqrt{\left\vert \det g\right\vert }=\left(  \det\left[
O^{\prime}\right]  \right)  Tr\left(  \left[  O\right]  \left[  \partial
_{\alpha}O^{\prime}\right]  \right) \label{E8}%
\end{equation}

Notice also the identity : $\frac{\partial}{\partial O_{\alpha}^{\prime i}%
}=\frac{\partial O_{j}^{\lambda}}{\partial O_{\alpha}^{\prime i}}%
\frac{\partial}{\partial O_{j}^{\lambda}}=-O_{i}^{\lambda}O_{j}^{\alpha}%
\frac{\partial}{\partial O_{j}^{\lambda}}$

\subsection{The other force fields}

We shall be brief as we do not try to specify the force fields considered. The
action of the force fields (other than gravitation) is represented through a
principal connection \textbf{A} over $U_{M}$\ , its connection form denoted
$\widehat{A}:U_{M}\rightarrow\Lambda_{1}\left(  TM^{\ast};T_{1}^{c}U\right)  $
and its potential $\grave{A}=\widehat{A}\left(  \varphi_{U}\left(
m,1_{U}\right)  \right)  .$ Notice that the connection is valued in the
complexified of the Lie algebra.

\subsection{The fiber bundle of force fields}

\paragraph{1)}

Both \textbf{G} and \textbf{A} can be defined as principal connection over
$Q_{M}.$ As equivariant connections they are essentially defined through their
form, which transforms in a gauge transformation as :

$q\left(  m\right)  =\varphi_{Q}\left(  m,\left(  1,1\right)  \right)
\rightarrow\widetilde{q}=\varphi_{Q}\left(  m,\left(  s,u\right)
^{-1}\right)  $

$G=\widehat{G}\left(  \varphi_{Q}\left(  m,\left(  1,1\right)  \right)
\right)  =\sum_{a,\alpha}G_{\alpha}^{a}dx^{\alpha}\otimes\overrightarrow
{\kappa}_{a}$

$\rightarrow\widetilde{G}=\widetilde{\widehat{G}}\left(  \varphi_{Q}\left(
m,\left(  s,u\right)  ^{-1}\right)  \right)  =\sum_{\alpha}\left(
Ad_{\mu\left(  s\right)  }\sum_{a}\left(  \widehat{G}_{\alpha}^{\alpha}%
-\kappa_{\alpha}^{a}\right)  \overrightarrow{\kappa}_{a}\right)  \otimes
dx^{\alpha}$

$\grave{A}=\widehat{\grave{A}}\left(  \varphi_{Q}\left(  m,\left(  1,1\right)
\right)  \right)  =\sum_{a,\alpha}\grave{A}_{\alpha}^{a}dx^{\alpha}%
\otimes\overrightarrow{\theta}_{a}$

$\rightarrow\widetilde{\grave{A}}=\widetilde{\widehat{\grave{A}}}\left(
\varphi_{Q}\left(  m,\left(  s,u\right)  ^{-1}\right)  \right)  =\sum_{\alpha
}\left(  Ad_{u}\sum_{a}\left(  \widehat{\grave{A}}_{\alpha}^{\alpha}%
-\theta_{\alpha}^{a}\right)  \overrightarrow{\theta}_{a}\right)  \otimes
dx^{\alpha}$

where

$\overrightarrow{\kappa}_{\alpha}=\kappa_{\alpha}^{a}\overrightarrow{\kappa
}_{a}\in o(3,1)$\ : $\frac{d\mu\left(  s\right)  }{dx}=L_{\mu\left(  s\right)
}^{\prime}(1)\kappa_{\alpha}^{a}dx^{\alpha}\otimes\overrightarrow{\kappa}_{a}$

$\overrightarrow{\theta}_{\alpha}=\theta_{\alpha}^{a}\overrightarrow{\theta
}_{a}\in T_{1}^{c}G:$\ $\frac{du}{dx}=L_{u}^{\prime}(1)\theta_{\alpha}%
^{a}dx^{\alpha}\otimes\overrightarrow{\theta}_{a}$

$\left(  \widehat{\grave{A}},\widehat{G}\right)  \rightarrow\widetilde{\left(
\widehat{\grave{A}},\widehat{G}\right)  }=\left(  Ad_{\mu\left(  s\right)
}\left(  \widehat{G}_{\alpha}-\overrightarrow{\kappa}_{\alpha}\right)
dx^{\alpha},Ad_{u}\left(  \widehat{\grave{A}}_{\alpha}-\theta_{\alpha}\right)
dx^{\alpha}\right)  $

Similar to the vector bundle describing the states of particles, there is a
fiber bundle describing the fields. But the relations above are affine and not
simply linear, so this fiber bundle is an affine bundle and not a\ vector bundle.

\paragraph{2)}

From the linear representations $\left(  o(3,1),Ad\right)  ,\left(
T_{1}U,Ad\right)  $ of the groups Spin(3,1),U over their own algebra one
builds the associated vector bundle : $F_{M}=Q_{M}\times_{Spin(3,1)\times
U}\left(  o(3,1)\times T_{1}U\right)  :$

$\varphi_{F}\left(  m,\left(  1_{Spin},1_{U}\right)  \right)  =\left(
\overrightarrow{\kappa}\left(  m\right)  ,\overrightarrow{\theta}\left(
m\right)  \right)  $

$\sim\varphi_{F}\left(  m,\left(  s,u\right)  ^{-1}\right)  =\left(
Ad_{s}\overrightarrow{\kappa}\left(  m\right)  ,Ad_{u}\overrightarrow{\theta
}\left(  m\right)  \right)  $

A local basis of this vector bundle is given by a couple of vectors

$\left(  \overrightarrow{\kappa}_{a}\left(  m\right)  ,\overrightarrow{\theta
}_{b}\left(  m\right)  \right)  $

This real vector bundle can be extended to a half-complexified vector bundle
to accomodate $T_{1}^{c}U.$

\paragraph{3)}

The 1-jet extension $J^{1}F_{M}$ of this vector bundle is a fiber bundle
coordinated by : $\left(  x^{\alpha},\left(  \kappa^{a},\theta^{b}\right)
,\left(  \kappa_{\alpha}^{a},\theta_{\beta}^{b}\right)  \right)  .$ It is a
vector bundle if restricted to the first two coordinates, meaning the bundle
$J^{1}F_{M}\rightarrow M$, and an affine bundle with the last coordinate
$J^{1}F_{M}\rightarrow F_{M}$.

Sections of the latter bundle can be seen as 1-form over M valued in $F_{M}:$

$\left(  x^{\alpha},\left(  \kappa_{\alpha}^{a}\left(  m\right)
,\theta_{\beta}^{b}\left(  m\right)  \right)  \right)  \leftrightarrow\left(
\kappa_{\alpha}^{a}\left(  m\right)  dx^{\alpha}\otimes\overrightarrow{\kappa
}_{a}\left(  m\right)  ,\theta_{\beta}^{b}\left(  m\right)  dx^{\alpha}%
\otimes\overrightarrow{\theta}_{b}\left(  m\right)  \right)  $

Force fields are described by connections and defined by their potential which
are 1-forms over M valued in the Lie algebras. There is a one-one
correspondance between principal connections over $Q_{M}$ and equivariant
sections over the affine bundle $J^{1}F_{M}\rightarrow F_{M}$ (Kol\`{a}r [14] IV.17).

\subsection{The covariant derivative over $E_{M}$}

\paragraph{1)}

The principal connections $\left(  \mathbf{G,A}\right)  $ over $Q_{M}$ induces
a covariant derivative denoted $\nabla$ acting on sections of the associated
vector bundle $E_{M}:$\ 

$\nabla\psi=\nabla\left(  q,\psi^{ij}e_{i}\otimes f_{j}\right)  $

$\nabla\psi=\left(  \partial_{\alpha}\psi^{ij}+\vartheta^{\prime
}(1_{Spin(3,1)},1_{U})\left(  G_{\alpha},\grave{A}_{\alpha}\right)
\psi\right)  dx^{\alpha}\otimes e_{i}\left(  m\right)  \otimes f_{j}\left(
m\right)  $

where :

$\vartheta^{\prime}(1_{Spin(3,1)},1_{U})\left(  G_{\alpha},\grave{A}_{\alpha
}\right)  \psi$

$=\sum_{j=1}^{4}\sum_{j=1}^{m}\left(  \sum_{k=1}^{4}\sum_{a=1}^{6}G_{\alpha
}^{a}\left[  \partial_{a}\rho\circ\Upsilon(1)\right]  _{k}^{i}\psi^{kj}%
+\sum_{k=1}^{m}\sum_{b}\grave{A}_{\alpha}^{b}\left[  \partial_{b}\chi\left(
1\right)  \right]  _{k}^{j}\psi^{ik}\right)  $ the index b running over the
dimension of U.

$\nabla\psi=\sum\left(  \partial_{\alpha}\psi^{ij}+G_{\alpha}^{a}\left[
\partial_{a}\rho\circ\Upsilon^{\prime}(1)\right]  _{k}^{i}\psi^{kj}+\grave
{A}_{\alpha}^{a}\left[  \partial_{a}\chi\left(  1\right)  \right]  _{k}%
^{j}\psi^{ik}\right)  dx^{\alpha}\otimes e_{i}\otimes f_{j}$

It has the following properties :

$\forall\psi\in\Lambda_{0}\left(  E_{M}\right)  ,\forall X,Y\in TM,\forall
\lambda,\mu\in%
\mathbb{R}
,f\in C(M;%
\mathbb{R}
):$

$\nabla_{\lambda X+\mu Y}\psi=\lambda\nabla_{X}\psi+\mu\nabla_{X}\psi
;\nabla_{fX}\psi=f\nabla_{X}\psi;$

$\nabla\left(  f\psi\right)  =f\left(  \nabla\psi\right)  +\left(
d_{M}f\right)  \otimes\psi$

In a gauge transformations we get :

$\psi=\left(  q,\psi^{ij}e_{i}\otimes f_{j}\right)  \simeq\left(  q\left(
s,u\right)  ^{-1},\widetilde{\psi}^{kl}e_{k}\otimes f_{l}\right)  $

$\psi^{ij}\rightarrow\widetilde{\psi}^{ij}=\left[  \rho\left(  s\right)
\right]  _{l}^{i}\left[  \chi\left(  g\right)  \right]  _{m}^{j}\psi^{lm}$

$G_{\alpha}\rightarrow\widetilde{G}_{\alpha}=Ad_{s}\left(  B_{\alpha}%
-\zeta_{\alpha}\right)  $

$\grave{A}_{\alpha}\rightarrow\widetilde{\grave{A}}_{\alpha}=Ad_{g}\left(
\grave{A}_{\alpha}-\xi_{\alpha}\right)  $

$\left(  q\left(  s,u\right)  ^{-1},\widetilde{\psi}^{kl}e_{k}\otimes
f_{l}\right)  $

$\simeq\left(  q,\left(  \partial_{\alpha}\psi^{ij}+\left[  \rho\circ
\Upsilon^{\prime}(1)G_{\alpha}\right]  _{k}^{i}\psi^{kj}+\left[  \chi^{\prime
}\left(  1\right)  \grave{A}_{\alpha}\right]  _{k}^{j}\psi^{ik}\right)
dx^{\alpha}\otimes e_{i}\otimes f_{j}\right)  $

\begin{notation}
:
\end{notation}

$\psi$ is a tensor, that will be conveniently represented as a matrix 4xm (it
is not square) : $\left[  \psi\right]  =\left[  \psi^{ij}\right]
_{j=1..m}^{i=0..3}$

We have previously seen the square 4x4 matrices :$\left[  \kappa_{a}\right]
=\left(  \rho\circ\Upsilon\right)  ^{\prime}(1)\left(  \widehat{\jmath}\left(
\overrightarrow{\kappa}_{a}\right)  \right)  .$ We will denote the square 4x4
matrices\ : $\sum_{a=1}^{4}\left[  \kappa_{a}\right]  G_{\alpha}^{a}=\left[
G_{\alpha}\right]  $. Notice that $\left[  \widetilde{G}_{\alpha}\right]
=\sum_{a}\left[  \widetilde{\kappa}_{a}\right]  G_{\alpha}^{a}\in o(3,1)$

$\chi^{\prime}(1):T_{1}^{c}U\rightarrow L_{C}\left(  W;W\right)  $ is a
complex linear map. $\partial_{a}\chi\left(  1_{G}\right)  \in L_{C}\left(
W;W\right)  $ is represented with the $\left(  f_{k}\right)  $\ basis by a
square mxm matrix with complex coefficients. We will denote the square mxm
matrices\ : $\partial_{a}\chi\left(  1_{G}\right)  =\chi^{\prime}\left(
1_{G}\right)  \overrightarrow{\theta}_{a}=\left[  \theta_{a}\right]  $.and
$\sum_{a}\left[  \theta_{a}\right]  \grave{A}_{\alpha}^{a}=\left[  \grave
{A}_{\alpha}\right]  $. The $\left(  W,\chi\right)  $ representation being
unitary : $\chi^{\prime}(1)=-\chi^{\prime}(1)^{\ast}$ and $\left[  \theta
_{a}\right]  =-\left[  \theta_{a}\right]  ^{\ast}.$%

\begin{equation}
\nabla\psi=\sum_{\alpha ijk}\left(  \partial_{\alpha}\psi^{ij}+\left[
\kappa_{a}\right]  _{k}^{i}G_{\alpha}^{a}\psi^{kj}+\left[  \theta_{a}\right]
_{k}^{j}\grave{A}_{\alpha}^{a}\psi^{ik}\right)  dx^{\alpha}\otimes
e_{i}\left(  m\right)  \otimes f_{j}\left(  m\right) \label{E9}%
\end{equation}

So the covariant derivative reads in matrix notation :%

\begin{equation}
\left[  \nabla_{\alpha}\psi\right]  \mathbf{=}\left[  \partial_{\alpha}%
\psi\right]  \mathbf{+}\left[  G_{\alpha}\right]  \left[  \psi\right]
\mathbf{+}\left[  \psi\right]  \left[  \grave{A}_{\alpha}\right]
^{t}\label{E10}%
\end{equation}

\paragraph{3)}

The covariant derivative gives a parallel transport over $E_{M}$ along a path
m(t) in M with the condition : $\nabla_{m^{\prime}(t)}\psi\left(  m\left(
t\right)  \right)  =0.$ Practically the observer must stay in a path such that
the effects of external fields do not change. Or equivalently two different
observers proceeding to the same experiment in similar conditions with a test
particle shall get equivariant measures. So, in principle, there is a way for
these two obervers to calibrate their instruments, that is to know where their
basis $e_{i}\left(  m\right)  \otimes f_{j}\left(  m\right)  $ stand
relatively to each other.

\paragraph{4)}

The covariant derivative acts on the section $\psi,$ the kinematic and the
physical characteristics. But as a 1-form on M it acts on the velocity, which
a vector in $T_{q}M.$ All these actions are local and linear, as expected. We
will have a better look at the mechanisms involved in the 4th part, until then
we will stay at a general level.

\subsection{Interaction Field/Field}

The force fields interact with each other. At this step we will\ not enter
into a precise description of the mechanisms involved, but just introduce one
key ingredient : the curvature. In the principle of least action picture we
need derivatives of the various quantities. For the states of particles which
are sections of associated bundle that is the covariant derivative. The force
fields are described as potential (G,\`{A}), which are 1-form over M, so we
need some kind of covariant derivative for forms.

\subsubsection{Exterior covariant derivative on principal bundles}

The force fields are fully described in the principal bundle picture, so only
these bundles are involved here.

\paragraph{1)}

The bracket of forms on M valued in a Lie algebra is defined as follows :

$\lambda\in\Lambda_{p}\left(  M;T_{1}U\right)  ,\mu\in\Lambda_{q}\left(
M;T_{1}^{c}U\right)  $

$\rightarrow\left[  \lambda,\mu\right]  =\sum_{a,b,c}C_{bc}^{a}\lambda
^{b}\wedge\mu^{c}\otimes\overrightarrow{\theta}_{a}\in\Lambda_{p+q}\left(
M;T_{1}U^{c}\right)  $

$\lambda\in\Lambda_{p}\left(  M;o(3,1)\right)  ,\mu\in\Lambda_{q}\left(
M;o(3,1)\right)  $

$\rightarrow\left[  \lambda,\mu\right]  =\sum_{a,b,c}G_{bc}^{a}\lambda
^{b}\wedge\mu^{c}\otimes\overrightarrow{\kappa}_{a}\in\Lambda_{p+q}\left(
M;o(3,1)\right)  $

where $C_{bc}^{a},G_{bc}^{a}$ are the structure coefficients of the algebras
(they are real numbers in both cases).

The exterior covariant derivative of a p-form on M valued in the Lie algebra
is defined as :

$\lambda\in\Lambda_{p}\left(  M;T_{1}^{c}U\right)  :\nabla_{e}\lambda
=d_{M}\lambda+\left[  \grave{A},\lambda\right]  $

$\lambda\in\Lambda_{p}\left(  M;o(3,1)\right)  :\nabla_{e}\lambda=d_{M}%
\lambda+\left[  G,\lambda\right]  $

where $d_{M}\lambda$ is the usual exterior derivative of the p-form on M.

\paragraph{2)}

The potential is a 1-form, so one can compute its exterior covariant
derivative :

$%
\mathcal{F}%
_{A}=\nabla_{e}\grave{A}=d_{V}\left(  \grave{A}\right)  +\left[  \grave
{A},\grave{A}\right]  _{T_{1}U}\in\Lambda_{2}\left(  M;T_{1}^{c}U\right)  $

$%
\mathcal{F}%
_{G}=\nabla_{e}G=d_{V}\left(  G\right)  +\left[  G,G\right]  _{o(3,1)}%
\in\Lambda_{2}\left(  M;o(3,1)\right)  $

They are 2-forms valued in the Lie algebra, expressed in components as :

$%
\mathcal{F}%
_{A}=\sum_{\left\{  \alpha,\beta\right\}  ,a}%
\mathcal{F}%
_{A\left\{  \alpha\beta\right\}  }^{a}dx^{\alpha}\wedge dx^{\beta}%
\otimes\overrightarrow{\theta}_{a}$

with $%
\mathcal{F}%
_{A\alpha\beta}^{a}=\partial_{\alpha}\grave{A}_{\beta}^{a}-\partial_{\beta
}\grave{A}_{\alpha}^{a}+\sum_{bc}C_{bc}^{a}\grave{A}_{\alpha}^{b}\grave
{A}_{\beta}^{c}$

$%
\mathcal{F}%
_{G}=\sum_{\left\{  \alpha,\beta\right\}  ,a}%
\mathcal{F}%
_{G\left\{  \alpha\beta\right\}  }^{a}dx^{\alpha}\wedge dx^{\beta}%
\otimes\overrightarrow{\kappa}_{a}$

with $%
\mathcal{F}%
_{G\alpha\beta}^{a}=\partial_{\alpha}G_{\beta}^{a}-\partial_{\beta}G_{\alpha
}^{a}+\sum_{bc}G_{bc}^{a}G_{\alpha}^{b}G_{\beta}^{c}$

and the usual notation $\left\{  \alpha,\beta\right\}  $ for an ordered set of indexes.

Their exterior covariant derivative is null : $\nabla_{e}%
\mathcal{F}%
_{A}=0;\nabla_{e}%
\mathcal{F}%
_{G}=0$

In a gauge transformation these forms transform as :

$%
\mathcal{F}%
_{A}\left(  q\left(  m\right)  \left(  s,u\right)  ^{-1}\right)  =Ad_{u}%
\mathcal{F}%
_{A}\left(  q\left(  m\right)  \right)  $

$%
\mathcal{F}%
_{G}\left(  q\left(  m\right)  \left(  s,u\right)  ^{-1}\right)  =Ad_{s}%
\mathcal{F}%
_{G}\left(  q\left(  m\right)  \right)  $

The definitions and names for these quantities vary in the litterature.\ We
will call them, in these definitions, the curvature forms.

\paragraph{3)}

They are the quantities (and possibly their derivatives) which should be put
in the lagrangian to account for the interactions between force fields. In
General Relativity it is usual to use the Riemann tensor and the scalar
curvature at this effect, so it is useful to see how these quantities are
related to our curvature forms.\ As previously with affine connections and
principal connexions the link goes through the associated vector bundle.

\subsubsection{Covariant exterior derivative on associated vector bundle}

\paragraph{1)}

For any covariant derivative $\nabla$\ on a vector bundle $G_{M}$ \footnote{We
take here the G$_{M}$ associated vector bundle but the procedure is general}
there is a unique extension as a linear operator

$\nabla_{e}:\Lambda_{r}(M;G_{M})\rightarrow\Lambda_{r+1}(M;G_{M})$\ on the
forms valued in\ $G_{M}$\ , such that :

$\forall\lambda\in\Lambda_{r}(M;%
\mathbb{R}
),\pi\in\Lambda_{r}(M;G_{M}):\nabla_{e}\left(  \lambda\wedge\pi\right)
=\left(  d_{M}\lambda\right)  \otimes\pi+(-1)^{r}\lambda\wedge\nabla_{e}\pi$

It is defined (Husemoller [10] 19.2) by :

$\pi\in\Lambda_{r}(M;G_{M}):\pi=\sum_{i}\sum_{\left\{  \alpha_{0}%
...\alpha_{r-1}\right\}  }\pi_{\left\{  \alpha_{0}...\alpha_{r-1}\right\}
}^{i}dx^{\alpha_{0}}\wedge..\wedge dx^{\alpha_{r-1}}\otimes\partial_{i}\left(
m\right)  $

$\nabla_{e}\pi=\sum_{i,\alpha}\sum_{\left\{  \alpha_{0}...\alpha
_{r-1}\right\}  }\left(  \partial_{\alpha}\pi_{\left\{  \alpha_{0}%
...\alpha_{r-1}\right\}  }^{i}+\left[  \widetilde{G}_{\alpha}\right]  _{k}%
^{i}\pi_{\left\{  \alpha_{0}...\alpha_{r-1}\right\}  }^{k}\right)  dx^{\alpha
}\wedge dx^{\alpha_{0}}\wedge..dx^{\alpha_{r-1}}\otimes\partial_{i}\left(
m\right)  $

$\Leftrightarrow\nabla_{e}\pi=\sum_{i}\left(  d_{M}\pi^{i}+\sum_{k}\left(
\sum_{\alpha}\left[  \widetilde{G}_{\alpha}\right]  _{k}^{i}dx^{\alpha
}\right)  \wedge\pi^{k}\right)  \otimes\partial_{i}$ where $d_{M}$\ is the
usual exterior differential on M.

\paragraph{2)}

If one applies two times this operator on the same form :

$\nabla_{e}^{2}\pi=\left(  \sum_{k}\left[  \digamma\right]  _{k}^{i}\wedge
\pi^{k}\right)  \otimes\partial_{i}$

where $\ \digamma=\jmath^{\prime}\left(  1\right)  \left(
\mathcal{F}%
_{G}\right)  =\sum_{\alpha<\beta}\sum_{a}%
\mathcal{F}%
_{G\left\{  \alpha\beta\right\}  }^{a}\left[  \widetilde{\kappa}_{a}\right]
dx^{\alpha}\wedge dx^{\beta}$ is a 2-form on M valued in the linear maps over
$G_{M}$\ and represented in the canonic basis of $%
\mathbb{R}
^{4}$ by matrices of o(3,1). $\digamma$ is nothing other than the curvature
form $%
\mathcal{F}%
_{G}$\ expressed in the orthonormal basis :

$\digamma=\frac{1}{2}\sum_{\alpha\beta}\sum_{a}%
\mathcal{F}%
_{G\alpha\beta}^{a}\left[  \widetilde{\kappa}_{a}\right]  dx^{\alpha}\wedge
dx^{\beta}=\sum_{\left\{  ij\right\}  kl}\digamma_{\left\{  ij\right\}  k}%
^{l}\partial^{i}\wedge\partial^{j}\otimes\partial^{k}\otimes\partial_{l}$

with $\digamma_{ijk}^{l}=\sum_{a\alpha\beta}%
\mathcal{F}%
_{G\alpha\beta}^{a}\left[  \widetilde{\kappa}_{a}\right]  _{k}^{l}%
O_{i}^{\alpha}O_{j}^{\beta}\Leftrightarrow\sum_{a\alpha\beta}%
\mathcal{F}%
_{G\alpha\beta}^{a}\left[  \widetilde{\kappa}_{a}\right]  _{k}^{l}%
=\sum\digamma_{ijk}^{l}O_{\alpha}^{\prime i}O_{\beta}^{\prime j}$

It can be shown that $\nabla_{e}\digamma_{i}=0$ where

$\digamma_{i}=\sum_{a\left\{  \alpha\beta\right\}  }%
\mathcal{F}%
_{G\alpha\beta}^{a}\left[  \widetilde{\kappa}_{a}\right]  _{i}^{j}dx^{\alpha
}\wedge dx^{\beta}\otimes\partial_{j}$

\paragraph{3)}

The same calculation can be done with any covariant derivative on a vector
bundle.\ With an affine connection on TM defined by the Christofell
coefficients $\Gamma_{\alpha\gamma}^{\beta}$ one gets :

$\nabla_{e}\left(  \sum_{\alpha,\left\{  \alpha_{1}...\alpha_{r}\right\}  }%
\pi_{\left\{  \alpha_{1}...\alpha_{r}\right\}  }^{\alpha}\left(
dx^{\alpha_{1}}\wedge...\wedge dx^{\alpha_{r}}\right)  \otimes\partial
_{\alpha}\right)  $

$=\sum_{\alpha}\left(  d\pi^{\alpha}+\sum_{\beta}\left(  \Gamma_{\gamma\beta
}^{\alpha}dx^{\gamma}\right)  \wedge\pi^{\beta}\right)  \otimes\partial
_{\alpha}$

$\nabla_{e}^{2}\left(  \sum_{\alpha,\left\{  \alpha_{1}...\alpha_{r}\right\}
}\pi_{\left\{  \alpha_{1}...\alpha_{r}\right\}  }^{\alpha}\left(
dx^{\alpha_{1}}\wedge...\wedge dx^{\alpha_{r}}\right)  \otimes\partial
_{\alpha}\right)  $

$=\sum_{\alpha}\sum_{\beta}\left(  d\left(  \Gamma_{\gamma\beta}^{\alpha
}dx^{\gamma}\right)  +\left(  \sum_{\gamma}\left(  \Gamma_{\lambda\gamma
}^{\alpha}dx^{\lambda}\right)  \wedge\left(  \Gamma_{\mu\beta}^{\gamma}%
dx^{\mu}\right)  \right)  \right)  \wedge\pi^{\beta}\otimes\partial_{\alpha}$

The quantity

$\sum_{\alpha}\sum_{\beta}\left(  d\left(  \Gamma_{\gamma\beta}^{\alpha
}dx^{\gamma}\right)  +\left(  \sum_{\gamma}\left(  \Gamma_{\lambda\gamma
}^{\alpha}dx^{\lambda}\right)  \wedge\left(  \Gamma_{\mu\beta}^{\gamma}%
dx^{\mu}\right)  \right)  \right)  \otimes\partial_{\alpha}$

$=\sum_{\alpha\beta}\sum_{\gamma<\eta}R_{\left\{  \gamma\eta\right\}  \beta
}^{\alpha}dx^{\gamma}\wedge dx^{\eta}\otimes dx^{\beta}\otimes\partial
_{\alpha}$ i

is the Riemann tensor : $R_{\gamma\eta\beta}^{\alpha}=\partial_{\gamma}%
\Gamma_{\eta\beta}^{\alpha}-\partial_{\eta}\Gamma_{\gamma\beta}^{\alpha
}+\Gamma_{\gamma\varepsilon}^{\alpha}\Gamma_{\eta\beta}^{\varepsilon}%
-\Gamma_{\eta\varepsilon}^{\alpha}\Gamma_{\gamma\beta}^{\varepsilon}$ which
can be seen as the curvature form of the affine connection \ and

$\nabla_{e}^{2}\left(  \pi\right)  =\sum_{\alpha\beta}\sum_{\gamma<\eta
}R_{\left\{  \gamma\eta\right\}  \beta}^{\alpha}\wedge\pi^{\beta}%
\otimes\partial_{\alpha}$.

From the Riemann tensor one deduces the Ricci tensor :

$Ric_{\gamma\beta}dx^{\gamma}\otimes dx^{\beta}=dx^{\alpha}\left(  R\left(
\partial_{\eta}\right)  \right)  ::Ric_{\alpha\beta}=\sum_{\gamma}%
R_{\alpha\gamma\beta}^{\gamma}$

Notice that these calculations can be done without any reference to a metric.

Now with a metric g there is the scalar curvature : $R=\sum_{\alpha\beta
}g^{\alpha\beta}Ric_{\alpha\beta}$

The Ricci tensor is symmetric if the connection is symmetric, but the scalar
curvature has a unique definition :

$R=\frac{1}{2}\sum_{\alpha\beta}g^{\alpha\beta}\left(  Ric_{\alpha\beta
}+Ric_{\beta\alpha}\right)  =\frac{1}{2}\left(  \sum_{\alpha\beta}%
g^{\alpha\beta}Ric_{\alpha\beta}+g^{\alpha\beta}Ric_{\beta\alpha}\right)  $

$=\frac{1}{2}\left(  \sum_{\alpha\beta}g^{\alpha\beta}Ric_{\alpha\beta
}+g^{\beta\alpha}Ric_{\beta\alpha}\right)  =\sum_{\alpha\beta}g^{\alpha\beta
}Ric_{\alpha\beta}$

\paragraph{4)}

We have seen previously that for a principal connection there is a unique
affine connection with $\left[  \Gamma_{\gamma}\right]  =\left(  \left[
O\right]  \left[  \widetilde{G}_{\gamma}\right]  -\partial_{\gamma}\left[
O\right]  \right)  \left[  O^{\prime}\right]  $ from which one can compute as
above from the connection \textbf{G} : :

\subparagraph{the Riemann tensor :}

$\left[  R_{\gamma\eta}\right]  $

$=\partial_{\gamma}\left[  \Gamma_{\eta}\right]  -\partial_{\eta}\left[
\Gamma_{\gamma}\right]  +\left[  \left[  \Gamma_{\gamma}\right]  ,\left[
\Gamma_{\eta}\right]  \right]  .$

$=\left[  O\right]  \left(  \left[  \widetilde{\partial_{\gamma}G_{\eta}%
}\right]  -\left[  \widetilde{\partial_{\eta}G_{\gamma}}\right]  +\left[
\widetilde{G_{\gamma}}\right]  \left[  \widetilde{G_{\eta}}\right]  -\left[
\widetilde{G_{\eta}}\right]  \left[  \widetilde{G_{\gamma}}\right]  \right)
\left[  O^{\prime}\right]  $

$-\left[  O\right]  \left[  \widetilde{G_{\eta}}\right]  \left[  O^{\prime
}\right]  \left[  \partial_{\gamma}O\right]  \left[  O^{\prime}\right]
+\left[  O\right]  \left[  \widetilde{G_{\eta}}\right]  \left[  O^{\prime
}\right]  \left[  \partial_{\gamma}O\right]  \left[  O^{\prime}\right]
+\left[  O\right]  \left[  \widetilde{G_{\gamma}}\right]  \left[  O^{\prime
}\right]  \left[  \partial_{\eta}O\right]  \left[  O^{\prime}\right]  $

$-\left[  O\right]  \left[  \widetilde{G_{\gamma}}\right]  \left[  O^{\prime
}\right]  \left[  \partial_{\eta}O\right]  \left[  O^{\prime}\right]  +\left[
\partial_{\eta}O\right]  \left[  O^{\prime}\right]  \left[  \partial_{\gamma
}O\right]  \left[  O^{\prime}\right]  -\left[  \partial_{\eta}O\right]
\left[  O^{\prime}\right]  \left[  \partial_{\gamma}O\right]  \left[
O^{\prime}\right]  $

$-\left[  \partial_{\gamma}O\right]  \left[  O^{\prime}\right]  \left[
\partial_{\eta}O\right]  \left[  O^{\prime}\right]  +\left[  \partial_{\gamma
}O\right]  \left[  O^{\prime}\right]  \left[  \partial_{\eta}O\right]  \left[
O^{\prime}\right]  $

$\left[  R_{\gamma\eta}\right]  =\left[  O\right]  \left[  \widetilde{%
\mathcal{F}%
_{G\gamma\eta}}\right]  \left[  O^{\prime}\right]  $

So the Riemann tensor associated to the connection \textbf{G} is :

$R_{\alpha\beta\gamma}^{\eta}=\sum_{ij}O_{i}^{\eta}\left[
\mathcal{F}%
_{G\alpha\beta}\right]  _{j}^{i}O_{\gamma}^{\prime j}\Leftrightarrow\left[
\mathcal{F}%
_{G\alpha\beta}\right]  _{j}^{i}=R_{\alpha\beta\gamma}^{\eta}O_{\eta}^{\prime
i}O_{j}^{\gamma}$

that is the tensor $\digamma$, which is the curvature form expressed in the
orthonormal basis.

\subparagraph{the Ricci tensor :}

$Ric=\sum_{\alpha\beta}Ric_{\alpha\beta}dx^{\alpha}\otimes dx^{\beta}%
=\sum_{ij\gamma}O_{i}^{\gamma}\left[  \widetilde{%
\mathcal{F}%
_{G\alpha\gamma}}\right]  _{j}^{i}O_{\beta}^{\prime j}dx^{\alpha}\otimes
dx^{\beta} $

$Ric_{\alpha\beta}=\sum_{ij\gamma}\left[  \widetilde{%
\mathcal{F}%
_{G\alpha\gamma}}\right]  _{j}^{i}O_{i}^{\gamma}O_{\beta}^{\prime j}%
=\sum_{aij\gamma}\left[  \widetilde{\kappa}_{a}\right]  _{j}^{i}O_{i}^{\gamma
}O_{\beta}^{\prime j}%
\mathcal{F}%
_{G\alpha\gamma}^{a}$

$=\sum_{ij\gamma}\left(  \delta^{ip_{a}}\delta_{jq_{a}}-\delta^{iq_{a}}%
\eta_{jp_{a}}\right)  O_{i}^{\gamma}O_{\beta}^{\prime j}%
\mathcal{F}%
_{G\alpha\gamma}^{a}$

$=\sum_{ij\gamma}\left(  O_{p_{a}}^{\gamma}O_{\beta}^{\prime q_{a}}%
-\eta_{p_{a}p_{a}}O_{q_{a}}^{\gamma}O_{\beta}^{\prime p_{a}}\right)
\mathcal{F}%
_{G\alpha\gamma}^{a}$

In the orthonormal frame the Ricci tensor from the connection \textbf{G}:
$Ric_{ij}=\sum_{k\alpha\gamma}\left[
\mathcal{F}%
_{G\alpha\gamma}\right]  _{j}^{k}O_{i}^{\alpha}O_{k}^{\gamma}=\sum_{k}%
\digamma_{ikj}^{k}$

\subparagraph{the scalar curvature :}

With the metric defined as : $g^{\alpha\beta}=\eta^{kl}O_{k}^{\alpha}%
O_{l}^{\beta}$ :

$R=\sum\eta^{kl}O_{k}^{\alpha}O_{l}^{\beta}\left[  \widetilde{%
\mathcal{F}%
_{G\alpha\gamma}}\right]  _{j}^{i}O_{i}^{\gamma}O_{\beta}^{\prime j}=\sum
\eta^{kj}\left[  \widetilde{%
\mathcal{F}%
_{G\alpha\beta}}\right]  _{j}^{i}O_{i}^{\beta}O_{k}^{\alpha}$

$=\sum_{a\alpha\beta ij}%
\mathcal{F}%
_{G\alpha\beta}^{a}\left(  \left[  \widetilde{\kappa}_{a}\right]  \left[
\eta\right]  \right)  _{j}^{i}O_{i}^{\beta}O_{j}^{\alpha}=\sum_{\alpha\beta
ij}%
\mathcal{F}%
_{G\alpha\beta}^{a}\left(  \delta^{ip_{a}}\delta_{jq_{a}}-\delta^{iq_{a}%
}\delta_{jp_{a}}\right)  O_{i}^{\beta}O_{j}^{\alpha}$

$=\sum_{a\alpha\beta}%
\mathcal{F}%
_{G\alpha\beta}^{a}\left(  O_{p_{a}}^{\beta}O_{q_{a}}^{\alpha}-O_{q_{a}%
}^{\beta}O_{p_{a}}^{\alpha}\right)  =2\sum_{a,\alpha<\beta}%
\mathcal{F}%
_{G\alpha\beta}^{a}\left(  O_{p_{a}}^{\beta}O_{q_{a}}^{\alpha}-O_{q_{a}%
}^{\beta}O_{p_{a}}^{\alpha}\right)  $%

\begin{equation}
\mathbf{R=}\sum_{a\alpha\beta}\mathbf{%
\mathcal{F}%
}_{G\alpha\beta}^{a}\left(  O_{p_{a}}^{\beta}O_{q_{a}}^{\alpha}-O_{q_{a}%
}^{\beta}O_{p_{a}}^{\alpha}\right)  \mathbf{=2}\sum_{a,\alpha<\beta}\mathbf{%
\mathcal{F}%
}_{G\alpha\beta}^{a}\left(  O_{p_{a}}^{\beta}O_{q_{a}}^{\alpha}-O_{q_{a}%
}^{\beta}O_{p_{a}}^{\alpha}\right) \label{E11}%
\end{equation}

In the orthonormal frame :

$R=\frac{1}{2}\sum_{akl\alpha\beta}\left(  \digamma_{pqj}^{q}\eta
^{pj}+\digamma_{pqj}^{q}\eta^{jp}\right)  =\sum_{pqj}\left(  \left[
\digamma_{pq}\right]  \left[  \eta\right]  \right)  _{p}^{q}=$ $\sum
_{ijk}\left[  \digamma\left(  \partial_{p},\partial_{q}\right)  \right]
_{j}^{q}\eta^{jp} $

$=\sum_{i}\left[  \digamma\left(  \partial_{1},\partial_{i}\right)  \right]
_{1}^{i}+\left[  \digamma\left(  \partial_{2},\partial_{i}\right)  \right]
_{2}^{i}+\left[  \digamma\left(  \partial_{3},\partial_{i}\right)  \right]
_{3}^{i}-\left[  \digamma\left(  \partial_{0},\partial_{i}\right)  \right]
_{0}^{i}$

If the connection \textbf{G}\ is symmetric this scalar curvature will be
identical to the usual quantity computed from g.

This quantity is preserved by a gauge transformation.

$\left[  O\right]  \rightarrow\left[  O\right]  \left[  \mu\left(  s\right)
\right]  ^{-1}$

$\left[  \widetilde{%
\mathcal{F}%
_{G\alpha\gamma}}\right]  \rightarrow Ad_{\mu\left(  s\right)  }\left[
\widetilde{%
\mathcal{F}%
_{B\alpha\gamma}}\right]  =\left[  \mu\left(  s\right)  \right]  \left[
\widetilde{%
\mathcal{F}%
_{G\alpha\gamma}}\right]  \left[  \mu\left(  s\right)  \right]  ^{-1}$

$\sum\left[  O\right]  _{i}^{\gamma}\left[  \widetilde{%
\mathcal{F}%
_{G\alpha\gamma}}\right]  _{j}^{i}\eta^{kj}\left[  O\right]  _{k}^{\alpha}$

$\rightarrow\sum\left[  O\right]  _{a}^{\gamma}\left[  \mu\left(  s\right)
^{-1}\right]  _{i}^{a}\left[  \mu\left(  s\right)  \right]  _{b}^{i}\left[
\widetilde{%
\mathcal{F}%
_{G\alpha\gamma}}\right]  _{c}^{b}\left[  \mu\left(  s\right)  ^{-1}\right]
_{j}^{c}\left[  \eta\right]  _{j}^{k}\left[  O\right]  _{d}^{\alpha}\left[
\mu\left(  s\right)  ^{-1}\right]  _{k}^{d}$

$=\sum\left[  O\right]  _{a}^{\gamma}\left[  \widetilde{%
\mathcal{F}%
_{G\alpha\gamma}}\right]  _{c}^{a}\left[  O\right]  _{d}^{\alpha}\left[
\mu\left(  s\right)  ^{-1}\right]  _{k}^{d}\left[  \eta\right]  _{j}%
^{k}\left[  \left(  \mu\left(  s\right)  ^{-1}\right)  ^{t}\right]  _{c}^{j}$

$=\sum\left[  O\right]  _{a}^{\gamma}\left[  \widetilde{%
\mathcal{F}%
_{G\alpha\gamma}}\right]  _{c}^{a}\left[  O\right]  _{d}^{\alpha}\eta^{dc}$

where we use the property of $\left[  \mu\left(  s\right)  ^{-1}\right]  \in
SO(3,1)\blacksquare$

\subsubsection{Torsion}

While we are in these calculations I take the opportunity to introduce the
torsion tensor and the structure coefficients which will be useful later on.

\paragraph{1)}

The fundamental form is the 1-form on M valued in $G_{M}:$ \ %

\begin{equation}
\Theta=\sum_{i=0}^{3}\partial^{i}\otimes\partial_{i}=\sum_{i\alpha}O_{\alpha
}^{\prime i}dx^{\alpha}\otimes\partial_{i}\in\Lambda_{1}\left(  M;G_{M}\right)
\label{E12}%
\end{equation}

This is a purely geometric quantity, independant from any connection.

\paragraph{2)}

The components of its exterior derivative $d_{M}\Theta$\ \ are the structure
coefficients $c_{lk}^{i}=\left[  \partial_{l},\partial_{k}\right]  ^{i}$ of
the algebra of vectors in the basis $\partial_{i}$ (the brackets are the
commutators of the vector fields $\partial_{i}=O_{i}^{\alpha}\partial_{\alpha
}$) :%

\begin{equation}
\mathbf{c}_{lk}^{i}\mathbf{=}\left[  \partial_{l},\partial_{k}\right]
^{i}\mathbf{=}\sum_{\alpha\beta}\mathbf{O}_{\beta}^{\prime i}\left(
O_{l}^{\alpha}\partial_{\alpha}O_{k}^{\beta}-O_{k}^{\alpha}\partial_{\alpha
}O_{l}^{\beta}\right) \label{E13}%
\end{equation}

$d\Theta=\sum_{r}\sum_{\alpha,\beta}\left(  \partial_{\beta}O_{\alpha}^{\prime
r}\right)  \partial^{\beta}\wedge\partial^{\alpha}\otimes\partial_{r}$

$=\sum_{r}\sum_{p,q}c_{pq}^{r}\partial^{p}\wedge\partial^{q}\otimes
\partial_{r}=2\sum_{r}\sum_{p<q}c_{pq}^{r}\partial^{p}\wedge\partial
^{q}\otimes\partial_{r}$

We have $c_{pq}^{r}=-c_{qp}^{r}$ so it is convenient to choose an order for
the indexes.\ Using the correspondance between the indexes in the basis of
o(3,1) and the couples (p,q) (see table 1) we will denote :%

\begin{align}
a  & =1..6,r=0,..3:c_{a}^{r}=c_{p_{a}q_{a}}^{r}:\label{E13a}\\
c_{1}^{r}  & =c_{32}^{r},c_{2}^{r}=c_{13}^{r},c_{3}^{r}=c_{21}^{r},c_{4}%
^{r}=c_{01}^{r},c_{5}^{r}=c_{02}^{r},c_{6}^{r}=c_{03}^{r}%
\end{align}

\bigskip

With this notation we have : $d\Theta=2\sum_{a=1}^{6}\sum_{r=0}^{3}c_{a}%
^{r}\partial^{p_{a}}\wedge\partial^{q_{a}}\otimes\partial_{r}$

\paragraph{3)}

The exterior covariant derivative of $\Theta$ is :

$\nabla_{e}\Theta=\sum_{r}\left(  d_{M}\Theta^{r}+\sum_{i}\left(  \left(
\sum_{\alpha}\left[  \widetilde{G}_{\alpha}\right]  _{i}^{r}dx^{\alpha
}\right)  \wedge\Theta^{i}\right)  \right)  \otimes\partial_{r}$

$=\left(  d_{M}\left(  \sum_{\alpha}O_{\alpha}^{\prime r}\partial^{\alpha
}\right)  +\sum_{i}\left(  \left(  \sum_{\alpha}\left[  \widetilde{G}_{\alpha
}\right]  _{i}^{r}dx^{\alpha}\right)  \wedge\left(  \sum_{\beta}O_{\beta
}^{\prime i}\partial^{\beta}\right)  \right)  \right)  \otimes\partial_{r}$

$=\sum_{r}\left(  \sum_{\alpha\beta}\left(  \partial_{\alpha}O_{\beta}^{\prime
r}+\left[  \widetilde{G}_{\alpha}\right]  _{i}^{r}O_{\beta}^{\prime i}\right)
dx^{\alpha}\wedge dx^{\beta}\otimes\partial_{r}\right)  $

$=2\sum_{r\gamma}\sum_{\alpha<\beta}\left(  \partial_{\alpha}O_{\beta}^{\prime
r}-\partial_{\beta}O_{\alpha}^{\prime r}+\left[  \widetilde{G}_{\alpha
}\right]  _{i}^{r}O_{\beta}^{\prime i}-\left[  \widetilde{G}_{\beta}\right]
_{i}^{r}O_{\alpha}^{\prime i}\right)  O_{r}^{\gamma}dx^{\alpha}\wedge
dx^{\beta}\otimes\partial_{\gamma}$

Expressed with $\left[  \Gamma_{\alpha}\right]  =\left(  \left[  O\right]
\left[  \widetilde{G}_{\alpha}\right]  -\left[  \partial_{\alpha}O\right]
\right)  \left[  O^{\prime}\right]  $ it gives:

$\nabla_{e}\Theta=\sum_{\alpha<\beta}\left(  \Gamma_{\alpha\beta}^{\gamma
}-\Gamma_{\beta\alpha}^{\gamma}\right)  dx^{\alpha}\wedge dx^{\beta}%
\otimes\partial_{\gamma}$

which is the usual torsion 2-form of the affine connection associated to
\textbf{G.}\ This connection is symmetric iff $\nabla_{e}\Theta=0$ and we have
the relation : $\nabla_{e}^{2}\Theta=\digamma\wedge\Theta.$

\paragraph{4)}

The torsion tensor can be expressed as :

$\nabla_{e}\Theta=d\Theta+\sum_{i\alpha\beta r}\left(  \left[  \widetilde
{G}_{\alpha}\right]  _{i}^{r}O_{\beta}^{\prime i}\right)  dx^{\alpha}\wedge
dx^{\beta}\otimes\partial_{r}$

with $d\Theta=2\sum_{r}\sum_{p<q}c_{pq}^{r}\partial^{p}\wedge\partial
^{q}\otimes\partial_{r}=2\sum_{a=1}^{6}\sum_{r=0}^{3}c_{a}^{r}\partial^{p_{a}%
}\wedge\partial^{q_{a}}\otimes\partial_{r}$

and

$\sum_{i\alpha\beta r}\left(  \left[  \widetilde{G}_{\alpha}\right]  _{i}%
^{r}O_{\beta}^{\prime i}\right)  dx^{\alpha}\wedge dx^{\beta}\otimes
\partial_{r}$

$=\sum_{\alpha\beta ijkr}\left[  \widetilde{G}_{\alpha}\right]  _{i}%
^{r}O_{\beta}^{\prime i}O_{j}^{\alpha}O_{k}^{\beta}\partial^{j}\wedge
\partial^{k}\otimes\partial_{r}$

$=\sum_{r}\sum_{ij}\left[  \widetilde{G}_{j}\right]  _{i}^{r}\partial
^{j}\wedge\partial^{i}\otimes\partial_{r}$

$=\sum_{a=1}^{6}\sum_{r=0}^{3}\left(  \left[  \widetilde{G}_{p_{a}}\right]
_{q_{a}}^{r}-\left[  \widetilde{G}_{q_{a}}\right]  _{p_{a}}^{r}\right)
\partial^{p_{a}}\wedge\partial^{q_{a}}\otimes\partial_{r}$

$=\sum_{a=1}^{6}\sum_{r=0}^{3}T^{ar}\partial^{p_{a}}\wedge\partial^{q_{a}%
}\otimes\partial_{r}$

with : $\left[  \widetilde{G}_{j}\right]  =\sum_{\alpha}\left[  \widetilde
{G}_{\alpha}\right]  O_{j}^{\alpha}.$

The table $T^{ar}$ \ indexed on a and r is computed with $\left[
\widetilde{\kappa}_{a}\right]  _{j}^{i}=\left(  \delta^{ip_{a}}\delta_{jq_{a}%
}-\delta^{iq_{a}}\eta_{jp_{a}}\right)  $ :

\bigskip

TABLE 2: $T^{ar}=$

$%
\begin{bmatrix}
a\setminus r & p & q & 0 & 1 & 2 & 3\\
1 & 3 & 2 & -G_{2}^{6}+G_{3}^{5} & -G_{2}^{2}-G_{3}^{3} & G_{2}^{1} &
G_{3}^{1}\\
2 & 1 & 3 & G_{1}^{6}-G_{3}^{4} & G_{1}^{2} & -G_{1}^{1}-G_{3}^{3} & G_{3}%
^{2}\\
3 & 2 & 1 & -G_{1}^{5}+G_{2}^{4} & G_{1}^{3} & G_{2}^{3} & -G_{1}^{1}%
-G_{2}^{2}\\
4 & 4 & 1 & G_{0}^{4} & -G_{1}^{4} & -G_{1}^{5}+G_{0}^{3} & -G_{0}^{2}%
-G_{1}^{6}\\
5 & 4 & 2 & G_{0}^{5} & -G_{2}^{4}-G_{0}^{3} & -G_{2}^{5} & G_{0}-G_{2}^{6}\\
6 & 4 & 3 & G_{0}^{6} & G_{0}^{2}-G_{3}^{4} & -G_{0}^{1}-G_{3}^{5} &
-G_{3}^{6}%
\end{bmatrix}
$

\bigskip

and we have :

$\nabla_{e}\Theta=\sum_{r,a}\left(  2c_{a}^{r}+T^{ar}\right)  \partial^{p_{a}%
}\wedge\partial^{q_{a}}\otimes\partial_{r}$

It should be noticed that the torsion, as the scalar curvature, are related to
a connection : they are not some geometrical properties of the manifold M.
They are computed composed either with $\left(  O_{p_{a}}^{\beta}O_{q_{a}%
}^{\alpha}-O_{q_{a}}^{\beta}O_{p_{a}}^{\alpha}\right)  $ or $O_{\beta}^{\prime
r}\left(  O_{p_{a}}^{\alpha}\partial_{\alpha}O_{q_{a}}^{\beta}-O_{q_{a}%
}^{\alpha}\partial_{\alpha}O_{p_{a}}^{\beta}\right)  $\ \ which are pure
geometrical quantities. As a manifold there are topological obstructions to
the existence of a structure of principal fiber bundle, but given a fiber
bundle there is no unique compatible connection. Indeed according to General
Relativity the connection is fixed through interactions with the content
(matter and fields) of the universe.\ Assuming that there is some intrinsic
scalar curvature or torsion of the universe would state that the vacuum has an
pre-existing physical structure.

\subsubsection{Remark}

In the tetrad method it is usual to introduce the quantities called :

a) the 1-connection 1-form : $\Omega:\Omega_{j}^{i}=\left[  \widetilde{G}%
_{j}\right]  _{k}^{i}\partial^{k}\in\Lambda(M;L(n))$

b) the 2-form torsion : $T^{i}=\dfrac{1}{2}T_{jk}^{i}\partial^{j}%
\Lambda\partial^{k}=\sum_{\left\{  jk\right\}  }T_{jk}^{i}\partial^{j}%
\Lambda\partial^{k};$

$T_{jk}^{i}=T(\partial_{j},\partial_{k})\partial^{i}=\left[  \widetilde{G}%
_{j}\right]  _{k}^{i}-\left[  \widetilde{G}_{k}\right]  _{j}^{i}-c_{jk}^{i}$

c) the 2-form curvature : $R_{j}^{i}=\dfrac{1}{2}R_{jkl}^{i}\partial
^{k}\Lambda\partial^{l}=\sum_{\left\{  kl\right\}  }R_{jkl}^{i}\partial
^{k}\Lambda\partial^{l}=\digamma$

And the following relations :

Cartan's structure equations : \ 

$d\partial^{i}+\Omega_{j}^{i}\Lambda\partial^{j}=T^{i}\Leftrightarrow
T=\nabla_{e}\Theta$

$d\Omega_{j}^{i}+\Omega_{k}^{i}\Lambda\Omega_{j}^{k}=R_{j}^{i}\Leftrightarrow%
\mathcal{F}%
_{G}=\nabla_{e}G$

Bianchi's identities :

$dT^{a}+\Omega_{j}^{a}\Lambda T^{j}=R_{j}^{a}\Lambda\partial^{j}\Rightarrow
dT^{a}\partial_{a}+\Omega_{j}^{a}\partial_{a}\Lambda T^{j}\partial_{j}=\left(
R_{j}^{a}\partial^{j}\otimes\partial_{a}\right)  \Lambda\partial
^{j}\Leftrightarrow\nabla_{e}T=\digamma\wedge\Theta=\nabla_{e}^{2}\Theta$

$dR_{j}^{a}+\Omega_{c}^{a}\Lambda R_{j}^{c}=R_{c}^{a}\Lambda\Omega_{j}%
^{c}\Leftrightarrow\nabla_{e}R=\nabla^{2}G=\digamma\wedge G$

\newpage

\part{LAGRANGIAN}

\section{PRINCIPLES}

\label{Lagrangian Principles}

\subsection{The point particle issue}

\paragraph{1)}

The system is described at each time t by the following quantities, measured
on the hypersurface S(t) :

- the geometry: the matrix $\left[  O\left(  m\right)  \right]  $ or
equivalently the fundamental form $\Theta$

- the state of the N particles : $\psi_{n}\left(  q_{n}\left(  t\right)
\right)  \in E_{M}$ located a some point $q_{n}\left(  t\right)  \in S\left(
t\right)  $

- the potential of the gravitation G and the other fields \`{A} on S(t)

For the complex variables the complex conjugate should also be involved. The
lagrangian is a real function, R-differentiable but not C-differentiable : if
it was holomorphic the partial derivatives with the conjugate variables would
be null.\ So we must consider separately the real and the imaginary part.

Let us denote all these variables $z^{j}$ where j runs over all the variables
and their coordinates. The lagrangian should also include their partial
derivatives : $\frac{\partial z^{i}}{\partial x^{\alpha}}.$ As the
interactions are described by first order connections it is sensible to limit
these partial derivatives to the first order also.

A general field model shall cover both the "vacuum" - no particles - and the
"free particles" - no force fields - cases. So the action can be split in one
part denoted $S_{M}$ addressing particles and interacting fields, and another
part $S_{F}$ addressing the interacting fields only. $\psi$ and its
derivatives figure in the $S_{M}$ part only.

\paragraph{2)}

The concept of point particle raises many difficulties in field theories. They
arise for the determination of the Lagrange equations (Poisson [21]) and the
trajectories (Quinn [22]). There are some ways to circumvent these problems,
but they are rather cumbersome and involve methods with which one cannot be
fully comfortable (such as fields propagation coming from the future). In fact
these issues appear in the case of a single particle interacting with its own
field, which is a simple model, but perhaps an unphysical one.\ Without
pretending to settle this issue I assume that the force fields are well
defined sections of the bundle $J^{1}F_{M}$ and the action $S_{F}$ reads :

$S_{F}=\int_{\Omega}L_{F}\left(  G_{\alpha}^{a},\partial_{\beta}G_{\alpha}%
^{a},\operatorname{Re}\grave{A}_{\alpha}^{a},\operatorname{Re}\partial_{\beta
}\grave{A}_{\alpha}^{a},\operatorname{Im}\grave{A}_{\alpha}^{a}%
,\operatorname{Im}\partial_{\beta}\grave{A}_{\alpha}^{a}z^{i}\left(  m\right)
,O_{\alpha}^{\prime i},\partial_{\beta}O_{\alpha}^{\prime i}\right)
\varpi_{4}$

with some real valued function $L_{F}.$ Notice this is $\varpi_{4}$\ in the
integral ($\varpi_{4}=\sqrt{\left\vert \det g\right\vert }\varpi_{0})$

\paragraph{3)}

The action $S_{M}$ should be some integral like :

$S_{M}=\sum_{k=1}^{N_{t}}$

$\int_{0}^{T}L_{M}\left(  v_{k}^{\alpha}\left(  t\right)  ,\operatorname{Re}%
\psi_{k}^{ij}\left(  t\right)  ,\operatorname{Im}\psi_{k}^{ij}\left(
t\right)  ,\operatorname{Re}\partial_{\alpha}\psi_{t}^{ij}\left(  t\right)
,\operatorname{Im}\partial_{\alpha}\psi_{k}^{ij}\left(  t\right)
,\operatorname{Re}\grave{A}_{\alpha}^{a}\left(  q_{k}\left(  t\right)
\right)  ,...\right)  dt$

with the sum over the $N_{t}$ particles present at t, their trajectories
$q_{k}\left(  t\right)  $ within $\Omega$ and their velocities : $v_{k}\left(
t\right)  =\frac{dq_{k}}{dt}.$

It is assumed that $\Omega$\ is large enough so that any particle entering
into the system stays within, or changes into another one (which is just a
change\ of the value of \ $\psi$) or is annihiliated : there is no particle
leaving or entering $\Omega$\ during the whole period of observation [0,T]
except at t=0 or t=T. Conversely particles can be created "from the vacuum".

\paragraph{4)}

Let us consider first the case where all the particles live over [0,T]. If the
problem has a "well posed" initial value formulation, that we will assume, the
principle of least action leads to equations such that the trajectories
$q_{k}\left(  t\right)  $ are uniquely determined from the initial values,
notably the positions $a_{k}$\ of the particle k at t=0. Implementing a
classical method attributed to Low, one can therefore assume that there is
some function $\widehat{f}:S(0)\times\left[  0,T\right]  \rightarrow\Omega$
such that the particle positioned at t=0 in $a\in S\left(  0\right)  $ is at
the time t at : $q=\widehat{f}\left(  a,t\right)  \in S(t).$ Of course the
function $\widehat{f}$\ depends itself on all the initial values, and is part
of the variables to be entered in the model.\ This is a strong assumption
indeed, as there is only one function for the whole system. The trajectory of
the particle k is : $q_{k}\left(  t\right)  =\widehat{f}\left(  a_{k}%
,t\right)  $ with the constant $a_{k}$ and its relative velocity is :
$v_{k}=\frac{\partial\widehat{f}\left(  a_{k},t\right)  }{\partial t}.$

Let us precise a key point : the value of the variables and their derivatives
are taken in $q=\widehat{f}\left(  a,t\right)  $, so :

$z^{i}=z^{i}\left(  \widehat{f}\left(  a,t\right)  \right)  ,$ $\partial
_{\alpha}z^{i}=\partial_{\alpha}z^{i}\left(  q\right)  =\frac{\partial z^{i}%
}{\partial x^{\alpha}}\left(  q\right)  \neq\frac{\partial z^{i}}%
{\partial\widehat{f}}\frac{\partial\widehat{f}}{\partial x^{\alpha}}.$

With these assumptions the action reads :

$S_{M}=\sum_{k=1}^{N}\int_{0}^{T}L_{M}\left(  \frac{\partial\widehat{f}\left(
a_{k},t\right)  }{\partial t},z^{i}\left(  \widehat{f}(a_{k},t)\right)
,z_{\alpha}^{i}\left(  \widehat{f}(a_{k},t)\right)  \right)  dt$

\paragraph{5)}

S(0) is a relatively compact riemanian manifold, so it is geodesically
complete and there is a unique Green function $\widetilde{N}:S(0)\times
S(0)\rightarrow%
\mathbb{R}
$ such that :

$\forall u\in L^{2}\left(  S(0)\right)  ,u|_{\partial S(0)}=0:$

$\int_{S(0)}\left(  \Delta_{a}\widetilde{N}(y,a)\right)  u\left(  a\right)
\varpi_{3}\left(  a\right)  =\int_{S(0)}\widetilde{N}(y,a)\Delta u\left(
x\right)  \varpi_{3}\left(  a\right)  =-u(y)$

where $\varpi_{3}$ is the induced euclidian metric on S(0) and $\Delta
=div\circ\nabla$ \ is the Laplace-Beltrami operator on S(0), incorporating
$\varpi_{3}$\ (Grigor'yan [6]). $\widetilde{N}$ is smooth outside x=y and
belongs to L%
${{}^2}$%
(S(0)). It is symmetric : $\widetilde{N}(a,y)=\widetilde{N}(y,a)$ ,positive on
S(0) and null on $\partial S\left(  0\right)  $. If S(0) is not too exotic
$\widetilde{V}$ is proportional to $\int_{d(a,y)}^{\infty}\frac{sds}{V(a,s)}$
where V(a,s) is the volume of the geodesic ball centered in x. $\widetilde{N}
$ is fully defined by S(0) and the induced metric on S(0).\ So it is a fixed
function in our problem.

The quantity $\phi\left(  y\right)  =\int_{0}^{T}L_{M}\left(  \frac
{\partial\widehat{f}\left(  y,t\right)  }{\partial t},z^{i}\left(  \widehat
{f}(y,t)\right)  ,z_{\alpha}^{i}\left(  \widehat{f}(y,t)\right)  \right)  dt$
is a function of $y\in S(0)$, null on $\partial S(0)$ if there is no particle
on the rim at t=0, and we can reasonably assume that $\phi\left(  x\right)
\in L^{2}\left(  S\left(  0\right)  \right)  $. So we can write :

$-\phi\left(  y\right)  =\int_{S\left(  0\right)  }\Delta_{x}\widetilde
{N}\left(  y,a\right)  \phi\left(  a\right)  \varpi_{3}\left(  a\right)  $

$-\phi\left(  a_{k}\right)  =\int_{S\left(  0\right)  }\Delta_{x}\widetilde
{N}\left(  a_{k},a\right)  \phi\left(  a\right)  \varpi_{3}\left(  a\right)  $

$S_{M}=-\sum_{k=1}^{N}\int_{S\left(  0\right)  }\left(  \Delta_{x}%
\widetilde{N}\left(  a_{k},a\right)  \right)  \phi\left(  a\right)  \varpi
_{3}\left(  a\right)  =$

$-\sum_{k=1}^{N}\int_{S\left(  0\right)  }\left(  \Delta_{x}\widetilde
{N}\left(  a_{k},a\right)  \right)  \int_{0}^{T}L_{M}\left(  \frac
{\partial\widehat{f}\left(  a,t\right)  }{\partial t},z^{i}\left(  \widehat
{f}(a,t)\right)  ,z_{\alpha}^{i}\left(  \widehat{f}(a,t)\right)  \right)
dt\varpi_{3}\left(  a\right)  $

\paragraph{6)}

$\Omega$ has the structure of a fiber bundle with base $%
\mathbb{R}
$ and typical fiber S(0), with trivialization : $\varphi_{\Omega}%
:S(0)\times\left[  0,T\right]  \rightarrow\Omega::m=\varphi_{\Omega}(a,t) $

So for each $\left(  a,t\right)  \in S(0)\times t$ the map $\varphi_{\Omega}$
defines a point in $\Omega,$ and conversely for each $m\in\Omega$ there are
unique coordinates $\left(  a,t\right)  \in S(0)\times t$ . Indeed this is
exactly how we have defined the chart. It works because the fiber bundle is trivial.

Let N be the function :

$N:\Omega\rightarrow%
\mathbb{R}
::$ $N\left(  \varphi_{\Omega}\left(  a,t)\right)  \right)  =-\sum_{k}%
\Delta_{S(0)}\widetilde{N}\left(  a_{k},\varphi_{\Omega}\left(  a,0)\right)
\right)  =-\sum_{k}\Delta_{x}\widetilde{N}\left(  a_{k},a\right)  $ .\ 

It is defined on $\Omega,$\ constant for all t, smooth and null on
$\partial\Omega.$ For a given system N should be fully known : it is included
in the initial values package.

Let be the map : $\widetilde{f}:\Omega\rightarrow\Omega::\widetilde{f}\left(
\varphi_{\Omega}\left(  a,t)\right)  \right)  =\widehat{f}\left(  a,t\right)
$ .\ It is defined on $\Omega$\ .\ The particle present at t in m if any would
have as coordinates : $m=\varphi_{\Omega}\left(  y,t)\right)  ,y\in S\left(
0\right)  .$ Its trajectory is : $q=\widehat{f}\left(  y,t\right)
=\widetilde{f}\left(  \varphi_{\Omega}\left(  y,t)\right)  \right)
=\widetilde{f}\left(  m\right)  .$ Its velocity in the chart is : $\frac
{dq}{dt}=\frac{\partial}{\partial t}\widehat{f}\left(  y,t\right)
=\partial_{0}\widetilde{f}\left(  \varphi_{\Omega}\left(  y,t)\right)
\right)  .\ $So we can define $V=\partial_{0}\widetilde{f}\left(  m\right)  $.
We will denote : $z^{0}=\widetilde{f}\Rightarrow V=\partial_{0}\widetilde
{f}=z_{0}^{0}$\ \ (but $z^{0}$\ itself cannot be explicitly in the lagrangian)
so the action reads :

$S_{M}=$

$\int_{S\left(  0\right)  }N\left(  \varphi_{\Omega}\left(  a,t)\right)
\right)  \int_{0}^{T}L_{M}\left(  z^{i}\left(  \widetilde{f}\left(
\varphi_{\Omega}\left(  a,t)\right)  \right)  \right)  ,z_{\alpha}^{i}\left(
\widetilde{f}\left(  \varphi_{\Omega}\left(  a,t)\right)  \right)  \right)
\right)  dt\varpi_{3}\left(  a\right)  $

$\Omega$ has the structure of a fiber bundle with base $%
\mathbb{R}
$ and typical fiber S(0). The volume measure on $\Omega$\ can be expressed as
: $\varpi_{4}=\varpi_{3}\otimes dt$ where dt is the Lebesgue measure on $%
\mathbb{R}
$\ (Lang [16] XV 6.4). Then:

$S_{M}=\int_{\Omega}N(m)L_{M}\left(  z^{i}\left(  \widetilde{f}(m)\right)
,z_{\alpha}^{i}\left(  \widetilde{f}(m)\right)  \right)  \varpi_{4}\left(
m\right)  $ and the action can be written : $S_{M}=\int_{\Omega}N\left(
m\right)  \widetilde{f}^{\ast}L_{M}\left(  z^{i},z_{\alpha}^{i}\right)
\varpi_{4}$

Remark : V is a vector field on $\Omega,$ which gives the velocity of the
particle that would be located at the same point.\ Thus we have a strong
analogy with a fluid mechanics model, but there are 2 differences. The V
vector field is "virtual" in the meaning that I do not assume that there is a
particle present \ : the lagrangian depends on $\widetilde{f},N$ and this has
important consequences. I\ do not assume any continuity of variables such as
density, and any conservation law should be deduced from the model.

\paragraph{7)}

Let us now consider the creation and annihiliation of particles. The presence
of a particle is felt in the action through $L_{M}.$ $E_{M}$ is a vector
bundle and 0 is a legitimate value for $\psi.$ After some adjustment if
necessary, it is possible to guarantee that $L_{M}=0$ whenever $\psi=0.$ The
previous construction stands with a variable number of particles and a map
$\widehat{f}$ defined over all S(0) with the convention that $:q=\widehat
{f}(a,t)$ corresponds to the trajectory of a particle iff $\psi\left(
\widehat{f}(a,t)\right)  \neq0.$

We have then a unique section $\psi:\Omega\rightarrow E_{M}$ for all of
$\Omega$\ and the particles. The continuity of such a section is questionable,
if the particles can be created and annihiliated, or change their physical
characteristics, but of course that is one of the main pending issues.

The physical vacuum means the absence of particles, that is $\psi\equiv0,$ and
the action is then restricted to $S_{F}.$ But the model stands as long as S(0)
is defined, with any function $N\left(  \varphi_{\Omega}\left(  \xi,t)\right)
\right)  =-\int_{S(0)}\Delta_{S(0)}\widetilde{N}\left(  x,\varphi_{\Omega
}\left(  a,0)\right)  \right)  p(a)\varpi_{3}\left(  a\right)  $ where p(a) is
an arbitrary function. One gets a section $\psi$\ which is not necessarily
null and defines some "fundamental state". We will come back later on this point.

\subsection{The configuration bundle}

\paragraph{1)}

In the following we keep the variables :

$\psi\in\Lambda_{0}\left(  E_{v}\right)  ,G\in\Lambda_{1}\left(
TV;o(3,1)\right)  ,\grave{A}\in\Lambda_{1}\left(  TV;T_{1}^{c}U\right)
,\Theta\in\Lambda_{1}\left(  TV;G_{M}\right)  $,

which are sections of the respective bundles, and the map $\widetilde{f}\in
C^{\infty}\left(  \Omega;\Omega\right)  .$

When it is useful we will denote : $Z_{i}^{\Diamond}=\widetilde{f}^{\ast}%
Z_{i},L_{M}^{\Diamond}=\widetilde{f}^{\ast}L_{M}.$

The action is :

$S=S_{M}+S_{F}$

$S_{M}=\int_{\Omega}NL_{M}(V^{\lozenge\alpha},\operatorname{Re}\psi^{\lozenge
ij},\operatorname{Im}\psi^{\lozenge ij},\operatorname{Re}\partial_{\alpha}%
\psi^{\lozenge ij},\operatorname{Im}\partial_{\alpha}\psi^{\lozenge
ij},\operatorname{Re}\grave{A}_{\alpha}^{\lozenge a},$

$\ \ \ \ \ \ \ \ \ \ \ \operatorname{Im}\grave{A}_{\alpha}^{\lozenge
a},\operatorname{Re}\partial_{\alpha}\grave{A}_{\beta}^{\lozenge
a},\operatorname{Im}\partial_{\alpha}\grave{A}_{\beta}^{\lozenge a},G_{\alpha
}^{\lozenge a},\partial_{\alpha}G_{\beta}^{\lozenge a},O_{k}^{\prime
\lozenge\alpha},\partial_{\beta}O_{k}^{\prime\lozenge\alpha})\varpi_{4}$

$S_{F}=\int_{\Omega}L_{F}\left(  \operatorname{Re}\grave{A}_{\alpha}%
^{a},\operatorname{Im}\grave{A}_{\alpha}^{a},\operatorname{Re}\partial
_{\alpha}\grave{A}_{\beta}^{a},\operatorname{Im}\partial_{\alpha}\grave
{A}_{\beta}^{a},G_{\alpha}^{a},\partial_{\alpha}G_{\beta}^{a},O_{k}%
^{\prime\alpha},\partial_{\beta}O_{k}^{\prime\alpha}\right)  \varpi_{4}$

\paragraph{2)}

There are 16m+36 variables

$z^{i}=\left(  V^{\alpha},\operatorname{Re}\psi^{ij},\operatorname{Im}%
\psi^{ij},\operatorname{Re}\grave{A}_{\alpha}^{a},\operatorname{Im}\grave
{A}_{\alpha}^{a},G_{\alpha}^{a},O_{k}^{\prime\alpha}\right)  $ ,

all real valued scalar functions. Let JZ be the vector bundle based over
$\Omega$\ modelled on the vector space spanned by $\left(  z^{i}\right)
$\ with the trivialization $Z=\varphi_{JZ}\left(  x^{\alpha},z^{i}\right)  $.
We will denote a vector of the tangent vector space to JZ : $v^{\alpha}%
\delta_{\alpha}+w^{i}\delta_{i}.$

A section of JZ is a map : $Z:\Omega\rightarrow JZ::Z\left(  m\right)
=\varphi_{JZ}\left(  x^{\alpha},z^{i}\left(  m\right)  \right)  .$ The first
jet extension $J^{1}Z$ of JZ is the set of the equivalence classes of the
sections on JZ with the same first order partial derivative (Kol\`{a}r [14]
IV). $J^{1}Z$ is coordinated by $j^{1}Z=\left(  x^{\alpha},z^{i},z_{\alpha
}^{i}\right)  $\ and $J^{1}Z\rightarrow JZ$\ is an affine bundle based over
JZ. $J^{1}Z$\ $\;$is identical to the set $JZ\otimes TM^{\ast}=L\left(
TM;JZ\right)  .$A section of $J^{1}Z$\ \ is a map : $j^{1}Z(m)=\varphi
_{J^{1}Z}\left(  x^{\alpha},z^{i}\left(  m\right)  ,\partial_{\alpha}%
z^{i}\left(  m\right)  \right)  $\ 

The configuration of the system is defined by a section $j^{1}Z$\ of $J^{1}%
Z$\ and a map $\widetilde{f}:\Omega\rightarrow\Omega$.

With these notations the action reads :%

\begin{equation}
\mathbf{S=S}_{M}\mathbf{+S}_{F}\mathbf{;\;S}_{M}\mathbf{=}\int_{\Omega
}\mathbf{N}\widetilde{f}^{\ast}\mathbf{j}^{1}\mathbf{Z}^{\ast}\mathbf{L}%
_{M}\left(  z^{i},z_{\alpha}^{i}\right)  \mathbf{\varpi}_{4}\mathbf{;\;S}%
_{F}\mathbf{=}\int_{\Omega}\mathbf{j}^{1}\mathbf{Z}^{\ast}\mathbf{L}%
_{F}\left(  z^{i},z_{\alpha}^{i}\right)  \mathbf{\varpi}_{4}\label{E14}%
\end{equation}

We will denote :

$%
\mathcal{L}%
_{M}=N\left(  m\right)  \widetilde{f}^{\ast}j^{1}Z^{\ast}\left(  L_{M}\left(
z^{i},z_{\alpha}^{i}\right)  \left(  \det O^{\prime}\right)  \right)  $

$%
\mathcal{L}%
_{F}=j^{1}Z^{\ast}\left(  L_{F}\left(  z^{i},z_{\alpha}^{i}\right)  \left(
\det O^{\prime}\right)  \right)  $

So $%
\mathcal{L}%
=%
\mathcal{L}%
_{M}^{\diamond}+%
\mathcal{L}%
_{F}$ and $S_{M}=\int_{\Omega}%
\mathcal{L}%
_{M}\varpi_{0};S_{F}=\int_{\Omega}%
\mathcal{L}%
_{F}\varpi_{0}$

\paragraph{3)}

The lagrangians $L_{M},L_{F}$ are real scalar functions on $J^{1}Z$ , which
together with the volume form $\varpi_{4}$\ define a 4-form on $\Omega.$

The specification of the lagrangian is a major issue in field theories.\ The
main road to set it out is by using the constraints imposed by covariance and
gauge equivariance.

The equations derived from the principle of least action lead to solutions
which shall be equivariant under a gauge transformation : observers with
different referentials shall be able to compare their results if they know how
to pass from one referential to the other.\ A general theorem states that this
is achieved if the lagrangian is invariant under a gauge transformation
(Giachetta [5] p.70).\ This will give us a first batch of relations to be met
by the lagrangian.

Covariance derives from the condition that the solutions, and thus the
mathematical objects involved in the model, should transform as expected in a
general change of chart on the manifold M : their coordinates are
representative of intrinsic geometrical objects. The action is the integral of
a 4-form over $\Omega.$\ So the functions $NL_{M},L_{F}$ must be invariant
under a change of chart.

We will address successively these two requirements.

We will prove that any lagrangian meeting the gauge and covariance conditions
must be of the form :

$\mathbf{L=}\left(  NL_{M}+L_{F}\right)  \det\mathbf{O}^{\prime}$

with

$\mathbf{L}_{M}\mathbf{=L}_{M}\left(  V^{\alpha},\operatorname{Re}\psi
\lozenge^{ij},\operatorname{Im}\psi\lozenge^{ij},\operatorname{Re}%
\nabla_{\alpha}\psi\lozenge^{ij},\operatorname{Im}\nabla_{\alpha}\psi
\lozenge^{ij},G_{\alpha}^{a\lozenge},O_{\alpha}^{\prime\lozenge i}%
,\partial_{\beta}O_{\alpha}^{\prime\lozenge i}\right)  $

$\mathbf{L}_{F}\mathbf{=L}_{F}\left(  \operatorname{Re}%
\mathcal{F}%
_{A\alpha\beta}^{a},\operatorname{Im}%
\mathcal{F}%
_{A\alpha\beta}^{a},%
\mathcal{F}%
_{B\alpha\beta}^{a},G_{\alpha}^{a},O_{\alpha}^{\prime i},\partial_{\beta
}O_{\alpha}^{\prime i}\right)  $

Furthermore G does not appear explicitly\ if the lagrangian does not depend on
$\partial_{\beta}O_{\alpha}^{\prime i}.$

We will prove that some of the partial derivatives of the lagrangian transform
as composants of tensors, thus they will be essential in definining "Noether currents".

But in the proof we will encounter many other mathematical quantities which
will be useful later.

\section{GAUGE EQUIVARIANCE}

\label{Guage equivariance}

Any physical measurement is done by the co-ordinated network of observers.\ So
it is not sufficient to know how each of them sets up its own apparatus, we
need to know how this set up changes as we move along the observers.\ A gauge
transformation is thus a continuous, and we will assume a diffentiable, map :
s(m)xu(m) on $Q_{M}$ extended to the bundles over M by use of the gauge
transformations rules. In fact the latter lead to parametrize the gauge
transformations by vectors of the Lie algebra, meaning using the fiber bundle
structure $F_{M}$ and its 1-jet extension introduced previously. This does not
concern $z^{0},V.$

\subsection{Gauge transformations on $Q_{M}$}

\paragraph{1)}

From a mathematical point of view a gauge transformation is a map :
$J:Q_{M}\rightarrow Q_{M}::J(q)=q\cdot f\left(  q\right)  $ where $\cdot$
stands for the right action on $Q_{M}$ and $f:Q_{M}\rightarrow\left(
Spin(3,1)\times U\right)  $ is such that : $f(q\cdot\left(  s,u\right)
)=\left(  s,u\right)  ^{-1}\times f(q)\times\left(  s,u\right)  .$ Applying
this formula to the section $q\left(  m\right)  =\varphi_{Q}\left(
m,1_{Spin}\times1_{U}\right)  $ gives : $f\left(  (\varphi_{Q}\left(
m,\left(  s,u\right)  \right)  \right)  =\left(  s,u\right)  ^{-1}\times
f(\varphi_{Q}\left(  m,1_{Spin}\times1_{U}\right)  )\times\left(  s,u\right)
$

A gauge transformation can thus be equivalently defined by a map :

$j:\Omega\rightarrow\left(  Spin(3,1)\times U\right)  $ : $\left(
j_{S}\left(  m\right)  ,j_{U}(m)\right)  =f\left(  \varphi_{Q}\left(
m,1_{Spin}\times1_{U}\right)  \right)  $

with the action on $Q_{M}:$

$J(\varphi_{Q}\left(  m,\left(  s,u\right)  \right)  )$

$=\varphi_{Q}\left(  m,\left(  s,u\right)  \right)  \cdot\left(  s,u\right)
^{-1}\left(  j_{S}\left(  m\right)  ,j_{U}(m)\right)  \left(  s,u\right)
=\varphi_{Q}\left(  m,\left(  j_{S}\left(  m\right)  s,j_{U}\left(  m\right)
u\right)  \right)  $

\paragraph{2)}

With pointwise product the set of gauge transformations has a group structure
(the gauge group). Among all these transformations we consider those which
form 1-parameter groups : the subsets of the gauge group parametrized by a
real scalar $\tau$ and such that :

$J_{\tau}\circ J_{\tau^{\prime}}=J_{\tau+\tau^{\prime}}\Leftrightarrow$

$\varphi_{Q}\left(  m,\left(  j_{\tau,S}\left(  m\right)  s,j_{\tau,A}\left(
m\right)  u\right)  \left(  j_{\tau^{\prime},S}\left(  m\right)
s,j_{\tau^{\prime},A}\left(  m\right)  u\right)  \right)  $

$=\varphi_{Q}\left(  m,\left(  j_{\tau+\tau^{\prime},S}\left(  m\right)
s,j_{\tau+\tau^{\prime},A}\left(  m\right)  u\right)  \right)  $

This condition is met with $\left(  j_{\tau,S}\left(  m\right)  ,j_{\tau
,A}\left(  m\right)  \right)  =\left(  \exp\tau\overrightarrow{\kappa}\left(
m\right)  ,\exp\tau\overrightarrow{\theta}\left(  m\right)  \right)  $ where
$\left(  \overrightarrow{\kappa}\left(  m\right)  ,\overrightarrow{\theta
}\left(  m\right)  \right)  $ is a map $\Omega\rightarrow\left(
o(3,1),T_{1}^{c}U\right)  $.

So we will focus on the gauge transformations such that :

$J_{\tau}(\varphi_{Q}\left(  m,\left(  s,u\right)  \right)  )=\varphi
_{Q}\left(  m,\left(  s,u\right)  \right)  \cdot\left(  s^{-1}\left(  \exp
\tau\overrightarrow{\kappa}\left(  m\right)  \right)  s,u^{-1}\left(  \exp
\tau\overrightarrow{\theta}\left(  m\right)  \right)  u\right)  $

$=\varphi_{Q}\left(  m,\left(  s,u\right)  \right)  \cdot\left(  \exp\tau
Ad_{s^{-1}}\overrightarrow{\kappa}\left(  m\right)  ,\exp\tau Ad_{u^{-1}%
}\overrightarrow{\theta}\left(  m\right)  \right)  $

$\frac{d}{d\tau}J_{\tau}(\varphi_{Q}\left(  m,\left(  s,u\right)  \right)
)|_{\tau=0}=$

$\left(  \left[  Ad_{s^{-1}}\right]  _{b}^{a}\kappa^{b}\left(  m\right)
\widehat{\kappa}_{a}(q\left(  \varphi_{Q}\left(  m,\left(  s,u\right)
\right)  \right)  ,\left[  Ad_{h^{-1}}\right]  _{c}^{a}\widehat{\theta}%
_{c}\left(  m\right)  \theta_{b}\left(  q\left(  \varphi_{Q}\left(  m,\left(
s,u\right)  \right)  \right)  \right)  \right)  $

where $\widehat{\kappa}_{a}(q),\widehat{\theta}_{b}\left(  q\right)  $ are the
fundamental vectors of $Q_{v}$

with $\overrightarrow{\kappa}\left(  m\right)  =\sum_{a}\kappa^{a}\left(
m\right)  \widehat{\kappa}_{a}\left(  m\right)  ,\overrightarrow{\theta
}\left(  m\right)  =\sum_{a}\theta^{a}\left(  m\right)  \widehat{\theta}%
_{a}\left(  m\right)  .$

Thus we have $\pi_{Q}\left(  J_{\tau}\left(  q\right)  \right)  =\pi
_{Q}\left(  q\right)  ;\frac{d}{d\tau}(J_{\tau}\left(  q\right)  )|_{\tau
=0}=Y(q)$ where Y is the equivariant vector field on $Q_{M}:$

$Y\in\Lambda_{0}\left(  TQ_{M}\right)  :Y\left(  q\right)  =\left(  \kappa
^{a}(q)\widehat{\kappa}_{a}(q),\theta^{b}\left(  q\right)  \widehat{\theta
}_{b}\left(  q\right)  \right)  $

with : $\left(  \kappa^{a}(\varphi_{Q}\left(  m,\left(  s,u\right)  \right)
),\theta^{b}\left(  \varphi_{Q}\left(  m,\left(  s,u\right)  \right)  \right)
\right)  =\left(  \left[  Ad_{s^{-1}}\right]  _{b}^{a}\kappa^{b}\left(
m\right)  ,\left[  Ad_{u^{-1}}\right]  _{c}^{a}\theta^{c}\left(  m\right)
\right)  $

Y is the infinitesimal generator of J$_{\tau}$ : $J_{\tau}\left(  q\right)
=\exp\tau Y\left(  q\right)  .$ Equivariant vector fields on $Q_{M}$\ are
described in the vector bundle $F_{M}:$

$\left(  \varphi_{Q}\left(  m,\left(  1,1\right)  \right)  ,\left(
\widehat{\kappa}\left(  m\right)  ,\widehat{\theta}\left(  m\right)  \right)
\right)  \simeq\left(  \varphi_{q}\left(  m,\left(  s,u\right)  ^{-1}\right)
,\left(  Ad_{s}\widehat{\kappa}\left(  m\right)  ,Ad_{u}\widehat{\theta
}\left(  m\right)  \right)  \right)  $

\paragraph{3)}

We have similar results for the gauge transformations on the fiber bundle
$J^{1}F_{M}\rightarrow F_{M}$ . A one parameter group of gauge transformations
is such that : $J_{\tau}\left(  G,\grave{A}\right)  \circ J_{\tau^{\prime}%
}\left(  G,\grave{A}\right)  =J_{\tau+\tau^{\prime}}\left(  G,\grave
{A}\right)  $ which is met by

$J_{\tau}\left(  G,\grave{A}\right)  =\left(  Ad_{\exp\tau\overrightarrow
{\kappa}}\left(  G_{\alpha}-\tau\partial_{\alpha}\overrightarrow{\kappa
}\right)  ,Ad_{\exp\tau\overrightarrow{\theta}}\left(  \grave{A}_{\alpha}%
-\tau\partial_{\alpha}\overrightarrow{\theta}\right)  \right)  $

$\frac{d}{d\tau}J_{\tau}(g\left(  m\right)  ,\grave{A}\left(  m\right)
)|_{\tau=0}$

$=(\left[  \frac{d}{d\tau}\exp\tau\overrightarrow{\kappa},G_{\alpha}%
-\tau\partial_{\alpha}\overrightarrow{\kappa}\right]  +\frac{d}{d\tau}\left(
G_{\alpha}-\tau\partial_{\alpha}\overrightarrow{\kappa}\right)  ,$

$\qquad\left[  \frac{d}{d\tau}\exp\tau\overrightarrow{\theta},\grave
{A}_{\alpha}-\tau\partial_{\alpha}\overrightarrow{\theta}\right]  +\frac
{d}{d\tau}\left(  \grave{A}_{\alpha}-\tau\partial_{\alpha}\overrightarrow
{\theta}\right)  )|_{\tau=0}$

$=\left(  \left[  \overrightarrow{\kappa}\left(  m\right)  ,G_{\alpha}\right]
-\partial_{\alpha}\overrightarrow{\kappa},\left[  \overrightarrow{\theta
}\left(  m\right)  ,\grave{A}_{\alpha}\right]  -\partial_{\alpha
}\overrightarrow{\theta}\right)  $

$=\left(  \left(  G_{bc}^{a}\kappa^{b}\left(  m\right)  G_{\alpha}%
^{c}-\partial_{\alpha}\kappa^{a}\right)  \widehat{\kappa}_{a},\left(
C_{bc}^{a}\theta^{b}\left(  m\right)  \grave{A}_{\alpha}^{c}-\partial_{\alpha
}\theta^{a}\right)  \widehat{\theta}_{a}\right)  $

where the brackets are on the respective Lie algebras.

Thus this kind of gauge transformation can be parametrized in $Q_{M}$\ and in
$F_{M}$\ by a section of $J^{1}F_{M}:$

$\zeta:\Omega\rightarrow J^{1}F_{M}:$:$\zeta\left(  m\right)  =\left(
\kappa^{a}\left(  m\right)  ,\theta^{b}\left(  m\right)  ,\partial_{\alpha
}\kappa^{a}\left(  m\right)  ,\partial_{\alpha}\theta^{b}\left(  m\right)
\right)  .$

\paragraph{4)}

The gauge transformations induced by $\kappa$\ act on \textbf{G} which is real
valued. But \`{A} is c-valued, so to be consistent one must allow $\theta$\ to
be c-valued, and extend $T_{1}U$\ to its complexified as well as the
representation $\left(  W,\chi\right)  $ to a representation of $U_{c}$.
$\theta$ and its partial derivatives are then complex valued and $\exp
\tau\theta$ is well defined.

It is clear that by proceeding this way one addresses specific gauge
transformations (only those that can be represented by one parameter group),
and so one does not cover all the constraints on the lagrangian.

\subsection{Gauge transformations in J$^{1}Z$}

The next step is to describe how these gauge transformations act on the
configuration space. At first we will describe the diffeomorphisms on $J^{1}Z
$, as it is a prerequisite for variational calculus.

\paragraph{1)}

A fibered isomorphism $\Phi:JZ\rightarrow JZ$\ is such that there is an
isomorphism $\Phi_{0}:\Omega\rightarrow\Omega$ with $\pi\circ\Phi=\Phi
_{0}\circ\pi$ where $\pi$ is the projection $JZ\rightarrow M$ . Its extension
$j^{1}\Phi:J^{1}Z\rightarrow J^{1}Z$ is defined by : $j^{1}\Phi\left(
j_{m}^{1}z\right)  =j_{\Phi_{0}\left(  m\right)  }^{1}\left(  \Phi\circ
z\circ\Phi_{0}^{-1}\left(  m\right)  \right)  $ for any section z on JZ. A
lagrangian is invariant by an automorphism $\Phi:JZ\rightarrow JZ$ if $\left(
j^{1}\Phi\right)  ^{\ast}\left(
\mathcal{L}%
\varpi_{0}\right)  =%
\mathcal{L}%
\varpi_{0}.$ Here $\varpi_{0}$ is the 4-form derived from a holonomic chart :
$\varpi_{0}=dx^{1}\wedge dx^{2}\wedge dx^{3}\wedge dx^{0}$

\paragraph{2)}

As above one focuses on one parameter groups of diffeomorphisms, with vector
fields generators. A vector field on JZ is written : $Y=Y^{\alpha}%
\delta_{\alpha}+Y^{i}\delta_{i}$ where the basis vectors $\delta_{\alpha
},\delta_{i}$\ and the components depend on the point $z\in JZ$. Y is a
projectable vector field if $\pi^{\prime}\left(  z\right)  Y$ is a vector
field on M.\ Its components $Y^{\alpha}$ depend on m only. Its flow is a
fibered diffeomorphism $\Phi_{\tau}^{Y}:JZ\rightarrow JZ::\frac{d}{d\tau}%
\Phi_{\tau}^{Y}\left(  z\right)  |_{\tau=0}=Y\left(  z\right)  $ (because :
$\pi\left(  \Phi_{\tau}^{Y}\left(  z\right)  \right)  =\exp\tau X\left(
\pi\left(  z\right)  \right)  )$ which can be extended on $J^{1}Z$ by the same
procedure as above : $j^{1}\Phi_{\tau}^{Y}\left(  j_{m}^{1}z\right)
=j_{\Phi_{\tau}^{X}\left(  m\right)  }^{1}\left(  \Phi_{\tau}^{Y}\circ
z\circ\Phi_{-\tau}^{X}\left(  m\right)  \right)  .$The one parameter group
$j^{1}\Phi_{\tau}^{Y}$ as for generator the vector field $j^{1}Y$ on $J^{1}Z $
defined by : $j^{1}Y\left(  j_{m}^{1}z\right)  =\frac{d}{d\tau}\left(
j^{1}\Phi_{\tau}^{v}\left(  j_{m}^{1}z\right)  \right)  |_{\tau=0}$ for any
section z on JZ. Its components are :

$j^{1}Y=Y^{\alpha}\delta_{\alpha}+Y^{i}\delta_{i}+\left(  \frac{\partial
Y^{i}}{\partial\xi^{\alpha}}+\frac{\partial Y^{i}}{\partial z^{j}}z_{\alpha
}^{j}-z_{\beta}^{i}\frac{\partial Y^{\beta}}{\partial\xi^{\alpha}}\right)
\delta_{i}^{\alpha}$ \ (Kol\`{a}r [14] p.360).

One can write : $j^{1}\Phi_{\tau}^{Y}\left(  j_{m}^{1}z\right)  =\Phi_{\tau
}^{j^{1}Y}\left(  j_{m}^{1}z\right)  $

$j^{1}z$ takes its value in $%
\mathbb{R}
^{16m+32}$\ and its variation is computed as :\ 

$\Phi_{\tau}^{j^{1}Y}\left(  j_{m}^{1}z\right)  -j_{m}^{1}z=\tau\left(
\frac{d}{d\tau}\Phi_{\tau}^{j^{1}Y}\left(  j_{m}^{1}z\right)  |_{\tau
=0}\right)  +\tau o\left(  \tau\right)  =\tau j^{1}Y\left(  j_{m}^{1}z\right)
+\tau o\left(  \tau\right)  $ with $\lim_{\tau\rightarrow0}o\left(
\tau\right)  =0$

The diffeomorphism $j^{1}\Phi_{\tau}^{Y}$ acting on $j^{1}z$ changes the value
of a lagrangian $%
\mathcal{L}%
\varpi_{0}$: $\left(  j^{1}z\right)  ^{\ast}%
\mathcal{L}%
\varpi_{0}\rightarrow\left(  \Phi_{\tau}^{j^{1}Y}\left(  j_{m}^{1}z\right)
\right)  ^{\ast}%
\mathcal{L}%
\varpi_{0}$

The derivative $\frac{d}{d\tau}\left(  \Phi_{\tau}^{j^{1}Y}\left(  j_{m}%
^{1}z\right)  ^{\ast}%
\mathcal{L}%
\varpi_{0}\right)  |_{\tau=0}=\left(  j^{1}z\right)  ^{\ast}\pounds _{j^{1}Y}%
\mathcal{L}%
\varpi_{0}$ is the Lie derivative $\pounds _{j^{1}Y}%
\mathcal{L}%
\varpi_{0}$ of $%
\mathcal{L}%
\varpi_{0}$\ along the vector field $J^{1}Y.$ A lagrangian $%
\mathcal{L}%
\varpi_{0} $ is invariant by $j^{1}\Phi_{\tau}^{Y}$ iff $\pounds _{j^{1}Y}%
\mathcal{L}%
\varpi_{0}=0.$

\paragraph{3)}

A one parameter group of diffeomorphism can be defined by any projectable
vector field Y on JZ (this is part of the basics of variational calculus) but
conversely if we are given a one parameter group of diffeomorphisms we can
compute its generator. We have seen previously that a one parameter group of
gauge transformations :

$q=\varphi_{Q}\left(  m,\left(  1,1\right)  \right)  \rightarrow\widehat
{q}=\varphi_{Q}\left(  m,\left(  \exp\tau\overrightarrow{\kappa}\left(
m\right)  ,\exp\tau\overrightarrow{\theta}\left(  m\right)  \right)
^{-1}\right)  .$on $Q_{M}$

can be parametrized by a section

$\zeta\in\Lambda_{0}\left(  J^{1}F_{M}\right)  $\ : $\zeta\left(  m\right)
=\left(  -\kappa^{a}\left(  m\right)  ,-\theta^{b}\left(  m\right)
,-\partial_{\alpha}\kappa^{a}\left(  m\right)  ,-\partial_{\alpha}\theta
^{b}\left(  m\right)  \right)  .$

We know how such a group acts on each variable. This is a fibered map
$\Phi_{\tau}^{Y}$ with a vertical vector field Y (the component along
$\delta_{\alpha}$\ is null) as generator,computed by : $Y\left(  Z\right)
=\frac{d}{d\tau}\Phi_{\tau}^{Y}\left(  Z\right)  |_{\tau=0}=\frac{d}{d\tau
}\widehat{Z}\left(  \tau\right)  |_{\tau=0}$

We will denote :

$Y=\left(  Y^{i}\delta_{i}\right)  $

$=\left(  Y^{\operatorname{Re}\psi^{ij}}\delta_{\operatorname{Re}\psi^{ij}%
},Y^{\operatorname{Im}\psi^{ij}}\delta_{\operatorname{Im}\psi^{ij}%
},Y^{G_{\alpha}^{a}}\delta_{G_{\alpha}^{a}},Y^{\operatorname{Re}\grave
{A}_{\alpha}^{a}}\delta_{\operatorname{Re}\grave{A}_{\alpha}^{a}%
},Y^{\operatorname{Re}\grave{A}_{\alpha}^{a}}\delta_{\operatorname{Im}%
\grave{A}_{\alpha}^{a}},Y^{O_{\alpha}^{\prime i}}\delta_{O_{\alpha}^{\prime
i}}\right)  ,$

$Y^{i}\left(  Z\right)  =\frac{d}{d\tau}\widehat{z}^{i}\left(  \tau\right)
|_{\tau=0}$

\paragraph{4)}

We have already seen the action of $\zeta$\ on the potentials :

$\sum_{a}\grave{A}_{\alpha}^{a}\overrightarrow{\theta}_{a}\rightarrow\sum
_{a}\widehat{\grave{A}}_{\alpha}\overrightarrow{\theta}_{a}=Ad_{\exp
\tau\overrightarrow{\theta}}\left(  \left(  \sum_{a}\grave{A}_{\alpha}%
^{a}-\tau\partial_{\alpha}\theta^{a}\right)  \overrightarrow{\theta}%
_{a}\right)  ;$

$\sum_{a}G_{\alpha}^{\alpha}\overrightarrow{\kappa}_{a}\rightarrow\sum
_{a}\widehat{G}_{\alpha}^{a}\overrightarrow{\kappa}_{a}=Ad_{\mu\left(
s\right)  }\left(  \sum_{a}\left(  G_{\alpha}^{a}-\tau\partial_{\alpha}%
\kappa^{a}\right)  \overrightarrow{\kappa}_{a}\right)  $

For G :

$\frac{d}{d\tau}\left(  \sum_{a}\widehat{G}_{\alpha}^{a}\overrightarrow
{\kappa}_{a}\right)  |_{\tau=0}$

$=\frac{d}{d\tau}\left(  Ad_{\mu\left(  s\right)  }\left(  \sum_{a}\left(
G_{\alpha}^{a}-\tau\partial_{\alpha}\kappa^{a}\right)  \overrightarrow{\kappa
}_{a}\right)  \right)  |_{\tau=0}=\left(  \left[  \overrightarrow{\kappa
},G_{\alpha}\right]  ^{a}-\partial_{\alpha}\kappa^{a}\right)  \overrightarrow
{\kappa}_{a}$

$Y^{G_{\alpha}^{a}}=\frac{d}{d\tau}\left(  \widehat{G}_{\alpha}^{a}\right)
|_{\tau=0}=\left[  \overrightarrow{\kappa},G_{\alpha}\right]  ^{a}%
-\partial_{\alpha}\kappa^{a}=G_{bc}^{a}\kappa^{b}G_{\alpha}^{c}-\partial
_{\alpha}\kappa^{a}$

For \`{A} we must compute the real and imaginary parts of the components (the
structure coefficients $C_{ac}^{b}$ are real) :

$\frac{d}{d\tau}\operatorname{Re}\widehat{A}_{\alpha}^{a}|_{\tau=0}=C_{bc}%
^{a}\left(  \operatorname{Re}\theta^{b}\operatorname{Re}\acute{A}_{\alpha}%
^{c}-\operatorname{Im}\theta^{b}\operatorname{Im}\acute{A}_{\alpha}%
^{c}\right)  -\operatorname{Re}\partial_{\alpha}\theta^{a}$

$\frac{d}{d\tau}\operatorname{Im}\widehat{A}_{\alpha}^{a}|_{\tau=0}=C_{bc}%
^{a}\left(  \operatorname{Re}\theta^{b}\operatorname{Im}\acute{A}_{\alpha}%
^{c}+\operatorname{Im}\theta^{b}\operatorname{Re}\acute{A}_{\alpha}%
^{c}\right)  -\operatorname{Im}\partial_{\alpha}\theta^{a}$

$Y^{\operatorname{Re}\grave{A}_{\alpha}^{a}}=\left(  C_{bc}^{a}\left(
\operatorname{Re}\theta^{b}\operatorname{Re}\acute{A}_{\alpha}^{c}%
-\operatorname{Im}\theta^{b}\operatorname{Im}\acute{A}_{\alpha}^{c}\right)
-\operatorname{Re}\partial_{\alpha}\theta^{a}\right)  $

$Y^{\operatorname{Im}\grave{A}_{\alpha}^{a}}=\left(  C_{bc}^{a}\left(
\operatorname{Re}\theta^{b}\operatorname{Im}\acute{A}_{\alpha}^{c}%
+\operatorname{Im}\theta^{b}\operatorname{Re}\acute{A}_{\alpha}^{c}\right)
-\operatorname{Im}\partial_{\alpha}\theta^{a}\right)  $

For O' :

$O_{\alpha}^{\prime i}\rightarrow\widehat{O^{\prime}}_{\alpha}^{i}=\left[
\left(  \mu\left(  \exp\tau\overrightarrow{\kappa}\right)  \right)  \right]
_{j}^{i}O_{\alpha}^{\prime j}$

$\frac{d}{d\tau}\widehat{O^{\prime}}_{\alpha}^{i}|_{\tau=0}=\left[
\jmath^{\prime}\circ\mu^{\prime}\left(  1\right)  \overrightarrow{\kappa
}\right]  _{j}^{i}O_{\alpha}^{\prime j}=\left[  \widetilde{\kappa}\right]
_{j}^{i}O_{\alpha}^{\prime j}$

$Y^{O_{\alpha}^{\prime i}}=\left(  \sum_{a}\kappa^{a}\left[  \widetilde
{\kappa}_{a}\right]  _{j}^{i}O_{\alpha}^{\prime j}\right)  $

\paragraph{5)}

The variables in $L_{M}$ depend on $\widetilde{f}$\ . The gauge transformation
acts with the values of the parameters as they are at the point $\widetilde
{f}\left(  m\right)  :$ $\kappa^{\Diamond}\left(  m\right)  =\kappa
\circ\widetilde{f}\left(  m\right)  ,\theta^{\Diamond}\left(  m\right)
=\theta\circ\widetilde{f}\left(  m\right)  $ and we have for $\psi^{ij}%
,\grave{A}_{\alpha}^{a},G_{\alpha}^{a},O_{\alpha}^{\prime i}:$

$Z\left(  \widetilde{f}\left(  m\right)  \right)  \rightarrow\widehat
{Z}\left(  \widetilde{f}\left(  m\right)  \right)  =\Phi_{\tau}^{Y}\left(
Z\left(  \widetilde{f}\left(  m\right)  \right)  \right)  $

$\Rightarrow Y\left(  Z\left(  \widetilde{f}\left(  m\right)  \right)
\right)  =\frac{d}{d\tau}\Phi_{\tau}^{Y}\left(  Z\left(  \widetilde{f}\left(
m\right)  \right)  \right)  |_{\tau=0}$

So the previous formulas stand if we consider the values $\widetilde
{f}Z=Z^{\diamond}$

For $\psi:$

$\psi^{ij\diamond}\rightarrow\widehat{\psi^{ij\diamond}}=\sum_{kl}\left[
\rho\left(  \exp\tau\overrightarrow{\kappa}^{\Diamond}\right)  \right]
_{k}^{i}\left[  \chi\left(  \exp\tau\overrightarrow{\theta}^{\Diamond}\right)
\right]  _{l}^{j}\psi^{\Diamond kl}$

$\frac{d}{d\tau}\widehat{\psi^{\diamond}}|_{\tau=0}=\left(  \rho^{\prime}%
\circ\Upsilon^{\prime}\left(  1\right)  \left(  \overrightarrow{\kappa
}^{\diamond}\right)  \otimes1+1\otimes\chi^{\prime}\left(  1\right)  \left(
\overrightarrow{\theta}^{\diamond}\right)  \right)  \psi^{\diamond}$

$\frac{d}{d\tau}\widehat{\psi^{\Diamond ij}}|_{\tau=0}=\left[  \kappa
^{\diamond}\right]  _{k}^{i}\psi^{\diamond kj}+\left[  \theta^{\diamond
}\right]  _{l}^{j}\psi^{\diamond il}$

$\frac{d}{d\tau}\operatorname{Re}\widehat{\psi^{\Diamond ij}}|_{\tau=0}$

$=\kappa^{\Diamond a}\operatorname{Re}\left(  \left[  \kappa_{a}\right]
_{k}^{i}\psi^{\Diamond kj}\right)  +\operatorname{Re}\theta^{\Diamond
a}\operatorname{Re}\left(  \left[  \theta_{a}\right]  _{k}^{j}\psi^{\Diamond
ik}\right)  -\operatorname{Im}\theta^{\Diamond a}\operatorname{Im}\left(
\left[  \theta_{a}\right]  _{k}^{j}\psi^{\Diamond ik}\right)  $

$\frac{d}{d\tau}\operatorname{Im}\widehat{\psi^{\Diamond ij}}|_{\tau=0}$

$=\kappa^{\Diamond a}\operatorname{Im}\left(  \left[  \kappa_{a}\right]
_{k}^{i}\psi^{\Diamond kj}\right)  +\operatorname{Re}\theta^{\Diamond
a}\operatorname{Im}\left(  \left[  \theta_{a}\right]  _{k}^{j}\psi^{\Diamond
ik}\right)  +\operatorname{Im}\theta^{\Diamond a}\operatorname{Re}\left(
\left[  \theta_{a}\right]  _{k}^{j}\psi^{\Diamond ik}\right)  $

$Y^{\operatorname{Re}\psi^{ij}}$

$=\sum_{a}\kappa^{\Diamond a}\operatorname{Re}\left(  \left[  \kappa
_{a}\right]  \left[  \psi^{\diamond}\right]  \right)  _{j}^{i}%
+\operatorname{Re}\theta^{\Diamond a}\operatorname{Re}\left(  \left[
\psi^{\Diamond}\right]  \left[  \theta_{a}\right]  ^{t}\right)  _{j}%
^{i}-\operatorname{Im}\theta^{\Diamond a}\operatorname{Im}\left(  \left[
\psi^{\Diamond}\right]  \left[  \theta_{a}\right]  ^{t}\right)  _{j}^{i}$

$Y^{\operatorname{Im}\psi^{ij}}$

$=\sum_{a}\kappa^{\Diamond a}\operatorname{Im}\left(  \left[  \kappa
_{a}\right]  \left[  \psi^{\diamond}\right]  \right)  _{j}^{i}%
+\operatorname{Re}\theta^{\Diamond a}\operatorname{Im}\left(  \left[
\psi^{\Diamond}\right]  \left[  \theta_{a}\right]  ^{t}\right)  _{j}%
^{i}+\operatorname{Im}\theta^{\Diamond a}\operatorname{Re}\left(  \left[
\psi^{\Diamond}\right]  \left[  \theta_{a}\right]  ^{t}\right)  _{j}^{i}$

\paragraph{6)}

The vector Y has an extention on $J^{1}Z:$ $j^{1}Y=\left(  Y,\partial_{\beta
}Y\right)  $ parametrized by $j^{2}\zeta=\left(  \zeta,\zeta_{\alpha}%
^{a},\zeta_{\alpha\beta}^{a}\right)  \in\Lambda_{0}J^{2}F_{M}.$\ Its
components can be computed from the general formula above, but here a direct
approach is easier. Let $\Phi_{\tau}^{Y}$ be the one parameter group generated
by Y.\ It is a vertical vector, so $\Phi_{0}=Id$\ and its extension is for a
section Z :$j^{1}\Phi_{\tau}^{Y}\left(  j_{m}^{1}Z\right)  =j_{m}^{1}\left(
\Phi_{\tau}^{Y}\circ Z\right)  =j_{m}^{1}\left(  \widehat{Z}\left(
\tau\right)  \right)  .$ So the components $\partial_{\beta}z$\ are computed
by the partial derivatives of the components of Y :

$Y^{i}\left(  Z\left(  m\right)  ,\zeta\left(  m\right)  \right)  =\frac
{d}{d\tau}\left(  \widehat{z}^{i}\left(  m,\tau\right)  \right)  |_{\tau=0}$

$Y_{\alpha}^{i}=\frac{d}{d\tau}\left(  \partial_{\alpha}\widehat{z}^{i}\left(
m,\tau\right)  \right)  |_{\tau=0}=\partial_{\alpha}\left(  \frac{d}{d\tau
}\left(  \widehat{z}^{i}\left(  m,\tau\right)  \right)  |_{\tau=0}\right)
=\partial_{\alpha}Y^{i}\left(  Z\left(  m\right)  ,\zeta\left(  m\right)
\right)  $

For the variables depending on $\widetilde{f}$\ the evaluation is still done
in $\widetilde{f}\left(  m\right)  $ so the partial derivatives of
$\widetilde{f}$\ are discarded.

\subparagraph{a) For \`{A} :}

$\partial_{\alpha}\widehat{\grave{A}}_{\beta}$

$=\partial_{\alpha}\left(  Ad_{u}\left(  \grave{A}_{\beta}-\tau\partial
_{\beta}\theta\right)  \right)  $

$=\left(  \partial_{\alpha}\left(  Ad_{u}\right)  \right)  \left(  \grave
{A}_{\beta}-\partial_{\beta}\theta\right)  +Ad_{u}\left(  \partial_{\alpha
}\grave{A}_{\beta}-\partial_{\alpha\beta}\theta\right)  $

$=Ad_{u}\left[  \tau\partial_{\alpha}\theta,\grave{A}_{\beta}-\tau
\partial_{\beta}\theta\right]  +Ad_{u}\left(  \partial_{\alpha}\grave
{A}_{\beta}-\tau\partial_{\alpha\beta}\theta\right)  $

with the general formula : $\partial_{\alpha}\left(  Ad_{\exp\tau
\overrightarrow{\theta}}\right)  =Ad_{\exp\tau\overrightarrow{\theta}}\circ
ad\left(  \partial_{\alpha}\left(  \tau\theta\right)  \right)  $

$\partial_{\alpha}\widehat{\grave{A}}_{\beta}=Ad_{\exp\tau\overrightarrow
{\theta}}\left(  \left[  \tau\partial_{\alpha}\theta,\grave{A}_{\alpha}%
-\tau\partial_{\beta}\theta\right]  +\partial_{\alpha}\grave{A}_{\alpha}%
-\tau\partial_{\alpha\beta}\theta\right)  $

$\frac{d}{d\tau}\partial_{\alpha}\widehat{A}_{\beta}|_{\tau=0}=\left[
\overrightarrow{\theta},\partial_{\alpha}\grave{A}_{\beta}\right]  +\left[
\overrightarrow{\partial_{\alpha}\theta},\grave{A}_{\beta}\right]
-\partial_{\alpha\beta}\theta$

$\frac{d}{d\tau}\partial_{\alpha}\widehat{A}_{\beta}^{a}|_{\tau=0}=C_{bc}%
^{a}\theta^{b}\partial_{\alpha}\grave{A}_{\beta}^{c}+C_{bc}^{a}\partial
_{\beta}\theta^{b}\grave{A}_{\beta}^{c}-\partial_{\alpha\beta}\theta^{a}$

$Y_{\alpha}^{\operatorname{Re}\grave{A}_{\beta}^{a}}=$

$C_{bc}^{a}\left(  \operatorname{Re}\partial_{\alpha}\theta^{b}%
\operatorname{Re}\grave{A}_{\beta}^{c}-\operatorname{Im}\partial_{\alpha
}\theta^{b}\operatorname{Im}\grave{A}_{\beta}^{c}+\operatorname{Re}\theta
^{b}\operatorname{Re}\partial_{\alpha}\grave{A}_{\beta}^{c}-\operatorname{Im}%
\theta^{b}\operatorname{Im}\partial_{\alpha}\grave{A}_{\beta}^{c}\right)
-\operatorname{Re}\partial_{\alpha\beta}\theta^{a}$

$Y_{\alpha}^{\operatorname{Im}\grave{A}_{\beta}^{a}}=$

$C_{ac}^{a}\left(  \operatorname{Re}\partial_{\alpha}\theta^{b}%
\operatorname{Im}\grave{A}_{\beta}^{c}+\operatorname{Im}\partial_{\alpha
}\theta^{c}\operatorname{Re}\grave{A}_{\beta}^{c}+\operatorname{Re}\theta
^{b}\operatorname{Im}\partial_{\alpha}\grave{A}_{\beta}^{c}+\operatorname{Im}%
\theta^{b}\operatorname{Re}\partial_{\alpha}\grave{A}_{\beta}^{c}\right)
-\operatorname{Im}\partial_{\alpha\beta}\theta^{a}$

\subparagraph{b) For G :}

$\widehat{\partial_{\alpha}G}_{\beta}=Ad_{\mu\left(  s\right)  }\left(
\left[  \tau\partial_{\alpha}\kappa,G_{\alpha}-\tau\partial_{\beta}%
\kappa\right]  +\partial_{\alpha}G_{\beta}-\tau\partial_{\alpha\beta}%
\kappa\right)  $

$Y_{\alpha}^{G_{\beta}^{a}}=\frac{d}{d\tau}\partial_{\alpha}\widehat{G}%
_{\beta}^{a}|_{\tau=0}=\kappa^{b}G_{bc}^{a}\partial_{\alpha}G_{\beta}%
^{c}+\partial_{\alpha}\kappa^{b}G_{bc}^{a}G_{\beta}^{c}-\partial_{\alpha\beta
}\kappa^{a}$

\subparagraph{c) For O' :}

$\partial_{\alpha}\widehat{O^{\prime}}_{\beta}^{i}=\left(  \tau\partial
_{\alpha}\kappa^{a}\left[  \left(  \mu\left(  \exp\tau\kappa\right)  \right)
\right]  _{k}^{i}\left[  \widetilde{\kappa}_{a}\right]  _{j}^{k}\right)
O_{\beta}^{\prime j}+\left[  \left(  \mu\left(  \exp\tau\kappa\right)
\right)  \right]  _{j}^{i}\partial_{\alpha}O_{\beta}^{\prime j}$

$Y_{\alpha}^{O_{\beta}^{\prime i}}=\frac{d}{d\tau}\partial_{\alpha}%
\widehat{O^{\prime}}_{\beta}^{i}|_{\tau=0}=\partial_{\alpha}\kappa^{a}\left[
\widetilde{\kappa}_{a}\right]  _{j}^{i}O_{\beta}^{\prime j}+\kappa^{a}\left[
\widetilde{\kappa}_{a}\right]  _{j}^{i}\partial_{\alpha}O_{\beta}^{\prime j}$

\subparagraph{d) For $\psi:$}

$\partial_{\alpha}\widehat{\psi}^{\Diamond ij}=\sum_{a}(\tau\left(
\partial_{\alpha}\kappa^{a}\right)  \left[  \rho\left(  \exp\tau
\kappa^{\Diamond}\right)  \right]  _{p}^{i}\left[  \kappa_{a}\right]  _{k}%
^{p}\left[  \chi\left(  \exp\tau\theta^{\Diamond}\right)  \right]  _{l}%
^{j}\psi^{kl}$

$+\tau\left(  \partial_{\alpha}\theta^{a}\right)  \left[  \rho\left(  \exp
\tau\kappa^{\Diamond}\right)  \right]  _{k}^{i}\left[  \chi\left(  \exp
\tau\theta^{\Diamond}\right)  \right]  _{p}^{j}\left[  \theta_{a}\right]
_{l}^{p}\psi^{kl}$

$+\left[  \rho\left(  \exp\tau\kappa^{\Diamond}\right)  \right]  _{k}%
^{i}\left[  \chi\left(  \exp\tau\theta^{\Diamond}\right)  \right]  _{l}%
^{j}\partial_{\alpha}\psi^{\Diamond kl})$

$Y_{\alpha}^{\operatorname{Re}\psi^{ij}}=\sum_{a}\kappa^{\Diamond
a}\operatorname{Re}\left(  \left[  \kappa_{a}\right]  \left[  \partial
_{\alpha}\psi^{\Diamond}\right]  \right)  _{j}^{i}+\partial_{\alpha}%
\kappa^{\Diamond a}\operatorname{Re}\left(  \left[  \kappa_{a}\right]  \left[
\partial_{\alpha}\psi^{\Diamond}\right]  \right)  _{j}^{i}$

$+\operatorname{Re}\theta^{\Diamond a}\operatorname{Re}\left(  \left[
\partial_{\alpha}\psi^{\Diamond}\right]  \left[  \theta_{a}\right]
^{t}\right)  +\operatorname{Re}\partial_{\alpha}\theta^{\Diamond
a}\operatorname{Re}\left(  \left[  \psi^{\Diamond}\right]  \left[  \theta
_{a}\right]  ^{t}\right)  _{j}^{i}$

$-\operatorname{Im}\theta^{\Diamond a}\operatorname{Im}\left(  \left[
\partial_{\alpha}\psi^{\Diamond}\right]  \left[  \theta_{a}\right]
^{t}\right)  _{j}^{i}-\operatorname{Im}\partial_{\alpha}\theta^{\Diamond
a}\operatorname{Im}\left(  \left[  \psi^{\Diamond}\right]  \left[  \theta
_{a}\right]  ^{t}\right)  _{j}^{i}$

$Y_{\alpha}^{\operatorname{Im}\psi^{ij}}=\sum_{a}\kappa^{\Diamond
a}\operatorname{Im}\left(  \left[  \kappa_{a}\right]  \left[  \partial
_{\alpha}\psi^{\Diamond}\right]  \right)  _{j}^{i}+\partial_{\alpha}%
\kappa^{\Diamond a}\operatorname{Im}\left(  \left[  \kappa_{a}\right]  \left[
\partial_{\alpha}\psi^{\Diamond}\right]  \right)  _{j}^{i}$

$+\operatorname{Re}\theta^{\Diamond a}\operatorname{Im}\left(  \left[
\partial_{\alpha}\psi^{\Diamond}\right]  \left[  \theta_{a}\right]
^{t}\right)  _{j}^{i}+\operatorname{Re}\partial_{\alpha}\theta^{\Diamond
a}\operatorname{Im}\left(  \left[  \psi^{\Diamond}\right]  \left[  \theta
_{a}\right]  ^{t}\right)  _{j}^{i}$

$+\operatorname{Im}\theta^{\Diamond a}\operatorname{Re}\left(  \left[
\partial_{\alpha}\psi^{\Diamond}\right]  \left[  \theta_{a}\right]
^{t}\right)  _{j}^{i}+\operatorname{Im}\partial_{\alpha}\theta^{\Diamond
a}\operatorname{Re}\left(  \left[  \psi^{\Diamond}\right]  \left[  \theta
_{a}\right]  ^{t}\right)  _{j}^{i}$

\subsection{Equivariance conditions}

The Lie derivative of the lagrangian must be null under a gauge
transformation.\ A direct computation gives :

$\pounds _{j^{2}\zeta}%
\mathcal{L}%
\varpi_{0}=\frac{d}{d\tau}%
\mathcal{L}%
\left(  \widehat{Z}\left(  j^{2}\zeta\right)  \right)  \varpi_{0}|_{\tau=0}$

$=\left(  \sum_{i}\frac{\partial%
\mathcal{L}%
_{M}}{\partial z^{i}}\frac{d\widehat{z}^{\Diamond i}}{d\tau}|_{\tau=0}%
+\frac{\partial%
\mathcal{L}%
_{M}}{\partial\partial_{\alpha}z^{i}}\frac{d\partial_{\alpha}\widehat
{z}^{\Diamond i}}{d\tau}|_{\tau=0}+\frac{\partial%
\mathcal{L}%
_{F}}{\partial Z^{i}}\frac{d\widehat{Z}^{i}}{d\tau}|_{\tau=0}+\frac{\partial%
\mathcal{L}%
_{F}}{\partial\partial_{\alpha}Z^{i}}\frac{d\partial_{\alpha}\widehat{Z}^{i}%
}{d\tau}|_{\tau=0}\right)  \varpi_{0}$

By definition :

$\frac{d\widehat{z}^{\Diamond i}}{d\tau}|_{\tau=0}=\frac{d\widehat{z}^{i}%
}{d\tau}|_{\tau=0}=Y^{i}\left(  Z,\zeta\right)  ,\frac{d\partial_{\alpha
}\widehat{z}^{\Diamond i}}{d\tau}|_{\tau=0}=\frac{d\partial_{\alpha}%
\widehat{z}^{i}}{d\tau}|_{\tau=0}=Y_{\alpha}^{i}\left(  Z,\zeta\right)  $

$%
\mathcal{L}%
_{M}$ depends on $f$ and not $%
\mathcal{L}%
_{F},$ so for all $j^{2}\zeta$\ we must have the identities :

$\sum_{i}\frac{\partial%
\mathcal{L}%
_{M}}{\partial z^{i}}Y^{i}\left(  Z,\zeta\right)  +\sum_{\alpha}\frac{\partial%
\mathcal{L}%
_{M}}{\partial\partial_{\alpha}z^{i}}Y_{\alpha}^{i}\left(  Z,\zeta\right)  =0$

$\sum_{i}\frac{\partial%
\mathcal{L}%
_{F}}{\partial z^{i}}Y^{i}\left(  Z,\zeta\right)  +\sum_{\alpha}\frac{\partial%
\mathcal{L}%
_{F}}{\partial\partial_{\alpha}z^{i}}Y_{\alpha}^{i}\left(  Z,\zeta\right)  =0$

Some of the variables can appear explicitly or through an other one. In order
to avoid confusion we will use the following conventions :

$\frac{d}{dz^{i}},\frac{d}{d\partial_{\alpha}z^{i}}$ such as $\frac
{d}{d\operatorname{Re}\psi^{ij}},\frac{d}{d\partial_{\alpha}\operatorname{Re}%
\psi^{ij}},...$ denotes the full partial derivatives with respect to the
variables $z^{i},z_{\alpha}^{i}$

$\frac{\partial}{\partial z^{i}},\frac{\partial}{\partial\partial_{\alpha
}z^{i}}$ such as $\frac{\partial}{\partial\operatorname{Re}\psi^{ij}}%
,\frac{\partial}{\partial\partial_{\alpha}\operatorname{Re}\psi^{ij}},...$
denotes the partial derivatives with respect to the variables $z^{i}%
,z_{\alpha}^{i}$ only when they appear explicitly

So the previous identities read :

$\sum_{i}\frac{d%
\mathcal{L}%
_{M}}{dz^{i}}Y^{i}\left(  Z,\zeta\right)  +\sum_{\alpha}\frac{d%
\mathcal{L}%
_{M}}{d\partial_{\alpha}z^{i}}Y_{\alpha}^{i}\left(  Z,\zeta\right)  =0$

$\sum_{i}\frac{d%
\mathcal{L}%
_{F}}{dz^{i}}Y^{i}\left(  Z,\zeta\right)  +\sum_{\alpha}\frac{d%
\mathcal{L}%
_{F}}{d\partial_{\alpha}z^{i}}Y_{\alpha}^{i}\left(  Z,\zeta\right)  =0$

\subsubsection{Lagrangien $%
\mathcal{L}%
_{F}$}

We address first the lagrangian $%
\mathcal{L}%
_{F}.$

\paragraph{1)}

\subparagraph{The terms in the second order in $j^{2}\zeta$ give :}%

\begin{align}
\forall\alpha,\beta,a  & :0=\frac{d%
\mathcal{L}%
_{F}}{d\partial_{\alpha}G_{\beta}^{a}}+\frac{d%
\mathcal{L}%
_{F}}{d\partial_{\beta}G_{\alpha}^{a}};0=\frac{d%
\mathcal{L}%
_{F}}{d\partial_{\beta}\operatorname{Re}\grave{A}_{\alpha}^{a}}+\frac{d%
\mathcal{L}%
_{F}}{d\partial_{\alpha}\operatorname{Re}\grave{A}_{\beta}^{a}};\label{E15}\\
0  & =\frac{d%
\mathcal{L}%
_{F}}{d\partial_{\beta}\operatorname{Im}\grave{A}_{\alpha}^{a}}+\frac{d%
\mathcal{L}%
_{F}}{d\partial_{\alpha}\operatorname{Im}\grave{A}_{\beta}^{a}}%
\end{align}

where we use : $\partial_{\beta\alpha}\kappa^{b}=\partial_{\alpha\beta}%
\kappa^{b},\partial_{\beta\alpha}\theta^{b}=\partial_{\alpha\beta}\theta^{b}$

So the partial derivatives are antisymmetric in $\partial_{\alpha}G_{\beta
}^{a},\partial_{\alpha}\grave{A}_{\beta}^{a}$ as could be expected.

\subparagraph{The terms in the first order in $\partial_{\alpha}\kappa^{a}$
give :}%

\begin{equation}
\forall a,\alpha:0=-\frac{d%
\mathcal{L}%
_{F}}{dG_{\alpha}^{a}}+\sum_{\beta b}\frac{d%
\mathcal{L}%
_{F}}{d\partial_{\alpha}G_{\beta}^{b}}\left[  \overrightarrow{\kappa}%
_{a},G_{\beta}\right]  ^{b}+\sum_{\beta i}\frac{d%
\mathcal{L}%
_{F}}{d\partial_{\alpha}O_{\beta}^{\prime i}}\left(  \left[  \widetilde
{\kappa}_{a}\right]  \left[  O^{\prime}\right]  \right)  _{\beta}%
^{i}\label{E16}%
\end{equation}

\subparagraph{The terms in the first order in $\partial_{\alpha}\theta^{a}$
give :}%

\begin{equation}
\forall\alpha,a:0=-\frac{d%
\mathcal{L}%
_{F}}{d\operatorname{Re}\grave{A}_{\alpha}^{a}}+\sum_{\beta}\frac{d%
\mathcal{L}%
_{F}}{d\operatorname{Re}\partial_{\alpha}\grave{A}_{\beta}^{b}}%
\operatorname{Re}\left[  \overrightarrow{\theta}_{a},\grave{A}_{\beta}\right]
^{b}+\frac{d%
\mathcal{L}%
_{F}}{d\operatorname{Im}\partial_{\alpha}\grave{A}_{\beta}^{b}}%
\operatorname{Im}\left[  \overrightarrow{\theta}_{a},\grave{A}_{\beta}\right]
^{b}\label{E17}%
\end{equation}

\begin{equation}
\forall\alpha,a:0=-\frac{d%
\mathcal{L}%
_{F}}{d\operatorname{Im}\grave{A}_{\alpha}^{a}}+\sum_{\beta}-\frac{d%
\mathcal{L}%
_{F}}{d\operatorname{Re}\partial_{\alpha}\grave{A}_{\beta}^{b}}%
\operatorname{Im}\left[  \overrightarrow{\theta}_{a},\grave{A}_{\beta}\right]
^{b}+\frac{d%
\mathcal{L}%
_{F}}{d\operatorname{Im}\partial_{\alpha}\grave{A}_{\beta}^{b}}%
\operatorname{Re}\left[  \overrightarrow{\theta}_{a},\grave{A}_{\beta}\right]
^{b}\label{E18}%
\end{equation}

where we use the fact that $\left(  \overrightarrow{\theta}_{a}\right)  $ is a
real basis and the structure coefficients $C_{bc}^{a}$ are real :

$\operatorname{Re}\left[  \overrightarrow{\theta}_{a},\grave{A}_{\beta
}\right]  ^{b}=\operatorname{Re}\sum_{c}\left(  C_{ac}^{b}\grave{A}_{\beta
}^{c}\right)  =\sum_{c}C_{ac}^{b}\operatorname{Re}\grave{A}_{\beta}^{c}$

\subparagraph{The terms in $\kappa^{a}$ give :}%

\begin{equation}
\label{E19}
\end{equation}

$\forall a:0=\sum_{\alpha\beta b}\frac{d%
\mathcal{L}%
_{F}}{dG_{\alpha}^{b}}\left[  \overrightarrow{\kappa}_{a},G_{\alpha}\right]
^{b}+\frac{d%
\mathcal{L}%
_{F}}{d\partial_{\beta}G_{\alpha}^{b}}\left[  \overrightarrow{\kappa}%
_{a},\partial_{\beta}G_{\alpha}\right]  ^{b}$

$\qquad+\sum_{\alpha\beta i}\frac{d%
\mathcal{L}%
_{F}}{dO_{\alpha}^{\prime i}}\left(  \left[  \widetilde{\kappa}_{a}\right]
\left[  O^{\prime}\right]  \right)  _{\alpha}^{i}+\frac{d%
\mathcal{L}%
_{F}}{d\partial_{\beta}O_{\alpha}^{\prime i}}\left(  \left[  \widetilde
{\kappa}_{a}\right]  \left[  \partial_{\beta}O^{\prime}\right]  \right)
_{\alpha}^{i}$

\bigskip

\subparagraph{The terms in $\theta^{a}$ give :}%

\begin{equation}
\label{E20}
\end{equation}

$\forall a:0=\sum_{\alpha b}\frac{d%
\mathcal{L}%
_{F}}{d\operatorname{Re}\grave{A}_{\alpha}^{b}}\operatorname{Re}\left[
\overrightarrow{\theta}_{a},\grave{A}_{\alpha}\right]  ^{b}+\frac{d%
\mathcal{L}%
_{F}}{d\operatorname{Im}\grave{A}_{\alpha}^{b}}\operatorname{Im}\left[
\overrightarrow{\theta}_{a},\grave{A}_{\alpha}\right]  ^{b}+$

$\qquad\sum_{\beta}\frac{\partial%
\mathcal{L}%
_{F}}{\partial\operatorname{Re}\partial_{\beta}\grave{A}_{\alpha}^{b}%
}\operatorname{Re}\left[  \overrightarrow{\theta}_{a},\partial_{\beta}%
\grave{A}_{\alpha}\right]  ^{b}+\frac{\partial%
\mathcal{L}%
_{F}}{\partial\operatorname{Im}\partial_{\beta}\grave{A}_{\alpha}^{b}%
}\operatorname{Im}\left[  \overrightarrow{\theta}_{a},\partial_{\beta}%
\grave{A}_{\alpha}\right]  ^{b}$%

\begin{equation}
\label{E20b}
\end{equation}

$\forall a:0=\sum_{\alpha b}-\frac{d%
\mathcal{L}%
_{F}}{d\operatorname{Re}\grave{A}_{\alpha}^{b}}\operatorname{Im}\left[
\overrightarrow{\theta}_{a},\grave{A}_{\alpha}\right]  ^{b}+\frac{d%
\mathcal{L}%
_{F}}{d\operatorname{Im}\grave{A}_{\alpha}^{b}}\operatorname{Re}\left[
\overrightarrow{\theta}_{a},\grave{A}_{\alpha}\right]  ^{b}$

$\qquad+\sum_{\beta}-\frac{\partial%
\mathcal{L}%
_{F}}{\partial\operatorname{Re}\partial_{\beta}\grave{A}_{\alpha}^{b}%
}\operatorname{Im}\left[  \overrightarrow{\theta}_{a},\partial_{\beta}%
\grave{A}_{\alpha}\right]  ^{b}+\frac{\partial%
\mathcal{L}%
_{F}}{\partial\operatorname{Im}\partial_{\beta}\grave{A}_{\alpha}^{b}%
}\operatorname{Re}\left[  \overrightarrow{\theta}_{a},\partial_{\beta}%
\grave{A}_{\alpha}\right]  ^{b}$

\paragraph{2)}

Let be $F_{G\alpha\beta}^{a}=\partial_{\alpha}G_{\beta}+\partial_{\beta
}G_{\alpha}$

By changing the variables :

$\partial_{\alpha}G_{\beta}\rightarrow\frac{1}{2}\left(
\mathcal{F}%
_{G\alpha\beta}^{a}-G_{bc}^{a}G_{\alpha}^{b}G_{\beta}^{c}+F_{G\alpha\beta}%
^{a}\right)  $

$\partial_{\beta}G_{\alpha}\rightarrow\frac{1}{2}\left(  -%
\mathcal{F}%
_{G\alpha\beta}^{a}+G_{bc}^{a}G_{\alpha}^{b}G_{\beta}^{c}+F_{G\alpha\beta}%
^{a}\right)  $

the equation \ref{E15} gives, when renaming $%
\mathcal{L}%
\prime_{F}$ the new lagrangian :

$\forall\alpha,\beta,b:\frac{d%
\mathcal{L}%
_{F}^{\prime}}{dF_{G\alpha\beta}^{b}}+\frac{d%
\mathcal{L}%
_{F}^{\prime}}{dF_{G\beta\alpha}^{b}}+\frac{d%
\mathcal{L}%
_{F}^{\prime}}{d%
\mathcal{F}%
_{G\alpha\beta}^{b}}-\frac{d%
\mathcal{L}%
_{F}^{\prime}}{d%
\mathcal{F}%
_{G\beta\alpha}^{b}}=0$

with the reversion of $\alpha,\beta:$

$\frac{d%
\mathcal{L}%
_{F}^{\prime}}{dF_{G\beta\alpha}^{b}}+\frac{d%
\mathcal{L}%
_{F}^{\prime}}{dF_{G\alpha\beta}^{b}}+\frac{d%
\mathcal{L}%
_{F}^{\prime}}{d%
\mathcal{F}%
_{G\beta\alpha}^{b}}-\frac{d%
\mathcal{L}%
_{F}^{\prime}}{d%
\mathcal{F}%
_{G\alpha\beta}^{b}}=0$

and adding the two : $2\left(  \frac{d%
\mathcal{L}%
_{F}^{\prime}}{dF_{G\beta\alpha}^{b}}+\frac{d%
\mathcal{L}%
_{F}^{\prime}}{dF_{G\alpha\beta}^{b}}\right)  =0$

$F_{G\alpha\beta}^{a}=F_{G\beta\alpha}^{a}\Rightarrow\frac{d%
\mathcal{L}%
_{F}^{\prime}}{dF_{G\alpha\beta}^{b}}=\frac{d%
\mathcal{L}%
_{F}^{\prime}}{dF_{G\beta\alpha}^{b}}=0$

We have a similar calculation for \`{A}. The first result is that the partial
derivatives of the potential G and \`{A} factorize through the curvature forms
$%
\mathcal{F}%
_{A},%
\mathcal{F}%
_{G}$.%

\begin{equation}
\frac{d%
\mathcal{L}%
_{F}}{d\partial_{\beta}G_{\alpha}^{a}}=-\frac{d%
\mathcal{L}%
_{F}}{d\partial_{\alpha}G_{\beta}^{a}}=2\frac{d%
\mathcal{L}%
_{F}}{d%
\mathcal{F}%
_{G\beta\alpha}^{a}}\label{E22a}%
\end{equation}

\begin{equation}
\frac{d%
\mathcal{L}%
_{F}}{d\operatorname{Re}\partial_{\alpha}\grave{A}_{\beta}^{a}}=-\frac{d%
\mathcal{L}%
_{F}}{d\operatorname{Re}\partial_{\beta}\grave{A}_{\alpha}^{a}}=2\frac{d%
\mathcal{L}%
_{F}}{d\operatorname{Re}%
\mathcal{F}%
_{\alpha\beta}^{a}};\frac{d%
\mathcal{L}%
_{F}}{d\operatorname{Im}\partial_{\alpha}\grave{A}_{\beta}^{a}}=-\frac{d%
\mathcal{L}%
_{F}}{d\operatorname{Im}\partial_{\beta}\grave{A}_{\alpha}^{a}}=2\frac{d%
\mathcal{L}%
_{F}}{d\operatorname{Im}%
\mathcal{F}%
_{\alpha\beta}^{a}}\label{E22b}%
\end{equation}

\paragraph{3)}

Equation \ref{E16} becomes :

$\forall a,\alpha:-\frac{d%
\mathcal{L}%
_{F}}{dG_{\alpha}^{a}}+2\sum_{\beta b}\frac{d%
\mathcal{L}%
_{F}}{d%
\mathcal{F}%
_{G\alpha\beta}^{a}}\left[  \overrightarrow{\kappa}_{a},G_{\beta}\right]
^{b}+\sum_{\beta i}\frac{d%
\mathcal{L}%
_{F}}{d\partial_{\alpha}O_{\beta}^{\prime i}}\left(  \left[  \widetilde
{\kappa}_{a}\right]  \left[  O^{\prime}\right]  \right)  _{\beta}^{i}=0$

But $%
\mathcal{F}%
_{G\lambda\mu}^{b}=\partial_{\lambda}G_{\mu}^{b}-\partial_{\mu}G_{\lambda}%
^{b}+G_{cd}^{b}G_{\lambda}^{c}G_{\mu}^{d} $ so

$\frac{d%
\mathcal{L}%
_{F}}{dG_{\alpha}^{a}}=\frac{\partial%
\mathcal{L}%
_{F}}{\partial G_{\alpha}^{a}}+\sum_{b\lambda\mu}\frac{d%
\mathcal{L}%
_{F}}{d%
\mathcal{F}%
_{G\lambda\mu}^{b}}\frac{\partial%
\mathcal{F}%
_{G\lambda\mu}^{b}}{\partial G_{\alpha}^{a}}=\frac{\partial%
\mathcal{L}%
_{F}}{\partial G_{\alpha}^{a}}+\sum_{bcd\lambda\mu}\frac{d%
\mathcal{L}%
_{F}}{d%
\mathcal{F}%
_{G\lambda\mu}^{b}}\left(  G_{cd}^{b}\delta_{c}^{a}\delta_{\lambda}^{\alpha
}G_{\mu}^{d}+G_{cd}^{b}G_{\lambda}^{c}\delta_{d}^{a}\delta_{\mu}^{\alpha
}\right)  $

$=\frac{\partial%
\mathcal{L}%
_{F}}{\partial G_{\alpha}^{a}}+\sum_{bcd\lambda\mu}\frac{d%
\mathcal{L}%
_{F}}{d%
\mathcal{F}%
_{G\alpha\mu}^{b}}\left(  G_{ad}^{b}G_{\mu}^{d}\right)  +\frac{d%
\mathcal{L}%
_{F}}{d%
\mathcal{F}%
_{G\lambda\alpha}^{b}}\left(  G_{cd}^{b}G_{\lambda}^{c}\right)  =\frac
{\partial%
\mathcal{L}%
_{F}}{\partial G_{\alpha}^{a}}+2\sum_{bc\beta}\frac{d%
\mathcal{L}%
_{F}}{d%
\mathcal{F}%
_{G\alpha\beta}^{b}}\left(  G_{ac}^{b}G_{\beta}^{c}\right)  $

$\frac{d%
\mathcal{L}%
_{F}}{dG_{\alpha}^{a}}=\frac{\partial%
\mathcal{L}%
_{F}}{\partial G_{\alpha}^{a}}+2\sum_{b\beta}\frac{d%
\mathcal{L}%
_{F}}{d%
\mathcal{F}%
_{G\alpha\beta}^{b}}\left[  \overrightarrow{\kappa}_{a},G_{\beta}\right]
^{b}$

Thus :%

\begin{equation}
\forall a,\alpha:0=-\frac{\partial%
\mathcal{L}%
_{F}}{\partial G_{\alpha}^{a}}+\sum_{\beta i}\frac{d%
\mathcal{L}%
_{F}}{d\partial_{\alpha}O_{\beta}^{\prime i}}\left(  \left[  \widetilde
{\kappa}_{a}\right]  \left[  O^{\prime}\right]  \right)  _{\beta}%
^{i}\label{E22}%
\end{equation}

So the second result is that $%
\mathcal{L}%
_{F}$ does not depend explicitly on G if it does not depend on the partial
derivatives $\partial_{\alpha}O_{\beta}^{\prime i}.$

Equation \ref{E17} becomes :

$\forall\alpha,a:-\frac{d%
\mathcal{L}%
_{F}}{d\operatorname{Re}\grave{A}_{\alpha}^{a}}+2\sum_{\beta}\frac{d%
\mathcal{L}%
_{F}}{d\operatorname{Re}%
\mathcal{F}%
_{\alpha\beta}^{b}}\operatorname{Re}\left[  \overrightarrow{\theta}_{a}%
,\grave{A}_{\beta}\right]  ^{b}+\frac{d%
\mathcal{L}%
_{F}}{d\operatorname{Im}%
\mathcal{F}%
_{\alpha\beta}^{b}}\operatorname{Im}\left[  \overrightarrow{\theta}_{a}%
,\grave{A}_{\beta}\right]  ^{b}=0$

But

$\operatorname{Re}%
\mathcal{F}%
_{A\lambda\mu}^{b}=\operatorname{Re}\partial_{\lambda}\grave{A}_{\mu}%
^{b}-\operatorname{Re}\partial_{\mu}\grave{A}_{\lambda}^{b}+C_{cd}^{b}\left(
\operatorname{Re}\grave{A}_{\lambda}^{c}\operatorname{Re}\grave{A}_{\mu}%
^{d}-\operatorname{Im}\grave{A}_{\lambda}^{c}\operatorname{Im}\grave{A}_{\mu
}^{d}\right)  $

$\operatorname{Im}%
\mathcal{F}%
_{A\lambda\mu}^{b}=\operatorname{Im}\partial_{\lambda}\grave{A}_{\mu}%
^{b}-\operatorname{Im}\partial_{\mu}\grave{A}_{\lambda}^{b}+C_{cd}^{b}\left(
\operatorname{Re}\grave{A}_{\lambda}^{c}\operatorname{Im}\grave{A}_{\mu}%
^{d}+\operatorname{Im}\grave{A}_{\lambda}^{c}\operatorname{Re}\grave{A}_{\mu
}^{d}\right)  $

with $\frac{d%
\mathcal{L}%
_{F}}{d\operatorname{Re}\grave{A}_{\alpha}^{a}}=\frac{\partial%
\mathcal{L}%
_{F}}{\partial\operatorname{Re}\grave{A}_{\alpha}^{a}}+\sum_{b\lambda\mu}%
\frac{d%
\mathcal{L}%
_{F}}{d\operatorname{Re}%
\mathcal{F}%
_{G\lambda\mu}^{b}}\frac{\partial\operatorname{Re}%
\mathcal{F}%
_{G\lambda\mu}^{b}}{\partial G_{\alpha}^{a}}+\frac{d%
\mathcal{L}%
_{F}}{d\operatorname{Im}%
\mathcal{F}%
_{G\lambda\mu}^{b}}\frac{\partial\operatorname{Im}%
\mathcal{F}%
_{G\lambda\mu}^{b}}{\partial G_{\alpha}^{a}}$

$\frac{d%
\mathcal{L}%
_{F}}{d\operatorname{Re}\grave{A}_{\alpha}^{a}}=\frac{\partial%
\mathcal{L}%
_{F}}{\partial\operatorname{Re}\grave{A}_{\alpha}^{a}}+2\sum_{b\beta}\frac{d%
\mathcal{L}%
_{F}}{d\operatorname{Re}%
\mathcal{F}%
_{G\alpha\beta}^{b}}C_{ac}^{b}\operatorname{Re}\grave{A}_{\beta}^{c}+\frac{d%
\mathcal{L}%
_{F}}{d\operatorname{Im}%
\mathcal{F}%
_{G\alpha\beta}^{b}}C_{ac}^{b}\operatorname{Im}\grave{A}_{\beta}^{c}$

$\frac{d%
\mathcal{L}%
_{F}}{d\operatorname{Re}\grave{A}_{\alpha}^{a}}=\frac{\partial%
\mathcal{L}%
_{F}}{\partial\operatorname{Re}\grave{A}_{\alpha}^{a}}+2\sum_{b\beta}\frac{d%
\mathcal{L}%
_{F}}{d\operatorname{Re}%
\mathcal{F}%
_{G\alpha\beta}^{b}}\operatorname{Re}\left[  \overrightarrow{\theta}%
_{a},\grave{A}_{\beta}\right]  ^{b}+\frac{d%
\mathcal{L}%
_{F}}{d\operatorname{Im}%
\mathcal{F}%
_{G\alpha\beta}^{b}}\operatorname{Im}\left[  \overrightarrow{\theta}%
_{a},\grave{A}_{\beta}\right]  ^{b}$

That is : $\forall\alpha,a:$ $\frac{\partial%
\mathcal{L}%
_{F}}{\partial\operatorname{Re}\grave{A}_{\alpha}^{a}}=0$

Similarly equation \ref{E18} gives :

$\forall\alpha,a:-\frac{d%
\mathcal{L}%
_{F}}{d\operatorname{Im}\grave{A}_{\alpha}^{a}}+2\sum_{\beta}-\frac{d%
\mathcal{L}%
_{F}}{d\operatorname{Re}%
\mathcal{F}%
_{\alpha\beta}^{b}}\operatorname{Im}\left[  \overrightarrow{\theta}_{a}%
,\grave{A}_{\beta}\right]  ^{b}+\frac{d%
\mathcal{L}%
_{F}}{d\operatorname{Im}%
\mathcal{F}%
_{\alpha\beta}^{b}}\operatorname{Re}\left[  \overrightarrow{\theta}_{a}%
,\grave{A}_{\beta}\right]  ^{b}=0$

$\frac{d%
\mathcal{L}%
_{F}}{d\operatorname{Im}\grave{A}_{\alpha}^{a}}=\frac{\partial%
\mathcal{L}%
_{F}}{\partial\operatorname{Im}\grave{A}_{\alpha}^{a}}+2\sum_{b\beta}-\frac{d%
\mathcal{L}%
_{F}}{d\operatorname{Re}%
\mathcal{F}%
_{G\alpha\beta}^{b}}\operatorname{Im}\left[  \overrightarrow{\theta}%
_{a},\grave{A}_{\beta}\right]  ^{b}+\frac{d%
\mathcal{L}%
_{F}}{d\operatorname{Im}%
\mathcal{F}%
_{G\alpha\beta}^{b}}\operatorname{Re}\left[  \overrightarrow{\theta}%
_{a},\grave{A}_{\beta}\right]  ^{b}$

$\Rightarrow\frac{\partial%
\mathcal{L}%
_{F}}{\partial\operatorname{Im}\grave{A}_{\alpha}^{a}}=0$

The third result is that \`{A}\ factorizes through the curvature form $%
\mathcal{F}%
_{A}.$%

\begin{equation}
\forall\alpha,a:\frac{\partial%
\mathcal{L}%
_{F}}{\partial\operatorname{Re}\grave{A}_{\alpha}^{a}}=0;\frac{\partial%
\mathcal{L}%
_{F}}{\partial\operatorname{Im}\grave{A}_{\alpha}^{a}}=0\label{E23a}%
\end{equation}

\paragraph{4)}

Equation \ref{E19} can be written :

$\forall a:0=\sum_{\alpha b}\left(  \frac{\partial%
\mathcal{L}%
_{F}}{\partial G_{\alpha}^{b}}+2\sum_{\beta c}\frac{d%
\mathcal{L}%
_{F}}{d%
\mathcal{F}%
_{G\alpha\beta}^{c}}\left[  \overrightarrow{\kappa}_{b},G_{\beta}\right]
^{c}\right)  \left[  \overrightarrow{\kappa}_{a},G_{\alpha}\right]  ^{b}%
+2\sum_{\alpha\beta b}\frac{d%
\mathcal{L}%
_{F}}{d%
\mathcal{F}%
_{G\alpha\beta}^{b}}\left[  \overrightarrow{\kappa}_{a},\partial_{\alpha
}G_{\beta}\right]  ^{b}$

$+\sum_{\alpha\beta i}\frac{d%
\mathcal{L}%
_{F}}{dO_{\alpha}^{\prime i}}\left(  \left[  \widetilde{\kappa}_{a}\right]
\left[  O^{\prime}\right]  \right)  _{\alpha}^{i}+\frac{d%
\mathcal{L}%
_{F}}{d\partial_{\beta}O_{\alpha}^{\prime i}}\left(  \left[  \widetilde
{\kappa}_{a}\right]  \left[  \partial_{\beta}O^{\prime}\right]  \right)
_{\alpha}^{i}$

$\forall a:0=2\sum_{\alpha\beta b}\frac{d%
\mathcal{L}%
_{F}}{d%
\mathcal{F}%
_{G\alpha\beta}^{b}}\left(  \left[  \left[  \overrightarrow{\kappa}%
_{a},G_{\alpha}\right]  ,G_{\beta}\right]  +\left[  \overrightarrow{\kappa
}_{a},\partial_{\alpha}G_{\beta}\right]  \right)  ^{b}+\sum_{\alpha b}%
\frac{\partial%
\mathcal{L}%
_{F}}{\partial G_{\alpha}^{b}}\left[  \overrightarrow{\kappa}_{a},G_{\alpha
}\right]  ^{b}$

$+\sum_{ij}\left[  \widetilde{\kappa}_{a}\right]  _{j}^{i}\sum_{\alpha}\left(
\frac{d%
\mathcal{L}%
_{F}}{dO_{\alpha}^{\prime i}}O_{\alpha}^{\prime j}+\sum_{\beta}\frac{d%
\mathcal{L}%
_{F}}{d\partial_{\beta}O_{\alpha}^{\prime i}}\partial_{\beta}O_{\alpha
}^{\prime j}\right)  $

$\sum_{\alpha\beta b}\frac{d%
\mathcal{L}%
_{F}}{d%
\mathcal{F}%
_{G\alpha\beta}^{b}}\left(  \left[  \left[  \overrightarrow{\kappa}%
_{a},G_{\alpha}\right]  ,G_{\beta}\right]  +\left[  \overrightarrow{\kappa
}_{a},\partial_{\alpha}G_{\beta}\right]  \right)  ^{b}$

$=\sum_{b,\alpha<\beta}\frac{d%
\mathcal{L}%
_{F}}{d%
\mathcal{F}%
_{G\alpha\beta}^{b}}\left(  \left[  \left[  \overrightarrow{\kappa}%
_{a},G_{\alpha}\right]  ,G_{\beta}\right]  +\left[  \overrightarrow{\kappa
}_{a},\partial_{\alpha}G_{\beta}\right]  -\left[  \left[  \overrightarrow
{\kappa}_{a},G_{\beta}\right]  ,G_{\alpha}\right]  -\left[  \overrightarrow
{\kappa}_{a},\partial_{\beta}G_{\alpha}\right]  \right)  ^{b}$

The brackets are computed in the Lie algebra.\ The Jacobi identities give :

$\left[  \left[  \overrightarrow{\kappa}_{a},G_{\alpha}\right]  ,G_{\beta
}\right]  +\left[  \left[  G_{\alpha},G_{\beta}\right]  ,\overrightarrow
{\kappa}_{a}\right]  +\left[  \left[  G_{\beta},\overrightarrow{\kappa}%
_{a}\right]  ,G_{\alpha}\right]  =0$

$\left[  \left[  \overrightarrow{\kappa}_{a},G_{\alpha}\right]  ,G_{\beta
}\right]  -\left[  \left[  \overrightarrow{\kappa}_{a},G_{\beta}\right]
,G_{\alpha}\right]  =\left[  \left[  \overrightarrow{\kappa}_{a},G_{\alpha
}\right]  ,G_{\beta}\right]  +\left[  \left[  G_{\beta},\overrightarrow
{\kappa}_{a}\right]  ,G_{\alpha}\right]  =-\left[  \left[  G_{\alpha}%
,G_{\beta}\right]  ,\overrightarrow{\kappa}_{a}\right]  $

$\left(  \left[  \left[  \overrightarrow{\kappa}_{a},G_{\alpha}\right]
,G_{\beta}\right]  +\left[  \overrightarrow{\kappa}_{a},\partial_{\alpha
}G_{\beta}\right]  -\left[  \left[  \overrightarrow{\kappa}_{a},G_{\beta
}\right]  ,G_{\alpha}\right]  -\left[  \overrightarrow{\kappa}_{a}%
,\partial_{\beta}G_{\alpha}\right]  \right)  $

$=\left[  \overrightarrow{\kappa}_{a},\left[  G_{\alpha},G_{\beta}\right]
\right]  +\left[  \overrightarrow{\kappa}_{a},\partial_{\alpha}G_{\beta
}\right]  -\left[  \overrightarrow{\kappa}_{a},\partial_{\beta}G_{\alpha
}\right]  =\left[  \overrightarrow{\kappa}_{a},%
\mathcal{F}%
_{G\alpha\beta}\right]  $

$\sum_{\alpha\beta b}\frac{d%
\mathcal{L}%
_{F}}{d%
\mathcal{F}%
_{G\alpha\beta}^{b}}\left(  \left[  \left[  \overrightarrow{\kappa}%
_{a},G_{\alpha}\right]  ,G_{\beta}\right]  +\left[  \overrightarrow{\kappa
}_{a},\partial_{\alpha}G_{\beta}\right]  \right)  ^{b}=\sum_{b,\alpha,\beta
}\frac{d%
\mathcal{L}%
_{F}}{d%
\mathcal{F}%
_{G\alpha\beta}^{b}}\left[  \overrightarrow{\kappa}_{a},%
\mathcal{F}%
_{G\alpha\beta}\right]  ^{b}$

So equation \ref{E19} reads :%

\begin{equation}
\label{E23b}
\end{equation}

$\forall a:0=\sum_{\alpha b}\left(  \frac{\partial%
\mathcal{L}%
_{F}}{\partial G_{\alpha}^{b}}\left[  \overrightarrow{\kappa}_{a},G_{\alpha
}\right]  ^{b}+2\sum_{\beta}\frac{d%
\mathcal{L}%
_{F}}{d%
\mathcal{F}%
_{G\alpha\beta}^{b}}\left[  \overrightarrow{\kappa}_{a},%
\mathcal{F}%
_{G\alpha\beta}\right]  ^{b}\right)  $

$\qquad+\sum_{ij}\left[  \widetilde{\kappa}_{a}\right]  _{j}^{i}\sum_{\alpha
}\left(  \frac{d%
\mathcal{L}%
_{F}}{dO_{\alpha}^{\prime i}}O_{\alpha}^{\prime j}+\sum_{\beta}\frac{d%
\mathcal{L}%
_{F}}{d\partial_{\beta}O_{\alpha}^{\prime i}}\partial_{\beta}O_{\alpha
}^{\prime j}\right)  $

\bigskip

Remark : with $\frac{d\det O^{\prime}}{dO_{\alpha}^{\prime i}}=O_{i}^{\alpha
}\det O^{\prime}$

$\sum_{\alpha\beta i}\frac{d%
\mathcal{L}%
_{F}}{dO_{\alpha}^{\prime i}}\left(  \left[  \widetilde{\kappa}_{a}\right]
\left[  O^{\prime}\right]  \right)  _{\alpha}^{i}=\sum_{\alpha\beta i}\left(
\det O^{\prime}\right)  \frac{dL_{F}}{dO_{\alpha}^{\prime i}}\left(  \left[
\widetilde{\kappa}_{a}\right]  \left[  O^{\prime}\right]  \right)  _{\alpha
}^{i}+L_{F}\left(  \left[  \widetilde{\kappa}_{a}\right]  \left[  O^{\prime
}\right]  \right)  _{\alpha}^{i}O_{i}^{\alpha}\det O^{\prime}$

$=\det O^{\prime}\sum_{\alpha\beta i}\frac{dL_{F}}{dO_{\alpha}^{\prime i}%
}\left(  \left[  \widetilde{\kappa}_{a}\right]  \left[  O^{\prime}\right]
\right)  _{\alpha}^{i}+L_{F}Tr\left[  \widetilde{\kappa}_{a}\right]  =\det
O^{\prime}\sum_{\alpha\beta i}\frac{dL_{F}}{dO_{\alpha}^{\prime i}}\left(
\left[  \widetilde{\kappa}_{a}\right]  \left[  O^{\prime}\right]  \right)
_{\alpha}^{i}$

(the matrices $\left[  \widetilde{\kappa}_{a}\right]  $\ are traceless) .So
the equation \ref{E23b} stands also with substituting $L_{F}$ to $%
\mathcal{L}%
_{F}.$

\paragraph{5)}

Equation \ref{E20} becomes :

$\forall a:0=2\sum_{\beta}\left(  \frac{\partial L_{F}}{\partial
\operatorname{Re}%
\mathcal{F}%
_{A\alpha\beta}^{c}}\operatorname{Re}\left[  \overrightarrow{\theta}%
_{b},\grave{A}_{\beta}\right]  ^{c}+\frac{\partial L_{F}}{\partial
\operatorname{Im}%
\mathcal{F}%
_{A\alpha\beta}^{c}}\operatorname{Im}\left[  \overrightarrow{\theta}%
_{b},\grave{A}_{\beta}\right]  ^{c}\right)  \operatorname{Re}\left[
\widehat{\theta}_{a},\grave{A}_{\alpha}\right]  ^{b}$

$+2\sum_{\beta}\left(  -\frac{\partial L_{F}}{\partial\operatorname{Re}%
\mathcal{F}%
_{A\alpha\beta}^{c}}\operatorname{Im}\left[  \overrightarrow{\theta}%
_{b},\grave{A}_{\beta}\right]  ^{c}+\frac{\partial L_{F}}{\partial
\operatorname{Im}%
\mathcal{F}%
_{A\alpha\beta}^{c}}\operatorname{Re}\left[  \overrightarrow{\theta}%
_{b},\grave{A}_{\beta}\right]  ^{c}\right)  \operatorname{Im}\left[
\overrightarrow{\theta}_{a},\grave{A}_{\alpha}\right]  ^{b}$

$-2\frac{\partial L_{F}}{\partial\operatorname{Re}%
\mathcal{F}%
_{A\alpha\beta}^{b}}\operatorname{Re}\left[  \overrightarrow{\theta}%
_{a},\partial_{\beta}\grave{A}_{\alpha}\right]  ^{b}-2\frac{\partial L_{F}%
}{\partial\operatorname{Im}%
\mathcal{F}%
_{A\alpha\beta}^{b}}\operatorname{Im}\left[  \overrightarrow{\theta}%
_{a},\partial_{\beta}\grave{A}_{\alpha}\right]  ^{b}$

that is :

$0=\sum_{\alpha\beta b}\frac{\partial L_{F}}{\partial\operatorname{Re}%
\mathcal{F}%
_{G\alpha\beta}^{b}}\operatorname{Re}\left(  \left[  \left[  \overrightarrow
{\theta}_{a},\grave{A}_{\alpha}\right]  ,\grave{A}_{\beta}\right]  -\left[
\overrightarrow{\theta}_{a},\partial_{\beta}\grave{A}_{\alpha}\right]
\right)  ^{b}$

$+\frac{\partial L_{F}}{\partial\operatorname{Im}%
\mathcal{F}%
_{B\alpha\beta}^{b}}\operatorname{Im}\left(  \left[  \left[  \overrightarrow
{\theta}_{a},\grave{A}_{\alpha}\right]  ,\grave{A}_{\beta}\right]  -\left[
\overrightarrow{\theta}_{a},\partial_{\beta}\grave{A}_{\alpha}\right]
\right)  ^{b}$

and the same calculation as previously gives :%

\begin{equation}
\forall a:0=\sum_{\alpha\beta b}\frac{\partial L_{F}}{\partial
\operatorname{Re}%
\mathcal{F}%
_{A\alpha\beta}^{b}}\operatorname{Re}\left(  \left[  \overrightarrow{\theta
}_{a},%
\mathcal{F}%
_{A\alpha\beta}\right]  \right)  ^{b}+\frac{\partial L_{F}}{\partial
\operatorname{Im}%
\mathcal{F}%
_{A\alpha\beta}^{b}}\operatorname{Im}\left(  \left[  \overrightarrow{\theta
}_{a},%
\mathcal{F}%
_{A\alpha\beta}\right]  \right)  ^{b}\label{E24}%
\end{equation}

Similar calculation with equation \ref{E20b} gives :%

\begin{equation}
\forall a:0=\sum_{\alpha\beta b}-\frac{\partial L_{F}}{\partial
\operatorname{Re}%
\mathcal{F}%
_{A\alpha\beta}^{b}}\operatorname{Im}\left(  \left[  \overrightarrow{\theta
}_{a},%
\mathcal{F}%
_{A\alpha\beta}\right]  \right)  ^{b}+\frac{\partial L_{F}}{\partial
\operatorname{Im}%
\mathcal{F}%
_{A\alpha\beta}^{b}}\operatorname{Re}\left(  \left[  \overrightarrow{\theta
}_{a},%
\mathcal{F}%
_{A\alpha\beta}\right]  \right)  ^{b}\label{E25}%
\end{equation}

\subsubsection{Lagrangian $%
\mathcal{L}%
_{M}$}

The partial derivatives of G and \`{A} appear only in the curvature forms, and
these only in the interactions fields/fields, so it is legitimate to assume
that $L_{M}$ does not depend on these variables :

$\frac{d%
\mathcal{L}%
_{M}}{d\partial_{\beta}G_{\alpha}^{a}}=0;\frac{d%
\mathcal{L}%
_{M}}{d\operatorname{Re}\partial_{\alpha}\grave{A}_{\beta}^{a}}=0;\frac{d%
\mathcal{L}%
_{M}}{d\operatorname{Im}\partial_{\alpha}\grave{A}_{\beta}^{a}}=0$

We proceed as above.

\paragraph{1)}

Terms in first order in $\partial_{\alpha}\kappa^{a}$ give :%

\begin{equation}
\label{E26}
\end{equation}

$\forall a,\alpha:0=\sum_{ij}\left(  \frac{d%
\mathcal{L}%
_{M}}{d\operatorname{Re}\partial_{\alpha}\psi^{ij}}\operatorname{Re}\left(
\left[  \kappa_{a}\right]  \left[  \psi^{\Diamond}\right]  \right)  _{j}%
^{i}+\frac{d%
\mathcal{L}%
_{M}}{d\operatorname{Im}\partial_{\alpha}\psi^{ij}}\operatorname{Im}\left(
\left[  \kappa_{a}\right]  \left[  \psi^{\Diamond}\right]  \right)  _{j}%
^{i}-\frac{d%
\mathcal{L}%
_{M}}{dG_{\alpha}^{a}}\right)  $

$\qquad+\sum_{\beta i}\frac{d%
\mathcal{L}%
_{M}}{d\partial_{\alpha}O_{\beta}^{\prime i}}\left(  \left[  \widetilde
{\kappa}_{a}\right]  \left[  O^{\prime}\right]  \right)  _{\beta}^{i}$

\bigskip

Terms in first order in $\partial_{\alpha}\theta^{a}$ give :%

\begin{equation}
\forall a,\alpha:0=\sum_{ij}\left(  \frac{d%
\mathcal{L}%
_{M}}{d\operatorname{Re}\partial_{\alpha}\psi^{ij}}\operatorname{Re}\left(
\left[  \psi^{\Diamond}\right]  \left[  \theta_{a}\right]  ^{t}\right)
_{j}^{i}+\frac{d%
\mathcal{L}%
_{M}}{d\operatorname{Im}\partial_{\alpha}\psi^{ij}}\operatorname{Im}\left(
\left[  \psi^{\Diamond}\right]  \left[  \theta_{a}\right]  ^{t}\right)
_{j}^{i}-\frac{d%
\mathcal{L}%
_{M}}{d\operatorname{Re}\grave{A}_{\alpha}^{a}}\right) \label{E27}%
\end{equation}

\begin{equation}
\forall a,\alpha:0=\sum_{ij}\left(  \frac{d%
\mathcal{L}%
_{M}}{d\operatorname{Re}\partial_{\alpha}\psi^{ij}}\operatorname{Im}\left(
\left[  \psi^{\Diamond}\right]  \left[  \theta_{a}\right]  ^{t}\right)
_{j}^{i}-\frac{d%
\mathcal{L}%
_{M}}{d\operatorname{Im}\partial_{\alpha}\psi^{ij}}\operatorname{Re}\left(
\left[  \psi^{\Diamond}\right]  \left[  \theta_{a}\right]  ^{t}\right)
_{j}^{i}+\frac{d%
\mathcal{L}%
_{M}}{d\operatorname{Im}\grave{A}_{\alpha}^{a}}\right) \label{E28}%
\end{equation}

Terms in $\kappa^{a}$ give :%

\begin{equation}
\end{equation}

$\forall a:0=\sum_{ij}\frac{d%
\mathcal{L}%
_{M}}{d\operatorname{Re}\psi^{ij}}\operatorname{Re}\left(  \left[  \kappa
_{a}\right]  \left[  \psi\right]  \right)  _{j}^{i}+\frac{d%
\mathcal{L}%
_{M}}{d\operatorname{Im}\psi^{ij}}\operatorname{Im}\left(  \left[  \kappa
_{a}\right]  \left[  \psi\right]  \right)  _{j}^{i}$

$+\sum_{\alpha ij}\frac{d%
\mathcal{L}%
_{M}}{d\operatorname{Re}\partial_{\alpha}\psi^{ij}}\operatorname{Re}\left(
\left[  \kappa_{a}\right]  \left[  \partial_{\alpha}\psi\right]  \right)
_{j}^{i}+\frac{d%
\mathcal{L}%
_{M}}{\partial\operatorname{Im}\partial_{\alpha}\psi^{ij}}\operatorname{Im}%
\left(  \left[  \kappa_{a}\right]  \left[  \partial_{\alpha}\psi\right]
\right)  _{j}^{i}$

$+\sum_{b\alpha}\frac{d%
\mathcal{L}%
_{M}}{dG_{\alpha}^{b}}\left[  \overrightarrow{\kappa}_{a},G_{\alpha}\right]
^{b}+\sum_{i\alpha}\frac{d%
\mathcal{L}%
_{M}}{dO_{\alpha}^{\prime i}}\left(  \left[  \widetilde{\kappa}_{a}\right]
\left[  O^{\prime}\right]  \right)  _{\alpha}^{i}+\frac{d%
\mathcal{L}%
_{M}}{d\partial_{\beta}O_{\alpha}^{\prime i}}\left(  \left[  \widetilde
{\kappa}_{a}\right]  \left[  O^{\prime}\right]  \right)  _{\alpha}^{i}$

$\bigskip$

Terms in $\theta^{a}$ give :%

\begin{equation}
\end{equation}

$\forall a:0=\sum_{ij}\frac{d%
\mathcal{L}%
_{M}}{d\operatorname{Re}\psi^{ij}}\operatorname{Re}\left(  \left[
\psi\right]  \left[  \theta_{a}\right]  ^{t}\right)  _{j}^{i}+\frac{d%
\mathcal{L}%
_{M}}{d\operatorname{Re}\partial_{\alpha}\psi^{ij}}\operatorname{Re}\left(
\left[  \partial_{\alpha}\psi\right]  \left[  \theta_{a}\right]  ^{t}\right)
_{j}^{i}$

$+\frac{d%
\mathcal{L}%
_{M}}{d\operatorname{Im}\psi^{ij}}\operatorname{Im}\left(  \left[
\psi\right]  \left[  \theta_{a}\right]  ^{t}\right)  _{j}^{i}+\frac{d%
\mathcal{L}%
_{M}}{d\operatorname{Im}\partial_{\alpha}\psi^{ij}}\operatorname{Im}\left(
\left[  \partial_{\alpha}\psi\right]  \left[  \theta_{a}\right]  ^{t}\right)
_{j}^{i}+$

$\sum_{b\alpha}\frac{d%
\mathcal{L}%
_{M}}{d\operatorname{Re}\grave{A}_{\alpha}^{b}}\operatorname{Re}\left[
\overrightarrow{\theta}_{a},\grave{A}_{\alpha}\right]  ^{b}+\frac{d%
\mathcal{L}%
_{M}}{d\operatorname{Im}\grave{A}_{\alpha}^{b}}\operatorname{Im}\left[
\overrightarrow{\theta}_{a},\grave{A}_{\alpha}\right]  ^{b}$

$\bigskip$%

\begin{equation}
\end{equation}

$\forall a:0=\sum_{ij}-\frac{d%
\mathcal{L}%
_{M}}{d\operatorname{Re}\psi^{ij}}\operatorname{Im}\left(  \left[
\psi\right]  \left[  \theta_{a}\right]  ^{t}\right)  _{j}^{i}-\frac{\partial%
\mathcal{L}%
_{M}}{\partial\operatorname{Re}\partial_{\alpha}\psi^{ij}}\operatorname{Im}%
\left(  \left[  \partial_{\alpha}\psi\right]  \left[  \theta_{a}\right]
^{t}\right)  _{j}^{i}$

$+\frac{\partial%
\mathcal{L}%
_{M}}{\partial\operatorname{Im}\psi^{ij}}\operatorname{Re}\left(  \left[
\psi\right]  \left[  \theta_{a}\right]  ^{t}\right)  _{j}^{i}+\frac{\partial%
\mathcal{L}%
_{M}}{\partial\operatorname{Im}\partial_{\alpha}\psi^{ij}}\operatorname{Re}%
\left(  \left[  \partial_{\alpha}\psi\right]  \left[  \theta_{a}\right]
^{t}\right)  _{j}^{i}$

$+\sum_{b\alpha}-\frac{\partial%
\mathcal{L}%
_{M}}{\partial\operatorname{Re}\grave{A}_{\alpha}^{b}}\operatorname{Im}\left[
\overrightarrow{\theta}_{a},\grave{A}_{\alpha}\right]  ^{b}+\frac{\partial%
\mathcal{L}%
_{M}}{\partial\operatorname{Im}\grave{A}_{\alpha}^{b}}\operatorname{Re}\left[
\overrightarrow{\theta}_{a},\grave{A}_{\alpha}\right]  ^{b}$

\bigskip

\paragraph{2)}

By changing the variables :

$\partial_{\alpha}\psi^{ij}\rightarrow\nabla_{\alpha}\psi-\left(  \left[
\kappa_{a}\right]  _{k}^{i}G_{\alpha}^{a}\psi^{kj}+\left[  \theta_{a}\right]
_{k}^{j}\grave{A}_{\alpha}^{a}\psi^{ik}\right)  $

$\operatorname{Re}\partial_{\alpha}\psi^{ij}\rightarrow\operatorname{Re}%
\nabla_{\alpha}\psi-\left(  G_{\alpha}^{a}\operatorname{Re}\left(  \left[
\kappa_{a}\right]  \left[  \psi\right]  \right)  _{j}^{i}+\operatorname{Re}%
\grave{A}_{\alpha}^{a}\operatorname{Re}\left(  \left[  \psi\right]  \left[
\theta_{a}\right]  ^{t}\right)  _{j}^{i}-\operatorname{Im}\grave{A}_{\alpha
}^{a}\operatorname{Im}\left(  \left[  \psi\right]  \left[  \theta_{a}\right]
^{t}\right)  _{j}^{i}\right)  $

$\operatorname{Im}\partial_{\alpha}\psi^{ij}\rightarrow\operatorname{Im}%
\nabla_{\alpha}\psi-\left(  G_{\alpha}^{a}\operatorname{Im}\left(  \left[
\kappa_{a}\right]  \left[  \psi\right]  \right)  _{j}^{i}+\operatorname{Re}%
\grave{A}_{\alpha}^{a}\operatorname{Im}\left(  \left[  \psi\right]  \left[
\theta_{a}\right]  ^{t}\right)  _{j}^{i}+\operatorname{Im}\grave{A}_{\alpha
}^{a}\operatorname{Re}\left(  \left[  \psi\right]  \left[  \theta_{a}\right]
^{t}\right)  _{j}^{i}\right)  $

and expressing $%
\mathcal{L}%
_{M}$\ \ as $%
\mathcal{L}%
\prime_{M}$ with these new arguments it comes :

$\frac{d%
\mathcal{L}%
_{M}}{d\operatorname{Re}\partial_{\alpha}\psi^{ij}}=\frac{d%
\mathcal{L}%
_{M}^{\prime}}{d\operatorname{Re}\nabla_{\alpha}\psi^{ij}};\frac{d%
\mathcal{L}%
_{M}}{d\operatorname{Im}\partial_{\alpha}\psi^{ij}}=\frac{d%
\mathcal{L}%
_{M}^{\prime}}{d\operatorname{Im}\nabla_{\alpha}\psi^{ij}}$

$\frac{d%
\mathcal{L}%
_{M}}{dG_{\alpha}^{a}}=\frac{\partial%
\mathcal{L}%
_{M}}{\partial G_{\alpha}^{a}}+\sum_{ij}\left(  \frac{d%
\mathcal{L}%
_{M}}{d\operatorname{Re}\partial_{\alpha}\psi^{ij}}\operatorname{Re}\left(
\left[  \kappa_{a}\right]  \left[  \psi^{\Diamond}\right]  \right)  _{j}%
^{i}+\frac{d%
\mathcal{L}%
_{M}}{d\operatorname{Im}\partial_{\alpha}\psi^{ij}}\operatorname{Im}\left(
\left[  \kappa_{a}\right]  \left[  \psi^{\Diamond}\right]  \right)  _{j}%
^{i}\right)  $

Equation \ref{E26} gives :%

\begin{equation}
\forall a,\alpha:\frac{\partial%
\mathcal{L}%
_{M}}{\partial G_{\alpha}^{a}}=\sum_{i\beta}\frac{d%
\mathcal{L}%
_{M}}{d\partial_{\alpha}O_{\beta}^{\prime i}}\left(  \left[  \widetilde
{\kappa}_{a}\right]  \left[  O^{\prime}\right]  \right)  _{\beta}%
^{i}\label{E32}%
\end{equation}

G factorizes through the covariant derivative if $%
\mathcal{L}%
_{M}$ does not depend on $\partial_{\beta}O_{\alpha}^{\prime i}.$

Equations \ref{E27} , \ref{E28} give :

$\frac{d%
\mathcal{L}%
_{M}}{d\operatorname{Re}\grave{A}_{\alpha}^{a}}=\sum_{ij}\frac{d%
\mathcal{L}%
_{M}}{d\operatorname{Re}\nabla_{\alpha}\psi^{ij}}\operatorname{Re}\left(
\left[  \psi^{\Diamond}\right]  \left[  \theta_{a}\right]  ^{t}\right)
_{j}^{i}+\frac{d%
\mathcal{L}%
_{M}}{d\operatorname{Im}\nabla_{\alpha}\psi^{ij}}\operatorname{Im}\left(
\left[  \psi^{\Diamond}\right]  \left[  \theta_{a}\right]  ^{t}\right)
_{j}^{i}$

$\frac{d%
\mathcal{L}%
_{M}}{d\operatorname{Im}\grave{A}_{\alpha}^{a}}=\sum_{ij}\left(  -\frac{d%
\mathcal{L}%
_{M}}{d\operatorname{Re}\nabla_{\alpha}\psi^{ij}}\operatorname{Im}\left(
\left[  \psi^{\Diamond}\right]  \left[  \theta_{a}\right]  ^{t}\right)
_{j}^{i}+\frac{d%
\mathcal{L}%
_{M}}{d\operatorname{Im}\nabla_{\alpha}\psi^{ij}}\operatorname{Re}\left(
\left[  \psi^{\Diamond}\right]  \left[  \theta_{a}\right]  ^{t}\right)
_{j}^{i}\right)  $%

\begin{equation}
\frac{\partial%
\mathcal{L}%
_{M}}{\partial\operatorname{Re}\grave{A}_{\alpha}^{a}}=0;\frac{\partial%
\mathcal{L}%
_{M}}{\partial\operatorname{Im}\grave{A}_{\alpha}^{a}}=0\label{E32a}%
\end{equation}

\`{A} factorizes through the covariant derivative.

\paragraph{3)}

With :

$\frac{d%
\mathcal{L}%
_{M}}{d\operatorname{Re}\psi^{ij}}=\frac{\partial%
\mathcal{L}%
_{M}}{\partial\operatorname{Re}\psi^{ij}}+\sum_{k}\frac{d%
\mathcal{L}%
_{M}}{d\operatorname{Re}\nabla_{\alpha}\psi^{kj}}\operatorname{Re}\left[
G_{\alpha}\right]  _{i}^{k}+\frac{d%
\mathcal{L}%
_{M}}{d\operatorname{Im}\nabla_{\alpha}\psi^{kj}}\operatorname{Im}\left[
G_{\alpha}\right]  _{i}^{k}$

$+\frac{d%
\mathcal{L}%
_{M}}{d\operatorname{Re}\nabla_{\alpha}\psi^{ik}}\operatorname{Re}\left[
\grave{A}_{\alpha}\right]  _{j}^{k}+\frac{d%
\mathcal{L}%
_{M}}{d\operatorname{Im}\nabla_{\alpha}\psi^{ik}}\operatorname{Im}\left[
\grave{A}_{\alpha}\right]  _{j}^{k}$

$\frac{d%
\mathcal{L}%
_{M}}{d\operatorname{Im}\psi^{ij}}=\frac{\partial%
\mathcal{L}%
_{M}}{\partial\operatorname{Im}\psi^{ij}}+\sum_{k}-\frac{d%
\mathcal{L}%
_{M}}{d\operatorname{Re}\nabla_{\alpha}\psi^{kj}}\operatorname{Im}\left[
G_{\alpha}\right]  _{i}^{k}+\frac{d%
\mathcal{L}%
_{M}}{d\operatorname{Im}\nabla_{\alpha}\psi^{kj}}\operatorname{Re}\left[
G_{\alpha}\right]  _{i}^{k}$

$-\frac{d%
\mathcal{L}%
_{M}}{d\operatorname{Re}\nabla_{\alpha}\psi^{ik}}\operatorname{Im}\left[
\grave{A}_{\alpha}\right]  _{j}^{k}+\frac{d%
\mathcal{L}%
_{M}}{d\operatorname{Im}\nabla_{\alpha}\psi^{ik}}\operatorname{Re}\left[
\grave{A}_{\alpha}\right]  _{j}^{k}$

Thus the 3 equations left give :

\bigskip

$\forall a:\sum_{ij}\frac{d%
\mathcal{L}%
_{M}}{d\operatorname{Re}\psi^{ij}}\operatorname{Re}\left(  \left[  \kappa
_{a}\right]  \left[  \psi\right]  \right)  _{j}^{i}+\frac{d%
\mathcal{L}%
_{M}}{d\operatorname{Im}\psi^{ij}}\operatorname{Im}\left(  \left[  \kappa
_{a}\right]  \left[  \psi\right]  \right)  _{j}^{i}$

$+\sum_{\alpha ij}\frac{d%
\mathcal{L}%
_{M}}{d\operatorname{Re}\partial_{\alpha}\psi^{ij}}\operatorname{Re}\left(
\left[  \kappa_{a}\right]  \left[  \partial_{\alpha}\psi\right]  \right)
_{j}^{i}+\frac{d%
\mathcal{L}%
_{M}}{\partial\operatorname{Im}\partial_{\alpha}\psi^{ij}}\operatorname{Im}%
\left(  \left[  \kappa_{a}\right]  \left[  \partial_{\alpha}\psi\right]
\right)  _{j}^{i}$

$+\sum_{b\alpha}\frac{d%
\mathcal{L}%
_{M}}{dB_{\alpha}^{b}}\left[  \overrightarrow{\kappa}_{a},G_{\alpha}\right]
^{b}+\sum_{i\alpha}\frac{d%
\mathcal{L}%
_{M}}{dO_{\alpha}^{\prime i}}\left(  \left[  \widetilde{\kappa}_{a}\right]
\left[  O^{\prime}\right]  \right)  _{\alpha}^{i}+\frac{d%
\mathcal{L}%
_{M}}{d\partial_{\beta}O_{\alpha}^{\prime i}}\left(  \left[  \widetilde
{\kappa}_{a}\right]  \left[  O^{\prime}\right]  \right)  _{\alpha}^{i}=0$%

\begin{equation}
\end{equation}

$\forall a:0=\sum_{ij\alpha}\frac{\partial L_{M}}{\partial\operatorname{Re}%
\psi^{ij}}\operatorname{Re}\left(  \left[  \kappa_{a}\right]  \left[
\psi\right]  \right)  _{j}^{i}+\frac{\partial L_{M}}{\partial\operatorname{Im}%
\psi^{ij}}\operatorname{Im}\left(  \left[  \kappa_{a}\right]  \left[
\psi\right]  \right)  _{j}^{i}$

$+\frac{dL_{M}}{d\operatorname{Re}\nabla_{\alpha}\psi^{ij}}\operatorname{Re}%
\left(  \left[  \kappa_{a}\right]  \left[  \nabla_{\alpha}\psi\right]
\right)  _{j}^{i}+\frac{dL_{M}}{d\operatorname{Im}\nabla_{\alpha}\psi^{ij}%
}\operatorname{Im}\left(  \left[  \kappa_{a}\right]  \left[  \nabla_{\alpha
}\psi\right]  \right)  _{j}^{i}$

$+\sum_{\alpha b}\frac{\partial L_{M}}{\partial G_{\alpha}^{b}}\left[
\overrightarrow{\kappa}_{a},G_{\alpha}\right]  ^{b}+\sum_{ij}\left[
\widetilde{\kappa}_{a}\right]  _{j}^{i}\sum_{i\alpha\beta}\frac{dL_{M}%
}{dO_{\alpha}^{\prime i}}O_{\alpha}^{\prime j}+\frac{dL_{M}}{d\partial_{\beta
}O_{\alpha}^{\prime i}}\partial_{\beta}O_{\alpha}^{\prime j}$

\bigskip%

\begin{equation}
\end{equation}

$\forall a:0=\sum_{ij\alpha}\frac{\partial L_{M}}{\partial\operatorname{Re}%
\psi^{ij}}\operatorname{Re}\left(  \left[  \psi\right]  \left[  \theta
_{a}\right]  ^{t}\right)  _{j}^{i}+\frac{\partial L_{M}}{\partial
\operatorname{Im}\psi^{ij}}\operatorname{Im}\left(  \left[  \psi\right]
\left[  \theta_{a}\right]  ^{t}\right)  _{j}^{i}$

$+\frac{dL_{M}}{d\operatorname{Re}\nabla_{\alpha}\psi^{ij}}\operatorname{Re}%
\left(  \left[  \nabla_{\alpha}\psi\right]  \left[  \theta_{a}\right]
^{t}\right)  _{j}^{i}+\frac{dL_{M}}{d\operatorname{Im}\nabla_{\alpha}\psi
^{ij}}\operatorname{Im}\left(  \left[  \nabla_{\alpha}\psi\right]  \left[
\theta_{a}\right]  ^{t}\right)  _{j}^{i}$

\bigskip%

\begin{equation}
\end{equation}

$\forall a:0=\sum_{ij\alpha}-\frac{\partial L_{M}}{\partial\operatorname{Re}%
\psi^{ij}}\operatorname{Im}\left(  \left[  \psi\right]  \left[  \theta
_{a}\right]  ^{t}\right)  _{j}^{i}+\frac{\partial L_{M}}{\partial
\operatorname{Im}\psi^{ij}}\operatorname{Re}\left(  \left[  \psi\right]
\left[  \theta_{a}\right]  ^{t}\right)  _{j}^{i}$

$-\frac{dL_{M}}{d\operatorname{Re}\nabla_{\alpha}\psi^{ij}}\operatorname{Im}%
\left(  \left[  \nabla_{\alpha}\psi\right]  \left[  \theta_{a}\right]
^{t}\right)  _{j}^{i}+\frac{dL_{M}}{d\operatorname{Im}\nabla_{\alpha}\psi
^{ij}}\operatorname{Re}\left(  \left[  \nabla_{\alpha}\psi\right]  \left[
\theta_{a}\right]  ^{t}\right)  _{j}^{i}$

\section{COVARIANCE}

\label{Covariance}

Covariance needs that the lagrangian be invariant under a change of the map of
the underlying manifold. The map $\widetilde{f}:\Omega\rightarrow\Omega$ is
not affected in such an operation.

One can proceed as in the previous section, but the group involved in
covariance is the group of general diffeomorphisms on M, it would be quite
restrictive to reduce it to some one parameter group.\ Moreover the
traditional way is simple and gives some results which will prove very useful.
So we follow the general method as presented in Lovelock [18].

\paragraph{1)}

A change of chart is a coordinates transformation $\xi\rightarrow\widehat{\xi
}=F\left(  \xi\right)  $ characterized by the jacobian $J_{\beta}^{\alpha
}\left(  m\right)  =\frac{\partial_{\alpha}}{\widehat{\partial}_{\beta}%
}=\left[  F^{\prime}\left(  m\right)  \right]  _{\beta}^{\alpha}$ whose matrix
is in GL(4).\ We denote its inverse matrix $K=J^{-1}.$ It induces the
following transformations :

$\partial_{\alpha}\rightarrow\widehat{\partial}_{\alpha}=J_{\alpha}^{\beta
}\partial_{\beta}$

$dx^{\alpha}\rightarrow\widehat{dx}^{\alpha}=K_{\beta}^{\alpha}dx^{\beta}$

$\varpi_{0}\rightarrow\widehat{\varpi}_{0}=\left(  \det K\right)  \varpi_{0}$

on vector fields on M : $X^{\alpha}\partial_{\alpha}=\widehat{X}^{\alpha
}\widehat{\partial}_{\alpha}\Rightarrow\widehat{X}^{\alpha}=K_{\beta}^{\alpha
}X^{\beta}$

on forms on M : $X_{\alpha}\partial^{\alpha}=\widehat{X}_{\alpha}%
\widehat{\partial}^{\alpha}\Rightarrow\widehat{X}_{\alpha}=J_{\alpha}^{\beta
}X_{\beta}$

$J^{1}Z$\ $\;$is identical to $JZ\otimes TV^{\ast}=L\left(  TV;JZ\right)  $ so
the transformations on $J^{1}Z$ are deduced from the transformations on TM.

$\psi$ is unchanged, all the other quantities are vectors and 1 or 2 forms :

$V^{\alpha}\rightarrow\widehat{V}^{\alpha}=K_{\beta}^{\alpha}V^{\beta}$

$O^{\prime}{}_{\alpha}^{i}\rightarrow\widehat{O}^{\prime}{}_{\alpha}%
^{i}=O_{\lambda}^{\prime i}J_{\alpha}^{\lambda}\Rightarrow\det O^{\prime
}\rightarrow\det\widehat{O}^{\prime}=\det J\det O^{\prime}$

$\partial_{\alpha}\psi^{ij}\rightarrow\widehat{\partial_{\alpha}\psi^{ij}%
}=J_{\alpha}^{\lambda}\partial_{\lambda}\psi^{ij}$

$\grave{A}_{\alpha}^{a}\rightarrow\widehat{\grave{A}}_{\alpha}^{a}=\grave
{A}_{\lambda}^{a}J_{\alpha}^{\lambda}$

$G_{\alpha}^{a}\rightarrow\widehat{G}_{\alpha}^{a}=J_{\alpha}^{\lambda
}G_{\lambda}^{a}$

$\partial_{\alpha}G_{\beta}^{a}\rightarrow\widehat{\partial_{\alpha}G_{\beta
}^{a}}=J_{\alpha}^{\lambda}J_{\beta}^{\mu}\partial_{\lambda}G_{\mu}^{a}$

$\partial_{\alpha}\grave{A}_{\beta}^{a}\rightarrow\widehat{\partial_{\alpha
}\grave{A}_{\beta}^{a}}=J_{\alpha}^{\lambda}J_{\beta}^{\mu}\partial_{\lambda
}\grave{A}_{\mu}^{a}$

$\partial_{\beta}O_{\alpha}^{\prime i}\rightarrow\widehat{\partial_{\beta
}O^{\prime}}_{\alpha}^{i}=J_{\alpha}^{\lambda}J_{\beta}^{\mu}\left(
\partial_{\lambda}O_{\mu}^{\prime i}\right)  $

and also :

$\nabla_{\alpha}\psi^{ij}\rightarrow\widehat{\nabla_{\alpha}\psi^{ij}%
}=J_{\alpha}^{\lambda}\nabla_{\lambda}\psi^{ij}$

$%
\mathcal{F}%
_{A,\alpha\beta}^{a}\rightarrow\widehat{%
\mathcal{F}%
_{A,\alpha\beta}^{a}}=J_{\alpha}^{\lambda}J_{\beta}^{\mu}%
\mathcal{F}%
_{A,\lambda\mu}^{a}$

$%
\mathcal{F}%
_{G,\alpha\beta}^{a}\rightarrow\widehat{%
\mathcal{F}%
_{G,\alpha\beta}^{a}}=J_{\alpha}^{\lambda}J_{\beta}^{\mu}%
\mathcal{F}%
_{G,\lambda\mu}^{a}$

$%
\mathcal{F}%
_{A,\alpha\beta}^{a}\rightarrow\widehat{%
\mathcal{F}%
_{A,\alpha\beta}^{a}}=J_{\alpha}^{\lambda}J_{\beta}^{\mu}%
\mathcal{F}%
_{A,\lambda\mu}^{a}$

J and K being real matrix all the formula stand for real and imaginary quantities.

$\widetilde{f}:\Omega\rightarrow\Omega$ is unchanged and the variables
$\left(  z^{i\Diamond},z_{\alpha}^{i\Diamond}\right)  =\widetilde{f}^{\ast
}\left(  z^{i},z_{\alpha}^{i}\right)  $ transform as $\left(  z^{i},z_{\alpha
}^{i}\right)  .$

\paragraph{2)}

We have for the 4-form on M :

$\left(  NL_{M}+L_{F}\right)  \varpi_{4}=\left(  NL_{M}+L_{F}\right)  \det
O^{\prime}\varpi_{0}$

$\rightarrow\left(  N\widehat{L}_{M}+\widehat{L}_{F}\right)  \det\widehat
{O}^{\prime}\widehat{\varpi}_{0}$

$=\left(  N\widehat{L}_{M}+\widehat{L}_{F}\right)  \det J\det K\det O^{\prime
}\varpi_{0}$

$=\left(  N\widehat{L}_{M}+\widehat{L}_{F}\right)  \det O^{\prime}\varpi_{0}$

$\Rightarrow\left(  NL_{M}+L_{F}\right)  =\left(  N\widehat{L}_{M}+\widehat
{L}_{F}\right)  $

$\widetilde{f}$ acts only on the first part, so we must have the two
identities :

$L_{M}\left(  z^{i},z_{\alpha}^{i}\right)  =\widehat{L}_{M}\left(  \widehat
{z}^{i},\widehat{z}_{\alpha}^{i}\right)  $

$L_{F}\left(  z^{i},z_{\alpha}^{i}\right)  =\widehat{L}_{F}\left(  \widehat
{z}^{i},\widehat{z}_{\alpha}^{i}\right)  $

$L_{M}\left(  V^{\alpha},\operatorname{Re}\psi^{ij},\operatorname{Im}\psi
^{ij},\operatorname{Re}\partial_{\alpha}\psi^{ij},\operatorname{Im}%
\partial_{\alpha}\psi^{ij},G_{\alpha}^{a},\operatorname{Re}\grave{A}_{\alpha
}^{a},\operatorname{Im}\grave{A}_{\alpha}^{a},O_{\alpha}^{i},\partial_{\beta
}O_{\alpha}^{\prime i}\right)  $

$=\widehat{L}_{M}(K_{\beta}^{\alpha}V^{\beta},\operatorname{Re}\psi
^{ij},\operatorname{Im}\psi^{ij},J_{\alpha}^{\lambda}\operatorname{Re}%
\partial_{\lambda}\psi^{ij},J_{\alpha}^{\lambda}\operatorname{Im}%
\partial_{\lambda}\psi^{ij},J_{\alpha}^{\lambda}G_{\lambda}^{a},$

$\qquad J_{\alpha}^{\lambda}\operatorname{Re}\grave{A}_{\lambda}^{a}%
,J_{\alpha}^{\lambda}\operatorname{Im}\grave{A}_{\lambda}^{a},O_{\lambda
}^{\prime i}J_{\alpha}^{\lambda},J_{\alpha}^{\lambda}J_{\beta}^{\mu}%
\partial_{\lambda}O_{\mu}^{\prime i})$

$L_{F}\left(  G_{\alpha}^{a},\partial_{\alpha}G_{\beta}^{a},\operatorname{Re}%
\grave{A}_{\alpha}^{a},\operatorname{Im}\grave{A}_{\alpha}^{a}%
,\operatorname{Re}\partial_{\alpha}\grave{A}_{\beta}^{a},\operatorname{Im}%
\partial_{\alpha}\grave{A}_{\beta}^{a},O_{i}^{\prime i},\partial_{\beta
}O_{\alpha}^{\prime i}\right)  $

$=\widehat{L}_{F}(J_{\alpha}^{\lambda}G_{\lambda}^{a},J_{\alpha}^{\lambda
}J_{\beta}^{\mu}\partial_{\lambda}G_{\mu}^{a},J_{\alpha}^{\lambda
}\operatorname{Re}\grave{A}_{\lambda}^{a},J_{\alpha}^{\lambda}%
\operatorname{Im}\grave{A}_{\lambda}^{a},J_{\alpha}^{\lambda}J_{\beta}^{\mu
}\operatorname{Re}\partial_{\lambda}\grave{A}_{\mu}^{a},$

$\qquad J_{\alpha}^{\lambda}J_{\beta}^{\mu}\operatorname{Im}\partial_{\lambda
}\grave{A}_{\mu}^{a},,O_{\lambda}^{\prime i}J_{\alpha}^{\lambda},J_{\alpha
}^{\lambda}J_{\beta}^{\mu}\partial_{\lambda}O_{\mu}^{\prime i})$

\paragraph{3)}

By differentiating with respect to $J_{\alpha}^{\beta}$ one gets the
identities :

with $\frac{d}{dJ_{\alpha}^{\beta}}=\sum_{\lambda\mu}\frac{d}{dK_{\mu
}^{\lambda}}\frac{dK_{\mu}^{\lambda}}{dJ_{\alpha}^{\beta}}=\sum_{\lambda\mu
}\left(  -K_{\beta}^{\lambda}K_{\mu}^{\alpha}\right)  \frac{d}{dK_{\mu
}^{\lambda}}$

$\forall\alpha,\beta:0=\sum_{\lambda\mu}\{\sum_{\gamma}\frac{d\widehat{L}_{M}%
}{d\widehat{V}^{\gamma}}\left(  -K_{\beta}^{\lambda}K_{\mu}^{\alpha}\right)
\frac{d\widehat{V}^{\gamma}}{dK_{\mu}^{\lambda}}$

$+\sum_{i,j}\frac{d\widehat{L}_{M}}{d\operatorname{Re}\widehat{\partial
_{\lambda}\psi^{ij}}}\frac{\partial J_{\lambda}^{\mu}\operatorname{Re}%
\partial_{\mu}\psi^{ij}}{\partial J_{\alpha}^{\beta}}+\frac{d\widehat{L}_{M}%
}{d\operatorname{Im}\widehat{\partial_{\lambda}\psi^{ij}}}\frac{\partial
J_{\lambda}^{\mu}\operatorname{Im}\partial_{\mu}\psi^{ij}}{\partial J_{\alpha
}^{\beta}}$

$+\sum_{a}\frac{d\widehat{L}_{M}}{d\widehat{G_{\lambda}^{a}}}\frac{\partial
J_{\lambda}^{\mu}G_{\mu}^{a}}{\partial J_{\alpha}^{\beta}}+\frac{d\widehat
{L}_{M}}{d\operatorname{Re}\widehat{\grave{A}_{\lambda}^{a}}}\frac{\partial
J_{\lambda}^{\mu}\operatorname{Re}\grave{A}_{\mu}^{a}}{\partial J_{\alpha
}^{\beta}}+\frac{d\widehat{L}_{M}}{d\operatorname{Im}\widehat{\grave
{A}_{\lambda}^{a}}}\frac{\partial J_{\lambda}^{\mu}\operatorname{Im}\grave
{A}_{\mu}^{a}}{\partial J_{\alpha}^{\beta}}$

$+\sum_{i}\frac{d\widehat{L}_{M}}{d\widehat{O_{\lambda}^{\prime i}}}%
\frac{\partial J_{\lambda}^{\mu}O_{\mu}^{\prime i}}{\partial J_{\alpha}%
^{\beta}}+\sum_{\xi\eta}\frac{d\widehat{L}_{M}}{d\widehat{\partial_{\lambda
}O_{\mu}^{i\prime}}}\frac{\partial J_{\lambda}^{\xi}J_{\mu}^{\eta}%
\partial_{\xi}O_{\eta}^{\prime i}}{\partial J_{\alpha}^{\beta}}\}$

$\forall\alpha,\beta:0=\sum_{\gamma}\frac{d\widehat{L}_{M}}{d\widehat
{V}^{\gamma}}\left(  -K_{\beta}^{\gamma}K_{\mu}^{\alpha}V^{\mu}\right)
+\sum_{i,j}\frac{d\widehat{L}_{M}}{d\operatorname{Re}\widehat{\partial
_{\alpha}\psi^{ij}}}\operatorname{Re}\partial_{\beta}\psi^{ij}+\frac
{d\widehat{L}_{M}}{d\operatorname{Im}\widehat{\partial_{\alpha}\psi^{ij}}%
}\operatorname{Im}\partial_{\beta}\psi^{ij}$

$+\sum_{a}\frac{d\widehat{L}_{M}}{d\widehat{G_{\alpha}^{a}}}G_{\beta}%
^{a}+\frac{d\widehat{L}_{M}}{d\operatorname{Re}\widehat{\grave{A}_{\alpha}%
^{a}}}\operatorname{Re}\grave{A}_{\beta}^{a}+\frac{d\widehat{L}_{M}%
}{d\operatorname{Im}\widehat{\grave{A}_{\alpha}^{a}}}\operatorname{Im}%
\grave{A}_{\beta}^{a}+\sum_{i}\frac{d\widehat{L}_{M}}{d\widehat{O_{\alpha
}^{\prime i}}}O_{\beta}^{\prime i}$

$+\sum_{i,\lambda,\mu}\frac{d\widehat{L}_{M}}{d\widehat{\partial_{\alpha
}O_{\lambda}^{i\prime}}}J_{\lambda}^{\mu}\partial_{\beta}O_{\mu}^{\prime
i}+\frac{d\widehat{L}_{M}}{d\widehat{\partial_{\lambda}O_{\alpha}^{i\prime}}%
}J_{\lambda}^{\mu}\partial_{\mu}O_{\beta}^{\prime i}$

and

$\forall\alpha,\beta:0=\sum_{\lambda\mu}\{\sum_{a}\frac{d\widehat{L}_{F}%
}{d\widehat{G_{\lambda}^{a}}}\frac{\partial J_{\lambda}^{\mu}G_{\mu}^{a}%
}{\partial J_{\alpha}^{\beta}}+\frac{d\widehat{L}_{F}}{d\operatorname{Re}%
\widehat{\grave{A}_{\lambda}^{a}}}\frac{\partial J_{\lambda}^{\mu
}\operatorname{Re}\grave{A}_{\mu}^{a}}{\partial J_{\alpha}^{\beta}}%
+\frac{d\widehat{L}_{F}}{d\operatorname{Im}\widehat{\grave{A}_{\lambda}^{a}}%
}\frac{\partial J_{\lambda}^{\mu}\operatorname{Im}\grave{A}_{\mu}^{a}%
}{\partial J_{\alpha}^{\beta}}$

$+\sum_{\xi\eta}\left(  \frac{d\widehat{L}_{F}}{d\operatorname{Re}%
\widehat{\partial_{\lambda}\grave{A}_{\mu}^{a}}}\frac{\partial J_{\lambda
}^{\xi}J_{\mu}^{\eta}\partial_{\xi}\operatorname{Re}\grave{A}_{\eta}^{a}%
}{\partial J_{\alpha}^{\beta}}+\frac{d\widehat{L}_{F}}{d\operatorname{Im}%
\widehat{\partial_{\lambda}\grave{A}_{\mu}^{a}}}\frac{\partial J_{\lambda
}^{\xi}J_{\mu}^{\eta}\partial_{\xi}\operatorname{Im}\grave{A}_{\eta}^{a}%
}{\partial J_{\alpha}^{\beta}}+\frac{d\widehat{L}_{F}}{d\widehat
{\partial_{\lambda}G_{\mu}^{a}}}\frac{\partial J_{\lambda}^{\xi}J_{\mu}^{\eta
}\partial_{\xi}G_{\eta}^{a}}{\partial J_{\alpha}^{\beta}}\right)  $

$+\sum_{i}\frac{d\widehat{L}_{F}}{d\widehat{O_{\lambda}^{\prime i}}}%
\frac{\partial J_{\lambda}^{\mu}O_{\mu}^{\prime i}}{\partial J_{\alpha}%
^{\beta}}+\sum_{\xi\eta}\frac{d\widehat{L}_{F}}{d\widehat{\partial_{\lambda
}O_{\mu}^{i\prime}}}\frac{\partial J_{\lambda}^{\xi}J_{\mu}^{\eta}%
\partial_{\xi}O_{\eta}^{\prime i}}{\partial J_{\alpha}^{\beta}}\}$

$\forall\alpha,\beta:0=\sum_{a}\frac{d\widehat{L}_{F}}{d\widehat{G_{\alpha
}^{a}}}G_{\beta}^{a}+\frac{d\widehat{L}_{F}}{d\operatorname{Re}\widehat
{\grave{A}_{\alpha}^{a}}}\operatorname{Re}\grave{A}_{\beta}^{a}+\frac
{d\widehat{L}_{F}}{d\operatorname{Im}\widehat{\grave{A}_{\alpha}^{a}}%
}\operatorname{Im}\grave{A}_{\beta}^{a}$

$+\sum_{a}\frac{d\widehat{L}_{F}}{d\widehat{O_{\alpha}^{\prime i}}}O_{\beta
}^{\prime i}+\sum_{i,\lambda,\mu}\frac{d\widehat{L}_{F}}{d\widehat
{\partial_{\alpha}O_{\lambda}^{i\prime}}}J_{\lambda}^{\mu}\partial_{\beta
}O_{\mu}^{\prime i}+\frac{d\widehat{L}_{F}}{d\widehat{\partial_{\lambda
}O_{\alpha}^{i\prime}}}J_{\lambda}^{\mu}\partial_{\mu}O_{\beta}^{\prime i}$

$+\sum_{\lambda\mu}(\frac{d\widehat{L}_{F}}{d\operatorname{Re}\widehat
{\partial_{\alpha}\grave{A}_{\lambda}^{a}}}J_{\lambda}^{\mu}\operatorname{Re}%
\left(  \partial_{\beta}\grave{A}_{\mu}^{a}-\partial_{\mu}\grave{A}_{\beta
}^{a}\right)  $

$+\frac{d\widehat{L}_{F}}{d\operatorname{Im}\widehat{\partial_{\alpha}%
\grave{A}_{\lambda}^{a}}}J_{\lambda}^{\mu}\operatorname{Im}\left(
\partial_{\beta}\grave{A}_{\mu}^{a}-\partial_{\mu}\grave{A}_{\beta}%
^{a}\right)  +\frac{d\widehat{L}_{F}}{d\widehat{\partial_{\alpha}G_{\lambda
}^{a}}}J_{\lambda}^{\mu}\left(  \partial_{\beta}G_{\mu}^{a}-\partial_{\mu
}G_{\beta}^{a}\right)  )$

\paragraph{4)}

By differentiating with respect to the original arguments one gets

\ $\frac{\partial L_{M}}{\partial z^{i}}=\frac{\partial\widehat{L}_{M}%
}{\partial\widehat{z}^{i}}\frac{\partial\widehat{z}^{i}}{\partial z^{i}}$
which reads :

$\frac{dL_{M}}{dV^{\alpha}}=\sum_{\beta}\frac{d\widehat{L}_{M}}{d\widehat
{V}^{\beta}}\frac{d\widehat{V}^{\beta}}{dV^{\alpha}}=\sum_{\beta}%
\frac{d\widehat{L}_{M}}{d\widehat{V}^{\beta}}K_{\alpha}^{\beta}\Leftrightarrow
\frac{d\widehat{L}_{M}}{d\widehat{V}^{\alpha}}=\sum_{\beta}\frac{dL_{M}%
}{dV^{\beta}}J_{\alpha}^{\beta}$

$\frac{dL_{M}}{d\operatorname{Re}\psi^{ij}}=\frac{d\widehat{L}_{M}}%
{d\widehat{\operatorname{Re}\psi}^{ij}};\frac{dL_{M}}{d\operatorname{Im}%
\psi^{ij}}=\frac{d\widehat{L}_{M}}{d\widehat{\operatorname{Im}\psi}^{ij}}$

$\frac{dL_{M}}{d\operatorname{Re}\partial_{\alpha}\psi^{ij}}=\sum_{\lambda
}J_{\lambda}^{\alpha}\frac{d\widehat{L}_{M}}{d\operatorname{Re}\widehat
{\partial_{\lambda}\psi}^{ij}};\frac{dL_{M}}{d\operatorname{Im}\partial
_{\alpha}\psi^{ij}}=\sum_{\lambda}J_{\lambda}^{\alpha}\frac{d\widehat{L}_{M}%
}{d\operatorname{Im}\widehat{\partial_{\lambda}\psi}^{ij}}$

$\frac{dL_{M}}{\partial G_{\alpha}^{a}}=J_{\lambda}^{\alpha}\frac{d\widehat
{L}_{M}}{d\widehat{G}_{\lambda}^{a}};\frac{dL_{F}}{\partial G_{\alpha}^{a}%
}=J_{\lambda}^{\alpha}\frac{d\widehat{L}_{F}}{d\widehat{G}_{\lambda}^{a}}$

$\frac{dL_{M}}{dO_{\alpha}^{\prime i}}=J_{\lambda}^{\alpha}\frac{d\widehat
{L}_{M}}{d\widehat{O}_{\lambda}^{\prime i}};\frac{dL_{F}}{dO_{\alpha}^{\prime
i}}=J_{\lambda}^{\alpha}\frac{d\widehat{L}_{F}}{d\widehat{O}_{\lambda}^{\prime
i}}$

$\frac{dL_{M}}{d\partial_{\beta}O_{\alpha}^{\prime i}}=J_{\lambda}^{\alpha
}J_{\mu}^{\beta}\frac{d\widehat{L}_{M}}{d\partial_{\mu}\widehat{O}_{\lambda
}^{\prime i}};\frac{dL_{F}}{d\partial_{\beta}O_{\alpha}^{\prime i}}%
=J_{\lambda}^{\alpha}J_{\mu}^{\beta}\frac{d\widehat{L}_{F}}{d\partial_{\mu
}\widehat{O}_{\lambda}^{\prime i}}$

$\frac{dL_{F}}{d\operatorname{Re}\partial_{\alpha}\grave{A}_{\beta}^{a}%
}=J_{\lambda}^{\alpha}J_{\mu}^{\beta}\frac{d\widehat{L}_{F}}%
{d\operatorname{Re}\widehat{\partial_{\lambda}\grave{A}_{\mu}^{a}}}%
;\frac{dL_{F}}{d\operatorname{Im}\partial_{\alpha}\grave{A}_{\beta}^{a}%
}=J_{\lambda}^{\alpha}J_{\mu}^{\beta}\frac{d\widehat{L}_{F}}%
{d\operatorname{Im}\widehat{\partial_{\lambda}\grave{A}_{\mu}^{a}}};$

\bigskip

\textbf{Some of these partial derivatives transform as composants of tensors},
and therefore we can introduce the corresponding tensorial objects :

$\frac{dL_{M}}{d\operatorname{Re}\psi^{ij}},\frac{dL_{M}}{d\operatorname{Im}%
\psi^{ij}}$ are functions over M

$\frac{dL_{M}}{dV^{\alpha}}$ are components of a one form field: $\sum
_{\alpha}\frac{dL_{M}}{dV^{\alpha}}dx^{\alpha}$

$\frac{dL_{M}}{d\operatorname{Re}\partial_{\alpha}\psi^{ij}},\frac{dL_{M}%
}{d\operatorname{Im}\partial_{\alpha}\psi^{ij}},\frac{dL_{M}}{dG_{\alpha}^{a}%
},\frac{dL_{F}}{dG_{\alpha}^{a}},\frac{dL_{M}}{dO_{\alpha}^{\prime i}}%
,\frac{dL_{F}}{dO_{\alpha}^{\prime i}}$ are components of vector fields :

$\sum_{\alpha}\frac{dL_{M}}{d\operatorname{Re}\partial_{\alpha}\psi^{ij}%
}\partial_{\alpha},\sum_{\alpha}\frac{dL_{M}}{d\operatorname{Im}%
\partial_{\alpha}\psi^{ij}}\partial_{\alpha},\sum_{\alpha}\frac{dL_{M}%
}{dG_{\alpha}^{a}}\partial_{\alpha},\sum_{\alpha}\frac{dL_{F}}{dG_{\alpha}%
^{a}}\partial_{\alpha},\sum_{\alpha}\frac{dL_{M}}{dO_{\alpha}^{\prime i}%
}\partial_{\alpha}$

$\frac{dL_{F}}{d\operatorname{Re}\partial_{\alpha}\grave{A}_{\beta}^{a}}%
,\frac{dL_{F}}{d\operatorname{Im}\partial_{\alpha}\grave{A}_{\beta}^{a}}%
,\frac{dL_{F}}{d\partial_{\alpha}G_{\beta}^{a}}$ are components of
anti-symmetric bi-vector fields :

$\sum_{\left\{  \alpha\beta\right\}  }\frac{dL_{F}}{d\operatorname{Re}%
\partial_{\alpha}\grave{A}_{\beta}^{a}}\partial_{\alpha}\wedge\partial_{\beta
},\sum_{\left\{  \alpha\beta\right\}  }\frac{dL_{F}}{d\operatorname{Im}%
\partial_{\alpha}\grave{A}_{\beta}^{a}}\partial_{\alpha}\wedge\partial_{\beta
},\sum_{\left\{  \alpha\beta\right\}  }\frac{dL_{F}}{d\partial_{\alpha
}G_{\beta}^{a}}\partial_{\alpha}\wedge\partial_{\beta},$

$\frac{dL_{M}}{d\partial_{\beta}O_{\alpha}^{\prime i}},\frac{dL_{F}}%
{d\partial_{\beta}O_{\alpha}^{\prime i}}$ are components of bi-tensor fields :
$\sum_{\alpha\beta}\frac{dL_{M}}{d\partial_{\alpha}O_{\beta}^{\prime i}%
}\partial_{\alpha}\otimes\partial_{\beta},\sum_{\alpha\beta}\frac{dL_{F}%
}{d\partial_{\alpha}O_{\beta}^{\prime i}}\partial_{\alpha}\otimes
\partial_{\beta}$

But the quantities such as $\frac{\partial L_{F}\det O^{\prime}}%
{\partial\operatorname{Re}%
\mathcal{F}%
_{A,\alpha\beta}^{a}}=\frac{\partial L_{F}}{\partial\operatorname{Re}%
\mathcal{F}%
_{A,\alpha\beta}^{a}}\det O^{\prime}$ are \textbf{not} tensorial.

\paragraph{5)}

By putting $J_{\beta}^{\alpha}=\delta_{\beta}^{\alpha}$ we see that the values
of the partial derivatives are unchanged. So the two previous identities give :%

\begin{equation}
\label{E37c}
\end{equation}

$\forall\alpha,\beta:0=-\frac{dL_{M}}{dV^{\beta}}V^{\alpha}+\sum_{i,j}%
\frac{dL_{M}}{d\operatorname{Re}\partial_{\alpha}\psi^{ij}}\operatorname{Re}%
\partial_{\beta}\psi^{ij}+\frac{dL_{M}}{d\operatorname{Im}\partial_{\alpha
}\psi^{ij}}\operatorname{Im}\partial_{\beta}\psi^{ij}$

$+\sum_{a}\frac{dL_{M}}{dG_{\alpha}^{a}}G_{\beta}^{a}+\frac{dL_{M}%
}{d\operatorname{Re}\grave{A}_{\alpha}^{a}}\operatorname{Re}\grave{A}_{\beta
}^{a}+\frac{dL_{M}}{d\operatorname{Im}\grave{A}_{\alpha}^{a}}\operatorname{Im}%
\grave{A}_{\beta}^{a}$

$+\sum_{i}\left(  \frac{dL_{M}}{dO_{\alpha}^{\prime i}}O_{\beta}^{\prime
i}+\sum_{\lambda}\frac{dL_{M}}{d\partial_{\lambda}O_{\alpha}^{i\prime}%
}\partial_{\lambda}O_{\beta}^{\prime i}+\frac{dL_{M}}{d\partial_{\alpha
}O_{\lambda}^{i\prime}}\partial_{\beta}O_{\lambda}^{\prime i}\right)  $

\bigskip%

\begin{equation}
\label{E37d}
\end{equation}

$\forall\alpha,\beta:0=\sum_{a}\frac{dL_{F}}{dG_{\alpha}^{a}}G_{\beta}%
^{a}+\frac{dL_{F}}{d\operatorname{Re}\grave{A}_{\alpha}^{a}}\operatorname{Re}%
\grave{A}_{\beta}^{a}+\frac{dL_{F}}{d\operatorname{Im}\grave{A}_{\alpha}^{a}%
}\operatorname{Im}\grave{A}_{\beta}^{a}$

$+\sum_{a,\lambda}(\frac{dL_{F}}{d\operatorname{Re}\partial_{\alpha}\grave
{A}_{\lambda}^{a}}\operatorname{Re}\left(  \partial_{\beta}\grave{A}_{\lambda
}^{a}-\partial_{\lambda}\grave{A}_{\beta}^{a}\right)  +\frac{dL_{F}%
}{d\operatorname{Im}\partial_{\alpha}\grave{A}_{\lambda}^{a}}\operatorname{Im}%
\left(  \partial_{\beta}\grave{A}_{\lambda}^{a}-\partial_{\lambda}\grave
{A}_{\beta}^{a}\right)  $

$+\frac{dL_{F}}{d\partial_{\alpha}G_{\lambda}^{a}}\left(  \partial_{\beta
}G_{\lambda}^{a}-\partial_{\lambda}G_{\beta}^{a}\right)  )$

$+\sum_{i}\left(  \frac{dL_{F}}{dO_{\alpha}^{\prime i}}O_{\beta}^{\prime
i}+\sum_{\lambda}\frac{dL_{F}}{d\partial_{\lambda}O_{\alpha}^{i\prime}%
}\partial_{\lambda}O_{\beta}^{\prime i}+\frac{dL_{F}}{d\partial_{\alpha
}O_{\lambda}^{i\prime}}\partial_{\beta}O_{\lambda}^{\prime i}\right)  $

\bigskip

\paragraph{6)}

By proceeding to the same calculations with the lagrangians :

$L_{M}\left(  \operatorname{Re}\psi^{ij},\operatorname{Im}\psi^{ij}%
,\operatorname{Re}\nabla_{\alpha}\psi^{ij},\operatorname{Im}\nabla_{\alpha
}\psi^{ij},G_{\alpha}^{a},O_{\alpha}^{i},\partial_{\beta}O_{\alpha}^{\prime
i}\right)  ,$

$L_{F}\left(  \operatorname{Re}%
\mathcal{F}%
_{A,\alpha\beta}^{a},\operatorname{Im}%
\mathcal{F}%
_{A,\alpha\beta}^{a},%
\mathcal{F}%
_{G,\alpha\beta}^{a},O_{i}^{\prime i},\partial_{\beta}O_{\alpha}^{\prime
i}\right)  $

one can check that the following quantities are tensorial :

\bigskip

$\frac{\partial L_{M}}{\partial\operatorname{Re}\nabla_{\alpha}\psi^{ij}%
},\frac{\partial L_{M}}{\partial\operatorname{Im}\nabla_{\alpha}\psi^{ij}}$
are components of vector fields : $\sum_{\alpha}\frac{\partial L_{M}}%
{\partial\operatorname{Re}\nabla_{\alpha}\psi^{ij}}\partial_{\alpha}%
,\sum_{\alpha}\frac{\partial L_{M}}{\partial\operatorname{Im}\nabla_{\alpha
}\psi^{ij}}\partial_{\alpha}$

$\frac{\partial L_{F}}{\partial\operatorname{Re}%
\mathcal{F}%
_{A,\alpha\beta}^{a}},\frac{\partial L_{F}}{\partial\operatorname{Im}%
\mathcal{F}%
_{A,\alpha\beta}^{a}},\frac{\partial L_{F}}{\partial%
\mathcal{F}%
_{G,\alpha\beta}^{a}}$ are components of anti-symmetric bi-vector fields :

$\sum_{\alpha\beta}\frac{\partial L_{F}}{\partial\operatorname{Re}%
\mathcal{F}%
_{A,\alpha\beta}^{a}}\partial_{\alpha}\wedge\partial_{\beta},\sum_{\alpha
\beta}\frac{\partial L_{F}}{\partial\operatorname{Im}%
\mathcal{F}%
_{A,\alpha\beta}^{a}}\partial_{\alpha}\wedge\partial_{\beta},\sum_{\alpha
\beta}\frac{\partial L_{F}}{\partial%
\mathcal{F}%
_{G,\alpha\beta}^{a}}\partial_{\alpha}\wedge\partial_{\beta}$

\paragraph{7)}

With the identities from the previous section the two equations \ref{E37c}%
,\ref{E37d} become :%

\begin{equation}
\label{E38a}
\end{equation}

$\forall\alpha,\beta:0=-\frac{dL_{M}}{dV^{\beta}}V^{\alpha}+\sum_{i,j}\left(
\frac{dL_{M}}{d\operatorname{Re}\nabla_{\alpha}\psi^{ij}}\operatorname{Re}%
\nabla_{\beta}\psi^{ij}+\frac{dL_{M}}{d\operatorname{Im}\nabla_{\alpha}%
\psi^{ij}}\operatorname{Im}\nabla_{\beta}\psi^{ij}\right)  $

$+\sum_{a}\frac{\partial L_{M}}{\partial G_{\alpha}^{a}}G_{\beta}^{a}+\sum
_{i}\frac{dL_{M}}{dO_{\alpha}^{\prime i}}O_{\beta}^{\prime i}+\sum_{i\lambda
}\frac{dL_{M}}{d\partial_{\lambda}O_{\alpha}^{\prime i}}\left(  \partial
_{\lambda}O_{\beta}^{\prime i}\right)  +\frac{dL_{M}}{d\partial_{\alpha
}O_{\lambda}^{\prime i}}\left(  \partial_{\beta}O_{\lambda}^{\prime i}\right)
$

\bigskip%

\begin{equation}
\label{E39a}
\end{equation}

$\forall\alpha,\beta:0=2\sum_{a\lambda}\left(  \frac{dL_{F}}%
{d\operatorname{Re}%
\mathcal{F}%
_{A,\alpha\lambda}^{a}}\operatorname{Re}%
\mathcal{F}%
_{A,\beta\lambda}^{a}+\frac{dL_{F}}{d\operatorname{Im}%
\mathcal{F}%
_{A,\alpha\lambda}^{a}}\operatorname{Im}%
\mathcal{F}%
_{A,\beta\lambda}^{a}+\frac{dL_{F}}{d%
\mathcal{F}%
_{G,\alpha\lambda}^{a}}%
\mathcal{F}%
_{G,\beta\lambda}^{a}\right)  $

$+\sum_{a}\frac{\partial L_{F}}{\partial G_{\alpha}^{a}}G_{\beta}^{a}+\sum
_{i}\frac{dL_{F}}{dO_{\alpha}^{\prime i}}O_{\beta}^{\prime i}+\sum_{i\lambda
}\frac{dL_{F}}{d\partial_{\lambda}O_{\alpha}^{i\prime}}\left(  \partial
_{\lambda}O_{\beta}^{\prime i}\right)  +\frac{dL_{F}}{d\partial_{\alpha
}O_{\lambda}^{i\prime}}\left(  \partial_{\beta}O_{\lambda}^{\prime i}\right)
$

\newpage

\part{LAGRANGE\ EQUATIONS}

\label{Lagrange equations principles}

\section{PRINCIPLES}

Functions for which the action is stationary are given by the standard
variational calculus, in the form of the Euler-Lagrange equations.\ We will
review them below.\ But our problem is more complicated due to the
$\widetilde{f}$\ map, which needs the use of functional derivatives techniques.

\subsection{Variational calculus}

\paragraph{1)}

We have seen previously how a projectable vector field on JZ is the generator
of a one parameter group $\Phi_{\tau}^{Y}$\ which can be extended to $J^{1}%
Z$\ .\ The group $\Phi_{\tau}^{J^{1}Y}$\ induces a deformation of a section
$j^{1}Z$ on $J^{1}Z$\ : $j^{1}Z\rightarrow\Phi_{\tau}^{J^{1}Y}\left(
j^{1}Z\right)  $ and of the value of the lagrangian and action :

$j^{1}Z^{\ast}%
\mathcal{L}%
\varpi_{0}\rightarrow\left(  \Phi_{\tau}^{J^{1}Y}\left(  j^{1}Z\right)
\right)  ^{\ast}%
\mathcal{L}%
\varpi_{0}$

$S\left(  Z\right)  =\int_{\Omega}\left(  j^{1}Z\right)  ^{\ast}%
\mathcal{L}%
\varpi_{0}$ $\rightarrow S\left(  \tau,Y\right)  =\int_{\Omega}\left(
\Phi_{\tau}^{J^{1}Y}\left(  j^{1}Z\right)  \right)  ^{\ast}%
\mathcal{L}%
\varpi_{0} $

For Y fixed $S\left(  \tau,Y\right)  $ is a function of the scalar $\tau.$ By
derivation with respect to $\tau$ in $\tau=0$ one gets the variational
derivative of
$\mathcal{L}$%
\ along Y :

$\frac{d}{d\tau}S(\tau,Y)|_{\tau=0}=\int_{\Omega}\frac{d}{d\tau}\left(
\left(  \Phi_{\tau}^{J^{1}Y}\left(  j^{1}Z\right)  \right)  ^{\ast}%
\mathcal{L}%
\varpi_{0}\right)  |_{\tau=0}=\int_{\Omega}\left(  j^{1}Z\right)  ^{\ast
}\pounds _{j^{1}Y}%
\mathcal{L}%
\varpi_{0}$ where $\pounds _{j^{1}Y}%
\mathcal{L}%
\varpi_{0}$\ is the Lie derivative of $%
\mathcal{L}%
\varpi_{0}$\ along the field $J^{1}Y.$

The solutions of the variational problem are taken as the sections Z such that
S is\ stationary for any projectable vector field Y with support included in
$\Omega$, that is : $\frac{d}{d\tau}S(\tau,Y)|_{\tau=0}=0$ or $\int_{\Omega
}\left(  j^{1}Z\right)  ^{\ast}\pounds _{j^{1}Y}%
\mathcal{L}%
\varpi_{0}$ $=0.$

\paragraph{2)}

The first variation formula of variational calculus gives the value of this
Lie derivative (Giachetta [5] p.75, Krupka [15]) :%

\begin{equation}
\pounds _{j^{1}Y}%
\mathcal{L}%
\varpi_{0}=\sum_{i}Y^{i}E_{i}\varpi_{0}+hd\left(  i_{j^{1}Y}\Theta_{L}\right)
\label{E40}%
\end{equation}

where

$\Theta_{L}=$\ \ $%
\mathcal{L}%
\varpi_{0}+\frac{\partial%
\mathcal{L}%
}{\partial z_{\alpha}^{i}}\varpi^{i}\wedge i_{\partial_{\alpha}}\varpi_{0} $
is the Poincar\'{e}-Cartan Lepage equivalent of the lagrangian, with
$\varpi^{i}=\left(  dz^{i}-z_{\alpha}^{i}d\xi^{\alpha}\right)  $

$E_{i}=\frac{\partial%
\mathcal{L}%
}{\partial z^{i}}-\sum_{\alpha=1}^{n}\frac{d}{d\xi^{\alpha}}\left(
\frac{\partial%
\mathcal{L}%
}{\partial z_{\alpha}^{i}}\right)  ;E=E_{i}\varpi^{i}\wedge\varpi_{0}$ is the
Euler-Lagrange form

h is the horizontalization, an exterior product preserving morphism :

$h:\Lambda_{q}(J^{r}Z;R)\rightarrow\Lambda_{q}^{H}(J^{r+1}Z;R)$ q%
$>$%
0,r%
$>$%
0

such that for a section $Z\subset\Lambda_{0}JZ$ and $\rho\in$\ $\Lambda
_{q}(J^{r}Z;R)$:

$\left(  j^{r}s\right)  ^{\ast}\rho=\left(  j^{r+1}s\right)  ^{\ast}h\rho$

\paragraph{3)}

Thus the variational derivative computes as :

$\frac{d}{d\tau}S(\tau,Y)|_{\tau=0}=\int_{\Omega}\left(  j^{1}Z\right)
^{\ast}\pounds _{j^{1}Y}%
\mathcal{L}%
\varpi_{0}=\int_{\Omega}\left(  j^{1}Z\right)  ^{\ast}\left(  Y^{i}E_{i}%
\varpi_{0}+hd\left(  i_{j^{1}Y}\Theta_{L}\right)  \right)  $

But $\int_{\Omega}\left(  j^{1}Z\right)  ^{\ast}hd\left(  i_{j^{1}Y}\Theta
_{L}\right)  =\int_{U}Z^{\ast}di_{j^{1}Y}\Theta_{L}=\int_{\partial U}Z^{\ast
}i_{j^{1}Y}\Theta_{L}=0$ with the Stockes theorem if Y is compactly supported

So for the solutions:

$\forall Y^{i}:\frac{d}{d\tau}S(\tau,Y)|_{\tau=0}=\int_{\Omega}\left(
j^{1}Z\right)  ^{\ast}Y^{i}E_{i}\varpi_{0}=0\Rightarrow E_{i}=\frac{\partial%
\mathcal{L}%
}{\partial z^{i}}-\sum_{\alpha=1}^{n}\frac{d}{d\xi^{\alpha}}\left(
\frac{\partial%
\mathcal{L}%
}{\partial z_{\alpha}^{i}}\right)  =0$

and we have the Euler-Lagrange equations:

$\frac{\partial%
\mathcal{L}%
}{\partial z^{i}}\mathbf{-}\sum_{\alpha=1}^{n}\frac{d}{d\xi^{\alpha}}\left(
\frac{\partial%
\mathcal{L}%
}{\partial z_{\alpha}^{i}}\right)  \mathbf{=0}$

\paragraph{4)}

This classical method can be implemented for the "field part" $%
\mathcal{L}%
_{F}$ of our lagrangian, but in the "matter part" $%
\mathcal{L}%
_{M}$ the map $\widetilde{f}$ does not fit well. So we tackle the problem
through the method of functional derivatives.

\subsection{Functional derivatives}

\paragraph{1)}

Let A be a set of scalar valued functions endowed with a Banach vector space
structure. A functional is a continuous operator : $S:A\rightarrow%
\mathbb{C}
.$ The general theory of derivatives can be fully implemented.\ The functional
derivative of S at f is a linear map :$\frac{dS}{df}\left(  f\right)
:A\rightarrow%
\mathbb{C}
$ such that for any infinitesimal $\delta f:S(f+\delta f)-S(f)=\frac{dS}%
{df}\left(  f\right)  \left(  \delta f\right)  +o\left(  \delta f\right)
\left\Vert \delta f\right\Vert $

It is computed easily by $\frac{\delta S}{\delta f}\left(  \delta f\right)
=\frac{d}{d\tau}S(f+\tau\delta f)|_{\tau=0}$ where $\delta f$\ is a compactly
supported function. $\frac{\delta S}{\delta f}$ is a distribution if it is
continuous. We have the usual theorems and properties of derivatives with some
caution because the product of two distributions is not defined. What matters
is not the domain where the function are defined, but the codomain, where they
take their value : the theory is legitimate as long as A is a Banach vector
space, such that functions can be added together.

\paragraph{2)}

If both F and f belong to A, the chain rule gives : $\frac{dS}{df}\left(
F\circ f\right)  =\frac{dS}{dF}\left(  F\circ f\right)  \frac{dF}{df}$ .\ As
the product of the function $\frac{dF}{df}$ by the distribution $\frac{\delta
S}{\delta F}\left(  F\right)  $ is well defined : $\frac{\delta S}{\delta
f}\left(  F\circ f\right)  \left(  \delta f\right)  =\frac{dF}{df}\frac{\delta
S}{\delta F}\left(  F\circ f\right)  \left(  \delta f\right)  .$ One can
compute the derivatives simultaneously with respect to F and f. Let us
consider the function : $\phi\left(  \tau_{1},\tau_{2}\right)  =S\left(
\left(  F+\tau_{1}\delta F\right)  \circ\left(  f+\tau_{2}\delta f\right)
\right)  $ and its partial derivatives with respect to $\tau_{1},\tau_{2}$ in
$\tau_{1}=0,\tau_{2}=0.$ That is :

$\delta S=S\left(  \left(  F+\tau_{1}\delta F\right)  \circ\left(  f+\tau
_{2}\delta f\right)  \right)  -S\left(  F\circ f\right)  =\phi\left(  \tau
_{1},\tau_{2}\right)  -\phi\left(  0,0\right)  $

$=\left(  \frac{\partial\phi}{\partial\tau_{1}}|_{\tau_{1},\tau_{2}=0}\right)
\tau_{1}+\left(  \frac{\partial\phi}{\partial\tau_{2}}|_{\tau_{1},\tau_{2}%
=0}\right)  \tau_{2}+o\left(  \tau_{1},\tau_{2}\right)  \left(  \left\vert
\tau_{1}\right\vert +\left\vert \tau_{2}\right\vert \right)  $

$\frac{\partial\phi}{\partial\tau_{1}}|_{\tau_{1}=0}=\frac{\delta S}{\delta
F}\left(  F\circ\left(  f+\tau_{2}\delta f\right)  \right)  \left(  \delta
F\circ\left(  f+\tau_{2}\delta f\right)  \right)  $

$\Rightarrow\frac{\partial\phi}{\partial\tau_{1}}|_{\tau_{1},\tau_{2}=0}%
=\frac{\delta S}{\delta F}\left(  F\circ f\right)  \delta\left(  F\circ
f\right)  $

$\frac{\partial\phi}{\partial\tau_{2}}|_{\tau_{2}=0}=\frac{dF}{df}\frac{\delta
S}{\delta F}\left(  \left(  F+\tau_{1}\delta F\right)  \circ f\right)  \delta
f$

$\Rightarrow\frac{\partial\phi}{\partial\tau_{2}}|_{\tau_{1},\tau_{2}=0}%
=\frac{dF}{df}\frac{\delta S}{\delta F}\left(  F\circ f\right)  \delta f $

Thus : $\delta S=\frac{\delta S}{\delta F}\left(  F\circ f\right)  \tau
_{1}\delta\left(  F\circ f\right)  +\frac{dF}{df}\frac{\delta S}{\delta
F}\left(  F\circ f\right)  \tau_{2}\left(  \delta f\right)  +o\left(  \tau
_{1},\tau_{2}\right)  \left(  \left\vert \tau_{1}\right\vert +\left\vert
\tau_{2}\right\vert \right)  $

The functional derivative of S with respect to $F\circ f$ is : $\frac{\delta
S}{\delta F}=\frac{\delta S}{\delta F}\left(  F\circ f\right)  ;$

and the functional derivative of S with respect to f is :$\ \frac{\delta
S}{\delta f}=\frac{dF}{df}\frac{\delta S}{\delta F}\left(  F\circ f\right)  $

Now if $S\left(  F\circ f,f\right)  $ we have by the same calculation :

$\delta S=S\left(  \left(  F+\tau_{1}\delta F\right)  \circ\left(  f+\tau
_{2}\delta f\right)  ,f+\tau_{2}\delta f\right)  -S\left(  F\circ f\right)  $

$\frac{\partial\phi}{\partial\tau_{1}}|_{\tau_{1}=0}=\frac{\delta S}{\delta
F}\left(  F\circ\left(  f+\tau_{2}\delta f\right)  \right)  \left(  \delta
F\circ\left(  f+\tau_{2}\delta f\right)  ,f+\tau_{2}\delta f\right)  $

$\Rightarrow\frac{\partial\phi}{\partial\tau_{1}}|_{\tau_{1},\tau_{2}=0}%
=\frac{\delta S}{\delta F}\left(  \left(  F\circ f\right)  ,f\right)
\delta\left(  F\circ f\right)  $

$\frac{\partial\phi}{\partial\tau_{2}}|_{\tau_{2}=0}=\frac{dF}{df}\frac{\delta
S}{\delta F}\left(  \left(  F+\tau_{1}\delta F\right)  \circ f,f\right)
\delta f+\frac{\delta S}{\delta f}\left(  \left(  F+\tau_{1}\delta F\right)
\circ f,f\right)  \delta f$

$\Rightarrow\frac{\partial\phi}{\partial\tau_{2}}|_{\tau_{1},\tau_{2}=0}%
=\frac{dF}{df}\frac{\delta S}{\delta F}\left(  F\circ f,f\right)  \delta
f+\frac{\delta S}{\delta f}\left(  F\circ f,f\right)  \delta f$

The functional derivative of S with respect to $F\circ f$ is : $\frac{\delta
S}{\delta F}=\frac{\delta S}{\delta F}\left(  F\circ f\right)  ;$ and the
functional derivative of S with respect to f is :$\ \frac{dF}{df}\frac{\delta
S}{\delta F}\left(  F\circ f,f\right)  +\frac{\delta S}{\delta f}\left(
F\circ f,f\right)  $ where the last term is the functional derivative for f as
a stand alone function.

\paragraph{3)}

Let us come back to our problem.\ JZ is a vector bundle, a section is valued
in $%
\mathbb{R}
^{16m+36},$ $J^{1}Z$ is also a vector space and we will assume that we
restrict ourselves to some set $H\subset\Lambda_{0}J^{1}Z$\ of bounded,
differentiable functions, endowed with a metric so that H is some Banach
vector space.\ The functional derivative in $j^{1}Z$ of a functional :
$S:H\rightarrow%
\mathbb{R}
$ is a continuous linear map : $U\in L(H;%
\mathbb{R}
)$ such that for any infinitesimal variation $j^{1}\delta Z=\left\{  \left(
\delta z^{i},\delta z_{\alpha}^{i}\right)  _{i,\alpha}\right\}  \in J^{1}Z$ :

$S(z^{i}+\delta z^{i},z_{\alpha}^{i}+\delta z_{\alpha}^{i})-S(z^{i},z_{\alpha
}^{i})=U\left(  \delta z^{i},\delta z_{\alpha}^{i}\right)  +o\left(  \delta
z^{i},\delta z_{\alpha}^{i}\right)  \left\Vert \left(  \delta z^{i},\delta
z_{\alpha}^{i}\right)  \right\Vert $

For a section $j^{1}Z\in H$ and the variation $\delta j^{1}Z=\left(  \delta
z^{i},\partial_{\alpha}\delta z^{i}\right)  =j^{1}\delta Z$ it reads :

$S(z^{i}+\delta z^{i},\partial_{\alpha}z^{i}+\partial_{\alpha}\delta
z^{i})-S(z^{i},\partial_{\alpha}z^{i})=U\left(  \delta z^{i},\partial_{\alpha
}\delta z^{i}\right)  +o\left(  \delta z^{i},\partial_{\alpha}\delta
z^{i}\right)  \left\Vert \left(  \delta z^{i},\partial_{\alpha}\delta
z^{i}\right)  \right\Vert $

that is : $S(j^{1}\delta Z)-S(j^{1}Z)=U\left(  j^{1}\delta Z\right)  +o\left(
j^{1}\delta Z\right)  \left\Vert \left(  j^{1}\delta Z\right)  \right\Vert $

If $\delta j^{1}Z$\ is defined by a projectable vector field in each point m,
we have for $\tau\in%
\mathbb{R}
:$

$j_{m}^{1}\delta Z=\Phi_{\tau}^{j^{1}Y}\left(  j_{m}^{1}Z\right)  -j_{m}%
^{1}Z=\tau\left(  \frac{d}{d\tau}\Phi_{\tau}^{j^{1}Y}\left(  j_{m}%
^{1}Z\right)  |_{\tau=0}\right)  +\tau o\left(  \tau\right)  =\tau
j^{1}Y\left(  j_{m}^{1}Z\right)  +\tau o\left(  \tau\right)  $

Therefore the functional derivative must meet the condition :

$S(\Phi_{\tau}^{j^{1}Y}\left(  j_{m}^{1}Z\right)  )-S(j_{m}^{1}Z)=\tau
Uj^{1}Y\left(  j_{m}^{1}Z\right)  +\tau o\left(  \tau\right)  $

The variational derivative is $\frac{d}{d\tau}S(\tau,Y)|_{\tau=0}=\int
_{\Omega}\left(  j^{1}Z\right)  ^{\ast}Y^{i}E_{i}\varpi_{0}$ thus it can be
associated with the functional derivative :

$\frac{\delta S}{\delta z^{i}}\left(  j^{1}Z\right)  =\left(  j^{1}Z\right)
^{\ast}E_{i}$ such that :

$S(\Phi_{\tau}^{j^{1}Y}\left(  j_{m}^{1}Z\right)  )-S(j_{m}^{1}Z)=\tau
\int_{\Omega}\left(  j^{1}Z\right)  ^{\ast}Y^{i}E_{i}\varpi_{0}+\tau o\left(
\tau\right)  $

The variational derivative is nothing but the value of the functional
derivative along a projectable vector field. Furthermore :

$\sum_{i}\frac{\delta S}{\delta z^{i}}\left(  j^{1}Z\right)  Y^{i}\varpi
_{0}=\left(  j^{1}Z\right)  ^{\ast}\pounds _{j^{1}Y}%
\mathcal{L}%
\varpi_{0}$

\paragraph{4)}

To implement this method we need to come back to functions with codomain $%
\mathbb{R}
^{n}.$ The trivialization of $\Omega:\varphi_{\Omega}:S(0)\times\left[
0,T\right]  \rightarrow\Omega$ stems from a chart : $\varphi_{U}:U_{0}%
\times\left[  0,T\right]  \rightarrow\Omega$ where $U_{0}$ is an open set in $%
\mathbb{R}
^{3}$. Let us define the map :

$f:U_{0}\times\left[  0,T\right]  \rightarrow U_{0}\times\left[  0,T\right]
::f=\varphi_{U}^{-1}\circ\widetilde{f}\circ\varphi_{U}$

$\alpha=0,1,23:\eta^{\alpha}=f^{\alpha}\left(  \xi^{1},\xi^{2},\xi
^{3},t\right)  $ are the coordinates at t of a particle whith coordinates
$\left(  \xi^{1},\xi^{2},\xi^{3},0\right)  $\ at t=0 and $V^{\alpha}%
=\frac{\partial f^{\alpha}}{\partial t}$

The matter part of the action is the functionnal (in putting V apart)

$S_{M}=\int_{\Omega}N\left(  m\right)  L_{M}\left(  V,z^{i}\circ\widetilde
{f},z_{\alpha}^{i}\circ\widetilde{f}\right)  \varpi_{4}=$

$\int_{U_{0}\times\left[  0,T\right]  }\left(  N\circ\varphi_{U}\right)
L_{M}\left(  \frac{\partial f^{\alpha}}{\partial t},z^{i}\widetilde{f}%
\varphi_{U},z_{\alpha}^{i}\widetilde{f}\varphi_{U}\right)  \left\vert \det
O^{\prime}\left(  \widetilde{f}\varphi_{U}\right)  \right\vert \left(
\xi\right)  d\xi^{1}\otimes d\xi^{2}\otimes d\xi^{3}\otimes dt$

$S_{M}=\int_{U_{0}\times\left[  0,T\right]  }\left(  N\circ\varphi_{U}\right)
L\left(  \frac{\partial f^{\alpha}}{\partial t},z^{i}\varphi_{U}f,z_{\alpha
}^{i}\varphi_{U}f\right)  \left\vert \det O^{\prime}\left(  \varphi
_{U}f\right)  \right\vert \left(  \xi\right)  d\xi^{1}\otimes d\xi^{2}\otimes
d\xi^{3}\otimes dt$

N is fixed and we have a functional of the composite function $j^{1}%
Z^{\diamond}=$ $j^{1}Z\circ f$ of the vector valued functions : $f\left(
\xi\right)  $ and $j^{1}Z=\left(  z^{i}\left(  \xi\right)  ,\frac{dz^{i}}%
{d\xi^{\alpha}}\left(  \xi\right)  \right)  $.

\paragraph{5)}

For the field part of the action, which does not depend on f, the functional
derivative is :

$\frac{\delta S_{F}}{\delta z^{i}}\left(  j^{1}Z\right)  =\frac{\partial%
\mathcal{L}%
_{F}}{\partial z^{i}}\left(  j^{1}Z\right)  -\sum_{\alpha}\frac{d}%
{d\xi^{\alpha}}\left(  \frac{\partial%
\mathcal{L}%
_{F}}{\partial z_{\alpha}^{i}}\left(  j^{1}Z\right)  \right)  =\frac{\partial%
\mathcal{L}%
_{F}}{\partial z^{i}}-\sum_{\alpha}\partial_{\alpha}\left(  \frac{\partial%
\mathcal{L}%
_{F}}{\partial z_{\alpha}^{i}}\left(  j^{1}Z\right)  \right)  $

\paragraph{6)}

For the matter part the functional derivatives are :

a) For i%
$>$%
0 : $\frac{\delta S_{M}}{\delta z^{i}}\left(  j^{1}Z^{\diamond}\right)
=\frac{\delta S_{M}}{\delta z^{i}}\left(  j^{1}Z^{\diamond}\right)
=E_{i}\left(  j^{1}Z^{\Diamond}\right)  =\frac{\partial%
\mathcal{L}%
_{M}}{\partial z^{i}}\left(  j^{1}Z^{\Diamond}\right)  -\sum_{\alpha}\frac
{d}{d\xi^{\alpha}}\left(  \frac{\partial%
\mathcal{L}%
_{M}}{\partial z_{\alpha}^{i}}\left(  j^{1}Z^{\Diamond}\right)  \right)  $

The quantities $\frac{d}{d\xi^{\alpha}}\left(  \frac{\partial%
\mathcal{L}%
_{M}}{\partial z_{\alpha}^{i}}\left(  j^{1}Z^{\Diamond}\right)  \right)  $
must be read with the total derivatives :

$\frac{d}{d\xi^{\alpha}}\left(  \frac{\partial%
\mathcal{L}%
_{M}}{\partial z_{\alpha}^{i}}\left(  j^{1}Z^{\Diamond}\right)  \right)
=\sum_{j}\frac{\partial%
\mathcal{L}%
_{M}}{\partial z^{j}\partial z_{\alpha}^{i}}\frac{dz^{j}}{d\xi^{\alpha}}%
+\sum_{j\beta}\frac{\partial%
\mathcal{L}%
_{M}}{\partial z_{\beta}^{j}\partial z_{\alpha}^{i}}\frac{dz_{\beta}^{j}}%
{d\xi^{\alpha}}$

b) for f :

as a stand alone function :

$\frac{\delta S_{M}}{\delta f^{\alpha}}=-\sum_{\beta}\frac{d}{d\xi^{\beta}%
}\left(  \frac{\partial%
\mathcal{L}%
_{M}}{\partial f_{\beta}^{\alpha}}\right)  =-\frac{d}{d\xi^{0}}\left(
\frac{\partial%
\mathcal{L}%
_{M}}{\partial V^{\alpha}}\right)  $

and composed with the other functions :

$\frac{\delta S_{M}}{\delta f^{\alpha}}\left(  j^{1}Z^{\diamond}\right)
=\sum_{i>0}\frac{dz^{i}}{df^{\alpha}}\frac{\delta S_{M}}{\delta z^{i}}\left(
j^{1}Z^{\diamond}\right)  -\frac{d}{d\xi^{0}}\left(  \frac{\partial%
\mathcal{L}%
_{M}}{\partial V^{\alpha}}\right)  $ with : $\frac{dz^{i}}{df^{\alpha}%
}=\partial_{\alpha}z^{i}$

\paragraph{7)}

The functional derivative of $S=S_{M}+S_{F}$ is the sum of the functional
derivatives :

i
$>$
0%

\begin{equation}
\frac{\delta S}{\delta z^{i}}\mathbf{=}\frac{\partial%
\mathcal{L}%
_{M}}{\partial z^{i}}\left(  j^{1}Z^{\Diamond}\right)  \mathbf{-}\sum_{\alpha
}\frac{d}{d\xi^{\alpha}}\left(  \frac{\partial%
\mathcal{L}%
_{M}}{\partial z_{\alpha}^{i}}\left(  j^{1}Z^{\Diamond}\right)  \right)
\mathbf{+}\frac{\partial%
\mathcal{L}%
_{F}}{\partial z^{i}}\left(  j^{1}Z\right)  \mathbf{-}\sum_{\alpha
}\mathbf{\partial}_{\alpha}\left(  \frac{\partial%
\mathcal{L}%
_{F}}{\partial z_{\alpha}^{i}}\left(  j^{1}Z\right)  \right) \label{E41}%
\end{equation}

and for f :%

\begin{equation}
\frac{\delta S}{\delta f^{\alpha}}\left(  j^{1}Z^{\diamond}\right)
\mathbf{=}\sum_{i>0}\left(  \partial_{\alpha}z^{i}\right)  \frac{\delta S_{M}%
}{\delta z^{i}}\left(  j^{1}Z^{\diamond}\right)  \mathbf{-}\frac{d}{d\xi^{0}%
}\left(  \frac{\partial%
\mathcal{L}%
_{M}}{\partial V^{\alpha}}\right) \label{E42}%
\end{equation}

We have furthermore for any projectable vector field Y:

$\sum_{i}\frac{\delta S}{\delta z^{i}}\left(  j^{1}s\right)  Y^{i}\varpi
_{0}=\left(  j^{1}s\right)  ^{\ast}\pounds _{j^{1}Y}\left(
\mathcal{L}%
\varpi_{0}\right)  =\left(  j^{1}s\right)  ^{\ast}\pounds _{j^{1}Y}\left(
\mathcal{L}%
_{M}^{\Diamond}+%
\mathcal{L}%
_{F}\right)  \varpi_{0}$

\section{LAGRANGE\ EQUATIONS}

\label{Lagrange equations}

\subsection{Equation of state}

The equations for the state $\psi$ are :

$\forall i,j:$ $\frac{\delta S}{\delta\operatorname{Re}\psi^{ij}}=N\left(
m\right)  \frac{dL_{M}\left(  \det O^{\prime}\right)  }{d\operatorname{Re}%
\psi^{ij}}\left(  Z^{\Diamond}\right)  -\sum_{\beta}\frac{d}{d\xi^{\beta}%
}\left(  N\left(  m\right)  \frac{dL_{M}\left(  \det O^{\prime}\right)
}{d\partial_{\beta}\operatorname{Re}\psi^{ij}}\left(  Z^{\Diamond}\right)
\right)  $

$\forall i,j:\frac{\delta S}{\delta\operatorname{Im}\psi^{ij}}=N\left(
m\right)  \frac{dL_{M}\left(  \det O^{\prime}\right)  }{d\operatorname{Im}%
\psi^{ij}}\left(  Z^{\Diamond}\right)  -\sum_{\beta}\frac{d}{d\xi^{\beta}%
}\left(  N\left(  m\right)  \frac{dL_{M}\left(  \det O^{\prime}\right)
}{d\partial_{\beta}\operatorname{Im}\psi^{ij}}\left(  Z^{\Diamond}\right)
\right)  $

With :

$\frac{d%
\mathcal{L}%
_{M}}{d\operatorname{Re}\psi^{ij}}=\frac{\partial%
\mathcal{L}%
_{M}}{\partial\operatorname{Re}\psi^{ij}}+\sum_{k}\frac{d%
\mathcal{L}%
_{M}}{d\operatorname{Re}\nabla_{\alpha}\psi^{kj}}\operatorname{Re}\left[
G_{\alpha}\right]  _{i}^{k}+\frac{d%
\mathcal{L}%
_{M}}{d\operatorname{Im}\nabla_{\alpha}\psi^{kj}}\operatorname{Im}\left[
G_{\alpha}\right]  _{i}^{k}$

$+\frac{d%
\mathcal{L}%
_{M}}{d\operatorname{Re}\nabla_{\alpha}\psi^{ik}}\operatorname{Re}\left[
\grave{A}_{\alpha}\right]  _{j}^{k}+\frac{d%
\mathcal{L}%
_{M}}{d\operatorname{Im}\nabla_{\alpha}\psi^{ik}}\operatorname{Im}\left[
\grave{A}_{\alpha}\right]  _{j}^{k}$

$\frac{d%
\mathcal{L}%
_{M}}{d\partial_{\beta}\operatorname{Re}\psi^{ij}}=\frac{d%
\mathcal{L}%
_{M}}{d\nabla_{\beta}\operatorname{Re}\psi^{ij}}$

$\frac{d%
\mathcal{L}%
_{M}}{d\operatorname{Im}\psi^{ij}}=\frac{\partial%
\mathcal{L}%
_{M}}{\partial\operatorname{Im}\psi^{ij}}+\sum_{k}-\frac{d%
\mathcal{L}%
_{M}}{d\operatorname{Re}\nabla_{\alpha}\psi^{kj}}\operatorname{Im}\left[
G_{\alpha}\right]  _{i}^{k}+\frac{d%
\mathcal{L}%
_{M}}{d\operatorname{Im}\nabla_{\alpha}\psi^{kj}}\operatorname{Re}\left[
G_{\alpha}\right]  _{i}^{k}$

$-\frac{d%
\mathcal{L}%
_{M}}{d\operatorname{Re}\nabla_{\alpha}\psi^{ik}}\operatorname{Im}\left[
\grave{A}_{\alpha}\right]  _{j}^{k}+\frac{d%
\mathcal{L}%
_{M}}{d\operatorname{Im}\nabla_{\alpha}\psi^{ik}}\operatorname{Re}\left[
\grave{A}_{\alpha}\right]  _{j}^{k}$

$\frac{d%
\mathcal{L}%
_{M}}{d\partial_{\beta}\operatorname{Im}\psi^{ij}}=\frac{d%
\mathcal{L}%
_{M}}{d\nabla_{\beta}\operatorname{Im}\psi^{ij}}$

we get the equations :%

\begin{equation}
\label{E43a}
\end{equation}

$\forall i,j:0=V\frac{\partial%
\mathcal{L}%
_{M}^{\Diamond}}{\partial\operatorname{Re}\psi^{ij}}+\sum_{\alpha k}\frac{d%
\mathcal{L}%
_{M}^{\Diamond}}{d\operatorname{Re}\nabla_{\alpha}\psi^{kj}}\operatorname{Re}%
\left[  G_{\alpha}\right]  _{i}^{k}+\frac{d%
\mathcal{L}%
_{M}^{\Diamond}}{d\operatorname{Im}\nabla_{\alpha}\psi^{kj}}\operatorname{Im}%
\left[  G_{\alpha}\right]  _{i}^{k}$

$+\frac{d%
\mathcal{L}%
_{M}^{\Diamond}}{d\operatorname{Re}\nabla_{\alpha}\psi^{ik}}\operatorname{Re}%
\left[  \grave{A}_{\alpha}\right]  _{j}^{k}+\frac{d%
\mathcal{L}%
_{M}^{\Diamond}}{d\operatorname{Im}\nabla_{\alpha}\psi^{ik}}\operatorname{Im}%
\left[  \grave{A}_{\alpha}\right]  _{j}^{k}\}-\sum_{\beta}\frac{d}{d\xi
^{\beta}}\left(  \frac{d%
\mathcal{L}%
_{M}^{\Diamond}}{d\operatorname{Re}\nabla_{\beta}\psi^{ij}}\right)  $

\bigskip%

\begin{equation}
\label{E44a}
\end{equation}

$\forall i,j:0=\frac{\partial%
\mathcal{L}%
_{M}^{\Diamond}}{\partial\operatorname{Im}\psi^{ij}}+\sum_{\alpha k}-\frac{d%
\mathcal{L}%
_{M}^{\Diamond}}{d\operatorname{Re}\nabla_{\alpha}\psi^{kj}}\operatorname{Im}%
\left[  G_{\alpha}\right]  _{i}^{k}+\frac{d%
\mathcal{L}%
_{M}^{\Diamond}}{d\operatorname{Im}\nabla_{\alpha}\psi^{kj}}\operatorname{Re}%
\left[  G_{\alpha}\right]  _{i}^{k}$

$-\frac{d%
\mathcal{L}%
_{M}^{\Diamond}}{d\operatorname{Re}\nabla_{\alpha}\psi^{ik}}\operatorname{Im}%
\left[  \grave{A}_{\alpha}\right]  _{j}^{k}+\frac{d%
\mathcal{L}%
_{M}^{\Diamond}}{d\operatorname{Im}\nabla_{\alpha}\psi^{ik}}\operatorname{Re}%
\left[  \grave{A}_{\alpha}\right]  _{j}^{k}-\sum_{\beta}\frac{d}{d\xi^{\beta}%
}\left(  \frac{d%
\mathcal{L}%
_{M}^{\Diamond}}{d\operatorname{Im}\nabla_{\alpha}\psi^{ij}}\right)  $

\subsection{Gravitational equations}

The equations for the gravitational potential G are :

$\forall a,\alpha:\frac{\delta S}{\delta G_{\alpha}^{a}}=\frac{d%
\mathcal{L}%
_{M}}{dG_{\alpha}^{a}}\left(  Z^{\Diamond}\right)  +\frac{d%
\mathcal{L}%
_{F}}{dG_{\alpha}^{a}}\left(  Z\left(  m\right)  \right)  -\sum_{\beta}%
\frac{d}{d\xi^{\beta}}\left(  \frac{d%
\mathcal{L}%
_{M}}{d\partial_{\beta}G_{\alpha}^{a}}\left(  Z^{\Diamond}\right)  +\frac{d%
\mathcal{L}%
_{F}}{d\partial_{\beta}G_{\alpha}^{a}}\left(  Z\left(  m\right)  \right)
\right)  $

$\frac{d%
\mathcal{L}%
_{M}}{dG_{\alpha}^{a}}=\frac{\partial%
\mathcal{L}%
_{M}}{\partial G_{\alpha}^{a}}+\sum_{ij}\left(  \frac{d%
\mathcal{L}%
_{M}}{d\operatorname{Re}\partial_{\alpha}\psi^{ij}}\operatorname{Re}\left(
\left[  \kappa_{a}\right]  \left[  \psi^{\Diamond}\right]  \right)  _{j}%
^{i}+\frac{d%
\mathcal{L}%
_{M}}{d\operatorname{Im}\partial_{\alpha}\psi^{ij}}\operatorname{Im}\left(
\left[  \kappa_{a}\right]  \left[  \psi^{\Diamond}\right]  \right)  _{j}%
^{i}\right)  $

$\frac{d%
\mathcal{L}%
_{F}}{dG_{\alpha}^{a}}=\frac{\partial%
\mathcal{L}%
_{F}}{\partial G_{\alpha}^{a}}+2\sum_{b\beta}\frac{d%
\mathcal{L}%
_{F}}{d%
\mathcal{F}%
_{G\alpha\beta}^{b}}\left[  \overrightarrow{\kappa}_{a},G_{\beta}\right]
^{b}$

$\frac{d%
\mathcal{L}%
_{M}}{d\partial_{\beta}G_{\alpha}^{a}}=0$

$\frac{d%
\mathcal{L}%
_{F}}{d\partial_{\beta}G_{\alpha}^{a}}=\sum_{\mu\lambda}\frac{dL_{F}}{d%
\mathcal{F}%
_{G\lambda\mu}^{b}}\frac{d%
\mathcal{F}%
_{G\lambda\mu}^{b}}{d\partial_{\beta}G_{\alpha}^{a}}=\sum_{\mu\lambda}%
\frac{dL_{F}}{d%
\mathcal{F}%
_{G\lambda\mu}^{b}}\frac{d\left(  \partial_{\lambda}G_{\mu}^{b}-\partial_{\mu
}G_{\lambda}^{b}\right)  }{d\partial_{\beta}G_{\alpha}^{a}}$

$\qquad=\frac{dL_{F}}{d%
\mathcal{F}%
_{G\beta\alpha}^{a}}-\frac{dL_{F}}{d%
\mathcal{F}%
_{G\alpha\beta}^{a}}=2\frac{d%
\mathcal{L}%
_{F}}{d%
\mathcal{F}%
_{G\beta\alpha}^{a}}$

$\frac{d}{d\xi^{\beta}}\frac{d%
\mathcal{L}%
_{F}}{d\partial_{\beta}G_{\alpha}^{a}}\left(  Z\left(  m\right)  \right)
=\partial_{\beta}\frac{d%
\mathcal{L}%
_{F}}{d\partial_{\beta}G_{\alpha}^{a}}$

which gives the equations :%

\begin{equation}
\label{E45}
\end{equation}

$\forall a,\alpha:0=\frac{\partial%
\mathcal{L}%
_{M}^{\Diamond}}{\partial G_{\alpha}^{a}}+\frac{\partial%
\mathcal{L}%
_{F}}{\partial G_{\alpha}^{a}}+\sum_{ij}\left(  \frac{\partial%
\mathcal{L}%
_{M}^{\Diamond}}{\partial\operatorname{Re}\nabla_{\alpha}\psi^{ij}%
}\operatorname{Re}\left(  \left[  \kappa_{a}\right]  \left[  \psi\right]
\right)  _{j}^{i}+\frac{\partial%
\mathcal{L}%
_{M}^{\Diamond}}{\partial\operatorname{Im}\nabla_{\alpha}\psi^{ij}%
}\operatorname{Im}\left(  \left[  \kappa_{a}\right]  \left[  \psi\right]
\right)  _{j}^{i}\right)  $

$+2\sum_{\beta}\left(  \sum_{b}\frac{\partial%
\mathcal{L}%
_{F}}{\partial%
\mathcal{F}%
_{G\alpha\beta}^{b}}\left[  \overrightarrow{\kappa}_{a},G_{\beta}\right]
^{b}+\partial_{\beta}\left(  \frac{\partial%
\mathcal{L}%
_{F}}{\partial%
\mathcal{F}%
_{G\alpha\beta}^{a}}\right)  \right)  $

\subsection{Equations for the other force fields}

The equations for the other potentials \`{A} are :

\paragraph{1)}

$\forall a,\alpha:\frac{\delta S}{\delta\operatorname{Re}\grave{A}_{\alpha
}^{a}}=\frac{d%
\mathcal{L}%
_{M}^{\Diamond}}{d\operatorname{Re}\grave{A}_{\alpha}^{a}}\left(  Z^{\Diamond
}\right)  +\frac{d%
\mathcal{L}%
_{F}}{d\operatorname{Re}\grave{A}_{\alpha}^{a}}\left(  Z\left(  m\right)
\right)  -\sum_{\beta}\frac{d}{d\xi^{\beta}}\left(  \frac{d%
\mathcal{L}%
_{F}}{d\partial_{\beta}\operatorname{Re}\grave{A}_{\alpha}^{a}}\left(
Z\left(  m\right)  \right)  \right)  =0$

with

$\frac{d%
\mathcal{L}%
_{M}}{d\operatorname{Re}\grave{A}_{\alpha}^{a}}=\sum_{ij}\frac{d%
\mathcal{L}%
_{M}}{d\operatorname{Re}\nabla_{\alpha}\psi^{ij}}\operatorname{Re}\left(
\left[  \psi^{\Diamond}\right]  \left[  \theta_{a}\right]  ^{t}\right)
_{j}^{i}+\frac{d%
\mathcal{L}%
_{M}}{d\operatorname{Im}\nabla_{\alpha}\psi^{ij}}\operatorname{Im}\left(
\left[  \psi^{\Diamond}\right]  \left[  \theta_{a}\right]  ^{t}\right)
_{j}^{i}$

$\frac{d%
\mathcal{L}%
_{F}}{d\operatorname{Re}\grave{A}_{\alpha}^{a}}=2\sum_{b\beta}\frac{d%
\mathcal{L}%
_{F}}{d\operatorname{Re}%
\mathcal{F}%
_{G\alpha\beta}^{b}}\operatorname{Re}\left[  \overrightarrow{\theta}%
_{a},\grave{A}_{\beta}\right]  ^{b}+\frac{d%
\mathcal{L}%
_{F}}{d\operatorname{Im}%
\mathcal{F}%
_{G\alpha\beta}^{b}}\operatorname{Im}\left[  \overrightarrow{\theta}%
_{a},\grave{A}_{\beta}\right]  ^{b}$

$\frac{d%
\mathcal{L}%
_{F}}{d\partial_{\beta}\operatorname{Re}\grave{A}_{\alpha}^{a}}=2\frac{d%
\mathcal{L}%
_{F}}{d\operatorname{Re}%
\mathcal{F}%
_{\beta\alpha}^{a}}$

we get the equation :%

\begin{equation}
\label{E46}
\end{equation}

$\forall a,\alpha:0=\sum_{ij}\left(  \frac{\partial%
\mathcal{L}%
_{M}^{\Diamond}}{\partial\operatorname{Re}\nabla_{\alpha}\psi^{ij}%
}\operatorname{Re}\left(  \left[  \psi\right]  \left[  \theta_{a}\right]
^{t}\right)  _{j}^{i}+\frac{\partial%
\mathcal{L}%
_{M}^{\Diamond}}{\partial\operatorname{Im}\nabla_{\alpha}\psi^{ij}%
}\operatorname{Im}\left(  \left[  \psi\right]  \left[  \theta_{a}\right]
^{t}\right)  _{j}^{i}\right)  $

$+2\sum_{\beta}\sum_{b}\left(  \frac{\partial%
\mathcal{L}%
_{F}}{\partial\operatorname{Re}%
\mathcal{F}%
_{A,\alpha\beta}^{b}}\operatorname{Re}\left[  \overrightarrow{\theta}%
_{a},\grave{A}_{\beta}\right]  ^{b}+\frac{\partial%
\mathcal{L}%
_{F}}{\partial\operatorname{Im}%
\mathcal{F}%
_{A,\alpha\beta}^{b}}\operatorname{Im}\left[  \overrightarrow{\theta}%
_{a},\grave{A}_{\beta}\right]  ^{b}\right)  +\partial_{\beta}\left(
\frac{\partial%
\mathcal{L}%
_{F}}{\partial\operatorname{Re}%
\mathcal{F}%
_{A,\alpha\beta}^{a}}\right)  $

\bigskip

\paragraph{2)}

The second set of equations is :

$\forall a,\alpha:\frac{\delta S}{\delta\operatorname{Im}\grave{A}_{\alpha
}^{a}}=\frac{\partial%
\mathcal{L}%
_{M}^{\Diamond}}{\partial\operatorname{Im}\grave{A}_{\alpha}^{a}}\left(
Z^{\Diamond}\right)  +\frac{\partial L_{F}\left(  \det O^{\prime}\right)
}{\partial\operatorname{Im}\grave{A}_{\alpha}^{a}}\left(  Z\left(  m\right)
\right)  -\sum_{\beta}\partial_{\beta}\left(  \frac{\partial L_{F}\left(  \det
O^{\prime}\right)  }{\partial\partial_{\beta}\operatorname{Im}\grave
{A}_{\alpha}^{a}}\left(  Z\left(  m\right)  \right)  \right)  =0$

$\frac{d%
\mathcal{L}%
_{F}}{d\operatorname{Im}\grave{A}_{\alpha}^{a}}=\frac{\partial%
\mathcal{L}%
_{F}}{\partial\operatorname{Im}\grave{A}_{\alpha}^{a}}+2\sum_{b\beta}-\frac{d%
\mathcal{L}%
_{F}}{d\operatorname{Re}%
\mathcal{F}%
_{G\alpha\beta}^{b}}\operatorname{Im}\left[  \overrightarrow{\theta}%
_{a},\grave{A}_{\beta}\right]  ^{b}+\frac{d%
\mathcal{L}%
_{F}}{d\operatorname{Im}%
\mathcal{F}%
_{G\alpha\beta}^{b}}\operatorname{Re}\left[  \overrightarrow{\theta}%
_{a},\grave{A}_{\beta}\right]  ^{b}$

$\frac{d%
\mathcal{L}%
_{M}}{d\operatorname{Im}\grave{A}_{\alpha}^{a}}=\sum_{ij}\left(  -\frac{d%
\mathcal{L}%
_{M}}{d\operatorname{Re}\nabla_{\alpha}\psi^{ij}}\operatorname{Im}\left(
\left[  \psi^{\Diamond}\right]  \left[  \theta_{a}\right]  ^{t}\right)
_{j}^{i}+\frac{d%
\mathcal{L}%
_{M}}{d\operatorname{Im}\nabla_{\alpha}\psi^{ij}}\operatorname{Re}\left(
\left[  \psi^{\Diamond}\right]  \left[  \theta_{a}\right]  ^{t}\right)
_{j}^{i}\right)  $

$\frac{d%
\mathcal{L}%
_{F}}{d\partial_{\beta}\operatorname{Im}\grave{A}_{\alpha}^{a}}=2\frac{d%
\mathcal{L}%
_{F}}{d\operatorname{Im}%
\mathcal{F}%
_{\beta\alpha}^{a}}$

and gives :%

\begin{equation}
\label{E47}
\end{equation}

$\forall a,\alpha:0=\sum_{ij}\left(  -\frac{\partial%
\mathcal{L}%
_{M}^{\Diamond}}{\partial\operatorname{Re}\nabla_{\alpha}\psi^{ij}%
}\operatorname{Im}\left(  \left[  \psi\right]  \left[  \theta_{a}\right]
^{t}\right)  _{j}^{i}+\frac{\partial%
\mathcal{L}%
_{M}^{\Diamond}}{\partial\operatorname{Im}\nabla_{\alpha}\psi^{ij}%
}\operatorname{Re}\left(  \left[  \psi\right]  \left[  \theta_{a}\right]
^{t}\right)  _{j}^{i}\right)  $

$+2\sum_{\beta}\sum_{b}\left(  -\frac{\partial%
\mathcal{L}%
_{F}}{\partial\operatorname{Re}%
\mathcal{F}%
_{A,\alpha\beta}^{b}}\operatorname{Im}\left[  \overrightarrow{\theta}%
_{a},\grave{A}_{\beta}\right]  ^{b}+\frac{\partial%
\mathcal{L}%
_{F}}{\partial\operatorname{Im}%
\mathcal{F}%
_{A,\alpha\beta}^{b}}\operatorname{Re}\left[  \overrightarrow{\theta}%
_{a},\grave{A}_{\beta}\right]  ^{b}\right)  +\partial_{\beta}\left(
\frac{\partial%
\mathcal{L}%
_{F}}{\partial\operatorname{Im}%
\mathcal{F}%
_{A,\alpha\beta}^{a}}\right)  $

\subsection{Frame equation}

The equations for O' are :

$\forall i,\alpha:\frac{\delta S}{\delta O_{i}^{\prime\alpha}}=\frac{d%
\mathcal{L}%
}{dO_{\alpha}^{\prime i}}-\sum_{\gamma}\frac{d}{d\xi^{\gamma}}\left(  \frac{d%
\mathcal{L}%
}{d\partial_{\gamma}O_{\alpha}^{\prime i}}\right)  =0$

with $%
\mathcal{L}%
=NL_{M}^{\Diamond}\det O^{\prime\diamond}+L_{F}\det O^{\prime}$

$\frac{\delta S}{\delta O_{i}^{\prime\alpha}}=N\left(  m\right)  \frac
{dL_{M}\left(  \det O^{\prime}\right)  }{dO_{\alpha}^{\prime i}}\left(
Z^{\Diamond}\right)  +\frac{dL_{F}\left(  \det O^{\prime}\right)  }%
{dO_{\alpha}^{\prime i}}\left(  Z\left(  m\right)  \right)  $

$-\sum_{\gamma}\frac{d}{d\xi^{\gamma}}\left(  N\frac{dL_{M}\left(  \det
O^{\prime}\right)  }{d\partial_{\gamma}O_{\alpha}^{\prime i}}\left(
Z^{\Diamond}\right)  +\frac{dL_{F}\left(  \det O^{\prime}\right)  }%
{d\partial_{\gamma}O_{\alpha}^{\prime i}}\left(  Z\left(  m\right)  \right)
\right)  $

with $\frac{\partial\left(  \det O^{\prime}\right)  }{\partial O_{\alpha
}^{\prime i}}=O_{i}^{\alpha}\left(  \det O^{\prime}\right)  ,L=VL_{M}%
^{\Diamond}+L_{F}$

$\frac{dL}{dO_{\alpha}^{\prime i}}+O_{i}^{\alpha}L-\frac{1}{\det O^{\prime}%
}\sum_{\gamma}\left(  \frac{d}{d\xi^{\beta}}\left(  V\frac{dL\det O^{\prime}%
}{d\partial_{\gamma}O_{\alpha}^{\prime i}}\right)  \right)  =0$

multiplying by $O_{\beta}^{\prime i}$ and adding :

$0=\sum_{i}\frac{dL}{dO_{\alpha}^{\prime i}}O_{\beta}^{\prime i}+\delta
_{\beta}^{\alpha}L-\frac{1}{\det O^{\prime}}\sum_{i\gamma}O_{\beta}^{\prime
i}\left(  \frac{d}{d\xi^{\gamma}}\left(  N\frac{dL\det O^{\prime}}%
{d\partial_{\gamma}O_{\alpha}^{\prime i}}\right)  \right)  $

$0=\sum_{i}\frac{dL}{dO_{\alpha}^{\prime i}}O_{\beta}^{\prime i}+\delta
_{\beta}^{\alpha}L-\frac{1}{\det O^{\prime}}O_{\beta}^{\prime i}\sum_{\gamma
}\left(  \frac{d}{d\xi^{\gamma}}\left(  N\frac{\partial L_{M}\left(  \det
O^{\prime}\right)  }{\partial\partial_{\gamma}O_{\alpha}^{\prime i}}\left(
Z^{\Diamond}\right)  \right)  +\partial_{\gamma}\left(  \frac{\partial
L_{F}\left(  \det O^{\prime}\right)  }{\partial\partial_{\gamma}O_{\alpha
}^{\prime i}}\right)  \right)  $

$0=\left(  \frac{dL}{dO_{\alpha}^{\prime i}}-\sum_{\gamma}\left(  \frac
{d}{d\xi^{\gamma}}\frac{NdL_{M}^{\diamond}}{d\partial_{\gamma}O_{\alpha
}^{\prime i}}+\partial_{\gamma}\frac{dL_{F}}{d\partial_{\gamma}O_{\alpha
}^{\prime i}}\right)  \right)  O_{\beta}^{\prime i}+\delta_{\beta}^{\alpha}L$

$-\frac{1}{\det O^{\prime}}O_{\beta}^{\prime i}\sum_{\gamma}\left(
\frac{NdL_{M}^{\diamond}}{d\partial_{\gamma}O_{\alpha}^{\prime i}}\frac{d\det
O^{\prime\diamond}}{d\xi^{\gamma}}+\frac{dL_{F}}{d\partial_{\gamma}O_{\alpha
}^{\prime i}}\partial_{\gamma}\det O^{\prime}\right)  $%

\begin{equation}
\label{E48}
\end{equation}

$\forall\alpha,\beta:0=\delta_{\beta}^{\alpha}L+\sum_{i}\left(  \frac
{dL}{dO_{\alpha}^{\prime i}}-\sum_{\gamma}\left(  \frac{d}{d\xi^{\gamma}}%
\frac{NdL_{M}^{\diamond}}{d\partial_{\gamma}O_{\alpha}^{\prime i}}%
+\partial_{\gamma}\frac{dL_{F}}{d\partial_{\gamma}O_{\alpha}^{\prime i}%
}\right)  \right)  O_{\beta}^{\prime i}$

$-\frac{1}{\det O^{\prime}}O_{\beta}^{\prime i}\sum_{\gamma}\left(
\frac{NdL_{M}^{\diamond}}{d\partial_{\gamma}O_{\alpha}^{\prime i}}\frac{d\det
O^{\prime\diamond}}{d\xi^{\gamma}}+\frac{dL_{F}}{d\partial_{\gamma}O_{\alpha
}^{\prime i}}\partial_{\gamma}\det O^{\prime}\right)  $

\bigskip

If the partial derivatives $\partial_{\gamma}O_{\alpha}^{\prime i}$ do not
appear in the lagrangian we have the simple equation:

$\forall\alpha,\beta:0=\sum_{i}\frac{dL}{dO_{\alpha}^{\prime i}}O_{\beta
}^{\prime i}+\delta_{\beta}^{\alpha}L$

\subsection{Trajectory}

\paragraph{1)}

The equation for f is :

$\forall\alpha:0=\frac{\delta S}{\delta f^{\alpha}}\left(  j^{1}Z^{\diamond
}\right)  =\sum_{i>0}\left(  \partial_{\alpha}z^{i}\right)  \frac{\delta
S_{M}}{\delta z^{i}}\left(  j^{1}Z^{\diamond}\right)  -\frac{d}{d\xi^{0}%
}\left(  \frac{\partial%
\mathcal{L}%
_{M}}{\partial V^{\alpha}}\right)  $

$=\sum_{ij}\frac{\delta S_{M}^{\diamond}}{\delta\operatorname{Re}\psi^{ij}%
}\frac{\partial\operatorname{Re}\psi^{ij}}{\partial f^{\alpha}}+\frac{\delta
S_{M}^{\diamond}}{\delta\operatorname{Im}\psi^{ij}}\frac{\partial
\operatorname{Im}\psi^{ij}}{\partial f^{\alpha}}$

$+\sum_{a,\beta}\frac{\delta S_{M}^{\diamond}}{\delta\operatorname{Re}%
\grave{A}_{\beta}^{a}}\frac{\partial\operatorname{Re}\grave{A}_{\beta}^{a}%
}{\partial f^{\alpha}}+\frac{\delta S_{M}^{\diamond}}{\delta\operatorname{Im}%
\grave{A}_{\beta}^{a}}\frac{\partial\operatorname{Im}\grave{A}_{\beta}^{a}%
}{\partial f^{\alpha}}+\frac{\delta S_{M}^{\diamond}}{\delta G_{\beta}^{a}%
}\frac{\partial G_{\beta}^{a}}{\partial f^{\alpha}}$

$+\sum_{i,\beta}\frac{\delta S_{M}^{\diamond}}{\delta O_{\beta}^{\prime i}%
}\frac{\partial O_{\beta}^{\prime i}}{\partial f^{\alpha}}-\frac{d}{d\xi^{0}%
}\left(  \frac{\partial%
\mathcal{L}%
_{M}}{\partial V^{\alpha}}\right)  $

$\frac{\partial\operatorname{Re}\psi^{ij}}{\partial f^{\alpha}}=\partial
_{\alpha}\operatorname{Re}\psi^{ij},\frac{\partial\operatorname{Im}\psi^{ij}%
}{\partial f^{\alpha}}=\partial_{\alpha}\operatorname{Im}\psi^{ij}%
,\frac{\partial\operatorname{Re}\grave{A}_{\beta}^{a}}{\partial f^{\alpha}%
}=\partial_{\alpha}\operatorname{Re}\grave{A}_{\beta}^{a},$

$\frac{\partial\operatorname{Im}\grave{A}_{\beta}^{a}}{\partial f^{\alpha}%
}=\partial_{\alpha}\operatorname{Im}\grave{A}_{\beta}^{a},\frac{\partial
G_{\beta}^{a}}{\partial f^{\alpha}}=\partial_{\alpha}G_{\beta}^{a}%
,\frac{\partial O_{\beta}^{\prime i}}{\partial f^{\alpha}}=\partial_{\alpha
}O_{\beta}^{\prime i}$

all these partial derivatives being evaluated at $j^{1}Z^{\diamond}%
=\widetilde{f}^{\ast}j^{1}Z$

On shell we have :

$\frac{\delta S_{M}^{\diamond}}{\delta\operatorname{Re}\psi^{ij}}%
=0,\frac{\delta S_{M}^{\diamond}}{\delta\operatorname{Im}\psi^{ij}}%
=0,\frac{\delta S_{M}^{\diamond}}{\delta\operatorname{Re}\grave{A}_{\beta}%
^{a}}+\frac{\delta S_{F}}{\delta\operatorname{Re}\grave{A}_{\beta}^{a}}=0,$

$\frac{\delta S_{M}^{\diamond}}{\delta\operatorname{Im}\grave{A}_{\beta}^{a}%
}+\frac{\delta S_{F}}{\delta\operatorname{Im}\grave{A}_{\beta}^{a}}%
=0,\frac{\delta S_{M}^{\diamond}}{\delta G_{\beta}^{a}}+\frac{\delta S_{F}%
}{\delta G_{\beta}^{a}}=0,\frac{\delta S_{M}^{\diamond}}{\delta O_{\beta
}^{\prime i}}+\frac{\delta S_{F}}{\delta O_{\beta}^{\prime i}}=0$

So we have two possible formulations for the equation. As $L_{F}$ does not
involve f, it is simpler to take its derivatives whenever useful.

\paragraph{2)}

$\frac{\delta S}{\delta f^{\alpha}}\left(  j^{1}Z^{\diamond}\right)  $

$=\left(  \sum_{a,\beta}\frac{\delta S_{M}^{\diamond}}{\delta\operatorname{Re}%
\grave{A}_{\beta}^{a}}\partial_{\alpha}\operatorname{Re}\grave{A}_{\beta}%
^{a}+\frac{\delta S_{M}^{\diamond}}{\delta\operatorname{Im}\grave{A}_{\beta
}^{a}}\partial_{\alpha}\operatorname{Im}\grave{A}_{\beta}^{a}+\frac{\delta
S_{M}^{\diamond}}{\delta G_{\beta}^{a}}\partial_{\alpha}G_{\beta}^{a}%
+\sum_{i,\beta}\frac{\delta S_{M}^{\diamond}}{\delta O_{\beta}^{\prime i}%
}\partial_{\alpha}O_{\beta}^{\prime i}\right)  \left(  j^{1}Z^{\diamond
}\right)  -\frac{d}{d\xi^{0}}\left(  \frac{\partial%
\mathcal{L}%
_{M}}{\partial V^{\alpha}}\right)  $

$=-\left(  \sum_{a,\beta}\frac{\delta S_{F}}{\delta\operatorname{Re}\grave
{A}_{\beta}^{a}}\partial_{\alpha}\operatorname{Re}\grave{A}_{\beta}^{a}%
+\frac{\delta S_{F}}{\delta\operatorname{Im}\grave{A}_{\beta}^{a}}%
\partial_{\alpha}\operatorname{Im}\grave{A}_{\beta}^{a}+\frac{\delta S_{F}%
}{\delta G_{\beta}^{a}}\partial_{\alpha}G_{\beta}^{a}+\sum_{i,\beta}%
\frac{\delta S_{F}}{\delta O_{\beta}^{\prime i}}\partial_{\alpha}O_{\beta
}^{\prime i}\right)  \left(  j^{1}Z\right)  -\frac{d}{d\xi^{0}}\left(
\frac{\partial%
\mathcal{L}%
_{M}}{\partial V^{\alpha}}\right)  $

$\frac{\delta S_{F}}{\delta\operatorname{Re}\grave{A}_{\beta}^{a}}%
=2\sum_{\gamma}\{\sum_{b}\left(  \frac{\partial%
\mathcal{L}%
_{F}}{\partial\operatorname{Re}%
\mathcal{F}%
_{A,\beta\gamma}^{b}}\operatorname{Re}\left[  \overrightarrow{\theta}%
_{a},\grave{A}_{\gamma}\right]  ^{b}+\frac{\partial%
\mathcal{L}%
_{F}}{\partial\operatorname{Im}%
\mathcal{F}%
_{A,\beta\gamma}^{b}}\operatorname{Im}\left[  \overrightarrow{\theta}%
_{a},\grave{A}_{\gamma}\right]  ^{b}\right)  $

$+\partial_{\gamma}\left(  \frac{\partial%
\mathcal{L}%
_{F}}{\partial\operatorname{Re}%
\mathcal{F}%
_{A,\beta\gamma}^{a}}\right)  \}$

$\frac{\delta S_{F}}{\delta\operatorname{Im}\grave{A}_{\beta}^{a}}%
=2\sum_{\gamma}\{\sum_{b}\left(  -\frac{\partial%
\mathcal{L}%
_{F}}{\partial\operatorname{Re}%
\mathcal{F}%
_{A,\beta\gamma}^{b}}\operatorname{Im}\left[  \overrightarrow{\theta}%
_{a},\grave{A}_{\gamma}\right]  ^{b}+\frac{\partial%
\mathcal{L}%
_{F}}{\partial\operatorname{Im}%
\mathcal{F}%
_{A,\beta\gamma}^{b}}\operatorname{Re}\left[  \overrightarrow{\theta}%
_{a},\grave{A}_{\gamma}\right]  ^{b}\right)  $

$+\partial_{\gamma}\left(  \frac{\partial%
\mathcal{L}%
_{F}}{\partial\operatorname{Im}%
\mathcal{F}%
_{A,\beta\gamma}^{b}}\right)  \}$

$\frac{\delta S_{F}}{\delta G_{\beta}^{a}}=\frac{\partial%
\mathcal{L}%
_{F}}{\partial G_{\beta}^{a}}+2\sum_{\gamma}\left(  \sum_{b}\frac{\partial%
\mathcal{L}%
_{F}}{\partial%
\mathcal{F}%
_{G\beta\gamma}^{b}}\left[  \overrightarrow{\kappa}_{a},G_{\gamma}\right]
^{b}+\partial_{\gamma}\left(  \frac{\partial%
\mathcal{L}%
_{F}}{\partial%
\mathcal{F}%
_{G\beta\gamma}^{a}}\right)  \right)  $

$\frac{\delta S_{F}}{\delta O_{\beta}^{\prime i}}=\frac{dL_{F}\left(  \det
O^{\prime}\right)  }{dO_{\beta}^{\prime i}}-\sum_{\gamma}\frac{d}{d\xi
^{\gamma}}\left(  \frac{dL_{F}\left(  \det O^{\prime}\right)  }{d\partial
_{\gamma}O_{\beta}^{\prime i}}\right)  $

So on shell the equation is :

$\forall\alpha:0=\frac{d}{d\xi^{0}}\left(  \frac{\partial%
\mathcal{L}%
_{M}}{\partial V^{\alpha}}\right)  $

$+2\sum_{a,\beta\gamma}\{\sum_{b}\left(  \frac{\partial%
\mathcal{L}%
_{F}}{\partial\operatorname{Re}%
\mathcal{F}%
_{A,\beta\gamma}^{b}}\operatorname{Re}\left[  \overrightarrow{\theta}%
_{a},\grave{A}_{\gamma}\right]  ^{b}+\frac{\partial%
\mathcal{L}%
_{F}}{\partial\operatorname{Im}%
\mathcal{F}%
_{A,\beta\gamma}^{b}}\operatorname{Im}\left[  \overrightarrow{\theta}%
_{a},\grave{A}_{\gamma}\right]  ^{b}\right)  $

$+\partial_{\gamma}\left(  \frac{\partial%
\mathcal{L}%
_{F}}{\partial\operatorname{Re}%
\mathcal{F}%
_{A,\beta\gamma}^{a}}\right)  \partial_{\alpha}\operatorname{Re}\grave
{A}_{\beta}^{a}\}$

$+2\sum_{a\beta\gamma}\{\sum_{b}\left(  -\frac{\partial%
\mathcal{L}%
_{F}}{\partial\operatorname{Re}%
\mathcal{F}%
_{A,\beta\gamma}^{b}}\operatorname{Im}\left[  \overrightarrow{\theta}%
_{a},\grave{A}_{\gamma}\right]  ^{b}+\frac{\partial%
\mathcal{L}%
_{F}}{\partial\operatorname{Im}%
\mathcal{F}%
_{A,\beta\gamma}^{b}}\operatorname{Re}\left[  \overrightarrow{\theta}%
_{a},\grave{A}_{\gamma}\right]  ^{b}\right)  $

$+\partial_{\gamma}\left(  \frac{\partial%
\mathcal{L}%
_{F}}{\partial\operatorname{Im}%
\mathcal{F}%
_{A,\beta\gamma}^{b}}\right)  \partial_{\alpha}\operatorname{Im}\grave
{A}_{\beta}^{a}\}$

$+\sum_{a\beta}\left(  \frac{\partial%
\mathcal{L}%
_{F}}{\partial G_{\beta}^{a}}+2\sum_{\gamma}\left(  \sum_{b}\frac{\partial%
\mathcal{L}%
_{F}}{\partial%
\mathcal{F}%
_{G\beta\gamma}^{b}}\left[  \overrightarrow{\kappa}_{a},G_{\gamma}\right]
^{b}+\partial_{\gamma}\left(  \frac{\partial%
\mathcal{L}%
_{F}}{\partial%
\mathcal{F}%
_{G\beta\gamma}^{a}}\right)  \right)  \right)  \partial_{\alpha}G_{\beta}^{a}$

$+\sum_{i\beta}\left(  \frac{dL_{F}\left(  \det O^{\prime}\right)  }%
{dO_{\beta}^{\prime i}}-\sum_{\gamma}\frac{d}{d\xi^{\gamma}}\left(
\frac{dL_{F}\left(  \det O^{\prime}\right)  }{d\partial_{\gamma}O_{\beta
}^{\prime i}}\right)  \right)  \partial_{\alpha}O_{\beta}^{\prime i}$

$\forall\alpha:0=\frac{d}{d\xi^{0}}\left(  \frac{\partial%
\mathcal{L}%
_{M}}{\partial V^{\alpha}}\right)  +2\sum_{a,\beta\gamma}\left(
\partial_{\alpha}\operatorname{Re}\grave{A}_{\beta}^{a}\right)  \partial
_{\gamma}\left(  \frac{\partial%
\mathcal{L}%
_{F}}{\partial\operatorname{Re}%
\mathcal{F}%
_{A,\beta\gamma}^{a}}\right)  +\left(  \partial_{\alpha}\operatorname{Im}%
\grave{A}_{\beta}^{a}\right)  \partial_{\gamma}\left(  \frac{\partial%
\mathcal{L}%
_{F}}{\partial\operatorname{Im}%
\mathcal{F}%
_{A,\beta\gamma}^{b}}\right)  $

$+2\sum_{b\beta\gamma}\frac{\partial%
\mathcal{L}%
_{F}}{\partial\operatorname{Re}%
\mathcal{F}%
_{A,\beta\gamma}^{b}}\operatorname{Re}\left(  \left[  \overrightarrow{\theta
}_{a},\grave{A}_{\gamma}\right]  ^{b}\partial_{\alpha}\grave{A}_{\beta}%
^{a}\right)  +\frac{\partial%
\mathcal{L}%
_{F}}{\partial\operatorname{Im}%
\mathcal{F}%
_{A,\beta\gamma}^{b}}\operatorname{Im}\left(  \left[  \overrightarrow{\theta
}_{a},\grave{A}_{\gamma}\right]  ^{b}\partial_{\alpha}\grave{A}_{\beta}%
^{a}\right)  \}$

$+\sum_{a\beta}\left(  \frac{\partial%
\mathcal{L}%
_{F}}{\partial G_{\alpha}^{a}}\partial_{\alpha}G_{\beta}^{a}+2\sum_{\gamma
}\left(  \left(  \partial_{\alpha}G_{\beta}^{a}\right)  \partial_{\gamma
}\left(  \frac{\partial%
\mathcal{L}%
_{F}}{\partial%
\mathcal{F}%
_{G\beta\gamma}^{a}}\right)  +\frac{\partial%
\mathcal{L}%
_{F}}{\partial%
\mathcal{F}%
_{G\beta\gamma}^{a}}\left[  \partial_{\alpha}G_{\beta},G_{\gamma}\right]
^{a}\right)  \right)  $

$+\sum_{i\beta}\left(  \frac{dL_{F}\left(  \det O^{\prime}\right)  }%
{dO_{\beta}^{\prime i}}-\sum_{\gamma}\frac{d}{d\xi^{\gamma}}\left(
\frac{dL_{F}\left(  \det O^{\prime}\right)  }{d\partial_{\gamma}O_{\beta
}^{\prime i}}\right)  \right)  \partial_{\alpha}O_{\beta}^{\prime i}$

Thus the equation is :%

\begin{equation}
\label{E49}
\end{equation}

$\forall\alpha:0=\frac{d}{d\xi^{0}}\left(  \frac{\partial%
\mathcal{L}%
_{M}}{\partial V^{\alpha}}\right)  +\sum_{i\beta}\left(  \frac{dL_{F}\left(
\det O^{\prime}\right)  }{dO_{\beta}^{\prime i}}-\sum_{\gamma}\frac{d}%
{d\xi^{\gamma}}\left(  \frac{dL_{F}\left(  \det O^{\prime}\right)  }%
{d\partial_{\gamma}O_{\beta}^{\prime i}}\right)  \right)  \partial_{\alpha
}O_{\beta}^{\prime i}$

$+2\sum_{a,\beta\gamma}(\left(  \partial_{\alpha}\operatorname{Re}\grave
{A}_{\beta}^{a}\right)  \partial_{\gamma}\left(  \frac{\partial%
\mathcal{L}%
_{F}}{\partial\operatorname{Re}%
\mathcal{F}%
_{A,\beta\gamma}^{a}}\right)  +\left(  \partial_{\alpha}\operatorname{Im}%
\grave{A}_{\beta}^{a}\right)  \partial_{\gamma}\left(  \frac{\partial%
\mathcal{L}%
_{F}}{\partial\operatorname{Im}%
\mathcal{F}%
_{A,\beta\gamma}^{b}}\right)  $

$+\frac{\partial%
\mathcal{L}%
_{F}}{\partial\operatorname{Re}%
\mathcal{F}%
_{A,\beta\gamma}^{a}}\operatorname{Re}\left[  \partial_{\alpha}\grave
{A}_{\beta},\grave{A}_{\gamma}\right]  ^{a}+\frac{\partial%
\mathcal{L}%
_{F}}{\partial\operatorname{Im}%
\mathcal{F}%
_{A,\beta\gamma}^{a}}\operatorname{Im}\left[  \partial_{\alpha}\grave
{A}_{\beta},\grave{A}_{\gamma}\right]  ^{a})$

$+\sum_{a\beta}\left(  \frac{\partial%
\mathcal{L}%
_{F}}{\partial G_{\alpha}^{a}}\partial_{\alpha}G_{\beta}^{a}+2\sum_{\gamma
}\left(  \left(  \partial_{\alpha}G_{\beta}^{a}\right)  \partial_{\gamma
}\left(  \frac{\partial%
\mathcal{L}%
_{F}}{\partial%
\mathcal{F}%
_{G\beta\gamma}^{a}}\right)  +\frac{\partial%
\mathcal{L}%
_{F}}{\partial%
\mathcal{F}%
_{G\beta\gamma}^{a}}\left[  \partial_{\alpha}G_{\beta},G_{\gamma}\right]
^{a}\right)  \right)  $

\section{NOETHER\ CURRENTS}

\label{Noether currents}

\subsection{Principles}

For a one parameter group of diffeomorphisms with projectable vector field Y
the Lie derivative is given by the formula \ref{E40} which is equivalent to
the following (Krupka [15] p.44):

\ $\pounds _{j^{1}Y}%
\mathcal{L}%
\varpi_{0}=\sum_{i}\left(  \left(  Y^{i}-z_{\beta}^{i}Y^{\beta}\right)
E_{i}+\frac{d}{d\xi^{\alpha}}\left(  Y^{\alpha}%
\mathcal{L}%
+\frac{\partial%
\mathcal{L}%
}{\partial z_{\alpha}^{i}}\left(  Y^{i}-X^{\beta}z_{\beta}^{i}\right)
\right)  \right)  \varpi_{0}$

that is if Y is vertical ($Y^{\alpha}=0):$

\ $\pounds _{j^{1}Y}%
\mathcal{L}%
\varpi_{0}=\sum_{i}\left(  Y^{i}E_{i}+\frac{d}{d\xi^{\alpha}}\left(
\frac{\partial%
\mathcal{L}%
}{\partial z_{\alpha}^{i}}Y^{i}\right)  \right)  \varpi_{0}$

were $E_{i}\left(
\mathcal{L}%
\right)  =\frac{\partial%
\mathcal{L}%
}{\partial z^{i}}-\sum_{\alpha}\frac{d}{d\xi^{\alpha}}\left(  \frac{\partial%
\mathcal{L}%
}{\partial z_{\alpha}^{i}}\right)  $ are the Lagrange forms

and $%
\mathcal{L}%
=\left(  VL_{M}+L_{F}\right)  \det O^{\prime}$.

On shell $E_{i}\left(
\mathcal{L}%
\right)  =0$ so :$\sum_{i,\alpha}\frac{d}{d\xi^{\alpha}}\left(  \frac{\partial%
\mathcal{L}%
}{\partial z_{\alpha}^{i}}Y^{i}\right)  \varpi_{0}=0$

The equivariance implies for any vertical vector field parametrized by a
section $\zeta$ of $J^{2}F_{M}:$

$\sum_{i}\frac{\partial%
\mathcal{L}%
_{M}}{\partial z^{i}}Y^{i}\left(  Z,\zeta\right)  +\frac{\partial%
\mathcal{L}%
_{M}}{\partial z_{\alpha}^{i}}Y_{\alpha}^{i}\left(  Z,\zeta\right)  =0$

$\sum_{i}\frac{\partial%
\mathcal{L}%
_{F}}{\partial z^{i}}Y^{i}\left(  Z,\zeta\right)  +\frac{\partial%
\mathcal{L}%
_{F}}{\partial z_{\alpha}^{i}}Y_{\alpha}^{i}\left(  Z,\zeta\right)  =0$

with $Y_{\alpha}^{i}\left(  Z,\zeta\right)  =\frac{d}{d\xi^{\alpha}}%
Y^{i}\left(  Z,\zeta\right)  $ that is : $\sum_{i}\frac{\partial%
\mathcal{L}%
}{\partial z^{i}}Y^{i}=-\sum_{i}\frac{\partial%
\mathcal{L}%
}{\partial z_{\alpha}^{i}}\frac{d}{d\xi^{\alpha}}Y^{i}$

Thus the quantity $\sum_{i}E_{i}\left(
\mathcal{L}%
\right)  Y^{i}=\sum_{i}\frac{\partial%
\mathcal{L}%
}{\partial z^{i}}Y^{i}-\sum_{i,\alpha}Y^{i}\frac{d}{d\xi^{\alpha}}\left(
\frac{\partial%
\mathcal{L}%
}{\partial z_{\alpha}^{i}}\right)  $ can be written :

$\sum_{i}E_{i}\left(
\mathcal{L}%
\right)  Y^{i}=-\sum_{i\alpha}\left(  \frac{\partial%
\mathcal{L}%
}{\partial z_{\alpha}^{i}}Y_{\alpha}^{i}+Y^{i}\frac{d}{d\xi^{\alpha}}\left(
\frac{\partial%
\mathcal{L}%
}{\partial z_{\alpha}^{i}}\right)  \right)  $

$=-\sum_{i\alpha}\left(  \frac{\partial%
\mathcal{L}%
}{\partial\partial_{\alpha}z^{i}}\frac{dY^{i}}{d\xi^{\alpha}}+Y^{i}\frac
{d}{d\xi^{\alpha}}\left(  \frac{\partial%
\mathcal{L}%
}{\partial z_{\alpha}^{i}}\right)  \right)  =-\sum_{\alpha}\frac{d}%
{d\xi^{\alpha}}\left(  \sum_{i}\frac{\partial%
\mathcal{L}%
}{\partial\partial_{\alpha}z^{i}}Y^{i}\right)  $

and on shell we get for any vertical field Y : $\sum_{\alpha}\frac{d}%
{d\xi^{\alpha}}\left(  \sum_{i}\frac{\partial%
\mathcal{L}%
}{\partial z_{\alpha}^{i}}Y^{i}\right)  =0.$. If one puts $\zeta=\left(
-\overrightarrow{\kappa}_{a},-\overrightarrow{\theta}_{b}\right)  =$Constant
then for each generator of the gauge group $\sum_{i}\frac{\partial%
\mathcal{L}%
}{\partial\partial_{\alpha}z^{i}}Y^{i}\left(  -\overrightarrow{\kappa}%
_{a},-\overrightarrow{\theta}_{b}\right)  $ is divergence free. But as we have
seen only the partial derivatives such that $\frac{\partial L_{M}}{\partial
z_{\lambda}^{i}},\frac{\partial L_{F}}{\partial z_{\lambda}^{i}}$ are
components of vector fields, so going to the conclusion is not so straightforward.

\bigskip

We will prove that both the gravitational equation and the equation for the
other fields can be written in purely geometrical manner of the kind :
$\varpi_{4}\left(  Y\right)  =\frac{1}{2}d\Pi$ where Y, the "Noether current",
is a vector and $\Pi$ , the "superpotential", is a 2-form.

\bigskip

\subsection{Noether currents for the gravitational field}

\paragraph{1)}

Let us fix $\overrightarrow{\kappa}=\sum_{a}\kappa^{a}\overrightarrow{\kappa
}_{a}=Ct$ then the vector Y has the components :

$Y^{\operatorname{Re}\psi^{ij}}=\sum_{a}\kappa^{a}\operatorname{Re}\left(
\left[  \kappa_{a}\right]  \left[  \psi\right]  \right)  _{j}^{i}$

$Y^{\operatorname{Im}\psi^{ij}}=\sum_{a}\kappa^{a}\operatorname{Im}\left(
\left[  \kappa_{a}\right]  \left[  \psi\right]  \right)  _{j}^{i}$

$Y^{G_{\beta}^{b}}=\sum_{a}\kappa^{a}\left[  \overrightarrow{\kappa}%
_{a},G_{\beta}\right]  ^{b}$

$Y^{O_{\beta}^{\prime i}}=\sum_{a}\kappa^{a}\left(  \left[  \widetilde{\kappa
}_{a}\right]  \left[  O^{\prime}\right]  \right)  _{\beta}^{i}$

V is not involved, so :

$\sum_{i}\frac{\partial%
\mathcal{L}%
}{\partial z_{\alpha}^{i}}Y^{i}\left(  \overrightarrow{\kappa}\right)
=\sum_{aij}\left(  \frac{d%
\mathcal{L}%
}{d\operatorname{Re}\partial_{\alpha}\psi^{ij}}\kappa^{a}\operatorname{Re}%
\left(  \left[  \kappa_{a}\right]  \left[  \psi\right]  \right)  _{j}%
^{i}+\frac{d%
\mathcal{L}%
}{d\operatorname{Im}\partial_{\alpha}\psi^{ij}}\kappa^{a}\operatorname{Im}%
\left(  \left[  \kappa_{a}\right]  \left[  \psi\right]  \right)  _{j}%
^{i}\right)  $

$+\sum_{ab\beta}\frac{d%
\mathcal{L}%
}{d\partial_{\alpha}G_{\beta}^{b}}\kappa^{a}\left[  \overrightarrow{\kappa
}_{a},G_{\beta}\right]  ^{b}+\sum_{ai\beta}\frac{d%
\mathcal{L}%
}{d\partial_{\alpha}O_{\beta}^{\prime i}}\kappa^{a}\left(  \left[
\widetilde{\kappa}_{a}\right]  \left[  O^{\prime}\right]  \right)  _{\beta
}^{i}$

$=\sum_{a}\kappa^{a}\left(  \det O^{\prime}\right)  \{\sum_{ij}\left(
\frac{d\left(  VL_{M}+L_{F}\right)  }{d\operatorname{Re}\partial_{\alpha}%
\psi^{ij}}\operatorname{Re}\left(  \left[  \kappa_{a}\right]  \left[
\psi\right]  \right)  _{j}^{i}+\frac{d\left(  VL_{M}+L_{F}\right)
}{d\operatorname{Im}\partial_{\alpha}\psi^{ij}}\operatorname{Im}\left(
\left[  \kappa_{a}\right]  \left[  \psi\right]  \right)  _{j}^{i}\right)  $

$+\sum_{b\beta}\frac{d\left(  VL_{M}+L_{F}\right)  }{d\partial_{\alpha
}G_{\beta}^{b}}\left[  \overrightarrow{\kappa}_{a},G_{\beta}\right]  ^{b}%
+\sum_{i\beta}\frac{d\left(  VL_{M}+L_{F}\right)  }{d\partial_{\alpha}%
O_{\beta}^{\prime i}}\left(  \left[  \widetilde{\kappa}_{a}\right]  \left[
O^{\prime}\right]  \right)  _{\beta}^{i}\}$

with $%
\mathcal{L}%
=\left(  NL_{M}+L_{F}\right)  \det O^{\prime}$

And :

$\frac{d\left(  NL_{M}+L_{F}\right)  }{d\operatorname{Re}\partial_{\alpha}%
\psi^{ij}}=N\frac{dL_{M}}{d\operatorname{Re}\nabla_{\alpha}\psi^{ij}}%
;\frac{dL}{d\operatorname{Im}\partial_{\alpha}\psi^{ij}}=N\frac{dL_{M}%
}{d\operatorname{Im}\nabla_{\alpha}\psi^{ij}}$

$\frac{d\left(  VL_{M}+L_{F}\right)  }{d\partial_{\alpha}G_{\beta}^{a}}%
=\frac{dL_{F}\left(  \det O^{\prime}\right)  }{d\partial_{\alpha}G_{\alpha
}^{a}}=2\frac{dL_{F}}{d%
\mathcal{F}%
_{G\alpha\beta}^{a}}$

$\frac{d\left(  NL_{M}+L_{F}\right)  }{d\partial_{\alpha}O_{\beta}^{\prime i}%
}=N\frac{dL_{M}}{d\partial_{\alpha}O_{\beta}^{\prime i}}+\frac{dL_{F}%
}{d\partial_{\alpha}O_{\beta}^{\prime i}}$

So :

$\sum_{i}\frac{\partial%
\mathcal{L}%
}{\partial z_{\alpha}^{i}}Y^{i}\left(  \overrightarrow{\kappa}\right)  $

$=\sum_{a}\kappa^{a}\left(  \det O^{\prime}\right)  \{\sum_{ij}N\left(
\frac{dL_{M}}{d\operatorname{Re}\nabla_{\alpha}\psi^{ij}}\operatorname{Re}%
\left(  \left[  \kappa_{a}\right]  \left[  \psi\right]  \right)  _{j}%
^{i}+\frac{dL_{M}}{d\operatorname{Im}\nabla_{\alpha}\psi^{ij}}%
\operatorname{Im}\left(  \left[  \kappa_{a}\right]  \left[  \psi\right]
\right)  _{j}^{i}\right)  $

$+2\sum_{a,\beta}\frac{dL_{F}}{d%
\mathcal{F}%
_{G\alpha\beta}^{b}}\left[  \overrightarrow{\kappa}_{a},G_{\beta}\right]
^{b}+\sum_{i\beta}\left(  N\frac{dL_{M}}{d\partial_{\alpha}O_{\beta}^{\prime
i}}+\frac{dL_{F}}{d\partial_{\alpha}O_{\beta}^{\prime i}}\right)  \left(
\left[  \widetilde{\kappa}_{a}\right]  \left[  O^{\prime}\right]  \right)
_{\beta}^{i}\}$

The identities \ref{E22},\ref{E32} from the gauge equivariance read :

$\forall a,\alpha:\frac{\partial L_{F}}{\partial G_{\alpha}^{a}}=\sum
_{ij}\left[  \widetilde{\kappa}_{a}\right]  _{j}^{i}\sum_{\beta}\frac{dL_{F}%
}{d\partial_{\alpha}O_{\beta}^{\prime i}}O_{\beta}^{\prime j}$

$\forall a,\alpha:\frac{\partial L_{M}}{\partial G_{\alpha}^{a}}=\sum
_{ij}\left[  \widetilde{\kappa}_{a}\right]  _{j}^{i}\sum_{\beta}\frac{dL_{M}%
}{d\partial_{\alpha}O_{\beta\alpha}^{\prime i}}O_{\beta}^{\prime j}$

$\sum_{i}\frac{\partial%
\mathcal{L}%
}{\partial z_{\alpha}^{i}}Y^{i}\left(  \overrightarrow{\kappa}\right)  $

$=\sum_{a}\kappa^{a}\left(  \det O^{\prime}\right)  \{\sum_{ij}N\left(
\frac{dL_{M}}{d\operatorname{Re}\nabla_{\alpha}\psi^{ij}}\operatorname{Re}%
\left(  \left[  \kappa_{a}\right]  \left[  \psi\right]  \right)  _{j}%
^{i}+\frac{dL_{M}}{d\operatorname{Im}\nabla_{\alpha}\psi^{ij}}%
\operatorname{Im}\left(  \left[  \kappa_{a}\right]  \left[  \psi\right]
\right)  _{j}^{i}\right)  $

$+2\sum_{b,\beta}\frac{dL_{F}}{d%
\mathcal{F}%
_{G\alpha\beta}^{b}}\left[  \overrightarrow{\kappa}_{a},G_{\beta}\right]
^{b}+V\frac{\partial L_{M}^{\Diamond}}{\partial G_{\alpha}^{a}}+\frac{\partial
L_{F}}{\partial G_{\alpha}^{a}}\}$

But from the definition of the partial derivatives :

$N\sum_{ij}\left(  \frac{dL_{M}^{\Diamond}}{d\operatorname{Re}\nabla_{\alpha
}\psi^{ij}}\operatorname{Re}\left(  \left[  \kappa_{a}\right]  \left[
\psi\right]  \right)  _{j}^{i}+\frac{dL_{M}^{\Diamond}}{d\operatorname{Im}%
\nabla_{\alpha}\psi^{ij}}\operatorname{Im}\left(  \left[  \kappa_{a}\right]
\left[  \psi\right]  \right)  _{j}^{i}\right)  +N\frac{\partial L_{M}%
^{\Diamond}}{\partial G_{\alpha}^{a}}=N\frac{dL_{M}^{\Diamond}}{dG_{\alpha
}^{a}}$

$\frac{dL_{F}}{dG_{\alpha}^{a}}=\frac{\partial L_{F}}{\partial G_{\alpha}^{a}%
}+\sum_{\mu\lambda}\frac{dL_{F}}{d%
\mathcal{F}%
_{G\lambda\mu}^{b}}\frac{d%
\mathcal{F}%
_{G\lambda\mu}^{b}}{dG_{\alpha}^{a}}$

$=\frac{\partial L_{F}}{\partial G_{\alpha}^{a}}+\sum_{\mu\lambda}\frac
{dL_{F}}{d%
\mathcal{F}%
_{G\lambda\mu}^{b}}\frac{d\left(  G_{cd}^{b}G_{\lambda}^{c}G_{\mu}^{d}\right)
}{dG_{\alpha}^{a}}=\frac{\partial L_{F}}{\partial G_{\alpha}^{a}}%
+2\sum_{b\beta}\frac{dL_{F}}{d%
\mathcal{F}%
_{G\alpha\beta}^{b}}\left[  \overrightarrow{\kappa}_{a},G_{\beta}\right]
^{b}$

So $\sum_{i}\frac{\partial%
\mathcal{L}%
}{\partial z_{\alpha}^{i}}Y^{i}\left(  \overrightarrow{\kappa}\right)
=\sum_{a}\kappa^{a}\left(  \det O^{\prime}\right)  \left(  N\frac
{dL_{M}^{\Diamond}}{dG_{\alpha}^{a}}+\frac{dL_{F}}{dG_{\alpha}^{a}}\right)
=\sum_{a}\kappa^{a}\frac{d%
\mathcal{L}%
}{dG_{\alpha}^{a}}$

The gravitational equation \ref{E45} reads:

$\forall a,\alpha:\frac{\delta S}{\delta G_{\alpha}^{a}}=\frac{d%
\mathcal{L}%
}{dG_{\alpha}^{a}}-\sum_{\beta}\frac{d}{d\xi^{\beta}}\left(  \frac{d%
\mathcal{L}%
}{d\partial_{\beta}G_{\alpha}^{a}}\right)  =0$

so $\sum_{i}\frac{\partial%
\mathcal{L}%
}{\partial z_{\alpha}^{i}}Y^{i}\left(  \overrightarrow{\kappa}\right)
=\sum_{a}\kappa^{a}\left(  \frac{\delta S}{\delta G_{\alpha}^{a}}+\sum_{\beta
}\frac{d}{d\xi^{\beta}}\left(  \frac{d%
\mathcal{L}%
}{d\partial_{\beta}G_{\alpha}^{a}}\right)  \right)  $ and on shell :

$\forall\overrightarrow{\kappa}:\sum_{\alpha}\frac{d}{d\xi^{\alpha}}\left(
\sum_{i}\frac{\partial%
\mathcal{L}%
}{\partial z_{\alpha}^{i}}Y^{i}\left(  \overrightarrow{\kappa}\right)
\right)  =\sum_{a}\kappa^{a}\sum_{\alpha\beta}\frac{d}{d\xi^{\alpha}}\frac
{d}{d\xi^{\beta}}\left(  \frac{d%
\mathcal{L}%
}{d\partial_{\beta}G_{\alpha}^{a}}\right)  =0$ because $\frac{d%
\mathcal{L}%
}{d\partial_{\alpha}G_{\beta}^{a}}+\frac{d%
\mathcal{L}%
}{d\partial_{\beta}G_{\alpha}^{a}}=0$

\paragraph{2)}

$\frac{d\left(  NL_{M}+L_{F}\right)  }{d\partial_{\beta}G_{\alpha}^{a}}%
=\frac{dL_{F}}{d\partial_{\beta}G_{\alpha}^{a}}=2\frac{dL_{F}}{d%
\mathcal{F}%
_{G,\beta\alpha}^{a}}$ are the components of the anti-symmetric bi-vector
field : $\ Z_{G}^{a}=\sum_{\left\{  \alpha\beta\right\}  }\frac{dL_{F}%
}{d\partial_{\alpha}G_{\beta}^{a}}\partial_{\alpha}\wedge\partial_{\beta}$ and
we can define the tensor : $Z_{G}=\sum_{a,\left\{  \alpha\beta\right\}  }%
\frac{dL_{F}}{d\partial_{\alpha}G_{\beta}^{a}}\partial_{\alpha}\wedge
\partial_{\beta}\otimes\overrightarrow{\kappa}_{a}=2\sum_{a,\left\{
\alpha\beta\right\}  }\frac{dL_{F}}{d%
\mathcal{F}%
_{G,\alpha\beta}^{a}}\partial_{\alpha}\wedge\partial_{\beta}\otimes
\overrightarrow{\kappa}_{a}.$ Beware that $Z_{G}$ here and in the following is
defined in respect with $\frac{dL_{F}}{d\partial_{\alpha}G_{\beta}^{a}}$ and
the 2 factor is needed when $Z_{G}$ is computed through $\frac{dL_{F}}{d%
\mathcal{F}%
_{G,\alpha\beta}^{a}}.$

We will compute the 2-form $\Pi_{G}=\varpi_{4}\left(  Z_{G}\right)  $ and its
exterior differential. For this we first establish several formulas which will
be extensively used in the following.

So let be the 2 antisymmetric bi-vector field : $\ Z=\sum_{\left\{
\alpha\beta\right\}  }Z^{\alpha\beta}\partial_{\alpha}\wedge\partial_{\beta} $

\subparagraph{a)}

$\Pi=\varpi_{4}\left(  Z\right)  =\varpi_{4}\left(  \sum_{\left\{  \alpha
\beta\right\}  }Z^{\alpha\beta}\partial_{\alpha}\wedge\partial_{\beta}\right)
$

$\varpi_{0}=\sum_{\alpha_{1}\alpha_{2}\alpha_{3}\alpha_{0}}\epsilon\left(
\alpha_{0},\alpha_{1},\alpha_{2},\alpha_{3}\right)  dx^{\alpha_{0}}\otimes
dx^{\alpha_{1}}\otimes dx^{\alpha_{2}}\otimes dx^{\alpha_{3}} $

$Z=\sum_{\alpha<\beta a}Z^{\alpha\beta}\left(  \partial_{\alpha}%
\otimes\partial_{\beta}-\partial_{\beta}\otimes\partial_{\alpha}\right)  $

$\varpi_{0}\left(  Z\right)  $

$=\sum_{\alpha_{1}\alpha_{2}\alpha_{3}\alpha_{0}}\sum_{\alpha<\beta}%
Z^{\alpha\beta}\epsilon\left(  \alpha_{0},\alpha_{1},\alpha_{2},\alpha
_{3}\right)  $

$\{\left(  dx^{\alpha_{0}}\otimes dx^{\alpha_{1}}\otimes dx^{\alpha_{2}%
}\left(  \partial_{\alpha}\right)  \otimes dx^{\alpha_{3}}\left(
\partial_{\beta}\right)  -dx^{\alpha_{0}}\otimes dx^{\alpha_{1}}\otimes
dx^{\alpha_{2}}\left(  \partial_{\beta}\right)  \otimes dx^{\alpha_{3}}\left(
\partial_{\alpha}\right)  \right)  \}$

$=\sum_{\alpha_{1}\alpha_{2}\alpha_{3}\alpha_{0}}\sum_{\alpha<\beta}%
Z^{\alpha\beta\alpha\beta}\epsilon\left(  \alpha_{0},\alpha_{1},\alpha
_{2},\alpha_{3}\right)  \left(  \delta_{\alpha}^{\alpha_{2}}\delta_{\beta
}^{\alpha_{3}}dx^{\alpha_{0}}\otimes dx^{\alpha_{1}}-\delta_{\beta}%
^{\alpha_{2}}\delta_{\alpha}^{\alpha_{3}}dx^{\alpha_{0}}\otimes dx^{\alpha
_{1}}\right)  $

$=\sum_{\alpha_{0}\alpha_{1}}\sum_{\alpha<\beta}Z^{\alpha\beta}\left(
\epsilon\left(  \alpha_{0},\alpha_{1},\alpha,\beta\right)  dx^{\alpha_{0}%
}\otimes dx^{\alpha_{1}}+\epsilon\left(  \alpha_{0},\alpha_{1},\alpha
,\beta\right)  dx^{\alpha_{0}}\otimes dx^{\alpha_{1}}\right)  $

$=2\sum_{\alpha<\beta}Z^{\alpha\beta}\sum_{\alpha_{0}\alpha_{1}}%
\epsilon\left(  \alpha_{0},\alpha_{1},\alpha,\beta\right)  dx^{\alpha_{0}%
}\otimes dx^{\alpha_{1}}$

$=2\sum_{\alpha<\beta}Z^{\alpha\beta}\left(  \sum_{\alpha_{0}<\alpha_{1}%
}\epsilon\left(  \alpha_{0},\alpha_{1},\alpha,\beta\right)  dx^{\alpha_{0}%
}\otimes dx^{\alpha_{1}}+\sum_{\alpha_{0}>\alpha_{1}}\epsilon\left(
\alpha_{0},\alpha_{1},\alpha,\beta\right)  dx^{\alpha_{0}}\otimes
dx^{\alpha_{1}}\right)  $

$=2\sum_{\alpha<\beta}Z^{\alpha\beta}\left(  \sum_{\alpha_{0}<\alpha_{1}%
}\epsilon\left(  \alpha_{0},\alpha_{1},\alpha,\beta\right)  dx^{\alpha_{0}%
}\otimes dx^{\alpha_{1}}+\sum_{\alpha_{1}>\alpha_{0}}\epsilon\left(
\alpha_{1},\alpha_{0},\alpha,\beta\right)  dx^{\alpha_{1}}\otimes
dx^{\alpha_{0}}\right)  $

$=2\sum_{\alpha<\beta}Z^{\alpha\beta}\sum_{\alpha_{0}<\alpha_{1}}\left(
\epsilon\left(  \alpha_{0},\alpha_{1},\alpha,\beta\right)  dx^{\alpha_{0}%
}\otimes dx^{\alpha_{1}}-\epsilon\left(  \alpha_{0},\alpha_{1},\alpha
,\beta\right)  dx^{\alpha_{1}}\otimes dx^{\alpha_{0}}\right)  $

$=2\sum_{\alpha<\beta}Z^{\alpha\beta}\sum_{\alpha_{0}<\alpha_{1}}%
\epsilon\left(  \alpha_{0},\alpha_{1},\alpha,\beta\right)  \left(
dx^{\alpha_{0}}\otimes dx^{\alpha_{1}}-dx^{\alpha_{1}}\otimes dx^{\alpha_{0}%
}\right)  $

$\varpi_{0}\left(  Z\right)  =2\sum_{\alpha<\beta}Z^{\alpha\beta}\sum
_{\alpha_{0}<\alpha_{1}}\epsilon\left(  \alpha_{0},\alpha_{1},\alpha
,\beta\right)  dx^{\alpha_{0}}\wedge dx^{\alpha_{1}}$

$\Pi=\varpi_{4}\left(  \sum_{\left\{  \alpha\beta\right\}  }Z^{\alpha\beta
}\partial_{\alpha}\wedge\partial_{\beta}\right)  $

$=2\left(  \det O^{\prime}\right)  \sum_{\lambda<\mu}\sum_{\alpha<\beta
,a}Z^{\alpha\beta}\epsilon\left(  \lambda,\mu,\alpha,\beta\right)
dx^{\lambda}\wedge dx^{\mu}$

Expressed in coordinates :

$\Pi=-2\left(  \det O^{\prime}\right)  \{Z^{32}dx^{0}\wedge dx^{1}%
+Z^{13}dx^{.0}\wedge dx^{2}+Z^{21}dx^{0}\wedge dx^{3}$

$+Z^{03}dx^{2}\wedge dx^{1}+Z^{02}dx^{1}\wedge dx^{3}+Z^{01}dx^{3}\wedge
dx^{2}\}$

Notice the choice of indexes : all the formulas are much simpler with this
one.\ Of course it is related to the table 1.

\subparagraph{b)}

Exterior derivative : $d\Pi$

$d\Pi=-2\{\partial_{2}\left(  Z^{32}\left(  \det O^{\prime}\right)  \right)
dx^{2}\wedge dx^{0}\wedge dx^{1}+\partial_{3}\left(  Z^{32}\left(  \det
O^{\prime}\right)  \right)  dx^{3}\wedge dx^{0}\wedge dx^{1}$

$+\partial_{1}\left(  Z^{13}\left(  \det O^{\prime}\right)  \right)
dx^{1}\wedge dx^{.0}\wedge dx^{2}+\partial_{3}\left(  Z^{13}\left(  \det
O^{\prime}\right)  \right)  dx^{3}\wedge dx^{.0}\wedge dx^{2}$

$+\partial_{1}\left(  Z^{21}\left(  \det O^{\prime}\right)  \right)
dx^{1}\wedge dx^{0}\wedge dx^{3}+\partial_{2}\left(  Z^{21}\left(  \det
O^{\prime}\right)  \right)  dx^{2}\wedge dx^{0}\wedge dx^{3}$

$+\partial_{0}\left(  Z^{03}\left(  \det O^{\prime}\right)  \right)
dx^{0}\wedge dx^{2}\wedge dx^{1}+\partial_{3}\left(  Z^{03}\left(  \det
O^{\prime}\right)  \right)  dx^{3}\wedge dx^{2}\wedge dx^{1}$

$+\partial_{0}\left(  Z^{02}\left(  \det O^{\prime}\right)  \right)
dx^{0}\wedge dx^{1}\wedge dx^{3}+\partial_{2}\left(  Z_{Ga}^{02}\left(  \det
O^{\prime}\right)  \right)  dx^{2}\wedge dx^{1}\wedge dx^{3}$

$+\partial_{0}\left(  Z^{01}\left(  \det O^{\prime}\right)  \right)
dx^{0}\wedge dx^{3}\wedge dx^{2}+\partial_{1}\left(  Z^{01}\left(  \det
O^{\prime}\right)  \right)  dx^{1}\wedge dx^{3}\wedge dx^{2}\}$

$=-2\{$

$-\left(  \partial_{3}\left(  Z^{03}\left(  \det O^{\prime}\right)  \right)
-\partial_{2}\left(  Z_{Ga}^{02}\left(  \det O^{\prime}\right)  \right)
-\partial_{1}\left(  Z^{01}\left(  \det O^{\prime}\right)  \right)  \right)
dx^{1}\wedge dx^{2}\wedge dx^{3}$

$+\left(  \partial_{3}\left(  Z^{13}\left(  \det O^{\prime}\right)  \right)
+\partial_{2}\left(  Z^{12}\left(  \det O^{\prime}\right)  \right)
+\partial_{0}\left(  Z^{10}\left(  \det O^{\prime}\right)  \right)  \right)
dx^{0}\wedge dx^{2}\wedge dx^{3}$

$-\left(  \partial_{3}\left(  Z^{23}\left(  \det O^{\prime}\right)  \right)
+\partial_{1}\left(  Z^{21}\left(  \det O^{\prime}\right)  \right)
+\partial_{0}\left(  Z^{20}\left(  \det O^{\prime}\right)  \right)  \right)
dx^{0}\wedge dx^{1}\wedge dx^{3}$

$+\left(  \partial_{2}\left(  Z^{32}\left(  \det O^{\prime}\right)  \right)
+\partial_{1}\left(  Z^{31}\left(  \det O^{\prime}\right)  \right)
+\partial_{0}\left(  Z^{30}\left(  \det O^{\prime}\right)  \right)  \right)
dx^{0}\wedge dx^{1}\wedge dx^{2}\}$

$d\Pi=-2\sum_{\alpha=0}^{3}\sum_{\beta=0}^{3}\left(  -1\right)  ^{\alpha
+1}\left(  \partial_{\beta}\left(  Z^{\alpha\beta}\det O^{\prime}\right)
\right)  dx^{0}\wedge...\wedge\widehat{dx^{\alpha}}\wedge...\wedge dx^{3}$

which is conveniently written as :

$d\Pi=d\left(  \varpi_{4}\left(  \sum_{\left\{  \alpha\beta\right\}
}Z^{\alpha\beta}\partial_{\alpha}\wedge\partial_{\beta}\right)  \right)
=-2\frac{1}{\det O^{\prime}}\sum_{a}\varpi_{4}\left(  \sum_{\alpha\beta
}\partial_{\beta}\left(  Z^{\alpha\beta}\det O^{\prime}\right)  \partial
_{\alpha}\right)  $

\subparagraph{c)}

Let us compute $\left(  \sum_{\gamma}X_{\gamma}dx^{\gamma}\right)  \wedge\Pi$
where $X=\sum_{\gamma}X_{\gamma}dx^{\gamma}\mathbf{\ }$is a one-form:

$\left(  \sum_{\gamma}X_{\gamma}dx^{\gamma}\right)  \wedge\Pi$

$=-2\left(  \det O^{\prime}\right)  \left(  \sum_{\gamma}X_{\gamma}dx^{\gamma
}\right)  \wedge\{Z^{32}dx^{0}\wedge dx^{1}+Z^{13}dx^{.0}\wedge dx^{2}$

$+Z^{21}dx^{0}\wedge dx^{3}+Z^{03}dx^{2}\wedge dx^{1}+Z^{02}dx^{1}\wedge
dx^{3}+Z^{01}dx^{3}\wedge dx^{2}\}$

$=-2\left(  \det O^{\prime}\right)  \{$

$-\left(  Z^{03}X_{3}+Z^{02}X_{2}+Z^{01}X_{1}\right)  dx^{1}\wedge
dx^{2}\wedge dx^{3}$

$+\left(  Z^{13}X_{3}+Z^{12}X_{2}+Z^{10}X_{0}\right)  dx^{0}\wedge
dx^{2}\wedge dx^{3}$

$-\left(  Z^{23}X_{3}+Z^{21}X_{1}+Z^{20}X_{0}\right)  dx^{0}\wedge
dx^{1}\wedge dx^{3}$

$+\left(  Z^{32}X_{2}+Z^{31}X_{1}+Z^{30}X_{0}\right)  dx^{0}\wedge
dx^{1}\wedge dx^{2}\}$

$=-2\left(  \det O^{\prime}\right)  \sum_{\alpha}\left(  -1\right)
^{\alpha+1}\sum_{\gamma}Z^{\alpha\gamma}X_{\gamma}dx^{0}\wedge...\wedge
\widehat{dx^{\alpha}}\wedge...\wedge dx^{3}$

where the symbol \ \symbol{94} over a variable denotes as usual that the
variable shall be omitted.\ 

On the other hand :

$\varpi_{4}\left(  \sum_{\alpha\gamma}Z^{\alpha\gamma}X_{\gamma}%
\partial_{\alpha}\right)  $

$=\left(  \det O^{\prime}\right)  \sum_{\alpha}\left(  -1\right)  ^{\alpha
+1}\sum_{\gamma}Z^{a,\alpha\gamma}X_{\gamma}dx^{0}\wedge...\wedge
\widehat{dx^{\alpha}}\wedge...\wedge dx^{3}$

So one can write :

$\varpi_{4}\left(  \sum_{\alpha\gamma}Z^{\alpha\gamma}X_{\gamma}%
\partial_{\alpha}\right)  =-\frac{1}{2}\left(  \sum_{\gamma}X_{\gamma
}dx^{\gamma}\right)  \wedge\Pi$

$=-\frac{1}{2}\left(  \sum_{\gamma}X_{\gamma}dx^{\gamma}\right)  \wedge
\varpi_{4}\left(  \sum_{\left\{  \alpha\beta\right\}  }Z^{\alpha\beta}%
\partial_{\alpha}\wedge\partial_{\beta}\right)  $

\paragraph{4)}

From these formulas we have :%

\begin{equation}
\Pi_{G}=\varpi_{4}\left(  \sum_{a,\left\{  \alpha\beta\right\}  }\frac{dL_{F}%
}{d\partial_{\alpha}G_{\beta}^{a}}\partial_{\alpha}\wedge\partial_{\beta
}\otimes\overrightarrow{\kappa}_{a}\right)  =2\sum_{\substack{\alpha<\beta,a
\\\lambda<\mu}}\epsilon\left(  \lambda,\mu,\alpha,\beta\right)  \frac{d%
\mathcal{L}%
_{F}}{d\partial_{\alpha}G_{\beta}^{a}}dx^{\lambda}\wedge dx^{\mu}%
\otimes\overrightarrow{\kappa}_{a}\label{E50}%
\end{equation}

This is a 2-form on M valued in o(3,1) called superpotential.

$\Pi_{G}=-2\left(  \det O^{\prime}\right)  \sum_{a}\{Z_{G}^{a,32}dx^{0}\wedge
dx^{1}+Z_{G}^{a,13}dx^{.0}\wedge dx^{2}+Z_{G}^{a,21}dx^{0}\wedge dx^{3}%
+Z_{G}^{a,03}dx^{2}\wedge dx^{1}$

$+Z_{G}^{a,02}dx^{1}\wedge dx^{3}+Z_{G}^{a,01}dx^{3}\wedge dx^{2}%
\}\otimes\overrightarrow{\kappa}_{a}$

And the exterior derivative :

$d\Pi_{G}=-2\frac{1}{\det O^{\prime}}\sum_{a}\varpi_{4}\left(  \sum
_{\alpha\beta}\partial_{\beta}\left(  \frac{d%
\mathcal{L}%
}{d\partial_{\alpha}G_{\beta}^{a}}\right)  \partial_{\alpha}\right)
\otimes\overrightarrow{\kappa}_{a}$

$=-2\sum_{a}\sum_{\alpha\beta=0}^{3}\left(  -1\right)  ^{\alpha+1}%
\partial_{\beta}\left(  \frac{d%
\mathcal{L}%
}{d\partial_{\alpha}G_{\beta}^{a}}\right)  dx^{0}\wedge...\wedge
\widehat{dx^{\alpha}}\wedge...\wedge dx^{3}\otimes\overrightarrow{\kappa}_{a}$

\paragraph{5)}

$\frac{dL}{dG_{\alpha}^{a}}=\sum_{ij}N\left(  \frac{dL_{M}}{d\operatorname{Re}%
\nabla_{\alpha}\psi^{ij}}\operatorname{Re}\left(  \left[  \kappa_{a}\right]
\left[  \psi\right]  \right)  _{j}^{i}+\frac{dL_{M}}{d\operatorname{Im}%
\nabla_{\alpha}\psi^{ij}}\operatorname{Im}\left(  \left[  \kappa_{a}\right]
\left[  \psi\right]  \right)  _{j}^{i}\right)  $

$+2\sum_{b,\beta}\frac{dL_{F}}{d%
\mathcal{F}%
_{G\alpha\beta}^{b}}\left[  \overrightarrow{\kappa}_{a},G_{\beta}\right]
^{b}+\frac{\partial\left(  VL_{M}^{\Diamond}+L_{F}\right)  }{\partial
G_{\alpha}^{a}}$

are the components of a tensor field :%

\begin{equation}
Y_{G}=\sum_{a,\alpha}\frac{d\left(  NL_{M}+L_{F}\right)  }{dG_{\alpha}^{a}%
}\partial_{\alpha}\otimes\overrightarrow{\kappa}_{a}\label{E51}%
\end{equation}

$Y_{G}$ is comprised of one part related to the particles $(L_{M})$ and one
part related to the gravitational field $\left(  L_{F}\right)  $.

\paragraph{6)}

We have :

$\varpi_{4}\left(  Y_{G}\right)  $

$=\sum_{a}\sum_{\alpha=0}^{3}\left(  -1\right)  ^{\alpha+1}\frac{d%
\mathcal{L}%
}{dG_{\alpha}^{a}}dx^{0}\wedge...\wedge\widehat{dx^{\alpha}}\wedge...\wedge
dx^{3}\otimes\overrightarrow{\kappa}_{a}$

$=\left(  \det O^{\prime}\right)  \sum_{\alpha}\sum_{\alpha_{1}\alpha
_{2}\alpha_{3}\alpha_{4}}Y_{G}^{a\alpha}\epsilon\left(  \alpha_{1},\alpha
_{2},\alpha_{3},\alpha_{4}\right)  dx^{\alpha_{1}}\otimes dx^{\alpha_{2}%
}\otimes dx^{\alpha_{3}}\otimes dx^{\alpha_{4}}\left(  \partial_{\alpha
}\right)  $

$=\left(  \det O^{\prime}\right)  \sum_{\alpha}\left(  -1\right)  ^{\alpha
+1}Y_{G}^{a\alpha}dx^{0}\wedge...\wedge\widehat{dx^{\alpha}}\wedge...\wedge
dx^{3}$

So the gravitational equation $\forall a,\alpha:0=\frac{d%
\mathcal{L}%
}{dG_{\alpha}^{a}}-\sum_{\beta}\frac{d}{d\xi^{\beta}}\left(  \frac{d%
\mathcal{L}%
}{d\partial_{\beta}G_{\alpha}^{a}}\right)  $ is equivalent to :

$\sum_{a}\sum_{\alpha=0}^{3}\left(  -1\right)  ^{\alpha+1}\frac{d%
\mathcal{L}%
}{dG_{\alpha}^{a}}dx^{0}\wedge...\wedge\widehat{dx^{\alpha}}\wedge...\wedge
dx^{3}\otimes\overrightarrow{\kappa}_{a}$

$=\sum_{a}\sum_{\alpha=0}^{3}\left(  -1\right)  ^{\alpha+1}\sum_{\beta
}\partial_{\beta}\left(  \frac{d%
\mathcal{L}%
}{d\partial_{\beta}G_{\alpha}^{a}}\right)  dx^{0}\wedge...\wedge
\widehat{dx^{\alpha}}\wedge...\wedge dx^{3}\otimes\overrightarrow{\kappa}_{a}$

$\varpi_{4}\left(  Y_{G}\right)  =-\sum_{a}\sum_{\alpha=0}^{3}\left(
-1\right)  ^{\alpha+1}\sum_{\beta}\partial_{\beta}\left(  \frac{d%
\mathcal{L}%
}{d\partial_{\alpha}G_{\beta}^{a}}\right)  dx^{0}\wedge...\wedge
\widehat{dx^{\alpha}}\wedge...\wedge dx^{3}\otimes\overrightarrow{\kappa}_{a}$%

\begin{equation}
\mathbf{\varpi}_{4}\left(  Y_{G}\right)  \mathbf{=}\frac{1}{2}\mathbf{d\Pi
}_{G}\label{E52}%
\end{equation}

As both quantities are tensors, this a fully geometric equation, which does
not involve coordinates and can be substituted to the gravitational equation.

\paragraph{7)}

The integral $\int_{\Omega\left(  t\right)  }d\varpi_{4}\left(  Y_{G}\right)
=\int_{\Omega\left(  t\right)  }\frac{1}{2}d^{2}\Pi_{G}$ over the region
delimited by S(0) and S(t) is null, but by Stockes theorem : $\int
_{\Omega\left(  t\right)  }d\varpi_{4}\left(  Y_{G}\right)  =0=\int
_{\partial\Omega\left(  t\right)  }\varpi_{4}\left(  Y_{G}\right)  $ .\ If
$Y_{G}=0$ on the rim of each S(t) the flux of the vector field $Y_{G}$ is
conserved : $\int_{S(0)}\varpi_{4}\left(  Y_{G}\right)  =\int_{S(t)}\varpi
_{4}\left(  Y_{G}\right)  .$

\paragraph{8)}

With the various gauge constraints :

$Y_{G}=\sum_{a\alpha}\{\sum_{ij}\left(  \frac{dNL_{M}}{d\operatorname{Re}%
\nabla_{\alpha}\psi^{ij}}\operatorname{Re}\left(  \left[  \kappa_{a}\right]
\left[  \psi\right]  \right)  _{j}^{i}+\frac{dNL_{M}}{d\operatorname{Im}%
\nabla_{\alpha}\psi^{ij}}\operatorname{Im}\left(  \left[  \kappa_{a}\right]
\left[  \psi\right]  \right)  _{j}^{i}\right)  $

$+2\sum_{b,\beta}\frac{dL_{F}}{d%
\mathcal{F}%
_{G\alpha\beta}^{b}}\left[  \overrightarrow{\kappa}_{a},G_{\beta}\right]
^{b}+\frac{\partial L}{\partial G_{\alpha}^{a}}\}\partial_{\alpha}%
\otimes\overrightarrow{\kappa}_{a}$

The equation can be written :

$\forall a:\varpi_{4}\left(  \sum_{\alpha ij}N\left(  \frac{dL_{M}%
}{d\operatorname{Re}\nabla_{\alpha}\psi^{ij}}\operatorname{Re}\left(  \left[
\kappa_{a}\right]  \left[  \psi\right]  \right)  _{j}^{i}+\frac{dL_{M}%
}{d\operatorname{Im}\nabla_{\alpha}\psi^{ij}}\operatorname{Im}\left(  \left[
\kappa_{a}\right]  \left[  \psi\right]  \right)  _{j}^{i}+\frac{\partial
L}{\partial G_{\alpha}^{a}}\right)  \partial_{\alpha}\right)  $

$=\frac{1}{2}d\Pi_{G}^{a}-\varpi_{4}\left(  2\sum_{b,\beta}\frac{dL_{F}}{d%
\mathcal{F}%
_{G\alpha\beta}^{b}}\left[  \overrightarrow{\kappa}_{a},G_{\beta}\right]
^{b}\partial_{\alpha}\right)  $

$\varpi_{4}\left(  2\sum_{b,\beta}\frac{dL_{F}}{d%
\mathcal{F}%
_{G\alpha\beta}^{b}}\left[  \overrightarrow{\kappa}_{a},G_{\beta}\right]
^{b}\partial_{\alpha}\right)  =\sum_{b}\varpi_{4}\left(  \sum_{\beta}%
Z_{G}^{b,\alpha\beta}\left[  \overrightarrow{\kappa}_{a},G_{\beta}\right]
^{b}\partial_{\alpha}\right)  $

$=-\frac{1}{2}\sum_{b}\left(  \sum_{\beta}\left[  \overrightarrow{\kappa}%
_{a},G_{\beta}\right]  ^{b}dx^{\beta}\right)  \wedge\Pi_{G}^{b}$

So the conservation equation reads :

$\forall a:$

$\varpi_{4}\left(  \sum_{\alpha ij}N\left(  \frac{dL_{M}}{d\operatorname{Re}%
\nabla_{\alpha}\psi^{ij}}\operatorname{Re}\left(  \left[  \kappa_{a}\right]
\left[  \psi\right]  \right)  _{j}^{i}+\frac{dL_{M}}{d\operatorname{Im}%
\nabla_{\alpha}\psi^{ij}}\operatorname{Im}\left(  \left[  \kappa_{a}\right]
\left[  \psi\right]  \right)  _{j}^{i}\right)  \partial_{\alpha}\right)
+\varpi_{4}\left(  \frac{\partial L}{\partial G_{\alpha}^{a}}\partial_{\alpha
}\right)  $

$=\frac{1}{2}d\Pi_{G}^{a}-\sum_{b}\left(  \sum_{\beta}\left[  \overrightarrow
{\kappa}_{a},G_{\beta}\right]  ^{b}dx^{\beta}\right)  \wedge\Pi_{G}^{b}$

\subsection{Noether currents for the other fields}

We will proceed in the same way as above, with just one complication coming
from the complex value of the quantities.

\paragraph{1)}

Let us fix $\overrightarrow{\theta}=\sum_{a}\theta^{a}\overrightarrow{\theta
}_{a}=Ct$ then the vector Y has the components :

$Y^{\operatorname{Re}\psi^{ij}}=\sum_{a}\operatorname{Re}\theta^{a}%
\operatorname{Re}\left(  \left[  \psi^{\Diamond}\right]  \left[  \theta
_{a}\right]  ^{t}\right)  _{j}^{i}-\operatorname{Im}\theta^{a}%
\operatorname{Im}\left(  \left[  \psi^{\Diamond}\right]  \left[  \theta
_{a}\right]  ^{t}\right)  _{j}^{i}$

$Y^{\operatorname{Im}\psi^{ij}}=\sum_{a}\operatorname{Re}\theta^{a}%
\operatorname{Im}\left(  \left[  \psi^{\Diamond}\right]  \left[  \theta
_{a}\right]  ^{t}\right)  _{j}^{i}+\operatorname{Im}\theta^{a}%
\operatorname{Re}\left(  \left[  \psi^{\Diamond}\right]  \left[  \theta
_{a}\right]  ^{t}\right)  _{j}^{i}$

$Y^{\operatorname{Re}\grave{A}_{\beta}^{a}}=\sum_{bc}C_{bc}^{a}\left(
\operatorname{Re}\theta^{b}\operatorname{Re}\acute{A}_{\beta}^{c}%
-\operatorname{Im}\theta^{b}\operatorname{Im}\acute{A}_{\beta}^{c}\right)  $

$Y^{\operatorname{Im}\grave{A}_{\beta}^{a}}=\sum_{bc}C_{bc}^{a}\left(
\operatorname{Re}\theta^{b}\operatorname{Im}\acute{A}_{\beta}^{c}%
+\operatorname{Im}\theta^{b}\operatorname{Re}\acute{A}_{\beta}^{c}\right)  $

And :

$\frac{d%
\mathcal{L}%
}{d\operatorname{Re}\partial_{\alpha}\psi^{ij}}=N\frac{dL_{M}\det O^{\prime}%
}{d\operatorname{Re}\nabla_{\alpha}\psi^{ij}};\frac{d%
\mathcal{L}%
}{d\operatorname{Im}\partial_{\alpha}\psi^{ij}}=N\frac{dL_{M}\det O^{\prime}%
}{d\operatorname{Im}\nabla_{\alpha}\psi^{ij}}$

$\frac{d%
\mathcal{L}%
}{d\operatorname{Re}\partial_{\alpha}\grave{A}_{\beta}^{a}}=2\frac{dL_{F}%
}{d\operatorname{Re}%
\mathcal{F}%
_{A,\alpha\beta}^{b}};\frac{d%
\mathcal{L}%
}{d\operatorname{Im}\partial_{\alpha}\grave{A}_{\beta}^{a}}=2\frac{dL_{F}%
}{d\operatorname{Im}%
\mathcal{F}%
_{A,\alpha\beta}^{b}}$

$\left(  \overrightarrow{\theta}_{a}\right)  $ is a real basis of the
complexified $T_{1}U^{c}=T_{1}U\oplus iT_{1}U$ with complex components. The
set $\left(  \overrightarrow{\theta}_{a},i\overrightarrow{\theta}_{a}\right)
$ is a real basis of the real vector space $T_{1}U^{c}$ with the real
components $\operatorname{Re}\theta^{a},\operatorname{Im}\theta^{a}.$

So we have to consider two quantities :

$\sum_{j}\frac{\partial%
\mathcal{L}%
}{\partial z_{\alpha}^{j}}Y^{j}\left(  \operatorname{Re}\overrightarrow
{\theta}\right)  $

$=\sum_{a}\{\sum_{ij}\frac{d%
\mathcal{L}%
}{d\operatorname{Re}\partial_{\alpha}\psi^{ij}}\operatorname{Re}\theta
^{a}\operatorname{Re}\left(  \left[  \psi^{\Diamond}\right]  \left[
\theta_{a}\right]  ^{t}\right)  _{j}^{i}+\frac{d%
\mathcal{L}%
}{d\operatorname{Im}\partial_{\alpha}\psi^{ij}}\operatorname{Re}\theta
^{a}\operatorname{Im}\left(  \left[  \psi^{\Diamond}\right]  \left[
\theta_{a}\right]  ^{t}\right)  _{j}^{i}$

$+\sum_{b\beta}\frac{d%
\mathcal{L}%
}{d\operatorname{Re}\partial_{\alpha}\grave{A}_{\beta}^{a}}C_{bc}%
^{a}\operatorname{Re}\theta^{b}\operatorname{Re}\grave{A}_{\beta}^{c}+\frac{d%
\mathcal{L}%
}{d\operatorname{Im}\partial_{\alpha}\grave{A}_{\beta}^{a}}C_{bc}%
^{a}\operatorname{Re}\theta^{b}\operatorname{Im}\grave{A}_{\beta}^{c}\}$

$=\sum_{a}\left(  \operatorname{Re}\theta^{a}\right)  \{\sum_{ij}N\frac
{dL_{M}\det O^{\prime}}{d\operatorname{Re}\nabla_{\alpha}\psi^{ij}%
}\operatorname{Re}\left(  \left[  \psi^{\Diamond}\right]  \left[  \theta
_{a}\right]  ^{t}\right)  _{j}^{i}+N\frac{dL_{M}\det O^{\prime}}%
{d\operatorname{Im}\nabla_{\alpha}\psi^{ij}}\operatorname{Im}\left(  \left[
\psi^{\Diamond}\right]  \left[  \theta_{a}\right]  ^{t}\right)  _{j}^{i}$

$+2\sum_{b\beta}\frac{dL_{F}\det O^{\prime}}{d\operatorname{Re}%
\mathcal{F}%
_{A,\alpha\beta}^{b}}\operatorname{Re}\left[  \overrightarrow{\theta}%
_{a},\grave{A}_{\beta}\right]  ^{b}+\frac{dL_{F}\det O^{\prime}}%
{d\operatorname{Im}%
\mathcal{F}%
_{A,\alpha\beta}^{b}}\operatorname{Im}\left[  \overrightarrow{\theta}%
_{a},\grave{A}_{\beta}\right]  ^{b}\}$

and

$\sum_{j}\frac{\partial%
\mathcal{L}%
}{\partial z_{\alpha}^{j}}Y^{j}\left(  \operatorname{Im}\overrightarrow
{\theta}\right)  $

$=\sum_{a}\{\sum_{ij}-\frac{d%
\mathcal{L}%
}{d\operatorname{Re}\partial_{\alpha}\psi^{ij}}\operatorname{Im}\theta
^{a}\operatorname{Im}\left(  \left[  \psi^{\Diamond}\right]  \left[
\theta_{a}\right]  ^{t}\right)  _{j}^{i}+\frac{d%
\mathcal{L}%
}{d\operatorname{Im}\partial_{\alpha}\psi^{ij}}\operatorname{Im}\theta
^{a}\operatorname{Re}\left(  \left[  \psi^{\Diamond}\right]  \left[
\theta_{a}\right]  ^{t}\right)  _{j}^{i}$

$+\sum_{b\beta}-\frac{d%
\mathcal{L}%
}{d\operatorname{Re}\partial_{\alpha}\grave{A}_{\beta}^{a}}C_{bc}%
^{a}\operatorname{Im}\theta^{b}\operatorname{Im}\grave{A}_{\beta}^{c}+\frac{d%
\mathcal{L}%
}{d\operatorname{Im}\partial_{\alpha}\grave{A}_{\beta}^{a}}C_{bc}%
^{a}\operatorname{Im}\theta^{b}\operatorname{Re}\grave{A}_{\beta}^{c}\}$

$=\sum_{a}\left(  \operatorname{Im}\theta^{a}\right)  \{N\sum_{ij}%
-\frac{dL_{M}\det O^{\prime}}{d\operatorname{Re}\nabla_{\alpha}\psi^{ij}%
}\operatorname{Im}\left(  \left[  \psi^{\Diamond}\right]  \left[  \theta
_{a}\right]  ^{t}\right)  _{j}^{i}+\frac{dL_{M}\det O^{\prime}}%
{d\operatorname{Im}\nabla_{\alpha}\psi^{ij}}\operatorname{Re}\left(  \left[
\psi^{\Diamond}\right]  \left[  \theta_{a}\right]  ^{t}\right)  _{j}^{i}$

$+2\sum_{b\beta}-\frac{dL_{F}\det O^{\prime}}{d\operatorname{Re}%
\mathcal{F}%
_{A,\alpha\beta}^{b}}\operatorname{Im}\left[  \overrightarrow{\theta}%
_{a},\grave{A}_{\beta}\right]  ^{b}+\frac{dL_{F}\det O^{\prime}}%
{d\operatorname{Im}%
\mathcal{F}%
_{A,\alpha\beta}^{b}}\operatorname{Re}\left[  \overrightarrow{\theta}%
_{a},\grave{A}_{\beta}\right]  ^{b}\}$

That is, with the gauge invariance identities :

$\sum_{j}\frac{\partial%
\mathcal{L}%
}{\partial z_{\alpha}^{j}}Y^{j}\left(  \operatorname{Re}\overrightarrow
{\theta}\right)  =\sum_{a}\left(  \operatorname{Re}\theta^{a}\right)  \frac{d%
\mathcal{L}%
}{d\operatorname{Re}\grave{A}_{\alpha}^{a}}$

$\sum_{j}\frac{\partial%
\mathcal{L}%
}{\partial z_{\alpha}^{j}}Y^{j}\left(  \operatorname{Im}\overrightarrow
{\theta}\right)  =\sum_{a}\left(  \operatorname{Im}\theta^{a}\right)  \frac{d%
\mathcal{L}%
}{d\operatorname{Im}\grave{A}_{\alpha}^{a}}$

The equations \ref{E46},\ref{E47} can be written as :

$\forall a,\alpha:\frac{\delta S}{\delta\operatorname{Re}\grave{A}_{\alpha
}^{a}}=\frac{d%
\mathcal{L}%
}{d\operatorname{Re}\grave{A}_{\alpha}^{a}}-\sum_{\beta}\frac{d}{d\xi^{\beta}%
}\left(  \frac{d%
\mathcal{L}%
}{d\partial_{\beta}\operatorname{Re}\grave{A}_{\alpha}^{a}}\right)  =0$

$\forall a,\alpha:\frac{\delta S}{\delta\operatorname{Im}\grave{A}_{\alpha
}^{a}}=\frac{d%
\mathcal{L}%
}{d\operatorname{Im}\grave{A}_{\alpha}^{a}}-\sum_{\beta}\frac{d}{d\xi^{\beta}%
}\left(  \frac{d%
\mathcal{L}%
}{d\partial_{\beta}\operatorname{Im}\grave{A}_{\alpha}^{a}}\right)  =0$

So :

$\sum_{j}\frac{\partial%
\mathcal{L}%
}{\partial z_{\alpha}^{j}}Y^{j}\left(  \operatorname{Re}\overrightarrow
{\theta}\right)  =\sum_{a}\left(  \operatorname{Re}\theta^{a}\right)  \left(
\frac{\delta S}{\delta\operatorname{Re}\grave{A}_{\alpha}^{a}}+\sum_{\beta
}\frac{d}{d\xi^{\beta}}\left(  \frac{d%
\mathcal{L}%
}{d\partial_{\beta}\operatorname{Re}\grave{A}_{\alpha}^{a}}\right)  \right)  $

$\sum_{j}\frac{\partial%
\mathcal{L}%
}{\partial z_{\alpha}^{j}}Y^{j}\left(  \operatorname{Im}\overrightarrow
{\theta}\right)  =\sum_{a}\left(  \operatorname{Im}\theta^{a}\right)  \left(
\frac{\delta S}{\delta\operatorname{Im}\grave{A}_{\alpha}^{a}}+\sum_{\beta
}\frac{d}{d\xi^{\beta}}\left(  \frac{d%
\mathcal{L}%
}{d\partial_{\beta}\operatorname{Im}\grave{A}_{\alpha}^{a}}\right)  \right)  $

and on shell : $\forall\overrightarrow{\theta}:\sum_{\alpha}\frac{d}%
{d\xi^{\alpha}}\left(  \sum_{i}\frac{\partial%
\mathcal{L}%
}{\partial z_{\alpha}^{i}}Y^{i}\left(  \operatorname{Re}\overrightarrow
{\theta}\right)  \right)  =\sum_{a}\left(  \operatorname{Re}\theta^{a}\right)
\sum_{\alpha\beta}\frac{d}{d\xi^{\alpha}}\frac{d}{d\xi^{\beta}}\left(  \frac{d%
\mathcal{L}%
}{d\partial_{\beta}\operatorname{Re}\grave{A}_{\alpha}^{a}}\right)  =0$
because $\frac{d%
\mathcal{L}%
}{d\partial_{\beta}\operatorname{Re}\grave{A}_{\alpha}^{a}}+\frac{d%
\mathcal{L}%
}{d\partial_{\alpha}\operatorname{Re}\grave{A}_{\beta}^{a}}=0$

\paragraph{2)}

The quantities : $\frac{d%
\mathcal{L}%
}{d\operatorname{Re}\partial_{\alpha}\grave{A}_{\beta}^{a}}=2\frac{dL_{F}%
}{d\operatorname{Re}%
\mathcal{F}%
_{A,\alpha\beta}^{a}},\frac{d%
\mathcal{L}%
}{d\operatorname{Im}\partial_{\alpha}\grave{A}_{\beta}^{a}}=2\frac{dL_{F}%
}{d\operatorname{Im}%
\mathcal{F}%
_{A,\alpha\beta}^{a}}$ are the components of the anti-symmetric 2-vector:

$Z_{AR}^{a}=\sum_{\left\{  \alpha\beta\right\}  }\frac{d%
\mathcal{L}%
}{d\operatorname{Re}\partial_{\alpha}\grave{A}_{\beta}^{a}}\partial_{\alpha
}\wedge\partial_{\beta},Z_{AI}^{a}=\sum_{\left\{  \alpha\beta\right\}  }%
\frac{d%
\mathcal{L}%
}{d\operatorname{Im}\partial_{\alpha}\grave{A}_{\beta}^{a}}\partial_{\alpha
}\wedge\partial_{\beta}.$

The 2-forms $\Pi_{AR}^{a}=\varpi_{4}\left(  Z_{AR}^{a}\right)  ,\Pi_{AI}%
^{a}=\varpi_{4}\left(  Z_{AI}^{a}\right)  $ can be computed as above :%

\begin{equation}
\Pi_{AR}^{a}=\varpi_{4}\left(  Z_{AR}^{a}\right)  =2\sum_{\lambda<\mu}%
\sum_{\alpha<\beta,a}\epsilon\left(  \lambda,\mu,\alpha,\beta\right)  \frac{d%
\mathcal{L}%
}{d\operatorname{Re}\partial_{\alpha}\grave{A}_{\beta}^{a}}dx^{\lambda}\wedge
dx^{\mu}\label{E53}%
\end{equation}

\begin{equation}
\Pi_{AI}^{a}=\varpi_{4}\left(  Z_{AR}^{a}\right)  =2\sum_{\lambda<\mu}%
\sum_{\alpha<\beta,a}\epsilon\left(  \lambda,\mu,\alpha,\beta\right)  \frac{d%
\mathcal{L}%
}{d\operatorname{Im}\partial_{\alpha}\grave{A}_{\beta}^{a}}dx^{\lambda}\wedge
dx^{\mu}\label{E54}%
\end{equation}

They are the superpotentiels of the field \`{A}. Their exterior derivative are :

$d\Pi_{AR}^{a}=-2\sum_{\alpha=0}^{3}\left(  \sum_{\beta=0}^{3}\left(
-1\right)  ^{\alpha+1}\partial_{\beta}\left(  \frac{d%
\mathcal{L}%
}{d\operatorname{Re}\partial_{\alpha}\grave{A}_{\beta}^{a}}\right)  \right)
dx^{0}\wedge...\wedge\widehat{dx^{\alpha}}\wedge...\wedge dx^{3}$

$=-2\frac{1}{\det O^{\prime}}\sum_{a}\varpi_{4}\left(  \sum_{\alpha\beta
}\partial_{\beta}\left(  \frac{d%
\mathcal{L}%
}{d\operatorname{Re}\partial_{\alpha}\grave{A}_{\beta}^{a}}\right)
\partial_{\alpha}\right)  $

$d\Pi_{AI}^{a}=-2\sum_{\alpha=0}^{3}\left(  \sum_{\beta=0}^{3}\left(
-1\right)  ^{\alpha+1}\partial_{\beta}\left(  \frac{d%
\mathcal{L}%
}{d\operatorname{Im}\partial_{\alpha}\grave{A}_{\beta}^{a}}\right)  \right)
dx^{0}\wedge...\wedge\widehat{dx^{\alpha}}\wedge...\wedge dx^{3}$

$=-2\frac{1}{\det O^{\prime}}\sum_{a}\varpi_{4}\left(  \sum_{\alpha\beta
}\partial_{\beta}\left(  \frac{d%
\mathcal{L}%
}{d\operatorname{Im}\partial_{\alpha}\grave{A}_{\beta}^{a}}\right)
\partial_{\alpha}\right)  $

\paragraph{3)}

The quantities :$Y_{AR}^{a,\alpha}=\frac{d\left(  NL_{M}+L_{F}\right)
}{d\operatorname{Re}\grave{A}_{\alpha}^{a}},Y_{AI}^{a,\alpha}=\frac{d\left(
VL_{M}+L_{F}\right)  }{d\operatorname{Im}\grave{A}_{\alpha}^{a}}$ are the
components of vector fields $Y_{AR}^{a}$,$Y_{AI}^{a}$ :%

\begin{equation}
\mathbf{Y}_{AR}^{a}\mathbf{=}\sum_{\alpha}\frac{d\left(  NL_{M}+L_{F}\right)
}{d\operatorname{Re}\grave{A}_{\alpha}^{a}}\mathbf{\}\partial}_{\alpha
}\mathbf{;Y}_{AI}^{a}\mathbf{=}\sum_{\alpha}\frac{d\left(  VL_{M}%
+L_{F}\right)  }{d\operatorname{Im}\grave{A}_{\alpha}^{a}}\mathbf{\}\partial
}_{\alpha}\label{E55}%
\end{equation}

and :

$\varpi_{4}\left(  Y_{AR}^{a}\right)  =\sum_{\alpha=0}^{3}\left(  -1\right)
^{\alpha+1}\left(  Y_{AR}^{a,\alpha}\right)  \left(  \det O^{\prime}\right)
dx^{0}\wedge...\wedge\widehat{dx^{\alpha}}\wedge...\wedge dx^{3}$

$\varpi_{4}\left(  Y_{AI}^{a}\right)  =\sum_{\alpha=0}^{3}\left(  -1\right)
^{\alpha+1}\left(  Y_{AI}^{a,\alpha}\right)  \left(  \det O^{\prime}\right)
dx^{0}\wedge...\wedge\widehat{dx^{\alpha}}\wedge...\wedge dx^{3}$

The equations \ref{E46},\ref{E47} are equivalent to :%

\begin{equation}
\mathbf{\varpi}_{4}\left(  Y_{AR}^{a}\right)  \mathbf{=}\frac{1}%
{2}\mathbf{d\Pi}_{AR}^{a}\mathbf{;\varpi}_{4}\left(  Y_{AI}^{a}\right)
\mathbf{=}\frac{1}{2}\mathbf{d\Pi}_{AI}^{a}\label{E57}%
\end{equation}

As seen previously the flow of the 2m vectors field $Y_{AR}^{a},Y_{AI}^{a}$ is conserved.

\paragraph{4)}

We can proceed to calculations similar as above :

$Y_{AR}^{\alpha a}=N\sum_{ij}\frac{dL_{M}}{d\operatorname{Re}\nabla_{\alpha
}\psi^{ij}}\operatorname{Re}\left(  \left[  \psi^{\Diamond}\right]  \left[
\theta_{a}\right]  ^{t}\right)  _{j}^{i}+\frac{dL_{M}}{d\operatorname{Im}%
\nabla_{\alpha}\psi^{ij}}\operatorname{Im}\left(  \left[  \psi^{\Diamond
}\right]  \left[  \theta_{a}\right]  ^{t}\right)  _{j}^{i}$

$+2\sum_{b\beta}\frac{dL_{F}}{d\operatorname{Re}%
\mathcal{F}%
_{A,\alpha\beta}^{b}}\operatorname{Re}\left[  \overrightarrow{\theta}%
_{a},\grave{A}_{\beta}\right]  ^{b}+\frac{dL_{F}}{d\operatorname{Im}%
\mathcal{F}%
_{A,\alpha\beta}^{b}}\operatorname{Im}\left[  \overrightarrow{\theta}%
_{a},\grave{A}_{\beta}\right]  ^{b}$

$Y_{AI}^{\alpha a}=N\sum_{ij}-\frac{dL_{M}}{d\operatorname{Re}\nabla_{\alpha
}\psi^{ij}}\operatorname{Im}\left(  \left[  \psi^{\Diamond}\right]  \left[
\theta_{a}\right]  ^{t}\right)  _{j}^{i}+\frac{dL_{M}}{d\operatorname{Im}%
\nabla_{\alpha}\psi^{ij}}\operatorname{Re}\left(  \left[  \psi^{\Diamond
}\right]  \left[  \theta_{a}\right]  ^{t}\right)  _{j}^{i}$

$+2\sum_{b\beta}-\frac{dL_{F}}{d\operatorname{Re}%
\mathcal{F}%
_{A,\alpha\beta}^{b}}\operatorname{Im}\left[  \overrightarrow{\theta}%
_{a},\grave{A}_{\beta}\right]  ^{b}+\frac{dL_{F}}{d\operatorname{Im}%
\mathcal{F}%
_{A,\alpha\beta}^{b}}\operatorname{Re}\left[  \overrightarrow{\theta}%
_{a},\grave{A}_{\beta}\right]  ^{b}$

$\varpi_{4}\left(  2\sum_{\beta}\frac{dL_{F}}{d\operatorname{Re}%
\mathcal{F}%
_{A\alpha\beta}^{b}}\operatorname{Re}\left[  \overrightarrow{\theta}%
_{a},\grave{A}_{\beta}\right]  ^{b}\partial_{\alpha}\right)  =\varpi
_{4}\left(  \sum_{\alpha\beta}Z_{AR}^{b\alpha\beta}\operatorname{Re}\left[
\overrightarrow{\theta}_{a},\grave{A}_{\beta}\right]  ^{b}\partial_{\alpha
}\right)  $

$=-\frac{1}{2}\left(  \sum_{\beta}\operatorname{Re}\left[  \overrightarrow
{\theta}_{a},\grave{A}_{\beta}\right]  ^{b}dx^{\beta}\right)  \wedge\Pi
_{AR}^{b}=-\frac{1}{2}\sum_{bc}C_{ac}^{b}\left(  \operatorname{Re}\grave
{A}^{c}\right)  \wedge\Pi_{AR}^{b}$

$\varpi_{4}\left(  2\sum_{\beta}\frac{dL_{F}}{d\operatorname{Im}%
\mathcal{F}%
_{A\alpha\beta}^{b}}\operatorname{Im}\left[  \overrightarrow{\theta}%
_{a},\grave{A}_{\beta}\right]  ^{b}\partial_{\alpha}\right)  =-\frac{1}%
{2}\left(  \sum_{\beta}\operatorname{Im}\left[  \overrightarrow{\theta}%
_{a},\grave{A}_{\beta}\right]  ^{b}dx^{\beta}\right)  \wedge\Pi_{AI}^{b}$

$=-\frac{1}{2}\sum_{bc}C_{ac}^{b}\left(  \operatorname{Im}\grave{A}%
^{c}\right)  \wedge\Pi_{AI}^{b}$

$\varpi_{4}\left(  2\sum_{\beta}\frac{dL_{F}}{d\operatorname{Re}%
\mathcal{F}%
_{A\alpha\beta}^{b}}\operatorname{Im}\left[  \overrightarrow{\theta}%
_{a},\grave{A}_{\beta}\right]  ^{b}\partial_{\alpha}\right)  =-\frac{1}%
{2}\left(  \sum_{\beta}\operatorname{Im}\left[  \overrightarrow{\theta}%
_{a},\grave{A}_{\beta}\right]  ^{b}dx^{\beta}\right)  \wedge\Pi_{AR}^{b}$

$=-\frac{1}{2}\sum_{bc}C_{ac}^{b}\left(  \operatorname{Im}\grave{A}%
^{c}\right)  \wedge\Pi_{AR}^{b}$

$\varpi_{4}\left(  2\sum_{\beta}\frac{dL_{F}}{d\operatorname{Im}%
\mathcal{F}%
_{A\alpha\beta}^{b}}\operatorname{Re}\left[  \overrightarrow{\theta}%
_{a},\grave{A}_{\beta}\right]  ^{b}\partial_{\alpha}\right)  =-\frac{1}%
{2}\left(  \sum_{\beta}\operatorname{Re}\left[  \overrightarrow{\theta}%
_{a},\grave{A}_{\beta}\right]  ^{b}dx^{\beta}\right)  \wedge\Pi_{AI}^{b}$

$=-\frac{1}{2}\sum_{bc}C_{ac}^{b}\left(  \operatorname{Re}\grave{A}%
^{c}\right)  \wedge\Pi_{AI}^{b}$

Thus the conservation equations read :

$\varpi_{4}\left(  N\sum_{ij}\left(  \frac{dL_{M}}{d\operatorname{Re}%
\nabla_{\alpha}\psi^{ij}}\operatorname{Re}\left(  \left[  \psi^{\Diamond
}\right]  \left[  \theta_{a}\right]  ^{t}\right)  _{j}^{i}+\frac{dL_{M}%
}{d\operatorname{Im}\nabla_{\alpha}\psi^{ij}}\operatorname{Im}\left(  \left[
\psi^{\Diamond}\right]  \left[  \theta_{a}\right]  ^{t}\right)  _{j}%
^{i}\right)  \partial_{\alpha}\right)  $

$=\frac{1}{2}\left(  d\Pi_{AR}^{a}+\sum_{bc}C_{ac}^{b}\left(
\operatorname{Re}\grave{A}^{c}\wedge\Pi_{AR}^{b}+\operatorname{Im}\grave
{A}^{c}\wedge\Pi_{AI}^{b}\right)  \right)  $

$\varpi_{4}\left(  N\sum_{ij}\left(  -\frac{dL_{M}}{d\operatorname{Re}%
\nabla_{\alpha}\psi^{ij}}\operatorname{Im}\left(  \left[  \psi^{\Diamond
}\right]  \left[  \theta_{a}\right]  ^{t}\right)  _{j}^{i}+\frac{dL_{M}%
}{d\operatorname{Im}\nabla_{\alpha}\psi^{ij}}\operatorname{Re}\left(  \left[
\psi^{\Diamond}\right]  \left[  \theta_{a}\right]  ^{t}\right)  _{j}%
^{i}\right)  \partial_{\alpha}\right)  $

$=\frac{1}{2}\left(  d\Pi_{AR}^{a}+\sum_{bc}C_{ac}^{b}\left(
-\operatorname{Im}\grave{A}^{c}\wedge\Pi_{AR}^{b}+\operatorname{Re}\grave
{A}^{c}\wedge\Pi_{AI}^{b}\right)  \right)  $

\paragraph{5)}

Remark : obviously we could combine both the real and the imaginary part, that
we will do later on, but so far it does not make the computations simpler.

\section{THE ENERGY MOMENTUM TENSOR}

\label{Energy momentum}

There are several ways to introduce the energy-momentum tensor. Because the
lagrangian does not depend explicitly of m (for covariance reason) the
Lagrange equations admit a first integral which is a conserved quantity. One
can also look for one parameter groups of diffeomorphisms over the cotangent
bundle in a way similar at what we have done for the gauge equivariance. But
here a more direct approach is simpler.\ In a first step we will prove
conservation laws of the kind encountered before, involving a "Noether-like"
current $Y_{H\beta}$\ and a super-potential.\ But from there it is possible to
prove a much stronger result, that we can call "super-conservation laws".

\subsection{Noether-like current}

\paragraph{1)}

Let $\widetilde{Y}_{H\beta}^{\alpha}$ be the quantities : \ $\widetilde
{Y}_{H\beta}^{\alpha}=\sum_{i>0}\frac{\partial L}{\partial z_{\alpha}^{i}%
}\partial_{\beta}z^{i}-\delta_{\beta}^{\alpha}L$ (without V) where
$L=NL_{M}+L_{F}.$ It is easily checked that they are the components of a
tensor.\ Indeed in a change of charts (see "covariance") :

$\widehat{\widetilde{Y}_{H\beta}^{\alpha}}=\sum_{\lambda\mu}\left(  \sum
_{i}K_{\lambda}^{\alpha}\frac{\partial L}{\partial z_{\alpha}^{i}}J_{\beta
}^{\mu}\partial_{\mu}z^{i}-K_{\lambda}^{\alpha}J_{\beta}^{\mu}\delta_{\mu
}^{\lambda}L\right)  =\sum_{\lambda\mu}K_{\lambda}^{\alpha}J_{\beta}^{\mu
}\widetilde{Y}_{H\mu}^{\gamma}$

$\widetilde{Y}_{H}=\sum_{\alpha\beta}\widetilde{Y}_{H\beta}^{\alpha}dx^{\beta
}\otimes\partial_{\alpha}$

So \textit{for }$\beta$\textit{\ fixed} we can consider the vector field

$Y_{H\beta}=\widetilde{Y}_{H\beta}^{\alpha}\partial_{\alpha}-\frac{dNL_{M}%
}{dV^{\beta}}V^{\alpha}\partial_{\alpha}=\sum_{\alpha}\left(  -\frac{dNL_{M}%
}{dV^{\beta}}V^{\alpha}+\sum_{i>1}\frac{\partial L}{\partial z_{\alpha}^{i}%
}\partial_{\beta}z^{i}-\delta_{\beta}^{\alpha}L\right)  \partial_{\alpha}$.

Its value is :%

\begin{equation}
\end{equation}

$Y_{H\beta}^{\alpha}=-\frac{dNL_{M}}{dV^{\beta}}V^{\alpha}+\sum_{ij}%
\frac{dNL_{M}}{d\operatorname{Re}\partial_{\alpha}\psi^{ij}}\operatorname{Re}%
\partial_{\beta}\psi^{ij}+\frac{dNL_{M}}{d\operatorname{Im}\partial_{\alpha
}\psi^{ij}}\operatorname{Im}\partial_{\beta}\psi^{ij}$

$+\sum_{a,\gamma}\frac{dL_{F}}{d\partial_{\alpha}G_{\gamma}^{a}}%
\partial_{\beta}G_{\gamma}^{a}+\frac{dL_{F}}{d\operatorname{Re}\partial
_{\alpha}\grave{A}_{\gamma}^{a}}\operatorname{Re}\partial_{\beta}\grave
{A}_{\gamma}^{a}+\frac{dL_{F}}{d\operatorname{Im}\partial_{\alpha}\grave
{A}_{\gamma}^{a}}\operatorname{Im}\partial_{\beta}\grave{A}_{\gamma}^{a}$

$+\sum_{i\gamma}\frac{dL}{d\partial_{\alpha}O_{\gamma}^{\prime i}}%
\partial_{\beta}O_{\gamma}^{\prime i}-\delta_{\beta}^{\alpha}L$

\bigskip

The index $\beta$\ plays for the "Noether-like" current $Y_{H\beta}$ a role
similar to the indexes "a" in the other Noether currents.\ 

\paragraph{2)}

Using the covariance identities \ref{E37c},\ref{E37d} we have :

$\forall\alpha,\beta:-\frac{dNL_{M}}{dV^{\beta}}V^{\alpha}+N\sum_{i,j}%
\frac{dL_{M}}{d\operatorname{Re}\partial_{\alpha}\psi^{ij}}\operatorname{Re}%
\partial_{\beta}\psi^{ij}+\frac{dL_{M}}{d\operatorname{Im}\partial_{\alpha
}\psi^{ij}}\operatorname{Im}\partial_{\beta}\psi^{ij}$

$=-N(\sum_{a}\frac{dL_{M}}{dG_{\alpha}^{a}}G_{\beta}^{a}+\frac{dL_{M}%
}{d\operatorname{Re}\grave{A}_{\alpha}^{a}}\operatorname{Re}\grave{A}_{\beta
}^{a}+\frac{dL_{M}}{d\operatorname{Im}\grave{A}_{\alpha}^{a}}\operatorname{Im}%
\grave{A}_{\beta}^{a}$

$+\sum_{i}\left(  \frac{dL_{M}}{dO_{\alpha}^{\prime i}}O_{\beta}^{\prime
i}+\sum_{\lambda}\frac{dL_{M}}{d\partial_{\lambda}O_{\alpha}^{i\prime}%
}\partial_{\lambda}O_{\beta}^{\prime i}+\frac{dL_{M}}{d\partial_{\alpha
}O_{\lambda}^{i\prime}}\partial_{\beta}O_{\lambda}^{\prime i}\right)  )$

$\forall\alpha,\beta:\sum_{a,\gamma}\left(  \frac{dL_{F}}{d\operatorname{Re}%
\partial_{\alpha}\grave{A}_{\gamma}^{a}}\operatorname{Re}\left(
\partial_{\beta}\grave{A}_{\gamma}^{a}\right)  +\frac{dL_{F}}%
{d\operatorname{Im}\partial_{\alpha}\grave{A}_{\gamma}^{a}}\operatorname{Im}%
\left(  \partial_{\beta}\grave{A}_{\gamma}^{a}\right)  +\frac{dL_{F}%
}{d\partial_{\alpha}G_{\gamma}^{a}}\left(  \partial_{\beta}G_{\gamma}%
^{a}\right)  \right)  $

$=\sum_{a,\gamma}\left(  \frac{dL_{F}}{d\operatorname{Re}\partial_{\alpha
}\grave{A}_{\gamma}^{a}}\operatorname{Re}\left(  \partial_{\gamma}\grave
{A}_{\beta}^{a}\right)  +\frac{dL_{F}}{d\operatorname{Im}\partial_{\alpha
}\grave{A}_{\gamma}^{a}}\operatorname{Im}\partial_{\gamma}\grave{A}_{\beta
}^{a}+\frac{dL_{F}}{d\partial_{\alpha}G_{\gamma}^{a}}\partial_{\gamma}%
G_{\beta}^{a}\right)  $

$-\sum_{a}\left(  \frac{dL_{F}}{dG_{\alpha}^{a}}G_{\beta}^{a}+\frac{dL_{F}%
}{d\operatorname{Re}\grave{A}_{\alpha}^{a}}\operatorname{Re}\grave{A}_{\beta
}^{a}+\frac{dL_{F}}{d\operatorname{Im}\grave{A}_{\alpha}^{a}}\operatorname{Im}%
\grave{A}_{\beta}^{a}\right)  $

$-\sum_{i}\left(  \frac{dL_{F}}{dO_{\alpha}^{\prime i}}O_{\beta}^{\prime
i}+\sum_{\lambda}\frac{dL_{F}}{d\partial_{\lambda}O_{\alpha}^{i\prime}%
}\partial_{\lambda}O_{\beta}^{\prime i}+\frac{dL_{F}}{d\partial_{\alpha
}O_{\lambda}^{i\prime}}\partial_{\beta}O_{\lambda}^{\prime i}\right)  $

So :

$Y_{H\beta}^{\alpha}=-\sum_{a}\left(  \frac{dNL_{M}}{dG_{\alpha}^{a}}G_{\beta
}^{a}+\frac{dNL_{M}}{d\operatorname{Re}\grave{A}_{\alpha}^{a}}%
\operatorname{Re}\grave{A}_{\beta}^{a}+\frac{dNL_{M}}{d\operatorname{Im}%
\grave{A}_{\alpha}^{a}}\operatorname{Im}\grave{A}_{\beta}^{a}\right)  $

$-\sum_{i}\left(  \frac{dNL_{M}}{dO_{\alpha}^{\prime i}}O_{\beta}^{\prime
i}+\sum_{\lambda}\frac{dNL_{M}}{d\partial_{\lambda}O_{\alpha}^{i\prime}%
}\partial_{\lambda}O_{\beta}^{\prime i}+\frac{dNL_{M}}{d\partial_{\alpha
}O_{\lambda}^{i\prime}}\partial_{\beta}O_{\lambda}^{\prime i}\right)  $

$+\sum_{a,\gamma}\left(  \frac{dL_{F}}{d\operatorname{Re}\partial_{\alpha
}\grave{A}_{\gamma}^{a}}\operatorname{Re}\left(  \partial_{\gamma}\grave
{A}_{\beta}^{a}\right)  +\frac{dL_{F}}{d\operatorname{Im}\partial_{\alpha
}\grave{A}_{\gamma}^{a}}\operatorname{Im}\partial_{\gamma}\grave{A}_{\beta
}^{a}+\frac{dL_{F}}{d\partial_{\alpha}G_{\gamma}^{a}}\partial_{\gamma}%
G_{\beta}^{a}\right)  $

$-\sum_{a}\left(  \frac{dL_{F}}{dG_{\alpha}^{a}}G_{\beta}^{a}+\frac{dL_{F}%
}{d\operatorname{Re}\grave{A}_{\alpha}^{a}}\operatorname{Re}\grave{A}_{\beta
}^{a}+\frac{dL_{F}}{d\operatorname{Im}\grave{A}_{\alpha}^{a}}\operatorname{Im}%
\grave{A}_{\beta}^{a}\right)  $

$-\sum_{i}\left(  \frac{dL_{F}}{dO_{\alpha}^{\prime i}}O_{\beta}^{\prime
i}+\sum_{\lambda}\frac{dL_{F}}{d\partial_{\lambda}O_{\alpha}^{i\prime}%
}\partial_{\lambda}O_{\beta}^{\prime i}+\frac{dL_{F}}{d\partial_{\alpha
}O_{\lambda}^{i\prime}}\partial_{\beta}O_{\lambda}^{\prime i}\right)
+\sum_{i\gamma}\frac{dL}{d\partial_{\alpha}O_{\gamma}^{\prime i}}%
\partial_{\beta}O_{\gamma}^{\prime i}-\delta_{\beta}^{\alpha}L$

$=-\sum_{a}\left(  \frac{dL}{dG_{\alpha}^{a}}G_{\beta}^{a}+\frac
{dL}{d\operatorname{Re}\grave{A}_{\alpha}^{a}}\operatorname{Re}\grave
{A}_{\beta}^{a}+\frac{dL}{d\operatorname{Im}\grave{A}_{\alpha}^{a}%
}\operatorname{Im}\grave{A}_{\beta}^{a}\right)  $

$+\sum_{a,\gamma}\left(  \frac{dL}{d\operatorname{Re}\partial_{\alpha}%
\grave{A}_{\gamma}^{a}}\operatorname{Re}\partial_{\gamma}\grave{A}_{\beta}%
^{a}+\frac{dL}{d\operatorname{Im}\partial_{\alpha}\grave{A}_{\gamma}^{a}%
}\operatorname{Im}\partial_{\gamma}\grave{A}_{\beta}^{a}+\frac{dL}%
{d\partial_{\alpha}G_{\gamma}^{a}}\partial_{\gamma}G_{\beta}^{a}\right)  $

$-\sum_{i}\left(  \frac{dL}{dO_{\alpha}^{\prime i}}O_{\beta}^{\prime i}%
+\sum_{\gamma}\frac{dL}{d\partial_{\gamma}O_{\alpha}^{i\prime}}\partial
_{\gamma}O_{\beta}^{\prime i}\right)  -\delta_{\beta}^{\alpha}L$

$Y_{H\beta}^{\alpha}\det O^{\prime}$

$=-\sum_{a}\left(  \frac{d%
\mathcal{L}%
}{dG_{\alpha}^{a}}G_{\beta}^{a}+\frac{d%
\mathcal{L}%
}{d\operatorname{Re}\grave{A}_{\alpha}^{a}}\operatorname{Re}\grave{A}_{\beta
}^{a}+\frac{d%
\mathcal{L}%
}{d\operatorname{Im}\grave{A}_{\alpha}^{a}}\operatorname{Im}\grave{A}_{\beta
}^{a}\right)  $

$+\sum_{a,\gamma}\left(  \frac{d%
\mathcal{L}%
}{d\operatorname{Re}\partial_{\alpha}\grave{A}_{\gamma}^{a}}\operatorname{Re}%
\partial_{\gamma}\grave{A}_{\beta}^{a}+\frac{d%
\mathcal{L}%
}{d\operatorname{Im}\partial_{\alpha}\grave{A}_{\gamma}^{a}}\operatorname{Im}%
\partial_{\gamma}\grave{A}_{\beta}^{a}+\frac{d%
\mathcal{L}%
}{d\partial_{\alpha}G_{\gamma}^{a}}\partial_{\gamma}G_{\beta}^{a}\right)  $

$-\sum_{i}\left(  \frac{d%
\mathcal{L}%
}{dO_{\alpha}^{\prime i}}O_{\beta}^{\prime i}+\sum_{\gamma}\frac{d%
\mathcal{L}%
}{d\partial_{\gamma}O_{\alpha}^{i\prime}}\partial_{\gamma}O_{\beta}^{\prime
i}\right)  $

where we used :

$\sum_{i}\frac{dL}{dO_{\alpha}^{\prime i}}O_{\beta}^{\prime i}\det O^{\prime
}=\sum_{i}\frac{d%
\mathcal{L}%
}{dO_{\alpha}^{\prime i}}O_{\beta}^{\prime i}-\sum_{i}\frac{d\det O^{\prime}%
}{dO_{\alpha}^{\prime i}}LO_{\beta}^{\prime i}$

$=\sum_{i}\frac{d%
\mathcal{L}%
}{dO_{\alpha}^{\prime i}}O_{\beta}^{\prime i}-\sum_{i}O_{i}^{\alpha}\left(
\det O^{\prime}\right)  LO_{\beta}^{\prime i}=\sum_{i}\frac{d%
\mathcal{L}%
}{dO_{\alpha}^{\prime i}}O_{\beta}^{\prime i}-\delta_{\beta}^{\alpha}%
\mathcal{L}%
$

$Y_{H\beta}^{\alpha}\det O^{\prime}$

$=-\sum_{a}\left(  \frac{d%
\mathcal{L}%
}{dG_{\alpha}^{a}}G_{\beta}^{a}+\frac{d%
\mathcal{L}%
}{d\operatorname{Re}\grave{A}_{\alpha}^{a}}\operatorname{Re}\grave{A}_{\beta
}^{a}+\frac{d%
\mathcal{L}%
}{d\operatorname{Im}\grave{A}_{\alpha}^{a}}\operatorname{Im}\grave{A}_{\beta
}^{a}\right)  $

$+\sum_{a,\gamma}\{\frac{d}{d\xi^{\gamma}}\left(  \frac{d%
\mathcal{L}%
}{d\operatorname{Re}\partial_{\alpha}\grave{A}_{\gamma}^{a}}\operatorname{Re}%
\grave{A}_{\beta}^{a}\right)  -\left(  \frac{d}{d\xi^{\gamma}}\frac{d%
\mathcal{L}%
}{d\operatorname{Re}\partial_{\alpha}\grave{A}_{\gamma}^{a}}\right)
\operatorname{Re}\grave{A}_{\beta}^{a}+\frac{d}{d\xi^{\gamma}}\left(  \frac{d%
\mathcal{L}%
}{d\operatorname{Im}\partial_{\alpha}\grave{A}_{\gamma}^{a}}\operatorname{Im}%
\grave{A}_{\beta}^{a}\right)  $

$-\left(  \frac{d}{d\xi^{\gamma}}\frac{d%
\mathcal{L}%
}{d\operatorname{Im}\partial_{\alpha}\grave{A}_{\gamma}^{a}}\right)
\operatorname{Im}\grave{A}_{\beta}^{a}+\frac{d}{d\xi^{\gamma}}\left(  \frac{d%
\mathcal{L}%
}{d\partial_{\alpha}G_{\gamma}^{a}}G_{\beta}^{a}\right)  -\left(  \frac
{d}{d\xi^{\gamma}}\frac{d%
\mathcal{L}%
}{d\partial_{\alpha}G_{\gamma}^{a}}\right)  G_{\beta}^{a}\}$

$-\sum_{i}\left(  \frac{d%
\mathcal{L}%
}{dO_{\alpha}^{\prime i}}O_{\beta}^{\prime i}+\sum_{\gamma}\frac{d}%
{d\xi^{\gamma}}\left(  \frac{d%
\mathcal{L}%
}{d\partial_{\gamma}O_{\alpha}^{i\prime}}O_{\beta}^{\prime i}\right)  -\left(
\frac{d}{d\xi^{\gamma}}\frac{d%
\mathcal{L}%
}{d\partial_{\gamma}O_{\alpha}^{i\prime}}\right)  O_{\beta}^{\prime i}\right)
$

$=-\sum_{a}\left(  \frac{d%
\mathcal{L}%
}{dG_{\alpha}^{a}}-\frac{d}{d\xi^{\gamma}}\frac{d%
\mathcal{L}%
}{d\partial_{\gamma}G_{\alpha}^{a}}\right)  G_{\beta}^{a}-\sum_{i}\left(
\left(  \frac{d%
\mathcal{L}%
}{dO_{\alpha}^{\prime i}}-\frac{d}{d\xi^{\gamma}}\frac{d%
\mathcal{L}%
}{d\partial_{\gamma}O_{\alpha}^{i\prime}}\right)  O_{\beta}^{\prime i}\right)
$

$+\sum_{a}\left(  \frac{d%
\mathcal{L}%
}{d\operatorname{Re}\grave{A}_{\alpha}^{a}}-\frac{d}{d\xi^{\gamma}}\frac{d%
\mathcal{L}%
}{d\operatorname{Re}\partial_{\gamma}\grave{A}_{\alpha}^{a}}\right)
\operatorname{Re}\grave{A}_{\beta}^{a}+\left(  \frac{d%
\mathcal{L}%
}{d\operatorname{Im}\grave{A}_{\alpha}^{a}}-\frac{d}{d\xi^{\gamma}}\frac{d%
\mathcal{L}%
}{d\operatorname{Im}\partial_{\gamma}\grave{A}_{\alpha}^{a}}\right)
\operatorname{Im}\grave{A}_{\beta}^{a}\}$

$+\sum_{\gamma}\frac{d}{d\xi^{\gamma}}\left(  \sum_{a}\frac{d%
\mathcal{L}%
}{d\operatorname{Re}\partial_{\alpha}\grave{A}_{\gamma}^{a}}\operatorname{Re}%
\grave{A}_{\beta}^{a}+\frac{d%
\mathcal{L}%
}{d\operatorname{Im}\partial_{\alpha}\grave{A}_{\gamma}^{a}}\operatorname{Im}%
\grave{A}_{\beta}^{a}+\frac{d%
\mathcal{L}%
}{d\partial_{\alpha}G_{\gamma}^{a}}G_{\beta}^{a}-\sum_{i}\frac{d%
\mathcal{L}%
}{d\partial_{\gamma}O_{\alpha}^{i\prime}}O_{\beta}^{\prime i}\right)  $

where we used :$\frac{d%
\mathcal{L}%
}{d\partial_{\gamma}G_{\alpha}^{a}}=-\frac{d%
\mathcal{L}%
}{d\partial_{\alpha}G_{\gamma}^{a}},\frac{d%
\mathcal{L}%
}{d\operatorname{Re}\partial_{\gamma}\grave{A}_{\alpha}^{a}}=-\frac{d%
\mathcal{L}%
}{d\operatorname{Re}\partial_{\alpha}\grave{A}_{\gamma}^{a}},\frac{d%
\mathcal{L}%
}{d\operatorname{Im}\partial_{\gamma}\grave{A}_{\alpha}^{a}}=-\frac{d%
\mathcal{L}%
}{d\operatorname{Im}\partial_{\alpha}\grave{A}_{\gamma}^{a}}$

$Y_{H\beta}^{\alpha}\det O^{\prime}=-\sum_{a}\left(  \frac{\delta S}{\delta
G_{\alpha}^{a}}G_{\beta}^{a}+\frac{\delta S}{\delta\operatorname{Re}\grave
{A}_{\alpha}^{a}}\operatorname{Re}\grave{A}_{\beta}^{a}+\frac{\delta%
\mathcal{L}%
}{\delta\operatorname{Im}\grave{A}_{\alpha}^{a}}\operatorname{Im}\grave
{A}_{\beta}^{a}\right)  -\sum_{i}\frac{\delta S}{\delta O_{\alpha}^{\prime i}%
}O_{\beta}^{\prime i}$

$+\sum_{\gamma}\frac{d}{d\xi^{\gamma}}\left(  \sum_{a}\frac{d%
\mathcal{L}%
}{d\operatorname{Re}\partial_{\gamma}\grave{A}_{\alpha}^{a}}\operatorname{Re}%
\grave{A}_{\beta}^{a}+\frac{d%
\mathcal{L}%
}{d\operatorname{Im}\partial_{\gamma}\grave{A}_{\alpha}^{a}}\operatorname{Im}%
\grave{A}_{\beta}^{a}+\frac{d%
\mathcal{L}%
}{d\partial_{\gamma}G_{\alpha}^{a}}G_{\beta}^{a}+\sum_{i}\frac{d%
\mathcal{L}%
}{d\partial_{\gamma}O_{\alpha}^{i\prime}}O_{\beta}^{\prime i}\right)  $

Thus on shell :

$Y_{H\beta}^{\alpha}\det O^{\prime}$

$=\sum_{\gamma}\frac{d}{d\xi^{\gamma}}\left(  \sum_{a}\frac{d%
\mathcal{L}%
}{d\operatorname{Re}\partial_{\gamma}\grave{A}_{\alpha}^{a}}\operatorname{Re}%
\grave{A}_{\beta}^{a}+\frac{d%
\mathcal{L}%
}{d\operatorname{Im}\partial_{\gamma}\grave{A}_{\alpha}^{a}}\operatorname{Im}%
\grave{A}_{\beta}^{a}+\frac{d%
\mathcal{L}%
}{d\partial_{\gamma}G_{\alpha}^{a}}G_{\beta}^{a}+\sum_{i}\frac{d%
\mathcal{L}%
}{d\partial_{\gamma}O_{\alpha}^{i\prime}}O_{\beta}^{\prime i}\right)  $

that hints at some kind of exterior derivative of a form. The trouble comes
from $\frac{d%
\mathcal{L}%
}{d\partial_{\gamma}O_{\alpha}^{i\prime}}$ which is not an antisymmetric 2-vector.

\subsection{Superpotential}

\paragraph{1)}

Let be the quantities

$Z_{H\beta}^{\alpha\gamma}=\sum_{a}\frac{dL}{d\operatorname{Re}\partial
_{\alpha}\grave{A}_{\gamma}^{a}}\operatorname{Re}\grave{A}_{\beta}^{a}%
+\frac{dL}{d\operatorname{Im}\partial_{\alpha}\grave{A}_{\gamma}^{a}%
}\operatorname{Im}\grave{A}_{\beta}^{a}+\frac{dL}{d\partial_{\alpha}G_{\gamma
}^{a}}G_{\beta}^{a}+\sum_{i}\left(  \frac{dL}{d\partial_{\alpha}O_{\gamma
}^{i\prime}}-\frac{dL}{d\partial_{\gamma}O_{\alpha}^{i\prime}}\right)
O_{\beta}^{\prime i}.$

and keep $\beta$\ fixed. They are the components of an antisymmetric 2-vector
field $Z_{H\beta}=\sum_{\left\{  \alpha\gamma\right\}  }Z_{H\beta}%
^{\alpha\gamma}\partial_{\alpha}\wedge\partial_{\gamma}$ on M.\ Indeed in a
change of chart we have (see section "covariance") :

$\widehat{Z_{H\beta}^{\alpha\gamma}}=\sum_{\lambda\mu\nu}\sum_{a}K_{\lambda
}^{\alpha}K_{\mu}^{\gamma}\frac{dL}{d\operatorname{Re}\partial_{\lambda}%
\grave{A}_{\mu}^{a}}\operatorname{Re}J_{\beta}^{\nu}\grave{A}_{\nu}%
^{a}+K_{\lambda}^{\alpha}K_{\mu}^{\gamma}\frac{dL}{d\operatorname{Im}%
\partial_{\lambda}\grave{A}_{\mu}^{a}}\operatorname{Im}J_{\beta}^{\nu}%
\grave{A}_{\nu}^{a}$

$+K_{\lambda}^{\alpha}K_{\mu}^{\gamma}\frac{dL}{d\partial_{\lambda}G_{\mu}%
^{a}}J_{\beta}^{\nu}G_{\nu}^{a}+\sum_{i}\left(  K_{\lambda}^{\alpha}K_{\mu
}^{\gamma}\frac{dL}{d\partial_{\lambda}O_{\mu}^{i\prime}}-K_{\lambda}^{\gamma
}K_{\mu}^{\alpha}\frac{dL}{d\partial_{\mu}O_{\lambda}^{i\prime}}\right)
J_{\beta}^{\nu}O_{\nu}^{\prime i}$

$\widehat{Z_{H\beta}^{\alpha\gamma}}=\sum_{\lambda\mu\nu}K_{\lambda}^{\alpha
}K_{\mu}^{\gamma}J_{\beta}^{\nu}Z_{H\nu}^{\lambda\mu}$

We have :

$\sum_{\left\{  \alpha\gamma\right\}  }Z_{H\beta}^{\alpha\gamma}%
\partial_{\alpha}\wedge\partial_{\gamma}$

$=\sum_{\left\{  \alpha\gamma\right\}  }\{\sum_{a}\left(  \frac{dL}%
{d\operatorname{Re}\partial_{\alpha}\grave{A}_{\gamma}^{a}}\partial_{\alpha
}\wedge\partial_{\gamma}\right)  \operatorname{Re}\grave{A}_{\beta}%
^{a}+\left(  \frac{dL}{d\operatorname{Im}\partial_{\alpha}\grave{A}_{\gamma
}^{a}}\partial_{\alpha}\wedge\partial_{\gamma}\right)  \operatorname{Im}%
\grave{A}_{\beta}^{a}$

$+\left(  \frac{dL}{d\partial_{\alpha}G_{\gamma}^{a}}\partial_{\alpha}%
\wedge\partial_{\gamma}\right)  G_{\beta}^{a}+\sum_{i}\left(  \left(
\frac{dL}{d\partial_{\alpha}O_{\gamma}^{i\prime}}-\frac{dL}{d\partial_{\gamma
}O_{\alpha}^{i\prime}}\right)  \partial_{\alpha}\wedge\partial_{\gamma
}\right)  O_{\beta}^{\prime i}\}$

$Z_{H\beta}=\sum_{a}Z_{AR}^{a}\operatorname{Re}\grave{A}_{\beta}^{a}%
+Z_{AI}^{a}\operatorname{Im}\grave{A}_{\beta}^{a}+Z_{G}^{a}G_{\beta}^{a}%
+\sum_{i}O_{\beta}^{\prime i}Z_{O}^{i}$

with $Z_{O}^{i}=\sum_{\left\{  \alpha\gamma\right\}  }\left(  \frac
{dL}{d\partial_{\alpha}O_{\gamma}^{i\prime}}-\frac{dL}{d\partial_{\gamma
}O_{\alpha}^{i\prime}}\right)  \partial_{\alpha}\wedge\partial_{\gamma} $

\paragraph{2)}

Compute the superpotential $\Pi_{H\beta}=\varpi_{4}\left(  Z_{H\beta}\right)
.$ By the same calculation as above we get :

$\Pi_{H\beta}=\varpi_{4}\left(  Z_{H\beta}\right)  =2\left(  \det O^{\prime
}\right)  \sum_{\lambda<\mu}\sum_{\alpha<\gamma}\epsilon\left(  \lambda
,\mu,\alpha,\gamma\right)  Z_{H\beta}^{\alpha\gamma}\left(  dx^{\lambda}\wedge
dx^{\mu}\right)  $ is a 2-form on M :

$\Pi_{H\beta}=-2\left(  \det O^{\prime}\right)  \sum_{a}\{Z_{H\beta}%
^{32}dx^{0}\wedge dx^{1}+Z_{H\beta}^{13}dx^{.0}\wedge dx^{2}+Z_{H\beta}%
^{21}dx^{0}\wedge dx^{3}+Z_{H\beta}^{03}dx^{2}\wedge dx^{1}+Z_{H\beta}%
^{02}dx^{1}\wedge dx^{3}+Z_{H\beta}^{01}dx^{3}\wedge dx^{2}\}$

$\Pi_{H\beta}=\sum_{a}\left(  \operatorname{Re}\grave{A}_{\beta}^{a}\right)
\Pi_{AR}^{a}+\left(  \operatorname{Im}\grave{A}_{\beta}^{a}\right)  \Pi
_{AI}^{a}+G_{\beta}^{a}\Pi_{G}^{a}+\sum_{i}O_{\beta}^{\prime i}\Pi_{O}^{i}$

with $\Pi_{O}^{i}=\varpi_{4}\left(  Z_{O}^{i}\right)  $

Its exterior derivative is a 3-form over M :

$d\Pi_{H\beta}$

$=-2\sum_{\alpha=0}^{3}\left(  -1\right)  ^{\alpha+1}\left(  \sum_{\gamma
=0}^{3}\partial_{\gamma}\left(  Z_{H\beta}^{\alpha\gamma}\det O^{\prime
}\right)  \right)  dx^{0}\wedge...\wedge\widehat{dx^{\alpha}}\wedge...\wedge
dx^{3}$

$=-2\frac{1}{\det O^{\prime}}\varpi_{4}\left(  \sum_{\alpha\gamma}\left(
\partial_{\gamma}\left(  Z_{H\beta}^{\alpha\gamma}\det O^{\prime}\right)
\right)  \partial_{\alpha}\right)  $

$=2\frac{1}{\det O^{\prime}}\varpi_{4}\left(  \sum_{\alpha\gamma}\left(
\partial_{\gamma}\left(  Z_{H\beta}^{\gamma\alpha}\det O^{\prime}\right)
\right)  \partial_{\alpha}\right)  $

$=2\sum_{\alpha=0}^{3}\left(  -1\right)  ^{\alpha+1}\left(  \sum_{\gamma
=0}^{3}\partial_{\gamma}\left(  Z_{H\beta}^{\gamma\alpha}\det O^{\prime
}\right)  \right)  dx^{0}\wedge...\wedge\widehat{dx^{\alpha}}\wedge...\wedge
dx^{3}$

that we can write : $d\Pi_{H\beta}=2\frac{1}{\det O^{\prime}}\varpi_{4}\left(
\sum_{\alpha\gamma}\left(  \frac{d}{d\xi^{\gamma}}\left(  Z_{H\beta}%
^{\gamma\alpha}\det O^{\prime}\right)  \right)  \partial_{\alpha}\right)  $
keeping in mind that when the derivative involves $L_{M}$ we must take the
composite function with f.

$d\Pi_{H\beta}=\frac{2}{\det O^{\prime}}\times$

$\varpi_{4}\left(  \sum_{\alpha\gamma}\frac{d}{d\xi^{\gamma}}\left(
\begin{array}
[c]{c}%
\sum_{a}\frac{d%
\mathcal{L}%
}{d\operatorname{Re}\partial_{\gamma}\grave{A}_{\alpha}^{a}}\operatorname{Re}%
\grave{A}_{\beta}^{a}+\frac{d%
\mathcal{L}%
}{d\operatorname{Im}\partial_{\gamma}\grave{A}_{\alpha}^{a}}\operatorname{Im}%
\grave{A}_{\beta}^{a}\\
+\frac{d%
\mathcal{L}%
}{d\partial_{\gamma}G_{\alpha}^{a}}G_{\beta}^{a}+\sum_{i}\left(  \frac{d%
\mathcal{L}%
}{d\partial_{\gamma}O_{\alpha}^{i\prime}}-\frac{d%
\mathcal{L}%
}{d\partial_{\alpha}O_{\gamma}^{i\prime}}\right)  O_{\beta}^{\prime i}%
\end{array}
\right)  \right)  dx^{0}\wedge..\widehat{dx^{\alpha}}..dx^{3}$

\subsection{Conservation law}

The value of $\varpi_{4}\left(  Y_{H\beta}\right)  $ is the 3-form over M :

$\varpi_{4}\left(  Y_{H\beta}\right)  =\sum_{\alpha=0}^{3}\left(  -1\right)
^{\alpha+1}Y_{H\beta}^{\alpha}\det O^{\prime}dx^{0}\wedge...\wedge
\widehat{dx^{\alpha}}\wedge...\wedge dx^{3}$

On shell we have :

$\sum_{\alpha=0}^{3}\left(  -1\right)  ^{\alpha+1}Y_{H\beta}^{\alpha}\left(
\det O^{\prime}\right)  dx^{0}\wedge...\widehat{dx^{\alpha}}..\wedge
dx^{3}=\varpi_{4}\left(  Y_{H\beta}\right)  $

$=\sum_{\alpha,\gamma=0}^{3}\left(  -1\right)  ^{\alpha+1}\frac{d}%
{d\xi^{\gamma}}(\sum_{a}\frac{d%
\mathcal{L}%
}{d\operatorname{Re}\partial_{\gamma}\grave{A}_{\alpha}^{a}}\operatorname{Re}%
\grave{A}_{\beta}^{a}+\frac{d%
\mathcal{L}%
}{d\operatorname{Im}\partial_{\gamma}\grave{A}_{\alpha}^{a}}\operatorname{Im}%
\grave{A}_{\beta}^{a} $

$+\frac{d%
\mathcal{L}%
}{d\partial_{\gamma}G_{\alpha}^{a}}G_{\beta}^{a}+\sum_{i}\frac{d%
\mathcal{L}%
}{d\partial_{\gamma}O_{\alpha}^{i\prime}}O_{\beta}^{\prime i})dx^{0}%
\wedge..\widehat{dx^{\alpha}}..dx^{3}$

$\varpi_{4}\left(  Y_{H\beta}\right)  $

$=\varpi_{4}\left(  \sum_{\alpha\gamma}\frac{d}{d\xi^{\gamma}}(\sum_{a}\frac{d%
\mathcal{L}%
}{d\operatorname{Re}\partial_{\gamma}\grave{A}_{\alpha}^{a}}\operatorname{Re}%
\grave{A}_{\beta}^{a}+\frac{d%
\mathcal{L}%
}{d\operatorname{Im}\partial_{\gamma}\grave{A}_{\alpha}^{a}}\operatorname{Im}%
\grave{A}_{\beta}^{a}+\frac{d%
\mathcal{L}%
}{d\partial_{\gamma}G_{\alpha}^{a}}G_{\beta}^{a}+\sum_{i}\frac{d%
\mathcal{L}%
}{d\partial_{\gamma}O_{\alpha}^{i\prime}}O_{\beta}^{\prime i})\partial
_{\alpha}\right)  $

$=\varpi_{4}(\sum_{\alpha\gamma}\frac{d}{d\xi^{\gamma}}(\sum_{a}\frac{d%
\mathcal{L}%
}{d\operatorname{Re}\partial_{\gamma}\grave{A}_{\alpha}^{a}}\operatorname{Re}%
\grave{A}_{\beta}^{a}+\frac{d%
\mathcal{L}%
}{d\operatorname{Im}\partial_{\gamma}\grave{A}_{\alpha}^{a}}\operatorname{Im}%
\grave{A}_{\beta}^{a}+\frac{d%
\mathcal{L}%
}{d\partial_{\gamma}G_{\alpha}^{a}}G_{\beta}^{a}$

$\qquad+\sum_{i}\left(  \frac{d%
\mathcal{L}%
}{d\partial_{\gamma}O_{\alpha}^{i\prime}}-\frac{d%
\mathcal{L}%
}{d\partial_{\alpha}O_{\gamma}^{i\prime}}\right)  O_{\beta}^{\prime
i})\partial_{\alpha})+\varpi_{4}\left(  \sum_{\alpha\gamma}\frac{d}%
{d\xi^{\gamma}}\left(  \sum_{i}\frac{d%
\mathcal{L}%
}{d\partial_{\alpha}O_{\gamma}^{i\prime}}O_{\beta}^{\prime i}\right)
\partial_{\alpha}\right)  $

$=\frac{1}{2}d\Pi_{H\beta}+\varpi_{4}\left(  \sum_{\alpha\gamma}\frac{d}%
{d\xi^{\gamma}}\left(  \sum_{i}\frac{d%
\mathcal{L}%
}{d\partial_{\alpha}O_{\gamma}^{i\prime}}O_{\beta}^{\prime i}\right)
\partial_{\alpha}\right)  $%

\begin{equation}
\varpi_{4}\left(  Y_{H\beta}\right)  =\frac{1}{2}\left(  d\Pi_{H\beta}%
+\Pi_{O\beta}\right) \label{E59}%
\end{equation}

with $\Pi_{O\beta}=\sum_{\alpha,\gamma=0}^{3}\left(  -1\right)  ^{\alpha
+1}\frac{d}{d\xi^{\gamma}}\left(  \sum_{i}\frac{d%
\mathcal{L}%
}{d\partial_{\alpha}O_{\gamma}^{i\prime}}O_{\beta}^{\prime i}\right)
dx^{0}\wedge...\wedge\widehat{dx^{\alpha}}\wedge...\wedge dx^{3}$

$=\frac{1}{\det O^{\prime}}\varpi_{4}\left(  \sum_{\alpha\gamma i}\left(
\frac{d}{d\xi^{\gamma}}\left(  \frac{d%
\mathcal{L}%
}{d\partial_{\alpha}O_{\gamma}^{i\prime}}O_{\beta}^{\prime i}\right)  \right)
\partial_{\alpha}\right)  $

If the equations related to the force fields are met, this latter equation is
equivalent to the "frame equation".

$d\left(  \varpi_{4}\left(  Y_{H\beta}\right)  \right)  =\left(  \sum
_{\alpha,\gamma=0}^{3}\frac{d^{2}}{d\xi^{\alpha}d\xi^{\gamma}}\left(  \sum
_{i}\frac{dL}{d\partial_{\alpha}O_{\gamma}^{i\prime}}O_{\beta}^{\prime i}\det
O^{\prime}\right)  \right)  \varpi_{0}$

The flow of the $Y_{H\beta}$ vector is conserved if this quantity is null,
which is met if L does not depend on $\partial_{\alpha}O_{\gamma}^{i\prime}%
$\ or if \ $\frac{dL}{d\partial_{\alpha}O_{\gamma}^{i\prime}}=-\frac
{dL}{d\partial_{\gamma}O_{\alpha}^{i\prime}}.$ But we will see a stronger result.

\subsection{The superconservation law}

It is intuitive that there is some relation between all the Noether currents
and potentials, looking like an energy conservation law.\ In addition neither
the term L in $Y_{H}$ or the $\Pi_{O}$ quantity in the latest equation are too
appealing. We can give a more convenient formula for this equation \ref{E59},
but that will require some work, that is, in some ways, a reverse engineering
of what has been done above.

We will prove that :

$\forall\alpha,\beta:Y_{H\beta}^{\alpha}=0$

$d\left(  \Pi_{H\beta}\right)  =-\frac{2}{\det O^{\prime}}\sum_{i}\varpi
_{4}\left(  \sum_{\alpha\gamma}\frac{d}{d\xi^{\gamma}}\left(  O_{\beta
}^{\prime i}\frac{d%
\mathcal{L}%
}{d\partial_{\alpha}O_{\gamma}^{i\prime}}\right)  \partial_{\alpha}\right)
=-2\Pi_{O\beta}$

meaning that the pertinent physical quantity is the energy-momentum tensor
$\delta_{\beta}^{\alpha}L$ . And the second equations gives, in the usual case
where the lagrangian does not depend of the derivatives $\partial_{\alpha
}O_{\gamma}^{i\prime}$ , a general law linking the gravitational and the other
force fields, without any involvement of the particles.

\paragraph{1)}

We will start by expliciting $\delta_{\beta}^{\alpha}L$ on shell$,$ that will
be useful later.

$\delta_{\beta}^{\alpha}L$ is given by the frame equation :

$\forall\alpha,\beta:\delta_{\beta}^{\alpha}L=-\sum_{i}\frac{dL}{dO_{\alpha
}^{\prime i}}O_{\beta}^{\prime i}+\frac{1}{\det O^{\prime}}\sum_{i\gamma
}O_{\beta}^{\prime i}\frac{d}{d\xi^{\gamma}}\left(  \frac{dL\det O^{\prime}%
}{d\partial_{\gamma}O_{\alpha}^{\prime i}}\right)  $

Using the covariance equations \ref{E37c},\ref{E37d} as above we get :

$\sum_{i}\frac{dL_{M}}{dO_{\alpha}^{\prime i}}O_{\beta}^{\prime i}%
=\frac{dL_{M}}{dV^{\beta}}V^{\alpha}-\sum_{i,j}\left(  \frac{dL_{M}%
}{d\operatorname{Re}\nabla_{\alpha}\psi^{ij}}\operatorname{Re}\nabla_{\beta
}\psi^{ij}+\frac{dL_{M}}{d\operatorname{Im}\nabla_{\alpha}\psi^{ij}%
}\operatorname{Im}\nabla_{\beta}\psi^{ij}\right)  $

$-\sum_{a}\frac{\partial L_{M}}{\partial G_{\alpha}^{a}}G_{\beta}^{a}%
-\sum_{i\lambda}\frac{dL_{M}}{d\partial_{\lambda}O_{\alpha}^{\prime i}}\left(
\partial_{\lambda}O_{\beta}^{\prime i}\right)  +\frac{dL_{M}}{d\partial
_{\alpha}O_{\lambda}^{\prime i}}\left(  \partial_{\beta}O_{\lambda}^{\prime
i}\right)  $

$\sum_{i}\frac{dL_{F}}{dO_{\alpha}^{\prime i}}O_{\beta}^{\prime i}%
=-2\sum_{a\lambda}\left(  \frac{dL_{F}}{d\operatorname{Re}%
\mathcal{F}%
_{A,\alpha\lambda}^{a}}\operatorname{Re}%
\mathcal{F}%
_{A,\beta\lambda}^{a}+\frac{dL_{F}}{d\operatorname{Im}%
\mathcal{F}%
_{A,\alpha\lambda}^{a}}\operatorname{Im}%
\mathcal{F}%
_{A,\beta\lambda}^{a}+\frac{dL_{F}}{d%
\mathcal{F}%
_{G,\alpha\lambda}^{a}}%
\mathcal{F}%
_{G,\beta\lambda}^{a}\right)  $

$-\sum_{a}\frac{\partial L_{F}}{\partial G_{\alpha}^{a}}G_{\beta}^{a}%
-\sum_{i\lambda}\frac{dL_{F}}{d\partial_{\lambda}O_{\alpha}^{i\prime}}\left(
\partial_{\lambda}O_{\beta}^{\prime i}\right)  +\frac{dL_{F}}{d\partial
_{\alpha}O_{\lambda}^{i\prime}}\left(  \partial_{\beta}O_{\lambda}^{\prime
i}\right)  $

$\sum_{i}\frac{dL}{dO_{\alpha}^{\prime i}}O_{\beta}^{\prime i}=\frac{dNL_{M}%
}{dV^{\beta}}V^{\alpha}-\sum_{i,j}\left(  \frac{dNL_{M}}{d\operatorname{Re}%
\nabla_{\alpha}\psi^{ij}}\operatorname{Re}\nabla_{\beta}\psi^{ij}%
+\frac{dNL_{M}}{d\operatorname{Im}\nabla_{\alpha}\psi^{ij}}\operatorname{Im}%
\nabla_{\beta}\psi^{ij}\right)  $

$-2\sum_{a\lambda}\left(  \frac{dL_{F}}{d\operatorname{Re}%
\mathcal{F}%
_{A,\alpha\lambda}^{a}}\operatorname{Re}%
\mathcal{F}%
_{A,\beta\lambda}^{a}+\frac{dL_{F}}{d\operatorname{Im}%
\mathcal{F}%
_{A,\alpha\lambda}^{a}}\operatorname{Im}%
\mathcal{F}%
_{A,\beta\lambda}^{a}+\frac{dL_{F}}{d%
\mathcal{F}%
_{G,\alpha\lambda}^{a}}%
\mathcal{F}%
_{G,\beta\lambda}^{a}\right)  $

$-\sum_{a}\frac{\partial L}{\partial G_{\alpha}^{a}}G_{\beta}^{a}%
-\sum_{i\lambda}\frac{dL}{d\partial_{\lambda}O_{\alpha}^{\prime i}}%
\partial_{\lambda}O_{\beta}^{\prime i}+\frac{dL}{d\partial_{\alpha}O_{\lambda
}^{\prime i}}\partial_{\beta}O_{\lambda}^{\prime i}$

$\delta_{\beta}^{\alpha}L=\sum_{i,j}\left(  \frac{dNL_{M}}{d\operatorname{Re}%
\nabla_{\alpha}\psi^{ij}}\operatorname{Re}\nabla_{\beta}\psi^{ij}%
+\frac{dNL_{M}}{d\operatorname{Im}\nabla_{\alpha}\psi^{ij}}\operatorname{Im}%
\nabla_{\beta}\psi^{ij}\right)  $

$+2\sum_{a\gamma}\left(  \frac{dL_{F}}{d\operatorname{Re}%
\mathcal{F}%
_{A,\alpha\gamma}^{a}}\operatorname{Re}%
\mathcal{F}%
_{A,\beta\gamma}^{a}+\frac{dL_{F}}{d\operatorname{Im}%
\mathcal{F}%
_{A,\alpha\gamma}^{a}}\operatorname{Im}%
\mathcal{F}%
_{A,\beta\gamma}^{a}+\frac{dL_{F}}{d%
\mathcal{F}%
_{G,\alpha\gamma}^{a}}%
\mathcal{F}%
_{G,\beta\gamma}^{a}\right)  $

$+\sum_{a}\frac{\partial L}{\partial G_{\alpha}^{a}}G_{\beta}^{a}%
+\sum_{i\gamma}\frac{dL}{d\partial_{\gamma}O_{\alpha}^{\prime i}}%
\partial_{\gamma}O_{\beta}^{\prime i}+\frac{dL}{d\partial_{\alpha}O_{\gamma
}^{\prime i}}\partial_{\beta}O_{\gamma}^{\prime i}\allowbreak+\frac{1}{\det
O^{\prime}}\sum_{i\gamma}O_{\beta}^{\prime i}\frac{d}{d\xi^{\gamma}}\left(
\frac{dL\det O^{\prime}}{d\partial_{\gamma}O_{\alpha}^{\prime i}}\right)
-\frac{dNL_{M}}{dV^{\beta}}V^{\alpha}$

Let us expand the first term.

$\sum_{i,j}\left(  \frac{dNL_{M}}{d\operatorname{Re}\nabla_{\alpha}\psi^{ij}%
}\operatorname{Re}\nabla_{\beta}\psi^{ij}+\frac{dNL_{M}}{d\operatorname{Im}%
\nabla_{\alpha}\psi^{ij}}\operatorname{Im}\nabla_{\beta}\psi^{ij}\right)  $

$=\sum_{i,j}\left(  \frac{dNL_{M}}{d\operatorname{Re}\nabla_{\alpha}\psi^{ij}%
}\operatorname{Re}\partial_{\beta}\psi^{ij}+\frac{dNL_{M}}{d\operatorname{Im}%
\nabla_{\alpha}\psi^{ij}}\operatorname{Im}\partial_{\beta}\psi^{ij}\right)  $

$+\sum_{aij}\frac{dNL_{M}}{d\operatorname{Re}\nabla_{\alpha}\psi^{ij}%
}\operatorname{Re}\left(  G_{\beta}^{a}\left[  \kappa_{a}\right]  \left[
\psi\right]  +\grave{A}_{\beta}^{a}\left[  \psi\right]  \left[  \theta
_{a}\right]  ^{t}\right)  ^{ij}$

$+\frac{dNL_{M}}{d\operatorname{Im}\nabla_{\alpha}\psi^{ij}}\operatorname{Im}%
\left(  G_{\beta}^{a}\left[  \kappa_{a}\right]  \left[  \psi\right]
+\grave{A}_{\beta}^{a}\left[  \psi\right]  \left[  \theta_{a}\right]
^{t}\right)  ^{ij}$

$=\sum_{i,j}\left(  \frac{dNL_{M}}{d\operatorname{Re}\nabla_{\alpha}\psi^{ij}%
}\operatorname{Re}\partial_{\beta}\psi^{ij}+\frac{dNL_{M}}{d\operatorname{Im}%
\nabla_{\alpha}\psi^{ij}}\operatorname{Im}\partial_{\beta}\psi^{ij}\right)  $

$+\sum_{a}G_{\beta}^{a}\sum_{ij}\frac{dNL_{M}}{d\operatorname{Re}%
\nabla_{\alpha}\psi^{ij}}\operatorname{Re}\left(  \left[  \kappa_{a}\right]
\left[  \psi\right]  \right)  ^{ij}+\frac{dNL_{M}}{d\operatorname{Im}%
\nabla_{\alpha}\psi^{ij}}\operatorname{Im}\left(  \left[  \kappa_{a}\right]
\left[  \psi\right]  \right)  ^{ij}$

$+\sum_{a}\left(  \operatorname{Re}\grave{A}_{\beta}^{a}\right)  \left(
\frac{dNL_{M}}{d\operatorname{Re}\nabla_{\alpha}\psi^{ij}}\operatorname{Re}%
\left(  \left[  \psi\right]  \left[  \theta_{a}\right]  ^{t}\right)
^{ij}+\frac{dNL_{M}}{d\operatorname{Im}\nabla_{\alpha}\psi^{ij}}%
\operatorname{Im}\left(  \left[  \psi\right]  \left[  \theta_{a}\right]
^{t}\right)  ^{ij}\right)  $

$+\left(  \operatorname{Im}\grave{A}_{\beta}^{a}\right)  \left(
-\frac{dNL_{M}}{d\operatorname{Re}\nabla_{\alpha}\psi^{ij}}\operatorname{Im}%
\left(  \left[  \psi\right]  \left[  \theta_{a}\right]  ^{t}\right)
^{ij}+\frac{dNL_{M}}{d\operatorname{Im}\nabla_{\alpha}\psi^{ij}}%
\operatorname{Re}\left(  \left[  \psi\right]  \left[  \theta_{a}\right]
^{t}\right)  ^{ij}\right)  $

And with the Noether currents it reads :

$\sum_{i,j}\left(  \frac{dNL_{M}}{d\operatorname{Re}\nabla_{\alpha}\psi^{ij}%
}\operatorname{Re}\nabla_{\beta}\psi^{ij}+\frac{dNL_{M}}{d\operatorname{Im}%
\nabla_{\alpha}\psi^{ij}}\operatorname{Im}\nabla_{\beta}\psi^{ij}\right)  $

$=\sum_{i,j}\left(  \frac{dNL_{M}}{d\operatorname{Re}\nabla_{\alpha}\psi^{ij}%
}\operatorname{Re}\partial_{\beta}\psi^{ij}+\frac{dNL_{M}}{d\operatorname{Im}%
\nabla_{\alpha}\psi^{ij}}\operatorname{Im}\partial_{\beta}\psi^{ij}\right)  $

$+\sum_{a}G_{\beta}^{a}\left(  Y_{G}^{a\alpha}-\frac{\partial L}{\partial
G_{\alpha}^{a}}-2\sum_{b,\gamma}\frac{dL_{F}}{d%
\mathcal{F}%
_{G\alpha\gamma}^{b}}\left[  \overrightarrow{\kappa}_{a},G_{\gamma}\right]
^{b}\right)  $

$+\sum_{a}\left(  \operatorname{Re}\grave{A}_{\beta}^{a}\right)  \left(
Y_{AR}^{\alpha a}-2\sum_{b\gamma}\frac{dL_{F}}{d\operatorname{Re}%
\mathcal{F}%
_{A,\alpha\gamma}^{b}}\operatorname{Re}\left[  \overrightarrow{\theta}%
_{a},\grave{A}_{\gamma}\right]  ^{b}+\frac{dL_{F}}{d\operatorname{Im}%
\mathcal{F}%
_{A,\alpha\gamma}^{b}}\operatorname{Im}\left[  \overrightarrow{\theta}%
_{a},\grave{A}_{\gamma}\right]  ^{b}\right)  $

$+\left(  \operatorname{Im}\grave{A}_{\beta}^{a}\right)  \left(
Y_{AI}^{\alpha a}-2\sum_{b\gamma}-\frac{dL_{F}}{d\operatorname{Re}%
\mathcal{F}%
_{A,\alpha\gamma}^{b}}\operatorname{Im}\left[  \overrightarrow{\theta}%
_{a},\grave{A}_{\gamma}\right]  ^{b}+\frac{dL_{F}}{d\operatorname{Im}%
\mathcal{F}%
_{A,\alpha\gamma}^{b}}\operatorname{Re}\left[  \overrightarrow{\theta}%
_{a},\grave{A}_{\gamma}\right]  ^{b}\right)  $

$=\sum_{i,j}\left(  \frac{dNL_{M}}{d\operatorname{Re}\nabla_{\alpha}\psi^{ij}%
}\operatorname{Re}\partial_{\beta}\psi^{ij}+\frac{dNL_{M}}{d\operatorname{Im}%
\nabla_{\alpha}\psi^{ij}}\operatorname{Im}\partial_{\beta}\psi^{ij}\right)
+\sum_{a}G_{\beta}^{a}\left(  Y_{G}^{a\alpha}-\frac{\partial L}{\partial
G_{\alpha}^{a}}\right)  $

$+Y_{AR}^{\alpha a}\operatorname{Re}\grave{A}_{\beta}^{a}+Y_{AI}^{\alpha
a}\operatorname{Im}\grave{A}_{\beta}^{a}-2\sum_{a,\gamma}\frac{dL_{F}}{d%
\mathcal{F}%
_{G\alpha\gamma}^{b}}\left[  G_{\beta},G_{\gamma}\right]  ^{a}$

$-2\sum_{a\gamma}\left(  \frac{dL_{F}}{d\operatorname{Re}%
\mathcal{F}%
_{A,\alpha\gamma}^{a}}\operatorname{Re}\left[  \grave{A}_{\beta},\grave
{A}_{\gamma}\right]  ^{a}+\frac{dL_{F}}{d\operatorname{Im}%
\mathcal{F}%
_{A,\alpha\gamma}^{a}}\operatorname{Im}\left[  \grave{A}_{\beta},\grave
{A}_{\gamma}\right]  ^{a}\right)  $

Thus :

$\delta_{\beta}^{\alpha}L=\sum_{i,j}\left(  \frac{dNL_{M}}{d\operatorname{Re}%
\nabla_{\alpha}\psi^{ij}}\operatorname{Re}\partial_{\beta}\psi^{ij}%
+\frac{dNL_{M}}{d\operatorname{Im}\nabla_{\alpha}\psi^{ij}}\operatorname{Im}%
\partial_{\beta}\psi^{ij}\right)  $

$+\sum_{a}G_{\beta}^{a}\left(  Y_{G}^{a\alpha}-\frac{\partial L}{\partial
G_{\alpha}^{a}}\right)  +Y_{AR}^{\alpha a}\operatorname{Re}\grave{A}_{\beta
}^{a}+Y_{AI}^{\alpha a}\operatorname{Im}\grave{A}_{\beta}^{a}-2\sum_{a,\gamma
}\frac{dL_{F}}{d%
\mathcal{F}%
_{G\alpha\gamma}^{b}}\left[  G_{\beta},G_{\gamma}\right]  ^{a}$

$-2\sum_{a\gamma}\left(  \frac{dL_{F}}{d\operatorname{Re}%
\mathcal{F}%
_{A,\alpha\gamma}^{a}}\operatorname{Re}\left[  \grave{A}_{\beta},\grave
{A}_{\gamma}\right]  ^{a}+\frac{dL_{F}}{d\operatorname{Im}%
\mathcal{F}%
_{A,\alpha\gamma}^{a}}\operatorname{Im}\left[  \grave{A}_{\beta},\grave
{A}_{\gamma}\right]  ^{a}\right)  $

$+2\sum_{a\gamma}\left(  \frac{dL_{F}}{d\operatorname{Re}%
\mathcal{F}%
_{A,\alpha\gamma}^{a}}\operatorname{Re}%
\mathcal{F}%
_{A,\beta\gamma}^{a}+\frac{dL_{F}}{d\operatorname{Im}%
\mathcal{F}%
_{A,\alpha\gamma}^{a}}\operatorname{Im}%
\mathcal{F}%
_{A,\beta\gamma}^{a}+\frac{dL_{F}}{d%
\mathcal{F}%
_{G,\alpha\gamma}^{a}}%
\mathcal{F}%
_{G,\beta\gamma}^{a}\right)  $

$+\sum_{a}\frac{\partial L}{\partial G_{\alpha}^{a}}G_{\beta}^{a}%
+\sum_{i\gamma}\frac{dL}{d\partial_{\gamma}O_{\alpha}^{\prime i}}%
\partial_{\gamma}O_{\beta}^{\prime i}+\frac{dL}{d\partial_{\alpha}O_{\gamma
}^{\prime i}}\partial_{\beta}O_{\gamma}^{\prime i}$

$+\frac{1}{\det O^{\prime}}\sum_{i\gamma}O_{\beta}^{\prime i}\frac{d}%
{d\xi^{\gamma}}\left(  \frac{dL\det O^{\prime}}{d\partial_{\gamma}O_{\alpha
}^{\prime i}}\right)  -\frac{dNL_{M}}{dV^{\beta}}V^{\alpha}$

$\delta_{\beta}^{\alpha}L=\sum_{i,j}\left(  \frac{dNL_{M}}{d\operatorname{Re}%
\nabla_{\alpha}\psi^{ij}}\operatorname{Re}\partial_{\beta}\psi^{ij}%
+\frac{dNL_{M}}{d\operatorname{Im}\nabla_{\alpha}\psi^{ij}}\operatorname{Im}%
\partial_{\beta}\psi^{ij}\right)  $

$+\sum_{a}G_{\beta}^{a}Y_{G}^{a\alpha}+Y_{AR}^{\alpha a}\operatorname{Re}%
\grave{A}_{\beta}^{a}+Y_{AI}^{\alpha a}\operatorname{Im}\grave{A}_{\beta}%
^{a}+2\sum_{a,\gamma}\frac{dL_{F}}{d%
\mathcal{F}%
_{G\alpha\gamma}^{a}}\left(
\mathcal{F}%
_{G,\beta\gamma}-\left[  G_{\beta},G_{\gamma}\right]  \right)  ^{a}$

$+2\sum_{a\gamma}\left(  \frac{dL_{F}}{d\operatorname{Re}%
\mathcal{F}%
_{A,\alpha\gamma}^{a}}\operatorname{Re}\left(
\mathcal{F}%
_{A,\beta\gamma}-\left[  \grave{A}_{\beta},\grave{A}_{\gamma}\right]  \right)
^{a}+\frac{dL_{F}}{d\operatorname{Im}%
\mathcal{F}%
_{A,\alpha\gamma}^{a}}\operatorname{Im}\left(
\mathcal{F}%
_{A,\beta\gamma}-\left[  \grave{A}_{\beta},\grave{A}_{\gamma}\right]  \right)
^{a}\right)  $

$+\sum_{i\gamma}\frac{dL}{d\partial_{\gamma}O_{\alpha}^{\prime i}}%
\partial_{\gamma}O_{\beta}^{\prime i}+\frac{dL}{d\partial_{\alpha}O_{\gamma
}^{\prime i}}\partial_{\beta}O_{\gamma}^{\prime i}+\frac{1}{\det O^{\prime}%
}\sum_{i\gamma}\frac{d}{d\xi^{\gamma}}\left(  \frac{dL\det O^{\prime}%
}{d\partial_{\gamma}O_{\alpha}^{\prime i}}\right)  -\frac{dNL_{M}}{dV^{\beta}%
}V^{\alpha}$

With the same conventions as above for derivation with composite functions :

$\sum_{i\gamma}\frac{dL}{d\partial_{\gamma}O_{\alpha}^{\prime i}}%
\partial_{\gamma}O_{\beta}^{\prime i}+\frac{1}{\det O^{\prime}}\sum_{i\gamma
}O_{\beta}^{\prime i}\frac{d}{d\xi^{\gamma}}\left(  \frac{dL\det O^{\prime}%
}{d\partial_{\gamma}O_{\alpha}^{\prime i}}\right)  $

$=\frac{1}{\det O^{\prime}}\sum_{i\gamma}\frac{d}{d\xi^{\gamma}}\left(
\frac{dL\det O^{\prime}}{d\partial_{\gamma}O_{\alpha}^{\prime i}}O_{\beta
}^{\prime i}\right)  =\frac{1}{\det O^{\prime}}\sum_{i\gamma}\frac{d}%
{d\xi^{\gamma}}\left(  \frac{d%
\mathcal{L}%
}{d\partial_{\gamma}O_{\alpha}^{\prime i}}O_{\beta}^{\prime i}\right)  $

$\delta_{\beta}^{\alpha}L$

$=\sum_{i,j}\left(  \frac{dNL_{M}}{d\operatorname{Re}\nabla_{\alpha}\psi^{ij}%
}\operatorname{Re}\partial_{\beta}\psi^{ij}+\frac{dNL_{M}}{d\operatorname{Im}%
\nabla_{\alpha}\psi^{ij}}\operatorname{Im}\partial_{\beta}\psi^{ij}\right)  $

$+\sum_{a}G_{\beta}^{a}Y_{G}^{a\alpha}+Y_{AR}^{\alpha a}\operatorname{Re}%
\grave{A}_{\beta}^{a}+Y_{AI}^{\alpha a}\operatorname{Im}\grave{A}_{\beta}^{a}$

$+2\sum_{a\gamma}(\frac{dL_{F}}{d%
\mathcal{F}%
_{G\alpha\gamma}^{b}}\left(  \partial_{\beta}G_{\gamma}-\partial_{\gamma
}G_{\beta}\right)  ^{a}+\frac{dL_{F}}{d\operatorname{Re}%
\mathcal{F}%
_{A,\alpha\gamma}^{a}}\operatorname{Re}\left(  \partial_{\beta}\grave
{A}_{\gamma}-\partial_{\gamma}\grave{A}_{\beta}\right)  ^{a}$

$+\frac{dL_{F}}{d\operatorname{Im}%
\mathcal{F}%
_{A,\alpha\gamma}^{a}}\operatorname{Im}\left(  \partial_{\beta}\grave
{A}_{\gamma}-\partial_{\gamma}\grave{A}_{\beta}\right)  ^{a})$

$+\sum_{i\gamma}\frac{dL}{d\partial_{\alpha}O_{\gamma}^{\prime i}}%
\partial_{\beta}O_{\gamma}^{\prime i}+\frac{1}{\det O^{\prime}}\sum_{i\gamma
}\frac{d}{d\xi^{\gamma}}\left(  \frac{d%
\mathcal{L}%
}{d\partial_{\gamma}O_{\alpha}^{\prime i}}O_{\beta}^{\prime i}\right)
-\frac{dNL_{M}}{dV^{\beta}}V^{\alpha}$

\paragraph{2)}

So on shell $Y_{H\beta}^{\alpha}$\ can be written :\ 

$Y_{H\beta}^{\alpha}=-\frac{dNL_{M}}{dV^{\beta}}V^{\alpha}+\sum_{ij}%
\frac{dNL_{M}}{d\operatorname{Re}\nabla_{\alpha}\psi^{ij}}\operatorname{Re}%
\partial_{\beta}\psi^{ij}+\frac{dNL_{M}}{d\operatorname{Im}\nabla_{\alpha}%
\psi^{ij}}\operatorname{Im}\partial_{\beta}\psi^{ij}$

$+2\sum_{a,\gamma}\frac{dL_{F}}{d%
\mathcal{F}%
_{G\alpha\gamma}^{a}}\partial_{\beta}G_{\gamma}^{a}+\frac{dL_{F}%
}{d\operatorname{Re}%
\mathcal{F}%
_{A\alpha\gamma}^{a}}\operatorname{Re}\partial_{\beta}\grave{A}_{\gamma}%
^{a}+\frac{dL_{F}}{d\operatorname{Im}%
\mathcal{F}%
_{A\alpha\gamma}^{a}}\operatorname{Im}\partial_{\beta}\grave{A}_{\gamma}%
^{a}+\sum_{i\gamma}\frac{dL}{d\partial_{\alpha}O_{\gamma}^{\prime i}}%
\partial_{\beta}O_{\gamma}^{\prime i}$

$-\sum_{i,j}\left(  \frac{dVL_{M}}{d\operatorname{Re}\nabla_{\alpha}\psi^{ij}%
}\operatorname{Re}\partial_{\beta}\psi^{ij}+\frac{dVL_{M}}{d\operatorname{Im}%
\nabla_{\alpha}\psi^{ij}}\operatorname{Im}\partial_{\beta}\psi^{ij}\right)  $

$-\sum_{a}G_{\beta}^{a}Y_{G}^{a\alpha}+Y_{AR}^{\alpha a}\operatorname{Re}%
\grave{A}_{\beta}^{a}+Y_{AI}^{\alpha a}\operatorname{Im}\grave{A}_{\beta}^{a}$

$-2\sum_{a\gamma}(\frac{dL_{F}}{d%
\mathcal{F}%
_{G\alpha\gamma}^{b}}\left(  \partial_{\beta}G_{\gamma}-\partial_{\gamma
}G_{\beta}\right)  ^{a}+\frac{dL_{F}}{d\operatorname{Re}%
\mathcal{F}%
_{A,\alpha\gamma}^{a}}\operatorname{Re}\left(  \partial_{\beta}\grave
{A}_{\gamma}-\partial_{\gamma}\grave{A}_{\beta}\right)  ^{a}$

$+\frac{dL_{F}}{d\operatorname{Im}%
\mathcal{F}%
_{A,\alpha\gamma}^{a}}\operatorname{Im}\left(  \partial_{\beta}\grave
{A}_{\gamma}-\partial_{\gamma}\grave{A}_{\beta}\right)  ^{a})$

$-\frac{dL}{d\partial_{\alpha}O_{\gamma}^{\prime i}}\partial_{\beta}O_{\gamma
}^{\prime i}-\frac{1}{\det O^{\prime}}\sum_{i\gamma}\frac{d}{d\xi^{\gamma}%
}\left(  \frac{d%
\mathcal{L}%
}{d\partial_{\gamma}O_{\alpha}^{\prime i}}O_{\beta}^{\prime i}\right)
+\frac{dNL_{M}}{dV^{\beta}}V^{\alpha}$

$Y_{H\beta}^{\alpha}=-\sum_{a}\left(  G_{\beta}^{a}Y_{G}^{a\alpha}%
+Y_{AR}^{\alpha a}\operatorname{Re}\grave{A}_{\beta}^{a}+Y_{AI}^{\alpha
a}\operatorname{Im}\grave{A}_{\beta}^{a}\right)  $

$+2\sum_{a\gamma}\left(  \frac{dL_{F}}{d%
\mathcal{F}%
_{G\alpha\gamma}^{b}}\partial_{\gamma}G_{\beta}^{a}+\frac{dL_{F}%
}{d\operatorname{Re}%
\mathcal{F}%
_{A,\alpha\gamma}^{a}}\operatorname{Re}\partial_{\gamma}\grave{A}_{\beta}%
^{a}+\frac{dL_{F}}{d\operatorname{Im}%
\mathcal{F}%
_{A,\alpha\gamma}^{a}}\operatorname{Im}\partial_{\gamma}\grave{A}_{\beta}%
^{a}\right)  $

$-\frac{1}{\det O^{\prime}}\sum_{i\gamma}\frac{d}{d\xi^{\gamma}}\left(
\frac{d%
\mathcal{L}%
}{d\partial_{\gamma}O_{\alpha}^{\prime i}}O_{\beta}^{\prime i}\right)  $

This formula will be improved.

\paragraph{3)}

Taking the value of $\varpi_{4}\left(  Y_{H\beta}\right)  $:

$\varpi_{4}\left(  Y_{H\beta}\right)  $

$=-\sum_{a}\left(  G_{\beta}^{a}\varpi_{4}\left(  Y_{G}^{a}\right)  +\left(
\operatorname{Re}\grave{A}_{\beta}^{a}\right)  \varpi_{4}\left(
Y_{AR}^{\alpha}\right)  +\left(  \operatorname{Im}\grave{A}_{\beta}%
^{a}\right)  \varpi_{4}\left(  Y_{AI}^{\alpha}\right)  \right)  $

$+\sum_{a}\varpi_{4}\left(  \left(  \sum_{\gamma}2\frac{dL}{d%
\mathcal{F}%
_{G\alpha\gamma}^{a}}\partial_{\gamma}G_{\beta}^{a}\right)  \partial_{\alpha
}\right)  +\varpi_{4}\left(  \left(  \sum_{\gamma}2\frac{dL_{F}}%
{d\operatorname{Re}%
\mathcal{F}%
_{A\alpha\gamma}^{a}}\operatorname{Re}\left(  \partial_{\gamma}\grave
{A}_{\beta}^{a}\right)  \right)  \partial_{\alpha}\right)  $

$+\varpi_{4}\left(  \left(  \sum_{\gamma}2\frac{dL_{F}}{d\operatorname{Im}%
\mathcal{F}%
_{A\alpha\gamma}^{a}}\operatorname{Im}\left(  \partial_{\gamma}\grave
{A}_{\beta}^{a}\right)  \right)  \partial_{\alpha}\right)  -\frac{1}{\det
O^{\prime}}\sum_{i}\varpi_{4}\left(  \sum_{\alpha\gamma}\left(  \frac{d}%
{d\xi^{\gamma}}\left(  \frac{d%
\mathcal{L}%
}{d\partial_{\gamma}O_{\alpha}^{\prime i}}O_{\beta}^{\prime i}\right)
\right)  \partial_{\alpha}\right)  $

With the usual algebraic calculation :

$\varpi_{4}\left(  \left(  \sum_{\gamma}2\frac{dL}{d%
\mathcal{F}%
_{G\alpha\gamma}^{a}}\partial_{\gamma}G_{\beta}^{a}\right)  \partial_{\alpha
}\right)  =\varpi_{4}\left(  \left(  \sum_{\gamma}Z_{G}^{a\alpha\gamma
}\partial_{\gamma}G_{\beta}^{a}\right)  \partial_{\alpha}\right)  $

$=-\frac{1}{2}\left(  \sum_{\gamma}\partial_{\gamma}G_{\beta}^{a}dx^{\gamma
}\right)  \wedge\Pi_{G}^{a}=-\frac{1}{2}d\left(  G_{\beta}^{a}\right)
\wedge\Pi_{G}^{a}$

$\varpi_{4}\left(  \left(  \sum_{\gamma}2\frac{dL_{F}}{d\operatorname{Re}%
\mathcal{F}%
_{A\alpha\gamma}^{a}}\operatorname{Re}\left(  \partial_{\gamma}\grave
{A}_{\beta}^{a}\right)  \right)  \partial_{\alpha}\right)  =-\frac{1}%
{2}d\left(  \operatorname{Re}\grave{A}_{\beta}^{a}\right)  \wedge\Pi_{AR}^{a}$

$\varpi_{4}\left(  \left(  \sum_{\gamma}2\frac{dL_{F}}{d\operatorname{Im}%
\mathcal{F}%
_{A\alpha\gamma}^{a}}\operatorname{Im}\left(  \partial_{\gamma}\grave
{A}_{\beta}^{a}\right)  \right)  \partial_{\alpha}\right)  =-\frac{1}%
{2}d\left(  \operatorname{Im}\grave{A}_{\beta}^{a}\right)  \wedge\Pi_{AI}^{a}$

$\varpi_{4}\left(  Y_{H\beta}\right)  =-\sum_{a}\left(  G_{\beta}^{a}%
\varpi_{4}\left(  Y_{G}^{a}\right)  +\left(  \operatorname{Re}\grave{A}%
_{\beta}^{a}\right)  \varpi_{4}\left(  Y_{AR}^{\alpha}\right)  +\left(
\operatorname{Im}\grave{A}_{\beta}^{a}\right)  \varpi_{4}\left(
Y_{AI}^{\alpha}\right)  \right)  $

$-\frac{1}{2}d\left(  G_{\beta}^{a}\right)  \wedge\Pi_{G}^{a}-\frac{1}%
{2}d\left(  \operatorname{Re}\grave{A}_{\beta}^{a}\right)  \wedge\Pi_{AR}%
^{a}-\frac{1}{2}d\left(  \operatorname{Im}\grave{A}_{\beta}^{a}\right)
\wedge\Pi_{AI}^{a}$

$-\frac{1}{\det O^{\prime}}\sum_{i}\varpi_{4}\left(  \sum_{\alpha\gamma
}\left(  \frac{d}{d\xi^{\gamma}}\left(  \frac{d%
\mathcal{L}%
}{d\partial_{\gamma}O_{\alpha}^{\prime i}}O_{\beta}^{\prime i}\right)
\right)  \partial_{\alpha}\right)  $

And on shell we have from the other conservation equations :

$\varpi_{4}\left(  Y_{G}^{a}\right)  =\frac{1}{2}d\Pi_{G}^{a},\varpi
_{4}\left(  Y_{AR}^{\alpha}\right)  =\frac{1}{2}d\Pi_{AR}^{a},\varpi
_{4}\left(  Y_{AI}^{\alpha}\right)  =\frac{1}{2}d\Pi_{AI}^{a}$

$\varpi_{4}\left(  Y_{H\beta}\right)  $

$=-\frac{1}{2}\sum_{a}G_{\beta}^{a}d\Pi_{G}^{a}+d\left(  G_{\beta}^{a}\right)
\wedge\Pi_{G}^{a}+\left(  \operatorname{Re}\grave{A}_{\beta}^{a}\right)
d\Pi_{AR}^{a}+d\left(  \operatorname{Re}\grave{A}_{\beta}^{a}\right)
\wedge\Pi_{AR}^{a}$

$+\left(  \operatorname{Im}\grave{A}_{\beta}^{a}\right)  d\Pi_{AI}%
^{a}+d\left(  \operatorname{Im}\grave{A}_{\beta}^{a}\right)  \wedge\Pi
_{AI}^{a}-\frac{1}{\det O^{\prime}}\sum_{i}\varpi_{4}\left(  \sum
_{\alpha\gamma}\left(  \frac{d}{d\xi^{\gamma}}\left(  \frac{d%
\mathcal{L}%
}{d\partial_{\gamma}O_{\alpha}^{\prime i}}O_{\beta}^{\prime i}\right)
\right)  \partial_{\alpha}\right)  $

$\varpi_{4}\left(  Y_{H\beta}\right)  =-\frac{1}{2}\sum_{a}d\left(  G_{\beta
}^{a}\Pi_{G}^{a}\right)  +d\left(  \left(  \operatorname{Re}\grave{A}_{\beta
}^{a}\right)  \Pi_{AR}^{a}\right)  +d\left(  \left(  \operatorname{Im}%
\grave{A}_{\beta}^{a}\right)  \Pi_{AI}^{a}\right)  $

$-\frac{1}{\det O^{\prime}}\sum_{i}\varpi_{4}\left(  \sum_{\alpha\gamma
}\left(  \frac{d}{d\xi^{\gamma}}\left(  \frac{d%
\mathcal{L}%
}{d\partial_{\gamma}O_{\alpha}^{\prime i}}O_{\beta}^{\prime i}\right)
\right)  \partial_{\alpha}\right)  $

\paragraph{4)}

The superpotential is :

$\Pi_{H\beta}=\sum_{a}\left(  \operatorname{Re}\grave{A}_{\beta}^{a}\right)
\Pi_{AR}^{a}+\left(  \operatorname{Im}\grave{A}_{\beta}^{a}\right)  \Pi
_{AI}^{a}+G_{\beta}^{a}\Pi_{G}^{a}+\sum_{i}O_{\beta}^{\prime i}\Pi_{O}^{i}$

with $\Pi_{O}^{i}=\varpi_{4}\left(  Z_{O}^{i}\right)  ,Z_{O}^{i}=\sum
_{i}\left(  \frac{dL}{d\partial_{\alpha}O_{\gamma}^{i\prime}}-\frac
{dL}{d\partial_{\gamma}O_{\alpha}^{i\prime}}\right)  \partial_{\alpha}%
\wedge\partial_{\gamma}$

and the conservation equation \ref{E59}\ 

$\varpi_{4}\left(  Y_{H\beta}\right)  =\frac{1}{2}d\Pi_{H\beta}+\frac{1}{\det
O^{\prime}}\varpi_{4}\left(  \sum_{\alpha\gamma i}\left(  \frac{d}%
{d\xi^{\gamma}}\left(  \frac{d%
\mathcal{L}%
}{d\partial_{\alpha}O_{\gamma}^{i\prime}}O_{\beta}^{\prime i}\right)  \right)
\partial_{\alpha}\right)  $

\ becomes :

$-\frac{1}{2}\sum_{a}d\left(  G_{\beta}^{a}\Pi_{G}^{a}\right)  +d\left(
\left(  \operatorname{Re}\grave{A}_{\beta}^{a}\right)  \Pi_{AR}^{a}\right)
+d\left(  \left(  \operatorname{Im}\grave{A}_{\beta}^{a}\right)  \Pi_{AI}%
^{a}\right)  $

$-\frac{1}{\det O^{\prime}}\sum_{i}\varpi_{4}\left(  \sum_{\alpha\gamma
}\left(  \frac{d}{d\xi^{\gamma}}\left(  \frac{d%
\mathcal{L}%
}{d\partial_{\gamma}O_{\alpha}^{\prime i}}O_{\beta}^{\prime i}\right)
\right)  \partial_{\alpha}\right)  $

$=\frac{1}{2}\sum_{a}d\left(  G_{\beta}^{a}\Pi_{G}^{a}\right)  +d\left(
\left(  \operatorname{Re}\grave{A}_{\beta}^{a}\right)  \Pi_{AR}^{a}\right)
+d\left(  \left(  \operatorname{Im}\grave{A}_{\beta}^{a}\right)  \Pi_{AI}%
^{a}\right)  $

$+\frac{1}{2}\sum_{i}d\left(  O_{\beta}^{\prime i}\Pi_{O}^{i}\right)
+\frac{1}{\det O^{\prime}}\varpi_{4}\left(  \sum_{\alpha\gamma i}\left(
\frac{d}{d\xi^{\gamma}}\left(  \frac{d%
\mathcal{L}%
}{d\partial_{\alpha}O_{\gamma}^{i\prime}}O_{\beta}^{\prime i}\right)  \right)
\partial_{\alpha}\right)  $

That is :

$\sum_{a}d\left(  G_{\beta}^{a}\Pi_{G}^{a}\right)  +d\left(  \left(
\operatorname{Re}\grave{A}_{\beta}^{a}\right)  \Pi_{AR}^{a}\right)  +d\left(
\left(  \operatorname{Im}\grave{A}_{\beta}^{a}\right)  \Pi_{AI}^{a}\right)  $

$=-\frac{1}{2}\sum_{i}d\left(  O_{\beta}^{\prime i}\Pi_{O}^{i}\right)
-\frac{1}{\det O^{\prime}}\sum_{i}\varpi_{4}\left(  \sum_{\alpha\gamma}%
\frac{d}{d\xi^{\gamma}}\left(  O_{\beta}^{\prime i}\left(  \frac{d%
\mathcal{L}%
}{d\partial_{\gamma}O_{\alpha}^{\prime i}}+\frac{d%
\mathcal{L}%
}{d\partial_{\alpha}O_{\gamma}^{i\prime}}\right)  \right)  \partial_{\alpha
}\right)  $

Or :

$d\left(  \Pi_{H\beta}\right)  =\sum_{i}\left(  \frac{1}{2}d\left(  O_{\beta
}^{\prime i}\Pi_{O}^{i}\right)  -\frac{1}{\det O^{\prime}}\varpi_{4}\left(
\sum_{\alpha\gamma}\frac{d}{d\xi^{\gamma}}\left(  O_{\beta}^{\prime i}\left(
\frac{d%
\mathcal{L}%
}{d\partial_{\gamma}O_{\alpha}^{\prime i}}+\frac{d%
\mathcal{L}%
}{d\partial_{\alpha}O_{\gamma}^{i\prime}}\right)  \right)  \partial_{\alpha
}\right)  \right)  $

But :

$d\left(  O_{\beta}^{\prime i}\Pi_{O}^{i}\right)  =\left(  dO_{\beta}^{\prime
i}\right)  \wedge\Pi_{O}^{i}+O_{\beta}^{\prime i}d\Pi_{O}^{i}$

$\varpi_{4}\left(  \sum_{\alpha\gamma}\left(  Z_{O}^{i\alpha\gamma}\frac
{d}{d\xi^{\gamma}}O_{\beta}^{\prime i}\right)  \partial_{\alpha}\right)  $

$=-\frac{1}{2}\left(  \sum_{\alpha\gamma}\left(  \frac{d}{d\xi^{\gamma}%
}O_{\beta}^{\prime i}\right)  dx^{\gamma}\right)  \wedge\Pi_{O}^{a}=-\frac
{1}{2}dO_{\beta}^{\prime i}\wedge\Pi_{O}^{a}$

$d\Pi_{O}^{i}=-2\frac{1}{\det O^{\prime}}\sum_{a}\varpi_{4}\left(
\sum_{\alpha\gamma}\frac{d}{d\xi^{\gamma}}\left(  Z_{O}^{i\alpha\gamma}\det
O^{\prime}\right)  \partial_{\alpha}\right)  $

$d\left(  \Pi_{H\beta}\right)  =\sum_{i}\{-\frac{1}{\det O^{\prime}}O_{\beta
}^{\prime i}\varpi_{4}\left(  \sum_{\alpha\gamma}\frac{d}{d\xi^{\gamma}%
}\left(  Z_{O}^{i\alpha\gamma}\det O^{\prime}\right)  \partial_{\alpha
}\right)  $

$-\varpi_{4}\left(  \sum_{\alpha\gamma}\left(  Z_{O}^{i\alpha\gamma}\frac
{d}{d\xi^{\gamma}}O_{\beta}^{\prime i}\right)  \partial_{\alpha}\right)  $

$-\frac{1}{\det O^{\prime}}\varpi_{4}\left(  \sum_{\alpha\gamma}\frac{d}%
{d\xi^{\gamma}}\left(  O_{\beta}^{\prime i}\left(  \frac{d%
\mathcal{L}%
}{d\partial_{\gamma}O_{\alpha}^{\prime i}}+\frac{d%
\mathcal{L}%
}{d\partial_{\alpha}O_{\gamma}^{i\prime}}\right)  \right)  \partial_{\alpha
}\right)  \}$

$=-\frac{1}{\det O^{\prime}}\sum_{i}\varpi_{4}\left(  \sum_{\alpha\gamma
}\left(
\begin{array}
[c]{c}%
O_{\beta}^{\prime i}\frac{d}{d\xi^{\gamma}}\left(  Z_{O}^{i\alpha\gamma}\det
O^{\prime}\right)  +Z_{O}^{i\alpha\gamma}\det O^{\prime}\frac{d}{d\xi^{\gamma
}}O_{\beta}^{\prime i}\\
+\frac{d}{d\xi^{\gamma}}\left(  O_{\beta}^{\prime i}\left(  \frac{d%
\mathcal{L}%
}{d\partial_{\gamma}O_{\alpha}^{\prime i}}+\frac{d%
\mathcal{L}%
}{d\partial_{\alpha}O_{\gamma}^{i\prime}}\right)  \right)
\end{array}
\right)  \partial_{\alpha}\right)  $

$=-\frac{1}{\det O^{\prime}}\sum_{i}\varpi_{4}\sum_{\alpha\gamma}\frac{d}%
{d\xi^{\gamma}}\left(  O_{\beta}^{\prime i}\left(  \frac{d%
\mathcal{L}%
}{d\partial_{\alpha}O_{\gamma}^{i\prime}}-\frac{d%
\mathcal{L}%
}{d\partial_{\gamma}O_{\alpha}^{\prime i}}+\frac{d%
\mathcal{L}%
}{d\partial_{\gamma}O_{\alpha}^{\prime i}}+\frac{d%
\mathcal{L}%
}{d\partial_{\alpha}O_{\gamma}^{i\prime}}\right)  \right)  \partial_{\alpha}$

$=-\frac{2}{\det O^{\prime}}\sum_{i}\varpi_{4}\left(  \sum_{\alpha\gamma}%
\frac{d}{d\xi^{\gamma}}\left(  O_{\beta}^{\prime i}\frac{d%
\mathcal{L}%
}{d\partial_{\alpha}O_{\gamma}^{i\prime}}\right)  \partial_{\alpha}\right)
=-2\Pi_{O\beta}$%

\begin{equation}
\mathbf{d}\left(  \Pi_{H\beta}\right)  \mathbf{=-}\frac{2}{\det O^{\prime}%
}\sum_{i}\mathbf{\varpi}_{4}\left(  \sum_{\alpha\gamma}\frac{d}{d\xi^{\gamma}%
}\left(  O_{\beta}^{\prime i}\frac{d%
\mathcal{L}%
}{d\partial_{\alpha}O_{\gamma}^{i\prime}}\right)  \partial_{\alpha}\right)
\mathbf{=-2\Pi}_{O\beta}\label{E59a}%
\end{equation}

\paragraph{5)}

Therefore :

$\varpi_{4}\left(  Y_{H\beta}\right)  =\frac{1}{2}d\Pi_{H\beta}+\Pi_{O\beta
}=0$

As :

$\varpi_{4}\left(  Y_{H\beta}\right)  =\sum_{\alpha=0}^{3}\left(  -1\right)
^{\alpha+1}Y_{H\beta}^{\alpha}\det O^{\prime}dx^{0}\wedge...\wedge
\widehat{dx^{\alpha}}\wedge...\wedge dx^{3}$

we have :%

\begin{equation}
\mathbf{\forall\alpha,\beta:Y}_{H\beta}^{\alpha}\mathbf{=0}\label{E59b}%
\end{equation}

\paragraph{6)}

And :

$\frac{1}{\det O^{\prime}}\sum_{i\gamma}\frac{d}{d\xi^{\gamma}}\left(  \frac{d%
\mathcal{L}%
}{d\partial_{\gamma}O_{\alpha}^{\prime i}}O_{\beta}^{\prime i}\right)
=-\sum_{a}\left(  G_{\beta}^{a}Y_{G}^{a\alpha}+Y_{AR}^{\alpha a}%
\operatorname{Re}\grave{A}_{\beta}^{a}+Y_{AI}^{\alpha a}\operatorname{Im}%
\grave{A}_{\beta}^{a}\right)  $

$+2\sum_{a\gamma}\left(  \frac{dL_{F}}{d%
\mathcal{F}%
_{G\alpha\gamma}^{b}}\partial_{\gamma}G_{\beta}^{a}+\frac{dL_{F}%
}{d\operatorname{Re}%
\mathcal{F}%
_{A,\alpha\gamma}^{a}}\operatorname{Re}\partial_{\gamma}\grave{A}_{\beta}%
^{a}+\frac{dL_{F}}{d\operatorname{Im}%
\mathcal{F}%
_{A,\alpha\gamma}^{a}}\operatorname{Im}\partial_{\gamma}\grave{A}_{\beta}%
^{a}\right)  $

So :

$\delta_{\beta}^{\alpha}L=\sum_{i,j}\left(  \frac{dVL_{M}}{d\operatorname{Re}%
\nabla_{\alpha}\psi^{ij}}\operatorname{Re}\partial_{\beta}\psi^{ij}%
+\frac{dVL_{M}}{d\operatorname{Im}\nabla_{\alpha}\psi^{ij}}\operatorname{Im}%
\partial_{\beta}\psi^{ij}\right)  $

$+\sum_{a}G_{\beta}^{a}Y_{G}^{a\alpha}+Y_{AR}^{\alpha a}\operatorname{Re}%
\grave{A}_{\beta}^{a}+Y_{AI}^{\alpha a}\operatorname{Im}\grave{A}_{\beta}^{a}$

$+2\sum_{a\gamma}(\frac{dL_{F}}{d%
\mathcal{F}%
_{G\alpha\gamma}^{b}}\left(  \partial_{\beta}G_{\gamma}-\partial_{\gamma
}G_{\beta}\right)  ^{a}+\frac{dL_{F}}{d\operatorname{Re}%
\mathcal{F}%
_{A,\alpha\gamma}^{a}}\operatorname{Re}\left(  \partial_{\beta}\grave
{A}_{\gamma}-\partial_{\gamma}\grave{A}_{\beta}\right)  ^{a}$

$+\frac{dL_{F}}{d\operatorname{Im}%
\mathcal{F}%
_{A,\alpha\gamma}^{a}}\operatorname{Im}\left(  \partial_{\beta}\grave
{A}_{\gamma}-\partial_{\gamma}\grave{A}_{\beta}\right)  ^{a})$

$+\sum_{i\gamma}\frac{dL}{d\partial_{\alpha}O_{\gamma}^{\prime i}}%
\partial_{\beta}O_{\gamma}^{\prime i}-\sum_{a}\left(  G_{\beta}^{a}%
Y_{G}^{a\alpha}+Y_{AR}^{\alpha a}\operatorname{Re}\grave{A}_{\beta}^{a}%
+Y_{AI}^{\alpha a}\operatorname{Im}\grave{A}_{\beta}^{a}\right)  $

$+2\sum_{a\gamma}\left(  \frac{dL_{F}}{d%
\mathcal{F}%
_{G\alpha\gamma}^{a}}\partial_{\gamma}G_{\beta}^{a}+\frac{dL_{F}%
}{d\operatorname{Re}%
\mathcal{F}%
_{A,\alpha\gamma}^{a}}\operatorname{Re}\partial_{\gamma}\grave{A}_{\beta}%
^{a}+\frac{dL_{F}}{d\operatorname{Im}%
\mathcal{F}%
_{A,\alpha\gamma}^{a}}\operatorname{Im}\partial_{\gamma}\grave{A}_{\beta}%
^{a}\right)  -\frac{dNL_{M}}{dV^{\beta}}V^{\alpha}$

We get a formula, valid on shell, for the quantity $\delta_{\beta}^{\alpha}%
L$\ that we name the energy-momentum tensor.\ \ %

\begin{equation}
\label{E59c}
\end{equation}

$\delta_{\beta}^{\alpha}L=-\frac{dNL_{M}}{dV^{\beta}}V^{\alpha}+\sum_{i\gamma
}\frac{dL}{d\partial_{\alpha}O_{\gamma}^{\prime i}}\partial_{\beta}O_{\gamma
}^{\prime i}+\sum_{i,j}\left(  \frac{dNL_{M}}{d\operatorname{Re}\nabla
_{\alpha}\psi^{ij}}\operatorname{Re}\partial_{\beta}\psi^{ij}+\frac{dNL_{M}%
}{d\operatorname{Im}\nabla_{\alpha}\psi^{ij}}\operatorname{Im}\partial_{\beta
}\psi^{ij}\right)  $

$\qquad+2\sum_{a\gamma}\left(  \frac{dL_{F}}{d%
\mathcal{F}%
_{G\alpha\gamma}^{a}}\partial_{\beta}G_{\gamma}^{a}+\frac{dL_{F}%
}{d\operatorname{Re}%
\mathcal{F}%
_{A,\alpha\gamma}^{a}}\operatorname{Re}\partial_{\beta}\grave{A}_{\gamma}%
^{a}+\frac{dL_{F}}{d\operatorname{Im}%
\mathcal{F}%
_{A,\alpha\gamma}^{a}}\operatorname{Im}\partial_{\beta}\grave{A}_{\gamma}%
^{a}\right)  $

\subsection{Energy}

\paragraph{1)}

It is useful to give some thoughts about the physical meaning of these results.

If in the equation \ref{E59c} we put $\alpha=\beta$ :

$L=-\frac{dNL_{M}}{dV^{\alpha}}V^{\alpha}+\sum_{i,j}\left(  \frac{dNL_{M}%
}{d\operatorname{Re}\nabla_{\alpha}\psi^{ij}}\operatorname{Re}\partial
_{\alpha}\psi^{ij}+\frac{dNL_{M}}{d\operatorname{Im}\nabla_{\alpha}\psi^{ij}%
}\operatorname{Im}\partial_{\alpha}\psi^{ij}\right)  $

$+2\sum_{a\gamma}\left(  \frac{dL_{F}}{d%
\mathcal{F}%
_{G\alpha\gamma}^{a}}\partial_{\alpha}G_{\gamma}^{a}+\frac{dL_{F}%
}{d\operatorname{Re}%
\mathcal{F}%
_{A,\alpha\gamma}^{a}}\operatorname{Re}\partial_{\alpha}\grave{A}_{\gamma}%
^{a}+\frac{dL_{F}}{d\operatorname{Im}%
\mathcal{F}%
_{A,\alpha\gamma}^{a}}\operatorname{Im}\partial_{\alpha}\grave{A}_{\gamma}%
^{a}\right)  +\sum_{i\gamma}\frac{dL}{d\partial_{\alpha}O_{\gamma}^{\prime i}%
}\partial_{\alpha}O_{\gamma}^{\prime i}$

As :

$\frac{dL_{M}}{d\operatorname{Re}\nabla_{\alpha}\psi^{ij}}=\frac
{dL}{d\operatorname{Re}\partial_{\alpha}\psi^{ij}},\frac{dL_{M}}%
{d\operatorname{Im}\nabla_{\alpha}\psi^{ij}}=\frac{dL}{d\operatorname{Im}%
\partial_{\alpha}\psi^{ij}},2\frac{dL_{F}}{d%
\mathcal{F}%
_{G\alpha\gamma}^{b}}=\frac{dL}{d\partial_{\alpha}G_{\gamma}^{b}},....$

this equation reads on shell : $\forall\alpha:L=-\frac{dL}{dV^{\alpha}%
}V^{\alpha}+\sum_{i}\frac{dL}{dz_{\alpha}^{i}}\partial_{\alpha}z^{i}$

The first term is a kind of "kinetic energy", the next six correspond to the
potential energy of the fields, but there is still the last one :
$\sum_{i\gamma}\frac{dL}{d\partial_{\alpha}O_{\gamma}^{\prime i}}%
\partial_{\beta}O_{\gamma}^{\prime i}$ which features the distorsion of the
space time. We know that energy in General Relativity is a difficult concept.
In some way we would expect that the gravitational field encompasses all the
effects on the geometry of the universe, and this is why usually one discards
the derivatives $\partial_{\gamma}O_{\alpha}^{\prime i},$ but as we see a more
open vision is perhaps necessary. Notice that this issue is not related to the
choice of \textbf{G} or g as key variable.\ In the traditional variational
version of General Relativity the fundamental term is given by the scalar
curvature which is, as seen before, nothing but the curvature form of
\textbf{G}.

\paragraph{2)}

The quantity $\int_{\Omega\left(  t\right)  }\varpi_{4}\left(  L\partial
_{\beta}\right)  $ is the energy flow through the borders of $\Omega\left(
t\right)  $ seen by an observer on the line $\partial_{\beta}.$The flow is
given by $\varpi_{4}\left(  L\partial_{\beta}\right)  :$

$\varpi_{4}\left(  L\partial_{\beta}\right)  =\varpi_{4}\left(  \left(
-\frac{dNL_{M}}{dV^{\beta}}V^{\alpha}+\sum_{i,j}\frac{dNL_{M}}%
{d\operatorname{Re}\nabla_{\alpha}\psi^{ij}}\operatorname{Re}\partial_{\beta
}\psi^{ij}+\frac{dNL_{M}}{d\operatorname{Im}\nabla_{\alpha}\psi^{ij}%
}\operatorname{Im}\partial_{\beta}\psi^{ij}\right)  \partial_{\alpha}\right)
$

$+\sum_{a}\varpi_{4}\left(  2\sum_{\alpha\gamma}\frac{dL_{F}}{d%
\mathcal{F}%
_{G,\alpha\gamma}^{a}}\partial_{\beta}G_{\gamma}^{a}\partial_{\alpha}\right)
+\varpi_{4}\left(  2\sum_{\alpha\gamma}\frac{dL_{F}}{d\operatorname{Re}%
\mathcal{F}%
_{A,\alpha\gamma}^{a}}\operatorname{Re}\partial_{\beta}\grave{A}_{\gamma}%
^{a}\partial_{\alpha}\right)  $

$+\varpi_{4}\left(  2\sum_{\alpha\gamma}\frac{dL_{F}}{d\operatorname{Im}%
\mathcal{F}%
_{A,\alpha\gamma}^{a}}\operatorname{Im}\partial_{\beta}\grave{A}_{\gamma}%
^{a}\partial_{\alpha}\right)  +\sum_{i}\varpi_{4}\left(  \sum_{\alpha\gamma
}\frac{dL}{d\partial_{\alpha}O_{\gamma}^{\prime i}}\partial_{\beta}O_{\gamma
}^{\prime i}\partial_{\alpha}\right)  $

$\varpi_{4}\left(  L\partial_{\beta}\right)  =\varpi_{4}\left(  \left(
-\frac{dNL_{M}}{dV^{\beta}}V^{\alpha}+\sum_{i,j}\frac{dNL_{M}}%
{d\operatorname{Re}\nabla_{\alpha}\psi^{ij}}\operatorname{Re}\partial_{\beta
}\psi^{ij}+\frac{dNL_{M}}{d\operatorname{Im}\nabla_{\alpha}\psi^{ij}%
}\operatorname{Im}\partial_{\beta}\psi^{ij}\right)  \partial_{\alpha}\right)
$

$+\sum_{i}\varpi_{4}\left(  \sum_{\alpha\gamma}\frac{dL}{d\partial_{\alpha
}O_{\gamma}^{\prime i}}\partial_{\beta}O_{\gamma}^{\prime i}\partial_{\alpha
}\right)  -\frac{1}{2}\sum_{a}\left(  \sum_{\gamma}\partial_{\beta}G_{\gamma
}^{a}dx^{\gamma}\right)  \wedge\Pi_{G}^{a}+\left(  \sum_{\gamma}%
\operatorname{Re}\partial_{\beta}\grave{A}_{\gamma}^{a}dx^{\gamma}\right)
\wedge\Pi_{AR}^{a}+\left(  \sum_{\gamma}\operatorname{Im}\partial_{\beta
}\grave{A}_{\gamma}^{a}dx^{\gamma}\right)  \wedge\Pi_{AR}^{a}$

\newpage

\part{\textbf{THE\ MODEL}}

So far we have shown the constraints imposed on a lagrangian and established
the lagrange equations for a not too specific model.\ In order to improve our
grasp of the problem, it is useful to go a step further, and to test our
concepts on some lagrangian, keeping it simple enough to enable calculations.
As we have seen, in the $L_{M}$ part of the lagrangian the quantities involved
are chiefly the velocity, the state tensors and their covariant derivatives,
and in the $L_{F}$ part they\ are the curvature forms
$\mathcal{F}$%
. The identities to be met hint at some kind of homogeneous function, so it is
legitimate to look for quadratic functions, and thus for scalar products. The
scalar products must be defined for the state tensors, on the vector space
$F\otimes W,$ and for the connections curvatures forms
$\mathcal{F}$%
. They must be invariant in a gauge transformation. If we want to compute a
scalar product involving the derivatives of the state tensors, we must find a
way to define a differential operator acting on the fiber bundle $E_{M}$, it
will be the Dirac operator. Eventually we can improve somewhat the definition
of the F vector space by introducing chirality.

\section{SCALAR PRODUCTS}

\label{Scalar products}

\subsection{Scalar products for the state tensors}

\paragraph{1)}

The first step is to define an hermitian scalar product $\left\langle
{}\right\rangle $ on the vector space F, invariant under a gauge
transformation by Spin(3,1). An hermitian scalar product on F is represented
in the basis $e_{i}$ by an hermitian matrix A, such that :

$\forall u,v\in F:\left\langle u,v\right\rangle =\left[  u\right]  ^{\ast
}\left[  A\right]  \left[  v\right]  ;\left[  A\right]  =\left[  A\right]
^{\ast}$

Spin(3,1) acts on F through : $\rho\circ\Upsilon:Spin(3,1)\rightarrow L(F;F)$
so A must be such that :

$\forall s\in Spin(3,1):\left[  \rho\left(  \Upsilon\left(  s\right)  \right)
\right]  ^{\ast}\left[  A\right]  \left[  \rho\left(  \Upsilon\left(
s\right)  \right)  \right]  =\left[  A\right]  $

\paragraph{2)}

The $\gamma$ matrices are defined up to conjugation by a constant matrix, and
any A hermitian matrix meeting the conditions will be defined up to
conjugation by an unitary matrix.\ We choose $A=\gamma_{0}.$ Let us prove that
it fits the constraints.

It is an hermitian matrix, as all the other $\gamma_{i}$ as we have assumed so far.

$s\in Spin(3,1)$ is the product of an even number of vectors of norm 1 in
Cl(3,1) :

$s=v_{1}\cdot.....\cdot v_{2r},$

$v_{k}=v_{1}\varepsilon_{1}+v_{2}\varepsilon_{2}+v_{3}\varepsilon_{3}%
+v_{0}\varepsilon_{0},v_{\alpha}\in R$

$v_{1}^{2}+v_{2}^{2}+v_{3}^{2}-v_{0}^{2}=1$

$\Upsilon\left(  v_{k}\right)  =v_{1}\varepsilon_{1}+v_{2}\varepsilon
_{2}+v_{3}\varepsilon_{3}+iv_{0}\varepsilon_{0}$

So $\rho\left(  \Upsilon\left(  s\right)  \right)  =\rho\left(  \Upsilon
\left(  v_{1}\right)  \right)  ...\rho\left(  \Upsilon\left(  v_{2r}\right)
\right)  $ and $\left[  \rho\left(  \Upsilon\left(  s\right)  \right)
\right]  ^{\ast}=\rho\left(  \Upsilon\left(  v_{2r}\right)  \right)  ^{\ast
}...\rho\left(  \Upsilon\left(  v_{1}\right)  \right)  ^{\ast}$

$\rho\left(  \Upsilon\left(  v_{k}\right)  \right)  =v_{1}\gamma_{1}%
+v_{2}\gamma_{2}+v_{3}\gamma_{3}+iv_{0}\gamma_{0}$

$\rho\left(  \Upsilon\left(  v_{k}\right)  \right)  ^{\ast}=v_{1}\gamma
_{1}+v_{2}\gamma_{2}+v_{3}\gamma_{3}-iv_{0}\gamma_{0}$ because all the
components are real and the $\gamma_{k}$ are hermitian

$\rho\left(  \Upsilon\left(  v_{k}\right)  \right)  ^{\ast}\gamma_{0}%
\rho\left(  \Upsilon\left(  v_{k}\right)  \right)  $

$=\left(  v_{1}\gamma_{1}+v_{2}\gamma_{2}+v_{3}\gamma_{3}-iv_{0}\gamma
_{0}\right)  \gamma_{0}\left(  v_{1}\gamma_{1}+v_{2}\gamma_{2}+v_{3}\gamma
_{3}+iv_{0}\gamma_{0}\right)  $

$=\left(  v_{1}\gamma_{1}\gamma_{0}+v_{2}\gamma_{2}\gamma_{0}+v_{3}\gamma
_{3}\gamma_{0}-iv_{0}\gamma_{0}\gamma_{0}\right)  \left(  v_{1}\gamma
_{1}+v_{2}\gamma_{2}+v_{3}\gamma_{3}+iv_{0}\gamma_{0}\right)  $

$=\left(  -v_{1}^{2}-v_{2}^{2}-v_{3}^{2}+v_{0}^{2}\right)  \gamma_{0}%
+v_{2}v_{1}\gamma_{0}\gamma_{1}\gamma_{2}-v_{1}v_{2}\gamma_{0}\gamma_{1}%
\gamma_{2}+v_{3}v_{1}\gamma_{0}\gamma_{1}\gamma_{3}$

$-v_{1}v_{3}\gamma_{0}\gamma_{1}\gamma_{3}+v_{3}v_{2}\gamma_{0}\gamma
_{2}\gamma_{3}-v_{2}v_{3}\gamma_{0}\gamma_{2}\gamma_{3}$

$+iv_{1}v_{0}\gamma_{1}-iv_{0}v_{1}\gamma_{1}+iv_{2}v_{0}\gamma_{2}%
-iv_{0}v_{2}\gamma_{2}+iv_{3}v_{0}\gamma_{3}-iv_{0}v_{3}\gamma_{3}$

$=-\gamma_{0}$

Thus $\left[  \rho\left(  \Upsilon\left(  s\right)  \right)  \right]  ^{\ast
}\gamma_{0}\left[  \rho\left(  \Upsilon\left(  s\right)  \right)  \right]
=\left(  -1\right)  ^{2r}\gamma_{0}=\gamma_{0}\blacksquare$

We will take as scalar product in F :

$u,v\in F:\left\langle u,v\right\rangle _{F}=\left[  u\right]  ^{\ast}%
\gamma_{0}\left[  v\right]  =\sum_{ij}\gamma_{0ij}\overline{u}^{i}v^{j}$

It is not degenerate, but not necessarily definite positive.

Notice that the scalar product is invariant by Spin(3,1), but not
Spin(4,C).\ The basis $\left(  e_{i}\right)  $ is not necessarily orthonormal
: $\left\langle e_{i},e_{k}\right\rangle =\left[  \gamma_{0}\right]  _{ik}$
and the representation $\left(  F,\rho\circ\Upsilon\right)  $ of Spin(3,1) is
not necessarily unitary.

From $\left[  \rho\left(  \Upsilon\left(  s\right)  \right)  \right]  ^{\ast
}\gamma_{0}\left[  \rho\left(  \Upsilon\left(  s\right)  \right)  \right]
=\gamma_{0}$ one deduces by differentiating with respect to s=1 :

$\forall\kappa\in o(3,1):\frac{d}{ds}\left(  \left[  \rho\left(
\Upsilon\left(  s\right)  \right)  \right]  ^{\ast}\gamma_{0}\left[
\rho\left(  \Upsilon\left(  s\right)  \right)  \right]  \right)  |_{s=1}=0$

$\left[  \rho\left(  \Upsilon\left(  s\right)  \right)  ^{\prime}%
\Upsilon^{\prime}(s)\kappa\right]  ^{\ast}\gamma_{0}\left[  \rho\left(
\Upsilon\left(  s\right)  \right)  \right]  +\left[  \rho\left(
\Upsilon\left(  s\right)  \right)  \right]  ^{\ast}\gamma_{0}\left[
\rho^{\prime}\left(  \Upsilon\left(  s\right)  \right)  \Upsilon^{\prime
}\left(  s\right)  \kappa\right]  =0$

$\left[  \rho\left(  1\right)  ^{\prime}\Upsilon^{\prime}(1)\kappa\right]
^{\ast}\gamma_{0}+\gamma_{4}\left[  \rho^{\prime}\left(  1\right)
\Upsilon^{\prime}\left(  1\right)  \kappa\right]  =0$

$\Upsilon^{\prime}(1)=Id$

$\left[  \rho\left(  1\right)  ^{\prime}\kappa\right]  ^{\ast}\gamma
_{0}+\gamma_{0}\left[  \rho^{\prime}\left(  1\right)  \kappa\right]  =0$

Thus with : $\rho^{\prime}(1)\overrightarrow{\kappa}_{a}=\left[  \kappa
_{a}\right]  :\left[  \kappa_{a}\right]  ^{\ast}\gamma_{0}+\gamma_{0}\left[
\kappa_{a}\right]  =0$%

\begin{equation}
\left[  \kappa_{a}\right]  ^{\ast}=-\left[  \gamma_{0}\right]  \left[
\kappa_{a}\right]  \left[  \gamma_{0}\right] \label{E60}%
\end{equation}

\paragraph{3)}

We assume that there is an hermitian scalar product on the vector space W,
invariant by $\chi:$

$\forall u\in U,\sigma,\sigma^{\prime}\in W:\left\langle \sigma,\sigma
^{\prime}\right\rangle =\left\langle \chi\left(  u\right)  \sigma,\chi\left(
u\right)  \sigma^{\prime}\right\rangle $

and that the basis $\left(  f_{j}\right)  $ is orthonormal $\left\langle
\sigma,\sigma^{\prime}\right\rangle =\sum_{i}\overline{\sigma}^{i}%
\sigma^{\prime i}$

Remark : on a complex vector space the signature of an hermitian form can be
set up at +.

\paragraph{4)}

From there we define an hermitian scalar product on $F\otimes W:$

$\left\langle \sigma_{1}^{i}\phi_{1}^{j}e_{i}\otimes f_{j},\sigma_{2}^{i}%
\phi_{2}^{j}e_{i}\otimes f_{j}\right\rangle =\left(  \sum_{ij}\gamma
_{0ij}\overline{\sigma_{1}}^{i}\sigma_{2}^{j}\right)  \left(  \sum
_{k}\overline{\phi}_{1}^{k}\phi_{2}^{k}\right)  =\sum_{ijk}\gamma
_{0ij}\overline{\sigma_{1}}^{i}\overline{\phi}_{1}^{k}\sigma_{2}^{j}\phi
_{2}^{k}$

So we define :

$\left\langle \psi_{1}^{ij}e_{i}\otimes f_{j},\psi_{2}^{ij}e_{i}\otimes
f_{j}\right\rangle =\sum_{ijk}\gamma_{0ij}\overline{\psi}_{1}^{ik}\psi
_{2}^{jk}=\left[  \psi_{1}^{\ast}\right]  _{k}^{i}\left[  \gamma_{0}\right]
_{j}^{i}\left[  \psi_{2}\right]  _{k}^{j}=Tr\left(  \left[  \psi_{1}\right]
^{\ast}\gamma_{0}\left[  \psi_{2}\right]  \right)  $

with the 4xm matrices : $\left[  \psi\right]  $

$Tr\overline{\left(  \left[  \psi_{1}\right]  ^{\ast}\gamma_{0}\left[
\psi_{2}\right]  \right)  }=Tr\left(  \overline{\left[  \psi_{1}\right]
}^{\ast}\overline{\gamma_{0}}\left[  \overline{\psi_{2}}\right]  \right)
=Tr\left(  \left[  \psi_{1}\right]  ^{t}\gamma_{4}^{t}\left[  \psi_{2}\right]
^{\ast t}\right)  =Tr\left(  \left[  \psi_{2}\right]  ^{\ast}\gamma_{0}\left[
\psi_{1}\right]  \right)  $

And the scalar product is extended on the vector bunle $E_{M}$\ $:$%

\begin{equation}
\mathbf{\psi}_{1}\mathbf{,\psi}_{2}\mathbf{\in\Lambda}_{0}\left(
E_{v}\right)  \mathbf{:}\left\langle \psi_{1},\psi_{2}\right\rangle
\mathbf{=Tr}\left(  \left[  \psi_{1}\right]  ^{\ast}\gamma_{0}\left[  \psi
_{2}\right]  \right)  \mathbf{=}\sum_{ijk}\mathbf{\gamma}_{0ij}\overline{\psi
}_{1}^{ik}\mathbf{\psi}_{2}^{jk}\label{E61}%
\end{equation}

It should be noticed that this scalar product, as any hermitian scalar
product, is invariant by the transformation : $\psi\rightarrow z\psi$ where z
is c-number valued function on M, with $\left\vert z\right\vert =1.$

\subsection{Scalar products for the curvature forms}

\paragraph{1)}

The connection and\ curvature forms are valued in the Lie algebras, and the
action of the gauge groups is the adjoint operator Ad, so we need a scalar
product on the Lie algebras invariant by the adjoint operator. On any Lie
algebra there is a bilinear symmetric form B (the Killing form) invariant by
Ad, but non necessarily positive definite. If the Lie algebra is semi-simple
(as o(3,1)) it is non degenerate. In the standard representation of o(3,1) :
$B(X,Y)=2Tr(\left[  X\right]  \left[  Y\right]  )$ and its signature is (+ + +
- - -) with the basis $\left(  \overrightarrow{\kappa}_{a}\right)  .$

We will not be so specific and just assume that there is an hermitian scalar
product $\left(  {}\right)  $, invariant by the adjoint action, on the Lie
algebras, not necessarily positive definite, for which the bases $\left(
\overrightarrow{\kappa}_{a}\right)  ,\left(  \overrightarrow{\theta}%
_{a}\right)  $\ are orthogonal\ :

- o(3,1) : as we use only the real form it is a symmetric real scalar product,
and we assume that $\left(  \overrightarrow{\kappa}_{a},\overrightarrow
{\kappa}_{b}\right)  =\eta_{ab}=\pm1$

- $T_{1}U^{c}:$the scalar product is assumed to be hermitian and the basis
$\left(  \overrightarrow{\theta}_{a}\right)  $\ orthonormal : \ $\left(
\overrightarrow{\theta}_{a},\overrightarrow{\theta}_{b}\right)  =\delta
_{ab}\Rightarrow\left(  \overrightarrow{\theta},\overrightarrow{\theta
}^{\prime}\right)  =\sum_{a}\overline{\theta}^{a}\theta^{\prime a}$

$\forall u\in U,\overrightarrow{\theta},\overrightarrow{\theta}^{\prime}\in
T_{1}U^{c}:\left(  Ad_{u}\overrightarrow{\theta},Ad_{u}\overrightarrow{\theta
}^{\prime}\right)  =\left(  \overrightarrow{\theta},\overrightarrow{\theta
}^{\prime}\right)  $

that implies : $u=\exp\tau\overrightarrow{\theta}:\frac{d}{d\tau}\left(
Ad_{\exp\tau\overrightarrow{\theta}}\overrightarrow{\theta}_{1},Ad_{\exp
\tau\overrightarrow{\theta}}\overrightarrow{\theta}_{2}\right)  |_{\tau
=0}=0=\left(  \left[  \overrightarrow{\theta},\overrightarrow{\theta}%
_{1}\right]  ,\overrightarrow{\theta}_{2}\right)  +\left(  \overrightarrow
{\theta}_{1},\left[  \overrightarrow{\theta},\overrightarrow{\theta}%
_{2}\right]  \right)  $

$\Rightarrow\forall\overrightarrow{\theta},\overrightarrow{\theta}%
_{1},\overrightarrow{\theta}_{2}\in T_{1}U^{c}:\left(  \left[  \overrightarrow
{\theta},\overrightarrow{\theta}_{1}\right]  ,\overrightarrow{\theta}%
_{2}\right)  =-\left(  \overrightarrow{\theta}_{1},\left[  \overrightarrow
{\theta},\overrightarrow{\theta}_{2}\right]  \right)  $

The space vector $T_{1}U^{c}$ endowed with this scalar product is a Hilbert space.

\paragraph{2)}

The potential and the forms are forms on the manifold M. M is endowed with the
lorentzian metric g. One defines the scalar product of two r-forms on M valued
in $%
\mathbb{C}
$\ \ by :

$G_{r}:\Lambda_{r}\left(  M;%
\mathbb{C}
\right)  \times\Lambda_{r}\left(  M;%
\mathbb{C}
\right)  \rightarrow%
\mathbb{C}
::$

$G_{r}\left(  \lambda,\mu\right)  =\sum_{\left\{  \alpha_{1}...\alpha
_{r}\right\}  ,\left\{  \beta_{1}...\beta_{r}\right\}  }\lambda_{\{\alpha
_{1}..\alpha_{r}\}}\mu_{\{\beta_{1}...\beta_{r}\}}\left[  \det g^{-1}%
(x)\right]  ^{\left\{  \alpha_{1}...\alpha_{r}\right\}  ,\left\{  \beta
_{1}...\beta_{r}\right\}  }$

$G_{r}\left(  \lambda,\mu\right)  $

$=\sum_{\left\{  \alpha_{1}...\alpha_{r}\right\}  }\lambda_{\{\alpha
_{1}...\alpha_{r}\}}\sum_{\beta_{1}...\beta_{r}}g^{\alpha_{1}\beta_{1}%
}...g^{\alpha_{r}\beta_{r}}\mu_{\beta_{1}...\beta_{r}}=\sum_{\left\{
\alpha_{1}...\alpha_{r}\right\}  }\lambda_{\{\alpha_{1}...\alpha_{r}\}}%
\mu^{\left\{  \alpha_{1}\alpha_{2}...\alpha_{r}\right\}  }$

where the indexes are uppered and lowered with g and $\left\{  \alpha\right\}
=\left\{  \alpha_{1}...\alpha_{r}\right\}  $ is an ordered set of r indexes.

The scalar product is symmetric, non degenerated and invariant under a change
of chart. It defines an isomorphism between the algebras of r and 4-r forms,
given by the Hodge dual.

The Hodge dual $\lambda^{\ast}$ of a r-form $\lambda\in\Lambda_{r}\left(  M;%
\mathbb{C}
\right)  $\ is the 4-r form $\lambda^{\ast}\in\Lambda_{4-r}\left(  M;%
\mathbb{C}
\right)  $\ such that :

$\forall\mu\in\Lambda_{r}\left(  M;%
\mathbb{R}
\right)  :\mu\wedge\ast\lambda=G_{r}\left(  \mu,\lambda\right)  \varpi_{4}$
where : $\varpi_{4}=\left(  \det O^{\prime}\right)  dx^{0}\wedge..\wedge
dx^{3}$ (Taylor [26] 5.8)

$\ast\left(  \sum_{\left\{  \alpha_{1}...\alpha_{r}\right\}  }\lambda
_{\{\alpha_{1}...\alpha_{r}\}}dx^{\alpha_{1}}..\wedge dx^{a_{r}}\right)  $

$=\sum_{\left\{  \alpha_{1}...\alpha_{n-r}\right\}  ,\left\{  \beta
_{1}...\beta_{r}\right\}  }\epsilon\left(  \beta_{1}..\beta_{r},\alpha
_{1},...\alpha_{n-r}\right)  u^{\beta_{1}...\beta_{r}}\left(  \det O^{\prime
}\right)  dx^{\alpha_{1}}..\wedge dx^{a_{4-r}}$

We have : $\ast\ast\lambda=(-1)^{r-1}\lambda$

$\ast\left(  \sum_{\alpha}\lambda_{\alpha}dx^{\alpha}\right)  =\sum_{\alpha
=0}^{4}\left(  -1\right)  ^{\alpha+1}g^{\alpha\beta}\lambda_{\beta}\left(
\det O^{\prime}\right)  dx^{0}\wedge..\widehat{dx^{\alpha}}\wedge...dx^{3}$

For a 2-form the formula is simple when one uses a convenient ordering of the components.

$\lambda\in\Lambda_{2}\left(  M;%
\mathbb{C}
\right)  :$

$\lambda=\{\lambda_{01}dx^{0}\wedge dx^{1}+\lambda_{02}dx^{0}\wedge
dx^{2}+\lambda_{03}dx^{0}\wedge dx^{3}$

$+\lambda_{32}dx^{3}\wedge dx^{2}+\lambda_{13}dx^{1}\wedge dx^{3}+\lambda
_{21}dx^{2}\wedge dx^{1}\}$%

\begin{equation}
\end{equation}

$\ast\lambda=-\{\lambda^{32}dx^{0}\wedge dx^{1}+\lambda^{13}dx^{0}\wedge
dx^{2}+\lambda^{21}dx^{0}\wedge dx^{3}$

$\qquad+\lambda^{01}dx^{3}\wedge dx^{2}+\lambda^{02}dx^{1}\wedge
dx^{3}+\lambda^{03}dx^{2}\wedge dx^{1}\}\det O^{\prime}$

\bigskip

Notice that this convenient ordering is deduced from the table 1.

In the orthonormal basis these formulas become :

$\lambda=\sum_{\left\{  i_{1}..i_{r}\right\}  }\lambda_{i_{1}...i_{r}}%
\partial^{i_{1}}\wedge..\wedge\partial^{i_{r}},\mu=\sum_{\left\{  i_{1}%
..i_{r}\right\}  }\mu_{i_{1}...i_{r}}\partial^{i_{1}}\wedge..\wedge
\partial^{i_{r}}$

$G_{r}(\lambda,\mu)=\sum_{\left\{  i_{1}..i_{p}\right\}  }\eta^{i_{1}i_{1}%
}...\eta^{i_{p}i_{p}}\lambda_{i_{1}..i_{p}}\mu_{i_{1}..i_{p}}$ where
$\eta^{i_{k}i_{k}}=\pm1$

$\ast(\sum_{\left\{  i_{1}..i_{p}\right\}  }\lambda_{i_{1}..i_{p}}%
\partial^{i_{1}}\wedge..\wedge\partial^{i_{p}})$

$=\sum_{\left\{  i_{1}..i_{p}\right\}  ,\left\{  j_{1}..j_{4-p}\right\}
}\epsilon\left(  \left\{  i_{1}..i_{p}\right\}  ,\left\{  j_{1}..j_{4-p}%
\right\}  \right)  \eta_{i_{1}i_{1}}...\eta_{i_{p}i_{p}}\lambda_{\{i_{1}%
..i_{p}\}}\partial^{j_{1}}\wedge...\wedge\partial^{j_{4-p}}$

$\ast\left(  \sum_{\left\{  i\right\}  }\lambda_{i}\partial^{i}\right)
=\sum_{j=0}^{3}\left(  -1\right)  ^{j+1}\eta_{jj}\lambda_{j}\partial^{0}%
\wedge.\widehat{\partial^{j}}..\wedge\partial^{3}$

$\ast\left(  \sum_{j=1}^{4}\lambda_{0..\widehat{j}...n}\partial^{1}%
\wedge.\widehat{\partial^{j}}..\wedge\partial^{3}\right)  =\sum_{j=0}%
^{3}\left(  -1\right)  ^{j+1}\eta_{jj}\lambda_{0..\widehat{j}...3}\partial
^{j}$

$\ast\sum_{\left\{  ij\right\}  }\lambda_{ij}\partial^{i}\wedge\partial^{j}$

$=-\lambda_{32}\partial^{0}\wedge\partial^{1}-\lambda_{13}\partial^{0}%
\wedge\partial^{2}-\lambda_{21}\partial^{0}\wedge\partial^{3}+\lambda
_{02}\partial^{1}\wedge\partial^{3}+\lambda_{01}\partial^{3}\wedge\partial
^{2}+\lambda_{03}\partial^{2}\wedge\partial^{1}$

$\ast\sum_{a}\lambda_{a}\partial^{p_{a}}\wedge\partial^{q_{a}}=\sum_{a=1}%
^{3}-\lambda_{a}\partial^{p_{a+3}}\wedge\partial^{q_{a+3}}+\lambda
_{a+3}\partial^{p_{a}}\wedge\partial^{q_{a}}$%

\begin{equation}
\end{equation}

$\ast\left(  \lambda_{01}\partial^{0}\wedge\partial^{1}+\lambda_{02}%
\partial^{0}\wedge\partial^{2}+\lambda_{03}\partial^{0}\wedge\partial
^{3}+\lambda_{32}\partial^{3}\wedge\partial^{2}+\lambda_{13}\partial^{1}%
\wedge\partial^{3}+\lambda_{21}\partial^{2}\wedge\partial^{1}\right)  $

$=\lambda_{01}\partial^{3}\wedge\partial^{2}+\lambda_{02}\partial^{1}%
\wedge\partial^{3}+\lambda_{03}\partial^{2}\wedge\partial^{1}-\lambda
_{32}\partial^{0}\wedge\partial^{1}-\lambda_{13}\partial^{0}\wedge\partial
^{2}-\lambda_{21}\partial^{0}\wedge\partial^{3}$

\bigskip

\paragraph{3)}

Let $H_{M}$ any vector bundle over M modelled on a vector space H endowed with
an hermitian product $\left(  {}\right)  .$ One defines the hermitian product
of 2 r-forms on M valued in $H_{M}$ by :

$\lambda,\mu\in\Lambda_{r}\left(  M;H_{M}\right)  :\lambda=\sum_{a}%
\sum_{\left\{  \alpha\right\}  }\lambda_{\{\alpha_{1}..\alpha_{r}\}}%
^{a}dx^{\alpha_{1}}\wedge..\wedge dx^{\alpha_{r}}\otimes\overrightarrow
{\varkappa}_{a}\left(  m\right)  $ where $\overrightarrow{\varkappa}%
_{a}\left(  m\right)  $\ is a local basis of H\ 

$\left\langle \lambda,\mu\right\rangle $

$=\sum_{a,b}G_{r}\left(  \overline{\lambda}^{a},\mu^{b}\right)  \left(
\overrightarrow{\varkappa}_{a},\overrightarrow{\varkappa}_{b}\right)  $

$=\sum_{a,b}\sum_{\left\{  \alpha_{1}...\alpha_{r}\right\}  }\left(
\overline{\lambda}_{\{\alpha_{1}...\alpha_{r}\}}^{a}\mu^{b,\alpha_{1}%
\alpha_{2}...\alpha_{r}}\right)  \left(  \overrightarrow{\varkappa}%
_{a},\overrightarrow{\varkappa}_{b}\right)  $

$=\sum_{\left\{  \alpha\right\}  ,\left\{  \beta\right\}  }\left(
\lambda_{\{\alpha\}}^{a}\overrightarrow{\varkappa}_{a},\mu_{\{\beta\}}%
^{b}\overrightarrow{\varkappa}_{b}\right)  _{H}\left[  \det g^{-1}(x)\right]
^{\{\alpha\},\{\beta\}}$

$=\sum_{a,b}\sum_{\left\{  \alpha_{1}...\alpha_{r}\right\}  }\left(
\overline{\lambda}_{\{\alpha_{1}...\alpha_{r}\}}^{a}\mu^{b,\alpha_{1}%
\alpha_{2}...\alpha_{r}}\right)  \left(  \overrightarrow{\varkappa}%
_{a},\overrightarrow{\varkappa}_{b}\right)  $

$=\sum_{a,b}\sum_{\left\{  \alpha_{1}...\alpha_{r}\right\}  }\left(
\overline{\lambda}^{a,\{\alpha_{1}...\alpha_{r}\}}\mu_{\{\alpha_{1}%
...\alpha_{r}\}}^{b}\right)  \left(  \overrightarrow{\varkappa}_{a}%
,\overrightarrow{\varkappa}_{b}\right)  $

It is hermitian, invariant by a chart transformation in M and does not depend
of the local basis. If $H_{M}$ is an associated bundle it is invariant by a
gauge transformation.

For the curvature form in an orthonormal basis of $T_{1}U^{c}:$%

\begin{equation}
\left\langle
\mathcal{F}%
_{A},%
\mathcal{F}%
_{A}\right\rangle \mathbf{=}\sum_{a}\sum_{\left\{  \alpha\beta\right\}
}\overline{%
\mathcal{F}%
}_{A\alpha\beta}^{a}\mathbf{%
\mathcal{F}%
}_{A}^{a,\alpha\beta}\label{E62}%
\end{equation}

$\left\langle
\mathcal{F}%
_{A},%
\mathcal{F}%
_{A}\right\rangle $

$=\sum_{a}\sum_{\left\{  \alpha\beta\right\}  }\overline{%
\mathcal{F}%
}_{A\alpha\beta}^{a}\sum_{\lambda\mu}g^{\alpha\lambda}g^{\beta\mu}%
\mathcal{F}%
_{A\lambda\mu}^{a}$

$=\frac{1}{2}\sum_{a}\sum_{\alpha\beta\lambda\mu}g^{\alpha\lambda}g^{\beta\mu
}\overline{%
\mathcal{F}%
}_{A\alpha\beta}^{a}%
\mathcal{F}%
_{A\lambda\mu}^{a}$

$=\frac{1}{2}\sum_{a}\sum_{\alpha\beta}\overline{%
\mathcal{F}%
}_{A\alpha\beta}^{a}%
\mathcal{F}%
_{A}^{a\alpha\beta}$

The Hodge dual is defined as : $\ast\left(  \sum_{a}\lambda^{a}\otimes
\overrightarrow{\varkappa}_{a}\right)  =\sum_{a}\ast\overline{\lambda}%
^{a}\otimes\overrightarrow{\varkappa}_{a}$ where $\ast\overline{\lambda}^{a}$
is the Hodge dual (in the previous meaning) of the C valued form
$\overline{\lambda}^{a}$

$\lambda,\mu\in\Lambda_{r}\left(  M;H_{M}\right)  :\left\langle \lambda
,\mu\right\rangle =\sum_{a,b}G_{r}\left(  \overline{\lambda}^{a},\mu
^{b}\right)  \left(  \overrightarrow{\varkappa}_{a},\overrightarrow{\varkappa
}_{b}\right)  $

$\mu^{b}\wedge\ast\overline{\lambda}^{a}=G_{r}\left(  \mu^{b},\overline
{\lambda}^{a}\right)  \varpi_{4}=G_{r}\left(  \overline{\lambda}^{a},\mu
^{b}\right)  =\mu^{b}\wedge\ast\overline{\lambda}^{a}$

$\left\langle \lambda,\mu\right\rangle \varpi_{4}=\sum_{a,b}G_{r}\left(
\overline{\lambda}^{a},\mu^{b}\right)  \left(  \overrightarrow{\varkappa}%
_{a},\overrightarrow{\varkappa}_{b}\right)  \varpi_{4}=\sum_{a,b}\left(
\mu^{b}\wedge\ast\overline{\lambda}^{a}\right)  \left(  \overrightarrow
{\varkappa}_{a},\overrightarrow{\varkappa}_{b}\right)  \varpi_{4}$

With an orthonormal basis : $\left\langle \lambda,\mu\right\rangle \varpi
_{4}=\sum_{a}\left(  \mu^{a}\wedge\ast\overline{\lambda}^{a}\right)
\varpi_{4}$

For the curvature form :%

\begin{equation}
\label{E63}
\end{equation}

$\ast\left(
\mathcal{F}%
_{A}\right)  =-\left(  \det O^{\prime}\right)  \sum_{a}\{\overline{%
\mathcal{F}%
}^{a,01}dx^{3}\wedge dx^{2}+\overline{%
\mathcal{F}%
}^{a,02}dx^{1}\wedge dx^{3}+\overline{%
\mathcal{F}%
}^{a,03}dx^{2}\wedge dx^{1}$

$\qquad+\overline{%
\mathcal{F}%
}^{a,32}dx^{0}\wedge dx^{1}+\overline{%
\mathcal{F}%
}^{a,13}dx^{0}\wedge dx^{2}+\overline{%
\mathcal{F}%
}^{a,21}dx^{0}\wedge dx^{3}\otimes\overrightarrow{\theta}_{a}\}$

\section{THE\ DIRAC\ OPERATOR}

\label{Dirac operator}

\paragraph{1)}

The fields change the state of particles and their velocity.\ The interaction
with the velocity is logically modelled by $\nabla_{V}\psi=\sum_{\alpha
}V^{\alpha}\nabla_{\alpha}\psi.$ But we can assume that there is also an
action on the state tensor itself, meaning independant of the dynamic
(represented by V). So it is sensible to look after some derivative operator :
$\Lambda_{0}E_{M}\rightarrow$ $\Lambda_{0}E_{M}.$ That is the Dirac operator,
which, in matrix form, is : $\left[  D\psi\right]  =\sum_{\alpha}O_{r}%
^{\alpha}\sum_{r=0}^{3}\left[  \gamma^{r}\right]  \left[  \nabla_{\alpha}%
\psi\right]  $ where $\left[  \gamma^{r}\right]  $ are matrices defined from
$\left[  \gamma_{r}\right]  .$ The construction is the following.

\paragraph{2)}

Spin(3,1)xU acts on $Cl(3,1)\times(F\otimes W)$ as : $\left(  s,u\right)
\times\left(  w,\psi\right)  =\left(  \mathbf{Ad}_{s}w,\vartheta\left(
s,u\right)  \psi\right)  $ so one can define the vector bundle $Q_{M}%
\times_{Spin(3,1)\times U}\left(  Cl(3,1)\times(F\otimes W)\right)  $
associated to $Q_{M}$ by the equivalence :

$\left(  q,\left(  w,\psi\right)  \right)  \simeq\left(  q\left(
s^{-1},u^{-1}\right)  ,\left(  \mathbf{Ad}_{s}w,\vartheta\left(  s,u\right)
\psi\right)  \right)  $

This action is linear with Cl(3,1) and $F\otimes W$\ so from there one can get
a vector bundle $D_{M}=Q_{M}\times_{Spin(3,1)\times U}\left(  Cl(3,1)\otimes
F\otimes W)\right)  $\ associated to $Q_{M}$\ modelled on $Cl(3,1)\otimes
F\otimes W$ (the tensor product is associative). The isomorphism :

$\left(  q\left(  m\right)  ,\left(  \varepsilon_{i}\otimes e_{j}\otimes
f_{k}\right)  \right)  \rightarrow\varphi_{C}\left(  m,\varepsilon_{i}\right)
\otimes e_{j}\left(  m\right)  \otimes f_{k}\left(  m\right)  =\partial
_{i}\left(  m\right)  \otimes e_{j}\left(  m\right)  \otimes f_{k}\left(
m\right)  $

is preserved by the action of Spin(3,1)xU:

$\left(  q\left(  m\right)  \left(  s^{-1},u^{-1}\right)  ,\left(  \left(
\mathbf{Ad}_{s}\varepsilon_{i}\right)  \otimes\rho\circ\Upsilon\left(
s\right)  \left(  e_{j}\right)  \otimes\chi\left(  u\right)  \left(
f_{k}\right)  \right)  \right)  $

$\rightarrow\left[  \jmath\circ\mu\left(  s\right)  \right]  _{i}^{l}%
\partial_{l}\otimes\rho\left(  \Upsilon\left(  s\right)  \right)  \left(
e_{j}\left(  m\right)  \right)  \otimes\chi\left(  u\right)  \left(
f_{k}\left(  m\right)  \right)  $

So one can pick $\partial_{i}\left(  m\right)  \otimes e_{j}\left(  m\right)
\otimes f_{k}\left(  m\right)  $ as a basis of this vector bundle $D_{M}$.

\paragraph{3)}

One defines the projection : $P:D_{M}\rightarrow E_{M}$ by :

$T^{ijk}\partial_{i}\left(  m\right)  \otimes e_{j}\left(  m\right)  \otimes
f_{k}\left(  m\right)  \rightarrow T^{ijk}\left(  \rho\circ\Upsilon\left(
\varepsilon_{i}\right)  \left(  e_{j}\right)  \right)  \left(  m\right)
\otimes f_{k}\left(  m\right)  $

extended by linearity. It is consistent because :

$T^{ijk}\partial_{i}\left(  m\right)  \otimes e_{j}\left(  m\right)  \otimes
f_{k}\left(  m\right)  $

$\simeq\left(  q\left(  m\right)  \left(  s^{-1},u^{-1}\right)  ,T^{ijk}%
\left(  Ad_{s}\varepsilon_{i}\otimes\rho\circ\Upsilon\left(  s\right)  \left(
e_{j}\right)  \otimes\chi\left(  u\right)  \left(  f_{k}\right)  \right)
\right)  $

$P\left(  \left(  q\left(  m\right)  \left(  s^{-1},u^{-1}\right)
,T^{ijk}\left(  \left(  Ad_{s}\varepsilon_{i}\right)  \otimes\rho\circ
\Upsilon\left(  s\right)  \left(  e_{j}\right)  \otimes\chi\left(  u\right)
\left(  f_{k}\right)  \right)  \right)  \right)  $

$=\left(  q\left(  m\right)  \left(  s^{-1},g^{-1}\right)  ,T^{ijk}\left(
\rho\circ\Upsilon\left(  Ad_{s}\varepsilon_{i}\right)  \rho\circ
\Upsilon\left(  s\right)  \left(  e_{j}\right)  \otimes\chi\left(  u\right)
\left(  f_{k}\right)  \right)  \right)  $

but

$\rho\circ\Upsilon\left(  Ad_{s}\varepsilon_{i}\right)  \rho\circ
\Upsilon\left(  s\right)  =\rho\circ\Upsilon\left(  Ad_{s}\varepsilon
_{i}.s\right)  $

$=\rho\circ\Upsilon\left(  s.\varepsilon_{i}.s^{-1}.s\right)  =\rho
\circ\Upsilon\left(  s.\varepsilon_{i}\right)  =\rho\circ\Upsilon\left(
s\right)  \rho\circ\Upsilon\left(  \varepsilon_{i}\right)  $

so :

$\left(  q\left(  m\right)  \left(  s^{-1},u^{-1}\right)  ,T^{ijk}\left(
\rho\circ\Upsilon\left(  s\right)  \rho\circ\Upsilon\left(  \varepsilon
_{i}\right)  \left(  e_{j}\right)  \otimes\chi\left(  u\right)  \left(
f_{k}\right)  \right)  \right)  $

$\simeq\left(  q\left(  m\right)  ,T^{ijk}\rho\circ\Upsilon\left(
\varepsilon_{i}\right)  e_{j}\otimes f_{k}\right)  =T^{ijk}\left(  \rho
\circ\Upsilon\left(  \varepsilon_{i}\right)  e_{j}\right)  \left(  m\right)
\otimes f_{k}\left(  m\right)  $

\paragraph{4)}

\ $\nabla\psi$ is a 1-form on TM*.\ With the metric we can go from TM* to TM :
$dx^{\alpha}\rightarrow g^{\alpha\beta}\partial_{\beta}$

$\left(  \nabla_{\alpha}\psi^{ij}\right)  dx^{\alpha}\otimes e_{i}\left(
m\right)  \otimes f_{i}(m)\rightarrow\left(  \nabla_{\alpha}\psi^{ij}\right)
g^{\alpha\beta}\partial_{\beta}\otimes e_{i}\left(  m\right)  \otimes
f_{j}\left(  m\right)  $

and from TM to $D_{M}:$

$\partial_{i}=O\left(  m\right)  _{i}^{\alpha}\partial_{\alpha}\Leftrightarrow
\partial_{\alpha}=O\left(  m\right)  _{\alpha}^{\prime i}\partial_{i}$

$\left(  \nabla_{\alpha}\psi^{ij}\right)  g^{\alpha\beta}\partial_{\beta
}\otimes e_{i}\left(  m\right)  \otimes f_{j}\left(  m\right)  $

$=\left(  \nabla_{\alpha}\psi^{ij}\right)  g^{\alpha\beta}O^{\prime}\left(
m\right)  _{\beta}^{l}\partial_{l}\otimes e_{i}\left(  m\right)  \otimes
f_{j}\left(  m\right)  $

$=\left(  \nabla_{r}\psi^{ij}\right)  \partial_{r}\left(  m\right)  \otimes
e_{i}\left(  m\right)  \otimes f_{j}\left(  m\right)  $

The resulting quantity is a section on $D_{M},$ that can be projected onto
$E_{M}.$

So, overall, the Dirac operator is the map $D:\Lambda_{0}\left(  E_{M}\right)
\rightarrow\Lambda_{0}\left(  E_{M}\right)  $

$D\psi=\sum_{\alpha ij}\left(  \nabla_{\alpha}\psi^{ij}\right)  \sum_{\beta
l}g^{\alpha\beta}O^{\prime}\left(  m\right)  _{\beta}^{l}\left(  \rho
\circ\Upsilon\left(  \varepsilon_{l}\right)  \left(  e_{i}\right)  \right)
\left(  m\right)  \otimes f_{j}\left(  m\right)  $

\paragraph{5)}

This quantity can be expressed in a better way. $\partial_{j}$ is orthonormal
so :

$\eta_{ij}=g_{\alpha\beta}\left[  O\right]  _{i}^{\alpha}\left[  O\right]
_{j}^{\beta}\Rightarrow g^{\alpha\beta}\left[  O^{\prime}\right]  _{\beta}%
^{l}=\eta^{lr}\left[  O\right]  _{r}^{\alpha}$

For : l=1,2,3: $\rho\circ\Upsilon\left(  \varepsilon_{l}\right)  =\gamma_{l};$
and $\rho\circ\Upsilon\left(  \varepsilon_{0}\right)  =i\gamma_{0}$

$\sum_{\beta l}g^{\alpha\beta}O^{\prime}\left(  m\right)  _{\beta}^{l}\left(
\rho\circ\Upsilon\left(  \varepsilon_{l}\right)  \left(  e_{i}\right)
\right)  \left(  m\right)  $

$=\sum_{rl}\eta^{lr}O_{r}^{\alpha}\left(  \rho\circ\Upsilon\left(
\varepsilon_{l}\right)  \right)  _{i}^{p}e_{p}\left(  m\right)  =\left(
\eta^{0r}i\left[  \gamma_{0}\right]  _{i}^{p}+\sum_{l=1}^{3}\left[  \gamma
_{l}\right]  _{i}^{p}\eta^{lr}\right)  e_{p}\left(  m\right)  $

We define the matrices : $\gamma^{r}$ (index up) such that : $\left[
\gamma^{r}\right]  _{i}^{p}=\eta^{0r}i\left[  \gamma_{0}\right]  _{i}^{p}%
+\sum_{l=1}^{3}\left[  \gamma_{l}\right]  _{i}^{p}\eta^{lr}.$ We have :%

\begin{equation}
\mathbf{\gamma}^{0}\mathbf{=-i\gamma}_{0}\mathbf{;r=1,2,3:\gamma}%
^{r}\mathbf{=\gamma}_{r}\label{E64}%
\end{equation}

\bigskip

$D\psi=\sum_{\alpha}\left(  \nabla_{\alpha}\psi^{ij}\right)  \sum_{r}\left[
O\right]  _{r}^{\alpha}\left[  \gamma^{r}\right]  _{i}^{p}e_{p}\left(
m\right)  \otimes f_{j}\left(  m\right)  $

We will denote :%

\begin{equation}
\left[  \gamma^{\alpha}\right]  \mathbf{=}\sum_{r=0}^{3}\left[  O\right]
_{r}^{\alpha}\left[  \gamma^{r}\right] \label{E65}%
\end{equation}

\begin{equation}
\sum_{\alpha=0}^{3}\left(  \nabla_{\alpha}\psi^{ij}\right)  \mathbf{O}%
_{r}^{\alpha}\mathbf{=\nabla}_{\partial_{r}}\mathbf{\psi}^{ij}\mathbf{=\nabla
}_{r}\mathbf{\psi}^{ij}\label{E66}%
\end{equation}

\bigskip

With these notations :

$D\psi$

$=\sum_{i,p=1}^{4}\sum_{j=1}^{m}\sum_{r=0}^{3}\left(  \nabla_{r}\psi
^{ij}\right)  \left[  \gamma^{r}\right]  _{i}^{p}e_{p}\left(  m\right)
\otimes f_{j}\left(  m\right)  $

$=\sum_{i,p=1}^{4}\sum_{j=1}^{m}\sum_{\alpha=0}^{3}\left(  \nabla_{\alpha}%
\psi^{ij}\right)  \left[  \gamma^{\alpha}\right]  _{i}^{p}e_{p}\left(
m\right)  \otimes f_{j}\left(  m\right)  $

Or in matrix notation :%

\begin{equation}
\left[  D\psi\right]  \mathbf{=}\sum_{r=0}^{3}\left[  \gamma^{r}\right]
\left[  \nabla_{r}\psi\right]  \mathbf{=}\sum_{\alpha=0}^{3}\left[
\gamma^{\alpha}\right]  \left[  \nabla_{\alpha}\psi\right] \label{E67}%
\end{equation}

We see on the relations above that the Dirac operator can be defined through
an orthonormal basis only, without any explicit reference to a metric.

\paragraph{6)}

The Dirac operator is well defined and linear.\ 

In a gauge transformation :

$\left(  \nabla_{\partial_{r}}\psi^{ij}\right)  e_{i}\left(  m\right)  \otimes
f_{j}\left(  m\right)  \simeq\left(  \widetilde{\nabla_{\widetilde{\partial
}_{r}}\psi^{ij}}\right)  \widetilde{e}_{i}\left(  m\right)  \otimes
\widetilde{f}_{j}\left(  m\right)  $

$\sum_{r}\left(  \nabla_{r}\psi^{ij}\right)  \left[  \gamma^{r}\right]
_{i}^{p}e_{p}\left(  m\right)  \otimes f_{j}\left(  m\right)  \simeq\sum
_{r}\left(  \widetilde{\nabla_{\widetilde{\partial}_{r}}\psi^{ij}}\right)
\left[  \gamma^{r}\right]  _{i}^{p}\widetilde{e}_{p}\left(  m\right)
\otimes\widetilde{f}_{j}\left(  m\right)  $

and it is invariant in a change of chart :

$\partial_{\alpha}\rightarrow\widetilde{\partial}_{\alpha}=J_{\alpha}^{\beta
}\partial_{\beta}$

$dx^{\alpha}\rightarrow\widetilde{dx^{\alpha}}=K_{\beta}^{\alpha}dx^{\beta}$
with $K_{\lambda}^{\alpha}J_{\beta}^{\lambda}=\delta_{\beta}^{\alpha}$

$\nabla_{\alpha}\psi^{ij}\rightarrow\widetilde{\nabla_{\alpha}\psi^{ij}%
}=J_{\alpha}^{\lambda}\nabla_{\lambda}\psi^{ij}$

$\left[  O\right]  _{r}^{\alpha}\rightarrow K_{\mu}^{\alpha}\left[  O\right]
_{r}^{\mu}$

$\nabla_{r}\psi^{ij}\rightarrow\sum_{\alpha=0}^{3}\left(  \widetilde
{\nabla_{\alpha}\psi^{ij}}\right)  \widetilde{O}_{r}^{\alpha}=\sum_{\alpha
=0}^{3}J_{\alpha}^{\lambda}\left(  \nabla_{\lambda}\psi^{ij}\right)  K_{\mu
}^{\alpha}\left[  O\right]  _{r}^{\mu}$

$=\sum_{\alpha=0}^{3}\left(  \nabla_{\lambda}\psi^{ij}\right)  \left[
O\right]  _{r}^{\lambda}=\nabla_{r}\psi^{ij}$

So $D\psi$ is invariant.$\blacksquare$

\section{CHIRALITY}

\label{Chirality}

We have left open the choice of the vector space F in the representation of
Cl(4,C). It can be made more precise, without no lost of generality, using a
feature of Clifford algebras. It is striking that this feature meets one
important characteristic of the physical world, that it distinguishes between
the "left" and the "right", that is chirality, and thus is fundamental in
particle physics.

\subsection{The splitting of Clifford algebras}

\paragraph{1)}

The ordered product $\varepsilon_{0}\cdot\varepsilon_{1}\cdot..\cdot
\varepsilon_{n}$ of the vectors of a direct orthonormal basis in a Clifford
algebra does not depend of the choice of this basis. In Cl(4,C) $\varepsilon
_{5}=\varepsilon_{0}\cdot\varepsilon_{1}\cdot\varepsilon_{2}\cdot
\varepsilon_{3}$ is such that $\left(  \varepsilon_{5}\right)  ^{2}=1.$

Let be :

$Cl^{+}\left(  4,C\right)  =\left\{  w\in Cl(4,C):\varepsilon_{5}\cdot
w=w\right\}  ,$

$Cl^{-}\left(  4,C\right)  =\left\{  w\in Cl(4,C):\varepsilon_{5}\cdot
w=-w\right\}  $

Remind that $Cl(4,C)=Cl_{0}\left(  4,C\right)  \oplus Cl_{1}\left(
4,C\right)  $ where $Cl_{0}\left(  4,C\right)  $ is the subalgebra of the
elements sum of an even product of vectors.

Let us prove that $Cl_{0}\left(  4,C\right)  =Cl_{0}^{+}\left(  4,C\right)
\oplus Cl_{0}^{-}\left(  4,C\right)  $ where $Cl_{0}^{+}\left(  4,C\right)
,Cl_{0}^{-}\left(  4,C\right)  $\ \ are two subalgebras which are isomorphic
and "orthogonal" in that : $\forall w\in Cl_{0}^{+}\left(  4,C\right)
,w^{\prime}\in Cl_{0}^{-}\left(  4,C\right)  :w\cdot w^{\prime}=0$

$Cl_{0}^{+}\left(  4,C\right)  =Cl_{0}\left(  4,C\right)  \cap Cl^{+}\left(
4,C\right)  $ and $Cl_{0}^{-}\left(  4,C\right)  =Cl_{0}\left(  4,C\right)
\cap Cl^{-}\left(  4,C\right)  $ are subspace of $Cl_{0}\left(  4,C\right)  $

$Cl_{0}\left(  4,C\right)  ,Cl^{+}\left(  4,C\right)  $ are subalgebras, so is
$Cl_{0}^{+}\left(  4,C\right)  $

if $w,w^{\prime}\in Cl_{0}^{-}\left(  4,C\right)  ,\varepsilon_{5}\cdot w\cdot
w^{\prime}=-w\cdot w^{\prime}\Leftrightarrow w\cdot w^{\prime}\in Cl_{0}%
^{-}\left(  4,C\right)  :$ $Cl_{0}^{-}\left(  4,C\right)  $ is a subalgebra

the only element common to the two subalgebras is 0, thus $Cl_{0}\left(
4,C\right)  =Cl_{0}^{+}\left(  4,C\right)  \oplus Cl_{0}^{-}\left(
4,C\right)  $

$\varepsilon_{5}$ commute with any element of $Cl_{0}\left(  4,C\right)  ,$
and anticommute with all elements of $Cl_{1}\left(  4,C\right)  $ so

If $w\in Cl_{0}^{+}\left(  4,C\right)  ,w^{\prime}\in Cl_{0}^{-}\left(
4,C\right)  :\varepsilon_{5}\cdot w=w,\varepsilon_{5}\cdot w^{\prime
}=-w^{\prime}$

$\varepsilon_{5}\cdot w\cdot\varepsilon_{5}\cdot w^{\prime}=w\cdot
\varepsilon_{5}\cdot\varepsilon_{5}\cdot w^{\prime}=w\cdot w^{\prime
}=-\varepsilon_{5}\cdot w\cdot w^{\prime}=-w\cdot w^{\prime}\Rightarrow w\cdot
w^{\prime}=0$

$\blacksquare$

Similarly : $Cl_{1}\left(  4,C\right)  =Cl_{1}^{+}\left(  4,C\right)  \oplus
Cl_{1}^{-}\left(  4,C\right)  $ (but they are not subalgebras)

So any element w\ of Cl(4,C) can be written : $w=w^{+}+w^{-}$ with $w^{+}\in
Cl^{+}\left(  4,C\right)  ,w^{-}\in Cl^{-}(4,C)$

\paragraph{2)}

Let be : $p_{\epsilon}=\frac{1}{2}\left(  1+\epsilon\varepsilon_{5}\right)
$\ with $\epsilon=\pm1.$\ $p_{+},p_{-}$ are the creation and annihiliation operators

There are the identities :

$p_{\epsilon}^{2}=p_{\epsilon},p_{+}\cdot p_{-}=p_{-}\cdot p_{+}=0,p_{+}%
+p_{-}=1$

For any $v\in C^{4}:p_{\epsilon}v=vp_{-\epsilon}$

For any $w=w^{+}+w^{-}\in Cl(4,C):p_{\epsilon}\cdot w=\frac{1}{2}\left(
\left(  1+\epsilon\right)  w^{+}+\left(  1-\epsilon\right)  w^{-}\right)  $

$\Rightarrow$

$p_{+}\cdot w=w^{+},p_{-}\cdot w=w^{-};p_{+}\cdot w^{+}=w^{+},$

$p_{-}\cdot w^{+}=0;p_{-}\cdot w^{-}=w^{-},p_{-}\cdot w^{+}=0$

If $w\in Cl_{0}\left(  4,C\right)  :w\cdot p_{\epsilon}=\frac{1}{2}\left(
\left(  1+\epsilon\right)  w^{+}+\left(  1-\epsilon\right)  w^{-}\right)  $

$\Rightarrow w^{+}.p_{+}=w^{+},w^{-}.p_{+}=0,w^{+}.p_{-}=0,w^{-}.p_{-}=w^{-}$

If $w\in Cl_{1}\left(  4,C\right)  :w\cdot p_{\epsilon}=w.\frac{1}{2}\left(
1+\epsilon\varepsilon_{5}\right)  =\frac{1}{2}\left(  w-\epsilon
\varepsilon_{5}.w\right)  $

$=\frac{1}{2}\left(  w^{+}+w^{-}-\epsilon w^{+}+\epsilon w^{-}\right)
=\frac{1}{2}\left(  \left(  1-\epsilon\right)  w^{+}+\left(  1+\epsilon
\right)  w^{-}\right)  $

$\Rightarrow w^{+}.p_{+}=0,w^{-}.p_{+}=w^{-},w^{+}.p_{-}=w^{+},w^{-}.p_{-}=0$

\paragraph{3)}

We have similarly in the $\left(  F,\rho\right)  $ representation of Cl(4,C) :

$\varepsilon_{5}\rightarrow\gamma_{5}=\gamma_{0}\gamma_{1}\gamma_{2}\gamma
_{3}$

$p_{\epsilon}\rightarrow\gamma_{\epsilon}=\frac{1}{2}\left(  I+\epsilon
\gamma_{5}\right)  $

We have the identities :

$\gamma_{5}^{2}=I;\gamma_{5}=\gamma_{5}^{\ast}$

$\gamma_{+}+\gamma_{-}=I;\gamma_{\epsilon}\gamma_{\epsilon}=\gamma_{\epsilon
};\gamma_{\epsilon}\gamma_{-\epsilon}=0$

k=0..3: $\gamma_{5}\gamma_{k}=-\gamma_{k}\gamma_{5};\gamma_{\epsilon}%
\gamma_{k}=\gamma_{k}\gamma_{-\epsilon};$

$\gamma_{+}\left[  \kappa_{a}\right]  =k\gamma_{+}\gamma_{p_{a}}\gamma_{q_{a}%
}=\frac{1}{2}k\gamma_{p_{a}}\gamma_{q_{a}}+\frac{1}{2}k\gamma_{5}\gamma
_{p_{a}}\gamma_{q_{a}}$

$=\frac{1}{2}k\gamma_{p_{a}}\gamma_{q_{a}}+\frac{1}{2}k\gamma_{p_{a}}%
\gamma_{q_{a}}\gamma_{5}=\left[  \kappa_{a}\right]  \gamma_{+}$

Let be the vector subspaces :$F^{+}=\gamma_{+}F,F^{-}=\gamma_{-}F$

$\gamma_{+}F^{-}=\gamma_{-}F^{+}=\left\{  0\right\}  \Rightarrow$\ $F^{+}\cap
F^{-}=\left\{  0\right\}  $

so : $F=F^{+}\oplus F^{-}$ and $\forall\phi\in F:\phi=\phi^{+}+\phi^{-}%
=\gamma_{+}\phi+\gamma_{-}\phi$

These two subspaces are isomorphic.\ Indeed for any non null vector v of Cl(4,C):

$p_{\epsilon}v=vp_{-\epsilon}\Rightarrow\gamma_{\epsilon}\rho\left(  v\right)
=\rho\left(  v\right)  \gamma_{-\epsilon}\Rightarrow\gamma_{\epsilon}%
\rho\left(  v\right)  \phi=\rho\left(  v\right)  \gamma_{-\epsilon}\phi$

$\Rightarrow\gamma_{-\epsilon}\phi=\rho\left(  v\right)  ^{-1}\gamma
_{\epsilon}\rho\left(  v\right)  \phi$

For any homogeneous element w of order k in Cl(4,C) :

$w=v_{1}.....v_{k}:$

$\rho\left(  w\right)  \gamma_{+}=\rho\left(  v_{1}.....v_{k}.p_{+}\right)
=\rho\left(  v_{1}.....v_{k-1}.p_{-}.v_{k}\right)  =\rho\left(  p_{\epsilon
}.v_{1}.....v_{k-1}v_{k}\right)  =\gamma_{\epsilon}\rho\left(  w\right)
$\ with $\epsilon=\left(  -1\right)  ^{k}$

and similarly for $\gamma_{-}.$ Thus if $w\in Cl_{0}\left(  4,C\right)  $ we
have $\rho\left(  w\right)  \gamma_{+}=\gamma_{+}\rho\left(  w\right)  $ and
$\rho\left(  w\right)  \gamma_{-}=\gamma_{-}\rho\left(  w\right)  $ : $F^{+}$
and $F^{-}$ are globally invariant for the $Cl_{0}\left(  4,C\right)  $
action.\ Conversely $Cl_{1}\left(  4,C\right)  $ exchanges $F^{+}$ and $F^{-}$.

The Spin(3,1) group image by $\Upsilon$\ is $Spin_{c}\left(  3,1\right)
\subset Cl_{0}\left(  4,C\right)  .$ So the splitting $F^{+},F^{-}$ is stable
under the Spin(3,1) action. $\left(  F,\rho\right)  $ is an irreducible
representation of the algebra Cl(4,C) but a reducible representation of
Spin(3,1) : $\left(  F,\rho\right)  =\left(  F^{+},\rho\right)  \oplus\left(
F^{-},\rho\right)  .$

\paragraph{4)}

This splitting is extended in $F\otimes W.$ Tensor product being associative
we have :

$F\otimes W=\left(  F^{+}\oplus F^{-}\right)  \otimes W=\left(  F^{+}\otimes
W\right)  \oplus\left(  F^{-}\otimes W\right)  $

and any tensor $\psi$\ can be splitted in the sum of two tensors, each one
comprised of 2 complex components over F:\ 

$\psi=\psi_{+}+\psi_{-}$ with $\gamma_{+}\psi_{+}=\psi_{+},\gamma_{-}\psi
_{-}=\psi_{-},\gamma_{\epsilon}\psi_{-\epsilon}=0$

$\psi=\sum_{ij}\psi^{ij}e_{i}\otimes f_{j}$

$\phi_{j}=\sum_{i}\psi^{ij}e_{i}=\gamma_{+}\phi_{j}+\gamma_{-}\phi_{j}$

$\psi=\sum_{j}\left(  \gamma_{+}\phi_{j}+\gamma_{-}\phi_{j}\right)  \otimes
f_{j}=\gamma_{+}\left(  \sum_{j}\phi_{j}\otimes f_{j}\right)  +\gamma
_{-}\left(  \sum_{j}\phi_{j}\otimes f_{j}\right)  =\gamma_{+}\psi+\gamma
_{-}\psi;$

$\psi^{ij}=\left[  \gamma_{+}\phi_{j}\right]  ^{i}+\left[  \gamma_{-}\phi
_{j}\right]  ^{i}=\left[  \gamma_{+}\right]  _{k}^{i}\phi_{j}^{k}+\left[
\gamma_{-}\right]  _{k}^{i}\phi_{j}^{k}=\left[  \gamma_{+}\right]  _{k}%
^{i}\psi^{kj}+\left[  \gamma_{-}\right]  _{k}^{i}\psi^{kj}$

The splitting is stable under $Cl_{0}\left(  4,C\right)  :$

$\forall w\in Cl_{0}\left(  4,C\right)  :\vartheta\left(  w,g\right)
\psi=\sum_{k,l}\psi^{kl}\left[  \rho\circ\Upsilon\left(  w\right)  \right]
_{k}^{i}\left[  \chi\left(  g\right)  \right]  _{l}^{j}e_{i}\otimes f_{j}$

$=\sum_{k,l,p}\left[  \gamma_{+}\right]  _{p}^{k}\psi^{pl}\left[  \rho
\circ\Upsilon\left(  w\right)  \right]  _{k}^{i}\left[  \chi\left(  g\right)
\right]  _{l}^{j}e_{i}\otimes f_{j}$

$+\sum_{k,l}\left[  \gamma_{-}\right]  _{p}^{k}\psi^{pl}\left[  \rho
\circ\Upsilon\left(  w\right)  \right]  _{k}^{i}\left[  \chi\left(  g\right)
\right]  _{l}^{j}e_{i}\otimes f_{j}$

$=\sum_{k,l,p}\psi^{pl}\left[  \rho\circ\Upsilon\left(  w\right)  \circ
\gamma_{+}\right]  _{p}^{i}\left[  \chi\left(  g\right)  \right]  _{l}%
^{j}e_{i}\otimes f_{j}$

$+\sum_{k,l}\psi^{pl}\left[  \rho\circ\Upsilon\left(  w\right)  \circ
\gamma_{-}\right]  _{p}^{i}\left[  \chi\left(  g\right)  \right]  _{l}%
^{j}e_{i}\otimes f_{j}$

$=\sum_{k,l,p}\psi^{pl}\left[  \gamma_{+}\circ\rho\circ\Upsilon\left(
w\right)  \right]  _{p}^{i}\left[  \chi\left(  g\right)  \right]  _{l}%
^{j}e_{i}\otimes f_{j}$

$+\sum_{k,l}\psi^{pl}\left[  \gamma_{-}\circ\rho\circ\Upsilon\left(  w\right)
\right]  _{p}^{i}\left[  \chi\left(  g\right)  \right]  _{l}^{j}e_{i}\otimes
f_{j}$

$=\sum_{k,l,p}\left[  \gamma_{+}\right]  _{k}^{i}\left[  \rho\circ
\Upsilon\left(  w\right)  \right]  _{p}^{k}\left[  \chi\left(  g\right)
\right]  _{l}^{j}\psi^{pl}e_{i}\otimes f_{j}$

$+\sum_{k,l}\left[  \gamma_{-}\right]  _{k}^{i}\left[  \rho\circ
\Upsilon\left(  w\right)  \right]  _{p}^{k}\left[  \chi\left(  g\right)
\right]  _{l}^{j}\psi^{pl}e_{i}\otimes f_{j}\blacksquare$

The two vector spaces are isotropic for the scalar product defined previously :

$\left\langle \psi_{1},\psi_{2}\right\rangle =Tr\left(  \left[  \psi
_{1}\right]  ^{\ast}\gamma_{0}\left[  \psi_{2}\right]  \right)  =\sum
_{ijk}\gamma_{4ij}\overline{\psi}_{1}^{ik}\psi_{2}^{jk}$

If $\left[  \psi_{2}\right]  =\gamma_{\epsilon}\psi_{2},\left[  \psi
_{1}\right]  =\gamma_{\epsilon^{\prime}}\psi_{1}:\left[  \psi_{1}\right]
^{\ast}\gamma_{0}\left[  \psi_{2}\right]  =\left[  \psi_{1}\right]  ^{\ast
}\gamma_{\epsilon}\gamma_{0}\gamma_{\epsilon^{\prime}}\left[  \psi_{2}\right]
=\left[  \psi_{1}\right]  ^{\ast}\gamma_{\epsilon}\gamma_{-\epsilon^{\prime}%
}\gamma_{0}\left[  \psi_{2}\right]  $

$\varepsilon=\varepsilon^{\prime}\Rightarrow\left\langle \psi_{1},\psi
_{2}\right\rangle =0\blacksquare$

\paragraph{5)}

$\varepsilon_{5}$ does not depend of the choice of an orthonormal basis, and
so for $\gamma_{\epsilon}.$The splitting can be lifted on the fiber bundle. At
each m the fiber over m can be splitted in $F^{+}\otimes W+F^{-}\otimes W.$

\subsection{A choice of the F space}

\paragraph{1)}

These features lead to precise the choice of a specific basis of the vector
space F : a basis which reflects the splitting in the direct sum of two 2
dimensional complex vector spaces.

Let be this basis such that $F^{+}$ corresponds to the two first vectors
$\left(  e_{1},e_{2}\right)  $\ and $F^{-}$\ to the last $\left(  e_{3}%
,e_{4}\right)  :$

$F^{+}=\left\{
\begin{bmatrix}
\phi_{1}\\
\phi_{2}\\
0\\
0
\end{bmatrix}
\right\}  ,F^{-}=\left\{
\begin{bmatrix}
0\\
0\\
\phi_{3}\\
\phi_{4}%
\end{bmatrix}
\right\}  $

The conditions : $F^{+}=\gamma_{+}F,F^{-}=\gamma_{-}F;\gamma_{+}+\gamma
_{-}=I;\gamma_{\epsilon}\gamma_{-\epsilon}=0$ lead to

\bigskip

$\gamma_{+}=%
\begin{bmatrix}
I_{2} & 0\\
0 & 0
\end{bmatrix}
,\gamma_{-}=%
\begin{bmatrix}
0 & 0\\
0 & I_{2}%
\end{bmatrix}
$

Thus $\gamma_{5}=\gamma_{+}-\gamma_{-}=%
\begin{bmatrix}
I & 0\\
0 & -I
\end{bmatrix}
$

\bigskip

The $\gamma_{k}$ matrices must meet : $\gamma_{i}\gamma_{j}+\gamma_{j}%
\gamma_{i}=2\delta_{ij}I_{4}$ and we assume as before that : $\gamma
_{k}=\gamma_{k}^{\ast}\Rightarrow\gamma_{k}\gamma_{k}^{\ast}=I$

So $\gamma_{k}=%
\begin{bmatrix}
A_{k} & B_{k}\\
B_{k}^{\ast} & D_{k}%
\end{bmatrix}
$

with $A_{k}=A_{k}^{\ast},D_{k}=D_{k}^{\ast},A_{k}^{2}+B_{k}B_{k}^{\ast
}=I=B_{k}^{\ast}B_{k}+D_{k}^{2},A_{k}B_{k}+B_{k}D_{k}=0$

The conditions k=0..3: $\gamma_{5}\gamma_{k}=-\gamma_{k}\gamma_{5}$ impose :
$A_{k}=D_{k}=0$

so $\gamma_{k}=%
\begin{bmatrix}
0 & B_{k}\\
B_{k}^{\ast} & 0
\end{bmatrix}
$ with $B_{k}B_{k}^{\ast}=I_{2}$

The condition $\gamma_{5}=\gamma_{0}\gamma_{1}\gamma_{2}\gamma_{3}$ imposes
:$B_{0}B_{1}^{\ast}B_{2}B_{3}^{\ast}=I_{2},B_{1}B_{0}^{\ast}B_{3}B_{2}^{\ast
}=-I_{2}$

\paragraph{2)}

There are not many choices left, and we come to the solution :

\bigskip

$\gamma_{1}=%
\begin{bmatrix}
0 & \sigma_{1}\\
\sigma_{1} & 0
\end{bmatrix}
;\gamma_{2}=%
\begin{bmatrix}
0 & \sigma_{2}\\
\sigma_{2} & 0
\end{bmatrix}
;\gamma_{3}=%
\begin{bmatrix}
0 & \sigma_{3}\\
\sigma_{3} & 0
\end{bmatrix}
;\gamma_{0}=i%
\begin{bmatrix}
0 & \sigma_{0}\\
-\sigma_{0} & 0
\end{bmatrix}
$

\bigskip

with the Pauli matrices :

$\sigma_{0}=%
\begin{bmatrix}
1 & 0\\
0 & 1
\end{bmatrix}
;\sigma_{1}=%
\begin{bmatrix}
0 & 1\\
1 & 0
\end{bmatrix}
;\sigma_{2}=i%
\begin{bmatrix}
0 & -1\\
1 & 0
\end{bmatrix}
;\sigma_{3}=%
\begin{bmatrix}
1 & 0\\
0 & -1
\end{bmatrix}
;$

\bigskip

$i,j=1,2,3:\sigma_{i}\sigma_{j}+\sigma_{j}\sigma_{i}=2\delta_{ij}\sigma
_{0};\sigma_{i}=\sigma_{i}^{\ast};\sigma_{1}\sigma_{2}=i\sigma_{3};\sigma
_{2}\sigma_{3}=i\sigma_{1};\sigma_{3}\sigma_{1}=i\sigma_{2}$

With this choice we have :

\bigskip

$\gamma^{r}=%
\begin{bmatrix}
0 & \sigma_{r}\\
\eta^{rr}\sigma_{r} & 0
\end{bmatrix}
$

a%
$<$%
4 : $\left[  \kappa_{a}\right]  =-i\frac{1}{2}%
\begin{bmatrix}
\sigma_{a} & 0\\
0 & \sigma_{a}%
\end{bmatrix}
;$ \ \ \ a%
$>$%
3 : $\left[  \kappa_{a}\right]  =\frac{1}{2}%
\begin{bmatrix}
\sigma_{a-3} & 0\\
0 & -\sigma_{a-3}%
\end{bmatrix}
$

\bigskip

\paragraph{3)}

The kinematic part of the state of particles is described in a two 2-complex
components vector : $\phi=%
\begin{bmatrix}
\phi_{R}\\
\phi_{L}%
\end{bmatrix}
,$ and $\phi_{R}\in F^{+},\phi_{L}\in F^{-}$ are Weyl spinors. The state is
described in a section $\psi=%
\begin{bmatrix}
\psi_{R}\\
\psi_{L}%
\end{bmatrix}
$ where $\psi_{R}=\psi_{R}^{ij},\psi_{L}=\psi_{L}^{ij},j=1..m,i=1,2$

We will denote $\left[  \psi_{R}\right]  ,\left[  \psi_{L}\right]  $ the 2xm matrices.

\section{LAGRANGIAN}

\label{Lagrangian Model}

The lagrangian is comprised of 3 parts, related to the particles alone, the
field forces alone, and the interactions.

\subsection{Particules}

\paragraph{1)}

For the particles the lagrangian cannot depend of the derivatives, which
depend on the fields. So the simplest choice is :

$a_{M}N\left\langle \psi,\psi\right\rangle =a_{M}NTr\left(  \left[
\psi\right]  ^{\ast}\gamma_{0}\left[  \psi\right]  \right)  $ with some real
scalar $a_{M}$

$\left\langle \psi,\psi\right\rangle =Tr\left(
\begin{bmatrix}
\psi_{R}^{\ast} & \psi_{L}^{\ast}%
\end{bmatrix}
i%
\begin{bmatrix}
0 & \sigma_{0}\\
-\sigma_{0} & 0
\end{bmatrix}%
\begin{bmatrix}
\psi_{R}\\
\psi_{L}%
\end{bmatrix}
\right)  =iTr\left(  \left[  \psi_{R}\right]  ^{\ast}\left[  \psi_{L}\right]
-\left[  \psi_{L}\right]  ^{\ast}\left[  \psi_{R}\right]  \right)  $

$\overline{Tr\left(  \left[  \psi_{R}\right]  ^{\ast}\left[  \psi_{L}\right]
\right)  }=Tr\left(  \left(  \left[  \psi_{R}\right]  ^{\ast}\left[  \psi
_{L}\right]  \right)  ^{\ast}\right)  =Tr\left(  \left[  \psi_{L}\right]
^{\ast}\left[  \psi_{R}\right]  \right)  $

$Tr\left(  \left[  \psi_{R}\right]  ^{\ast}\left[  \psi_{L}\right]  -\left[
\psi_{L}\right]  ^{\ast}\left[  \psi_{R}\right]  \right)  =2i\operatorname{Im}%
Tr\left(  \left[  \psi_{R}\right]  ^{\ast}\left[  \psi_{L}\right]  \right)  $

$\left\langle \psi,\psi\right\rangle =-2\operatorname{Im}Tr\left(  \left[
\psi_{R}\right]  ^{\ast}\left[  \psi_{L}\right]  \right)  $

This quantity does not depend of the jauge or the chart : this a function on
$\Omega,$ evaluated at $\widetilde{f}\left(  m\right)  $ .

\paragraph{2)}

It would be legitimate to add some dynamic part like $\sum_{\alpha}V^{\alpha
}V_{\alpha}$ but it raises two issues.\ First we should put some "mass",which
figures in $\psi.$ Second the mass is the "charge" associated to the
gravitation field, so it makes sense to keep the dynamic part linked with the
covariant derivative $\nabla\psi.$ We will see how.

\subsection{Fields}

\paragraph{1)}

The lagrangian depends on the curvature forms and cannot involve $\psi$ or
explicitely \`{A}.\ We will assume that the derivatives $\partial_{\alpha
}O^{\prime}$ are not present, thus G does not appear and the lagrangian
depends uniquely on the curvature forms.

\paragraph{2)}

The simplest solution is to take the scalar product defined before. For
$\mathcal{F}$%
$_{A}$ it reads :

$a_{F}\left\langle
\mathcal{F}%
_{A},%
\mathcal{F}%
_{A}\right\rangle =a_{F}\sum_{a}\sum_{\left\{  \alpha\beta\right\}  }%
\overline{%
\mathcal{F}%
}_{A\alpha\beta}^{a}%
\mathcal{F}%
_{A}^{a,\alpha\beta}$

$=a_{F}\sum_{a}\sum_{\left\{  \alpha\beta\right\}  }\overline{%
\mathcal{F}%
}_{A\alpha\beta}^{a}\sum_{\lambda\mu}g^{\alpha\lambda}g^{\beta\mu}%
\mathcal{F}%
_{A\lambda\mu}^{a}=a_{F}\frac{1}{2}\sum_{a}\sum_{\alpha\beta\lambda\mu
}g^{\alpha\lambda}g^{\beta\mu}\overline{%
\mathcal{F}%
}_{A\alpha\beta}^{a}%
\mathcal{F}%
_{A\lambda\mu}^{a}$

This a real quantity because this scalar product is hermitian.

\paragraph{3)}

The same choice for gravitation would lead to $a_{G}\sum_{\left\{  \alpha
\beta\right\}  }B\left(
\mathcal{F}%
_{G\alpha\beta},%
\mathcal{F}%
_{G}^{\alpha\beta}\right)  $ \ with some scalar product on o(3,1). The only
natural choice for this scalar product is the Killing form $B(X,Y)=2Tr(\left[
\widetilde{X}\right]  \left[  \widetilde{Y}\right]  )$ which gives with the
Riemann tensor :

$\frac{1}{4}a_{G}\sum_{\alpha\beta\lambda\mu}g^{\alpha\lambda}g^{\beta\mu
}B\left(
\mathcal{F}%
_{B\alpha\beta},%
\mathcal{F}%
_{B\lambda\mu}\right)  $

$=\frac{1}{2}a_{G}\sum_{\alpha\beta\lambda\mu}g^{\alpha\lambda}g^{\beta\mu
}Tr\left(  \left[  \widetilde{%
\mathcal{F}%
}_{B\alpha\beta}\right]  \left[  \widetilde{%
\mathcal{F}%
}_{B\lambda\mu}\right]  \right)  $

$=\frac{1}{2}a_{G}\sum_{\alpha\beta\lambda\mu}g^{\alpha\lambda}g^{\beta\mu
}Tr\left(  O^{-1}\left[  R_{\alpha\beta}\right]  OO^{-1}\left[  R_{\lambda\mu
}\right]  O\right)  $

$=\frac{1}{2}a_{G}\sum_{\alpha\beta\lambda\mu}g^{\alpha\lambda}g^{\beta\mu
}Tr\left(  \left[  R_{\alpha\beta}\right]  \left[  R_{\lambda\mu}\right]
\right)  $

$=\frac{1}{2}a_{G}\sum_{\alpha\beta\lambda\mu}g^{\alpha\lambda}g^{\beta\mu
}\sum_{cd}R_{\alpha\beta d}^{c}R_{\lambda\mu c}^{d}$

$=\frac{1}{2}a_{G}\sum_{\alpha\beta\lambda\mu}g^{\alpha\lambda}g^{\beta\mu
}Tr\left(  \left[  R_{\alpha\beta}\right]  \left[  R_{\lambda\mu}\right]
\right)  $

The trouble is that the Killing form is not positive definite. It is possible
to turn over this problem by using the "compact real form" of o(3,1) (Knapp
[12] VI.1), in fact treating gravitation on the same footing as the other
fields, and considering a complex valued connection. But this would move us
further away from the traditional affine connections.

The alternate option is use the Palatini action, as in General Relativity. It
can be computed with our variables, as seen before :

The scalar curvature is (equation \ref{E11}):

$R=\sum_{\alpha\beta ijk}\left[  \widetilde{%
\mathcal{F}%
_{G\alpha\beta}}\right]  _{k}^{i}\eta^{kj}O_{i}^{\beta}O_{j}^{\alpha}%
=\sum_{a,\alpha\beta}%
\mathcal{F}%
_{G\alpha\beta}^{a}\left(  O_{p_{a}}^{\beta}O_{q_{a}}^{\alpha}-O_{q_{a}%
}^{\beta}O_{p_{a}}^{\alpha}\right)  $

and the lagrangian is $a_{G}\sum_{a\alpha\beta ij}%
\mathcal{F}%
_{G\alpha\beta}^{a}\left(  \left[  \widetilde{\kappa}_{a}\right]  \left[
\eta\right]  \right)  _{j}^{i}O_{i}^{\beta}O_{j}^{\alpha}$ with a real scalar
$a_{G}.$

So explicitly :

$L_{F}=a_{F}\left\langle
\mathcal{F}%
_{A},%
\mathcal{F}%
_{A}\right\rangle +a_{G}\sum_{a,\alpha\beta}%
\mathcal{F}%
_{G\alpha\beta}^{a}\left(  O_{p_{a}}^{\beta}O_{q_{a}}^{\alpha}-O_{q_{a}%
}^{\beta}O_{p_{a}}^{\alpha}\right)  $

$L_{F}=\frac{1}{2}a_{F}\sum_{a\alpha\beta\lambda\mu ijkl}\eta^{ji}\eta
^{lk}O_{j}^{\lambda}O_{i}^{\alpha}O_{l}^{\mu}O_{k}^{\beta}%
\mathcal{F}%
_{A\alpha\beta}^{a}\overline{%
\mathcal{F}%
}_{A\lambda\mu}^{a}$

$\qquad+a_{G}\sum_{a,\alpha\beta}%
\mathcal{F}%
_{G\alpha\beta}^{a}\left(  O_{p_{a}}^{\beta}O_{q_{a}}^{\alpha}-O_{q_{a}%
}^{\beta}O_{p_{a}}^{\alpha}\right)  $

Remark : one could introduce a cosmological constant $\Lambda$\ such that the
lagrangian becomes : \ $a_{G}\left(  \Lambda+R\right)  $ and $\Lambda$\ acts
in the action through the density $\sqrt{\left\vert \det g\right\vert }. $\ 

\subsection{Interactions}

As said before we have to address two interactions.

\paragraph{1) The "static" part :}

It models the pure action on the state of particles. This is for what we
defined the Dirac operator. So the simplest choices are : $\left\langle
D\psi,D\psi\right\rangle ,\left\langle D\psi,\psi\right\rangle ,\left\langle
\psi,D\psi\right\rangle .$ The first one leads to quadratic terms in the
derivatives, but the two last ones cannot guarantee to deliver a real scalar.
So we have two options :

$\frac{1}{2}\left(  \left\langle D\psi,\psi\right\rangle +\left\langle
\psi,D\psi\right\rangle \right)  =\operatorname{Re}\left\langle \psi
,D\psi\right\rangle $

$\frac{1}{2}i\left(  \left\langle D\psi,\psi\right\rangle -\left\langle
\psi,D\psi\right\rangle \right)  =\operatorname{Im}\left\langle \psi
,D\psi\right\rangle $

Clearly the term in $\partial_{\alpha}\psi$ in the lagrangian is
crucial.\ That is

$\sum_{r}Tr\left(  \left[  \psi^{\ast}\right]  \gamma_{0}\gamma^{r}\left[
\partial_{\alpha}\psi\right]  \right)  =iTr\left(  \eta^{rr}\left[  \psi
_{R}\right]  ^{\ast}\sigma_{r}\left[  \partial_{r}\psi_{R}\right]  -\left[
\psi_{L}\right]  ^{\ast}\sigma_{r}\left[  \partial_{r}\psi_{L}\right]
\right)  $

with $\operatorname{Re}\left\langle \psi,D\psi\right\rangle $ we have

$\sum_{r}\operatorname{Re}Tr\left(  \left[  \psi^{\ast}\right]  \gamma
_{0}\gamma^{r}\left[  \partial_{\alpha}\psi\right]  \right)  =$ $\sum
_{r}\{-\operatorname{Im}Tr\left(  \eta^{rr}\left[  \psi_{R}\right]  ^{\ast
}\sigma_{r}\left[  \partial_{r}\psi_{R}\right]  -\left[  \psi_{L}\right]
^{\ast}\sigma_{r}\left[  \partial_{r}\psi_{L}\right]  \right)  $

with $\operatorname{Im}\left\langle \psi,D\psi\right\rangle $ we have

$\sum_{r}\operatorname{Im}Tr\left(  \left[  \psi^{\ast}\right]  \gamma
_{0}\gamma^{r}\left[  \partial_{\alpha}\psi\right]  \right)  =\sum
_{r}\{\operatorname{Re}Tr\left(  \eta^{rr}\left[  \psi_{R}\right]  ^{\ast
}\sigma_{r}\left[  \partial_{r}\psi_{R}\right]  -\left[  \psi_{L}\right]
^{\ast}\sigma_{r}\left[  \partial_{r}\psi_{L}\right]  \right)  $

and we have the identities : $\operatorname{Re}Tr\left(  \left[  \psi\right]
^{\ast}\sigma_{r}\left[  \partial_{\alpha}\psi\right]  \right)  =\frac{1}%
{2}\partial_{\alpha}Tr\left(  \left[  \psi\right]  ^{\ast}\sigma_{r}\left[
\psi\right]  \right)  $

Indeed :

$Tr\left[  \psi\right]  ^{\ast}\sigma_{r}\left[  \partial_{r}\psi\right]  $

$=\sum_{\alpha}O_{r}^{\prime\alpha}\sum_{klj}\sigma_{r}^{kl}\overline{\psi
}^{kj}\partial_{\alpha}\psi^{lj}$

$=\sum_{\alpha}O_{r}^{\prime\alpha}\sum_{j}\sigma_{r}^{11}\overline{\psi}%
^{1j}\partial_{\alpha}\psi^{1j}+\sigma_{r}^{12}\overline{\psi}^{1j}%
\partial_{\alpha}\psi^{2j}+\sigma_{r}^{21}\overline{\psi}^{2j}\partial
_{\alpha}\psi^{1j}+\sigma_{r}^{22}\overline{\psi}^{2j}\partial_{\alpha}%
\psi^{2j}$

$Tr\left[  \psi\right]  ^{\ast}\sigma_{0}\left[  \partial_{r}\psi\right]
=\frac{1}{2}\partial_{r}\sum_{ij}\left\vert \psi^{ij}\right\vert ^{2}$

$Tr\left[  \psi\right]  ^{\ast}\sigma_{1}\left[  \partial_{r}\psi\right]
=\sum_{j}\partial_{r}\left(  \psi^{1j}\psi^{2j}\right)  $

$Tr\left[  \psi\right]  ^{\ast}\sigma_{2}\left[  \partial_{r}\psi\right]
=i\sum_{j}-\overline{\psi}^{1j}\partial_{r}\psi^{2j}+\overline{\psi}%
^{2j}\partial_{r}\psi^{1j}$

$Tr\left[  \psi\right]  ^{\ast}\sigma_{3}\left[  \partial_{r}\psi\right]
=\frac{1}{2}\sum_{j}\partial_{r}\left(  \left\vert \psi^{1j}\right\vert
^{2}-\left\vert \psi^{2j}\right\vert ^{2}\right)  $

As we see if we take the imaginary part of the above expressions il would be
null for $r=0,3$ thus making two privileged directions, and that does not
happen with the real part. It is a rather weak argument, but let us say that
this option has been tested before (Giachetta [5]).

So we choose $\epsilon=-1$ and the following lagrangian :

$a_{I}\frac{1}{2}i\left(  \left\langle D\psi,\psi\right\rangle -\left\langle
\psi,D\psi\right\rangle \right)  =a_{I}\operatorname{Im}\left\langle
\psi,D\psi\right\rangle $

\paragraph{2) The "dynamic part"}

It involves the fields, the state of particles and their velocity.\ The
simplest choice is :

$\left\langle \psi,\sum_{\alpha}V^{\alpha}\nabla_{\alpha}\psi\right\rangle $

V is a vector field, so this quantity is covariant, and invariant in a change
of gauge. As above we need a real quantity. There is no obvious reason for one
or the other option. Let us say that after testing both, the most physically
meaningful is the same as above.\ So I take :

$a_{D}\sum_{\alpha}\frac{1}{2}V^{\alpha}i\left(  \left\langle \nabla_{\alpha
}\psi,\psi\right\rangle +\left\langle \psi,\nabla_{\alpha}\psi\right\rangle
\right)  =a_{D}\sum_{\alpha}V^{\alpha}\operatorname{Im}\left\langle
\psi,\nabla_{\alpha}\psi\right\rangle $

with a real scalar constant $a_{D}$

\subsection{Summary}

The full lagrangian of this model is :%

\begin{equation}
\end{equation}

$%
\mathcal{L}%
_{M}=N\left(  a_{M}\left\langle \psi,\psi\right\rangle +a_{I}\operatorname{Im}%
\left\langle \psi,D\psi\right\rangle +a_{D}\sum_{\alpha}V^{\alpha
}\operatorname{Im}\left\langle \psi,\nabla_{\alpha}\psi\right\rangle \right)
\det O^{\prime}$

$%
\mathcal{L}%
_{F}=\left(  a_{F}\left\langle
\mathcal{F}%
_{A},%
\mathcal{F}%
_{A}\right\rangle +a_{G}R\right)  \det O^{\prime}$

\bigskip

The lagrangian is defined through intrinsic quantities so it is invariant
under gauge transformations or change of charts.

It does not involve the partial derivatives $\partial_{\alpha}O_{\beta
}^{\prime i}$ of the tetrad. There is no obvious need to introduce them, as
the scalar curvature answers to the interaction between the geometry and the
gravitational field (as it appears very clearly in the formula $R=$
$\sum_{ijk}\left[  \digamma\left(  \partial_{p},\partial_{q}\right)  \right]
_{j}^{q}\eta^{jp})$ . An additional item should be purely geometrical in
nature.\ In some ways a cosmological constant (which acts through the volume
form) is a tentative solution, but "purely scalar" in that it does not involve
any space-time distorsion other than dilation. Eventually what is missing is
some equivalent of the term $a_{M}\left\langle \psi,\psi\right\rangle ,$
meaning that space-time itself is more than a container, and possesses some
intrinsic property independantly from matter and force fields, all this
without reinventing the aether...So for the simple purpose I have in mind I
will keep the simplest solution and stick to this basic lagrangian.

The matter part of the lagrangian writes :

$%
\mathcal{L}%
_{M}=N\left(  \det O^{\prime}\right)  \left(  a_{M}\left\langle \psi
,\psi\right\rangle +a_{I}\operatorname{Im}\left\langle \psi,D\psi\right\rangle
+a_{D}\sum_{\alpha}V^{\alpha}\operatorname{Im}\left\langle \psi,\nabla
_{\alpha}\psi\right\rangle \right)  $

$=N\left(  \det O^{\prime}\right)  \left(  a_{M}\left\langle \psi
,\psi\right\rangle +\operatorname{Im}\left\langle \psi,\sum_{r}a_{I}%
O_{r}^{\alpha}\left[  \gamma^{r}\right]  \left[  \nabla_{\alpha}\psi\right]
\right\rangle +\operatorname{Im}\left\langle \psi,\sum_{\alpha}V^{\alpha}%
a_{D}\nabla_{\alpha}\psi\right\rangle \right)  $

$=N\left(  \det O^{\prime}\right)  \left(  a_{M}\left\langle \psi
,\psi\right\rangle +\operatorname{Im}\left\langle \psi,\left(  \sum_{r}%
a_{I}O_{r}^{\alpha}\left[  \gamma^{r}\right]  +\sum_{\alpha}V^{\alpha}%
a_{D}I\right)  \left[  \nabla_{\alpha}\psi\right]  \right\rangle \right)  $

with $I$\ \ the unitary 4x4 matrix, and it will be convenient to denote the
operator :%

\begin{equation}
D_{M}^{\alpha}=\left(  \sum_{r}a_{I}O_{r}^{\alpha}\left[  \gamma^{r}\right]
+\sum_{\alpha}V^{\alpha}a_{D}I\right) \label{E71b}%
\end{equation}

so : $%
\mathcal{L}%
_{M}=N\left(  \det O^{\prime}\right)  \left(  a_{M}\left\langle \psi
,\psi\right\rangle +\operatorname{Im}\left\langle \psi,\sum_{\alpha}%
D_{M}^{\alpha}\left[  \nabla_{\alpha}\psi\right]  \right\rangle \right)  $

With this notation it is obvious that $a_{I}\operatorname{Im}\left\langle
\psi,D\psi\right\rangle $ accounts for the kinematic part (rotations
\textit{in} the tetrad) and $a_{D}\sum_{\alpha}V^{\alpha}\operatorname{Im}%
\left\langle \psi,\nabla_{\alpha}\psi\right\rangle $ for the dynamic part
(displacement within $\Omega).$

\section{LAGRANGE\ EQUATIONS}

\label{Model Lagrange equations}

The Lagrange equations are just the transcription of the previous ones but, as
expected, one can get more insightful results.

\subsection{Gravitation}

We will define two moments : P, which can be seen as a "linear momentum", and
J, as an "angular momentum", both computed from the state tensor, but gauge
and chart invariant. The most important result is that the gravitational
potential G can be explicitly computed from the structure coefficients and
these two moments. The torsion is then easily computed and it appears that
usually the gravitational connection would not be torsionfree.

\subsubsection{Noether currents}

\paragraph{1)}

The Noether current reads (equation \ref{E51}):

$Y_{G}=\sum_{a,\alpha}\frac{d\left(  VL_{M}+L_{F}\right)  }{dG_{\alpha}^{a}%
}\partial_{\alpha}\otimes\overrightarrow{\kappa}_{a}$

with $Y_{G}^{a\alpha}=V\sum_{ij}\left(  \frac{dL_{M}^{\Diamond}}%
{d\operatorname{Re}\nabla_{\alpha}\psi^{ij}}\operatorname{Re}\left(  \left[
\kappa_{a}\right]  \left[  \psi\right]  \right)  _{j}^{i}+\frac{dL_{M}%
^{\Diamond}}{d\operatorname{Im}\nabla_{\alpha}\psi^{ij}}\operatorname{Im}%
\left(  \left[  \kappa_{a}\right]  \left[  \psi\right]  \right)  _{j}%
^{i}\right)  $

$+V\frac{\partial L_{M}^{\Diamond}}{\partial G_{\alpha}^{a}}+\frac{\partial
L_{F}}{\partial G_{\alpha}^{a}}+2\sum_{b,\beta}\frac{dL_{F}}{d%
\mathcal{F}%
_{G\alpha\beta}^{b}}\left[  \overrightarrow{\kappa}_{a},G_{\beta}\right]
^{b}$

$V\frac{\partial L_{M}^{\Diamond}}{\partial G_{\alpha}^{a}}=\frac{\partial
L_{F}}{\partial G_{\alpha}^{a}}=0$

$\frac{dL_{F}}{d%
\mathcal{F}%
_{G\alpha\beta}^{b}}=a_{G}\left(  O_{p_{b}}^{\beta}O_{q_{b}}^{\alpha}%
-O_{q_{b}}^{\beta}O_{p_{b}}^{\alpha}\right)  $

Let us compute the partial derivatives with respect to $\operatorname{Re}%
\nabla_{\alpha}\psi^{ij},\operatorname{Im}\nabla_{\alpha}\psi^{ij}:$

$L_{M}=N\left(  a_{M}\left\langle \psi,\psi\right\rangle +\operatorname{Im}%
\left\langle \psi,\sum_{\alpha}D_{M}^{\alpha}\left[  \nabla_{\alpha}%
\psi\right]  \right\rangle \right)  $

$\operatorname{Im}\left\langle \psi,\sum_{\alpha}D_{M}^{\alpha}\left[
\nabla_{\alpha}\psi\right]  \right\rangle $

$=\sum_{\alpha pq}\operatorname{Im}\left(  \left[  \psi\right]  ^{\ast}\left[
\gamma_{0}\right]  \left[  D_{M}^{\alpha}\right]  \right)  _{q}^{p}\left[
\nabla_{\alpha}\psi\right]  ^{qp}$

$=\sum_{\alpha pq}\operatorname{Re}\left(  \left[  \psi\right]  ^{\ast}\left[
\gamma_{0}\right]  \left[  D_{M}^{\alpha}\right]  \right)  _{q}^{p}%
\operatorname{Im}\left[  \nabla_{\alpha}\psi\right]  ^{qp}+\operatorname{Im}%
\left(  \left[  \psi\right]  ^{\ast}\left[  \gamma_{0}\right]  \left[
D_{M}^{\alpha}\right]  \right)  _{q}^{p}\operatorname{Re}\left[
\nabla_{\alpha}\psi\right]  ^{qp}$

$\frac{d\operatorname{Im}\left\langle \psi,\sum_{\alpha}D_{M}^{\alpha}\left[
\nabla_{\alpha}\psi\right]  \right\rangle }{d\operatorname{Re}\nabla_{\alpha
}\psi^{ij}}=\sum_{pq}\operatorname{Im}\left(  \left[  \psi\right]  ^{\ast
}\left[  \gamma_{0}\right]  \left[  D_{M}^{\alpha}\right]  \right)  _{q}%
^{p}\delta_{q}^{i}\delta_{p}^{j}=\operatorname{Im}\left(  \left[  \psi\right]
^{\ast}\left[  \gamma_{0}\right]  \left[  D_{M}^{\alpha}\right]  \right)
_{i}^{j}$

$\frac{d\operatorname{Im}\left\langle \psi,\sum_{\alpha}D_{M}^{\alpha}\left[
\nabla_{\alpha}\psi\right]  \right\rangle }{d\operatorname{Im}\nabla_{\alpha
}\psi^{ij}}=\sum_{pq}\operatorname{Re}\left(  \left[  \psi\right]  ^{\ast
}\left[  \gamma_{0}\right]  \left[  D_{M}^{\alpha}\right]  \right)  _{q}%
^{p}\delta_{q}^{i}\delta_{p}^{j}=\operatorname{Re}\left(  \left[  \psi\right]
^{\ast}\left[  \gamma_{0}\right]  \left[  D_{M}^{\alpha}\right]  \right)
_{i}^{j}$

$\frac{dL_{M}}{d\operatorname{Re}\nabla_{\alpha}\psi^{ij}}=N\operatorname{Im}%
\left(  \left[  \psi\right]  ^{\ast}\left[  \gamma_{0}\right]  \left[
D_{M}^{\alpha}\right]  \right)  _{i}^{j}$

$\frac{dL_{M}}{d\operatorname{Im}\nabla_{\alpha}\psi^{ij}}=N\operatorname{Re}%
\left(  \left[  \psi\right]  ^{\ast}\left[  \gamma_{0}\right]  \left[
D_{M}^{\alpha}\right]  \right)  _{i}^{j}$

So :

$Y_{G}^{a\alpha}=N\sum_{ij}\left(  \operatorname{Im}\left(  \left[
\psi\right]  ^{\ast}\left[  \gamma_{0}\right]  \left[  D_{M}^{\alpha}\right]
\right)  _{i}^{j}\operatorname{Re}\left(  \left[  \kappa_{a}\right]  \left[
\psi\right]  \right)  _{j}^{i}+\operatorname{Re}\left(  \left[  \psi\right]
^{\ast}\left[  D_{M}^{\alpha}\right]  \right)  _{i}^{j}\operatorname{Im}%
\left(  \left[  \kappa_{a}\right]  \left[  \psi\right]  \right)  _{j}%
^{i}\right)  $

$+2\sum_{b,\beta}a_{G}\left(  O_{p_{b}}^{\beta}O_{q_{b}}^{\alpha}-O_{q_{b}%
}^{\beta}O_{p_{b}}^{\alpha}\right)  \left[  \overrightarrow{\kappa}%
_{a},G_{\beta}\right]  ^{b}$

$Y_{G}^{a\alpha}=N\operatorname{Im}Tr\left[  \psi\right]  ^{\ast}\left[
\gamma_{0}\right]  \left[  D_{M}^{\alpha}\right]  \left(  \left[  \kappa
_{a}\right]  \left[  \psi\right]  \right)  +2\sum_{b,\beta}a_{G}\left(
O_{p_{b}}^{\beta}O_{q_{b}}^{\alpha}-O_{q_{b}}^{\beta}O_{p_{b}}^{\alpha
}\right)  \left[  \overrightarrow{\kappa}_{a},G_{\beta}\right]  ^{b}$

In the orthonormal basis :

$Y_{G}=(\sum_{a,r}N\operatorname{Im}Tr\left[  \psi\right]  ^{\ast}\left[
D_{M}^{r}\right]  \left(  \left[  \kappa_{a}\right]  \left[  \psi\right]
\right)  $

$+2a_{G}\sum_{b,\alpha\beta}\left(  O_{p_{b}}^{\beta}O_{q_{b}}^{\alpha
}-O_{q_{b}}^{\beta}O_{p_{b}}^{\alpha}\right)  \left[  \overrightarrow{\kappa
}_{a},G_{\beta}\right]  ^{b}O_{\alpha}^{\prime r})\partial_{r}\otimes
\overrightarrow{\kappa}_{a}$

with $\left[  D_{M}^{r}\right]  =D_{M}^{\alpha}=\left(  a_{I}\left[
\gamma^{r}\right]  +\sum_{r\alpha}O_{\alpha}^{\prime r}V^{\alpha}%
a_{D}I\right)  =\left(  a_{I}\left[  \gamma^{r}\right]  +V^{r}a_{D}I\right)  $

$Y_{G}=\sum_{a,r}\{N\left(  a_{I}\left(  \operatorname{Im}Tr\left(  \left[
\psi\right]  ^{\ast}\left[  \gamma_{0}\gamma^{r}\right]  \left[  \kappa
_{a}\right]  \left[  \psi\right]  \right)  +a_{D}V^{r}\operatorname{Im}%
Tr\left(  \left[  \psi^{\ast}\right]  \left[  \gamma_{0}\right]  \left[
\kappa_{a}\right]  \left[  \psi\right]  \right)  \right)  \right)  $

$+2a_{G}\sum_{b,\alpha\beta}\left(  O_{p_{b}}^{\beta}O_{q_{b}}^{\alpha
}-O_{q_{b}}^{\beta}O_{p_{b}}^{\alpha}\right)  \left[  \overrightarrow{\kappa
}_{a},G_{\beta}\right]  ^{b}O_{\alpha}^{\prime r}\}\partial_{r}\otimes
\overrightarrow{\kappa}_{a}$

\paragraph{2)}

We have for the first term: $a_{I}\operatorname{Im}Tr\left(  \left[
\psi\right]  ^{\ast}\left[  \gamma_{0}\gamma^{r}\right]  \left[  \kappa
_{a}\right]  \left[  \psi\right]  \right)  $

a%
$<$%
4 : $Tr\left(  \left[  \psi\right]  ^{\ast}\left[  \gamma_{0}\gamma
^{r}\right]  \left[  \kappa_{a}\right]  \left[  \psi\right]  \right)
=\frac{1}{2}Tr\left(  \eta^{rr}\psi_{R}^{\ast}\sigma_{r}\sigma_{a}\psi
_{R}-\psi_{L}^{\ast}\sigma_{r}\sigma_{a}\psi_{L}\right)  $

a%
$>$%
3 : $Tr\left(  \left[  \psi\right]  ^{\ast}\left[  \gamma_{0}\gamma
^{r}\right]  \left[  \kappa_{a}\right]  \left[  \psi\right]  \right)
=i\frac{1}{2}Tr\left(  \eta^{rr}\psi_{R}^{\ast}\sigma_{r}\sigma_{a-3}\psi
_{R}+\psi_{L}^{\ast}\sigma_{r}\sigma_{a-3}\psi_{L}\right)  $

with $\sigma_{1}\sigma_{2}=i\sigma_{3};\sigma_{2}\sigma_{3}=i\sigma_{1}%
;\sigma_{3}\sigma_{1}=i\sigma_{2}$ \ the quantity $Tr\left(  \left[
\psi\right]  ^{\ast}\left[  \gamma_{0}\gamma^{r}\right]  \left[  \kappa
_{a}\right]  \left[  \psi\right]  \right)  $\ \ is given by the table :

\bigskip

$%
\begin{bmatrix}
a\backslash r & 0 & 1\\
1 & -\frac{1}{2}Tr\left(  \psi_{R}^{\ast}\sigma_{1}\psi_{R}+\psi_{L}^{\ast
}\sigma_{1}\psi_{L}\right)  & \frac{1}{2}Tr\left(  \psi_{R}^{\ast}\psi
_{R}-\psi_{L}^{\ast}\psi_{L}\right) \\
2 & -\frac{1}{2}Tr\left(  \psi_{R}^{\ast}\sigma_{2}\psi_{R}+\psi_{L}^{\ast
}\sigma_{2}\psi_{L}\right)  & i\frac{1}{2}Tr\left(  \psi_{R}^{\ast}\sigma
_{3}\psi_{R}-\psi_{L}^{\ast}\sigma_{3}\psi_{L}\right) \\
3 & -\frac{1}{2}Tr\left(  \psi_{R}^{\ast}\sigma_{3}\psi_{R}+\psi_{L}^{\ast
}\sigma_{3}\psi_{L}\right)  & -i\frac{1}{2}Tr\left(  \psi_{R}^{\ast}\sigma
_{2}\psi_{R}-\psi_{L}^{\ast}\sigma_{2}\psi_{L}\right) \\
4 & -i\frac{1}{2}Tr\left(  \psi_{R}^{\ast}\sigma_{1}\psi_{R}-\psi_{L}^{\ast
}\sigma_{1}\psi_{L}\right)  & i\frac{1}{2}Tr\left(  \psi_{R}^{\ast}\psi
_{R}+\psi_{L}^{\ast}\psi_{L}\right) \\
5 & -i\frac{1}{2}Tr\left(  \psi_{R}^{\ast}\sigma_{2}\psi_{R}-\psi_{L}^{\ast
}\sigma_{2}\psi_{L}\right)  & -\frac{1}{2}Tr\left(  \psi_{R}^{\ast}\sigma
_{3}\psi_{R}+\psi_{L}^{\ast}\sigma_{3}\psi_{L}\right) \\
6 & -i\frac{1}{2}Tr\left(  \psi_{R}^{\ast}\sigma_{3}\psi_{R}-\psi_{L}^{\ast
}\sigma_{3}\psi_{L}\right)  & \frac{1}{2}Tr\left(  \psi_{R}^{\ast}\sigma
_{2}\psi_{R}+\psi_{L}^{\ast}\sigma_{2}\psi_{L}\right)
\end{bmatrix}
$

$%
\begin{bmatrix}
a\backslash r & 2 & 3\\
1 & -i\frac{1}{2}Tr\left(  \psi_{R}^{\ast}\sigma_{3}\psi_{R}-\psi_{L}^{\ast
}\sigma_{3}\psi_{L}\right)  & i\frac{1}{2}Tr\left(  \psi_{R}^{\ast}\sigma
_{2}\psi_{R}-\psi_{L}^{\ast}\sigma_{2}\psi_{L}\right) \\
2 & \frac{1}{2}Tr\left(  \psi_{R}^{\ast}\psi_{R}-\psi_{L}^{\ast}\psi
_{L}\right)  & -i\frac{1}{2}Tr\left(  \psi_{R}^{\ast}\sigma_{1}\psi_{R}%
-\psi_{L}^{\ast}\sigma_{1}\psi_{L}\right) \\
3 & i\frac{1}{2}Tr\left(  \psi_{R}^{\ast}\sigma_{1}\psi_{R}-\psi_{L}^{\ast
}\sigma_{1}\psi_{L}\right)  & \frac{1}{2}Tr\left(  \psi_{R}^{\ast}\psi
_{R}-\psi_{L}^{\ast}\psi_{L}\right) \\
4 & \frac{1}{2}Tr\left(  \psi_{R}^{\ast}\sigma_{3}\psi_{R}+\psi_{L}^{\ast
}\sigma_{3}\psi_{L}\right)  & -\frac{1}{2}Tr\left(  \psi_{R}^{\ast}\sigma
_{2}\psi_{R}+\psi_{L}^{\ast}\sigma_{2}\psi_{L}\right) \\
5 & i\frac{1}{2}Tr\left(  \psi_{R}^{\ast}\psi_{R}+\psi_{L}^{\ast}\psi
_{L}\right)  & \frac{1}{2}Tr\left(  \psi_{R}^{\ast}\sigma_{1}\psi_{R}+\psi
_{L}^{\ast}\sigma_{1}\psi_{L}\right) \\
6 & -\frac{1}{2}Tr\left(  \psi_{R}^{\ast}\sigma_{1}\psi_{R}+\psi_{L}^{\ast
}\sigma_{1}\psi_{L}\right)  & i\frac{1}{2}Tr\left(  \psi_{R}^{\ast}\psi
_{R}+\psi_{L}^{\ast}\psi_{L}\right)
\end{bmatrix}
$

\bigskip

As $Tr\left(  \psi^{\ast}\sigma_{k}\psi\right)  \in%
\mathbb{R}
$ the quantity $\operatorname{Im}Tr\left(  \left[  \psi\right]  ^{\ast}\left[
\gamma_{0}\gamma^{r}\right]  \left[  \kappa_{a}\right]  \left[  \psi\right]
\right)  $ is given by the table :

\bigskip

$\operatorname{Im}Tr\left(  \left[  \psi\right]  ^{\ast}\left[  \gamma
_{0}\gamma^{r}\right]  \left[  \kappa_{a}\right]  \left[  \psi\right]
\right)  =%
\begin{bmatrix}
a\backslash r & 0 & 1 & 2 & 3\\
1 & 0 & 0 & J^{3} & -J^{2}\\
2 & 0 & -J^{3} & 0 & J^{1}\\
3 & 0 & J^{2} & -J^{1} & 0\\
4 & J^{1} & J^{0} & 0 & 0\\
5 & J^{2} & 0 & J^{0} & 0\\
6 & J^{3} & 0 & 0 & J^{0}%
\end{bmatrix}
$

\bigskip

with%

\begin{equation}
\mathbf{J}_{k}\mathbf{=-}\frac{1}{2}\mathbf{Tr}\left(  \eta^{kk}\psi_{R}%
^{\ast}\sigma_{k}\psi_{R}-\psi_{L}^{\ast}\sigma_{k}\psi_{L}\right) \label{E72}%
\end{equation}

One can check that :

$\operatorname{Im}Tr\left(  \left[  \psi\right]  ^{\ast}\left[  \gamma
_{0}\gamma^{r}\right]  \left[  \kappa_{a}\right]  \left[  \psi\right]
\right)  =\sum_{k}J_{k}\left[  \widetilde{\kappa}_{a}\right]  _{r}^{k}$

We denote by $\left[  J\right]  $ the 1x4 row matrix $J_{k}.$ So :%

\begin{equation}
\operatorname{Im}\left\langle \psi,\left[  \gamma^{r}\right]  \left[
\kappa_{a}\right]  \left[  \psi\right]  \right\rangle \mathbf{=}%
\operatorname{Im}\mathbf{Tr}\left(  \left[  \psi\right]  ^{\ast}\left[
\gamma_{0}\gamma^{r}\right]  \left[  \kappa_{a}\right]  \left[  \psi\right]
\right)  \mathbf{=}\left(  \left[  J\right]  \left[  \widetilde{\kappa}%
_{a}\right]  \right)  _{r}\label{E72b}%
\end{equation}

and :

$\sum_{a,r}\operatorname{Im}Tr\left(  \left[  \psi\right]  ^{\ast}\left[
\gamma_{0}\gamma^{r}\right]  \left[  \kappa_{a}\right]  \left[  \psi\right]
\right)  \partial_{r}\otimes\overrightarrow{\kappa}_{a}$

$=\sum_{a,r}\left(  J_{p_{a}}\partial_{q_{a}}-\eta^{p_{a}p_{a}}J_{q_{a}%
}\partial_{p_{a}}\right)  \otimes\overrightarrow{\kappa}_{a}=\sum_{a,r}\left(
\left[  J\right]  \left[  \widetilde{\kappa}_{a}\right]  \right)  _{r}%
\partial_{r}\otimes\overrightarrow{\kappa}_{a}$

The quantities $J_{k}$ are real scalar functions, invariant in a gauge
transformations, and so are not components of a vector field on M.\ We see
that the geometrical pertinent quantity is $\sum_{a,r}\left(  \left[
J\right]  \left[  \widetilde{\kappa}_{a}\right]  \right)  _{r}\partial_{r}$
which is a vector, similar to a relativistic angular momentum.

Remark : one can check :

$\left[  \psi\right]  ^{\ast}\left[  \gamma_{0}\gamma^{r}\right]  \left[
\psi\right]  =i\left(  \eta^{kk}\psi_{R}^{\ast}\sigma_{k}\psi_{R}-\psi
_{L}^{\ast}\sigma_{k}\psi_{L}\right)  $

So $\left\langle \psi,\gamma^{r}\psi\right\rangle =Tr\left(  \left[
\psi\right]  ^{\ast}\left[  \gamma_{0}\gamma^{r}\right]  \left[  \psi\right]
\right)  =iTr\left(  \eta^{kk}\psi_{R}^{\ast}\sigma_{k}\psi_{R}-\psi_{L}%
^{\ast}\sigma_{k}\psi_{L}\right)  =-2J_{r}$

\paragraph{3)}

The second item : $a_{D}V^{r}\operatorname{Im}Tr\left(  \left[  \psi^{\ast
}\right]  \left[  \gamma_{0}\right]  \left[  \kappa_{a}\right]  \left[
\psi\right]  \right)  $

$Tr\left(  \left[  \psi^{\ast}\right]  \left[  \gamma_{0}\right]  \left[
\kappa_{a}\right]  \left[  \psi\right]  \right)  $ is a pure imaginary scalar
as one can check :

$\overline{Tr\left(  \left[  \psi^{\ast}\right]  \left[  \gamma_{0}\right]
\left[  \kappa_{a}\right]  \left[  \psi\right]  \right)  }=Tr\left(  \left[
\psi^{\ast}\right]  \left[  \gamma_{0}\right]  \left[  \kappa_{a}\right]
\left[  \psi\right]  \right)  ^{\ast}$

$=Tr\left(  \left[  \psi^{\ast}\right]  \left[  \kappa_{a}\right]  ^{\ast
}\left[  \gamma_{0}\right]  \left[  \psi\right]  \right)  =-Tr\left(  \left[
\psi^{\ast}\right]  \left[  \gamma_{0}\right]  \left[  \kappa_{a}\right]
\left[  \gamma_{0}\right]  \left[  \gamma_{0}\right]  \left[  \psi\right]
\right)  =-Tr\left(  \left[  \psi^{\ast}\right]  \left[  \gamma_{0}\right]
\left[  \kappa_{a}\right]  \left[  \psi\right]  \right)  $

So let be $P_{a}\in%
\mathbb{R}
:Tr\left(  \left[  \psi^{\ast}\right]  \left[  \gamma_{0}\right]  \left[
\kappa_{a}\right]  \left[  \psi\right]  \right)  =iP_{a}$

a%
$<$%
4 : $Tr\left(  \left[  \psi^{\ast}\right]  \left[  \gamma_{0}\right]  \left[
\kappa_{a}\right]  \left[  \psi\right]  \right)  =\frac{1}{2}Tr\left(  \left[
\psi_{R}^{\ast}\right]  \left[  \sigma_{a}\right]  \left[  \psi_{L}\right]
-\left[  \psi_{L}^{\ast}\right]  \left[  \sigma_{a}\right]  \left[  \psi
_{R}\right]  \right)  $

a%
$>$%
3 :$Tr\left(  \left[  \psi^{\ast}\right]  \left[  \gamma_{0}\right]  \left[
\kappa_{a}\right]  \left[  \psi\right]  \right)  =-i\frac{1}{2}Tr\left(
\left[  \psi_{R}^{\ast}\right]  \left[  \sigma_{a-3}\right]  \left[  \psi
_{L}\right]  +\left[  \psi_{L}^{\ast}\right]  \left[  \sigma_{a-3}\right]
\left[  \psi_{R}\right]  \right)  $

So :%

\begin{equation}
\left\langle \psi,\left[  \kappa_{a}\right]  \left[  \psi\right]
\right\rangle \mathbf{=Tr}\left(  \left[  \psi^{\ast}\right]  \left[
\gamma_{0}\right]  \left[  \kappa_{a}\right]  \left[  \psi\right]  \right)
\mathbf{=iP}_{a}\label{E72c}%
\end{equation}

The quantities $P_{a}$ are real scalar functions, invariant in a gauge
transformations. The geometrical pertinent quantity is $\sum_{a,r}V^{r}%
P_{a}\partial_{r}\otimes\overrightarrow{\kappa}_{a}$ .

One can check that :

$\sum_{a=1}^{6}\left(  P_{a}\right)  ^{2}$

$=\sum_{a=1}^{3}\left(  -i\frac{1}{2}Tr\left(  \left[  \psi_{R}^{\ast}\right]
\left[  \sigma_{a}\right]  \left[  \psi_{L}\right]  -\left[  \psi_{L}^{\ast
}\right]  \left[  \sigma_{a}\right]  \left[  \psi_{R}\right]  \right)
\right)  ^{2}$

$\qquad+\left(  -\frac{1}{2}Tr\left(  \left[  \psi_{R}^{\ast}\right]  \left[
\sigma_{a}\right]  \left[  \psi_{L}\right]  +\left[  \psi_{L}^{\ast}\right]
\left[  \sigma_{a}\right]  \left[  \psi_{R}\right]  \right)  \right)  ^{2}$

$=\sum_{a=1}^{3}\left\vert Tr\left(  \left[  \psi_{R}^{\ast}\right]  \left[
\sigma_{a}\right]  \left[  \psi_{L}\right]  \right)  \right\vert ^{2}>0$

\paragraph{4)}

The third term is $\sum_{b,\alpha\beta}\left(  O_{p_{b}}^{\beta}O_{q_{b}%
}^{\alpha}-O_{q_{b}}^{\beta}O_{p_{b}}^{\alpha}\right)  \left[  \overrightarrow
{\kappa}_{a},G_{\beta}\right]  ^{b}O_{\alpha}^{\prime r}$

$\sum_{b,\alpha\beta}\left(  O_{p_{b}}^{\beta}O_{q_{b}}^{\alpha}-O_{q_{b}%
}^{\beta}O_{p_{b}}^{\alpha}\right)  \left[  \overrightarrow{\kappa}%
_{a},G_{\beta}\right]  ^{b}O_{\alpha}^{\prime r}$

$=\sum_{b,\beta}\left(  \delta_{q_{b}}^{r}O_{p_{b}}^{\beta}-\delta_{p_{b}}%
^{r}O_{q_{b}}^{\beta}\right)  \left[  \overrightarrow{\kappa}_{a},G_{\beta
}\right]  ^{b}$

$=\sum_{b}\delta_{q_{b}}^{r}\left[  \overrightarrow{\kappa}_{a},G_{p_{b}%
}\right]  ^{b}-\delta_{p_{b}}^{r}\left[  \overrightarrow{\kappa}_{a},G_{q_{b}%
}\right]  ^{b}$

$=\sum_{bc=1}^{6}G_{ac}^{b}\left(  \delta_{q_{b}}^{r}G_{p_{b}}^{c}%
-\delta_{p_{b}}^{r}G_{q_{b}}^{c}\right)  $

$\sum_{bc=1}^{6}G_{ac}^{b}\left(  \delta_{q_{b}}^{r}G_{p_{b}}^{c}%
-\delta_{p_{b}}^{r}G_{q_{b}}^{c}\right)  $ can be computed explicitly :

a=1 : $G_{11}^{b}\left(  \delta_{q_{b}}^{r}G_{p_{b}}^{1}-\delta_{p_{b}}%
^{r}G_{q_{b}}^{1}\right)  +G_{12}^{b}\left(  \delta_{q_{b}}^{r}G_{p_{b}}%
^{2}-\delta_{p_{b}}^{r}G_{q_{b}}^{2}\right)  +G_{13}^{b}\left(  \delta_{q_{b}%
}^{r}G_{p_{b}}^{3}-\delta_{p_{b}}^{r}G_{q_{b}}^{3}\right)  $

$+G_{14}^{b}\left(  \delta_{q_{b}}^{r}G_{p_{b}}^{4}-\delta_{p_{b}}^{r}%
G_{q_{b}}^{4}\right)  +G_{15}^{b}\left(  \delta_{q_{b}}^{r}G_{p_{b}}%
^{5}-\delta_{p_{b}}^{r}G_{q_{b}}^{5}\right)  +G_{16}^{b}\left(  \delta_{q_{b}%
}^{r}G_{p_{b}}^{6}-\delta_{p_{b}}^{r}G_{q_{b}}^{6}\right)  $

a=2 :

$G_{21}^{b}\left(  \delta_{q_{b}}^{r}G_{p_{b}}^{1}-\delta_{p_{b}}^{r}G_{q_{b}%
}^{1}\right)  +G_{22}^{b}\left(  \delta_{q_{b}}^{r}G_{p_{b}}^{2}-\delta
_{p_{b}}^{r}G_{q_{b}}^{2}\right)  +G_{23}^{b}\left(  \delta_{q_{b}}%
^{r}G_{p_{b}}^{3}-\delta_{p_{b}}^{r}G_{q_{b}}^{3}\right)  $

$+G_{24}^{b}\left(  \delta_{q_{b}}^{r}G_{p_{b}}^{4}-\delta_{p_{b}}^{r}%
G_{q_{b}}^{4}\right)  +G_{25}^{b}\left(  \delta_{q_{b}}^{r}G_{p_{b}}%
^{5}-\delta_{p_{b}}^{r}G_{q_{b}}^{5}\right)  +G_{26}^{b}\left(  \delta_{q_{b}%
}^{r}G_{p_{b}}^{6}-\delta_{p_{b}}^{r}G_{q_{b}}^{6}\right)  $

a=3 :

$G_{31}^{b}\left(  \delta_{q_{b}}^{r}G_{p_{b}}^{1}-\delta_{p_{b}}^{r}G_{q_{b}%
}^{1}\right)  +G_{32}^{b}\left(  \delta_{q_{b}}^{r}G_{p_{b}}^{2}-\delta
_{p_{b}}^{r}G_{q_{b}}^{2}\right)  +G_{33}^{b}\left(  \delta_{q_{b}}%
^{r}G_{p_{b}}^{3}-\delta_{p_{b}}^{r}G_{q_{b}}^{3}\right)  $

$+G_{34}^{b}\left(  \delta_{q_{b}}^{r}G_{p_{b}}^{4}-\delta_{p_{b}}^{r}%
G_{q_{b}}^{4}\right)  +G_{35}^{b}\left(  \delta_{q_{b}}^{r}G_{p_{b}}%
^{5}-\delta_{p_{b}}^{r}G_{q_{b}}^{5}\right)  +G_{36}^{b}\left(  \delta_{q_{b}%
}^{r}G_{p_{b}}^{6}-\delta_{p_{b}}^{r}G_{q_{b}}^{6}\right)  $

a=4 :

$G_{41}^{b}\left(  \delta_{q_{b}}^{r}G_{p_{b}}^{1}-\delta_{p_{b}}^{r}G_{q_{b}%
}^{1}\right)  +G_{42}^{b}\left(  \delta_{q_{b}}^{r}G_{p_{b}}^{2}-\delta
_{p_{b}}^{r}G_{q_{b}}^{2}\right)  +G_{43}^{b}\left(  \delta_{q_{b}}%
^{r}G_{p_{b}}^{3}-\delta_{p_{b}}^{r}G_{q_{b}}^{3}\right)  $

$+G_{44}^{b}\left(  \delta_{q_{b}}^{r}G_{p_{b}}^{4}-\delta_{p_{b}}^{r}%
G_{q_{b}}^{4}\right)  +G_{45}^{b}\left(  \delta_{q_{b}}^{r}G_{p_{b}}%
^{5}-\delta_{p_{b}}^{r}G_{q_{b}}^{5}\right)  +G_{46}^{b}\left(  \delta_{q_{b}%
}^{r}G_{p_{b}}^{6}-\delta_{p_{b}}^{r}G_{q_{b}}^{6}\right)  $

a=5 :

$G_{51}^{b}\left(  \delta_{q_{b}}^{r}G_{p_{b}}^{1}-\delta_{p_{b}}^{r}G_{q_{b}%
}^{1}\right)  +G_{52}^{b}\left(  \delta_{q_{b}}^{r}G_{p_{b}}^{2}-\delta
_{p_{b}}^{r}G_{q_{b}}^{2}\right)  +G_{53}^{b}\left(  \delta_{q_{b}}%
^{r}G_{p_{b}}^{3}-\delta_{p_{b}}^{r}G_{q_{b}}^{3}\right)  $

$+G_{54}^{b}\left(  \delta_{q_{b}}^{r}G_{p_{b}}^{4}-\delta_{p_{b}}^{r}%
G_{q_{b}}^{4}\right)  +G_{55}^{b}\left(  \delta_{q_{b}}^{r}G_{p_{b}}%
^{5}-\delta_{p_{b}}^{r}G_{q_{b}}^{5}\right)  +G_{56}^{b}\left(  \delta_{q_{b}%
}^{r}G_{p_{b}}^{6}-\delta_{p_{b}}^{r}G_{q_{b}}^{6}\right)  $

a=6 :

$G_{61}^{b}\left(  \delta_{q_{b}}^{r}G_{p_{b}}^{1}-\delta_{p_{b}}^{r}G_{q_{b}%
}^{1}\right)  +G_{62}^{b}\left(  \delta_{q_{b}}^{r}G_{p_{b}}^{2}-\delta
_{p_{b}}^{r}G_{q_{b}}^{2}\right)  +G_{63}^{b}\left(  \delta_{q_{b}}%
^{r}G_{p_{b}}^{3}-\delta_{p_{b}}^{r}G_{q_{b}}^{3}\right)  $

$+G_{64}^{b}\left(  \delta_{q_{b}}^{r}G_{p_{b}}^{4}-\delta_{p_{b}}^{r}%
G_{q_{b}}^{4}\right)  +G_{65}^{b}\left(  \delta_{q_{b}}^{r}G_{p_{b}}%
^{5}-\delta_{p_{b}}^{r}G_{q_{b}}^{5}\right)  +G_{66}^{b}\left(  \delta_{q_{b}%
}^{r}G_{p_{b}}^{6}-\delta_{p_{b}}^{r}G_{q_{b}}^{6}\right)  $

With the chosen basis in o(3,1) the structure coefficients are :

$G_{12}^{b}=\delta_{3}^{b};G_{13}^{b}=-\delta_{2}^{b};G_{14}^{b}=0;G_{15}%
^{b}=\delta_{6}^{b};G_{16}^{b}=-\delta_{5}^{b}$

$G_{23}^{b}=\delta_{1}^{b};G_{24}^{b}=-\delta_{6}^{b};G_{25}^{b}=0;G_{26}%
^{b}=\delta_{4}^{b}$

$G_{34}^{b}=\delta_{5}^{b};G_{35}^{b}=-\delta_{4}^{b};G_{36}^{b}=0$

$G_{45}^{b}=-\delta_{3}^{b};G_{46}^{b}=\delta_{2}^{b}$

$G_{56}^{b}=-\delta_{1}^{b}$

So :

a=1 : $\delta_{3}^{b}\left(  \delta_{q_{b}}^{r}G_{p_{b}}^{2}-\delta_{p_{b}%
}^{r}G_{q_{b}}^{2}\right)  -\delta_{2}^{b}\left(  \delta_{q_{b}}^{r}G_{p_{b}%
}^{3}-\delta_{p_{b}}^{r}G_{q_{b}}^{3}\right)  +\delta_{6}^{b}\left(
\delta_{q_{b}}^{r}G_{p_{b}}^{5}-\delta_{p_{b}}^{r}G_{q_{b}}^{5}\right)
-\delta_{5}^{b}\left(  \delta_{q_{b}}^{r}G_{p_{b}}^{6}-\delta_{p_{b}}%
^{r}G_{q_{b}}^{6}\right)  $

a=2 : $-\delta_{3}^{b}\left(  \delta_{q_{b}}^{r}G_{p_{b}}^{1}-\delta_{p_{b}%
}^{r}G_{q_{b}}^{1}\right)  +\delta_{1}^{b}\left(  \delta_{q_{b}}^{r}G_{p_{b}%
}^{3}-\delta_{p_{b}}^{r}G_{q_{b}}^{3}\right)  -\delta_{6}^{b}\left(
\delta_{q_{b}}^{r}G_{p_{b}}^{4}-\delta_{p_{b}}^{r}G_{q_{b}}^{4}\right)
+\delta_{4}^{b}\left(  \delta_{q_{b}}^{r}G_{p_{b}}^{6}-\delta_{p_{b}}%
^{r}G_{q_{b}}^{6}\right)  $

a=3 : $\delta_{2}^{b}\left(  \delta_{q_{b}}^{r}G_{p_{b}}^{1}-\delta_{p_{b}%
}^{r}G_{q_{b}}^{1}\right)  -\delta_{1}^{b}\left(  \delta_{q_{b}}^{r}G_{p_{b}%
}^{2}-\delta_{p_{b}}^{r}G_{q_{b}}^{2}\right)  +\delta_{5}^{b}\left(
\delta_{q_{b}}^{r}G_{p_{b}}^{4}-\delta_{p_{b}}^{r}G_{q_{b}}^{4}\right)
-\delta_{4}^{b}\left(  \delta_{q_{b}}^{r}G_{p_{b}}^{5}-\delta_{p_{b}}%
^{r}G_{q_{b}}^{5}\right)  $

a=4 : $\delta_{6}^{b}\left(  \delta_{q_{b}}^{r}G_{p_{b}}^{2}-\delta_{p_{b}%
}^{r}G_{q_{b}}^{2}\right)  -\delta_{5}^{b}\left(  \delta_{q_{b}}^{r}G_{p_{b}%
}^{3}-\delta_{p_{b}}^{r}G_{q_{b}}^{3}\right)  -\delta_{3}^{b}\left(
\delta_{q_{b}}^{r}G_{p_{b}}^{5}-\delta_{p_{b}}^{r}G_{q_{b}}^{5}\right)
+\delta_{2}^{b}\left(  \delta_{q_{b}}^{r}G_{p_{b}}^{6}-\delta_{p_{b}}%
^{r}G_{q_{b}}^{6}\right)  $

a=5 : $-\delta_{6}^{b}\left(  \delta_{q_{b}}^{r}G_{p_{b}}^{1}-\delta_{p_{b}%
}^{r}G_{q_{b}}^{1}\right)  +\delta_{4}^{b}\left(  \delta_{q_{b}}^{r}G_{p_{b}%
}^{3}-\delta_{p_{b}}^{r}G_{q_{b}}^{3}\right)  +\delta_{3}^{b}\left(
\delta_{q_{b}}^{r}G_{p_{b}}^{4}-\delta_{p_{b}}^{r}G_{q_{b}}^{4}\right)
-\delta_{1}^{b}\left(  \delta_{q_{b}}^{r}G_{p_{b}}^{6}-\delta_{p_{b}}%
^{r}G_{q_{b}}^{6}\right)  $

a=6 : $\delta_{5}^{b}\left(  \delta_{q_{b}}^{r}G_{p_{b}}^{1}-\delta_{p_{b}%
}^{r}G_{q_{b}}^{1}\right)  -\delta_{4}^{b}\left(  \delta_{q_{b}}^{r}G_{p_{b}%
}^{2}-\delta_{p_{b}}^{r}G_{q_{b}}^{2}\right)  -\delta_{2}^{b}\left(
\delta_{q_{b}}^{r}G_{p_{b}}^{4}-\delta_{p_{b}}^{r}G_{q_{b}}^{4}\right)
+\delta_{1}^{b}\left(  \delta_{q_{b}}^{r}G_{p_{b}}^{5}-\delta_{p_{b}}%
^{r}G_{q_{b}}^{5}\right)  $

One gets $\sum_{bc=1}^{6}G_{ac}^{b}\left(  \delta_{q_{b}}^{r}G_{p_{b}}%
^{c}-\delta_{p_{b}}^{r}G_{q_{b}}^{c}\right)  =\left[  T_{G}\right]  ^{ar}%
$\ \ with the table $T_{G}$ :

\bigskip

$%
\begin{bmatrix}
a\backslash r & 0 & 1 & 2 & 3\\
1 & \left(  G_{2}^{6}-G_{3}^{5}\right)  & \left(  G_{2}^{2}+G_{3}^{3}\right)
& \left(  -G_{1}^{2}-G_{0}^{6}\right)  & \left(  G_{0}^{5}-G_{1}^{3}\right) \\
2 & \left(  G_{3}^{4}-G_{1}^{6}\right)  & \left(  G_{0}^{6}-G_{2}^{1}\right)
& \left(  G_{1}^{1}+G_{3}^{3}\right)  & \left(  -G_{2}^{3}-G_{0}^{4}\right) \\
3 & \left(  G_{1}^{5}-G_{2}^{4}\right)  & \left(  -G_{0}^{5}-G_{3}^{1}\right)
& \left(  G_{0}^{4}-G_{3}^{2}\right)  & \left(  G_{2}^{2}+G_{1}^{1}\right) \\
4 & \left(  G_{2}^{3}-G_{3}^{2}\right)  & \left(  -G_{2}^{5}-G_{3}^{6}\right)
& \left(  G_{1}^{5}-G_{0}^{3}\right)  & \left(  G_{1}^{6}+G_{0}^{2}\right) \\
5 & \left(  G_{3}^{1}-G_{1}^{3}\right)  & \left(  G_{0}^{3}+G_{2}^{4}\right)
& \left(  -G_{1}^{4}-G_{3}^{6}\right)  & \left(  G_{2}^{6}-G_{0}^{1}\right) \\
6 & \left(  G_{1}^{2}-G_{2}^{1}\right)  & \left(  G_{3}^{4}-G_{0}^{2}\right)
& \left(  G_{0}^{1}+G_{3}^{5}\right)  & \left(  -G_{1}^{4}-G_{2}^{5}\right)
\end{bmatrix}
$

\bigskip%

\begin{equation}
\left[  T_{G}\right]  ^{ar}=\sum_{bc=1}^{6}G_{ac}^{b}\left(  \delta_{q_{b}%
}^{r}G_{p_{b}}^{c}-\delta_{p_{b}}^{r}G_{q_{b}}^{c}\right) \label{E72a}%
\end{equation}

\paragraph{5)}

The matter part of the Noether current $Y_{G}$ is :

$N\sum_{a,r}\left(  a_{I}\left(  \left[  J\right]  \left[  \widetilde{\kappa
}_{a}\right]  \right)  _{r}+V^{r}P_{a}\right)  \partial_{r}\otimes
\overrightarrow{\kappa}_{a}$ and :%

\begin{equation}
Y_{G}=\sum_{a,r}\left(  N\left(  a_{I}\left(  \left[  J\right]  \left[
\widetilde{\kappa}_{a}\right]  \right)  _{r}+a_{D}V^{r}P_{a}\right)
+2a_{G}\left[  T_{G}\right]  _{r}^{a}\right)  \partial_{r}\otimes
\overrightarrow{\kappa}_{a}\label{E73}%
\end{equation}

\subsubsection{Superpotential}

\paragraph{1)}

The superpotential $\varpi_{4}\left(  Z_{G}\right)  $\ is given by the
equation \ref{E50} with

$Z_{G}=\sum_{a\left\{  \alpha\beta\right\}  }\frac{dL_{F}}{d\partial_{\alpha
}G_{\beta}^{a}}\partial_{\alpha}\wedge\partial_{\beta}\otimes\overrightarrow
{\kappa}_{a}=2\sum_{a\left\{  \alpha\beta\right\}  }\frac{dL_{F}}{d%
\mathcal{F}%
_{\alpha\beta}^{a}}\partial_{\alpha}\wedge\partial_{\beta}\otimes
\overrightarrow{\kappa}_{a}$

It will be more convenient to isolate the constant $a_{G}$ and denote from now
on : $\varpi_{4}\left(  Z_{G}\right)  =a_{G}\Pi_{G}$ so that :

$Z_{G}=2\sum_{a\left\{  \alpha\beta\right\}  }\left(  O_{p_{a}}^{\beta
}O_{q_{a}}^{\alpha}-O_{q_{a}}^{\beta}O_{p_{a}}^{\alpha}\right)  \partial
_{\alpha}\wedge\partial_{\beta}\otimes\overrightarrow{\kappa}_{a}$

$\Pi_{G}=\varpi_{4}\left(  2\sum_{a\left\{  \alpha\beta\right\}  }\left(
O_{p_{a}}^{\beta}O_{q_{a}}^{\alpha}-O_{q_{a}}^{\beta}O_{p_{a}}^{\alpha
}\right)  \partial_{\alpha}\wedge\partial_{\beta}\right)  \otimes
\overrightarrow{\kappa}_{a}$

\paragraph{2)}

Its exterior derivative is :

$d\Pi_{G}$

$=-4\sum_{\alpha\beta=0}^{3}\left(  -1\right)  ^{\alpha+1}\partial_{\beta
}\left(  \left(  O_{p_{a}}^{\beta}O_{q_{a}}^{\alpha}-O_{q_{a}}^{\beta}%
O_{p_{a}}^{\alpha}\right)  \det O^{\prime}\right)  dx^{0}..\wedge
\widehat{dx^{\alpha}}..dx^{3}\otimes\overrightarrow{\kappa}_{a}$

$\sum_{\beta=0}^{3}\partial_{\beta}\left(  \left(  O_{p_{a}}^{\beta}O_{q_{a}%
}^{\alpha}-O_{q_{a}}^{\beta}O_{p_{a}}^{\alpha}\right)  \det O^{\prime}\right)
$

$=\sum_{\beta}\partial_{\beta}\left(  \left(  O_{p_{a}}^{\beta}O_{q_{a}%
}^{\alpha}-O_{q_{a}}^{\beta}O_{p_{a}}^{\alpha}\right)  \left(  \det O^{\prime
}\right)  \right)  +\left(  O_{p_{a}}^{\beta}O_{q_{a}}^{\alpha}-O_{q_{a}%
}^{\beta}O_{p_{a}}^{\alpha}\right)  \partial_{\beta}\left(  \left(  \det
O^{\prime}\right)  \right)  $

$=\left(  \det O^{\prime}\right)  \sum_{\beta}O_{p_{a}}^{\beta}\partial
_{\beta}O_{q_{a}}^{\alpha}-O_{q_{a}}^{\beta}\partial_{\beta}O_{p_{a}}^{\alpha
}+O_{q_{a}}^{\alpha}\left(  \partial_{\beta}O_{p_{a}}^{\beta}+O_{p_{a}}%
^{\beta}Tr\left(  O\partial_{\beta}O^{\prime}\right)  \right)  \allowbreak
-O_{p_{a}}^{\alpha}\left(  \partial_{\beta}O_{q_{a}}^{\beta}+O_{q_{a}}^{\beta
}Tr\left(  O\partial_{\beta}O^{\prime}\right)  \right)  $

$=\left(  \sum_{ra}O_{r}^{\alpha}c_{a}^{r}+\left(  O_{q_{a}}^{\alpha}D_{p_{a}%
}-O_{p_{a}}^{\alpha}D_{q_{a}}\right)  \right)  \left(  \det O^{\prime}\right)
$ with $c_{p_{a}q_{a}}^{r}=c_{a}^{r}$

Where :

$\sum_{\alpha\beta}O_{\alpha}^{\prime r}\left(  O_{p}^{\beta}\partial_{\beta
}O_{q}^{\alpha}-O_{q}^{\beta}\partial_{\beta}O_{p}^{\alpha}\right)  =\left[
\partial_{p},\partial_{q}\right]  ^{r}=c_{pq}^{r}$ are the structure
coefficients of the basis $\partial_{i}$\ 

$D_{i}=\sum_{\beta}\partial_{\beta}O_{i}^{\beta}+O_{i}^{\beta}Tr\left(
O\partial_{\beta}O^{\prime}\right)  =\frac{1}{\det O^{\prime}}\sum_{\beta
}\partial_{\beta}\left(  O_{i}^{\beta}\left(  \det O^{\prime}\right)  \right)
$ is the divergence of the vector field $\partial_{i}:D_{i}\varpi
_{4}=\pounds _{\partial_{i}}\varpi_{4}\Leftrightarrow D_{i}=Div\left(
\partial_{i}\right)  $

It is easy to check that : i=0,..3: $D_{i}=\sum_{j=0}^{3}c_{ji}^{j}$

$D_{0}=c_{10}^{1}+c_{20}^{2}+c_{30}^{3}=-c_{4}^{1}-c_{5}^{2}-c_{6}^{3}$

$D_{1}=c_{01}^{0}+c_{21}^{2}+c_{31}^{3}=c_{4}^{0}+c_{3}^{2}-c_{2}^{3}$

$D_{2}=c_{02}^{0}+c_{12}^{1}+c_{32}^{3}=c_{5}^{0}-c_{3}^{1}+c_{1}^{3}$

$D_{0}=c_{03}^{0}+c_{13}^{1}+c_{23}^{2}=c_{6}^{1}+c_{2}^{1}-c_{1}^{2}$

So :

$d\Pi_{G}=-4\sum_{\alpha,r=0}^{3}\left(  -1\right)  ^{\alpha+1}\left(
O_{r}^{\alpha}c_{p_{a}q_{a}}^{r}+O_{q_{a}}^{\alpha}D_{p_{a}}-O_{p_{a}}%
^{\alpha}D_{q_{a}}\right)  \det O^{\prime}$

$\qquad dx^{0}\wedge.\wedge\widehat{dx^{\alpha}}..\wedge dx^{3}\otimes
\overrightarrow{\kappa}_{a}$%

\begin{equation}
d\Pi_{G}=-4\sum_{\alpha=0}^{3}\left(  -1\right)  ^{\alpha+1}\sum_{r}\left(
c_{a}^{r}+\delta_{q_{a}}^{r}D_{p_{a}}-\delta_{p_{a}}^{r}D_{q_{a}}\right)
O_{r}^{\alpha}\det O^{\prime}dx^{0}\wedge..\wedge\widehat{dx^{\alpha}}..\wedge
dx^{3}\otimes\overrightarrow{\kappa}_{a}\label{E74}%
\end{equation}

\paragraph{3)}

Remark :$\varpi_{4}\left(  Z_{G}\right)  $ can be computed directly with :
$\varpi_{4}=\partial^{0}\wedge\partial^{1}\wedge\partial^{2}\wedge\partial
^{3}$ :

$\varpi_{4}\left(  \sum_{a}\partial_{p_{a}}\wedge\partial_{q_{a}}%
\otimes\overrightarrow{\kappa}_{a}\right)  $

$=\left(  \partial^{1}\wedge\partial^{2}\wedge\partial^{3}\wedge\partial
^{4}\right)  \left(  \sum_{a}\partial_{p_{a}}\wedge\partial_{q_{a}}%
\otimes\overrightarrow{\kappa}_{a}\right)  $

$=\sum_{a}\left(  \left(  \partial^{1}\wedge\partial^{2}\wedge\partial
^{3}\wedge\partial^{4}\right)  \left(  \partial_{p_{a}}\wedge\partial_{q_{a}%
}\right)  \right)  \otimes\overrightarrow{\kappa}_{a}$

$\varpi_{4}\left(  Z_{G}\right)  =-4a_{G}\varpi_{4}\left(  \sum_{a}%
\partial_{p_{a}}\wedge\partial_{q_{a}}\otimes\overrightarrow{\kappa}%
_{a}\right)  $

$=8a_{G}(\partial^{0}\wedge\partial^{1}\otimes\overrightarrow{\kappa}%
_{1}+\partial^{0}\wedge\partial^{2}\otimes\overrightarrow{\kappa}_{2}%
+\partial^{0}\wedge\partial^{3}\otimes\overrightarrow{\kappa}_{3}$

$+\partial^{3}\wedge\partial^{2}\otimes\overrightarrow{\kappa}_{4}%
+\partial^{1}\wedge\partial^{3}\otimes\overrightarrow{\kappa}_{5}+\partial
^{2}\wedge\partial^{1}\otimes\overrightarrow{\kappa}_{6})$

$=8a_{G}\left(  \sum_{a=1}^{3}\partial^{p_{a}}\wedge\partial^{q_{a}}%
\otimes\overrightarrow{\kappa}_{a+3}+\partial^{p_{a+3}}\wedge\partial
^{q_{a+3}}\otimes\overrightarrow{\kappa}_{a}\right)  $

This quantity can be expressed with the Hodge dual of : $\sum_{a=1}^{6}%
\eta_{p_{a}p_{a}}\partial^{p_{a}}\wedge\partial^{q_{a}}\otimes\overrightarrow
{\kappa}_{a}$ :

$\varpi_{4}\left(  Z_{G}\right)  =8a_{G}\ast(-\partial^{0}\wedge\partial
^{1}\otimes\overrightarrow{\kappa}_{1}-\partial^{0}\wedge\partial^{2}%
\otimes\overrightarrow{\kappa}_{2}-\partial^{0}\wedge\partial^{3}%
\otimes\overrightarrow{\kappa}_{3}$

$+\partial^{3}\wedge\partial^{2}\otimes\overrightarrow{\kappa}_{4}%
+\partial^{1}\wedge\partial^{3}\otimes\overrightarrow{\kappa}_{5}+\partial
^{2}\wedge\partial^{1}\otimes\overrightarrow{\kappa}_{6})$

$\varpi_{4}\left(  Z_{G}\right)  =8a_{G}\ast\left(  \sum_{a=1}^{6}\eta
_{p_{a}p_{a}}\partial^{p_{a}}\wedge\partial^{q_{a}}\otimes\overrightarrow
{\kappa}_{a}\right)  $

\subsubsection{Equation}

This equation is linear in $G_{r}^{a}$ and thus can be solved explicitly.

\paragraph{1)}

The gravitational equation reads :

$\varpi_{4}\left(  Y_{G}\right)  =\frac{1}{2}a_{G}d\Pi_{G}$ or $\forall
a,\alpha:Y_{G}^{a\alpha}=-2\sum_{\beta b}\frac{1}{\det O^{\prime}}%
\partial_{\beta}\left(  \frac{dL_{F}\left(  \det O^{\prime}\right)  }{d%
\mathcal{F}%
_{G\alpha\beta}^{a}}\right)  $

with $Y_{G}^{ar}=\sum_{a,r}\left(  N\left(  a_{I}\left(  \left[  J\right]
\left[  \widetilde{\kappa}_{a}\right]  \right)  _{r}+a_{D}V^{r}P_{a}\right)
+2a_{G}\left[  T_{G}\right]  ^{ar}\right)  $

$\left(  N\left(  a_{I}\left(  \left[  J\right]  \left[  \widetilde{\kappa
}_{a}\right]  \right)  _{r}+a_{D}V^{r}P_{a}\right)  +2a_{G}\left[
T_{G}\right]  ^{ar}\right)  O_{r}^{\alpha}$

$=-2a_{G}\sum_{r}\left(  c_{p_{a}q_{a}}^{r}+\delta_{q_{a}}^{r}D_{p_{a}}%
-\delta_{p_{a}}^{r}D_{q_{a}}\right)  O_{r}^{\alpha}$

That is :%

\begin{equation}
\left[  T_{G}\right]  ^{ar}=-c_{a}^{r}+\delta_{p_{a}}^{r}D_{q_{a}}%
-\delta_{q_{a}}^{r}D_{p_{a}}-\frac{N}{2a_{G}}\left(  a_{I}\left(  \left[
J\right]  \left[  \widetilde{\kappa}_{a}\right]  \right)  _{r}+a_{D}V^{r}%
P_{a}\right) \label{E75}%
\end{equation}

where $\left[  T_{G}\right]  $ is the previous table. We have 24 linear
equations which are with%

\begin{equation}
K_{a}^{r}=\frac{N}{2a_{G}}\left(  a_{I}\left(  \left[  J\right]  \left[
\widetilde{\kappa}_{a}\right]  \right)  _{r}+a_{D}V^{r}P_{a}\right)
\label{E75b}%
\end{equation}

$\sum_{bc=1}^{6}G_{ac}^{b}\left(  \delta_{q_{b}}^{r}G_{p_{b}}^{c}%
-\delta_{p_{b}}^{r}G_{q_{b}}^{c}\right)  =-c_{a}^{r}+\delta_{p_{a}}%
^{r}D_{q_{a}}-\delta_{q_{a}}^{r}D_{p_{a}}-K_{a}^{r}$

\newpage

$\bigskip\bigskip\bigskip$

$\;%
\begin{bmatrix}
a & r & LHS & RHS\\
1 & 1 & G_{2}^{2}+G_{3}^{3} & -c_{1}^{1}-K_{1}^{1}\\
1 & 2 & -G_{1}^{2}-G_{0}^{6} & -D_{3}-c_{1}^{2}-K_{1}^{2}\\
1 & 3 & -G_{1}^{3}+G_{0}^{5} & D_{2}-c_{1}^{3}-K_{1}^{3}\\
1 & 0 & G_{2}^{6}-G_{3}^{5} & -c_{1}^{0}-K_{1}^{0}\\
2 & 1 & -G_{2}^{1}+G_{0}^{6} & -c_{2}^{1}+D_{3}-K_{2}^{1}\\
2 & 2 & G_{1}^{1}+G_{3}^{3} & -c_{2}^{2}-K_{2}^{2}\\
2 & 3 & -G_{2}^{3}-G_{0}^{4} & -D_{1}-c_{2}^{3}-K_{2}^{3}\\
2 & 0 & -G_{1}^{6}+G_{3}^{4} & -c_{2}^{0}-K_{2}^{0}\\
3 & 1 & -G_{3}^{1}-G_{0}^{5} & -c_{3}^{1}-D_{2}-K_{3}^{1}\\
3 & 2 & -G_{3}^{2}+G_{0}^{4} & D_{1}-c_{3}^{2}-K_{3}^{2}\\
3 & 3 & G_{1}^{1}+G_{2}^{2} & -c_{3}^{3}-K_{3}^{3}\\
3 & 0 & G_{1}^{5}-G_{2}^{4} & -c_{3}^{0}-K_{3}^{0}\\
4 & 1 & -G_{2}^{5}-G_{3}^{6} & -c_{4}^{1}-D_{0}-K_{4}^{1}\\
4 & 2 & G_{1}^{5}-G_{0}^{3} & -c_{4}^{2}-K_{4}^{2}\\
4 & 3 & G_{4}^{2}+G_{1}^{6} & -c_{4}^{3}-K_{4}^{3}\\
4 & 0 & G_{2}^{3}-G_{3}^{2} & D_{1}-c_{4}^{0}-K_{4}^{0}\\
5 & 1 & G_{2}^{4}+G_{4}^{3} & -c_{5}^{1}-K_{5}^{1}\\
5 & 2 & -G_{1}^{4}-G_{3}^{6} & -D_{0}-c_{5}^{2}-K_{5}^{2}\\
5 & 3 & -G_{4}^{1}+G_{2}^{6} & -c_{5}^{3}-K_{5}^{3}\\
5 & 0 & G_{3}^{1}-G_{1}^{3} & D_{2}-c_{5}^{0}-K_{5}^{0}\\
6 & 1 & -G_{4}^{2}+G_{3}^{4} & -c_{6}^{1}-K_{6}^{1}\\
6 & 2 & G_{4}^{1}+G_{3}^{5} & -c_{6}^{2}-K_{6}^{2}\\
6 & 3 & -G_{1}^{4}-G_{2}^{5} & -D_{0}-c_{6}^{3}-K_{6}^{3}\\
6 & 0 & -G_{2}^{1}+G_{1}^{2} & D_{3}-c_{6}^{0}-K_{6}^{0}%
\end{bmatrix}
$

\newpage

and the solution is : $2G_{r}^{a}=$

\bigskip

$%
\begin{bmatrix}
2G_{r}^{a} & r=0 &  & r=1\\
G_{0}^{1} & -K_{1}^{0}+K_{5}^{3}-K_{6}^{2}-c_{1}^{0}+c_{5}^{3}-c_{6}^{2} &
G_{1}^{1} & K_{1}^{1}+c_{1}^{1}-K_{2}^{2}-K_{3}^{3}-c_{2}^{2}-c_{3}^{3}\\
G_{0}^{2} & K_{6}^{1}+c_{6}^{1}-K_{2}^{0}-K_{4}^{3}-c_{2}^{0}-c_{4}^{3} &
G_{1}^{2} & K_{2}^{1}+2c_{2}^{1}+K_{1}^{2}-K_{6}^{0}\\
G_{0}^{3} & -K_{5}^{1}-c_{5}^{1}+K_{4}^{2}-K_{3}^{0}+c_{4}^{2}-c_{3}^{0} &
G_{1}^{3} & K_{3}^{1}+2c_{3}^{1}+K_{1}^{3}+K_{5}^{0}\\
G_{0}^{4} & K_{2}^{3}-K_{3}^{2}+K_{4}^{0}+2c_{4}^{0} & G_{1}^{4} & -K_{4}%
^{1}-2c_{4}^{1}+K_{5}^{2}+K_{6}^{3}\\
G_{0}^{5} & K_{3}-K_{1}^{3}+K_{5}^{0}+2c_{5}^{0} & G_{1}^{5} & -K_{5}%
^{1}-c_{5}^{1}-K_{4}^{2}-K_{3}^{0}-c_{4}^{2}-c_{3}^{0}\\
G_{0}^{6} & -K_{2}+K_{1}^{2}+K_{6}^{0}+2c_{6}^{0} & G_{1}^{6} & -K_{6}%
^{1}-c_{6}^{1}+K_{2}^{0}-K_{4}^{3}+c_{2}^{0}-c_{4}^{3}%
\end{bmatrix}
$

$%
\begin{bmatrix}
& r=2 &  & r=3\\
G_{2}^{1} & K_{2}^{1}+K_{1}^{2}+K_{6}^{0}+2c_{1}^{2} & G_{3}^{1} & K_{3}%
+K_{1}^{3}-K_{5}^{0}+2c_{1}^{3}\\
G_{2}^{2} & -K_{1}^{1}-c_{1}^{1}+K_{2}^{2}-K_{3}^{3}+c_{2}^{2}-c_{3}^{3} &
G_{3}^{2} & K_{2}^{3}+K_{3}^{2}+K_{4}^{0}+2c_{2}^{3}\\
G_{2}^{3} & K_{2}^{3}+K_{3}^{2}-K_{4}^{0}+2c_{3}^{2} & G_{3}^{3} & -K_{1}%
^{1}-c_{1}^{1}-K_{2}^{2}+K_{3}^{3}-c_{2}^{2}+c_{3}^{3}\\
G_{2}^{4} & -K_{5}^{1}-c_{5}^{1}-K_{4}^{2}+K_{3}^{0}-c_{4}^{2}+c_{3}^{0} &
G_{3}^{4} & -K_{6}^{1}-c_{6}^{1}-K_{2}^{0}-K_{4}^{3}-c_{2}^{0}-c_{4}^{3}\\
G_{2}^{5} & K_{4}^{1}-K_{5}^{2}+K_{6}^{3}-2c_{5}^{2} & G_{3}^{5} & K_{1}%
^{0}-K_{5}^{3}-K_{6}^{2}+c_{1}^{0}-c_{5}^{3}-c_{6}^{2}\\
G_{2}^{6} & -K_{1}^{0}-K_{5}^{3}-K_{6}^{2}-c_{1}^{0}-c_{5}^{3}-c_{6}^{2} &
G_{3}^{6} & K_{4}^{1}+K_{5}^{2}-K_{6}^{3}-2c_{6}^{3}%
\end{bmatrix}
$

\bigskip

Notice that in this equation ony the momentum K appears, and not $\psi$ per se.

In the vacuum the gravitation field is not null (in accordance with the
General Relativity) and entirely given by the structure coefficients which
thus fully represent the geometry of the universe.

\paragraph{2)}

The Noether current : \ $Y_{G}^{ar}=-2a_{G}\sum_{r}\left(  c_{p_{a}q_{a}}%
^{r}+\delta_{q_{a}}^{r}D_{p_{a}}-\delta_{p_{a}}^{r}D_{q_{a}}\right)  $ is
expressed with the structure coefficients only.

We have the table :

\bigskip

$Y_{G}^{ar}=2a_{G}%
\begin{bmatrix}
a\backslash r & 0 & 1 & 2 & 3\\
1 & -c_{1}^{0} & -c_{1}^{1} & -\left(  c_{2}^{1}+c_{6}^{0}\right)  & \left(
-c_{3}^{1}+c_{5}^{0}\right) \\
2 & -c_{2}^{0} & \left(  -c_{1}^{2}+c_{6}^{0}\right)  & -c_{2}^{2} & -\left(
c_{3}^{2}+c_{4}^{0}\right) \\
3 & -c_{3}^{0} & -\left(  c_{1}^{3}+c_{5}^{0}\right)  & \left(  -c_{2}%
^{3}+c_{4}^{0}\right)  & -c_{3}^{3}\\
4 & \left(  c_{3}^{2}-c_{2}^{3}\right)  & \left(  c_{5}^{2}+c_{6}^{3}\right)
& -c_{4}^{2} & -c_{4}^{3}\\
5 & \left(  -c_{3}^{1}+c_{1}^{3}\right)  & -c_{5}^{1} & \left(  c_{4}%
^{1}+c_{6}^{3}\right)  & -c_{5}^{3}\\
6 & \left(  c_{2}^{1}-c_{1}^{2}\right)  & -c_{6}^{1} & -c_{6}^{2} & \left(
c_{4}^{1}+c_{5}^{2}\right)
\end{bmatrix}
$

\bigskip

The flux of this vector through the S(t) hypersurfaces is conserved. That is
$\int_{S(0)}\varpi_{4}\left(  Y_{G}\right)  =\int_{S(t)}\varpi_{4}\left(
Y_{G}\right)  $

\subsubsection{Symmetry}

The value of the torsion tensor $\nabla_{e}\Theta=\sum_{r,a}\left(
2c_{p_{a}q_{a}}^{r}+T^{ar}\right)  \partial^{p_{a}}\wedge\partial^{q_{a}%
}\otimes\partial_{r}$ has been calculated previously (table 2).

Using the results above we get the table:

\bigskip

$\left(  \nabla_{e}\Theta\right)  _{p_{a}q_{a}}^{r}=3\left[  c_{a}^{r}\right]
+\left[  K\right]  _{a}^{r}+\frac{1}{2}%
\begin{bmatrix}
0 & 0 & \Theta_{3} & -\Theta_{2}\\
0 & -\Theta_{3} & 0 & \Theta_{1}\\
0 & \Theta_{2} & -\Theta_{1} & 0\\
-\Theta_{1} & \Theta_{0} & 0 & 0\\
-\Theta_{2} & 0 & \Theta_{0} & 0\\
-\Theta_{3} & 0 & 0 & \Theta_{0}%
\end{bmatrix}
$

\bigskip

with : $\Theta_{0}=-\left(  K_{4}^{1}+K_{5}^{2}+K_{6}^{3}\right)  ;\Theta
_{1}=-K_{2}^{3}+K_{3}^{2}+K_{4}^{0};\Theta_{2}=-K_{3}^{1}+K_{1}^{3}+K_{5}%
^{0};\Theta_{3}=-K_{1}^{2}+K_{2}^{1}+K_{6}^{0}$

$\Theta_{0}=-\sum_{a=1}^{3}K_{a+3}^{a};k=1,2,3:\Theta_{k}=K_{p_{k}}^{q_{k}%
}-K_{q_{k}}^{p_{k}}+K_{k+3}^{0}$

So in the vacuum the torsion is given by the structure coefficients.\ And the
connection is torsion free iff the particles have some specific distribution.

\subsection{Other force fields}

We will define two moments : the "charge" $\rho_{a}$ and the "magnetic moment"
$\overrightarrow{\mu}_{a}$, both computed from the state tensor, gauge and
chart invariant. The law for the fields take then a simple, geometric form,
independant from the gravitational field:

$N\varpi_{4}\left(  a_{D}\rho_{a}\overrightarrow{V}-ia_{I}\overrightarrow{\mu
}_{a}\right)  =\nabla_{e}\ast\overline{%
\mathcal{F}%
}_{A}^{a}$

\subsubsection{Noether currents}

\paragraph{1)}

The Noether currents are : $Y_{AR}^{a}=\sum_{\alpha}Y_{AR}^{\alpha a}%
\partial_{\alpha},Y_{AI}^{a}=\sum_{\alpha}Y_{AI}^{\alpha a}\partial_{\alpha}$
with :

$Y_{AR}^{\alpha a}=V\sum_{ij}\frac{dL_{M}}{d\operatorname{Re}\nabla_{\alpha
}\psi^{ij}}\operatorname{Re}\left(  \left[  \psi^{\Diamond}\right]  \left[
\theta_{a}\right]  ^{t}\right)  _{j}^{i}+\frac{dL_{M}}{d\operatorname{Im}%
\nabla_{\alpha}\psi^{ij}}\operatorname{Im}\left(  \left[  \psi^{\Diamond
}\right]  \left[  \theta_{a}\right]  ^{t}\right)  _{j}^{i}$

$+2\sum_{b\beta}\frac{dL_{F}}{d\operatorname{Re}%
\mathcal{F}%
_{A,\alpha\beta}^{b}}\operatorname{Re}\left[  \overrightarrow{\theta}%
_{a},\grave{A}_{\beta}\right]  ^{b}+\frac{dL_{F}}{d\operatorname{Im}%
\mathcal{F}%
_{A,\alpha\beta}^{b}}\operatorname{Im}\left[  \overrightarrow{\theta}%
_{a},\grave{A}_{\beta}\right]  ^{b}$

$Y_{AI}^{\alpha a}=V\sum_{ij}-\frac{dL_{M}}{d\operatorname{Re}\nabla_{\alpha
}\psi^{ij}}\operatorname{Im}\left(  \left[  \psi^{\Diamond}\right]  \left[
\theta_{a}\right]  ^{t}\right)  _{j}^{i}+\frac{dL_{M}}{d\operatorname{Im}%
\nabla_{\alpha}\psi^{ij}}\operatorname{Re}\left(  \left[  \psi^{\Diamond
}\right]  \left[  \theta_{a}\right]  ^{t}\right)  _{j}^{i}$

$+2\sum_{b\beta}-\frac{dL_{F}}{d\operatorname{Re}%
\mathcal{F}%
_{A,\alpha\beta}^{b}}\operatorname{Im}\left[  \overrightarrow{\theta}%
_{a},\grave{A}_{\beta}\right]  ^{b}+\frac{dL_{F}}{d\operatorname{Im}%
\mathcal{F}%
_{A,\alpha\beta}^{b}}\operatorname{Re}\left[  \overrightarrow{\theta}%
_{a},\grave{A}_{\beta}\right]  ^{b}$

We have already :

$\frac{dL_{M}}{d\operatorname{Re}\nabla_{\alpha}\psi^{ij}}=N\operatorname{Im}%
\left(  \left[  \psi\right]  ^{\ast}\left[  \gamma_{0}\right]  \left[
D_{M}^{\alpha}\right]  \right)  _{i}^{j}$

$\frac{dL_{M}}{d\operatorname{Im}\nabla_{\alpha}\psi^{ij}}=N\operatorname{Re}%
\left(  \left[  \psi\right]  ^{\ast}\left[  \gamma_{0}\right]  \left[
D_{M}^{\alpha}\right]  \right)  _{i}^{j}$

Computation of the derivatives with respect to $\operatorname{Re}%
\mathcal{F}%
_{A,\alpha\beta}^{b},\operatorname{Im}%
\mathcal{F}%
_{A,\alpha\beta}^{b}$

$L_{F}=a_{F}\frac{1}{2}\sum_{a}\sum_{\eta\xi\lambda\mu}g^{\eta\lambda}%
g^{\xi\mu}\overline{%
\mathcal{F}%
}_{A\eta\xi}^{a}%
\mathcal{F}%
_{A\lambda\mu}^{a}+a_{G}R$

$=a_{F}\frac{1}{2}\sum_{a}\sum_{\eta\xi\lambda\mu}g^{\eta\lambda}g^{\xi\mu
}\left(  \operatorname{Re}%
\mathcal{F}%
_{A\eta\xi}^{a}-i\operatorname{Im}%
\mathcal{F}%
_{A\eta\xi}^{a}\right)  \left(  \operatorname{Re}%
\mathcal{F}%
_{A\lambda\mu}^{a}+i\operatorname{Im}%
\mathcal{F}%
_{A\lambda\mu}^{a}\right)  +a_{G}R$

$\frac{dL_{F}}{d\operatorname{Re}%
\mathcal{F}%
_{A,\alpha\beta}^{b}}=\frac{1}{2}a_{F}\sum_{\eta\xi\lambda\mu}g^{\eta\lambda
}g^{\xi\mu}\left(  \delta_{\eta}^{\alpha}\delta_{\xi}^{\beta}\overline{%
\mathcal{F}%
}_{A\lambda\mu}^{b}+%
\mathcal{F}%
_{A\eta\xi}^{b}\delta_{\lambda}^{\alpha}\delta_{\mu}^{\beta}\right)  =\frac
{1}{2}a_{F}\left(  \overline{%
\mathcal{F}%
}_{A}^{b\alpha\beta}+%
\mathcal{F}%
_{A}^{b,\alpha\beta}\right)  $

$\frac{dL_{F}}{d\operatorname{Re}%
\mathcal{F}%
_{A,\alpha\beta}^{b}}=a_{F}\operatorname{Re}%
\mathcal{F}%
_{A}^{b,\alpha\beta}$

$\frac{dL_{F}}{d\operatorname{Im}%
\mathcal{F}%
_{A,\alpha\beta}^{b}}=\frac{1}{2}a_{F}\sum_{\eta\xi\lambda\mu}g^{\eta\lambda
}g^{\xi\mu}\left(  -i\delta_{\eta}^{\alpha}\delta_{\xi}^{\beta}%
\mathcal{F}%
_{A\lambda\mu}^{b}+i\overline{%
\mathcal{F}%
}_{A\eta\xi}^{a}\delta_{\lambda}^{\alpha}\delta_{\mu}^{\beta}\right)  $

$=\frac{1}{2}ia_{F}\sum_{\eta\xi\lambda\mu}\left(  -g^{\alpha\lambda}%
g^{\beta\mu}%
\mathcal{F}%
_{A\lambda\mu}^{b}+g^{\eta\alpha}g^{\xi\beta}\overline{%
\mathcal{F}%
}_{A\eta\xi}^{a}\right)  $

$=i\frac{1}{2}a_{F}\left(  -%
\mathcal{F}%
_{A}^{b\alpha\beta}+\overline{%
\mathcal{F}%
}_{A}^{b,\alpha\beta}\right)  =i\frac{1}{2}a_{F}\left(  -2i\operatorname{Im}%
\mathcal{F}%
_{A}^{b,\alpha\beta}\right)  $

$\frac{dL_{F}}{d\operatorname{Im}%
\mathcal{F}%
_{A,\alpha\beta}^{b}}=a_{F}\operatorname{Im}%
\mathcal{F}%
_{A}^{b,\alpha\beta}$

So the equations are :

$Y_{AR}^{\alpha a}=N\sum_{ij}\operatorname{Im}\left(  \left[  \psi\right]
^{\ast}\left[  \gamma_{0}\right]  \left[  D_{M}^{\alpha}\right]  \right)
_{i}^{j}\operatorname{Re}\left(  \left[  \psi\right]  \left[  \theta
_{a}\right]  ^{t}\right)  _{j}^{i}+\operatorname{Re}\left(  \left[
\psi\right]  ^{\ast}\left[  \gamma_{0}\right]  \left[  D_{M}^{\alpha}\right]
\right)  _{i}^{j}\operatorname{Im}\left(  \left[  \psi^{\Diamond}\right]
\left[  \theta_{a}\right]  ^{t}\right)  _{j}^{i}$

$+2a_{F}\sum_{b\beta}\operatorname{Re}%
\mathcal{F}%
_{A}^{b,\alpha\beta}\operatorname{Re}\left[  \overrightarrow{\theta}%
_{a},\grave{A}_{\beta}\right]  ^{b}+\operatorname{Im}%
\mathcal{F}%
_{A}^{b,\alpha\beta}\operatorname{Im}\left[  \overrightarrow{\theta}%
_{a},\grave{A}_{\beta}\right]  ^{b}$

$=N\operatorname{Im}Tr\left[  \psi\right]  ^{\ast}\left[  \gamma_{0}\right]
\left[  D_{M}^{\alpha}\right]  \left[  \psi\right]  \left[  \theta_{a}\right]
^{t}+2a_{F}\sum_{b\beta}\operatorname{Re}\left(
\mathcal{F}%
_{A}^{b,\alpha\beta}\overline{\left[  \overrightarrow{\theta}_{a},\grave
{A}_{\beta}\right]  }^{b}\right)  $

$Y_{AR}^{\alpha a}=N\operatorname{Im}Tr\left[  \psi\right]  ^{\ast}\left[
\gamma_{0}\right]  \left[  D_{M}^{\alpha}\right]  \left[  \psi\right]  \left[
\theta_{a}\right]  ^{t}+2a_{F}\sum_{b\beta}\operatorname{Re}\left(  \left[
\overrightarrow{\theta}_{a},\grave{A}_{\beta}\right]  ,%
\mathcal{F}%
_{A}^{\alpha\beta}\right)  $

$Y_{AI}^{\alpha a}=N\sum_{ij}-\operatorname{Im}\left(  \left[  \psi\right]
^{\ast}\left[  \gamma_{0}\right]  \left[  D_{M}^{\alpha}\right]  \right)
_{i}^{j}\operatorname{Im}\left(  \left[  \psi\right]  \left[  \theta
_{a}\right]  ^{t}\right)  _{j}^{i}+\operatorname{Re}\left(  \left[
\psi\right]  ^{\ast}\left[  \gamma_{0}\right]  \left[  D_{M}^{\alpha}\right]
\right)  _{i}^{j}\operatorname{Re}\left(  \left[  \psi\right]  \left[
\theta_{a}\right]  ^{t}\right)  _{j}^{i}$

$+2a_{F}\sum_{b\beta}-\operatorname{Re}%
\mathcal{F}%
_{A}^{b,\alpha\beta}\operatorname{Im}\left[  \overrightarrow{\theta}%
_{a},\grave{A}_{\beta}\right]  ^{b}+\operatorname{Im}%
\mathcal{F}%
_{A}^{b,\alpha\beta}\operatorname{Re}\left[  \overrightarrow{\theta}%
_{a},\grave{A}_{\beta}\right]  ^{b}$

$=N\operatorname{Re}Tr\left[  \psi\right]  ^{\ast}\left[  \gamma_{0}\right]
\left[  D_{M}^{\alpha}\right]  \left[  \psi\right]  \left[  \theta_{a}\right]
^{t}+2a_{F}\sum_{b\beta}\operatorname{Im}\left(
\mathcal{F}%
_{A}^{b,\alpha\beta}\overline{\left[  \overrightarrow{\theta}_{a},\grave
{A}_{\beta}\right]  }^{b}\right)  $

$Y_{AR}^{\alpha a}=N\operatorname{Re}Tr\left[  \psi\right]  ^{\ast}\left[
\gamma_{0}\right]  \left[  D_{M}^{\alpha}\right]  \left[  \psi\right]  \left[
\theta_{a}\right]  ^{t}+2a_{F}\sum_{b\beta}\operatorname{Im}\left(  \left[
\overrightarrow{\theta}_{a},\grave{A}_{\beta}\right]  ,%
\mathcal{F}%
_{A}^{\alpha\beta}\right)  $

$\sum_{b}%
\mathcal{F}%
_{A}^{b,\alpha\beta}\overline{\left[  \overrightarrow{\theta}_{a},\grave
{A}_{\beta}\right]  }^{b}$ is the scalar product $\left(  \left[
\overrightarrow{\theta}_{a},\grave{A}_{\beta}\right]  ,%
\mathcal{F}%
_{A}^{\alpha\beta}\right)  $ in $T_{1}U^{c}$\ and with our assumptions that
the scalar product is preserved by the adjoint operator : $\forall
\overrightarrow{\theta},\overrightarrow{\theta}_{1},\overrightarrow{\theta
}_{2}\in T_{1}U^{c}:\left(  \left[  \overrightarrow{\theta},\overrightarrow
{\theta}_{1}\right]  ,\overrightarrow{\theta}_{2}\right)  =-\left(
\overrightarrow{\theta}_{1},\left[  \overrightarrow{\theta},\overrightarrow
{\theta}_{2}\right]  \right)  $

$\left(  \left[  \overrightarrow{\theta}_{a},\grave{A}_{\beta}\right]  ,%
\mathcal{F}%
_{A}^{\alpha\beta}\right)  =-\left(  \left[  \grave{A}_{\beta},\overrightarrow
{\theta}_{a}\right]  ,%
\mathcal{F}%
_{A}^{\alpha\beta}\right)  =\left(  \overrightarrow{\theta}_{a},\left[
\grave{A}_{\beta},%
\mathcal{F}%
_{A}^{\alpha\beta}\right]  \right)  =\left[  \grave{A}_{\beta},%
\mathcal{F}%
_{A}^{\alpha\beta}\right]  ^{a}$

Thus :

$Y_{AR}^{\alpha a}=N\operatorname{Im}Tr\left[  \psi\right]  ^{\ast}\left[
\gamma_{0}\right]  \left[  D_{M}^{\alpha}\right]  \left[  \psi\right]  \left[
\theta_{a}\right]  ^{t}+2a_{F}\sum_{\beta}\operatorname{Re}\left[  \grave
{A}_{\beta},%
\mathcal{F}%
_{A}^{\alpha\beta}\right]  ^{a}$

$Y_{AI}^{\alpha a}=N\operatorname{Re}Tr\left[  \psi\right]  ^{\ast}\left[
\gamma_{0}\right]  \left[  D_{M}^{\alpha}\right]  \left[  \psi\right]  \left[
\theta_{a}\right]  ^{t}+2a_{F}\sum_{\beta}\operatorname{Im}\left[  \grave
{A}_{\beta},%
\mathcal{F}%
_{A}^{\alpha\beta}\right]  ^{a}$

\paragraph{2)}

Let us compute the first terms : $\operatorname{Im}Tr\left[  \psi\right]
^{\ast}\left[  D_{M}^{\alpha}\right]  \left[  \psi\right]  \left[  \theta
_{a}\right]  ^{t},\operatorname{Re}Tr\left[  \psi\right]  ^{\ast}\left[
D_{M}^{\alpha}\right]  \left[  \psi\right]  \left[  \theta_{a}\right]  ^{t}$

$Tr\left[  \psi\right]  ^{\ast}\left[  D_{M}^{\alpha}\right]  \left[
\psi\right]  \left[  \theta_{a}\right]  ^{t}$

$=a_{I}\sum_{r}O_{r}^{\alpha}\operatorname{Im}Tr\left(  \left[  \psi\right]
^{\ast}\left[  \gamma_{0}\gamma^{r}\right]  \left[  \psi\right]  \left[
\theta_{a}\right]  ^{t}\right)  +a_{D}V^{\alpha}\operatorname{Im}Tr\left(
\left[  \psi^{\ast}\right]  \left[  \gamma_{0}\right]  \left[  \psi\right]
\left[  \theta_{a}\right]  ^{t}\right)  $

\subparagraph{a)}

$\left[  \psi\right]  ^{\ast}\left[  \gamma_{0}\gamma^{r}\right]  \left[
\psi\right]  \left[  \theta_{a}\right]  ^{t}=i\left(  \eta^{rr}\left[
\psi_{R}\right]  ^{\ast}\sigma_{r}\left[  \psi_{R}\right]  \left[  \theta
_{a}\right]  ^{t}-\left[  \psi_{L}\right]  ^{\ast}\sigma_{r}\left[  \psi
_{L}\right]  \left[  \theta_{a}\right]  ^{t}\right)  $

So : $Tr\left(  \left[  \psi\right]  ^{\ast}\left[  \gamma_{0}\gamma
^{r}\right]  \left[  \psi\right]  \left[  \theta_{a}\right]  ^{t}\right)
=iTr\left(  \left(  \eta^{rr}\left[  \psi_{R}\right]  ^{\ast}\sigma_{r}\left[
\psi_{R}\right]  \left[  \theta_{a}\right]  ^{t}-\left[  \psi_{L}\right]
^{\ast}\sigma_{r}\left[  \psi_{L}\right]  \left[  \theta_{a}\right]
^{t}\right)  \right)  $

We have assumed that the representation $\left(  W,\chi\right)  $\ is unitary
so $\left[  \theta_{a}\right]  $\ is antihermitian :$\left[  \theta
_{a}\right]  ^{\ast}=-\left[  \theta_{a}\right]  ,\left[  \theta_{a}\right]
^{t}=-\overline{\left[  \theta_{a}\right]  }$ and :

$\overline{Tr\left(  \left[  \psi_{R}\right]  ^{\ast}\sigma_{r}\left[
\psi_{R}\right]  \left[  \theta_{a}\right]  ^{t}\right)  }=Tr\left(  \left[
\psi_{R}\right]  ^{\ast}\sigma_{r}\left[  \psi_{R}\right]  \left[  \theta
_{a}\right]  ^{t}\right)  ^{\ast}=-Tr\left(  \left[  \psi_{R}\right]
^{t}\sigma_{r}^{t}\overline{\left[  \psi_{R}\right]  }\left[  \theta
_{a}\right]  \right)  $

$=-Tr\left(  \left[  \theta_{a}\right]  ^{t}\left[  \psi_{R}\right]  ^{\ast
}\sigma_{r}\left[  \psi_{R}\right]  \right)  =-Tr\left(  \left[  \psi
_{R}\right]  ^{\ast}\sigma_{r}\left[  \psi_{R}\right]  \left[  \theta
_{a}\right]  ^{t}\right)  $

Each quantity $Tr\left(  \left[  \psi_{R}\right]  ^{\ast}\sigma_{r}\left[
\psi_{R}\right]  \left[  \theta_{a}\right]  ^{t}\right)  ,Tr\left(  \left[
\psi_{L}\right]  ^{\ast}\sigma_{r}\left[  \psi_{L}\right]  \left[  \theta
_{a}\right]  ^{t}\right)  $ is an imaginary scalar.

Let be :

$\left\langle \psi,\left[  \gamma_{0}\gamma^{r}\right]  \left[  \psi\right]
\left[  \theta_{a}\right]  ^{t}\right\rangle =iTr\left(  \eta^{rr}\left[
\psi_{R}\right]  ^{\ast}\sigma_{r}\left[  \psi_{R}\right]  \left[  \theta
_{a}\right]  ^{t}-\left[  \psi_{L}\right]  ^{\ast}\sigma_{r}\left[  \psi
_{L}\right]  \left[  \theta_{a}\right]  ^{t}\right)  =-\left[  \mu\right]
_{a}^{r}$

$\left[  \mu\right]  _{a}^{r}=-iTr\left(  \eta^{rr}\left[  \psi_{R}\right]
^{\ast}\sigma_{r}\left[  \psi_{R}\right]  \left[  \theta_{a}\right]
^{t}\right)  +iTr\left(  \left[  \psi_{L}\right]  ^{\ast}\sigma_{r}\left[
\psi_{L}\right]  \left[  \theta_{a}\right]  ^{t}\right)  $%

\begin{equation}
\left\langle \psi,\left[  \gamma_{0}\gamma^{r}\right]  \left[  \psi\right]
\left[  \theta_{a}\right]  ^{t}\right\rangle \mathbf{=iTr}\left(  \eta
^{rr}\left[  \psi_{R}\right]  ^{\ast}\sigma_{r}\left[  \psi_{R}\right]
\left[  \theta_{a}\right]  ^{t}-\left[  \psi_{L}\right]  ^{\ast}\sigma
_{r}\left[  \psi_{L}\right]  \left[  \theta_{a}\right]  ^{t}\right)
\mathbf{=-}\left[  \mu\right]  _{a}^{r}%
\end{equation}

\subparagraph{b)}

$\left[  \psi^{\ast}\right]  \left[  \gamma_{0}\right]  \left[  \psi\right]
\left[  \theta_{a}\right]  ^{t}=i\left(  \left[  \psi_{R}\right]  ^{\ast
}\left[  \psi_{L}\right]  \left[  \theta_{a}\right]  ^{t}-\left[  \psi
_{L}\right]  ^{\ast}\left[  \psi_{R}\right]  \left[  \theta_{a}\right]
^{t}\right)  $

So : $Tr\left(  \left[  \psi^{\ast}\right]  \left[  \gamma_{0}\right]  \left[
\psi\right]  \left[  \theta_{a}\right]  ^{t}\right)  =iTr\left(  \left[
\psi_{R}\right]  ^{\ast}\left[  \psi_{L}\right]  \left[  \theta_{a}\right]
^{t}-\left[  \psi_{L}\right]  ^{\ast}\left[  \psi_{R}\right]  \left[
\theta_{a}\right]  ^{t}\right)  $

With the same assumption about $\left[  \theta_{a}\right]  $ as above :

$\overline{Tr\left[  \psi_{R}\right]  ^{\ast}\left[  \psi_{L}\right]  \left[
\theta_{a}\right]  ^{t}}=Tr\left(  \left(  \left[  \psi_{R}\right]  ^{\ast
}\left[  \psi_{L}\right]  \left[  \theta_{a}\right]  ^{t}\right)  ^{\ast
}\right)  $

$=Tr\left(  \overline{\left[  \theta_{a}\right]  }\left[  \psi_{L}\right]
^{\ast}\left[  \psi_{R}\right]  \right)  =-Tr\left(  \left[  \theta
_{a}\right]  ^{t}\left[  \psi_{L}\right]  ^{\ast}\left[  \psi_{R}\right]
\right)  =-Tr\left(  \left[  \psi_{L}\right]  ^{\ast}\left[  \psi_{R}\right]
\left[  \theta_{a}\right]  ^{t}\right)  $

So one cannot tell much about the individual quantities

$Tr\left(  \left[  \psi_{L}\right]  ^{\ast}\left[  \psi_{R}\right]  \left[
\theta_{a}\right]  ^{t}\right)  ,Tr\left(  \left[  \psi_{R}\right]  ^{\ast
}\left[  \psi_{L}\right]  \left[  \theta_{a}\right]  ^{t}\right)  $

but that they are complex conjugates of each other. So the difference is a
real number:

$\overline{Tr\left(  \left[  \psi_{R}\right]  ^{\ast}\left[  \psi_{L}\right]
\left[  \theta_{a}\right]  ^{t}-\left[  \psi_{L}\right]  ^{\ast}\left[
\psi_{R}\right]  \left[  \theta_{a}\right]  ^{t}\right)  }$

$=\overline{Tr\left(  \left[  \psi_{R}\right]  ^{\ast}\left[  \psi_{L}\right]
\left[  \theta_{a}\right]  ^{t}\right)  }-\overline{Tr\left(  \left[  \psi
_{L}\right]  ^{\ast}\left[  \psi_{R}\right]  \left[  \theta_{a}\right]
^{t}\right)  }$

$=-Tr\left(  \left[  \psi_{L}\right]  ^{\ast}\left[  \psi_{R}\right]  \left[
\theta_{a}\right]  ^{t}+\left[  \psi_{R}\right]  ^{\ast}\left[  \psi
_{L}\right]  \left[  \theta_{a}\right]  ^{t}\right)  $

$=-Tr\left(  \left[  \psi_{L}\right]  ^{\ast}\left[  \psi_{R}\right]  \left[
\theta_{a}\right]  ^{t}-\left[  \psi_{R}\right]  ^{\ast}\left[  \psi
_{L}\right]  \left[  \theta_{a}\right]  ^{t}\right)  $

$=Tr\left(  \left[  \psi_{R}\right]  ^{\ast}\left[  \psi_{L}\right]  \left[
\theta_{a}\right]  ^{t}-\left[  \psi_{L}\right]  ^{\ast}\left[  \psi
_{R}\right]  \left[  \theta_{a}\right]  ^{t}\right)  \in%
\mathbb{R}
$

So let be :%

\begin{equation}
\left\langle \psi,\left[  \gamma_{0}\gamma^{r}\right]  \left[  \psi\right]
\left[  \theta_{a}\right]  ^{t}\right\rangle \mathbf{=i\rho}_{a}%
\mathbf{=iTr}\left(  \left[  \psi_{R}\right]  ^{\ast}\left[  \psi_{L}\right]
\left[  \theta_{a}\right]  ^{t}-\left[  \psi_{L}\right]  ^{\ast}\left[
\psi_{R}\right]  \left[  \theta_{a}\right]  ^{t}\right)
\end{equation}

\paragraph{3)}

The currents read :

$Y_{AR}^{\alpha a}=Na_{D}V^{\alpha}\rho_{a}+2a_{F}\sum_{\beta}%
\operatorname{Re}\left[  \grave{A}_{\beta},%
\mathcal{F}%
_{A}^{\alpha\beta}\right]  ^{a}$

$Y_{AI}^{\alpha a}=-Na_{I}\sum_{r}O_{r}^{\alpha}\left[  \mu_{R}-\mu
_{L}\right]  _{a}^{r}+2a_{F}\sum_{\beta}\operatorname{Im}\left[  \grave
{A}_{\beta},%
\mathcal{F}%
_{A}^{\alpha\beta}\right]  ^{a}$

We can combine both currents $Y_{A}^{a}=\sum_{\alpha}\left(  Y_{AR}^{a\alpha
}+iY_{AI}^{a\alpha}\right)  \partial_{\alpha}$\ \ and denote :%

\begin{equation}
Y_{A}^{a}=N\left(  a_{D}\sum_{\alpha}V^{\alpha}\rho_{a}\partial_{\alpha
}-ia_{I}\sum_{r}\left[  \mu_{A}\right]  _{a}^{r}\partial_{r}\right)
+2a_{F}\sum_{\alpha\beta}\left[  \grave{A}_{\beta},%
\mathcal{F}%
_{A}^{\alpha\beta}\right]  ^{a}\partial_{\alpha}\label{E79}%
\end{equation}

$\rho_{a}$ is similar to a "charge" of the particle, and $\mu_{A}$ \ to a
"magnetic moment". The a index is related to the kind of force. These
quantities are functions on M and therefore invariant by a change of gauge.
The matter part of the Noether current is :

$N\sum_{\alpha}\left(  a_{D}V^{\alpha}\rho_{a}-ia_{I}\sum_{r}O_{r}^{\alpha
}\left[  \mu_{A}\right]  _{a}^{r}\right)  \partial_{\alpha}$

The Noether current is a geometric quantity.\ As we see in the formula above
the pertinent geometric quantities, pertaining to the particle, are the
4-vectors :

- the "charge current" linked with the velocity : $\rho_{a}\overrightarrow
{V}=\sum_{\alpha}V^{\alpha}\rho_{a}\partial_{\alpha}$

- the "magnetic moment" linked with the tetrad : $\overrightarrow{\mu}%
_{a}=\sum_{r}\left[  \mu_{A}\right]  _{a}^{r}\partial_{r}$

Remark : these equations do no involve gravitation.\ They can be seen as
relating the fields to the sources.\ The gravitational field comes back
through the trajectories of the particles.

\subsubsection{Superpotential}

Implementing the previous definitions :

$\Pi_{AR}^{a}=4a_{F}\left(  \det O^{\prime}\right)  \sum_{\lambda<\mu}%
\sum_{\alpha<\beta,a}\epsilon\left(  \lambda,\mu,\alpha,\beta\right)
\operatorname{Re}%
\mathcal{F}%
_{A}^{a,\alpha\beta}dx^{\lambda}\wedge dx^{\mu}$

$\Pi_{AI}^{a}=4a_{F}\left(  \det O^{\prime}\right)  \sum_{\lambda<\mu}%
\sum_{\alpha<\beta,a}\epsilon\left(  \lambda,\mu,\alpha,\beta\right)
\operatorname{Im}%
\mathcal{F}%
_{A}^{a,\alpha\beta}dx^{\lambda}\wedge dx^{\mu}$

Let us introduce :

$\Pi_{A}^{a}=\Pi_{AR}^{a}+i\Pi_{AI}^{a}=4\left(  \det O^{\prime}\right)
a_{F}\sum_{\lambda<\mu}\sum_{\alpha<\beta,a}\epsilon\left(  \lambda,\mu
,\alpha,\beta\right)
\mathcal{F}%
_{A}^{a\alpha\beta}dx^{\lambda}\wedge dx^{\mu}$

$\Pi_{A}^{a}=-4\left(  \det O^{\prime}\right)  a_{F}\{%
\mathcal{F}%
^{a,01}dx^{3}\wedge dx^{2}+%
\mathcal{F}%
^{a,02}dx^{1}\wedge dx^{3}+%
\mathcal{F}%
^{a,03}dx^{2}\wedge dx^{1}$

$+%
\mathcal{F}%
^{a,32}dx^{0}\wedge dx^{1}+%
\mathcal{F}%
^{a,13}dx^{0}\wedge dx^{2}+%
\mathcal{F}%
^{a,21}dx^{0}\wedge dx^{3}\}$

We can recognize the Hodge dual of the conjugate of the curvature form (cf
\ref{E63}):

So we can write :%

\begin{equation}
\Pi_{A}^{a}=4a_{F}\left(  \ast\overline{%
\mathcal{F}%
}_{A}^{a}\right) \label{E80}%
\end{equation}

\subsubsection{Equations}

\paragraph{1)}

The equations \ref{E57} take here the simple form :

$\varpi_{4}\left(  Y_{AR}^{a}\right)  =\frac{1}{2}d\Pi_{AR}^{a}%
=\operatorname{Re}\varpi_{4}\left(  Y_{A}^{a}\right)  =\frac{1}{2}%
d\operatorname{Re}\Pi_{AR}^{a};\varpi_{4}\left(  Y_{AI}^{a}\right)  =\frac
{1}{2}d\Pi_{AI}^{a}=\operatorname{Im}\varpi_{4}\left(  Y_{AI}^{a}\right)
=\frac{1}{2}d\operatorname{Im}\Pi_{AI}^{a}$%

\begin{equation}
\varpi_{4}\left(  Y_{A}^{a}\right)  =2a_{F}d\left(  \ast\overline{%
\mathcal{F}%
}_{A}^{a}\right) \label{E81}%
\end{equation}

\paragraph{2)}

The equations read :

$\forall a,\alpha:Y_{AR}^{\alpha a}=-2\frac{1}{\det O^{\prime}}\sum_{\beta
}\partial_{\beta}\left(  \frac{dL_{F}\left(  \det O^{\prime}\right)
}{d\operatorname{Re}%
\mathcal{F}%
_{A,\alpha\beta}^{a}}\right)  =-2\frac{1}{\det O^{\prime}}a_{F}\sum_{\beta
}\operatorname{Re}\partial_{\beta}\left(
\mathcal{F}%
_{A}^{a\alpha\beta}\left(  \det O^{\prime}\right)  \right)  $

$\forall a,\alpha:Y_{AI}^{\alpha a}=-2\frac{1}{\det O^{\prime}}\sum_{\beta
}\partial_{\beta}\left(  \frac{\partial L_{F}\left(  \det O^{\prime}\right)
}{\partial\operatorname{Im}%
\mathcal{F}%
_{A,\alpha\beta}^{a}}\right)  =-2\frac{1}{\det O^{\prime}}a_{F}\sum_{\beta
}\operatorname{Im}\partial_{\beta}\left(
\mathcal{F}%
_{A}^{a\alpha\beta}\left(  \det O^{\prime}\right)  \right)  $

That we can write :

$\forall a,\alpha:Y_{A}^{\alpha a}=-2\frac{1}{\det O^{\prime}}a_{F}\sum
_{\beta}\partial_{\beta}\left(
\mathcal{F}%
_{A}^{a\alpha\beta}\left(  \det O^{\prime}\right)  \right)  $

Or :%

\begin{equation}
\label{E82}
\end{equation}

$\forall a,\alpha:N\left(  a_{D}V^{\alpha}\rho_{a}-ia_{I}\sum_{r}O_{r}%
^{\alpha}\left[  \mu_{A}\right]  _{a}^{r}\right)  \det O^{\prime}$

$\qquad=-2a_{F}\sum_{\beta}\left(  \left[  \grave{A}_{\beta},%
\mathcal{F}%
_{A}^{\alpha\beta}\det O^{\prime}\right]  ^{a}+\partial_{\beta}\left(
\mathcal{F}%
_{A}^{a\alpha\beta}\left(  \det O^{\prime}\right)  \right)  \right)  $

\bigskip

\paragraph{3)}

This equation can also be written as :

$\varpi_{4}\left(  Y_{A}^{a}\right)  =2a_{F}d\left(  \ast%
\mathcal{F}%
_{A}^{a}\right)  $

$=N\varpi_{4}\left(  a_{D}\sum_{\alpha}V^{\alpha}\rho_{a}\partial_{\alpha
}-ia_{I}\sum_{r}\left[  \mu_{R}-\mu_{L}\right]  _{a}^{r}\partial_{r}\right)
+2a_{F}\varpi_{4}\left(  \sum_{\beta}\left[  \grave{A}_{\beta},%
\mathcal{F}%
_{A}^{\alpha\beta}\right]  ^{a}\partial_{\alpha}\right)  $

$N\varpi_{4}\left(  a_{D}\sum_{\alpha}V^{\alpha}\rho^{a}\partial_{\alpha
}-ia_{I}\sum_{r}\left[  \mu_{R}-\mu_{L}\right]  _{a}^{r}\partial_{r}\right)  $

$=2a_{F}\left(  d\left(  \ast\overline{%
\mathcal{F}%
}_{A}^{a}\right)  -\varpi_{4}\left(  \sum_{\beta}\left[  \grave{A}_{\beta},%
\mathcal{F}%
_{A}^{\alpha\beta}\right]  ^{a}\partial_{\alpha}\right)  \right)  $

and :

$\varpi_{4}\left(  \sum_{\beta}\left[  \grave{A}_{\beta},%
\mathcal{F}%
_{A}^{\alpha\beta}\right]  ^{a}\partial_{\alpha}\right)  $

$=\sum_{\alpha}\left(  -1\right)  ^{\alpha+1}\sum_{\beta}\left[  \grave
{A}_{\beta},%
\mathcal{F}%
_{A}^{\alpha\beta}\right]  ^{a}\det O^{\prime}dx^{0}\wedge..\widehat
{dx^{\alpha}}..\wedge dx^{3}$

$=\{-\sum_{\beta}\left[  \grave{A}_{\beta},%
\mathcal{F}%
_{A}^{0\beta}\right]  ^{a}dx^{1}\wedge dx^{2}\wedge dx^{3}+\sum_{\beta}\left[
\grave{A}_{\beta},%
\mathcal{F}%
_{A}^{1\beta}\right]  ^{a}dx^{0}\wedge dx^{2}\wedge dx^{3}$

$-\sum_{\beta}\left[  \grave{A}_{\beta},%
\mathcal{F}%
_{A}^{2\beta}\right]  ^{a}dx^{0}\wedge dx^{1}\wedge dx^{3}+\sum_{\beta}\left[
\grave{A}_{\beta},%
\mathcal{F}%
_{A}^{3\beta}\right]  ^{a}dx^{0}\wedge dx^{1}\wedge dx^{2}\}\left(  \det
O^{\prime}\right)  $

On the other hand :

$\left[  \grave{A},\ast\overline{%
\mathcal{F}%
}_{A}\right]  ^{a}=\sum_{\lambda<\mu}\sum_{\beta}\left[  \grave{A}_{\beta
},\left(  \ast\overline{%
\mathcal{F}%
}_{A}\right)  _{\lambda\mu}\right]  ^{a}dx^{\beta}\wedge dx^{\lambda}\wedge
dx^{\mu}$

with :

$\ast\overline{%
\mathcal{F}%
}_{A}^{a}=-\{%
\mathcal{F}%
^{a,01}dx^{3}\wedge dx^{2}+%
\mathcal{F}%
^{a,02}dx^{1}\wedge dx^{3}+%
\mathcal{F}%
^{a,03}dx^{2}\wedge dx^{1}$

$+%
\mathcal{F}%
^{a,32}dx^{0}\wedge dx^{1}+%
\mathcal{F}%
^{a,13}dx^{0}\wedge dx^{2}+%
\mathcal{F}%
^{a,21}dx^{0}\wedge dx^{3}\}$

$\left[  \grave{A},\ast\overline{%
\mathcal{F}%
}_{A}\right]  ^{a}=$

$-\sum_{\beta}\left[  \grave{A}_{\beta},%
\mathcal{F}%
^{01}\right]  ^{a}dx^{\beta}\wedge dx^{3}\wedge dx^{2}+\left[  \grave
{A}_{\beta},%
\mathcal{F}%
^{02}\right]  ^{a}dx^{\beta}\wedge dx^{1}\wedge dx^{3}+\left[  \grave
{A}_{\beta},%
\mathcal{F}%
^{03}\right]  ^{a}dx^{\beta}\wedge dx^{2}\wedge dx^{1}$

$+\left[  \grave{A}_{\beta},%
\mathcal{F}%
^{32}\right]  ^{a}dx^{\beta}\wedge dx^{0}\wedge dx^{1}+\left[  \grave
{A}_{\beta},%
\mathcal{F}%
^{13}\right]  ^{a}dx^{\beta}\wedge dx^{0}\wedge dx^{2}+\left[  \grave
{A}_{\beta},%
\mathcal{F}%
^{21}\right]  ^{a}dx^{\beta}\wedge dx^{0}\wedge dx^{3}$

$=-\{-\left(  \left[  \grave{A}_{1},%
\mathcal{F}%
^{01}\right]  ^{a}+\left[  \grave{A}_{2},%
\mathcal{F}%
^{02}\right]  ^{a}+\left[  \grave{A}_{3},%
\mathcal{F}%
^{03}\right]  ^{a}\right)  dx^{1}\wedge dx^{2}\wedge dx^{3}$

$+\left(  \left[  \grave{A}_{0},%
\mathcal{F}%
^{10}\right]  ^{a}+\left[  \grave{A}_{3},%
\mathcal{F}%
^{13}\right]  ^{a}+\left[  \grave{A}_{2},%
\mathcal{F}%
^{12}\right]  ^{a}\right)  dx^{0}\wedge dx^{2}\wedge dx^{3}\}$

$-\left(  \left[  \grave{A}_{0},%
\mathcal{F}%
^{20}\right]  ^{a}+\left[  \grave{A}_{3},%
\mathcal{F}%
^{23}\right]  ^{a}+\left[  \grave{A}_{1},%
\mathcal{F}%
^{21}\right]  ^{a}\right)  dx^{0}\wedge dx^{1}\wedge dx^{3}$

$+\left(  \left[  \grave{A}_{0},%
\mathcal{F}%
^{30}\right]  ^{a}+\left[  \grave{A}_{2},%
\mathcal{F}%
^{32}\right]  ^{a}+\left[  \grave{A}_{1},%
\mathcal{F}%
^{31}\right]  ^{a}\right)  dx^{0}\wedge dx^{1}\wedge dx^{2}\}$

$=-\sum_{\alpha}\left(  -1\right)  ^{\alpha+1}\left(  \sum_{\beta}\left[
\grave{A}_{\beta},%
\mathcal{F}%
_{A}^{\alpha\beta}\right]  ^{a}\right)  dx^{0}\wedge...\wedge\widehat
{dx^{\alpha}}\wedge...\wedge dx^{3}$

So : $\left[  \grave{A},\ast\overline{%
\mathcal{F}%
}_{A}\right]  ^{a}=-\varpi_{4}\left(  \sum_{\beta}\left[  \grave{A}_{\beta},%
\mathcal{F}%
_{A}^{\alpha\beta}\right]  ^{a}\partial_{\alpha}\right)  $

Thus : $N\varpi_{4}\left(  a_{D}\sum_{\alpha}V^{\alpha}\rho^{a}\partial
_{\alpha}-ia_{I}\sum_{r}\left[  \mu_{A}\right]  _{a}^{r}\partial_{r}\right)
=d\left(  \ast\overline{%
\mathcal{F}%
}_{A}^{a}\right)  +\left[  \grave{A},\ast\overline{%
\mathcal{F}%
}_{A}\right]  ^{a}$ where we recognize the exterior covariant derivative.

And we have the geometric form of the equation :%

\begin{equation}
\mathbf{N\varpi}_{4}\left(  a_{D}\rho_{a}\overrightarrow{V}-ia_{I}%
\overrightarrow{\mu}_{a}\right)  \mathbf{=\nabla}_{e}\mathbf{\ast}\overline{%
\mathcal{F}%
}_{A}^{a}\label{E83}%
\end{equation}

On the right hand side we have the moments, evaluated at a point through f,
and at the right hand side the curvature form evaluated without f. Notice that
the tensor $\psi$ does not appear per se.

We see that the real part of the field acts through the "charge current" and
the imaginary part through the "magnetic moment". So one can guess that the
first impacts the velocity and the second the "rotation" of the particle.

\subsection{Energy-momentum tensor}

According to equations established in the previous part, the energy-momentum
tensor $\delta_{\beta}^{\alpha}L$ can be expressed in two different equivalent
ways from the moments and the force fields. We have a general, simple,
equation which links all the force fields connections. It is then possible to
get a simple equation for the scalar curvature, which does not require the
explicit computation of the gravitational 2-form $%
\mathcal{F}%
_{G}.$

\subsubsection{Moments}

Now we have defined all the moments that we needed.\ As they are crucial in
all the calculations it is convenient to sum up here their definitions and
properties. They are all real scalar functions, invariant by gauge or chart changes.

\paragraph{1) The kinematic moments :}

\subparagraph{a) the "linear momentum" P :}

$P_{a}=\operatorname{Im}Tr\left(  \left[  \psi^{\ast}\right]  \left[
\gamma_{0}\right]  \left[  \kappa_{a}\right]  \left[  \psi\right]  \right)
=\operatorname{Im}\left\langle \psi,\left[  \kappa_{a}\right]  \left[
\psi\right]  \right\rangle $

$\left\langle \psi,\left[  \kappa_{a}\right]  \left[  \psi\right]
\right\rangle =Tr\left(  \left[  \psi^{\ast}\right]  \left[  \gamma
_{0}\right]  \left[  \kappa_{a}\right]  \left[  \psi\right]  \right)  =iP_{a}$

$a<4:P_{a}=-i\frac{1}{2}Tr\left(  \left[  \psi_{R}^{\ast}\right]  \left[
\sigma_{a}\right]  \left[  \psi_{L}\right]  -\left[  \psi_{L}^{\ast}\right]
\left[  \sigma_{a}\right]  \left[  \psi_{R}\right]  \right)  ;$

$a>3:P_{a}=-\frac{1}{2}Tr\left(  \left[  \psi_{R}^{\ast}\right]  \left[
\sigma_{a-3}\right]  \left[  \psi_{L}\right]  +\left[  \psi_{L}^{\ast}\right]
\left[  \sigma_{a-3}\right]  \left[  \psi_{R}\right]  \right)  $

It depends only on 3 complex scalars :

a=1,2,3 : $p_{a}=Tr\left(  \left[  \psi_{R}^{\ast}\right]  \left[  \sigma
_{a}\right]  \left[  \psi_{L}\right]  \right)  ;\overline{p}_{a}=Tr\left(
\left[  \psi_{L}^{\ast}\right]  \left[  \sigma_{a}\right]  \left[  \psi
_{R}\right]  \right)  $

a%
$<$%
4 :$P_{a}=-i\frac{1}{2}\left(  p^{a}-\overline{p}^{a}\right)  $

a%
$>$%
3 :$P_{a}=-\frac{1}{2}\left(  p^{a-3}+\overline{p}^{a-3}\right)  $

It is never null :

$\sum_{a=1}^{6}\left(  P_{a}\right)  ^{2}=\sum_{a=1}^{3}\left\vert Tr\left(
\left[  \psi_{R}^{\ast}\right]  \left[  \sigma_{a}\right]  \left[  \psi
_{L}\right]  \right)  \right\vert ^{2}>0$

The physical quantity is the tensor :$\sum_{a\alpha}V^{\alpha}P_{a}%
\partial_{\alpha}\otimes\overrightarrow{\kappa}_{a}$

\subparagraph{b) the angular momentum J :}

$J_{r}=-\frac{1}{2}Tr\left(  \eta^{rr}\psi_{R}^{\ast}\sigma_{r}\psi_{R}%
-\psi_{L}^{\ast}\sigma_{r}\psi_{L}\right)  $

It can be null :

$\sum_{k=0}^{3}J_{k}^{2}$

$=\frac{1}{4}\left(  \sum_{k=0}^{3}\left(  Tr\left(  \psi_{R}^{\ast}\sigma
_{k}\psi_{R}\right)  \right)  ^{2}+\left(  Tr\left(  \psi_{L}^{\ast}\sigma
_{k}\psi_{L}\right)  \right)  ^{2}-2\eta^{kk}Tr\left(  \psi_{R}^{\ast}%
\sigma_{k}\psi_{R}\right)  Tr\left(  \psi_{L}^{\ast}\sigma_{k}\psi_{L}\right)
\right)  $

$\left\langle \psi,\gamma^{r}\psi\right\rangle =Tr\left(  \left[  \psi\right]
^{\ast}\left[  \gamma_{0}\gamma^{r}\right]  \left[  \psi\right]  \right)
=iTr\left(  \eta^{rr}\psi_{R}^{\ast}\sigma_{r}\psi_{R}-\psi_{L}^{\ast}%
\sigma_{r}\psi_{L}\right)  =-2iJ_{r}$

$\operatorname{Im}\left\langle \psi,\left[  \gamma^{r}\right]  \left[
\kappa_{a}\right]  \psi\right\rangle =\operatorname{Im}Tr\left(  \left[
\psi\right]  ^{\ast}\left[  \gamma_{0}\gamma^{r}\right]  \left[  \kappa
_{a}\right]  \left[  \psi\right]  \right)  =\left(  \left[  J\right]  \left[
\widetilde{\kappa}_{a}\right]  \right)  _{r}$

The physical quantity is the tensor :

$\sum_{a,r}\operatorname{Im}Tr\left(  \left[  \psi\right]  ^{\ast}\left[
\gamma_{0}\gamma^{r}\right]  \left[  \kappa_{a}\right]  \left[  \psi\right]
\right)  \partial_{r}\otimes\overrightarrow{\kappa}_{a}=\sum_{a,r}\left(
\left[  J\right]  \left[  \widetilde{\kappa}_{a}\right]  \right)  _{r}%
\partial_{r}\otimes\overrightarrow{\kappa}_{a}$ where $\left[  J\right]  $ is
a 1x4 row matrix

\subparagraph{c) one can add the function :}

$\left\langle \psi,\psi\right\rangle =-2\operatorname{Im}Tr\left(  \left[
\psi_{R}\right]  ^{\ast}\left[  \psi_{L}\right]  \right)  =iTr\left(  \left[
\psi_{R}\right]  ^{\ast}\left[  \psi_{L}\right]  -\left[  \psi_{L}\right]
^{\ast}\left[  \psi_{R}\right]  \right)  $

\subparagraph{d)}

There is an important property of the partial derivative.\ 

As it is easy to check the derivation commutes with the trace operator

$Tr\left(  \partial_{\beta}\left(  \left[  \psi_{R}\right]  ^{\ast}\sigma
_{r}\left[  \psi_{R}\right]  \right)  \right)  =\sum_{j=1}^{m}\partial_{\beta
}\left(  \left[  \psi_{R}\right]  ^{\ast}\sigma_{r}\left[  \psi_{R}\right]
\right)  _{j}^{j}$

$=\partial_{\beta}\sum_{j=1}^{m}\left(  \left[  \psi_{R}\right]  ^{\ast}%
\sigma_{r}\left[  \psi_{R}\right]  \right)  _{j}^{j}=\partial_{\beta}Tr\left(
\left[  \psi_{R}\right]  ^{\ast}\sigma_{r}\left[  \psi_{R}\right]  \right)  $

b) so :

$\partial_{\beta}Tr\left(  \left[  \psi_{R}\right]  ^{\ast}\sigma_{r}\left[
\psi_{R}\right]  \right)  =Tr\left(  \partial_{\beta}\left(  \left[  \psi
_{R}\right]  ^{\ast}\sigma_{r}\left[  \psi_{R}\right]  \right)  \right)  $

$=Tr\left(  \left[  \partial_{\beta}\psi_{R}\right]  ^{\ast}\sigma_{r}\left[
\psi_{R}\right]  \right)  +Tr\left(  \left[  \psi_{R}\right]  ^{\ast}%
\sigma_{r}\left[  \partial_{\beta}\psi_{R}\right]  \right)  $

but :

$\overline{Tr\left(  \left[  \partial_{\beta}\psi_{R}\right]  ^{\ast}%
\sigma_{r}\left[  \psi_{R}\right]  \right)  }=Tr\left(  \left(  \left[
\partial_{\beta}\psi_{R}\right]  ^{\ast}\sigma_{r}\left[  \psi_{R}\right]
\right)  ^{\ast}\right)  =Tr\left(  \left[  \psi_{R}\right]  ^{\ast}\sigma
_{r}\left[  \partial_{\beta}\psi_{R}\right]  \right)  $

so :

$\partial_{\beta}Tr\left(  \left[  \psi_{R}\right]  ^{\ast}\sigma_{r}\left[
\psi_{R}\right]  \right)  $

$=Tr\left(  \left[  \psi_{R}\right]  ^{\ast}\sigma_{r}\left[  \partial_{\beta
}\psi_{R}\right]  \right)  +\overline{Tr\left(  \left[  \psi_{R}\right]
^{\ast}\sigma_{r}\left[  \partial_{\beta}\psi_{R}\right]  \right)  }$

$=2\operatorname{Re}Tr\left(  \left[  \psi_{R}\right]  ^{\ast}\sigma
_{r}\left[  \partial_{\beta}\psi_{R}\right]  \right)  $

$\Rightarrow\operatorname{Re}Tr\left(  \left[  \psi_{R}\right]  ^{\ast}%
\sigma_{r}\left[  \partial_{\beta}\psi_{R}\right]  \right)  =\frac{1}%
{2}Tr\left(  \partial_{\beta}\left(  \left[  \psi_{R}\right]  ^{\ast}%
\sigma_{r}\left[  \psi_{R}\right]  \right)  \right)  $

and we have the identity :

$\operatorname{Re}Tr\left(  \eta^{rr}\left[  \psi_{R}\right]  ^{\ast}%
\sigma_{r}\left[  \partial_{\beta}\psi_{R}\right]  -\left[  \psi_{L}\right]
^{\ast}\sigma_{r}\left[  \partial_{\beta}\psi_{L}\right]  \right)  $

$=\frac{1}{2}\partial_{\beta}\operatorname{Re}Tr\left(  \left(  \eta
^{rr}\left[  \psi_{R}\right]  ^{\ast}\sigma_{r}\left[  \psi_{R}\right]
-\left[  \psi_{L}\right]  ^{\ast}\sigma_{r}\left[  \psi_{L}\right]  \right)
\right)  $

$=\frac{1}{2}\partial_{\beta}Tr\left(  \left(  \eta^{rr}\left[  \psi
_{R}\right]  ^{\ast}\sigma_{r}\left[  \psi_{R}\right]  -\left[  \psi
_{L}\right]  ^{\ast}\sigma_{r}\left[  \psi_{L}\right]  \right)  \right)
=-\partial_{\beta}J_{r}$

And :

$\left\langle \psi,\gamma^{r}\partial_{\beta}\psi\right\rangle =Tr\left(
\left[  \psi\right]  ^{\ast}\left[  \gamma_{0}\gamma^{r}\right]  \left[
\partial_{\beta}\psi\right]  \right)  =iTr\left(  \eta^{rr}\left[  \psi
_{R}\right]  ^{\ast}\sigma_{r}\left[  \partial_{\beta}\psi_{R}\right]
-\left[  \psi_{L}\right]  ^{\ast}\sigma_{r}\left[  \partial_{\beta}\psi
_{L}\right]  \right)  $

So

$\operatorname{Im}\left\langle \psi,\gamma^{r}\partial_{\beta}\psi
\right\rangle =\operatorname{Im}Tr\left(  \left[  \psi\right]  ^{\ast}\left[
\gamma_{0}\gamma^{r}\right]  \left[  \partial_{\beta}\psi\right]  \right)  $

$=\operatorname{Re}Tr\left(  \eta^{rr}\left[  \psi_{R}\right]  ^{\ast}%
\sigma_{r}\left[  \partial_{\beta}\psi_{R}\right]  -\left[  \psi_{L}\right]
^{\ast}\sigma_{r}\left[  \partial_{\beta}\psi_{L}\right]  \right)
=-\partial_{\beta}J_{r}$

\paragraph{2) The force fields moments :}

\subparagraph{a) the "charges"}

$\rho_{a}=Tr\left(  \left[  \psi_{R}\right]  ^{\ast}\left[  \psi_{L}\right]
\left[  \theta_{a}\right]  ^{t}-\left[  \psi_{L}\right]  ^{\ast}\left[
\psi_{R}\right]  \left[  \theta_{a}\right]  ^{t}\right)  $

$\left\langle \psi,\left[  \psi\right]  \left[  \theta_{a}\right]
^{t}\right\rangle =Tr\left(  \left[  \psi^{\ast}\right]  \left[  \gamma
_{0}\right]  \left[  \psi\right]  \left[  \theta_{a}\right]  ^{t}\right)
=iTr\left(  \left[  \psi_{R}\right]  ^{\ast}\left[  \psi_{L}\right]  \left[
\theta_{a}\right]  ^{t}-\left[  \psi_{L}\right]  ^{\ast}\left[  \psi
_{R}\right]  \left[  \theta_{a}\right]  ^{t}\right)  =i\rho_{a}$

and the "charge current" : linked with the velocity : $\rho_{a}\overrightarrow
{V}=\sum_{\alpha}V^{\alpha}\rho_{a}\partial_{\alpha}$

\subparagraph{b) the magnetic moment :}

$\left[  \mu_{A}\right]  _{a}^{r}=-\left\langle \psi,\gamma^{r}\left[
\psi\right]  \left[  \theta_{a}\right]  ^{t}\right\rangle =-Tr\left(  \left[
\psi\right]  ^{\ast}\left[  \gamma_{0}\gamma^{r}\right]  \left[  \psi\right]
\left[  \theta_{a}\right]  ^{t}\right)  $

$=-iTr\left(  \eta^{rr}\left[  \psi_{R}\right]  ^{\ast}\sigma_{r}\left[
\psi_{R}\right]  \left[  \theta_{a}\right]  ^{t}-\left[  \psi_{L}\right]
^{\ast}\sigma_{r}\left[  \psi_{L}\right]  \left[  \theta_{a}\right]
^{t}\right)  $

and the "magnetic moment" : linked with the tetrad : $\overrightarrow{\mu}%
_{a}=\sum_{r}\left[  \mu_{A}\right]  _{a}^{r}\partial_{r}$

\paragraph{3)}

The state tensor is the sum of 2 right and left components : $\psi=\psi
_{R}+\psi_{L}.$\ If each of these components is decomposable : $\psi=\psi
_{R}\otimes\sigma_{R}+\psi_{L}\otimes\sigma_{L}$ one can write the matrix

$\left[  \psi\right]  =%
\begin{bmatrix}
\psi_{R}\\
\psi_{L}%
\end{bmatrix}
=%
\begin{bmatrix}
\left[  \phi_{R}\right]  \left[  \sigma_{R}\right] \\
\left[  \phi_{L}\right]  \left[  \sigma_{L}\right]
\end{bmatrix}
$

where $\left[  \phi_{R}\right]  ,\left[  \phi_{L}\right]  $ are 2x1 column
matrices and $\left[  \sigma_{R}\right]  ,\left[  \sigma_{L}\right]  $ are 1xm
row matrices. The previous formulas are simpler.

with any $\left[  \mu\right]  $ matrix : $Tr\left(  \left[  \psi_{1}^{\ast
}\right]  \left[  \mu\right]  \left[  \psi_{2}\right]  \right)  =Tr\left(
\left[  \sigma_{1}\right]  ^{\ast}\left[  \phi_{1}^{\ast}\right]  \left[
\mu\right]  \left[  \phi_{2}\right]  \left[  \sigma_{2}\right]  \right)
=\left(  \left[  \phi_{1}^{\ast}\right]  \left[  \mu\right]  \left[  \phi
_{2}\right]  \right)  Tr\left(  \left[  \sigma_{1}\right]  ^{\ast}\left[
\sigma_{2}\right]  \right)  =\left(  \left[  \phi_{1}^{\ast}\right]  \left[
\mu\right]  \left[  \phi_{2}\right]  \right)  \left(  \left[  \sigma
_{2}\right]  ^{t}\overline{\left[  \sigma_{1}\right]  }\right)  $

$Tr\left(  \left[  \psi_{1}^{\ast}\right]  \left[  \mu\right]  \left[
\psi_{2}\right]  \left[  \theta_{a}\right]  ^{t}\right)  =Tr\left(  \left[
\sigma_{1}\right]  ^{\ast}\left[  \phi_{1}^{\ast}\right]  \left[  \mu\right]
\left[  \phi_{2}\right]  \left[  \sigma_{2}\right]  \left[  \theta_{a}\right]
^{t}\right)  =\left(  \left[  \phi_{1}^{\ast}\right]  \left[  \mu\right]
\left[  \phi_{2}\right]  \right)  \left(  \left[  \sigma_{2}\right]  \left[
\theta_{a}\right]  ^{t}\left[  \sigma_{1}\right]  ^{\ast}\right)  $

a) $P_{a}=\operatorname{Im}Tr\left(  \left[  \psi^{\ast}\right]  \left[
\gamma_{0}\right]  \left[  \kappa_{a}\right]  \left[  \psi\right]  \right)  :$

$a<4:P_{a}=\operatorname{Im}\left(  \left(  \left[  \phi_{R}^{\ast}\right]
\left[  \sigma_{a}\right]  \left[  \phi_{L}\right]  \right)  \left(  \left[
\sigma_{L}\right]  ^{t}\overline{\left[  \sigma_{R}\right]  }\right)  \right)
$

$a>3:P_{a}=-\operatorname{Re}\left(  \left(  \left[  \phi_{R}^{\ast}\right]
\left[  \sigma_{a-3}\right]  \left[  \phi_{L}\right]  \right)  \left(  \left[
\sigma_{L}\right]  ^{t}\overline{\left[  \sigma_{R}\right]  }\right)  \right)
$

b) $J_{r}=-\frac{1}{2}Tr\left(  \eta^{rr}\psi_{R}^{\ast}\sigma_{r}\psi
_{R}-\psi_{L}^{\ast}\sigma_{r}\psi_{L}\right)  $

$=-\frac{1}{2}\left(  \eta^{rr}\left(  \left[  \phi_{R}^{\ast}\right]  \left[
\sigma_{a}\right]  \left[  \phi_{R}\right]  \right)  \left(  \left[
\sigma_{R}\right]  ^{t}\overline{\left[  \sigma_{R}\right]  }\right)  -\left(
\left[  \phi_{L}^{\ast}\right]  \left[  \sigma_{a}\right]  \left[  \phi
_{L}\right]  \right)  \left(  \left[  \sigma_{L}\right]  ^{t}\overline{\left[
\sigma_{L}\right]  }\right)  \right)  $

c) $\left\langle \psi,\psi\right\rangle =-2\operatorname{Im}Tr\left(  \left[
\psi_{R}\right]  ^{\ast}\left[  \psi_{L}\right]  \right)  =-2\operatorname{Im}%
\left(  \left(  \left[  \phi_{R}\right]  ^{\ast}\left[  \phi_{L}\right]
\right)  \left(  \left[  \sigma_{L}\right]  ^{t}\overline{\left[  \sigma
_{R}\right]  }\right)  \right)  $

d) $\rho^{a}=Tr\left(  \left[  \psi_{R}\right]  ^{\ast}\left[  \psi
_{L}\right]  \left[  \theta_{a}\right]  ^{t}-\left[  \psi_{L}\right]  ^{\ast
}\left[  \psi_{R}\right]  \left[  \theta_{a}\right]  ^{t}\right)  $

$=\left(  \left[  \phi_{R}^{\ast}\right]  \left[  \phi_{L}\right]  \right)
\left(  \left[  \sigma_{L}\right]  \left[  \theta_{a}\right]  ^{t}\left[
\sigma_{R}\right]  ^{\ast}\right)  -\left(  \left[  \phi_{L}^{\ast}\right]
\left[  \phi_{R}\right]  \right)  \left(  \left[  \sigma_{R}\right]  \left[
\theta_{a}\right]  ^{t}\left[  \sigma_{L}\right]  ^{\ast}\right)  $

$\overline{\left(  \left[  \sigma_{L}\right]  \left[  \theta_{a}\right]
^{t}\left[  \sigma_{R}\right]  ^{\ast}\right)  }=\left(  \left[  \sigma
_{L}\right]  \left[  \theta_{a}\right]  ^{t}\left[  \sigma_{R}\right]  ^{\ast
}\right)  ^{\ast}=\left(  \left[  \sigma_{R}\right]  ^{\ast}\overline{\left[
\theta_{a}\right]  }\left[  \sigma_{L}\right]  \right)  =-\left(  \left[
\sigma_{R}\right]  ^{\ast}\left[  \theta_{a}\right]  ^{t}\left[  \sigma
_{L}\right]  \right)  $

$\overline{\left(  \left[  \phi_{R}^{\ast}\right]  \left[  \phi_{L}\right]
\right)  }=\left(  \left[  \phi_{R}^{\ast}\right]  \left[  \phi_{L}\right]
\right)  ^{\ast}=\left(  \left[  \phi_{L}^{\ast}\right]  \left[  \phi
_{R}\right]  \right)  $

$\rho^{a}=2\operatorname{Re}\left(  \left(  \left[  \phi_{R}^{\ast}\right]
\left[  \phi_{L}\right]  \right)  \left(  \left[  \sigma_{L}\right]  \left[
\theta_{a}\right]  ^{t}\left[  \sigma_{R}\right]  ^{\ast}\right)  \right)  $

e) $\left[  \mu_{A}\right]  _{a}^{r}=-i\left(  \eta^{rr}\left[  \phi_{R}%
^{\ast}\right]  \sigma_{r}\left[  \phi_{R}\right]  \left(  \left[  \sigma
_{R}\right]  \left[  \theta_{a}\right]  ^{t}\left[  \sigma_{R}\right]  ^{\ast
}\right)  -\left[  \psi_{L}\right]  ^{\ast}\sigma_{r}\left[  \psi_{L}\right]
\left[  \theta_{a}\right]  ^{t}\left(  \left[  \sigma_{L}\right]  \left[
\theta_{a}\right]  ^{t}\left[  \sigma_{L}\right]  ^{\ast}\right)  \right)  $

\paragraph{4)}

The matter lagrangian can be expressed with respect to the moments :

a) $\operatorname{Im}\left\langle \psi,\nabla_{\alpha}\psi\right\rangle
=\operatorname{Im}\left\langle \psi,\partial_{\alpha}\psi\right\rangle
+\sum_{a}G_{\alpha}^{a}\operatorname{Im}\left\langle \psi,\left[  \kappa
_{a}\right]  \left[  \psi\right]  \right\rangle +\operatorname{Im}\grave
{A}_{\alpha}^{a}\left\langle \psi,\left[  \psi\right]  \left[  \theta
_{a}\right]  ^{t}\right\rangle $

$\operatorname{Im}\left\langle \psi,\nabla_{\alpha}\psi\right\rangle
=\operatorname{Im}\left\langle \psi,\partial_{\alpha}\psi\right\rangle
+\sum_{a}G_{\alpha}^{a}P_{a}+\rho_{a}\operatorname{Re}\grave{A}_{\alpha}^{a}$

b) $\operatorname{Im}\left\langle \psi,\gamma^{r}\nabla_{\alpha}%
\psi\right\rangle =\operatorname{Im}\left\langle \psi,\gamma^{r}%
\partial_{\alpha}\psi\right\rangle +\sum_{a}G_{\alpha}^{a}\operatorname{Im}%
\left\langle \psi,\gamma^{r}\left[  \kappa_{a}\right]  \left[  \psi\right]
\right\rangle +\operatorname{Im}\grave{A}_{\alpha}^{a}\left\langle \psi
,\gamma^{r}\left[  \psi\right]  \left[  \theta_{a}\right]  ^{t}\right\rangle $

$\operatorname{Im}\left\langle \psi,\gamma^{r}\nabla_{\alpha}\psi\right\rangle
=-\partial_{\alpha}J_{r}+\sum_{a}G_{\alpha}^{a}\left(  \left[  J\right]
\left[  \widetilde{\kappa}_{a}\right]  \right)  _{r}-\left[  \mu_{R}-\mu
_{L}\right]  _{a}^{r}\operatorname{Im}\grave{A}_{\alpha}^{a}$

c) $\operatorname{Im}\left\langle \psi,D_{M}^{\alpha}\nabla_{\alpha}%
\psi\right\rangle =\operatorname{Im}\left\langle \psi,\left(  \sum_{r}%
a_{I}O_{r}^{\alpha}\left[  \gamma^{r}\right]  +\sum_{\alpha}V^{\alpha}%
a_{D}I\right)  \nabla_{\alpha}\psi\right\rangle $

$\operatorname{Im}\left\langle \psi,D_{M}^{\alpha}\nabla_{\alpha}%
\psi\right\rangle =a_{I}\left(  -\sum_{\alpha,r}O_{r}^{\alpha}\partial
_{\alpha}J_{r}+\sum_{ar}G_{r}^{a}\left(  \left[  J\right]  \left[
\widetilde{\kappa}_{a}\right]  \right)  _{r}-\left[  \mu_{R}-\mu_{L}\right]
_{a}^{r}\operatorname{Im}\grave{A}_{r}^{a}\right)  +a_{D}\sum_{\alpha
}V^{\alpha}\left(  \operatorname{Im}\left\langle \psi,\partial_{\alpha}%
\psi\right\rangle +\sum_{a}\left(  G_{\alpha}^{a}P_{a}+\rho_{a}%
\operatorname{Re}\grave{A}_{\alpha}^{a}\right)  \right)  $

d) $L_{M}=a_{M}\left\langle \psi,\psi\right\rangle +a_{I}\left(  -\sum
_{\alpha,r}O_{r}^{\alpha}\partial_{\alpha}J_{r}+\sum_{ar}G_{r}^{a}\left(
\left[  J\right]  \left[  \widetilde{\kappa}_{a}\right]  \right)  _{r}-\left[
\mu_{R}-\mu_{L}\right]  _{a}^{r}\operatorname{Im}\grave{A}_{r}^{a}\right)
+a_{D}\sum_{\alpha}V^{\alpha}\left(  \operatorname{Im}\left\langle
\psi,\partial_{\alpha}\psi\right\rangle +\sum_{a}\left(  G_{\alpha}^{a}%
P_{a}+\rho_{a}\operatorname{Re}\grave{A}_{\alpha}^{a}\right)  \right)  $

Notice that when the sections are composed with f then the full derivatives
must be used :

$\operatorname{Im}\left\langle \psi,\nabla_{\alpha}\psi\right\rangle
=\operatorname{Im}\left\langle \psi,\frac{d\psi}{d\xi^{\alpha}}\right\rangle
+\sum_{a}G_{\alpha}^{a}P_{a}+\rho_{a}\operatorname{Re}\grave{A}_{\alpha}^{a}$

$\operatorname{Im}\left\langle \psi,\gamma^{r}\nabla_{\alpha}\psi\right\rangle
=-\frac{dJ_{r}}{d\xi^{\alpha}}+\sum_{a}G_{\alpha}^{a}\left(  \left[  J\right]
\left[  \widetilde{\kappa}_{a}\right]  \right)  _{r}-\left[  \mu_{R}-\mu
_{L}\right]  _{a}^{r}\operatorname{Im}\grave{A}_{\alpha}^{a}$

\subsubsection{The energy-momentum tensor}

Back to the energy momentum tensor. We have two ways to compute $\delta
_{\beta}^{\alpha}L$ which will be both useful.

\paragraph{1)}

Equation \ref{E59c} reads :

$\delta_{\beta}^{\alpha}L=-\frac{dNL_{M}}{dV^{\beta}}V^{\alpha}+\sum
_{i,j}\left(  \frac{dNL_{M}}{d\operatorname{Re}\nabla_{\alpha}\psi^{ij}%
}\operatorname{Re}\partial_{\beta}\psi^{ij}+\frac{dNL_{M}}{d\operatorname{Im}%
\nabla_{\alpha}\psi^{ij}}\operatorname{Im}\partial_{\beta}\psi^{ij}\right)  $

$+2\sum_{a\gamma}\left(  \frac{dL_{F}}{d%
\mathcal{F}%
_{G\alpha\gamma}^{a}}\partial_{\beta}G_{\gamma}^{a}+\frac{dL_{F}%
}{d\operatorname{Re}%
\mathcal{F}%
_{A,\alpha\gamma}^{a}}\operatorname{Re}\partial_{\beta}\grave{A}_{\gamma}%
^{a}+\frac{dL_{F}}{d\operatorname{Im}%
\mathcal{F}%
_{A,\alpha\gamma}^{a}}\operatorname{Im}\partial_{\beta}\grave{A}_{\gamma}%
^{a}\right)  +\sum_{i\gamma}\frac{dL}{d\partial_{\alpha}O_{\gamma}^{\prime i}%
}\partial_{\beta}O_{\gamma}^{\prime i}$

$\frac{dL_{M}}{d\operatorname{Re}\nabla_{\alpha}\psi^{ij}}=N\operatorname{Im}%
\left(  \left[  \psi\right]  ^{\ast}\left[  \gamma_{0}\right]  \left[
D_{M}^{\alpha}\right]  \right)  _{i}^{j};\frac{dL_{M}}{d\operatorname{Im}%
\nabla_{\alpha}\psi^{ij}}=N\operatorname{Re}\left(  \left[  \psi\right]
^{\ast}\left[  \gamma_{0}\right]  \left[  D_{M}^{\alpha}\right]  \right)
_{i}^{j}$

$\frac{dL_{F}}{d%
\mathcal{F}%
_{G\alpha\gamma}^{a}}=a_{G}\left(  O_{p_{a}}^{\gamma}O_{q_{a}}^{\alpha
}-O_{q_{a}}^{\gamma}O_{p_{a}}^{\alpha}\right)  ;\frac{dL_{F}}%
{d\operatorname{Re}%
\mathcal{F}%
_{A,\alpha\gamma}^{a}}=a_{F}\operatorname{Re}%
\mathcal{F}%
_{A}^{a,\alpha\gamma};\frac{dL_{F}}{d\operatorname{Im}%
\mathcal{F}%
_{A,\alpha\gamma}^{a}}=a_{F}\operatorname{Im}%
\mathcal{F}%
_{A}^{a,\alpha\gamma}$

$\frac{dL}{d\partial_{\alpha}O_{\gamma}^{\prime i}}=0$

$\frac{dNL_{M}}{dV^{\beta}}=a_{D}N\operatorname{Im}\left\langle \psi
,\nabla_{\beta}\psi\right\rangle $

$\delta_{\beta}^{\alpha}L=-a_{D}NV^{\alpha}\operatorname{Im}\left\langle
\psi,\nabla_{\beta}\psi\right\rangle +\sum_{i,j}\operatorname{Im}\left(
\left[  \psi\right]  ^{\ast}\left[  \gamma_{0}\right]  \left[  D_{M}^{\alpha
}\right]  \right)  _{i}^{j}\operatorname{Re}\partial_{\beta}\psi^{ij}$

$+\operatorname{Re}\left(  \left[  \psi\right]  ^{\ast}\left[  \gamma
_{0}\right]  \left[  D_{M}^{\alpha}\right]  \right)  _{i}^{j}\operatorname{Im}%
\partial_{\beta}\psi^{ij}+2\sum_{a,\gamma}a_{G}\left(  O_{p_{b}}^{\gamma
}O_{q_{b}}^{\alpha}-O_{q_{b}}^{\gamma}O_{p_{b}}^{\alpha}\right)
\partial_{\beta}G_{\gamma}^{a}$

$+a_{F}\operatorname{Re}%
\mathcal{F}%
_{A}^{a,\alpha\gamma}\operatorname{Re}\partial_{\beta}\grave{A}_{\gamma}%
^{a}+a_{F}\operatorname{Im}%
\mathcal{F}%
_{A}^{a,\alpha\gamma}\operatorname{Im}\partial_{\beta}\grave{A}_{\gamma}^{a}$

$\delta_{\beta}^{\alpha}L=-a_{D}NV^{\alpha}\operatorname{Im}\left\langle
\psi,\nabla_{\beta}\psi\right\rangle +N\operatorname{Im}\left\langle
\psi,\left[  \gamma_{0}\right]  \left[  D_{M}^{\alpha}\right]  \left[
\partial_{\beta}\psi\right]  \right\rangle $

$+2\sum_{a,\gamma}a_{G}\left(  O_{p_{b}}^{\gamma}O_{q_{b}}^{\alpha}-O_{q_{b}%
}^{\gamma}O_{p_{b}}^{\alpha}\right)  \partial_{\beta}G_{\gamma}^{a}%
+a_{F}\left(  \operatorname{Re}%
\mathcal{F}%
_{A}^{a,\alpha\gamma}\overline{\partial_{\beta}\grave{A}_{\gamma}^{a}}\right)
$

So the equation becomes :%

\begin{equation}
\end{equation}

$\delta_{\beta}^{\alpha}L=-N\left(  a_{I}\sum_{r}O_{r}^{\alpha}\frac{dJ_{r}%
}{d\xi^{\alpha}}+a_{D}V^{\alpha}\sum_{a}\left(  G_{\beta}^{a}P_{a}+\rho
^{a}\operatorname{Re}\grave{A}_{\beta}^{a}\right)  \right)  $

$\qquad+2\sum_{a,\gamma}\left(  a_{G}\left(  O_{p_{a}}^{\gamma}O_{q_{a}%
}^{\alpha}-O_{q_{a}}^{\gamma}O_{p_{a}}^{\alpha}\right)  \partial_{\beta
}G_{\gamma}^{a}+a_{F}\operatorname{Re}\left(  \partial_{\beta}\grave
{A}_{\gamma},%
\mathcal{F}%
_{A}^{\alpha\gamma}\right)  \right)  $

\bigskip

\paragraph{2)}

We can get a more convenient equation.\ We go back to the equation \ref{E48}
which reads here :

$\forall\alpha,\beta:0=\sum_{i,}\frac{dL}{dO_{\alpha}^{\prime i}}O_{\beta
}^{\prime i}+\delta_{\beta}^{\alpha}L$

expressed with respect to O :

$\frac{dL}{dO_{\alpha}^{\prime i}}=\sum_{j\beta}\frac{dL}{dO_{j}^{\beta}}%
\frac{dO_{j}^{\beta}}{dO_{\alpha}^{\prime i}}=-\sum_{j\beta}O_{i}^{\beta}%
O_{j}^{\alpha}\frac{dL}{dO_{j}^{\beta}}$

a) Let us compute the derivatives :

$L_{M}=a_{M}\left\langle \psi,\psi\right\rangle +a_{I}\operatorname{Im}%
\sum_{\lambda r}O_{r}^{\lambda}\left\langle \psi,\gamma^{r}\nabla_{\lambda
}\psi\right\rangle +a_{D}\sum_{\alpha}V^{\alpha}\left\langle \psi
,\nabla_{\alpha}\psi\right\rangle $

$\frac{dL_{M}}{dO_{j}^{\beta}}=a_{I}\operatorname{Im}\left\langle \psi
,\gamma^{j}\nabla_{\beta}\psi\right\rangle $

$\frac{dL_{M}}{dO_{\alpha}^{\prime i}}=-\sum_{j\beta}O_{i}^{\beta}%
O_{j}^{\alpha}a_{I}\operatorname{Im}\left\langle \psi,\gamma^{j}\nabla_{\beta
}\psi\right\rangle $

$L_{F}=\frac{1}{2}a_{F}\sum_{a\lambda\mu\gamma\xi pqkl}\eta^{qp}\eta^{lk}%
O_{q}^{\lambda}O_{p}^{\gamma}O_{l}^{\mu}O_{k}^{\xi}%
\mathcal{F}%
_{A\gamma\xi}^{a}\overline{%
\mathcal{F}%
}_{A\lambda\mu}^{a}$

$+a_{G}\sum_{a,\lambda\mu}%
\mathcal{F}%
_{G\lambda\mu}^{a}\left(  O_{p_{a}}^{\mu}O_{q_{a}}^{\lambda}-O_{q_{a}}^{\mu
}O_{p_{a}}^{\lambda}\right)  $

$\frac{\partial L_{F}}{\partial O_{j}^{\beta}}=$

$\frac{1}{2}a_{F}\sum\eta^{pq}\eta^{lk}%
\mathcal{F}%
_{A\gamma\xi}^{a}\overline{%
\mathcal{F}%
}_{A\lambda\mu}^{a}\left(  \delta_{\lambda}^{\beta}\delta_{q}^{j}O_{p}%
^{\gamma}O_{l}^{\mu}O_{k}^{\xi}+O_{q}^{\lambda}\delta_{\gamma}^{\beta}%
\delta_{p}^{j}O_{l}^{\mu}O_{k}^{\xi}+O_{q}^{\lambda}O_{p}^{\gamma}\delta_{\mu
}^{\beta}\delta_{l}^{j}O_{k}^{\xi}+O_{q}^{\lambda}O_{p}^{\gamma}O_{l}^{\mu
}\delta_{\xi}^{\beta}\delta_{k}^{j}\right)  $

$+a_{G}\sum_{a,\lambda\mu}%
\mathcal{F}%
_{G\lambda\mu}^{a}\left(  \delta_{\mu}^{\beta}\delta_{p_{a}}^{j}O_{q_{a}%
}^{\lambda}+O_{p_{a}}^{\mu}\delta_{\lambda}^{\beta}\delta_{q_{a}}^{j}%
-\delta_{\mu}^{\beta}\delta_{q_{a}}^{j}O_{p_{a}}^{\lambda}-O_{q_{a}}^{\mu
}\delta_{\lambda}^{\beta}\delta_{p_{a}}^{j}\right)  $

$=\frac{1}{2}a_{F}\sum(\eta^{pj}g^{\mu\lambda}%
\mathcal{F}%
_{A\gamma\lambda}^{a}\overline{%
\mathcal{F}%
}_{A\beta\mu}^{a}O_{p}^{\gamma}+\eta^{jq}O_{q}^{\lambda}g^{\gamma\mu}%
\mathcal{F}%
_{A\beta\mu}^{a}\overline{%
\mathcal{F}%
}_{A\lambda\gamma}^{a}-g^{\mu\gamma}\eta^{jk}O_{k}^{\lambda}%
\mathcal{F}%
_{A\gamma\lambda}^{a}\overline{%
\mathcal{F}%
}_{A\beta\mu}^{a} $

$-g^{\lambda\mu}\eta^{lj}O_{l}^{\gamma}%
\mathcal{F}%
_{A\beta\mu}^{a}\overline{%
\mathcal{F}%
}_{A\lambda\gamma}^{a})+2a_{G}\sum_{a,\lambda\mu}%
\mathcal{F}%
_{G\beta\lambda}^{a}\left(  O_{p_{a}}^{\lambda}\delta_{q_{a}}^{j}-O_{q_{a}%
}^{\lambda}\delta_{p_{a}}^{j}\right)  $

$\frac{dL_{F}}{dO_{\alpha}^{\prime i}}=-\sum_{j\beta}O_{i}^{\beta}%
O_{j}^{\alpha}\frac{dL}{dO_{j}^{\beta}}$

$=-\frac{1}{2}a_{F}O_{i}^{\beta}\sum(\eta^{pj}O_{p}^{\gamma}O_{j}^{\alpha
}g^{\mu\lambda}%
\mathcal{F}%
_{A\gamma\lambda}^{a}\overline{%
\mathcal{F}%
}_{A\beta\mu}^{a}+\eta^{jq}O_{q}^{\lambda}O_{j}^{\alpha}g^{\gamma\mu}%
\mathcal{F}%
_{A\beta\mu}^{a}\overline{%
\mathcal{F}%
}_{A\lambda\gamma}^{a}$

$-g^{\mu\gamma}\eta^{jk}O_{k}^{\lambda}O_{j}^{\alpha}%
\mathcal{F}%
_{A\gamma\lambda}^{a}\overline{%
\mathcal{F}%
}_{A\beta\mu}^{a}-g^{\lambda\mu}\eta^{lj}O_{l}^{\gamma}O_{j}^{\alpha}%
\mathcal{F}%
_{A\beta\mu}^{a}\overline{%
\mathcal{F}%
}_{A\lambda\gamma}^{a})-2a_{G}O_{i}^{\beta}\sum%
\mathcal{F}%
_{G\beta\lambda}^{a}\left(  O_{p_{a}}^{\lambda}\delta_{q_{a}}^{j}O_{j}%
^{\alpha}-O_{q_{a}}^{\lambda}\delta_{p_{a}}^{j}O_{j}^{\alpha}\right)  $

$\frac{dL_{F}}{dO_{\alpha}^{\prime i}}=-\frac{1}{2}a_{F}O_{i}^{\beta}%
\sum\left(
\mathcal{F}%
_{A}^{a\alpha\gamma}\overline{%
\mathcal{F}%
}_{A\beta\gamma}^{a}+%
\mathcal{F}%
_{A\beta\gamma}^{a}\overline{%
\mathcal{F}%
}_{A}^{a\alpha\gamma}+%
\mathcal{F}%
_{A}^{a\alpha\gamma}\overline{%
\mathcal{F}%
}_{A\beta\gamma}^{a}+%
\mathcal{F}%
_{A\beta\gamma}^{a}\overline{%
\mathcal{F}%
}_{A}^{a\alpha\gamma}\right)  $

$-2a_{G}O_{i}^{\beta}\sum%
\mathcal{F}%
_{G\beta\gamma}^{a}\left(  O_{p_{a}}^{\gamma}O_{q_{a}}^{\alpha}-O_{q_{a}%
}^{\gamma}O_{p_{a}}^{\alpha}\right)  $

$\frac{dL_{F}}{dO_{\alpha}^{\prime i}}=-2a_{F}\sum_{\beta}O_{i}^{\beta}%
\sum_{\gamma}\operatorname{Re}\left(
\mathcal{F}%
_{A}^{\alpha\gamma},%
\mathcal{F}%
_{A\beta\gamma}\right)  -2a_{G}O_{i}^{\beta}\sum_{a\gamma}%
\mathcal{F}%
_{G\beta\gamma}^{a}\left(  O_{p_{a}}^{\gamma}O_{q_{a}}^{\alpha}-O_{q_{a}%
}^{\gamma}O_{p_{a}}^{\alpha}\right)  $

$\frac{dL}{dO_{\alpha}^{\prime i}}$

$=-\sum_{\beta}O_{i}^{\beta}a_{I}\sum_{r}O_{r}^{\alpha}\operatorname{Im}%
\left\langle \psi,\gamma^{r}\nabla_{\beta}\psi\right\rangle $

$+2a_{F}\sum_{\gamma}\operatorname{Re}\left(
\mathcal{F}%
_{A}^{\alpha\gamma},%
\mathcal{F}%
_{A\beta\gamma}\right)  +2a_{G}\sum_{a\gamma}%
\mathcal{F}%
_{G\beta\gamma}^{a}\left(  O_{p_{a}}^{\gamma}O_{q_{a}}^{\alpha}-O_{q_{a}%
}^{\gamma}O_{p_{a}}^{\alpha}\right)  $

b) Thus :%

\begin{equation}
\label{E87}
\end{equation}

$\forall\alpha,\beta:\delta_{\beta}^{\alpha}L=a_{I}N\operatorname{Im}%
\left\langle \psi,\gamma^{\alpha}\nabla_{\beta}\psi\right\rangle +2a_{F}%
\sum_{\gamma}\operatorname{Re}\left(
\mathcal{F}%
_{A}^{\alpha\gamma},%
\mathcal{F}%
_{A\beta\gamma}\right)  $

$+2a_{G}\sum_{a\lambda}%
\mathcal{F}%
_{G\beta\lambda}^{a}\left(  O_{p_{a}}^{\gamma}O_{q_{a}}^{\alpha}-O_{q_{a}%
}^{\gamma}O_{p_{a}}^{\alpha}\right)  $

\bigskip

\paragraph{3)}

Using the moments we get \ for equation \ref{E87} :

$\forall\alpha,\beta:\delta_{\beta}^{\alpha}L=a_{I}N\sum_{\alpha r}%
O_{r}^{\alpha}\operatorname{Im}\left\langle \psi,\gamma^{r}\nabla_{\beta}%
\psi\right\rangle +2a_{F}\sum_{\gamma}\operatorname{Re}\left(
\mathcal{F}%
_{A}^{\alpha\gamma},%
\mathcal{F}%
_{A\beta\gamma}\right)  +2a_{G}\sum_{a\lambda}%
\mathcal{F}%
_{G\beta\lambda}^{a}\left(  O_{p_{a}}^{\gamma}O_{q_{a}}^{\alpha}-O_{q_{a}%
}^{\gamma}O_{p_{a}}^{\alpha}\right)  $%

\begin{equation}
\label{E89}
\end{equation}

$\forall\alpha,\beta:\delta_{\beta}^{\alpha}L=Na_{I}\sum_{r}O_{r}^{\alpha
}\left(  -\frac{dJ_{r}}{d\xi^{\beta}}+\sum_{a}\left(  G_{\beta}^{a}\left(
\left[  J\right]  \left[  \widetilde{\kappa}_{a}\right]  \right)  _{r}-\left[
\mu_{A}\right]  _{a}^{r}\operatorname{Im}\grave{A}_{\beta}^{a}\right)
\right)  $

$\qquad+2a_{F}\sum_{\gamma}\operatorname{Re}\left(
\mathcal{F}%
_{A}^{\alpha\gamma},%
\mathcal{F}%
_{A\beta\gamma}\right)  +2a_{G}\sum_{a\lambda}%
\mathcal{F}%
_{G\beta\lambda}^{a}\left(  O_{p_{a}}^{\gamma}O_{q_{a}}^{\alpha}-O_{q_{a}%
}^{\gamma}O_{p_{a}}^{\alpha}\right)  $

\bigskip

\subsubsection{Superpotential}

The superpotential is here :

$\Pi_{H\beta}=\sum_{a}\left(  \operatorname{Re}\grave{A}_{\beta}^{a}\right)
\Pi_{AR}^{a}+\left(  \operatorname{Im}\grave{A}_{\beta}^{a}\right)  \Pi
_{AI}^{a}+a_{G}G_{\beta}^{a}\Pi_{G}^{a}$

$\Pi_{G}^{a}=4\left(  \det O^{\prime}\right)  \sum_{\lambda<\mu}\sum
_{\alpha<\gamma}\epsilon\left(  \lambda,\mu,\alpha,\gamma\right)  \left(
O_{p_{a}}^{\gamma}O_{q_{a}}^{\alpha}-O_{q_{a}}^{\gamma}O_{p_{a}}^{\alpha
}\right)  dx^{\lambda}\wedge dx^{\mu}$

$\Pi_{AR}^{a}=4a_{F}\left(  \det O^{\prime}\right)  \sum_{\lambda<\mu}%
\sum_{\alpha<\gamma}\epsilon\left(  \lambda,\mu,\alpha,\gamma\right)
\operatorname{Re}%
\mathcal{F}%
_{A}^{a,\alpha\beta}dx^{\lambda}\wedge dx^{\mu}$

$\Pi_{AI}^{a}=4a_{F}\left(  \det O^{\prime}\right)  \sum_{\lambda<\mu}%
\sum_{\alpha<\gamma}\epsilon\left(  \lambda,\mu,\alpha,\gamma\right)
\operatorname{Im}%
\mathcal{F}%
_{A}^{a,\alpha\gamma}dx^{\lambda}\wedge dx^{\mu}$

$\left(  \operatorname{Re}\grave{A}_{\beta}^{a}\right)  \Pi_{AR}^{a}+\left(
\operatorname{Im}\grave{A}_{\beta}^{a}\right)  \Pi_{AI}^{a}$

$=4a_{F}\left(  \det O^{\prime}\right)  \sum_{\lambda<\mu}\sum_{\alpha<\gamma
}\epsilon\left(  \lambda,\mu,\alpha,\gamma\right)  \left(  \operatorname{Re}%
\mathcal{F}%
_{A}^{a,\alpha\gamma}\left(  \operatorname{Re}\grave{A}_{\beta}^{a}\right)
+\left(  \operatorname{Im}\grave{A}_{\beta}^{a}\right)  \operatorname{Im}%
\mathcal{F}%
_{A}^{a,\alpha\gamma}\right)  dx^{\lambda}\wedge dx^{\mu}$

$=4a_{F}\left(  \det O^{\prime}\right)  \sum_{\lambda<\mu}\sum_{\alpha
<\gamma,a}\epsilon\left(  \lambda,\mu,\alpha,\gamma\right)  \operatorname{Re}%
\left(  \overline{\grave{A}}_{\beta}^{a}%
\mathcal{F}%
_{A}^{a,\alpha\gamma}\right)  dx^{\lambda}\wedge dx^{\mu}$

and :%

\begin{equation}
\label{E84c}
\end{equation}

$\Pi_{H\beta}=4\det O^{\prime}\sum_{a}\sum_{\substack{\lambda<\mu
\\\alpha<\gamma}}$

$\epsilon\left(  \lambda,\mu,\alpha,\gamma\right)  \left(  a_{G}G_{\beta}%
^{a}\left(  O_{p_{a}}^{\gamma}O_{q_{a}}^{\alpha}-O_{q_{a}}^{\gamma}O_{p_{a}%
}^{\alpha}\right)  +a_{F}\operatorname{Re}\left(  \overline{\grave{A}}_{\beta
}^{a}%
\mathcal{F}%
_{A}^{a,\alpha\gamma}\right)  \right)  dx^{\lambda}\wedge dx^{\mu}$

\subsubsection{The equation}

The equation \ref{E59a} reads here : $d\left(  \Pi_{H\beta}\right)  =0$

That is :

$d\Pi_{H\beta}$

$=-4\sum_{a}\varpi_{4}\left(  \sum_{\alpha\gamma}\partial_{\gamma}\left(
\sum_{a}\left(
\begin{array}
[c]{c}%
a_{G}G_{\beta}^{a}\left(  O_{p_{a}}^{\gamma}O_{q_{a}}^{\alpha}-O_{q_{a}%
}^{\gamma}O_{p_{a}}^{\alpha}\right) \\
+a_{F}\operatorname{Re}\left(  \overline{\grave{A}}_{\beta}^{a}%
\mathcal{F}%
_{A}^{a,\alpha\gamma}\right)
\end{array}
\right)  \det O^{\prime}\right)  \partial_{\alpha}\right)  $

$=-4\sum_{\alpha\gamma}\left(  -1\right)  ^{\alpha+1}\partial_{\gamma}\left(
\sum_{a}\left(
\begin{array}
[c]{c}%
a_{G}G_{\beta}^{a}\left(  O_{p_{a}}^{\gamma}O_{q_{a}}^{\alpha}-O_{q_{a}%
}^{\gamma}O_{p_{a}}^{\alpha}\right) \\
+a_{F}\operatorname{Re}\left(  \overline{\grave{A}}_{\beta}^{a}%
\mathcal{F}%
_{A}^{a,\alpha\gamma}\right)
\end{array}
\right)  \det O^{\prime}\right)  $

$\qquad\qquad\qquad\times dx^{0}\wedge..\widehat{dx^{\alpha}}\wedge...\wedge
dx^{3}$

So $:$%

\begin{equation}
\mathbf{\forall\alpha,\beta:0=}\sum_{\gamma}\mathbf{\partial}_{\gamma}\left(
\left(  a_{F}\operatorname{Re}\left(  \grave{A}_{\beta},%
\mathcal{F}%
_{A}^{\alpha\gamma}\right)  +a_{G}\sum_{a}\left(  O_{p_{a}}^{\gamma}O_{q_{a}%
}^{\alpha}-O_{q_{a}}^{\gamma}O_{p_{a}}^{\alpha}\right)  G_{\beta}^{a}\right)
\det O^{\prime}\right) \label{E90}%
\end{equation}

One can check that this equation is equivalent to the equality of the two
previous expressions for $\delta_{\beta}^{\alpha}L.$This equation involves
neither f (all functions come from $L_{F})$ or the state tensor $\psi.$A
change in the gravitational field should entail a change in the other fields,
whatever the presence of particles, and conversely.

\subsubsection{Scalar curvature}

If we put $\alpha=\beta$ in equation \ref{E87}\ we get :

$\forall\alpha:L=Na_{I}\operatorname{Im}\left\langle \psi,\gamma^{\alpha
}\nabla_{\alpha}\psi\right\rangle +2a_{F}\sum_{\lambda}\operatorname{Re}%
\left(
\mathcal{F}%
_{A\alpha\lambda},%
\mathcal{F}%
_{A}^{\alpha\lambda}\right)  $

$\qquad+2a_{G}\sum_{a\lambda}%
\mathcal{F}%
_{G\alpha\lambda}^{a}\left(  O_{q_{a}}^{\alpha}O_{p_{a}}^{\lambda}-O_{p_{a}%
}^{\alpha}O_{q_{a}}^{\lambda}\right)  $

and by adding over $\alpha:$

$4L=Na_{I}\operatorname{Im}\left\langle \psi,D\psi\right\rangle +2a_{F}%
\sum_{\alpha\lambda}\left(
\mathcal{F}%
_{A\alpha\lambda}^{a},%
\mathcal{F}%
_{A}^{a\alpha\lambda}\right)  $

$\qquad+2a_{G}\sum_{\alpha\lambda a}%
\mathcal{F}%
_{G\alpha\lambda}^{a}\left(  O_{q_{a}}^{\alpha}O_{p_{a}}^{\lambda}-O_{p_{a}%
}^{\alpha}O_{q_{a}}^{\lambda}\right)  $

That is :

$4L=Na_{I}\operatorname{Im}\left\langle \psi,D\psi\right\rangle +4a_{F}%
\left\langle
\mathcal{F}%
_{A},%
\mathcal{F}%
_{A}\right\rangle +2R$

with :

$a_{F}\left\langle
\mathcal{F}%
_{A},%
\mathcal{F}%
_{A}\right\rangle =a_{F}\sum_{a}\sum_{\left\{  \alpha\beta\right\}  }%
\overline{%
\mathcal{F}%
}_{A\alpha\beta}^{a}%
\mathcal{F}%
_{A}^{a,\alpha\beta}$

$=\frac{1}{2}a_{F}\sum_{a}\sum_{\alpha\beta}\overline{%
\mathcal{F}%
}_{A\alpha\beta}^{a}%
\mathcal{F}%
_{A}^{a,\alpha\beta}=\frac{1}{2}a_{F}\sum_{\alpha\beta}\left(
\mathcal{F}%
_{A\alpha\beta},%
\mathcal{F}%
_{A}^{\alpha\beta}\right)  $

$a_{G}R=a_{G}\sum_{\alpha\lambda}%
\mathcal{F}%
_{G\alpha\lambda}^{a}\left(  O_{q_{a}}^{\alpha}O_{p_{a}}^{\lambda}-O_{p_{a}%
}^{\alpha}O_{q_{a}}^{\lambda}\right)  $

But :

$L=Na_{M}\left\langle \psi,\psi\right\rangle +Na_{I}\operatorname{Im}%
\left\langle \psi,D\psi\right\rangle $

$\qquad+Na_{D}\sum_{\alpha}\operatorname{Im}\left\langle \psi,V^{\alpha}%
\nabla_{\alpha}\right\rangle +a_{F}\left\langle
\mathcal{F}%
_{A},%
\mathcal{F}%
_{A}\right\rangle +a_{G}R$

So equation \ref{E87}\ implies :

$Na_{I}\operatorname{Im}\left\langle \psi,D\psi\right\rangle +4a_{F}%
\left\langle
\mathcal{F}%
_{A},%
\mathcal{F}%
_{A}\right\rangle +2R=4Na_{M}\left\langle \psi,\psi\right\rangle
+4Na_{I}\operatorname{Im}\left\langle \psi,D\psi\right\rangle $

$\qquad+4Na_{D}\sum_{\alpha}\operatorname{Im}\left\langle \psi,V^{\alpha
}\nabla_{\alpha}\right\rangle +4a_{F}\left\langle
\mathcal{F}%
_{A},%
\mathcal{F}%
_{A}\right\rangle +4a_{G}R$

So we get the formula for the scalar curvature :%

\begin{equation}
R=-\frac{N}{a_{G}}\left(  2a_{M}\left\langle \psi,\psi\right\rangle +\frac
{3}{2}a_{I}\operatorname{Im}\left\langle \psi,D\psi\right\rangle +2a_{D}%
\sum_{\alpha}\operatorname{Im}\left\langle \psi,V^{\alpha}\nabla_{\alpha
}\right\rangle \right) \label{E88}%
\end{equation}

It depends on the particles and the other fields, as expected, but it is null
if there is no particle. The curvature being small, the constant $a_{G}$
should be much larger than $a_{M},a_{I},a_{D}.$ As the sign of N is difficult
to predict one cannot guess anything about the signs of the terms. All the
second order (the torsion and the scalar curvature) gravitational quantities
are thus easily computed, without need of the curvature form $%
\mathcal{F}%
_{G}$ which does not appear in the equations.

\subsection{Equation of state}

The equations of state is written from the results of the previous part.\ It
is not too complicated and involves the derivatives of the velocity. But from
it one can prove two striking conservation laws for the moments: one related
to the density of particles and the other to the "particles energy":

$\sum_{\alpha}\frac{d}{d\xi^{\alpha}}\left(  \left(  N\det O^{\prime}\right)
V^{\alpha}\left\langle \psi,\psi\right\rangle \right)  =0$

$N\left(  \det O^{\prime}\right)  L_{M}+a_{I}\sum_{\alpha r}\frac{d}%
{d\xi^{\alpha}}\left(  N\left(  \det O^{\prime}\right)  O_{r}^{\alpha}%
J_{r}\right)  =0$

\subsubsection{Equation}

The equations \ref{E43a},\ref{E44a} reads here :

$\forall i,j:0=N\{\sum_{\alpha k}\frac{dL_{M}^{\Diamond}}{d\operatorname{Re}%
\nabla_{\alpha}\psi^{kj}}\operatorname{Re}\left[  G_{\alpha}\right]  _{i}%
^{k}+\frac{dL_{M}^{\Diamond}}{d\operatorname{Im}\nabla_{\alpha}\psi^{kj}%
}\operatorname{Im}\left[  G_{\alpha}\right]  _{i}^{k}$

$+\frac{dL_{M}^{\Diamond}}{d\operatorname{Re}\nabla_{\alpha}\psi^{ik}%
}\operatorname{Re}\left[  \grave{A}_{\alpha}\right]  _{j}^{k}+\frac
{dL_{M}^{\Diamond}}{d\operatorname{Im}\nabla_{\alpha}\psi^{ik}}%
\operatorname{Im}\left[  \grave{A}_{\alpha}\right]  _{j}^{k}\}$

$-\sum_{\beta}\frac{1}{\left(  \det O^{\prime}\right)  }\frac{d}{d\xi^{\beta}%
}\left(  \frac{d%
\mathcal{L}%
_{M}}{d\operatorname{Re}\nabla_{\beta}\psi^{ij}}\right)  +N\frac{\partial
L_{M}^{\Diamond}}{\partial\operatorname{Re}\psi^{ij}}$

$\forall i,j:0=N\{\sum_{\alpha k}-\frac{dL_{M}^{\Diamond}}{d\operatorname{Re}%
\nabla_{\alpha}\psi^{kj}}\operatorname{Im}\left[  G_{\alpha}\right]  _{i}%
^{k}+\frac{dL_{M}^{\Diamond}}{d\operatorname{Im}\nabla_{\alpha}\psi^{kj}%
}\operatorname{Re}\left[  G_{\alpha}\right]  _{i}^{k}$

$-\frac{dL_{M}^{\Diamond}}{d\operatorname{Re}\nabla_{\alpha}\psi^{ik}%
}\operatorname{Im}\left[  \grave{A}_{\alpha}\right]  _{j}^{k}+\frac
{dL_{M}^{\Diamond}}{d\operatorname{Im}\nabla_{\alpha}\psi^{ik}}%
\operatorname{Re}\left[  \grave{A}_{\alpha}\right]  _{j}^{k}$

$-\sum_{\beta}\frac{1}{\left(  \det O^{\prime}\right)  }\frac{d}{d\xi^{\beta}%
}\left(  \frac{d%
\mathcal{L}%
_{M}}{d\operatorname{Im}\nabla_{\alpha}\psi^{ij}}\right)  +N\frac{\partial
L_{M}^{\Diamond}}{\partial\operatorname{Im}\psi^{ij}}$

\paragraph{1)}

Computation of the derivatives $\frac{\partial L_{M}^{\Diamond}}%
{\partial\operatorname{Re}\psi^{ij}},\frac{\partial L_{M}^{\Diamond}}%
{\partial\operatorname{Im}\psi^{ij}}$ .\ Notice that only the terms in $\psi
$\ (and not $\nabla_{\alpha}\psi)$\ are involved here.

$L_{M}=N\left(  a_{M}\left\langle \psi,\psi\right\rangle +\operatorname{Im}%
\left\langle \psi,\sum_{\alpha}D_{M}^{\alpha}\left[  \nabla_{\alpha}%
\psi\right]  \right\rangle \right)  $

$\left\langle \psi,\psi\right\rangle =\sum_{pqk}\left[  \overline{\psi
}\right]  ^{kp}\left[  \gamma_{0}\right]  _{q}^{k}\left[  \psi\right]  ^{qp}$

$\frac{\partial\left\langle \psi,\psi\right\rangle }{\partial\operatorname{Re}%
\psi^{ij}}=\sum\delta_{k}^{i}\delta_{p}^{j}\left[  \gamma_{0}\right]  _{q}%
^{k}\left[  \psi\right]  ^{qp}+\left[  \overline{\psi}\right]  ^{kp}\left[
\gamma_{0}\right]  _{q}^{k}\delta_{q}^{i}\delta_{p}^{j}$

$=\sum_{pqk}\left[  \gamma_{0}\right]  _{q}^{i}\left[  \psi\right]
^{qj}+\left[  \overline{\psi}\right]  ^{kj}\left[  \gamma_{0}\right]  _{i}%
^{k}=\left(  \left[  \gamma_{0}\right]  \left(  \left[  \psi\right]  -\left[
\overline{\psi}\right]  \right)  \right)  _{j}^{i}$

$\frac{\partial\left\langle \psi,\psi\right\rangle }{\partial\operatorname{Re}%
\psi^{ij}}=2\left(  i\left[  \gamma_{0}\right]  \operatorname{Im}\left[
\psi\right]  \right)  _{j}^{i}=2\operatorname{Im}\left(  i\left[  \gamma
_{0}\right]  \left[  \psi\right]  \right)  _{j}^{i}$

$\frac{\partial\left\langle \psi,\psi\right\rangle }{\partial\operatorname{Im}%
\psi^{ij}}=\sum-i\delta_{k}^{i}\delta_{p}^{j}\left[  \gamma_{0}\right]
_{q}^{k}\left[  \psi\right]  ^{qp}+i\left[  \overline{\psi}\right]
^{kp}\left[  \gamma_{0}\right]  _{q}^{k}\delta_{q}^{i}\delta_{p}^{j}$

$=\sum_{pqk}-i\left[  \gamma_{0}\right]  _{q}^{i}\left[  \psi\right]
^{qj}+i\left[  \overline{\psi}\right]  ^{kj}\left[  \gamma_{0}\right]
_{i}^{k}=-i\left(  \left[  \gamma_{0}\right]  \left(  \left[  \psi\right]
+\left[  \overline{\psi}\right]  \right)  \right)  _{j}^{i}$

$\frac{\partial\left\langle \psi,\psi\right\rangle }{\partial\operatorname{Im}%
\psi^{ij}}=-2i\left(  \left[  \gamma_{0}\right]  \operatorname{Re}\left[
\psi\right]  \right)  _{j}^{i}=-2\operatorname{Re}\left(  i\left[  \gamma
_{0}\right]  \left[  \psi\right]  \right)  _{j}^{i}$

$\operatorname{Im}\left\langle \psi,\sum_{\alpha}D_{M}^{\alpha}\left[
\nabla_{\alpha}\psi\right]  \right\rangle =\sum_{\alpha pq}\operatorname{Im}%
\left(  \overline{\left[  \psi\right]  }^{qp}\left(  \left[  \gamma
_{0}\right]  \left[  D_{M}^{\alpha}\right]  \left[  \nabla_{\alpha}%
\psi\right]  \right)  ^{qp}\right)  $

$=\sum_{\alpha pq}\operatorname{Im}\left(  \overline{\left[  \psi\right]
}^{qp}\right)  \operatorname{Re}\left(  \left[  \gamma_{0}\right]  \left[
D_{M}^{\alpha}\right]  \left[  \nabla_{\alpha}\psi\right]  \right)
^{qp}+\operatorname{Re}\left(  \overline{\left[  \psi\right]  }^{qp}\right)
\operatorname{Im}\left(  \left[  \gamma_{0}\right]  \left[  D_{M}^{\alpha
}\right]  \left[  \nabla_{\alpha}\psi\right]  \right)  ^{qp}$

$=\sum_{\alpha pq}-\operatorname{Im}\left(  \left[  \psi\right]  ^{qp}\right)
\operatorname{Re}\left(  \left[  \gamma_{0}\right]  \left[  D_{M}^{\alpha
}\right]  \left[  \nabla_{\alpha}\psi\right]  \right)  ^{qp}+\operatorname{Re}%
\left(  \left[  \psi\right]  ^{qp}\right)  \operatorname{Im}\left(  \left[
\gamma_{0}\right]  \left[  D_{M}^{\alpha}\right]  \left[  \nabla_{\alpha}%
\psi\right]  \right)  ^{qp}$

$\frac{\partial\operatorname{Im}\left\langle \psi,\sum_{\alpha}D_{M}^{\alpha
}\left[  \nabla_{\alpha}\psi\right]  \right\rangle }{\partial\operatorname{Re}%
\psi^{ij}}=\sum_{\alpha pq}\delta_{q}^{i}\delta_{p}^{j}\operatorname{Im}%
\left(  \left[  \gamma_{0}\right]  \left[  D_{M}^{\alpha}\right]  \left[
\nabla_{\alpha}\psi\right]  \right)  ^{qp}=\sum_{\alpha}\operatorname{Im}%
\left(  \left[  \gamma_{0}\right]  \left[  D_{M}^{\alpha}\right]  \left[
\nabla_{\alpha}\psi\right]  \right)  _{j}^{i}$

$\frac{\partial\operatorname{Im}\left\langle \psi,\sum_{\alpha}D_{M}^{\alpha
}\left[  \nabla_{\alpha}\psi\right]  \right\rangle }{\partial\operatorname{Im}%
\psi^{ij}}=\sum_{\alpha pq}-\delta_{q}^{i}\delta_{p}^{j}\operatorname{Re}%
\left(  \left[  \gamma_{0}\right]  \left[  D_{M}^{\alpha}\right]  \left[
\nabla_{\alpha}\psi\right]  \right)  ^{qp}=-\sum_{\alpha}\operatorname{Re}%
\left(  \left[  \gamma_{0}\right]  \left[  D_{M}^{\alpha}\right]  \left[
\nabla_{\alpha}\psi\right]  \right)  _{j}^{i}$

$\frac{\partial L_{M}^{\Diamond}}{\partial\operatorname{Re}\psi^{ij}}%
=2a_{M}\operatorname{Im}\left(  i\left[  \gamma_{0}\right]  \left[
\psi\right]  \right)  _{j}^{i}+\sum_{\alpha}\operatorname{Im}\left(  \left[
\gamma_{0}\right]  \left[  D_{M}^{\alpha}\right]  \left[  \nabla_{\alpha}%
\psi\right]  \right)  _{j}^{i}$

$=\operatorname{Im}\left(  2a_{M}i\left[  \gamma_{0}\right]  \left[
\psi\right]  +\sum_{\alpha}\left[  \gamma_{0}\right]  \left[  D_{M}^{\alpha
}\right]  \left[  \nabla_{\alpha}\psi\right]  \right)  _{j}^{i}$

$\frac{\partial L_{M}^{\Diamond}}{\partial\operatorname{Im}\psi^{ij}}%
=-2a_{M}\operatorname{Re}\left(  i\left[  \gamma_{0}\right]  \left[
\psi\right]  \right)  _{j}^{i}-\sum_{\alpha}\operatorname{Re}\left(  \left[
\gamma_{0}\right]  \left[  D_{M}^{\alpha}\right]  \left[  \nabla_{\alpha}%
\psi\right]  \right)  _{j}^{i}$

$=-\operatorname{Re}\left(  2a_{M}i\left[  \gamma_{0}\right]  \left[
\psi\right]  +\sum_{\alpha}\left[  \gamma_{0}\right]  \left[  D_{M}^{\alpha
}\right]  \left[  \nabla_{\alpha}\psi\right]  \right)  _{j}^{i}$

\paragraph{2)}

We have already

$\frac{dL_{M}}{d\operatorname{Re}\nabla_{\alpha}\psi^{ij}}=N\operatorname{Im}%
\left(  \left[  \psi\right]  ^{\ast}\left[  \gamma_{0}\right]  \left[
D_{M}^{\alpha}\right]  \right)  _{i}^{j}$

$\frac{dL_{M}}{d\operatorname{Im}\nabla_{\alpha}\psi^{ij}}=N\operatorname{Re}%
\left(  \left[  \psi\right]  ^{\ast}\left[  \gamma_{0}\right]  \left[
D_{M}^{\alpha}\right]  \right)  _{i}^{j}$

So the equations read:

$0=N\{\sum_{\alpha k}\operatorname{Im}\left(  \left[  \psi\right]  ^{\ast
}\left[  \gamma_{0}\right]  \left[  D_{M}^{\alpha}\right]  \right)  _{k}%
^{j}\operatorname{Re}\left[  G_{\alpha}\right]  _{i}^{k}+\operatorname{Re}%
\left(  \left[  \psi\right]  ^{\ast}\left[  \gamma_{0}\right]  \left[
D_{M}^{\alpha}\right]  \right)  _{k}^{j}\operatorname{Im}\left[  G_{\alpha
}\right]  _{i}^{k}$

$+\operatorname{Im}\left(  \left[  \psi\right]  ^{\ast}\left[  \gamma
_{0}\right]  \left[  D_{M}^{\alpha}\right]  \right)  _{i}^{k}\operatorname{Re}%
\left[  \grave{A}_{\alpha}\right]  _{j}^{k}+\operatorname{Re}\left(  \left[
\psi\right]  ^{\ast}\left[  \gamma_{0}\right]  \left[  D_{M}^{\alpha}\right]
\right)  _{i}^{k}\operatorname{Im}\left[  \grave{A}_{\alpha}\right]  _{j}%
^{k}\}$

$-\sum_{\alpha}\frac{1}{\left(  \det O^{\prime}\right)  }\frac{d}{d\xi
^{\alpha}}\left(  N\left(  \det O^{\prime}\right)  \operatorname{Im}\left(
\left[  \psi\right]  ^{\ast}\left[  \gamma_{0}\right]  \left[  D_{M}^{\alpha
}\right]  \right)  _{i}^{j}\right)  $

$+N\operatorname{Im}\left(  2a_{M}i\left[  \gamma_{0}\right]  \left[
\psi\right]  +\sum_{\alpha}\left[  \gamma_{0}\right]  \left[  D_{M}^{\alpha
}\right]  \left[  \nabla_{\alpha}\psi\right]  \right)  _{j}^{i}$

$0=N\{\sum_{\alpha k}-\operatorname{Im}\left(  \left[  \psi\right]  ^{\ast
}\left[  \gamma_{0}\right]  \left[  D_{M}^{\alpha}\right]  \right)  _{k}%
^{j}\operatorname{Im}\left[  G_{\alpha}\right]  _{i}^{k}+\operatorname{Re}%
\left(  \left[  \psi\right]  ^{\ast}\left[  \gamma_{0}\right]  \left[
D_{M}^{\alpha}\right]  \right)  _{k}^{j}\operatorname{Re}\left[  G_{\alpha
}\right]  _{i}^{k}$

$-\operatorname{Im}\left(  \left[  \psi\right]  ^{\ast}\left[  \gamma
_{0}\right]  \left[  D_{M}^{\alpha}\right]  \right)  _{i}^{k}\operatorname{Im}%
\left[  \grave{A}_{\alpha}\right]  _{j}^{k}+\operatorname{Re}\left(  \left[
\psi\right]  ^{\ast}\left[  \gamma_{0}\right]  \left[  D_{M}^{\alpha}\right]
\right)  _{i}^{k}\operatorname{Re}\left[  \grave{A}_{\alpha}\right]  _{j}^{k}$

$-\sum_{\alpha}\frac{1}{\left(  \det O^{\prime}\right)  }\frac{d}{d\xi
^{\alpha}}\left(  N\left(  \det O^{\prime}\right)  \operatorname{Re}\left(
\left[  \psi\right]  ^{\ast}\left[  \gamma_{0}\right]  \left[  D_{M}^{\alpha
}\right]  \right)  _{i}^{j}\right)  $

$-N\operatorname{Re}\left(  2a_{M}i\left[  \gamma_{0}\right]  \left[
\psi\right]  +\sum_{\alpha}\left[  \gamma_{0}\right]  \left[  D_{M}^{\alpha
}\right]  \left[  \nabla_{\alpha}\psi\right]  \right)  _{j}^{i}$

$0=N\sum_{\alpha}\operatorname{Im}\left(  \left[  \psi\right]  ^{\ast}\left[
\gamma_{0}\right]  \left[  D_{M}^{\alpha}\right]  \left[  G_{\alpha}\right]
+\left[  \grave{A}_{\alpha}\right]  ^{t}\left[  \psi\right]  ^{\ast}\left[
\gamma_{0}\right]  \left[  D_{M}^{\alpha}\right]  \right)  _{i}^{j}$

$-\sum_{\alpha}\frac{1}{\left(  \det O^{\prime}\right)  }\frac{d}{d\xi
^{\alpha}}\left(  N\left(  \det O^{\prime}\right)  \operatorname{Im}\left(
\left[  \psi\right]  ^{\ast}\left[  \gamma_{0}\right]  \left[  D_{M}^{\alpha
}\right]  \right)  _{i}^{j}\right)  $

$+N\operatorname{Im}\left(  2a_{M}i\left[  \gamma_{0}\right]  \left[
\psi\right]  +\sum_{\alpha}\left[  \gamma_{0}\right]  \left[  D_{M}^{\alpha
}\right]  \left[  \nabla_{\alpha}\psi\right]  \right)  _{j}^{i}$

$0=N\sum_{\alpha k}\operatorname{Re}\left(  \left[  \psi\right]  ^{\ast
}\left[  \gamma_{0}\right]  \left[  D_{M}^{\alpha}\right]  \left[  G_{\alpha
}\right]  +\left[  \grave{A}_{\alpha}\right]  ^{t}\left[  \psi\right]  ^{\ast
}\left[  \gamma_{0}\right]  \left[  D_{M}^{\alpha}\right]  \right)  _{i}^{j}$

$-\sum_{\alpha}\frac{1}{\left(  \det O^{\prime}\right)  }\frac{d}{d\xi
^{\alpha}}\left(  N\left(  \det O^{\prime}\right)  \operatorname{Re}\left(
\left[  \psi\right]  ^{\ast}\left[  \gamma_{0}\right]  \left[  D_{M}^{\alpha
}\right]  \right)  _{i}^{j}\right)  $

$-N\operatorname{Re}\left(  2a_{M}i\left[  \gamma_{0}\right]  \left[
\psi\right]  +\sum_{\alpha}\left[  \gamma_{0}\right]  \left[  D_{M}^{\alpha
}\right]  \left[  \nabla_{\alpha}\psi\right]  \right)  _{j}^{i}$

\paragraph{3)}

We have two real equations, that we can combine in the complex matrix equation :

$-N\operatorname{Re}\left(  2a_{M}i\left[  \gamma_{0}\right]  \left[
\psi\right]  +\sum_{\alpha}\left[  \gamma_{0}\right]  \left[  D_{M}^{\alpha
}\right]  \left[  \nabla_{\alpha}\psi\right]  \right)  _{j}^{i}$

$+iN\operatorname{Im}\left(  2a_{M}i\left[  \gamma_{0}\right]  \left[
\psi\right]  +\sum_{\alpha}\left[  \gamma_{0}\right]  \left[  D_{M}^{\alpha
}\right]  \left[  \nabla_{\alpha}\psi\right]  \right)  _{j}^{i}$

$=-N\left(  \left(  2a_{M}i\left[  \gamma_{0}\right]  \left[  \psi\right]
+\sum_{\alpha}\left[  \gamma_{0}\right]  \left[  D_{M}^{\alpha}\right]
\left[  \nabla_{\alpha}\psi\right]  \right)  ^{\ast}\right)  _{i}^{j}$

$0=N\sum_{\alpha}\left[  \psi\right]  ^{\ast}\left[  \gamma_{0}\right]
\left[  D_{M}^{\alpha}\right]  \left[  G_{\alpha}\right]  +\left[  \grave
{A}_{\alpha}\right]  ^{t}\left[  \psi\right]  ^{\ast}\left[  \gamma
_{0}\right]  \left[  D_{M}^{\alpha}\right]  $

$-N\left(  2a_{M}i\left[  \gamma_{0}\right]  \left[  \psi\right]
+\sum_{\alpha}\left[  \gamma_{0}\right]  \left[  D_{M}^{\alpha}\right]
\left[  \nabla_{\alpha}\psi\right]  \right)  ^{\ast}$

$-\sum_{\alpha}\frac{1}{\left(  \det O^{\prime}\right)  }\frac{d}{d\xi
^{\alpha}}\left(  N\left(  \det O^{\prime}\right)  \left[  \psi\right]
^{\ast}\left[  \gamma_{0}\right]  \left[  D_{M}^{\alpha}\right]  \right)  $

By conjugate transpose :

$0=-2a_{M}iN\left[  \gamma_{0}\right]  \left[  \psi\right]  -\sum_{\alpha
}\frac{1}{\left(  \det O^{\prime}\right)  }\frac{d}{d\xi^{\alpha}}\left(
N\left(  \det O^{\prime}\right)  \left[  D_{M}^{\alpha}\right]  ^{\ast}\left[
\gamma_{0}\right]  \left[  \psi\right]  \right)  $

$+N\sum_{\alpha}\left(  \left[  G_{\alpha}\right]  ^{\ast}\left[
D_{M}^{\alpha}\right]  ^{\ast}\left[  \gamma_{0}\right]  \left[  \psi\right]
+\left[  D_{M}^{\alpha}\right]  ^{\ast}\left[  \gamma_{0}\right]  \left[
\psi\right]  \overline{\left[  \grave{A}_{\alpha}\right]  }-\left[  \gamma
_{0}\right]  \left[  D_{M}^{\alpha}\right]  \left[  \nabla_{\alpha}%
\psi\right]  \right)  $

$0=-2a_{M}iN\left[  \gamma_{0}\right]  \left[  \psi\right]  -\sum_{\alpha
}\frac{1}{\left(  \det O^{\prime}\right)  }\frac{d}{d\xi^{\alpha}}\left(
N\left(  \det O^{\prime}\right)  \left[  D_{M}^{\alpha}\right]  ^{\ast}\left[
\gamma_{0}\right]  \left[  \psi\right]  \right)  $

$+N\sum_{\alpha}\left(  G_{\alpha}^{a}\left[  \kappa_{a}\right]  ^{\ast
}\left[  D_{M}^{\alpha}\right]  ^{\ast}\left[  \gamma_{0}\right]  \left[
\psi\right]  +\overline{\grave{A}}_{\alpha}^{a}\left[  D_{M}^{\alpha}\right]
^{\ast}\left[  \gamma_{0}\right]  \left[  \psi\right]  \overline{\left[
\theta_{a}\right]  }\psi-\left[  \gamma_{0}\right]  \left[  D_{M}^{\alpha
}\right]  \left[  \nabla_{\alpha}\psi\right]  \right)  $

with $\overline{\left[  \theta_{a}\right]  }=-\left[  \theta_{a}\right]
^{t},\left[  \kappa_{a}\right]  ^{\ast}=-\left[  \gamma_{0}\right]  \left[
\kappa_{a}\right]  \left[  \gamma_{0}\right]  $

$0=-2a_{M}iN\left[  \gamma_{0}\right]  \left[  \psi\right]  -\sum_{\alpha
}\frac{1}{\left(  \det O^{\prime}\right)  }\frac{d}{d\xi^{\alpha}}\left(
N\left(  \det O^{\prime}\right)  \left[  D_{M}^{\alpha}\right]  ^{\ast}\left[
\gamma_{0}\right]  \left[  \psi\right]  \right)  $

$-N\sum_{\alpha}\left(  G_{\alpha}^{a}\left[  \gamma_{0}\right]  \left[
\kappa_{a}\right]  \left[  \gamma_{0}\right]  \left[  D_{M}^{\alpha}\right]
^{\ast}\left[  \gamma_{0}\right]  \left[  \psi\right]  +\overline{\grave{A}%
}_{\alpha}^{a}\left[  D_{M}^{\alpha}\right]  ^{\ast}\left[  \gamma_{0}\right]
\left[  \psi\right]  \left[  \theta_{a}\right]  ^{t}\psi+\left[  \gamma
_{0}\right]  \left[  D_{M}^{\alpha}\right]  \left[  \nabla_{\alpha}%
\psi\right]  \right)  $

By left multiplication by $\left[  \gamma_{0}\right]  $

$0=-2a_{M}iN\left[  \psi\right]  -\sum_{\alpha}\frac{1}{\left(  \det
O^{\prime}\right)  }\frac{d}{d\xi^{\alpha}}\left(  N\left(  \det O^{\prime
}\right)  \left[  \gamma_{0}\right]  \left[  D_{M}^{\alpha}\right]  ^{\ast
}\left[  \gamma_{0}\right]  \left[  \psi\right]  \right)  $

$-N\sum_{\alpha}\left(  G_{\alpha}^{a}\left[  \kappa_{a}\right]  \left[
\gamma_{0}\right]  \left[  D_{M}^{\alpha}\right]  ^{\ast}\left[  \gamma
_{0}\right]  \left[  \psi\right]  +\overline{\grave{A}}_{\alpha}^{a}\left[
\gamma_{0}\right]  \left[  D_{M}^{\alpha}\right]  ^{\ast}\left[  \gamma
_{0}\right]  \left[  \psi\right]  \left[  \theta_{a}\right]  ^{t}+\left[
D_{M}^{\alpha}\right]  \left[  \nabla_{\alpha}\psi\right]  \right)  $

$\left[  \gamma_{0}\right]  \left[  D_{M}^{\alpha}\right]  ^{\ast}\left[
\gamma_{0}\right]  $

$=\left[  \gamma_{0}\right]  \left(  \sum_{r}a_{I}O_{r}^{\alpha}\left[
\gamma^{r}\right]  +\sum_{\alpha}V^{\alpha}a_{D}I\right)  ^{\ast}\left[
\gamma_{0}\right]  $

$=\left(  \sum_{r}a_{I}O_{r}^{\alpha}\left[  \gamma_{0}\right]  \left[
\gamma^{r}\right]  ^{\ast}\left[  \gamma_{0}\right]  +\sum_{\alpha}V^{\alpha
}a_{D}I\right)  $

$=\left(  \sum_{r}-a_{I}O_{r}^{\alpha}\left[  \gamma^{r}\right]  +\sum
_{\alpha}V^{\alpha}a_{D}I\right)  $

$=\left[  D_{M}^{\prime\alpha}\right]  $ with $\left[  \gamma_{0}\right]
\left[  \gamma^{r}\right]  ^{\ast}\left[  \gamma_{0}\right]  =-\left[
\gamma^{r}\right]  $

That is :%

\begin{equation}
\label{E91b}
\end{equation}

$\sum_{\alpha}\frac{d}{d\xi^{\alpha}}\left(  N\det O^{\prime}\left[
D_{M}^{\prime\alpha}\right]  \left[  \psi\right]  \right)  $

$\mathbf{=-N}\det\mathbf{O}^{\prime}\left(  2a_{M}i\left[  \psi\right]
+\sum_{\alpha a}\left[  D_{M}^{\alpha}\right]  \left[  \nabla_{\alpha}%
\psi\right]  +\left[  G_{\alpha}\right]  \left[  D_{M}^{\prime\alpha}\right]
\left[  \psi\right]  -\left[  D_{M}^{\prime\alpha}\right]  \left[
\psi\right]  \overline{\left[  \grave{A}_{\alpha}\right]  }\right)  $

\bigskip

The scalar product for the state tensor, and consequently the moments, are
unchanged by multiplication by a c-number z : $\left\vert z\right\vert =1.$
But as one can see in the equation z must be constant, so we are not fully
allowed to normalize $\psi$.

\paragraph{4)}

One can expand the derivative. In this equation all variables, but N, are
valued at a point $m\left(  f\right)  .$ So:

$\sum_{\alpha}\frac{d}{d\xi^{\alpha}}\left(  N\left(  \det O^{\prime}\right)
\left[  D_{M}^{\prime\alpha}\right]  \left[  \psi\right]  \right)  $

$=\sum_{\alpha}\frac{d}{d\xi^{\alpha}}\left(  N\left(  \det O^{\prime}\right)
\left(  \sum_{r}-a_{I}O_{r}^{\alpha}\left[  \gamma^{r}\right]  \left[
\psi\right]  +\sum_{\alpha}V^{\alpha}a_{D}\left[  \psi\right]  \right)
\right)  $

$=-a_{I}\sum_{\alpha r}\frac{dN\left(  \det O^{\prime}\right)  O_{r}^{\alpha}%
}{d\xi^{\alpha}}\left[  \gamma^{r}\right]  \left[  \psi\right]  +a_{D}%
\sum_{\alpha}\frac{dN\left(  \det O^{\prime}\right)  V^{\alpha}}{d\xi^{\alpha
}}\left[  \psi\right]  +N\left(  \det O^{\prime}\right)  \sum_{\alpha}\left[
D_{M}^{\prime\alpha}\right]  \frac{d\psi}{d\xi^{\alpha}}$

and the equation becomes :

$0=-a_{I}\sum_{\alpha r}\frac{dN\left(  \det O^{\prime}\right)  O_{r}^{\alpha
}}{d\xi^{\alpha}}\left[  \gamma^{r}\right]  \left[  \psi\right]  +a_{D}%
\sum_{\alpha}\frac{dN\left(  \det O^{\prime}\right)  V^{\alpha}}{d\xi^{\alpha
}}\left[  \psi\right]  +N\left(  \det O^{\prime}\right)  \sum_{\alpha}\left[
D_{M}^{\prime\alpha}\right]  \frac{d\psi}{d\xi^{\alpha}}$

$+\left(  N\det O^{\prime}\right)  \left(  2a_{M}i\left[  \psi\right]
+\sum_{\alpha}\left[  D_{M}^{\alpha}\right]  \left[  \nabla_{\alpha}%
\psi\right]  +\left(  \left[  G_{\alpha}\right]  \left[  D_{M}^{\prime\alpha
}\right]  \left[  \psi\right]  -\left[  D_{M}^{\prime\alpha}\right]  \left[
\psi\right]  \overline{\left[  \grave{A}_{\alpha}\right]  }\right)  \right)  $

$0=-a_{I}\sum_{\alpha r}\frac{dN\left(  \det O^{\prime}\right)  O_{r}^{\alpha
}}{d\xi^{\alpha}}\left[  \gamma^{r}\right]  \left[  \psi\right]  +a_{D}%
\sum_{\alpha}\frac{dN\left(  \det O^{\prime}\right)  V^{\alpha}}{d\xi^{\alpha
}}\left[  \psi\right]  +2a_{M}i\left(  N\det O^{\prime}\right)  \left[
\psi\right]  $

$+\left(  N\det O^{\prime}\right)  \sum_{\alpha}\left(  \left[  D_{M}^{\alpha
}\right]  \left[  \nabla_{\alpha}\psi\right]  +\left[  G_{\alpha}\right]
\left[  D_{M}^{\prime\alpha}\right]  \left[  \psi\right]  -\left[
D_{M}^{\prime\alpha}\right]  \left[  \psi\right]  \overline{\left[  \grave
{A}_{\alpha}\right]  }+\left[  D_{M}^{\prime\alpha}\right]  \left[
\frac{d\psi}{d\xi^{\alpha}}\right]  \right)  $

Let us expand the last term :

$\left[  D_{M}^{\alpha}\right]  \left[  \nabla_{\alpha}\psi\right]  +\left[
G_{\alpha}\right]  \left[  D_{M}^{\prime\alpha}\right]  \left[  \psi\right]
-\left[  D_{M}^{\prime\alpha}\right]  \left[  \psi\right]  \overline{\left[
\grave{A}_{\alpha}\right]  }+\left[  D_{M}^{\prime\alpha}\right]  \left[
\frac{d\psi}{d\xi^{\alpha}}\right]  $

$=\left[  D_{M}^{\alpha}\right]  \left(  \left[  \frac{d\psi}{d\xi^{\alpha}%
}\right]  +\left[  G_{\alpha}\right]  \left[  \psi\right]  +\left[
\psi\right]  \left[  \grave{A}_{\alpha}\right]  ^{t}\right)  +\left[
G_{\alpha}\right]  \left[  D_{M}^{\prime\alpha}\right]  \left[  \psi\right]
-\left[  D_{M}^{\prime\alpha}\right]  \left[  \psi\right]  \overline{\left[
\grave{A}_{\alpha}\right]  }+\left[  D_{M}^{\prime\alpha}\right]  \left[
\frac{d\psi}{d\xi^{\alpha}}\right]  $

$=\left(  \left[  D_{M}^{\alpha}\right]  +\left[  D_{M}^{\prime\alpha}\right]
\right)  \left[  \frac{d\psi}{d\xi^{\alpha}}\right]  +G_{\alpha}\left(
\left[  D_{M}^{\alpha}\right]  \left[  \kappa_{a}\right]  +\left[  \kappa
_{a}\right]  \left[  D_{M}^{\prime\alpha}\right]  \right)  \left[
\psi\right]  $

$+\grave{A}_{\alpha}^{a}\left[  D_{M}^{\alpha}\right]  \left[  \psi\right]
\left[  \theta_{a}\right]  ^{t}+\overline{\grave{A}_{\alpha}^{a}}\left[
D_{M}^{\prime\alpha}\right]  \left[  \psi\right]  \left[  \theta_{a}\right]
^{t}$

$=2\sum_{\alpha}V^{\alpha}a_{D}\left[  \frac{d\psi}{d\xi^{\alpha}}\right]
+G_{\alpha}\left(  \left[  D_{M}^{\alpha}\right]  \left[  \kappa_{a}\right]
+\left[  \kappa_{a}\right]  \left[  D_{M}^{\prime\alpha}\right]  \right)
\left[  \psi\right]  $

$+\left(  \grave{A}_{\alpha}^{a}\left[  D_{M}^{\alpha}\right]  +\overline
{\grave{A}_{\alpha}^{a}}\left[  D_{M}^{\prime\alpha}\right]  \right)  \left[
\psi\right]  \left[  \theta_{a}\right]  ^{t}$

$=2\sum_{\alpha}V^{\alpha}a_{D}\left[  \frac{d\psi}{d\xi^{\alpha}}\right]
+G_{\alpha}\left(  a_{I}\sum_{r}O_{r}^{\alpha}\left(  \left[  \gamma
^{r}\right]  \left[  \kappa_{a}\right]  -\left[  \kappa_{a}\right]  \left[
\gamma^{r}\right]  \right)  +2a_{D}V^{\alpha}\left[  \kappa_{a}\right]
\right)  \left[  \psi\right]  $

$+\left(  a_{I}\sum_{r}\left(  \grave{A}_{\alpha}^{a}-\overline{\grave
{A}_{\alpha}^{a}}\right)  O_{r}^{\alpha}\left[  \gamma^{r}\right]
+\sum_{\alpha}a_{D}V^{\alpha}\left(  \grave{A}_{\alpha}^{a}+\overline
{\grave{A}_{\alpha}^{a}}\right)  I\right)  \left[  \psi\right]  \left[
\theta_{a}\right]  ^{t}$

$=2\sum_{\alpha}V^{\alpha}a_{D}\left[  \frac{d\psi}{d\xi^{\alpha}}\right]
+G_{\alpha}\left(  a_{I}\sum_{r}O_{r}^{\alpha}\left(  \left[  \gamma
^{r}\right]  \left[  \kappa_{a}\right]  -\left[  \kappa_{a}\right]  \left[
\gamma^{r}\right]  \right)  +2a_{D}V^{\alpha}\left[  \kappa_{a}\right]
\right)  \left[  \psi\right]  $

$+2\left(  ia_{I}\sum_{r}\left(  \operatorname{Im}\grave{A}_{\alpha}%
^{a}\right)  O_{r}^{\alpha}\left[  \gamma^{r}\right]  +\sum_{\alpha}%
a_{D}V^{\alpha}\left(  \operatorname{Re}\grave{A}_{\alpha}^{a}\right)
I\right)  \left[  \psi\right]  \left[  \theta_{a}\right]  ^{t}$

$=a_{I}\sum_{r}\left(  G_{r}^{a}\left(  \left[  \gamma^{r}\right]  \left[
\kappa_{a}\right]  -\left[  \kappa_{a}\right]  \left[  \gamma^{r}\right]
\right)  \left[  \psi\right]  +2i\left(  \operatorname{Im}\grave{A}_{r}%
^{a}\right)  \left[  \gamma^{r}\right]  \left[  \psi\right]  \left[
\theta_{a}\right]  ^{t}\right)  $

$+2a_{D}\sum_{\alpha}V^{\alpha}\left(  \left[  \frac{d\psi}{d\xi^{\alpha}%
}\right]  +\left[  G_{\alpha}\right]  \left[  \psi\right]  +\left(
\operatorname{Re}\grave{A}_{\alpha}^{a}\right)  \left[  \psi\right]  \left[
\theta_{a}\right]  ^{t}\right)  $

The expanded equation reads:

$0=-a_{I}\sum_{\alpha r}\frac{dN\left(  \det O^{\prime}\right)  O_{r}^{\alpha
}}{d\xi^{\alpha}}\left[  \gamma^{r}\right]  \left[  \psi\right]  +a_{D}%
\sum_{\alpha}\frac{dN\left(  \det O^{\prime}\right)  V^{\alpha}}{d\xi^{\alpha
}}\left[  \psi\right]  +2a_{M}i\left(  N\det O^{\prime}\right)  \left[
\psi\right]  $

$+\left(  N\det O^{\prime}\right)  a_{I}\sum_{r}\left(  G_{r}^{a}\left(
\left[  \gamma^{r}\right]  \left[  \kappa_{a}\right]  -\left[  \kappa
_{a}\right]  \left[  \gamma^{r}\right]  \right)  \left[  \psi\right]
+2i\left(  \operatorname{Im}\grave{A}_{r}^{a}\right)  \left[  \gamma
^{r}\right]  \left[  \psi\right]  \left[  \theta_{a}\right]  ^{t}\right)  $

$+2a_{D}\sum_{\alpha}V^{\alpha}\left(  \left[  \frac{d\psi}{d\xi^{\alpha}%
}\right]  +\left[  G_{\alpha}\right]  \left[  \psi\right]  +\left(
\operatorname{Re}\grave{A}_{\alpha}^{a}\right)  \left[  \psi\right]  \left[
\theta_{a}\right]  ^{t}\right)  $

$0=2a_{M}i\left(  N\det O^{\prime}\right)  \left[  \psi\right]  -\sum_{\alpha
}\frac{dN\left(  \det O^{\prime}\right)  O_{r}^{\alpha}}{d\xi^{\alpha}}\left[
\gamma^{r}\right]  \left[  \psi\right]  )$

$+a_{I}\sum_{r}(\left(  N\det O^{\prime}\right)  \left(  G_{r}^{a}\left(
\left[  \gamma^{r}\right]  \left[  \kappa_{a}\right]  -\left[  \kappa
_{a}\right]  \left[  \gamma^{r}\right]  \right)  \left[  \psi\right]
\allowbreak+2i\left(  \operatorname{Im}\grave{A}_{r}^{a}\right)  \left[
\gamma^{r}\right]  \left[  \psi\right]  \left[  \theta_{a}\right]
^{t}\right)  $

$+a_{D}\sum_{\alpha}\left(  \frac{dN\left(  \det O^{\prime}\right)  V^{\alpha
}}{d\xi^{\alpha}}\left[  \psi\right]  +2\left(  N\det O^{\prime}\right)
V^{\alpha}\left(  \left[  \frac{d\psi}{d\xi^{\alpha}}\right]  +\left[
G_{\alpha}\right]  \left[  \psi\right]  +\left(  \operatorname{Re}\grave
{A}_{\alpha}^{a}\right)  \left[  \psi\right]  \left[  \theta_{a}\right]
^{t}\right)  \right)  $

It will be most useful latter.

\paragraph{4)}

One can compute the equation with respect to $\psi_{R},\psi_{L}$

With : $\left[  \nabla_{\alpha}\psi\right]  =$ $%
\begin{bmatrix}
\partial_{\alpha}\psi_{R}+\frac{1}{2}\sum_{1}^{3}\left(  G_{\alpha}%
^{a+3}-iG_{\alpha}^{a}\right)  \sigma_{a}\psi_{R}+\psi_{R}\left[  \grave
{A}_{\alpha}\right]  ^{t}\\
\partial_{\alpha}\psi_{L}-\frac{1}{2}\sum_{1}^{3}\left(  G_{\alpha}%
^{a+3}+iG_{\alpha}^{a}\right)  \sigma_{a}\psi_{L}+\psi_{L}\left[  \grave
{A}_{\alpha}\right]  ^{t}%
\end{bmatrix}
$

the result is the two matrix equations, which are not simple...%

\begin{equation}
\end{equation}

$\frac{1}{N\det O^{\prime}}\sum_{\alpha}\frac{d}{d\xi^{\alpha}}\left(  \left(
-a_{D}V^{\alpha}\psi_{R}+a_{I}\sum_{r}O_{r}^{\alpha}\sigma_{r}\psi_{L}\right)
\left(  N\det O^{\prime}\right)  \right)  $

$=2ia_{M}\psi_{R}+a_{D}\sum_{\alpha}V^{\alpha}\left(  \partial_{\alpha}%
\psi_{R}+\sum_{a=1}^{3}\left(  G_{\alpha}^{a+3}-iG_{\alpha}^{a}\right)
\sigma_{a}\psi_{R}+2\psi_{R}\left[  \operatorname{Re}\grave{A}_{\alpha
}\right]  ^{t}\right)  $

$+a_{I}\sum_{r}\sigma_{r}\left(  \sum_{\alpha}O_{r}^{\alpha}\partial_{\alpha
}\psi_{L}-\sum_{a=1}^{3}\left(  \delta_{r}^{a}\left(  G_{r}^{a+3}-iG_{r}%
^{a}\right)  \sigma_{r}+iG_{r}^{a}\sigma_{a}\right)  \psi_{L}+2i\psi
_{L}\left[  \operatorname{Im}\grave{A}_{\alpha}\right]  ^{t}\right)  $%

\begin{equation}
\end{equation}

$\frac{1}{N\det O^{\prime}}\sum_{\alpha}\frac{d}{d\xi^{\alpha}}\left(  \left(
a_{I}\sum_{r}O_{r}^{\alpha}\eta^{rr}\sigma_{r}\psi_{R}-a_{D}V^{\alpha}\psi
_{L}\right)  \left(  N\det O^{\prime}\right)  \right)  $

$=a_{I}\sum_{r}\eta^{rr}\sigma_{r}\left(  \partial_{\alpha}\psi_{R}+\sum
_{a=1}^{3}\left(  \delta_{r}^{a}\left(  G_{r}^{a+3}+iG_{r}^{a}\right)
\sigma_{r}-iG_{r}^{a}\sigma_{a}\right)  \psi_{R}+2i\psi_{R}\left[
\operatorname{Im}\grave{A}_{\alpha}\right]  ^{t}\right)  $

$+2ia_{M}\psi_{L}+a_{D}\sum_{\alpha}V^{\alpha}\left(  \partial_{\alpha}%
\psi_{L}-\sum_{a=1}^{3}\left(  G_{\alpha}^{a+3}+iG_{\alpha}^{a}\right)
\sigma_{a}\psi_{L}+2\psi_{L}\left[  \operatorname{Re}\grave{A}_{\alpha
}\right]  ^{t}\right)  $

\bigskip

Usually in this kind of equations the Dirac operator brings trouble. Indeed it
exchanges the two subspaces $F^{+},F^{-},$ so one cannot have $\psi_{R}%
\equiv0$\ or $\psi_{L}\equiv0$\ without $\psi=0.$ It does not happen here
thanks to the introduction of V.

\subsubsection{Moments}

From these equation we can deduce new equations for the moments. We will use
the expanded equation :

$0=2a_{M}i\left(  N\det O^{\prime}\right)  \left[  \psi\right]  $

$+a_{I}\sum_{r}(\left(  N\det O^{\prime}\right)  \left(  G_{r}^{a}\left(
\left[  \gamma^{r}\right]  \left[  \kappa_{a}\right]  -\left[  \kappa
_{a}\right]  \left[  \gamma^{r}\right]  \right)  \left[  \psi\right]
+2i\left(  \operatorname{Im}\grave{A}_{r}^{a}\right)  \left[  \gamma
^{r}\right]  \left[  \psi\right]  \left[  \theta_{a}\right]  ^{t}\right)
-\sum_{\alpha}\frac{dN\left(  \det O^{\prime}\right)  O_{r}^{\alpha}}%
{d\xi^{\alpha}}\left[  \gamma^{r}\right]  \left[  \psi\right]  )$

$+a_{D}\sum_{\alpha}\left(  \frac{dN\left(  \det O^{\prime}\right)  V^{\alpha
}}{d\xi^{\alpha}}\left[  \psi\right]  +2\left(  N\det O^{\prime}\right)
V^{\alpha}\left(  \left[  \frac{d\psi}{d\xi^{\alpha}}\right]  +\left[
G_{\alpha}\right]  \left[  \psi\right]  +\left(  \operatorname{Re}\grave
{A}_{\alpha}^{a}\right)  \left[  \psi\right]  \left[  \theta_{a}\right]
^{t}\right)  \right)  $

\paragraph{1)}

Taking the scalar product on the left with $\psi:$

$0=2a_{M}i\left(  N\det O^{\prime}\right)  \left\langle \psi,\psi\right\rangle
-a_{I}\sum_{\alpha}\frac{dN\left(  \det O^{\prime}\right)  O_{r}^{\alpha}%
}{d\xi^{\alpha}}\left\langle \psi,\left[  \gamma^{r}\right]  \left[
\psi\right]  \right\rangle $

$+a_{D}\sum_{\alpha}\frac{dN\left(  \det O^{\prime}\right)  V^{\alpha}}%
{d\xi^{\alpha}}\left\langle \psi,\psi\right\rangle +$

$+a_{I}\left(  N\det O^{\prime}\right)  \sum_{r}\left(  G_{r}^{a}\left\langle
\psi,\left(  \left[  \gamma^{r}\right]  \left[  \kappa_{a}\right]  -\left[
\kappa_{a}\right]  \left[  \gamma^{r}\right]  \right)  \left[  \psi\right]
\right\rangle +2i\left(  \operatorname{Im}\grave{A}_{r}^{a}\right)
\left\langle \psi,\left[  \gamma^{r}\right]  \left[  \psi\right]  \left[
\theta_{a}\right]  ^{t}\right\rangle \right)  $

$+a_{D}2\left(  N\det O^{\prime}\right)  \sum_{\alpha}V^{\alpha}\left(
\left\langle \psi,\left[  \frac{d\psi}{d\xi^{\alpha}}\right]  \right\rangle
+\left\langle \psi,\left[  G_{\alpha}\right]  \left[  \psi\right]
\right\rangle +\left(  \operatorname{Re}\grave{A}_{\alpha}^{a}\right)
\left\langle \psi,\left[  \psi\right]  \left[  \theta_{a}\right]
^{t}\right\rangle \right)  $

It comes:

$0=\left(  2a_{M}i\left(  N\det O^{\prime}\right)  +a_{D}\sum_{\alpha}%
\frac{dN\left(  \det O^{\prime}\right)  V^{\alpha}}{d\xi^{\alpha}}\right)
\left\langle \psi,\psi\right\rangle $

$+a_{D}\sum_{\alpha}\left(  2\left(  N\det O^{\prime}\right)  V^{\alpha
}\left(  \left\langle \psi,\frac{d\psi}{d\xi^{\alpha}}\right\rangle
+G_{\alpha}^{a}iP_{a}+\left(  \operatorname{Re}\grave{A}_{\alpha}^{a}\right)
i\rho_{a}\right)  \right)  $

$+a_{I}\sum_{r}\left(  N\det O^{\prime}\right)  \sum_{a}(\left(
\begin{array}
[c]{c}%
G_{r}^{a}\left(  \left\langle \psi,\left[  \gamma^{r}\right]  \left[
\kappa_{a}\right]  \left[  \psi\right]  \right\rangle -\left\langle
\psi,\left[  \kappa_{a}\right]  \left[  \gamma^{r}\right]  \left[
\psi\right]  \right\rangle \right) \\
-2i\left(  \operatorname{Im}\grave{A}_{r}^{a}\right)  \left[  \mu_{A}\right]
_{a}^{r}%
\end{array}
\right)  $

$+2i\sum_{\alpha}\frac{dN\left(  \det O^{\prime}\right)  O_{r}^{\alpha}}%
{d\xi^{\alpha}}J_{r})$

with :

$\left\langle \psi,\left[  \psi\right]  \left[  \theta_{a}\right]
^{t}\right\rangle =i\rho_{a}$

$\left\langle \psi,\gamma^{r}\left[  \psi\right]  \left[  \theta_{a}\right]
^{t}\right\rangle =-\left[  \mu_{A}\right]  _{a}^{r}$

$\left\langle \psi,\left[  \kappa_{a}\right]  \left[  \psi\right]
\right\rangle =iP_{a}$

$\left\langle \psi,\gamma^{r}\psi\right\rangle =-2iJ_{r}$

\paragraph{2)}

We are left with : $\left\langle \psi,\left[  \gamma^{r}\right]  \left[
\kappa_{a}\right]  \left[  \psi\right]  \right\rangle -\left\langle
\psi,\left[  \kappa_{a}\right]  \left[  \gamma^{r}\right]  \left[
\psi\right]  \right\rangle $

But : $\left\langle \psi,\left[  \kappa_{a}\right]  \left[  \gamma^{r}\right]
\left[  \psi\right]  \right\rangle =Tr\left(  \left[  \psi\right]  ^{\ast
}\left[  \gamma_{0}\right]  \left[  \kappa_{a}\right]  \left[  \gamma
^{r}\right]  \left[  \psi\right]  \right)  $

$\overline{Tr\left(  \left[  \psi\right]  ^{\ast}\left[  \gamma_{0}\right]
\left[  \kappa_{a}\right]  \left[  \gamma^{r}\right]  \left[  \psi\right]
\right)  }=Tr\left(  \left(  \left[  \psi\right]  ^{\ast}\left[  \gamma
_{0}\right]  \left[  \kappa_{a}\right]  \left[  \gamma^{r}\right]  \left[
\psi\right]  \right)  ^{\ast}\right)  $

$=Tr\left(  \left[  \psi\right]  ^{\ast}\left[  \gamma^{r}\right]  ^{\ast
}\left[  \kappa_{a}\right]  ^{\ast}\left[  \gamma_{0}\right]  \left[
\psi\right]  \right)  =Tr\left(  \left[  \psi\right]  ^{\ast}\left[
\gamma_{0}\right]  \left[  \gamma^{r}\right]  \left[  \gamma_{0}\right]
\left[  \gamma_{0}\right]  \left[  \kappa_{a}\right]  \left[  \gamma
_{0}\right]  \left[  \gamma_{0}\right]  \left[  \psi\right]  \right)  $

$=Tr\left(  \left[  \psi\right]  ^{\ast}\left[  \gamma_{0}\right]  \left[
\gamma^{r}\right]  \left[  \kappa_{a}\right]  \left[  \psi\right]  \right)
=\left\langle \psi,\left[  \gamma^{r}\right]  \left[  \kappa_{a}\right]
\left[  \psi\right]  \right\rangle $

with $\left[  \gamma_{0}\right]  \left[  \gamma^{r}\right]  ^{\ast}\left[
\gamma_{0}\right]  =-\left[  \gamma^{r}\right]  ,\left[  \kappa_{a}\right]
^{\ast}=-\left[  \gamma_{0}\right]  \left[  \kappa_{a}\right]  \left[
\gamma_{0}\right]  $

So $\left\langle \psi,\left[  \kappa_{a}\right]  \left[  \gamma^{r}\right]
\left[  \psi\right]  \right\rangle =\overline{\left\langle \psi,\left[
\gamma^{r}\right]  \left[  \kappa_{a}\right]  \left[  \psi\right]
\right\rangle }$

$\left\langle \psi,\left[  \gamma^{r}\right]  \left[  \kappa_{a}\right]
\left[  \psi\right]  \right\rangle -\left\langle \psi,\left[  \kappa
_{a}\right]  \left[  \gamma^{r}\right]  \left[  \psi\right]  \right\rangle $

$=\left\langle \psi,\left[  \gamma^{r}\right]  \left[  \kappa_{a}\right]
\left[  \psi\right]  \right\rangle -\overline{\left\langle \psi,\left[
\gamma^{r}\right]  \left[  \kappa_{a}\right]  \left[  \psi\right]
\right\rangle }$

$=2i\operatorname{Im}\left\langle \psi,\left[  \gamma^{r}\right]  \left[
\kappa_{a}\right]  \left[  \psi\right]  \right\rangle =2i\left(  \left[
J\right]  \left[  \widetilde{\kappa}_{a}\right]  \right)  _{r}$

with : $\operatorname{Im}\left\langle \psi,\left[  \gamma^{r}\right]  \left[
\kappa_{a}\right]  \psi\right\rangle =\left(  \left[  J\right]  \left[
\widetilde{\kappa}_{a}\right]  \right)  _{r}$

and : $\left\langle \psi,\frac{d\psi}{d\xi^{\alpha}}\right\rangle $

But : $\frac{d}{d\xi^{\alpha}}\left\langle \psi,\psi\right\rangle
=\left\langle \frac{d\psi}{d\xi^{\alpha}},\psi\right\rangle +\left\langle
\psi,\frac{d\psi}{d\xi^{\alpha}}\right\rangle =2\operatorname{Re}\left\langle
\psi,\frac{d\psi}{d\xi^{\alpha}}\right\rangle $

$\left\langle \psi,\frac{d\psi}{d\xi^{\alpha}}\right\rangle =\frac{1}{2}%
\frac{d}{d\xi^{\alpha}}\left\langle \psi,\psi\right\rangle +i\operatorname{Im}%
\left\langle \psi,\frac{d\psi}{d\xi^{\alpha}}\right\rangle $

The equation reads:

$0=\left(  2a_{M}i\left(  N\det O^{\prime}\right)  +a_{D}\sum_{\alpha}%
\frac{dN\left(  \det O^{\prime}\right)  V^{\alpha}}{d\xi^{\alpha}}\right)
\left\langle \psi,\psi\right\rangle $

$+a_{D}\sum_{\alpha}\left(  2\left(  N\det O^{\prime}\right)  V^{\alpha
}\left(  \frac{1}{2}\frac{d}{d\xi^{\alpha}}\left\langle \psi,\psi\right\rangle
+i\operatorname{Im}\left\langle \psi,\frac{d\psi}{d\xi^{\alpha}}\right\rangle
+G_{\alpha}^{a}iP_{a}+\left(  \operatorname{Re}\grave{A}_{\alpha}^{a}\right)
i\rho_{a}\right)  \right)  $

$+2ia_{I}\sum_{r}\left(  \left(  N\det O^{\prime}\right)  \sum_{a}\left(
G_{r}^{a}\left(  \left[  J\right]  \left[  \widetilde{\kappa}_{a}\right]
\right)  _{r}-\left(  \operatorname{Im}\grave{A}_{r}^{a}\right)  \left[
\mu_{R}-\mu_{L}\right]  _{a}^{r}\right)  +\sum_{\alpha}\frac{dN\left(  \det
O^{\prime}\right)  O_{r}^{\alpha}}{d\xi^{\alpha}}J_{r}\right)  $

\paragraph{3)}

Taking the real part :

$0=a_{D}\sum_{\alpha}\frac{dN\left(  \det O^{\prime}\right)  V^{\alpha}}%
{d\xi^{\alpha}}\left\langle \psi,\psi\right\rangle +a_{D}\left(  N\det
O^{\prime}\right)  \sum_{\alpha}V^{\alpha}\frac{d}{d\xi^{\alpha}}\left\langle
\psi,\psi\right\rangle $%

\begin{equation}
\sum_{\alpha}\frac{d}{d\xi^{\alpha}}\left(  \left(  N\det O^{\prime}\right)
V^{\alpha}\left\langle \psi,\psi\right\rangle \right)  \mathbf{=0}\label{E92b}%
\end{equation}

Remind the convention about the creation and annihiliation of particles :
$\psi\left(  m\right)  =0$ means no particle .\ So $\left(  N\det O^{\prime
}\right)  \left\langle \psi,\psi\right\rangle $ can be seen roughly as a
density of particles, this equation expresses a conservation law of the flow
of moving particles.

\paragraph{4)}

Taking the imaginary part :

$0=2a_{M}\left(  N\det O^{\prime}\right)  \left\langle \psi,\psi\right\rangle
+2\left(  N\det O^{\prime}\right)  a_{D}\sum_{\alpha}V^{\alpha}\left(
\operatorname{Im}\left\langle \psi,\frac{d\psi}{d\xi^{\alpha}}\right\rangle
+G_{\alpha}^{a}P_{a}+\left(  \operatorname{Re}\grave{A}_{\alpha}^{a}\right)
\rho_{a}\right)  $

$+2a_{I}\sum_{r}\left(  \left(  N\det O^{\prime}\right)  \sum_{a}\left(
G_{r}^{a}\left(  \left[  J\right]  \left[  \widetilde{\kappa}_{a}\right]
\right)  _{r}-\left(  \operatorname{Im}\grave{A}_{r}^{a}\right)  \left[
\mu_{A}\right]  _{a}^{r}\right)  +\sum_{\alpha}\frac{dN\left(  \det O^{\prime
}\right)  O_{r}^{\alpha}}{d\xi^{\alpha}}J_{r}\right)  $

$0=a_{M}\left\langle \psi,\psi\right\rangle +a_{D}\sum_{\alpha}V^{\alpha
}\left(  \operatorname{Im}\left\langle \psi,\frac{d\psi}{d\xi^{\alpha}%
}\right\rangle +\sum_{a}\left(  G_{\alpha}^{a}P_{a}+\rho_{a}\operatorname{Re}%
\grave{A}_{\alpha}^{a}\right)  \right)  +a_{I}\left(  \sum_{\alpha r}%
\frac{dN\left(  \det O^{\prime}\right)  O_{r}^{\alpha}}{N\left(  \det
O^{\prime}\right)  d\xi^{\alpha}}J_{r}+\left(  \left[  J\right]  \left[
G_{r}\right]  \right)  _{r}-\sum_{a}\left(  \operatorname{Im}\grave{A}_{r}%
^{a}\right)  \left[  \mu_{A}\right]  _{a}^{r}\right)  $

With $\sum_{\alpha}V^{\alpha}\left(  \operatorname{Im}\left\langle \psi
,\frac{d\psi}{d\xi^{\alpha}}\right\rangle \right)  =\operatorname{Im}%
\left\langle \psi,\sum_{\alpha}V^{\alpha}\frac{d\psi}{d\xi^{\alpha}%
}\right\rangle =\operatorname{Im}\left\langle \psi,\frac{d\psi}{d\xi^{0}%
}\right\rangle $%

\begin{equation}
\end{equation}

$0=a_{M}\left\langle \psi,\psi\right\rangle +a_{D}\operatorname{Im}%
\left\langle \psi,\frac{d\psi}{d\xi^{0}}\right\rangle +a_{D}\sum_{\alpha
a}V^{\alpha}\left(  G_{\alpha}^{a}P_{a}+\rho_{a}\operatorname{Re}\grave
{A}_{\alpha}^{a}\right)  $

$\qquad+a_{I}\left(  \sum_{\alpha r}\frac{dN\left(  \det O^{\prime}\right)
O_{r}^{\alpha}}{N\left(  \det O^{\prime}\right)  d\xi^{\alpha}}J_{r}+\left(
\left[  J\right]  \left[  G_{r}\right]  \right)  _{r}-\sum_{a}\left(
\operatorname{Im}\grave{A}_{r}^{a}\right)  \left[  \mu_{A}\right]  _{a}%
^{r}\right)  $

As we have :

$L_{M}=a_{M}\left\langle \psi,\psi\right\rangle +a_{I}\left(  -\sum_{\alpha
,r}O_{r}^{\alpha}\frac{dJ_{r}}{d\xi^{\alpha}}+\left(  \left[  J\right]
\left[  G_{r}\right]  \right)  _{r}-\sum_{a}\left(  \operatorname{Im}\grave
{A}_{r}^{a}\right)  \left[  \mu_{A}\right]  _{a}^{r}\right)  $

$+a_{D}\sum_{\alpha}V^{\alpha}\left(  \operatorname{Im}\left\langle \psi
,\frac{d\psi}{d\xi^{\alpha}}\right\rangle +\sum_{a}\left(  G_{\alpha}^{a}%
P_{a}+\rho_{a}\operatorname{Re}\grave{A}_{\alpha}^{a}\right)  \right)  $

$0=L_{M}+a_{I}\left(  \sum_{\alpha r}\frac{dN\left(  \det O^{\prime}\right)
O_{r}^{\alpha}}{N\left(  \det O^{\prime}\right)  d\xi^{\alpha}}J_{r}+\left(
\left[  J\right]  \left[  G_{r}\right]  \right)  _{r}-\sum_{a}\left(
\operatorname{Im}\grave{A}_{r}^{a}\right)  \left[  \mu_{A}\right]  _{a}%
^{r}\right)  $

$-a_{I}\left(  -\sum_{\alpha,r}O_{r}^{\alpha}\frac{dJ_{r}}{d\xi^{\alpha}%
}+\left(  \left[  J\right]  \left[  G_{r}\right]  \right)  _{r}-\sum
_{a}\left(  \operatorname{Im}\grave{A}_{r}^{a}\right)  \left[  \mu_{A}\right]
_{a}^{r}\right)  $

$0=L_{M}+a_{I}\sum_{\alpha r}\left(  \frac{dN\left(  \det O^{\prime}\right)
O_{r}^{\alpha}}{N\left(  \det O^{\prime}\right)  d\xi^{\alpha}}J_{r}%
+O_{r}^{\alpha}\frac{dJ_{r}}{d\xi^{\alpha}}\right)  $%

\begin{equation}
\mathbf{0=N}\left(  \det O^{\prime}\right)  \mathbf{L}_{M}\mathbf{+a}_{I}%
\sum_{\alpha r}\frac{d}{d\xi^{\alpha}}\left(  N\left(  \det O^{\prime}\right)
O_{r}^{\alpha}J_{r}\right) \label{E92d}%
\end{equation}

$N\left(  \det O^{\prime}\right)  L_{M}$ can be seen roughly as a density of
the energy of the particles, and $N\left(  \det O^{\prime}\right)
O_{r}^{\alpha}J_{r}$ as the "internal energy" , so this equation can be seen
as a conservation of energy.\ Notice that there is no clear equivalent of rest
mass (but for $J_{0}$ ?).

\paragraph{5)}

Remind the equation for the scalar curvature :

$R=-\frac{N}{a_{G}}\left(  2a_{M}\left\langle \psi,\psi\right\rangle +\frac
{3}{2}a_{I}\operatorname{Im}\left\langle \psi,D\psi\right\rangle +2a_{D}%
\sum_{\alpha}\operatorname{Im}\left\langle \psi,V^{\alpha}\nabla_{\alpha
}\right\rangle \right)  $

It reads :

$a_{G}R\left(  \det O^{\prime}\right)  =-2NL_{M}\left(  \det O^{\prime
}\right)  +\frac{1}{2}N\left(  \det O^{\prime}\right)  a_{I}\operatorname{Im}%
\left\langle \psi,D\psi\right\rangle $

$=2a_{I}\sum_{\alpha r}\frac{dN\left(  \det O^{\prime}\right)  O_{r}^{\alpha
}J_{r}}{d\xi^{\alpha}}+\frac{1}{2}N\left(  \det O^{\prime}\right)
a_{I}\operatorname{Im}\left\langle \psi,D\psi\right\rangle $%

\begin{equation}
\mathbf{a}_{G}\mathbf{R}\left(  \det O^{\prime}\right)  \mathbf{=a}_{I}\left(
2\sum_{\alpha r}\frac{dN\left(  \det O^{\prime}\right)  O_{r}^{\alpha}J_{r}%
}{d\xi^{\alpha}}+\frac{1}{2}N\left(  \det O^{\prime}\right)  \operatorname{Im}%
\left\langle \psi,D\psi\right\rangle \right) \label{E92f}%
\end{equation}

The scalar curvature is entirely linked to the kinematic part of the particles
: are involved neither the field (but they are involved in the covariant
derivative) or the velocity.

\subsection{Trajectory}

The equation for the trajectory of particles reads :

$\forall\alpha:a_{D}\sum_{\beta}V^{\beta}\partial_{\beta}\left(
\operatorname{Im}\left\langle \psi,\nabla_{\alpha}\psi\right\rangle \det
O^{\prime}\right)  $

$=L_{M}\partial_{\alpha}\det O^{\prime}+\left(  \det O^{\prime}\right)
a_{D}\sum_{a\beta}\left(  V^{\beta}\left(  \partial_{\alpha}G_{\beta}%
^{a}\right)  P_{a}+V^{\beta}\rho_{a}\left(  \partial_{\alpha}\operatorname{Re}%
\grave{A}_{\beta}^{a}\right)  \right)  $

$+\left(  \det O^{\prime}\right)  a_{I}\sum_{\beta r}\left(  -\frac{dJ_{r}%
}{d\xi^{\alpha}}\left(  \partial_{\alpha}O_{r}^{\beta}\right)  +\sum
_{a}\left(  \partial_{\alpha}G_{r}^{a}\right)  \left(  \left[  J\right]
\left[  \widetilde{\kappa}_{a}\right]  \right)  _{r}-\left[  \mu_{A}\right]
_{a}^{r}\operatorname{Im}\left(  \partial_{\alpha}\grave{A}_{r}^{\alpha
}\right)  \right)  $

$L_{M}\partial_{\alpha}\det O^{\prime}$ is obtained through one previous
equation :

$0=N\left(  \det O^{\prime}\right)  L_{M}+a_{I}\sum_{\alpha r}\frac{dN\left(
\det O^{\prime}\right)  O_{r}^{\alpha}J_{r}}{d\xi^{\alpha}}$

So the equation links the derivative $\partial_{\beta}\left(
\operatorname{Im}\left\langle \psi,\nabla_{\alpha}\psi\right\rangle \det
O^{\prime}\right)  $ to the moments and the velocity V (the derivatives of V
figures in the state equation).

\paragraph{1)}

The equation \ref{E49} reads :

$\forall\alpha:0=\frac{d}{d\xi^{0}}\left(  \frac{\partial%
\mathcal{L}%
_{M}}{\partial V^{\alpha}}\right)  $

$+2\sum_{a,\beta\gamma}\{\left(  \partial_{\alpha}\operatorname{Re}\grave
{A}_{\beta}^{a}\right)  \partial_{\gamma}\left(  \frac{\partial%
\mathcal{L}%
_{F}}{\partial\operatorname{Re}%
\mathcal{F}%
_{A,\beta\gamma}^{a}}\right)  +\left(  \partial_{\alpha}\operatorname{Im}%
\grave{A}_{\beta}^{a}\right)  \partial_{\gamma}\left(  \frac{\partial%
\mathcal{L}%
_{F}}{\partial\operatorname{Im}%
\mathcal{F}%
_{A,\beta\gamma}^{b}}\right)  $

$+\frac{\partial%
\mathcal{L}%
_{F}}{\partial\operatorname{Re}%
\mathcal{F}%
_{A,\beta\gamma}^{a}}\operatorname{Re}\left[  \partial_{\alpha}\grave
{A}_{\beta},\grave{A}_{\gamma}\right]  ^{a}+\frac{\partial%
\mathcal{L}%
_{F}}{\partial\operatorname{Im}%
\mathcal{F}%
_{A,\beta\gamma}^{a}}\operatorname{Im}\left[  \partial_{\alpha}\grave
{A}_{\beta},\grave{A}_{\gamma}\right]  ^{a}\}$

$+\sum_{a\beta}\left(  \frac{\partial%
\mathcal{L}%
_{F}}{\partial G_{\alpha}^{a}}\partial_{\alpha}G_{\beta}^{a}+2\sum_{\gamma
}\left(  \left(  \partial_{\alpha}G_{\beta}^{a}\right)  \partial_{\gamma
}\left(  \frac{\partial%
\mathcal{L}%
_{F}}{\partial%
\mathcal{F}%
_{G\beta\gamma}^{a}}\right)  +\frac{\partial%
\mathcal{L}%
_{F}}{\partial%
\mathcal{F}%
_{G\beta\gamma}^{a}}\left[  \partial_{\alpha}G_{\beta},G_{\gamma}\right]
^{a}\right)  \right)  $

$+\sum_{i\beta}\left(  \frac{dL_{F}\left(  \det O^{\prime}\right)  }%
{dO_{\beta}^{\prime i}}-\sum_{\gamma}\frac{d}{d\xi^{\gamma}}\left(
\frac{dL_{F}\left(  \det O^{\prime}\right)  }{d\partial_{\gamma}O_{\beta
}^{\prime i}}\right)  \right)  \partial_{\alpha}O_{\beta}^{\prime i}$

We have the derivatives (all derivatives in $L_{F}$ do not involve f) :

$\frac{\partial%
\mathcal{L}%
_{F}}{\partial%
\mathcal{F}%
_{G\beta\gamma}^{a}}=a_{G}\left(  O_{p_{a}}^{\gamma}O_{q_{a}}^{\beta}%
-O_{q_{a}}^{\gamma}O_{p_{a}}^{\beta}\right)  \det O^{\prime}$

$\frac{dL_{F}}{d\operatorname{Re}%
\mathcal{F}%
_{A,\beta\gamma}^{a}}=a_{F}\operatorname{Re}%
\mathcal{F}%
_{A}^{a,\beta\gamma};\frac{dL_{F}}{d\operatorname{Im}%
\mathcal{F}%
_{A,\beta\gamma}^{a}}=a_{F}\operatorname{Im}%
\mathcal{F}%
_{A}^{a,\beta\gamma}$

$\frac{dL}{d\partial_{\alpha}O_{\gamma}^{\prime i}}=0$

$\frac{dNL_{M}}{dV^{\alpha}}=a_{D}N\operatorname{Im}\left\langle \psi
,\nabla_{\alpha}\psi\right\rangle $

$\frac{dL_{F}}{dO_{\beta}^{\prime i}}=-2\sum_{\gamma}O_{i}^{\gamma}\left(
a_{F}\sum_{\mu}\operatorname{Re}\left(
\mathcal{F}%
_{A}^{\beta\mu},%
\mathcal{F}%
_{A\gamma\mu}\right)  +a_{G}\sum_{a\mu}%
\mathcal{F}%
_{G\gamma\mu}^{a}\left(  O_{p_{a}}^{\mu}O_{q_{a}}^{\beta}-O_{q_{a}}^{\mu
}O_{p_{a}}^{\beta}\right)  \right)  $

$\frac{dL_{F}\left(  \det O^{\prime}\right)  }{dO_{\beta}^{\prime i}}%
=\frac{dL_{F}}{dO_{\beta}^{\prime i}}\left(  \det O^{\prime}\right)
+L_{F}\frac{d\det O^{\prime}}{dO_{\beta}^{\prime i}}$

$=-2\left(  \det O^{\prime}\right)  \sum_{\gamma}O_{i}^{\gamma}\left(
a_{F}\sum_{\mu}\operatorname{Re}\left(
\mathcal{F}%
_{A}^{\beta\mu},%
\mathcal{F}%
_{A\gamma\mu}\right)  +a_{G}\sum_{a\mu}%
\mathcal{F}%
_{G\gamma\mu}^{a}\left(  O_{p_{a}}^{\mu}O_{q_{a}}^{\beta}-O_{q_{a}}^{\mu
}O_{p_{a}}^{\beta}\right)  \right)  +L_{F}O_{i}^{\beta}\left(  \det O^{\prime
}\right)  $

with $\frac{\partial\left(  \det O^{\prime}\right)  }{\partial O_{\beta
}^{\prime i}}=O_{i}^{\beta}\left(  \det O^{\prime}\right)  $

The equation reads :

$\forall\alpha:0=\frac{d}{d\xi^{0}}\left(  a_{D}N\left(  \det O^{\prime
}\right)  \operatorname{Im}\left\langle \psi,\nabla_{\alpha}\psi\right\rangle
\right)  $

$+2a_{F}\sum_{a,\beta\gamma}\{\left(  \partial_{\alpha}\operatorname{Re}%
\grave{A}_{\beta}^{a}\right)  \partial_{\gamma}\left(  \operatorname{Re}%
\mathcal{F}%
_{A}^{a,\beta\gamma}\det O^{\prime}\right)  +\left(  \partial_{\alpha
}\operatorname{Im}\grave{A}_{\beta}^{a}\right)  \partial_{\gamma}\left(
\operatorname{Im}%
\mathcal{F}%
_{A}^{a,\beta\gamma}\det O^{\prime}\right)  $

$+\operatorname{Re}%
\mathcal{F}%
_{A}^{a,\beta\gamma}\operatorname{Re}\left[  \partial_{\alpha}\grave{A}%
_{\beta},\grave{A}_{\gamma}\right]  ^{a}\det O^{\prime}+\operatorname{Im}%
\mathcal{F}%
_{A}^{a,\beta\gamma}\operatorname{Im}\left[  \partial_{\alpha}\grave{A}%
_{\beta},\grave{A}_{\gamma}\right]  ^{a}\det O^{\prime}\}$

$+2a_{G}\sum_{a\beta\gamma}\{\partial_{\alpha}G_{\beta}^{a}\partial_{\gamma
}\left(  \left(  O_{p_{a}}^{\gamma}O_{q_{a}}^{\beta}-O_{q_{a}}^{\gamma
}O_{p_{a}}^{\beta}\right)  \det O^{\prime}\right)  $

$\qquad+\left(  O_{p_{a}}^{\gamma}O_{q_{a}}^{\beta}-O_{q_{a}}^{\gamma}%
O_{p_{a}}^{\beta}\right)  \det O^{\prime}\left[  \partial_{\alpha}G_{\beta
},G_{\gamma}\right]  ^{a}\}$

$+\sum_{i\beta}\{-2\left(  \det O^{\prime}\right)  \sum_{\gamma}O_{i}^{\gamma
}\left(  a_{F}\sum_{\mu}\operatorname{Re}\left(
\mathcal{F}%
_{A}^{\beta\mu},%
\mathcal{F}%
_{A\gamma\mu}\right)  +a_{G}\sum_{a\mu}%
\mathcal{F}%
_{G\gamma\mu}^{a}\left(  O_{p_{a}}^{\mu}O_{q_{a}}^{\beta}-O_{q_{a}}^{\mu
}O_{p_{a}}^{\beta}\right)  \right)  $

$+L_{F}O_{i}^{\beta}\left(  \det O^{\prime}\right)  \}\partial_{\alpha
}O_{\beta}^{\prime i}$

$\forall\alpha:0=\frac{d}{d\xi^{0}}\left(  a_{D}N\left(  \det O^{\prime
}\right)  \operatorname{Im}\left\langle \psi,\nabla_{\alpha}\psi\right\rangle
\right)  $

$+L_{F}\partial_{\alpha}\det O^{\prime}+2a_{F}\sum_{a,\beta\gamma
}\operatorname{Re}\left(  \overline{\left(  \partial_{\alpha}\grave{A}_{\beta
}^{a}\right)  }\partial_{\gamma}\left(
\mathcal{F}%
_{A}^{a,\beta\gamma}\det O^{\prime}\right)  +%
\mathcal{F}%
_{A}^{a,\beta\gamma}\overline{\left[  \partial_{\alpha}\grave{A}_{\beta
},\grave{A}_{\gamma}\right]  }^{a}\det O^{\prime}\right)  $

$+2a_{G}\sum_{a\beta\gamma}\{\partial_{\alpha}G_{\beta}^{a}\partial_{\gamma
}\left(  \left(  O_{p_{a}}^{\gamma}O_{q_{a}}^{\beta}-O_{q_{a}}^{\gamma
}O_{p_{a}}^{\beta}\right)  \det O^{\prime}\right)  $

$\qquad+\left(  O_{p_{a}}^{\gamma}O_{q_{a}}^{\beta}-O_{q_{a}}^{\gamma}%
O_{p_{a}}^{\beta}\right)  \det O^{\prime}\left[  \partial_{\alpha}G_{\beta
},G_{\gamma}\right]  ^{a}\}$

$-2\left(  \det O^{\prime}\right)  \sum_{i\beta\gamma\mu}O_{i}^{\gamma}\left(
\partial_{\alpha}O_{\beta}^{\prime i}\right)  \left(  a_{F}\operatorname{Re}%
\left(
\mathcal{F}%
_{A}^{\beta\mu},%
\mathcal{F}%
_{A\gamma\mu}\right)  +a_{G}%
\mathcal{F}%
_{G\gamma\mu}^{a}\left(  O_{p_{a}}^{\mu}O_{q_{a}}^{\beta}-O_{q_{a}}^{\mu
}O_{p_{a}}^{\beta}\right)  \right)  $

\paragraph{2)}

Let us compute the third term :

$2a_{F}\sum_{\beta\gamma}\operatorname{Re}\left(  \left(  \partial_{\alpha
}\grave{A}_{\beta},\partial_{\gamma}\left(
\mathcal{F}%
_{A}^{a,\beta\gamma}\det O^{\prime}\right)  \right)  +\left(  \left[
\partial_{\alpha}\grave{A}_{\beta},\grave{A}_{\gamma}\right]  ,%
\mathcal{F}%
_{A}^{\beta\gamma}\det O^{\prime}\right)  \right)  $

Using the identity in $T_{1}U^{c}:$

$\forall\overrightarrow{\theta},\overrightarrow{\theta}_{1},\overrightarrow
{\theta}_{2}\in T_{1}U^{c}:\left(  \left[  \overrightarrow{\theta
},\overrightarrow{\theta}_{1}\right]  ,\overrightarrow{\theta}_{2}\right)
=-\left(  \overrightarrow{\theta}_{1},\left[  \overrightarrow{\theta
},\overrightarrow{\theta}_{2}\right]  \right)  $

$\left(  \left[  \partial_{\alpha}\grave{A}_{\beta},\grave{A}_{\gamma}\right]
,%
\mathcal{F}%
_{A}^{\beta\gamma}\det O^{\prime}\right)  $

$=-\left(  \left[  \grave{A}_{\gamma},\partial_{\alpha}\grave{A}_{\beta
},\right]  ,%
\mathcal{F}%
_{A}^{\beta\gamma}\det O^{\prime}\right)  $

$=\left(  \partial_{\alpha}\grave{A}_{\beta},\left[  \grave{A}_{\gamma},%
\mathcal{F}%
_{A}^{\beta\gamma}\det O^{\prime}\right]  \right)  $

$2a_{F}\sum_{\beta\gamma}\left(  \left(  \partial_{\alpha}\grave{A}_{\beta
},\partial_{\gamma}\left(
\mathcal{F}%
_{A}^{a,\beta\gamma}\det O^{\prime}\right)  \right)  +\left(  \left[
\partial_{\alpha}\grave{A}_{\beta},\grave{A}_{\gamma}\right]  ,%
\mathcal{F}%
_{A}^{\beta\gamma}\det O^{\prime}\right)  \right)  $

$=2a_{F}\sum_{\beta\gamma}\operatorname{Re}\left(  \partial_{\alpha}\grave
{A}_{\beta},\left[  \grave{A}_{\gamma},%
\mathcal{F}%
_{A}^{\beta\gamma}\det O^{\prime}\right]  +\partial_{\gamma}\left(
\mathcal{F}%
_{A}^{a,\beta\gamma}\det O^{\prime}\right)  \right)  $

$=2a_{F}\sum_{\beta\gamma}\operatorname{Re}\left(  \left[  \grave{A}_{\gamma},%
\mathcal{F}%
_{A}^{\beta\gamma}\det O^{\prime}\right]  +\partial_{\gamma}\left(
\mathcal{F}%
_{A}^{a,\beta\gamma}\det O^{\prime}\right)  ,\partial_{\alpha}\grave{A}%
_{\beta}\right)  $

$=2a_{F}\sum_{\beta\gamma}\sum_{a}\operatorname{Re}\left(  \overline{\left(
\left[  \grave{A}_{\gamma},%
\mathcal{F}%
_{A}^{\beta\gamma}\det O^{\prime}\right]  +\partial_{\gamma}\left(
\mathcal{F}%
_{A}^{a,\beta\gamma}\det O^{\prime}\right)  \right)  }^{a}\left(
\partial_{\alpha}\grave{A}_{\beta}^{a}\right)  \right)  $

with equation \ref{E82} on shell :

$N\left(  a_{D}V^{\beta}\rho_{a}-ia_{I}\sum_{r}O_{r}^{\beta}\left[  \mu
_{A}\right]  _{a}^{r}\right)  \det O^{\prime}$

$=-2a_{F}\sum_{\gamma}\left(  \left[  \grave{A}_{\beta},%
\mathcal{F}%
_{A}^{\beta\gamma}\det O^{\prime}\right]  ^{a}+\partial_{\gamma}\left(
\mathcal{F}%
_{A}^{a\beta\gamma}\left(  \det O^{\prime}\right)  \right)  \right)  $

So :

$2a_{F}\sum_{\beta\gamma}\left(  \left(  \partial_{\alpha}\grave{A}_{\beta
},\partial_{\gamma}\left(
\mathcal{F}%
_{A}^{a,\beta\gamma}\det O^{\prime}\right)  \right)  +\left(  \left[
\partial_{\alpha}\grave{A}_{\beta},\grave{A}_{\gamma}\right]  ,%
\mathcal{F}%
_{A}^{\beta\gamma}\det O^{\prime}\right)  \right)  $

$=-N\left(  \det O^{\prime}\right)  \sum_{a\beta}\operatorname{Re}\left(
\overline{\left(  a_{D}V^{\beta}\rho_{a}-ia_{I}\sum_{r}O_{r}^{\beta}\left[
\mu_{A}\right]  _{a}^{r}\right)  }\left(  \partial_{\alpha}\grave{A}_{\beta
}^{a}\right)  \right)  $

$=-N\left(  \det O^{\prime}\right)  \sum_{a\beta}\operatorname{Re}\left(
\left(  a_{D}V^{\beta}\rho_{a}+ia_{I}\sum_{r}O_{r}^{\beta}\left[  \mu
_{A}\right]  _{a}^{r}\right)  \left(  \partial_{\alpha}\grave{A}_{\beta}%
^{a}\right)  \right)  $

$=-N\left(  \det O^{\prime}\right)  \left(  a_{D}\sum_{a\beta}V^{\beta}%
\rho_{a}\operatorname{Re}\left(  \partial_{\alpha}\grave{A}_{\beta}%
^{a}\right)  -a_{I}\sum_{r}O_{r}^{\beta}\left[  \mu_{A}\right]  _{a}%
^{r}\operatorname{Im}\left(  \partial_{\alpha}\grave{A}_{\beta}^{a}\right)
\right)  $

\paragraph{3)}

The fourth term :

$2a_{G}\sum_{a\beta\gamma}\left(
\begin{array}
[c]{c}%
\left(  \partial_{\alpha}G_{\beta}^{a}\right)  \partial_{\gamma}\left(
\left(  O_{p_{a}}^{\gamma}O_{q_{a}}^{\beta}-O_{q_{a}}^{\gamma}O_{p_{a}}%
^{\beta}\right)  \det O^{\prime}\right) \\
+\left(  O_{p_{a}}^{\gamma}O_{q_{a}}^{\beta}-O_{q_{a}}^{\gamma}O_{p_{a}%
}^{\beta}\right)  \det O^{\prime}\left[  \partial_{\alpha}G_{\beta},G_{\gamma
}\right]  ^{a}%
\end{array}
\right)  $

\subparagraph{a)}

We have seen (equation \ref{E75}\ \ ) that :

$\sum_{\beta=0}^{3}\partial_{\beta}\left(  \left(  O_{p_{a}}^{\beta}O_{q_{a}%
}^{\alpha}-O_{q_{a}}^{\beta}O_{p_{a}}^{\alpha}\right)  \det O^{\prime}\right)
=\left(  \sum_{r}O_{r}^{\alpha}c_{a}^{r}+\left(  O_{q_{a}}^{\alpha}D_{p_{a}%
}-O_{p_{a}}^{\alpha}D_{q_{a}}\right)  \right)  \left(  \det O^{\prime}\right)
$

So :

$\sum_{a\beta}\left(  \partial_{\alpha}G_{\beta}^{a}\right)  \left(
\sum_{\gamma}\partial_{\gamma}\left(  \left(  O_{p_{a}}^{\gamma}O_{q_{a}%
}^{\beta}-O_{q_{a}}^{\gamma}O_{p_{a}}^{\beta}\right)  \det O^{\prime}\right)
\right)  $

$=\sum_{a\beta r}\left(  \partial_{\alpha}G_{\beta}^{a}\right)  \left(
O_{r}^{\beta}c_{a}^{r}+\left(  O_{q_{a}}^{\beta}D_{p_{a}}-O_{p_{a}}^{\beta
}D_{q_{a}}\right)  \right)  \left(  \det O^{\prime}\right)  $

\subparagraph{b)}

$\sum_{a\beta\gamma}\left(  O_{p_{a}}^{\gamma}O_{q_{a}}^{\beta}-O_{q_{a}%
}^{\gamma}O_{p_{a}}^{\beta}\right)  \left[  \partial_{\alpha}G_{\beta
},G_{\gamma}\right]  ^{a}=\sum_{\beta\gamma b}\left(  O_{p_{b}}^{\gamma
}O_{q_{b}}^{\beta}-O_{q_{b}}^{\gamma}O_{p_{b}}^{\beta}\right)  \left[
\partial_{\alpha}G_{\beta},G_{\gamma}\right]  ^{b}$

$=\sum_{ab\beta\gamma r}O_{r}^{\beta}\left(  \partial_{\alpha}G_{\beta}%
^{a}\right)  \left(  O_{p_{b}}^{\gamma}\delta_{q_{b}}^{r}-O_{q_{b}}^{\gamma
}\delta_{p_{b}}^{r}\right)  \left[  \overrightarrow{\kappa}_{a},G_{\gamma
}\right]  ^{b}$

$=\sum_{a\beta r}O_{r}^{\beta}\left(  \partial_{\alpha}G_{\beta}^{a}\right)
\sum_{b}\left(  O_{p_{b}}^{\gamma}\delta_{q_{b}}^{r}-O_{q_{b}}^{\gamma}%
\delta_{p_{b}}^{r}\right)  \left[  \overrightarrow{\kappa}_{a},G_{\gamma
}\right]  ^{b}$

$\sum_{b}\left(  O_{p_{b}}^{\gamma}\delta_{q_{b}}^{r}-O_{q_{b}}^{\gamma}%
\delta_{p_{b}}^{r}\right)  \left[  \overrightarrow{\kappa}_{a},G_{\gamma
}\right]  ^{b}$

$=\sum_{b}\delta_{q_{b}}^{r}\left[  \overrightarrow{\kappa}_{a},G_{p_{b}%
}\right]  ^{b}-\delta_{p_{b}}^{r}\left[  \overrightarrow{\kappa}_{a},G_{q_{b}%
}\right]  ^{b}$

$=\sum_{bc=1}^{6}G_{ac}^{b}\left(  \delta_{q_{b}}^{r}G_{p_{b}}^{c}%
-\delta_{p_{b}}^{r}G_{q_{b}}^{c}\right)  $

We have seen (equation \ref{E72a} )\ that :

$\left[  T_{G}\right]  ^{ar}=\sum_{bc=1}^{6}G_{ac}^{b}\left(  \delta_{q_{b}%
}^{r}G_{p_{b}}^{c}-\delta_{p_{b}}^{r}G_{q_{b}}^{c}\right)  $

So : $\sum_{a\beta\gamma}\left(  O_{p_{a}}^{\gamma}O_{q_{a}}^{\beta}-O_{q_{a}%
}^{\gamma}O_{p_{a}}^{\beta}\right)  \left[  \partial_{\alpha}G_{\beta
},G_{\gamma}\right]  ^{a}=\sum_{a\beta r}O_{r}^{\beta}\left(  \partial
_{\alpha}G_{\beta}^{a}\right)  \left[  T_{G}\right]  ^{ar}$

\subparagraph{c)}

$2a_{G}\sum_{a\beta\gamma}\{\left(  \partial_{\alpha}G_{\beta}^{a}\right)
\partial_{\gamma}\left(  \left(  O_{p_{a}}^{\gamma}O_{q_{a}}^{\beta}-O_{q_{a}%
}^{\gamma}O_{p_{a}}^{\beta}\right)  \det O^{\prime}\right)  $

$+\left(  O_{p_{a}}^{\gamma}O_{q_{a}}^{\beta}-O_{q_{a}}^{\gamma}O_{p_{a}%
}^{\beta}\right)  \det O^{\prime}\left[  \partial_{\alpha}G_{\beta},G_{\gamma
}\right]  ^{a}\}$

$=2a_{G}\left(  \sum_{a\beta r}\left(  \partial_{\alpha}G_{\beta}^{a}\right)
\left(  O_{r}^{\beta}c_{a}^{r}+\left(  O_{q_{a}}^{\beta}D_{p_{a}}-O_{p_{a}%
}^{\beta}D_{q_{a}}\right)  \right)  +O_{r}^{\beta}\left(  \partial_{\alpha
}G_{\beta}^{a}\right)  \left[  T_{G}\right]  ^{ar}\right)  \det O^{\prime}$

$=2a_{G}\left(  \sum_{a\beta r}\left(  \partial_{\alpha}G_{\beta}^{a}\right)
O_{r}^{\beta}\left(  c_{a}^{r}+\delta_{q_{a}}^{r}D_{p_{a}}-\delta_{p_{a}}%
^{r}D_{q_{a}}+\left[  T_{G}\right]  ^{ar}\right)  \right)  \det O^{\prime}$

But from the equation \ref{E75}:

$\left[  T_{G}\right]  ^{ar}=-c_{a}^{r}+\delta_{p_{a}}^{r}D_{q_{a}}%
-\delta_{q_{a}}^{r}D_{p_{a}}-\frac{N}{2a_{G}}\left(  a_{I}\left(  \left[
J\right]  \left[  \widetilde{\kappa}_{a}\right]  \right)  _{r}+a_{D}V^{r}%
P_{a}\right)  $

$2a_{G}\sum_{a\beta\gamma}\left(  \left(  \partial_{\alpha}G_{\beta}%
^{a}\right)  \partial_{\gamma}\left(  \left(  O_{p_{a}}^{\gamma}O_{q_{a}%
}^{\beta}-O_{q_{a}}^{\gamma}O_{p_{a}}^{\beta}\right)  \det O^{\prime}\right)
+\left(  O_{p_{a}}^{\gamma}O_{q_{a}}^{\beta}-O_{q_{a}}^{\gamma}O_{p_{a}%
}^{\beta}\right)  \det O^{\prime}\left[  \partial_{\alpha}G_{\beta},G_{\gamma
}\right]  ^{a}\right)  $

$=2a_{G}\left(  \sum_{a\beta r}\left(  \partial_{\alpha}G_{\beta}^{a}\right)
O_{r}^{\beta}\left(  -\frac{N}{2a_{G}}\left(  a_{I}\left(  \left[  J\right]
\left[  \widetilde{\kappa}_{a}\right]  \right)  _{r}+a_{D}V^{r}P_{a}\right)
\right)  \right)  \det O^{\prime}$

$=-N\sum_{a\beta r}\left(  \partial_{\alpha}G_{\beta}^{a}\right)  O_{r}%
^{\beta}\left(  a_{I}\left(  \left[  J\right]  \left[  \widetilde{\kappa}%
_{a}\right]  \right)  _{r}+a_{D}V^{r}P_{a}\right)  \det O^{\prime}$

\paragraph{4)}

The last term :

$2\sum_{i\beta\gamma\mu}O_{i}^{\gamma}\left(  \partial_{\alpha}O_{\beta
}^{\prime i}\right)  \left(  a_{F}\operatorname{Re}\left(
\mathcal{F}%
_{A}^{\beta\mu},%
\mathcal{F}%
_{A\gamma\mu}\right)  +a_{G}%
\mathcal{F}%
_{G\gamma\mu}^{a}\left(  O_{p_{a}}^{\mu}O_{q_{a}}^{\beta}-O_{q_{a}}^{\mu
}O_{p_{a}}^{\beta}\right)  \right)  \left(  \det O^{\prime}\right)  $

From equation \ref{E87} we have :

$\forall\alpha,\beta:\delta_{\beta}^{\alpha}L$

$=a_{I}N\operatorname{Im}\left\langle \psi,\gamma^{\alpha}\nabla_{\beta}%
\psi\right\rangle +2a_{F}\sum_{\gamma}\operatorname{Re}\left(
\mathcal{F}%
_{A}^{\alpha\gamma},%
\mathcal{F}%
_{A\beta\gamma}\right)  +2a_{G}\sum_{a\lambda}%
\mathcal{F}%
_{G\beta\gamma}^{a}\left(  O_{p_{a}}^{\gamma}O_{q_{a}}^{\alpha}-O_{q_{a}%
}^{\gamma}O_{p_{a}}^{\alpha}\right)  $

$2\left(  a_{F}\sum_{\gamma}\operatorname{Re}\left(
\mathcal{F}%
_{A}^{\alpha\gamma},%
\mathcal{F}%
_{A\beta\gamma}\right)  +a_{G}\sum_{a\lambda}%
\mathcal{F}%
_{G\beta\gamma}^{a}\left(  O_{p_{a}}^{\gamma}O_{q_{a}}^{\alpha}-O_{q_{a}%
}^{\gamma}O_{p_{a}}^{\alpha}\right)  \right)  $

$=\left(  \delta_{\beta}^{\alpha}L-a_{I}N\operatorname{Im}\left\langle
\psi,\gamma^{\alpha}\nabla_{\beta}\psi\right\rangle \right)  $

So :

$2\sum_{i\beta\gamma\mu}O_{i}^{\gamma}\left(  \partial_{\alpha}O_{\beta
}^{\prime i}\right)  \left(  a_{F}\operatorname{Re}\left(
\mathcal{F}%
_{A}^{\beta\mu},%
\mathcal{F}%
_{A\gamma\mu}\right)  +a_{G}%
\mathcal{F}%
_{G\gamma\mu}^{a}\left(  O_{p_{a}}^{\mu}O_{q_{a}}^{\beta}-O_{q_{a}}^{\mu
}O_{p_{a}}^{\beta}\right)  \right)  \left(  \det O^{\prime}\right)  $

$=\sum_{i\beta\gamma}O_{i}^{\gamma}\left(  \partial_{\alpha}O_{\beta}^{\prime
i}\right)  \left(  \left(  \delta_{\gamma}^{\beta}L-a_{I}N\operatorname{Im}%
\left\langle \psi,\gamma^{\beta}\nabla_{\gamma}\psi\right\rangle \right)
\right)  \left(  \det O^{\prime}\right)  $

$=\left(  \partial_{\alpha}\det O^{\prime}\right)  L+a_{I}N\left(  \det
O^{\prime}\right)  \sum_{i\beta\gamma}O_{i}^{\gamma}O_{\beta}^{\prime
i}\left(  \partial_{\alpha}O_{r}^{\beta}\right)  \operatorname{Im}\left\langle
\psi,\gamma^{r}\nabla_{\gamma}\psi\right\rangle $

$=\left(  \partial_{\alpha}\det O^{\prime}\right)  L+a_{I}N\left(  \det
O^{\prime}\right)  \sum_{\beta}\left(  \partial_{\alpha}O_{r}^{\beta}\right)
\operatorname{Im}\left\langle \psi,\gamma^{r}\nabla_{\beta}\psi\right\rangle $

\paragraph{5)}

The equation becomes :

$\forall\alpha:\frac{d}{d\xi^{0}}\left(  a_{D}N\left(  \det O^{\prime}\right)
\operatorname{Im}\left\langle \psi,\nabla_{\alpha}\psi\right\rangle \right)
=Na_{D}\det O^{\prime}\sum_{a\beta}V^{\beta}\left(  \left(  \partial_{\alpha
}G_{\beta}^{a}\right)  P_{a}+\rho_{a}\left(  \partial_{\alpha}%
\operatorname{Re}\grave{A}_{\beta}^{a}\right)  \right)  +$

$+a_{I}\det O^{\prime}N\sum_{r}\{\left(  \partial_{\alpha}G_{\beta}%
^{a}\right)  O_{r}^{\beta}\left(  \left[  J\right]  \left[  \widetilde{\kappa
}_{a}\right]  \right)  _{r}-O_{r}^{\beta}\left[  \mu_{A}\right]  _{a}%
^{r}\left(  \partial_{\alpha}\operatorname{Im}\grave{A}_{\beta}^{a}\right)  $

$-O_{i}^{\gamma}\left(  \partial_{\alpha}O_{\beta}^{\prime i}\right)
\operatorname{Im}\left\langle \psi,\gamma^{\beta}\nabla_{\gamma}%
\psi\right\rangle +NL_{M}\partial_{\alpha}\det O^{\prime}\}$

That is:

$\frac{d}{d\xi^{0}}\left(  a_{D}N\left(  \det O^{\prime}\right)
\operatorname{Im}\left\langle \psi,\nabla_{\alpha}\psi\right\rangle \right)  $

$=NL_{M}\partial_{\alpha}\det O^{\prime}+\left(  N\det O^{\prime}\right)
a_{D}\sum_{a\beta}V^{\beta}\left(  \left(  \partial_{\alpha}G_{\beta}%
^{a}\right)  P_{a}+\rho_{a}\left(  \partial_{\alpha}\operatorname{Re}\grave
{A}_{\beta}^{a}\right)  \right)  +$

$+N\det O^{\prime}a_{I}\sum_{\beta r}\left(  \partial_{\alpha}O_{r}^{\beta
}\operatorname{Im}\left\langle \psi,\gamma^{r}\nabla_{\beta}\psi\right\rangle
+\sum_{a}O_{r}^{\beta}\left(  \partial_{\alpha}G_{\beta}^{a}\left(  \left[
J\right]  \left[  \widetilde{\kappa}_{a}\right]  \right)  _{r}-\left[  \mu
_{A}\right]  _{a}^{r}\partial_{\alpha}\operatorname{Im}\grave{A}_{\beta}%
^{a}\right)  \right)  $

We have seen that :

$\left\langle \psi,\gamma^{r}\nabla_{\beta}\psi\right\rangle =-\frac{dJ_{r}%
}{d\xi^{\beta}}+\sum_{a}G_{\beta}^{a}\left(  \left[  J\right]  \left[
\widetilde{\kappa}_{a}\right]  \right)  _{r}-\left[  \mu_{A}\right]  _{a}%
^{r}\operatorname{Im}\grave{A}_{\beta}^{\alpha}$

$\sum_{\beta r}\left(  \left(  \partial_{\alpha}O_{r}^{\beta}\right)
\operatorname{Im}\left\langle \psi,\gamma^{r}\nabla_{\beta}\psi\right\rangle
+\sum_{a}O_{r}^{\beta}\left(  \left(  \partial_{\alpha}G_{\beta}^{a}\right)
\left(  \left[  J\right]  \left[  \widetilde{\kappa}_{a}\right]  \right)
_{r}-\left[  \mu_{A}-\mu_{L}\right]  _{a}^{r}\left(  \partial_{\alpha
}\operatorname{Im}\grave{A}_{\beta}^{a}\right)  \right)  \right)  $

$=\sum_{\beta r}\{\left(  \partial_{\alpha}O_{r}^{\beta}\right)  \left(
-\frac{dJ_{r}}{d\xi^{\beta}}+\sum_{a}G_{\beta}^{a}\left(  \left[  J\right]
\left[  \widetilde{\kappa}_{a}\right]  \right)  _{r}-\left[  \mu_{R}-\mu
_{L}\right]  _{a}^{r}\operatorname{Im}\grave{A}_{\beta}^{\alpha}\right)  $

$+\sum_{a}O_{r}^{\beta}\left(  \left(  \partial_{\alpha}G_{\beta}^{a}\right)
\left(  \left[  J\right]  \left[  \widetilde{\kappa}_{a}\right]  \right)
_{r}-\left[  \mu_{A}-\mu_{L}\right]  _{a}^{r}\left(  \partial_{\alpha
}\operatorname{Im}\grave{A}_{\beta}^{a}\right)  \right)  \}$

$=\sum_{\beta r}\left(  \left(  -\frac{dJ_{r}}{d\xi^{\beta}}\left(
\partial_{\alpha}O_{r}^{\beta}\right)  +\sum_{a}\left(  \partial_{\alpha}%
O_{r}^{\beta}G_{\beta}^{a}\right)  \left(  \left[  J\right]  \left[
\widetilde{\kappa}_{a}\right]  \right)  _{r}-\left[  \mu_{A}\right]  _{a}%
^{r}\operatorname{Im}\left(  \partial_{\alpha}O_{r}^{\beta}\grave{A}_{\beta
}^{\alpha}\right)  \right)  \right)  $

$=\sum_{\beta r}\left(  \left(  -\frac{dJ_{r}}{d\xi^{\beta}}\left(
\partial_{\alpha}O_{r}^{\beta}\right)  +\sum_{a}\left(  \partial_{\alpha}%
G_{r}^{a}\right)  \left(  \left[  J\right]  \left[  \widetilde{\kappa}%
_{a}\right]  \right)  _{r}-\left[  \mu_{A}\right]  _{a}^{r}\operatorname{Im}%
\left(  \partial_{\alpha}\grave{A}_{r}^{\alpha}\right)  \right)  \right)  $

$\frac{d}{d\xi^{0}}\left(  a_{D}N\left(  \det O^{\prime}\right)
\operatorname{Im}\left\langle \psi,\nabla_{\alpha}\psi\right\rangle \right)  $

$=NL_{M}\partial_{\alpha}\det O^{\prime}+\left(  N\det O^{\prime}\right)
a_{D}\sum_{a\beta}V^{\beta}\left(  \left(  \partial_{\alpha}G_{\beta}%
^{a}\right)  P_{a}+\rho_{a}\left(  \partial_{\alpha}\operatorname{Re}\grave
{A}_{\beta}^{a}\right)  \right)  +$

$+\left(  N\det O^{\prime}\right)  a_{I}\sum_{\beta r}\left(  \left(
-\frac{dJ_{r}}{d\xi^{\beta}}\left(  \partial_{\alpha}O_{r}^{\beta}\right)
+\sum_{a}\left(  \partial_{\alpha}G_{r}^{a}\right)  \left(  \left[  J\right]
\left[  \widetilde{\kappa}_{a}\right]  \right)  _{r}-\left[  \mu_{A}\right]
_{a}^{r}\operatorname{Im}\left(  \partial_{\alpha}\grave{A}_{r}^{\alpha
}\right)  \right)  \right)  $

\paragraph{6)}

The derivative reads:

$\frac{d}{d\xi^{0}}\left(  a_{D}\left(  \det O^{\prime}\right)
N\operatorname{Im}\left\langle \psi,\nabla_{\alpha}\psi\right\rangle \right)
$

$=a_{D}N\frac{d}{d\xi^{0}}\left(  \left(  \det O^{\prime}\right)
\operatorname{Im}\left\langle \psi,\nabla_{\alpha}\psi\right\rangle \right)  $
as N is constant along $\xi^{0}$

$a_{D}N\frac{d}{d\xi^{0}}\left(  \left(  \det O^{\prime}\right)
\operatorname{Im}\left\langle \psi,\nabla_{\alpha}\psi\right\rangle \right)
=a_{D}N\sum_{\beta}\frac{df^{\beta}}{d\xi^{0}}\partial_{\beta}\left(
\operatorname{Im}\left\langle \psi,\nabla_{\alpha}\psi\right\rangle \det
O^{\prime}\right)  $

$=a_{D}N\sum_{\beta}V^{\beta}\partial_{\beta}\left(  \operatorname{Im}%
\left\langle \psi,\nabla_{\alpha}\psi\right\rangle \det O^{\prime}\right)  $

so, assuming that N$\neq0$ we have the equation :%

\begin{equation}
\label{E93}
\end{equation}

$\forall\alpha:a_{D}\sum_{\beta}V^{\beta}\partial_{\beta}\left(
\operatorname{Im}\left\langle \psi,\nabla_{\alpha}\psi\right\rangle \det
O^{\prime}\right)  $

$=L_{M}\partial_{\alpha}\det O^{\prime}+\left(  \det O^{\prime}\right)
a_{D}\sum_{a\beta}\left(  V^{\beta}\left(  \partial_{\alpha}G_{\beta}%
^{a}\right)  P_{a}+V^{\beta}\rho_{a}\left(  \partial_{\alpha}\operatorname{Re}%
\grave{A}_{\beta}^{a}\right)  \right)  $

$+\left(  \det O^{\prime}\right)  a_{I}\sum_{\beta r}\left(  -\frac{dJ_{r}%
}{d\xi^{\alpha}}\left(  \partial_{\alpha}O_{r}^{\beta}\right)  +\sum
_{a}\left(  \partial_{\alpha}G_{r}^{a}\right)  \left(  \left[  J\right]
\left[  \widetilde{\kappa}_{a}\right]  \right)  _{r}-\left[  \mu_{A}\right]
_{a}^{r}\operatorname{Im}\left(  \partial_{\alpha}\grave{A}_{r}^{\alpha
}\right)  \right)  $

\bigskip

\section{CHOOSING\ A\ GAUGE}

\label{Choosing a gauge}

We have, all together 16m+36 first degree partial differential equations for
16m+36 variables. But the 16 tetrad coefficients O' are defined within a
SO(3,1) matrix and so we could fix 10 parameters. Now we intend to use this
gauge freedom, by choosing a chart and a tetrad.

\paragraph{1)}

The most physical choice for the chart is just that we have built in the
beginning :

- the time vector n is taken as the 0 basis vector, both for the tetrad and
the chart.\ In this section $\left(  t,\xi=\left(  \xi^{1},\xi^{2},\xi
^{3}\right)  \right)  $ are the coordinates along n on one hand, and in a
chart of S(t) on the other hand. So we have%

\begin{equation}
O_{0}^{\alpha}=\delta_{0}^{\alpha},O_{0}^{\prime i}=\delta_{0}^{i}\label{E94}%
\end{equation}

$\frac{d}{dt}O_{i}^{\alpha}\left(  \xi,t\right)  =n^{\beta}\partial_{\beta
}O\left(  \xi,t\right)  =\partial_{4}O_{i}^{\alpha}\left(  \xi,t\right)
=\partial_{t}O_{i}^{\alpha}\left(  \xi,t\right)  $

- the vectors $\partial^{i}$ are parallel transported along a gravitational
geodesic :

$\forall i:\widehat{\nabla}\partial_{i}\left(  n\right)  =0=G_{0}^{a}\left[
\widetilde{\kappa}_{a}\right]  _{i}^{j}\partial_{j}=G_{0}^{a}\left(  \left(
\delta^{jp}\eta_{iq}-\delta^{jq}\eta_{ip}\right)  \right)  \partial_{j}%
=G_{0}^{a}\eta_{iq}\partial_{p}-B_{0}^{a}\eta_{ip}\partial_{q}=0$

$G_{0}^{a}\eta_{pp}\partial_{q}=0;G_{0}^{a}\partial_{p}=0$

$\Rightarrow$%

\begin{equation}
\forall a:G_{0}^{a}=0\label{E95}%
\end{equation}

- the world lines of the particles are therefore such that $f^{0}%
=t\Rightarrow\frac{df^{0}}{d\xi^{\alpha}}=\delta_{\alpha}^{0}$

- the function N does not depend on t

\paragraph{2)}

As we have seen the structure coefficients are key variables in the model :

$c_{pq}^{r}=\sum_{\alpha\beta}O_{\alpha}^{\prime r}\left(  O_{p}^{\beta
}\partial_{\beta}O_{q}^{\alpha}-O_{q}^{\beta}\partial_{\beta}O_{p}^{\alpha
}\right)  =\left[  \partial_{p},\partial_{q}\right]  ^{r}$

There are 24 coefficients denoted $c_{a}^{r}=c_{p_{a}q_{a}}^{r}$ which \ are
not independant.\ Indeed there are the Jacobi identities coming from the
commutator rules :

$\forall\left(  a,b,c\right)  :$ $\left[  \partial_{a},\left[  \partial
_{b},\partial_{c}\right]  \right]  +\left[  \partial_{b},\left[  \partial
_{c},\partial_{a}\right]  \right]  +\left[  \partial_{c},\left[  \partial
_{a},\partial_{b}\right]  \right]  =0$

$\Rightarrow\forall\left(  a,b,c\right)  ,d=0..3:\sum_{i=1}^{4}\left(
c_{bc}^{i}c_{ai}^{d}+c_{ca}^{i}c_{bi}^{d}+c_{ab}^{i}c_{ci}^{d}\right)  =0$

With the present assumptions :

$c_{pq}^{0}=\sum_{\alpha\beta}\left(  O_{p}^{\beta}\partial_{\beta}O_{q}%
^{0}-O_{q}^{\beta}\partial_{\beta}O_{p}^{0}\right)  =0,c_{0q}^{r}=\sum
_{\alpha}O_{\alpha}^{\prime r}\partial_{t}O_{q}^{\alpha},$

and we have 18 non null coefficients linked by 12 identities (with d=1,2,3),
so only 6 of them can be considered as independant.

The Jacobi identities can be conveniently put on the matrix form:

\bigskip

$%
\begin{bmatrix}
\left(  c_{3}^{2}-c_{2}^{3}\right)  & \left(  c_{3}^{1}+c_{1}^{3}\right)  &
\left(  c_{2}^{1}-c_{1}^{2}\right)  & 0 & 0 & 0\\
\left(  c_{6}^{3}+c_{5}^{2}\right)  & c_{5}^{1} & -c_{6}^{1} & -c_{1}^{1} &
-c_{1}^{2} & -c_{1}^{3}\\
c_{4}^{2} & -\left(  c_{4}^{1}+c_{6}^{3}\right)  & c_{6}^{2} & c_{2}^{1} &
c_{2}^{2} & c_{2}^{3}\\
-c_{4}^{3} & c_{5}^{3} & c_{4}^{1}+c_{5}^{2} & -c_{3}^{1} & -c_{3}^{2} &
-c_{3}^{3}%
\end{bmatrix}%
\begin{bmatrix}
c_{1}^{1} & c_{1}^{2} & c_{1}^{3}\\
c_{2}^{1} & c_{2}^{2} & c_{2}^{3}\\
c_{3}^{1} & c_{3}^{2} & c_{3}^{3}\\
c_{4}^{1} & c_{4}^{2} & c_{4}^{3}\\
c_{5}^{1} & c_{5}^{2} & c_{5}^{3}\\
c_{6}^{1} & c_{6}^{2} & c_{6}^{3}%
\end{bmatrix}
=0$

\bigskip

The divergence D of the tetrad takes here the following value :

$D_{1}=\left(  c_{3}^{2}-c_{2}^{3}\right)  $

$D_{2}=\left(  -c_{3}^{1}+c_{1}^{3}\right)  $

$D_{3}=\left(  c_{2}^{1}-c_{1}^{2}\right)  $

$D_{0}=-\left(  c_{4}^{1}+c_{5}^{2}+c_{6}^{3}\right)  $

The gravitational Noether current $Y_{G}$ is conserved, and it is directly
related to the structure coefficients.\ Thus it should be doable to take these
coefficients as constant in a first approximation.

\paragraph{3)}

From $G_{0}^{a}=0$ we deduce :

$-K_{1}^{0}+K_{5}^{3}-K_{6}^{2}=c_{1}^{4}-c_{5}^{3}+c_{6}^{2}$

$-K_{2}^{0}+K_{6}^{1}-K_{4}^{3}=-c_{6}^{1}+c_{2}^{0}+c_{4}^{3}$

$-K_{3}^{0}-K_{5}^{1}+K_{4}^{2}=c_{5}^{1}-c_{4}^{2}+c_{3}^{0}$

$K_{4}^{0}+K_{2}^{3}-K_{3}^{2}=-2c_{4}^{0}$

$K_{5}^{0}+K_{3}-K_{1}^{3}=-2c_{5}^{0}$

$K_{6}^{0}-K_{2}+K_{1}^{2}=-2c_{6}^{0}$

In most of the problems where the coefficients structure are either assumed
constants, or linked by some symmetry, it is convenient to keep K as the
variables, and the $K_{a}^{0}$ are given by the previous relations. The
gravitational potential becomes :

\bigskip

$2G_{r}^{a}=$

$%
\begin{bmatrix}
r=1\\
K_{1}^{1}+c_{1}^{1}-K_{2}^{2}-c_{2}^{2}-K_{3}^{3}-c_{3}^{3}\\
2K_{1}^{2}+2c_{2}^{1}+2c_{6}^{0}\\
2K_{1}^{3}+2c_{3}^{1}-2c_{5}^{0}\\
-K_{4}^{1}-2c_{4}^{1}+K_{5}^{2}+K_{6}^{3}\\
-2K_{4}^{2}-2c_{4}^{2}\\
-2K_{4}^{3}-2c_{4}^{3}%
\end{bmatrix}
$

$%
\begin{bmatrix}
r=2\\
2K_{2}^{1}+2c_{1}^{2}-2c_{6}^{0}\\
-K_{1}^{1}-c_{1}^{1}+K_{2}^{2}+c_{2}^{2}-K_{3}^{3}-c_{3}^{3}\\
2K_{2}^{3}+2c_{3}^{2}+2c_{4}^{0}\\
-2K_{5}^{1}-2c_{5}^{1}\\
K_{4}^{1}-K_{5}^{2}-2c_{5}^{2}+K_{6}^{3}\\
-2K_{5}^{3}-2c_{5}^{3}%
\end{bmatrix}
$

$%
\begin{bmatrix}
r=3\\
2K_{3}^{1}+2c_{1}^{3}+2c_{5}^{0}\\
+2K_{3}^{2}+2c_{2}^{3}-2c_{4}^{0}\\
-K_{1}^{1}-c_{1}^{1}-K_{2}^{2}-c_{2}^{2}+K_{3}^{3}+c_{3}^{3}\\
-2K_{6}^{1}-2c_{6}^{1}\\
-2K_{6}^{2}-2c_{5}^{3}\\
K_{4}^{1}+K_{5}^{2}-K_{6}^{3}-2c_{6}^{3}%
\end{bmatrix}
$

\bigskip

\paragraph{4)}

In the Special Relativity picture all the structure coefficient are null, but
the connection \textbf{G}\ , depending on V and J, is not null. It can be seen
as the stress-tensor of the system. But of course a lagrangian with the scalar
curvature R is questionable in this picture, at the least.\ However the fields
equations \ref{E82} are still fully valid.

\newpage

\part{\textbf{SYMMETRIES}}

Symmetries are everywhere in physics, but the same word is used with many
different meanings. They arouse some of the most difficult questions in
physics, so they cannot be dealt with lightly. According to the relativity
principle physical reality does not depend on the mathematics that we used,
and the measures of two observers for the same system should be
\textit{equivariant}, meaning that they can be deduced from each other with
only the knowledge of the mathematical rules to pass from one observer to the
other. We have used abondantly this principle in this paper. If these measures
are \textit{identical} we shall conclude that the system itself is special :
there is a physical symmetry. In both cases the mathematics involved are
similar, they rely on group theory, and physical symmetries are identified by
a departure from the general rule of equivariance, which must be set up first.
But their physical meaning is very different. Equivariance is a consistency
requirement, assuming that the right mathematical structure has been picked up
to describe a set of measures. A physical symmetry should be an experimental
outcome, requiring changes in a pre-existent model by adding assumptions about
the configuration of the system or the mechanisms that it tries to describe.

It can happen that all observers get the same measures for a physical quantity
: it can be a special case of equivariance, requiring that the quantity should
be described by scalar functions (in differential geometry), or that the
configuration of the system is isotropic. But most often physical symmetries
can be seen by a specific class of observers only and it is convenient to
characterize the symmetry by this class. For instance a cylinder looks the
same for the observers located in the same plane orthogonal to its axis, and
one concludes that the body has a "cylindrical symmetry". This is convenient,
and authorized by the usual duality active / passive measures, but to some
extent only, as we will see later.

Symmetries are often a specific characteristic of the system itself : there
are not an issue, just a good mean to alleviate the computations. In
theoretical physics, and in the kind of model that is involved here, one looks
for symmetries that occur in any system, whatever its initial configuration,
possibly for certain kinds of physical objets, as a way to classify these
objects. They have been extensively studied in particles physics, where one
discerns 3 "symmetry modes" (Guidry [7]), that we will address in several ways :

- the "Wigner mode" : the lagrangian is symmetric and the vacuum is invariant

- the "Goldstone mode" : there is a global symmetry for the lagrangian but the
vacuum is not invariant

- the "Higgs mode" : there is a local symmetry for the lagrangian, and the
vacuum is not invariant

The "vacuum" is essentially a quantum concept. In our picture one can see the
vacuum as the value of the section $\psi:M\rightarrow E_{M},$ without the map
f, which defines the initial state of the system. This section has a life on
its own, closely linked to the physical objects involved. To say that the
vacuum is symmetric is a strong assumption about the true nature of the
particles and the fields. Notably the choice of the group to represent $\psi
$\ depends on how we see the particles. One has a "Wigner symmetry" if there
is a subgroup H of $Spin(3,1)\times U$ such that 2 states $\psi,\psi^{\prime}$
related by a gauge in H look identical. One has this kind of symmetry whenever
one uses an hermitian scalar product : if $\psi^{\prime}=z\psi$ with z a
c-number of module 1 one gets the same value for the lagrangian. This symmetry
is usually seen as a mathematical artefact that the physicists discard by a
normalization of the state vectors.

The change of orientation of space-time (CPT invariance) or of the signature
of the metric are related to the "Goldstone mode".

We will address also with more details the spatial and physical symmetries,
which are of particular importance.

We will also give a general picture of the "symmetry breakdown", which is a
fundamental mechanism in particles physics to give a mass to the bosons.

\section{CPT INVARIANCE}

\label{CPT}

In the first section we noticed that, if the universe is orientable, a change
of gauge that does not preserve the orientation cannot be acceptable. And it
is easy to check in the model that the equations are significantly altered in
a change of space (P) or time (T) orientation. But a global change of both the
space and the time orientation could be acceptable, because it does keep the
orientation of the 4 dimensional universe. Such a change is indeed a change of
chart, that we have studied in the "covariance" section. It is easy to check
that the equations are not altered, and it is just the consequence of their
covariance. Notice that the Green function N keeps its sign (it is positive on S(0)).

But there is a significant change with the Noether currents : the moments J,
P,$\rho,\mu$ are scalar functions, invariant by a change of chart or
gauge.\ And in the model they act by multiplication with vectors (usually
$\partial_{r}$ or $V^{\alpha})$\ . So if the orientation of this vector is
reversed, one should change the sign of J, P,$\rho,\mu$ to stay consistent.
This operation, denoted as "C symmetry" (for "charge"), is not included in the
"gauge transform" package, it must be dealt with separately.

The usual interpretation is that the label (positive, negative,..) that we
give to particles is purely conventional and linked with the choice made to
name "direct" or "indirect" the geometric basis. The interesting point is that
the gravitational charges are equally affected.

\section{SIGNATURE}

\label{Signature}

\paragraph{1)}

If the signature becomes (- + + + ) -%
$>$
(+ - - - ) : $\eta^{rr}$\ takes the new values $\eta^{00}=+1,r>0:\eta
^{rr}=-1.$ We have seen that the Clifford algebras Cl(1,3) and Cl(3,1) are not
isomorphic, but the representation $\left(  F,\rho\right)  $ of Cl(4,C) is not
affected, so the $\gamma$ matrices and the split of F are unchanged. The
change of the signature\ impacts only the quantities defined through the map
$\Upsilon$, that is :

- the matrices $\left[  \kappa_{a}\right]  :a<4:\left[  \kappa_{a}^{\prime
}\right]  =-\left[  \kappa_{a}\right]  $

- the matrices $\left[  \gamma^{r}\right]  :\left[  \gamma^{\prime r}\right]
^{\ast}=-i\eta^{rr}\left[  \gamma^{r}\right]  ^{\ast}$

The scalar product is unchanged.

One goes from one signature to the other by :

$\psi\rightarrow\psi^{\prime}=\sum_{jk}-i\eta^{jj}\psi^{jk}e_{j}\otimes
f_{k}\Leftrightarrow\left[  \psi^{\prime}\right]  =-i\left[  \eta\right]
\left[  \psi\right]  $

\paragraph{2)}

In the choice of a lagrangian the key argument was the imaginary and real
parts of $\sum_{r}Tr\left(  \left[  \psi^{\ast}\right]  \gamma_{0}\gamma
^{r}\left[  \partial_{\alpha}\psi\right]  \right)  $. With the other signature
it becomes

$\sum_{r}Tr\left(  \left[  \psi^{\prime\ast}\right]  \gamma_{0}\gamma^{\prime
r}\left[  \partial_{\alpha}\psi^{\prime}\right]  \right)  $

$=\sum_{r}Tr\left(  i\left[  \psi\right]  ^{\ast}\left[  \eta\right]  \left[
\gamma_{0}\right]  \left(  i\eta^{rr}\right)  \left[  \gamma^{r}\right]
\left(  -i\left[  \eta\right]  \left[  \partial_{\alpha}\psi\right]  \right)
\right)  $

$=\sum_{r}i\eta^{rr}Tr\left(  \left[  \psi\right]  ^{\ast}\left(  \left[
\eta\right]  \gamma_{0}\left[  \gamma^{r}\right]  \left[  \eta\right]
\right)  \left[  \partial_{\alpha}\psi\right]  \right)  $

So we would be lead to take the opposite choice for the lagrangian.

\paragraph{3)}

Therefore check the consequences of the choice of signature is not a simple
thing...Intuitively all the equations which can be expressed with the moments
should not be impacted, but for a change of definition for the momenta.\ This
leaves the equation of state which depends heavily of the coordinates. I will
let this issue open.\ If ever this model was pertinent, this change would,
perhaps, be a mean to test the physical significance of the signature. After
all we have so far no sensible explanation for the imbalance between matter
and anti-matter.

\section{SPATIAL\ SYMMETRIES}

\label{Spatial symmetries}

Spatial symmetry is a subtle matter.\ We will first define what could be a
spatial symmetry.\ In the general framework used to describe the state of
particles we will characterize "symmetric states".\ It will be done without
any model, even without the principle of least action.\ Eventually, with the
simple model built previously we will try to bring some light to the results.

\subsection{Definition}

\paragraph{1)}

We will say that there is a spatial symmetry if the measures done by a class
of observers, using different spatial gauges, are identical for any system, at
least for some kinds of particles. Mathematically there is :

- a set of state tensors, described as the product $F_{S}\otimes W_{S},$ of
two vector subspaces $W_{S}\subset W$\ ,$F_{S}\subset F$ ,

- a subgroup S of of Spin(3,1),

such that two observers (located a the same point as the particle) whose
frames $\left(  \partial_{r}\right)  $ differ by $s\in S$ get the same
measures on the system whenever $\psi\in$\ $F_{S}\otimes W_{S}$

\paragraph{2)}

Let us assume that there is a symmetry on the Wigner mode.\ So there is a
group H and an action $\varkappa:H\rightarrow L(F\otimes W;F\otimes W)$\ such
that the states $\psi$ and $\widehat{\psi}=\varkappa\left(  h\right)
\psi,\forall h\in H$ are imperceptible.

The spatial symmetry is then fully defined by the condition :

$\forall\psi\in F_{S}\otimes W_{S},\forall s\in S,\forall u\in U,\exists h\in
H:$

$\vartheta\left(  s,u\right)  \psi=\varkappa\left(  h\right)  \vartheta\left(
1,u\right)  \psi\Leftrightarrow\forall\psi\in F_{S}\otimes W_{S}%
,\vartheta\left(  S,U\right)  \psi\subset\varkappa\left(  H\right)  \psi$

which implies

$\forall\psi\in F_{S}\otimes W_{S},\forall\left(  s,u\right)  \in S\times
U:\vartheta\left(  s,u\right)  \psi\in F_{S}\otimes W_{S}$

because $\vartheta\left(  s^{\prime},u^{\prime}\right)  \vartheta\left(
s,u\right)  \psi=\vartheta\left(  s^{\prime}s,u^{\prime}u\right)
\psi=\varkappa\left(  h^{\prime}h\right)  \vartheta\left(  1,u^{\prime
}u\right)  \psi$

Therefore $\left(  F_{S}\otimes W_{S},\vartheta\right)  $ is a linear
representation of SxU.

\paragraph{3)}

We will focus on the important case where H=U(1) : two states $\psi
,\psi^{\prime}$ are imperceptible if there is a c-complex number such that
$\psi^{\prime}=z\psi,\left\vert z\right\vert =1.$ It happens whenever the
lagrangian is defined from hermitian scalar products. So the condition reads :

$\forall\psi\in F_{S}\otimes W_{S},\forall s\in S,\forall u\in U,\exists
\theta\in%
\mathbb{R}
:\vartheta\left(  s,u\right)  \psi=e^{i\theta}\vartheta\left(  1,u\right)
\psi$

$\theta$ can depend on s, u and $\psi.$

Notice that this definition, per se, is quite general, and does not involve
the principle of least action.

\subsection{The space of spacially symmetric states}

\paragraph{1)}

Let be the m vectors of F such that :

$\phi_{j}=\sum_{i=1}^{4}\psi^{ij}e_{i}$ so $\psi=\sum_{j=1}^{m}\phi_{j}\otimes
f_{j}$

$\vartheta\left(  s,u\right)  \psi=\sum_{k,l}\psi^{kl}\left[  \rho
\circ\Upsilon\left(  s\right)  \right]  _{k}^{i}\left[  \chi\left(  u\right)
\right]  _{l}^{j}e_{i}\otimes f_{j}$

So if $\psi\in F_{S}\otimes W_{S}:\forall s\in S,\forall u\in U:$

$\sum_{j=1}^{m}\vartheta\left(  s,u\right)  \left(  \phi_{j}\otimes
f_{j}\right)  =\sum_{j=1}^{m}e^{i\theta\left(  s\right)  }\phi_{j}\otimes
f_{j}=\sum_{j=1}^{m}\left(  \rho\circ\Upsilon\left(  s\right)  \phi
_{j}\right)  \otimes\left(  \chi\left(  u\right)  f_{j}\right)  $

Let u=1 : $\sum_{j=1}^{m}\left(  \rho\circ\Upsilon\left(  s\right)  \phi
_{j}\right)  \otimes f_{j}=\sum_{j=1}^{m}\left(  e^{i\theta\left(  s\right)
}\phi_{j}\right)  \otimes f_{j}$

$\Rightarrow\forall\psi\in F_{S}\otimes W_{S},\forall s\in S:\rho\circ
\Upsilon\left(  s\right)  \phi_{j}=e^{i\theta\left(  s\right)  }\phi_{j}$

So for each j $\left(  \phi_{j},\rho\circ\Upsilon|_{S}\right)  $ is a linear
representation of S, of complex dimension 1. Besides the scalar matrices the
only subgroups \ S of Spin(3,1) which admit such a representation are the
abelian groups generated by $\left(  \overrightarrow{\kappa}_{a}\right)
_{a=1}^{3}$ on one hand, $\left(  \overrightarrow{\kappa}_{a}\right)
_{a=4}^{6}$ on the other.

In the standard representation of SO(3,1) the first is the subgroup of
rotations with fixed axis of components $\left(  0,y_{1},y_{2},y_{3}\right)  $
in an orthonormal basis with $y_{1}^{2}+y_{2}^{2}+y_{3}^{2}=1$, generated by
the matrices $\jmath\circ\mu\left(  \exp\tau\left(  y_{1}\overrightarrow
{\kappa}_{1}+y_{2}\overrightarrow{\kappa}_{2}+y_{3}\overrightarrow{\kappa}%
_{3}\right)  \right)  .$

The second is the subgroup of space-time rotations of matrices $\ $

$\jmath\circ\mu\left(  \exp\tau\left(  z_{1}\overrightarrow{\kappa}_{4}%
+z_{2}\overrightarrow{\kappa}_{5}+z_{3}\overrightarrow{\kappa}_{6}\right)
\right)  ,$ with $z_{1}^{2}+z_{2}^{2}+z_{3}^{2}=1.$

They correspond to the coordinate change between two observers moving at the
spatial speed $\left(  z_{1}c\tanh\tau,z_{2}c\tanh\tau,z_{3}c\tanh\tau\right)
.$

\paragraph{2)}

The action of these two subgroups can be described by $\rho\circ
\Upsilon\left(  \sum r^{a}\overrightarrow{\kappa}_{a}\right)  $ where the
components $r^{a}$ are fixed and the index a runs from 1 to 3 for the first
and 4 to 6 for the second subgroup.

So :\ 

$\forall\tau\in%
\mathbb{R}
:\vartheta\left(  \exp\tau\sum r^{a}\overrightarrow{\kappa}_{a},1_{U}\right)
\psi$

$=\sum_{i,k,j}\psi^{kj}\left[  \rho\circ\Upsilon\left(  \exp\tau\sum
r^{a}\overrightarrow{\kappa}_{a}\right)  \right]  _{k}^{i}e_{i}\otimes f_{j}$

$=\sum_{j}\left[  \rho\circ\Upsilon\left(  \exp\tau\sum r^{a}\overrightarrow
{\kappa}_{a}\right)  \right]  \left(  \phi_{j}\right)  \otimes f_{j}$

$\vartheta\left(  \exp\tau\sum r^{a}\overrightarrow{\kappa}_{a},1_{G}\right)
\psi=\sum_{i,j}e^{i\theta\left(  \tau\right)  }\psi^{ij}e_{i}\otimes
f_{j}=\sum_{j}e^{i\theta\left(  \tau\right)  }\phi_{j}\otimes f_{j}$

$\Leftrightarrow\forall_{j}:\left[  \rho\circ\Upsilon\left(  \exp\tau\sum
r^{a}\overrightarrow{\kappa}_{a}\right)  \right]  \left(  \phi_{j}\right)
=e^{i\theta\left(  \tau\right)  }\phi_{j}$

By differenciating with respect to $\tau$\ in $\tau=0$ :

$\left(  \rho\circ\Upsilon\right)  ^{\prime}(1)\left(  \sum r^{a}%
\overrightarrow{\kappa}_{a}\right)  \phi_{j}=i\theta^{\prime}\left(  0\right)
\phi_{j}=\sum_{a}r^{a}\left[  \kappa_{a}\right]  \phi_{j}=\left[  \kappa
_{S}\right]  \phi_{j}$

So a necessary condition is that the vectors $\phi_{j}$ are eigen vectors of
the matrices $\left[  \kappa_{S}\right]  =\sum r^{a}\left[  \kappa_{a}\right]
$ with the same imaginary eigen value. Conversely $\left(  \rho\circ
\Upsilon,F\right)  $ is a representation of Spin(3,1), so $\left(
\rho^{\prime}\circ\Upsilon^{\prime}(1),F\right)  $ is a representation of
o(3,1), and the matrices of the representation of Spin(3,1) in F are the
exponential of the matrices of the representation of o(3,1), which are here
the matrices $\left[  \kappa_{S}\right]  .$ Thus : $\rho\circ\Upsilon\left(
\exp\tau\sum r_{a}^{a}\overrightarrow{\kappa}\right)  =\exp\left(  \left[
\kappa_{S}\right]  \right)  .$If $\phi_{j}$ is eigen vector of $\left[
\kappa_{S}\right]  $ with the eigen value $i\lambda,\lambda\in%
\mathbb{R}
$\ it is an eigen vector of $\exp\tau\left[  \kappa_{S}\right]  $ with eigen
value $\exp\tau i\lambda$ and this condition is also sufficient.

\paragraph{3)}

The matrices $\left[  \kappa_{a}\right]  $ are built from the matrices
$\gamma,$ which are defined within conjugation and so their eigen values do
not depend of the choice of the $\gamma$ matrices and we can take those
defined previously. The eigen values are $\pm\frac{1}{2}i$\ for the first
subgroup and $\pm\frac{1}{2}$ for the second.\ Therefore the spatial rotations
are the only possible case, and $\theta\left(  s\right)  =\frac{\epsilon}%
{2},\epsilon=\pm1$. The $\left[  \kappa_{S}\right]  $\ matrices are normal, so
diagonalizable and each of their eigen spaces $F_{\epsilon}\left(  r\right)  $
is 2 dimensional.

The axis of the rotation is fixed by the eigen vector r (which is space like)
and the direction of the rotation by $\epsilon.$

\paragraph{4)}

Let $\left(  u_{\epsilon}\left(  r\right)  ,v_{\epsilon}\left(  r\right)
\right)  $ be a basis of $F_{\epsilon}\left(  r\right)  .$For each symmetric
state $\psi$ there is some r such that:

$\psi=\sum_{j=1}^{m}\phi_{j}\otimes f_{j},\phi_{j}\in F_{\epsilon}\left(
r\right)  $

that we can write : $\ \phi_{j}=a_{j}u_{\epsilon}\left(  r\right)
+b_{j}v_{\epsilon}\left(  r\right)  ,\left(  a_{j},b_{j}\right)  \in%
\mathbb{C}
$

So the set of symmetric states is the subspace of $F\otimes W:$

$\psi=u_{\epsilon}\left(  r\right)  \otimes\left(  \sum_{j=1}^{m}a_{j}%
f_{j}\right)  +v_{\epsilon}\left(  r\right)  \otimes\left(  \sum_{j=1}%
^{m}b_{j}f_{j}\right)  =u_{\epsilon}\left(  r\right)  \otimes\sigma
_{1}+v_{\epsilon}\left(  r\right)  \otimes\sigma_{2}$ with some $\sigma
_{1},\sigma_{2}\in W$

\paragraph{5)}

If $\phi\in F_{\epsilon}\left(  r\right)  :\phi=\gamma_{+}\phi+\gamma_{-}\phi$

$\sum_{a=1}^{3}r^{a}\left[  \kappa_{a}\right]  \left(  \gamma_{\epsilon}%
\phi\right)  =\gamma_{\epsilon}\sum_{a=1}^{3}r^{a}\left[  \kappa_{a}\right]
\phi=\epsilon\frac{1}{2}i\gamma_{\epsilon}\phi$ with $\gamma_{\epsilon}\left[
\kappa_{a}\right]  =\left[  \kappa_{a}\right]  \gamma_{\epsilon}$

thus $\gamma_{+}\phi$ and $\gamma_{-}\phi$ are still eigen vectors with the
same eigen value, and $F_{\epsilon}\left(  r\right)  $ split between
$F^{+},F^{-}$ and the states can be written :%

\begin{equation}
\psi=\phi_{+}\otimes\sigma_{+}+\phi_{-}\otimes\sigma_{-}\label{E96}%
\end{equation}

$\Rightarrow\left[  \psi\right]  =\left[  \phi_{+}\right]  \left[  \sigma
_{+}\right]  ^{t}+\left[  \phi_{-}\right]  \left[  \sigma_{-}\right]  ^{t}$
\ with $\gamma_{+}\phi_{+}=\phi_{+};\gamma_{-}\phi_{-}=\phi_{-}$%

\begin{equation}
\sum_{a=1}^{3}r^{a}\left[  \kappa_{a}\right]  \phi_{+}=\epsilon\frac{1}%
{2}i\phi_{+};\sum_{a=1}^{3}r^{a}\left[  \kappa_{a}\right]  \phi_{-}%
=\epsilon\frac{1}{2}i\phi_{-}\label{E97}%
\end{equation}

The tensors $\psi$\ are eigen vectors of the operator : $\left(  \sum
_{a=1}^{3}r^{a}\left[  \kappa_{a}\right]  \right)  \otimes\left(
1_{U}\right)  $ in the following meaning :

$\left(  \sum_{a=1}^{3}r^{a}\left[  \kappa_{a}\right]  \right)  \otimes\left(
1_{U}\right)  \left(  \psi\right)  =\left(  \sum_{a=1}^{3}r^{a}\left[
\kappa_{a}\right]  \right)  \otimes\left(  1_{U}\right)  \left(  \phi
_{+}\otimes\sigma_{+}\right)  +\left(  \sum_{a=1}^{3}r^{a}\left[  \kappa
_{a}\right]  \right)  \otimes\left(  1_{G}\right)  \left(  \phi_{-}%
\otimes\sigma_{-}\right)  $

$=\left(  \sum_{a=1}^{3}r^{a}\left[  \kappa_{a}\right]  \right)  \left(
\phi_{+}\right)  \otimes\left(  1_{U}\right)  \left(  \sigma_{+}\right)
+\left(  \sum_{a=1}^{3}r^{a}\left[  \kappa_{a}\right]  \right)  \left(
\phi_{-}\right)  \otimes\left(  1_{U}\right)  \left(  \sigma_{-}\right)  $

$=i\epsilon\left(  \phi_{+}\right)  \otimes\left(  1_{U}\right)  \left(
\sigma_{+}\right)  +i\epsilon\left(  \phi_{-}\right)  \otimes\left(
1_{U}\right)  \left(  \sigma_{-}\right)  =\frac{1}{2}i\epsilon\psi$

Under a change of gauge these states tensors transform as usual.\ If the
change is of the kind : $\left(  \rho\circ\Upsilon\left(  \exp\tau\sum
r_{a}^{a}\overrightarrow{\kappa}\right)  ,1_{U}\right)  $ we get:

$\widehat{\psi}=\vartheta\left(  \rho\circ\Upsilon\left(  \exp\tau\sum
r_{a}^{a}\overrightarrow{\kappa}\right)  ,1_{U}\right)  \psi=\left(
\exp\epsilon\tau i\frac{1}{2}\right)  \psi$

6) With the $\gamma$\ previously defined one can compute the matrices $\left[
\kappa_{a}\right]  $.\ Let us fix the real scalars $\left(  r_{1},r_{2}%
,r_{3}\right)  $ and compute the eigen vectors of the matrix $\left[
\kappa_{S}\right]  =\sum_{a}r^{a}\left[  \kappa_{a}\right]  $

We get the following\ with $\sum_{j=1}^{3}\left(  r_{j}\right)  ^{2}=1:$

a) If $r_{3}\neq1:$

eigen value : $\epsilon\frac{1}{2}i:\phi_{-}=%
\begin{bmatrix}
0\\
0\\
\epsilon\left(  -r_{1}+ir_{2}\right) \\
-1+\epsilon r_{3}%
\end{bmatrix}
\in F^{-},\phi_{+}=%
\begin{bmatrix}
\epsilon\left(  -r_{1}+ir_{2}\right) \\
-1+\epsilon r_{3}\\
0\\
0
\end{bmatrix}
\in F^{+},$

b) If $r_{3}=1:$

eigen value : $\frac{1}{2}i:\phi_{-}=%
\begin{bmatrix}
0\\
0\\
0\\
1
\end{bmatrix}
\in F^{-},\phi_{+}=%
\begin{bmatrix}
0\\
1\\
0\\
0
\end{bmatrix}
\in F^{+};$

eigen value : $-\frac{1}{2}i:\phi_{-}=%
\begin{bmatrix}
0\\
0\\
1\\
0
\end{bmatrix}
\in F^{-},\phi_{+}=%
\begin{bmatrix}
1\\
0\\
0\\
0
\end{bmatrix}
\in F^{+};$

Conversely for any vector of F such as : $\phi=%
\begin{bmatrix}
ue^{i\alpha}\\
ve^{i\beta}\\
ue^{i\alpha}\\
ve^{i\beta}%
\end{bmatrix}
\neq0,u,v\in%
\mathbb{R}
\Leftrightarrow\phi_{R}=\phi_{L}$ \ one can find $\left(  r_{1},r_{2}%
,r_{3}\right)  $ such that $\phi$ is an eigen vector of $\sum_{a}r^{a}\left[
\kappa_{a}\right]  :r_{1}=-2\epsilon\frac{uv}{\left(  u^{2}+v^{2}\right)
}\cos\left(  \nu\right)  ;r_{2}=2\epsilon\frac{uv}{\left(  u^{2}+v^{2}\right)
}\sin\left(  \nu\right)  ;r_{3}=-\frac{\epsilon\left(  u^{2}-v^{2}\right)
}{\left(  u^{2}+v^{2}\right)  }$

The set of such vectors is the set of invariant vectors under $-i\gamma
_{1}\gamma_{2}\gamma_{3}=%
\begin{bmatrix}
0 & \sigma_{0}\\
\sigma_{0} & 0
\end{bmatrix}
.$ It is a 2 dimensional subspace of F, but it is not invariant by the action
of Spin(3,1).\ Indeed $-i\gamma_{1}\gamma_{2}\gamma_{3}=-i\rho\left(
\varepsilon_{1}\varepsilon_{2}\varepsilon_{3}\right)  $ and there are elements
of Spin(3,1) which do not commute with $\varepsilon_{1}\varepsilon
_{2}\varepsilon_{3}$ as it is not too difficult to see in Cl(4,C).\ So if a
state is symmetric for some frame, it is no longer symmetric for at least
another frame : for these particles there are privileged frames, and so
privileged directions in the universe.

The spatially symmetric states are : $\psi=\phi_{+}\otimes\sigma_{R}+\phi
_{-}\otimes\sigma_{L}$ in matrix form:

$\left[  \psi\right]  =%
\begin{bmatrix}
\psi_{R}\\
\psi_{L}%
\end{bmatrix}
=%
\begin{bmatrix}
\phi\sigma_{R}\\
\phi\sigma_{L}%
\end{bmatrix}
$ with $\left[  \phi\right]  =%
\begin{bmatrix}
ue^{i\alpha}\\
ve^{i\beta}%
\end{bmatrix}
$ 2x1 matrix, $\left[  \sigma_{R}\right]  ,\left[  \sigma_{L}\right]  $ 1xm matrix

7) It is easy to compute the moments with these states :

$\left[  \phi\right]  ^{\ast}\left[  \phi\right]  =u^{2}+v^{2}$

$\left[  \phi\right]  ^{\ast}\sigma_{1}\left[  \phi\right]  =2uv\cos\left(
\nu\right)  =-\epsilon r_{1}\left(  u^{2}+v^{2}\right)  $

$\left[  \phi\right]  ^{\ast}\sigma_{2}\left[  \phi\right]  =-2uv\sin\left(
\nu\right)  =-\epsilon r_{2}\left(  u^{2}+v^{2}\right)  $

$\left[  \phi\right]  ^{\ast}\sigma_{3}\left[  \phi\right]  =u^{2}%
-v^{2}=-\epsilon r_{3}\left(  u^{2}+v^{2}\right)  $

$\sum_{1}^{3}\left(  \left[  \phi\right]  ^{\ast}\sigma_{a}\left[
\phi\right]  \right)  ^{2}=$ $\left(  u^{2}+v^{2}\right)  ^{2}$

All these quantities depend on 3 real scalars only. One cannot have $\left[
\phi\right]  ^{\ast}\sigma_{0}\left[  \phi\right]  =0$ or r%
$>$%
0: $\left[  \phi\right]  ^{\ast}\sigma_{r}\left[  \phi\right]  =0$

With the formulas given previously the moments are the following :

a)

$a<4:P_{a}=-\epsilon r_{a}\left(  u^{2}+v^{2}\right)  \operatorname{Im}\left(
\left[  \sigma_{L}\right]  ^{t}\overline{\left[  \sigma_{R}\right]  }\right)
$

$a>3:P_{a}=\epsilon r_{a-3}\left(  u^{2}+v^{2}\right)  \operatorname{Re}%
\left(  \left[  \sigma_{L}\right]  ^{t}\overline{\left[  \sigma_{R}\right]
}\right)  $

b)

$J_{0}=-\frac{1}{2}\left(  \allowbreak u^{2}+v^{2}\right)  \left(  \left[
\sigma_{R}\right]  ^{t}\overline{\left[  \sigma_{R}\right]  }+\left[
\sigma_{L}\right]  ^{t}\overline{\left[  \sigma_{L}\right]  }\right)  ;$

$r>0:$ $J_{r}=\frac{1}{2}\left(  \allowbreak u^{2}+v^{2}\right)  \epsilon
r_{r}\left(  \left[  \sigma_{R}\right]  ^{t}\overline{\left[  \sigma
_{R}\right]  }-\left[  \sigma_{L}\right]  ^{t}\overline{\left[  \sigma
_{L}\right]  }\right)  $

$\sum_{0}^{3}J_{r}^{2}=\frac{1}{2}\left(  \allowbreak u^{2}+v^{2}\right)
^{2}\left(  \left(  \left[  \sigma_{R}\right]  ^{t}\overline{\left[
\sigma_{R}\right]  }\right)  ^{2}+\left(  \left[  \sigma_{L}\right]
^{t}\overline{\left[  \sigma_{L}\right]  }\right)  ^{2}\right)  >0$

c) $\left\langle \psi,\psi\right\rangle =-2\left(  \allowbreak u^{2}%
+v^{2}\right)  \operatorname{Im}\left(  \left[  \sigma_{L}\right]
^{t}\overline{\left[  \sigma_{R}\right]  }\right)  $

d) $\rho^{a}=2\left(  \allowbreak u^{2}+v^{2}\right)  \operatorname{Re}\left(
\left[  \sigma_{L}\right]  \left[  \theta_{a}\right]  ^{t}\left[  \sigma
_{R}\right]  ^{\ast}\right)  $

e)

$\left[  \mu_{A}\right]  _{a}^{0}=-i\left(  u^{2}+v^{2}\right)  \left(
\left[  \sigma_{R}\right]  \left[  \theta_{a}\right]  ^{t}\left[  \sigma
_{R}\right]  ^{\ast}\right)  +\left(  \left[  \sigma_{L}\right]  \left[
\theta_{a}\right]  ^{t}\left[  \sigma_{L}\right]  ^{\ast}\right)  $

$\left[  \mu_{A}\right]  _{a}^{j}=i\epsilon r_{j}\left(  u^{2}+v^{2}\right)
\left(  \left[  \sigma_{R}\right]  \left[  \theta_{a}\right]  ^{t}\left[
\sigma_{R}\right]  ^{\ast}\right)  -\left(  \left[  \sigma_{L}\right]  \left[
\theta_{a}\right]  ^{t}\left[  \sigma_{L}\right]  ^{\ast}\right)  $

The kinematic moments depend on 3 real scalars (for $\phi)$ and 2 real
($\left[  \sigma_{R}\right]  \left[  \sigma_{R}\right]  ^{\ast},\left[
\sigma_{L}\right]  \left[  \sigma_{L}\right]  ^{\ast})$ and 1 complex scalar
$\left(  \left[  \sigma_{L}\right]  \left[  \sigma_{R}\right]  ^{\ast}\right)
$ that is 7 degrees of freedom. The physical moments depend on 3 real scalars
(for $\phi)$ and mx2 reals scalars ($\left[  \sigma_{R}\right]  \left[
\theta_{a}\right]  ^{t}\left[  \sigma_{R}\right]  ^{\ast},\left(  \left[
\sigma_{L}\right]  \left[  \theta_{a}\right]  ^{t}\left[  \sigma_{L}\right]
^{\ast}\right)  $ imaginary) and m complex numbers $\left(  \left[  \sigma
_{R}\right]  \left[  \theta_{a}\right]  ^{t}\left[  \sigma_{L}\right]  ^{\ast
}\right)  .$

The moments are gauge and chart invariant, so their value is the same for all
the observers.

\subsection{Physical meaning}

\paragraph{1)}

Let us pause to think about the physical meaning of these results. Starting
from general assumptions, we have proven that the only spatial symmetry, as
defined, are rotations around a space like vector r, and that the state tensor
is then an eigen vector of an operator $\left(  \sum_{a=1}^{3}r^{a}\left[
\kappa_{a}\right]  \right)  \otimes\left(  1_{U}\right)  .$ But of course this
outcome could be some mathematical artefact without physical meaning and only
experiments could tell if such symmetries exist.

\paragraph{2)}

If particles have this kind of feature that means that for them some specific
geometric directions are privileged, and we should expect that they behave
accordingly. This is not an issue of quantization, but of the existence of
something like an magnetic moment. And we know that it is the case for most of
the elementary particles.

Indeed\ the equation \ref{E83} reads :

$N\varpi_{4}\left(  a_{D}\rho^{a}\overrightarrow{V}-ia_{I}\overrightarrow{\mu
}_{a}\right)  =\nabla_{e}\ast\overline{%
\mathcal{F}%
}_{A}^{a}$

And for a symmetric state the "magnetic moment" is the 4-vector

$\overrightarrow{\mu}=\sum_{r}\left[  \mu_{A}\right]  _{a}^{r}\partial
_{r}=\left(  u^{2}+v^{2}\right)  \sum_{\alpha}\left(  O_{0}^{\alpha}\sigma
^{2}-\epsilon\left(  r_{1}O_{1}^{\alpha}+r_{2}O_{2}^{\alpha}+r_{3}%
O_{3}^{\alpha}\right)  \sigma^{3}\right)  \partial_{\alpha}$

with :

$\sigma^{1}=2\operatorname{Im}\left(  \left[  \sigma_{R}\right]  \left[
\theta_{a}\right]  ^{t}\left[  \sigma_{L}\right]  ^{\ast}\right)  $

$i\sigma^{2}=\left(  \left[  \sigma_{R}\right]  \left[  \theta_{a}\right]
^{t}\left[  \sigma_{R}\right]  ^{\ast}\right)  +\left(  \left[  \sigma
_{L}\right]  \left[  \theta_{a}\right]  ^{t}\left[  \sigma_{L}\right]  ^{\ast
}\right)  $

$i\sigma^{3}=\left(  \left[  \sigma_{R}\right]  \left[  \theta_{a}\right]
^{t}\left[  \sigma_{R}\right]  ^{\ast}\right)  -\left(  \left[  \sigma
_{L}\right]  \left[  \theta_{a}\right]  ^{t}\left[  \sigma_{L}\right]  ^{\ast
}\right)  $

This equation is geometric : all the quantities are defined as vectors or
tensors so, as written, it stands for any observer, it is fully gauge
equivariant and covariant. The moments do not depend of the gauge, so the
quantities $r_{1},r_{2},r_{3}$ are not the components of a vector expressed in
a frame : they are invariant (it is the consequence of the invariance of the
scalar product). The magnetic moment is defined as the \textit{sum of vectors} :

$\overrightarrow{\mu}=\left(  u^{2}+v^{2}\right)  \sum_{\alpha}\left(
\sigma^{2}\overrightarrow{\partial_{0}}-\epsilon\left(  r_{1}\overrightarrow
{\partial_{1}}+r_{2}\overrightarrow{\partial_{2}}+r_{3}\overrightarrow
{\partial_{3}}\right)  \sigma^{3}\right)  $

If one changes the tetrad the vectors $\left(  \overrightarrow{\partial}%
_{i}\right)  $ change and $\overrightarrow{\mu}$ changes accordingly. This
feature is common to all the moments that we have defined.

\paragraph{3)}

So it seems that we have a paradox, or an inconsistency : the vector
\ $\overrightarrow{\mu}$ plays clearly some specific role, it is a well
defined vector, but it changes with the observer.

Let us first clarify a point about the previous demonstration. We have
characterized the set of observers $O_{S}$ , but all the reasoning above has
been done from a frame belonging to $O_{S}$\ , so if we could know the
components of the vector r we do not know in which frame they are measured.
Indeed our question was : "what are the symmetric states in a tetrad ?" and
the answer is right, but it is true for any tetrad, and it is what we checked
for the magnetic moment.

So from the above equation it is probable that any observer can find tetrads
in which the states are symmetric, and check that the direction of the
4-vector $\overrightarrow{\mu}$\ is specific, but these directions change with
the observer. That the symmetry is geometrically defined does not entail that
some particles, and over more their kinematic mode, have a symmetry of a
geometric nature (which is anyway difficult to apprehend for pointlike objects).

\paragraph{4)}

Furthermore we do have a privileged orientation in our model : the velocity of
the particles. As one can see in the equation above the net impact of the
field depends of a sum of $\overrightarrow{V}$ and $\overrightarrow{\mu}_{a} $
and, even if the real and imaginary parts of the fields act differently, the
net impact should depend of the relative orientations of the two vectors. The
spin of a particle is measured with respect to its velocity. It would be of
interest to investigate the equations listed here for particles which have
moments of the symmetric mode.

\section{PHYSICAL\ SYMMETRIES}

\label{Contant physical characteristics}

One of the main features of particles is that their "physical
characteristics", modelled in the space vector W, are constant. So we are lead
to study the symmetries occuring in relation with the group U.\ They come in
two flavours :

- families of particles sharing the same constant physical characteristics,
and corresponding to subgroups of U : they define elementary particles

- specificities of the vacuum which force the states of the particles, as we
see them, to belong to some representation of a subgroup of U. This is the
symmetry breackdown mechanism

\subsection{Families of particles}

The simplest way to define families of particles is to proceed as above and
look for symmetries related to the U group.

\paragraph{1)}

Two states $\psi,\psi^{\prime}$ are physically imperceptible if there are a
vector subspace $W_{S}\subset W,$ and a closed subgroup $U_{S}\subset U$ such
that :

$\forall\psi\in F\otimes W_{S},\forall s\in Spin(3,1),\forall u\in
U_{S},\exists\theta\in%
\mathbb{R}
:\vartheta\left(  s,u\right)  \psi=e^{i\theta}\vartheta\left(  s,1\right)
\psi$

$\theta$ can depend on s, u and $\psi.$

\paragraph{2)}

Let be the 4 vectors $\sigma_{i}$ of W such that : $\sigma_{i}=\sum_{j=1}%
^{m}\psi^{ij}f_{j}$ so $\psi=\sum_{i=1}^{4}e_{i}\otimes\sigma_{i}$

So if $\psi\in F\otimes W_{S}:\forall s\in Spin(3,1),\forall u\in U_{S}:$

$\sum_{i=1}^{4}\vartheta\left(  s,u\right)  \left(  e_{i}\otimes\sigma
_{i}\right)  =\sum_{i=1}^{4}e^{i\theta\left(  u\right)  }e_{i}\otimes
\sigma_{i}=\sum_{i=1}^{4}\left(  \rho\circ\Upsilon\left(  s\right)
e_{i}\right)  \otimes\left(  \chi\left(  u\right)  \sigma_{i}\right)  $

Let s=1 : $\sum_{i=1}^{4}\vartheta\left(  1,u\right)  \left(  e_{i}%
\otimes\sigma_{i}\right)  =\sum_{i=1}^{4}e_{i}\otimes\left(  \chi\left(
u\right)  \sigma_{i}\right)  =\sum_{i=1}^{4}e_{i}\otimes e^{i\theta\left(
u\right)  }\sigma_{i}$

$\Rightarrow\forall\psi\in F\otimes W_{S},\forall u\in U_{S}:\chi\left(
u\right)  \sigma_{i}=e^{i\theta\left(  u\right)  }\sigma_{i}$

So for each i $\left(  \sigma_{i},\chi|_{U_{S}}\right)  $ is a linear
representation of $U_{S}$\ , of complex dimension 1. $\left(
\mathbb{C}
,\chi^{\prime}\left(  1\right)  \right)  $ is a linear 1-dimensional
representation of the Lie algebra of $U_{S}.$ The only compact groups with non
trivial 1 dimensional representations are abelian, so $U_{S}$ is an abelian
subgroup of U.\ 

\paragraph{3)}

Any compact Lie group has abelian subgroups, which are tori. The maximal tori
are p dimensional with p=rank of U, all conjugated to each others. One can
always choose a basis in $T_{1}U$ such that the first p vectors
$\overrightarrow{\theta_{j}}$ belong to a maximal torus of the algebra.\ They
are orthonormal in the hermitian, Ad invariant, scalar product on $T_{1}^{c}U$\ .

The irreducible representations $\left(  W_{\lambda},\chi_{\lambda}\right)  $
of the compact group U\ are indexed on the highest weight $\lambda$, The
corresponding vector space $W_{\lambda}$ contains a vector $\sigma_{\lambda}$
which is an eigen-vector of $\chi_{\lambda}\left(  u\right)  $ for any element
of a maximal torus $U_{S}$:

$\forall u\in U_{S}:\chi_{\lambda}\left(  u\right)  \sigma_{\lambda
}=e^{i\lambda\left(  u\right)  }\sigma_{\lambda},\lambda\left(  u\right)  \in%
\mathbb{R}
$

The other vectors of $W_{\lambda}$ are generated as successive applications of
some elements of G. So these vectors $\sigma_{\lambda}$ characterize distinct
families of particles.

Any representation of U is the sum of irreducible representations, so W has a
collection of such vectors $u_{\lambda}$.

\paragraph{4)}

For a particle belonging to a family one should expect that its tensor state
$\psi$ is such that its physical components $\sigma_{\lambda}$ stays within
$W_{\lambda}.$ But one cannot exclude more complicated states, involving more
than one family of particles.

So the whole story is about the definition of irreducible representations of
compact groups, and finding, through experiments, the representations which
are found in the real world. This is at the foundation of gauge theories of
particles physics.

\section{SYMMETRY\ BREAKDOWN}

\label{Symmetry breakdown}

\subsubsection{Principle}

\paragraph{1)}

The invariance of the lagrangian implies that the potentials \`{A} must
factorize through the covariant derivative, or the curvature form for their
derivatives. In quantum theory of fields a particle - a boson - must be
associated to each field and therefore this boson must be massless,
contradictory to the experiments. Symmetry breakdown is first a mechanism to
turn over this issue, so far critical to the consistency of the standard model
of particles.

\paragraph{2)}

The "Higgs mechanism" can be summarized as follow (Bednyakov [2]). The Euler
Lagrange equations give only necessary conditions, and it could happen that
the solutions are not unique. This can be a mathematical artefact but, in many
physical situations, a system can possibly follow several paths, and the
actual choice depends on the initial conditions or an outside action. In these
cases it is logical to reparametrize the model, generally by introducing
discrepancy variables. \ In the Higgs mechanism it is the fundamental state,
corresponding to the vacuum, which is assumed to offer several paths (the
vacuum state is degenerate). The most common explanation to this phenomenon is
cosmological : the specificities of the present vacuum would come from the
initial conditions of the "big bang". Some kinds of particles would have been
privileged, and the situation hereafter would have been frozen, as if a phase
transition had occured. Thus to account for this specific initial values
conditions, one proceeds to a change of variables, evidencing the discrepancy
with the actual vacuum. The new variables take the form of fields (Goldstone
bosons and Higgs field) which interact with the existing matter and fields and
give a mass to some bosons. This correction to the gauge model is
phenomenological : we see one privileged solution among others, and the basic
theory cannot forecast which one the system takes, so additional variables are
needed, that only experiments can fixed.

\paragraph{3)}

In particle physics all this happens in the framework of quantum theory of
fields, but the breakdown of symmetry is actually a fairly common phenomenon,
met in classical situations, such as ferromagnetism or phase transition, and
therefore it is in the scope of classical field theory. In the general picture
used here:

a) H is a subgroup of the group U of "internal symmetries" (the kinematic part
is not involved here).\ A member of U can be written as $u=xh$ with $x\in
X=U/H.$ The quotient space X acts as an intermediary level in the gauge group
and there is a fiber bundle $X_{M}$ over M modelled on U/H.

b) The system is still modelled as previously but there is some "fundamental
state" of the universe, pre-existing to any system, and characterized by a
section $\varkappa\left(  m\right)  $\ of the fiber bundle $X_{M}$, similar to
the "Higgs field". This field interacts with the force fields (other than
gravitation) and therefore constrains their value and conversely this
interaction fixes $\varkappa.$ A transition phase has occurred.

c) Therefore the gauge group is reduced in that the only visible gauge
transformations are those in the equivalence class of $\varkappa,$ that is of
the form $u=xh,$ with H as apparent gauge group : the action of U is "hidden"
by the Higgs field. The states $\psi$\ are still described in a fiber bundle
associated to U, and the principle of least action still stands, but one has
to account for the pre-existing Higgs field $\varkappa$

So, this is not simple...I found it better to proceed step by step, and use
long but basic mathematical developments than to race through highly
specialized short-cuts. We have to address successively the additional
mathematical structures for U, the fiber bundle $U_{M},$ the connection
\textbf{A}\ , and the fiber bundle of fields.

\subsubsection{The fiber bundle U}

The basic rule is that any element in U can be written as : $u=\Lambda\left(
x,h\right)  $ where $x\in U/H,h\in H$

\paragraph{1)}

H is assumed to be a closed non discret subgroup of U, it is therefore a
compact Lie group. The quotient space X=U/H (called homogeneous space) is
defined by the equivalence relation :

$x\sim y\in U\Leftrightarrow\exists h\in H:x=yh\Leftrightarrow y^{-1}x\in H$

\paragraph{2)}

Under this assumption U is a \textit{principal fiber bundle over U/H of group
H} (Kolar [ ] 10.5) which implies :

a) U/H\ is a smooth metrisable manifold

b) There is a projection $\pi_{H}:U\rightarrow U/H$ such that $\forall s\sim
s^{\prime}\in U:\forall u\in U:\pi_{H}(us)=\pi_{H}(us^{\prime})$ and therefore :

$h\in H:\pi_{H}\left(  h\right)  =\pi_{H}\left(  1_{U}\right)  =\pi_{H}\left(
1_{H}\right)  $

$\pi_{H}$ is onto : $\forall\xi_{x}\in T_{x}U/H,\exists\overrightarrow{\theta
}\in T_{1}U:\pi_{H}^{\prime}\left(  x\right)  \overrightarrow{\theta}=\xi_{x}$

c) There is an open cover $\left(  U_{i}\right)  $\ of U and trivializations :

$\Lambda_{i}:\pi_{H}^{-1}\left(  U_{i}\right)  \times H\rightarrow
U::\Lambda_{i}\left(  x,h\right)  \in U$ with the usual \textit{right} action
of H on U :$\Lambda_{i}\left(  x,h\right)  \times h^{\prime}=\Lambda
_{i}\left(  x,hh^{\prime}\right)  $

The trivializations are defined by the values $\Lambda_{i}\left(  x,1\right)
$

d) The fundamental vectors $\Lambda_{h}^{\prime}(x,h)R_{h}^{\prime}\left(
1\right)  \xi^{h},\xi^{h}\in T_{1}H$ are the generators of the vertical space
isomorphic to $T_{1}H$ and:

$\pi_{H}^{\prime}\left(  \Lambda_{i}\left(  x,h\right)  \right)  \Lambda
_{ih}^{\prime}(x,h)R_{h}^{\prime}\left(  1\right)  \xi^{h}=0$

$\pi_{H}\left(  \Lambda_{i}\left(  x,h\right)  \right)  =x\Rightarrow\pi
_{H}^{\prime}\left(  \Lambda_{i}\left(  x,h\right)  \right)  \Lambda
_{ix}^{\prime}(x,h)\xi^{x}=\xi^{x}$

e) There is a \textit{left} action of U on U/H which is denoted $\lambda
\left(  u,x\right)  :$

$\pi_{H}\left(  s\right)  =x:\lambda\left(  u,x\right)  =\pi_{H}\left(
us\right)  \Leftrightarrow\lambda\left(  u,\pi_{H}\left(  s\right)  \right)
=\pi_{H}\left(  us\right)  $

\paragraph{2)}

In the standard model U is the direct product of the compact groups
$U=SU(3)\times SU(2)\times U\left(  1\right)  $ and H is the projection of U
on SU(2)xU(1) or U(1). Without being too specific but with the purpose to be
simple we will assume the following:

a) $T_{1}U=h_{0}\oplus l_{0}$ with $h_{0}$ a r dimensional sub-algebra and
$l_{0}$ a m-r vector subspace$.$ The basis of $T_{1}U$ is comprised of r
vectors $\overrightarrow{\theta}_{a}\in h_{0}$ (a=1..r) and m-r vectors
$\overrightarrow{\theta}_{a}\in l_{0}$ (a=r+1...m)

b) $h_{0}$ is the Lie algebra of the compact subgroup H, which is generated by
$\exp\left(  \sum_{a=1}^{r}r^{a}\overrightarrow{\theta}_{a}\right)  $

c) The map : $H\times l_{0}\rightarrow U::u=\exp l\times h$ is a
diffeomorphism. One can identify $x$ with $\Lambda\left(  x,1\right)  $ ,X=U/H
with a subset of U : $X=U/H=\left\{  \exp l,l\in l_{0}\right\}  $ and
$\Lambda(x,h)=\Lambda\left(  \exp l,h\right)  =\left(  \exp l\right)  h$ . The
fiber bundle $U\left(  X,H,\pi_{H}\right)  $ is trivial.

d) The bracket on $T_{1}G$ is such as : $\left[  h_{0},h_{0}\right]  \subset
h_{0},\left[  l_{0},l_{0}\right]  \subset h_{0},\left[  h_{0},l_{0}\right]
\subset l_{0}$

e) There is a bilinear symmetric scalar form on $T_{1}U,$ invariant by the
adjoint operator, for which the subspaces $h_{0},l_{0}$ are oorthogonal, and
positive definite on $T_{1}H.$

These conditions are met if U and H are linked in a Cartan decomposition
(Knapp [ ] 6.31). All semi-simple Lie groups have such decompositions. The
conditions d) and e) will not be used in the following but are part of the
definition of a Cartan decomposition.

Remark : property c is usually written as: $u=h\exp l.$ Both formulations are equivalent.

\begin{proof}
\ Indeed :
\end{proof}

$Ad_{h}\left(  h_{0}\right)  \subset h_{0}$ because $h_{0}$\ is the Lie
algebra of H, $h_{0}$ is\ Ad$_{h}$ invariant, so is its orthogonal complement
$l_{0}.$

Thus: $\forall h\in H,l\in l_{0},\exists l^{\prime}\in l_{0}:Ad_{h}%
l=l^{\prime}\Rightarrow\exp Ad_{h}l=h\left(  \exp l\right)  h^{-1}=\exp
l^{\prime}\Rightarrow\exp\left(  -r\right)  \exp l^{\prime}\exp r=\exp l$

$u=\exp r\exp l=\exp r\exp\left(  -r\right)  \exp l^{\prime}\exp r=\exp
l^{\prime}\exp r\blacksquare$

\subsubsection{Principal fiber bundles on M}

The splitting U/H, H is prolonged in the principal fiber bundles on M. The
principal fiber bundle $U_{M}\rightarrow M$ split in : $X_{M}\rightarrow M$
corresponding to U/H and $\widetilde{U}_{M}\rightarrow X_{M}$\ corresponding
to H. The splitting is attributable to the Higgs field, materialized by a
section $\varkappa$\ \ \ \ on $X_{M}.$

\paragraph{1)}

We still have the same principal fiber bundle $U_{M}$\ base M, group U, with
the projection $\pi:U_{M}\rightarrow M$\ , and the trivializations on an open
cover: $\varphi_{Ui}:\pi^{-1}\left(  U_{Mi}\right)  \times U\rightarrow
U_{M}::p=\varphi_{Ui}\left(  m,u\right)  $ and we denote $\widehat{p}%
_{i}=\varphi_{Ui}\left(  m,1\right)  $,so $p=u_{i}\widehat{p}_{i}=\widehat
{p}_{i}u_{i}$

As a manifold $U_{M}$ has charts deduced from $\varphi_{Ui}\left(
m,\Lambda\left(  x,h\right)  \right)  =p$ and one can construct additional structures.

\paragraph{2)}

The \textit{associated fiber bundle }$X_{M}=U_{M}\times_{U}X$ with typical
fiber X=U/H associated with $U_{M}$ through the U action :

$\left(  U_{M}\times X\right)  \times G\rightarrow\left(  U_{M}\times
X\right)  ::\left(  p,x\right)  \times u\rightarrow\left(  pu^{-1}%
,\lambda\left(  u,x\right)  \right)  $

$X_{M}$ is a U-fiber bundle (but not a principal fiber bundle), with base M, trivializations:

$\pi^{-1}\left(  U_{Mi}\right)  \times U/H\rightarrow X_{M}:\varphi
_{Xi}\left(  m,x\right)  =\varphi_{Ui}\left(  m,\Lambda\left(  x,1\right)
\right)  $

and U left action : $u\times\varphi_{Xi}\left(  m,x\right)  =\varphi
_{Xi}\left(  m,\lambda\left(  u,x\right)  \right)  $

With a Cartan decomposition $X_{M}$ can be seen as a sub-bundle, embedded in
$U_{M}:\varphi_{Ui}\left(  m,\exp l\right)  $

A section on $X_{M}$ is a map : $\chi\left(  m\right)  =\varphi_{Xi}\left(
m,\varkappa\left(  m\right)  \right)  =\varphi_{Ui}\left(  m,\Lambda\left(
x\left(  m\right)  ,1\right)  \right)  $ .The Higgs fields are such sections
:they fix the state of the vacuum, which is characterized by an equivalence
class of U/H. With the Cartan decomposition : $\chi\left(  m\right)
=\varphi_{Ui}\left(  m,\exp l\left(  m\right)  \right)  $ where
$l:M\rightarrow l_{0}$

\paragraph{3)}

$U_{M}$\ is endowed with the \textit{principal fiber bundle }%
structure\textit{\ }$\widetilde{U}_{M}$\textit{\ with base }$X_{M}$\textit{\ ,
group H} and :

projection : $\widetilde{\pi}\left(  \varphi_{iU}\left(  m,u\right)  \right)
=\varphi_{iX}\left(  m,\pi_{H}\left(  u\right)  \right)  $

open cover : $X_{Mi}=U_{Mi}\cap X_{M}$

trivializations : $\widetilde{\varphi}_{Ui}:X_{Mi}\times H\rightarrow
U_{M}::\widetilde{\varphi}_{Ui}\left(  q,h\right)  =qh$

$\Leftrightarrow\widetilde{\varphi}_{Ui}\left(  \varphi_{Ui}\left(
m,\Lambda\left(  x,1\right)  \right)  ,h\right)  =\varphi_{Ui}\left(
m,\Lambda\left(  x,h\right)  \right)  $

H action : $h\times p=h\times\widetilde{\varphi}_{Ui}\left(  q,h^{\prime
}\right)  =\varphi_{Ui}\left(  \pi\left(  p\right)  ,\Lambda(x,hh^{\prime
})\right)  $

A section on $\widetilde{U}_{M}$ is : $s=\widetilde{\varphi}_{U}\left(
q,h_{s}\left(  q\right)  \right)  $

One has a 2-levels composite fiber bundle $:U_{M}\equiv\widetilde{U}%
_{M}\overset{H}{\rightarrow}X_{M}\overset{U}{\rightarrow}M$

\paragraph{4)}

The composition of a section $\varkappa$ on $X_{M}$ and a section $s$ on
$\widetilde{U}_{M}$ is a section $S$\ on $U_{M}.$ Conversely a section $s$ on
$\widetilde{U}_{M}$ has for image $s(X_{M})$ a sub-manifold embedded in
$U_{M}.$ Any section $S$ in $U_{M}$ is the composite of $\varkappa\left(
m\right)  =\widetilde{\pi}\left(  S\left(  m\right)  \right)  $ on $X_{M}$ and
a section $s$ on $\widetilde{U}_{M}:M\overset{\varkappa}{\rightarrow}%
X_{M}\overset{s}{\rightarrow}U_{M}:S\left(  m\right)  =s\circ\varkappa\left(
m\right)  $.\ Remember that the physical characteristics $\sigma$ are sections
of $U_{M}.$ They are now defined in two steps : the first with $\varkappa,$
the second with H.

A local gauge transformation is given either by a section on $U_{M}$\ or by
the composite:

$S\left(  m\right)  =s\circ\varkappa\left(  m\right)  =\widetilde{\varphi}%
_{U}\left(  \varkappa\left(  m\right)  ,h_{s}\left(  m\right)  \right)
;\widetilde{\pi}\left(  S\left(  m\right)  \right)  =\varkappa\left(
m\right)  $

With Cartan decomposition : $S\left(  m\right)  =\varphi_{U}\left(  m,\exp
l\left(  m\right)  \exp r\left(  m\right)  \right)  $

\paragraph{5)}

There is a bijection between the \textit{principal bundle structures} $H_{M}$
\textit{with base M and group H} on one hand, and the global sections
$\varkappa\left(  m\right)  $ on $X_{M}$ \ (Kol\={a}r [14] 10.13) :
$\varkappa=\varphi_{X}\left(  m,x\left(  m\right)  \right)  $. The
trivialization is : $\varphi_{H}\left(  m,h\right)  =\widetilde{\varphi}%
_{U}\left(  \varkappa\left(  m\right)  ,h\right)  =\varphi_{U}\left(
m,\Lambda(x\left(  m\right)  ,h)\right)  $ . With Cartan decomposition :
$\varkappa=\varphi_{G}\left(  m,\exp l\left(  m\right)  \right)
\rightarrow\varphi_{H}\left(  m,h\right)  =\varphi_{U}\left(  m,\exp l\left(
m\right)  \times h\right)  .$

Such $H_{M}$ \ fiber bundles are not necessarily isomorphic. So its definition
requires both $U_{M}$ and $\chi$.

\subsubsection{The induced connections}

A connection is a projection from the tangent space on the vertical space. The
tangent space of $U_{M}$ splits and a connection \textbf{A} induces a
connection on $X_{M},$\ but it induces a connection on $H_{M}$ \ iff
$\nabla\varkappa=0.$ Let us first define the tangent spaces.

\paragraph{1)}

\subparagraph{a)}

The tangent space of $U\rightarrow U/H$ can be defined through the
trivialization :

$\left(  \xi^{x},\xi^{h}\right)  \in T_{x}X\times T_{h}H\rightarrow
\overrightarrow{\xi}^{u}=\Lambda^{\prime}\left(  x,h\right)  \left(  \xi
^{x},\xi^{h}\right)  \in T_{u}U,$

$\overrightarrow{\xi}^{u}=\Lambda_{x}^{\prime}\left(  x,h\right)  \xi
^{x}+\Lambda_{h}^{\prime}\left(  x,h\right)  \xi^{h}=\left(  R_{h}^{\prime
}x\right)  \xi^{x}+\left(  L_{x}^{\prime}h\right)  \xi^{h}\Leftrightarrow
\Lambda_{h}^{\prime}\left(  x,h\right)  =L_{x}^{\prime}h;\Lambda_{x}^{\prime
}\left(  x,h\right)  =R_{h}^{\prime}x$

X=U/H can be identified with a subset of U, so the tangent space $T_{u}U$
splits :

$T_{u}U$ $=\left\{  L_{u}^{\prime}T_{1}U\right\}  =\left\{  R_{u}^{\prime
}T_{1}U\right\}  $ Notice that the map works on right and left, we will need both

$\overrightarrow{\theta}^{h}=\sum_{a=1}^{r}l^{a}\widehat{\theta}_{a}\in
h_{0}\rightarrow\xi^{h}=\left(  L_{h}^{\prime}1\right)  \overrightarrow
{\theta}^{h},$

$\overrightarrow{\theta}^{l}=\sum_{a=r+1}^{m}l^{a}\widehat{\theta}_{a}\in
l_{0}\rightarrow\xi^{x}=\left(  R_{x}^{\prime}1\right)  \overrightarrow
{\theta}^{l},$

$\overrightarrow{\xi}^{u}=\left(  R_{h}^{\prime}x\right)  \left(
R_{x}^{\prime}1\right)  \overrightarrow{\theta}^{l}+\left(  L_{x}^{\prime
}h\right)  \left(  L_{h}^{\prime}1\right)  \overrightarrow{\theta}^{h}%
=R_{u}^{\prime}1\overrightarrow{\theta}^{l}+L_{u}^{\prime}1\overrightarrow
{\theta}^{h}$

$\overrightarrow{\xi}^{u}=L_{u}^{\prime}1\overrightarrow{\theta}^{\prime
}=R_{u}^{\prime}1\overrightarrow{\theta}"\Rightarrow\overrightarrow{\theta
}"=R_{u^{-1}}^{\prime}\left(  u\right)  L_{u}^{\prime}1\overrightarrow{\theta
}^{\prime}=Ad_{u}\overrightarrow{\theta}^{\prime}$

The vectors $\left(  L_{u}^{\prime}1\right)  \overrightarrow{\theta}%
^{h},\overrightarrow{\theta}^{h}\in l_{0}$ are the generators of the vertical
space :

$\pi_{H}^{\prime}\left(  u\right)  \left(  L_{u}^{\prime}1\right)
\overrightarrow{\theta}^{h}=0$

$\pi_{H}^{\prime}\left(  u\right)  \overrightarrow{\xi}^{u}=R_{u}^{\prime
}1\overrightarrow{\theta}^{l}$

and we have $\lambda\left(  u,x\right)  =\pi_{H}\left(  ux\right)
\Rightarrow$

$\lambda_{u}^{\prime}\left(  u,x\right)  \xi^{u}+\lambda_{x}^{\prime}\left(
u,x\right)  \xi^{x}=\pi_{H}^{\prime}\left(  ux\right)  \left(  R_{x}^{\prime
}u\xi^{u}+L_{u}^{\prime}x\xi^{x}\right)  $

$\lambda_{u}^{\prime}\left(  u,x\right)  =\pi_{H}^{\prime}\left(  ux\right)
R_{x}^{\prime}u;$

$\lambda_{x}^{\prime}\left(  u,x\right)  =\pi_{H}^{\prime}\left(  ux\right)
L_{u}^{\prime}x$

\subparagraph{b)}

The tangent space $T_{p}U_{M}$ can be defined by :

$\left(  \xi^{m},\xi^{u}\right)  \in T_{m}M\times T_{u}U\rightarrow\xi
_{p}=\varphi_{iUm}^{\prime}\left(  m,u\right)  \xi^{m}+\varphi_{iUu}^{\prime
}\left(  m,u\right)  \xi^{u},$

The vertical space splits : $\varphi_{iUu}^{\prime}\left(  m,u\right)  \xi
^{u}=\varphi_{iUu}^{\prime}\left(  m,u\right)  R_{u}^{\prime}1\overrightarrow
{\theta}^{l}+\varphi_{iUu}^{\prime}\left(  m,u\right)  L_{u}^{\prime
}1\overrightarrow{\theta}^{h}$ \ and we have the basis :

a=1,..r : $\delta_{a}\left(  p\right)  =\varphi_{iUu}^{\prime}\left(
m,u\right)  L_{u}^{\prime}1\overrightarrow{\theta}_{a}$

a=r+1,...m: $\delta_{a}\left(  p\right)  =\varphi_{iUu}^{\prime}\left(
m,u\right)  R_{u}^{\prime}1\overrightarrow{\theta}_{a}$

\subparagraph{c)}

The tangent space $T_{q}X_{M}$\ can be defined by :

$q=\varphi_{iX}\left(  m,x\right)  =\varphi_{iU}\left(  m,\Lambda\left(
x,1\right)  \right)  $

$\xi_{q}=\varphi_{iX}^{\prime}\left(  m,x\right)  \left(  \xi^{m},\xi
^{x}\right)  =\varphi_{iUm}^{\prime}\left(  m,\Lambda\left(  x,1\right)
\right)  \xi^{m}+\varphi_{iUu}^{\prime}\left(  m,u\right)  R_{x}^{\prime
}1\overrightarrow{\theta}^{l}$

the vertical space isomorphic to $T_{x}X$ is generated by $:\varphi
_{iUu}^{\prime}\left(  m,u\right)  R_{x}^{\prime}1\overrightarrow{\theta}^{l}$

\subparagraph{d)}

The tangent space $T_{q}\widetilde{U}_{M}$\ can be defined by :

$q=\widetilde{\varphi}_{iU}\left(  \varphi_{iU}\left(  m,\Lambda\left(
x,1\right)  \right)  ,h\right)  =\varphi_{iU}\left(  m,\Lambda\left(
x,h\right)  \right)  $

$\left(  \xi^{m},\xi^{x},\xi^{h}\right)  \in T_{m}M\times T_{x}X\times T_{h}H$

$\rightarrow\xi_{q}=\varphi_{iUm}^{\prime}\left(  m,\Lambda\left(  x,1\right)
\right)  \xi^{m}+\varphi_{iUu}^{\prime}\left(  m,\Lambda\left(  x,1\right)
\right)  \left(  \Lambda_{x}^{\prime}\left(  x,h\right)  \xi^{x}+\Lambda
_{h}^{\prime}\left(  x,h\right)  \xi^{h}\right)  $

$=\varphi_{Um}^{\prime}\left(  m,\Lambda\left(  x,1\right)  \right)  \xi
^{m}+\varphi_{Uu}^{\prime}\left(  m,\Lambda\left(  x,1\right)  \right)
R_{h}^{\prime}x\xi^{x}+\varphi_{Uu}^{\prime}\left(  m,\Lambda\left(
x,1\right)  \right)  L_{x}^{\prime}h\xi^{h}$

$=\varphi_{Um}^{\prime}\left(  m,\Lambda\left(  x,1\right)  \right)  \xi
^{m}+\varphi_{Uu}^{\prime}\left(  m,\Lambda\left(  x,1\right)  \right)
R_{xh}^{\prime}1\overrightarrow{\theta}^{l}+\varphi_{Uu}^{\prime}\left(
m,\Lambda\left(  x,1\right)  \right)  L_{xh}^{\prime}1\overrightarrow{\theta
}^{h}$

So it can be identified with $T_{\Lambda\left(  x,h\right)  }U_{M}$\ but the
vertical space is generated here by $\varphi_{Uu}^{\prime}\left(
m,\Lambda\left(  x,1\right)  \right)  L_{xh}^{\prime}1\overrightarrow{\theta
}^{h}.$

\subparagraph{e)}

The tangent space $T_{q}H_{M}$\ with the section $\chi\in\Lambda_{0}X_{v}$
\ can be defined by :

$q=\varphi_{H}\left(  m,h\right)  =\varphi_{Ui}\left(  m,\Lambda(x\left(
m\right)  ,h)\right)  $

$\left(  \xi^{m},\xi^{h}\right)  \in T_{m}M\times T_{h}H\rightarrow\xi
_{q}=\varphi_{Hm}^{\prime}\left(  m,h\right)  \xi^{m}+\varphi_{Hh}^{\prime
}\left(  m,h\right)  \xi^{h}$

$\xi_{q}=\varphi_{Um}^{\prime}\left(  m,\Lambda(\varkappa\left(  m\right)
,h)\right)  \xi^{m}$

$+\varphi_{Uu}^{\prime}\left(  m,\Lambda(\varkappa\left(  m\right)
,h)\right)  \left(  \Lambda_{x}^{\prime}(\varkappa\left(  m\right)
,h)\varkappa^{\prime}\left(  m\right)  \xi^{m}+\Lambda_{h}^{\prime}%
(\varkappa\left(  m\right)  ,h)\xi^{h}\right)  $

$=\left(  \varphi_{Um}^{\prime}\left(  m,\Lambda(\varkappa\left(  m\right)
,h)\right)  +\varphi_{Uu}^{\prime}\left(  m,\Lambda(\varkappa\left(  m\right)
,h)\right)  \left(  R_{h}^{\prime}\varkappa\right)  \varkappa^{\prime}\left(
m\right)  \right)  \xi^{m}$

$+\varphi_{Uu}^{\prime}\left(  m,\Lambda(\varkappa\left(  m\right)
,h)\right)  \left(  L_{\varkappa\left(  m\right)  }^{\prime}h\right)  \xi^{h}$

$\varphi_{Hh}^{\prime}\left(  m,h\right)  \xi^{h}=\varphi_{Uu}^{\prime}\left(
m,\Lambda(\varkappa\left(  m\right)  ,h)\right)  \left(  L_{\varkappa\left(
m\right)  }^{\prime}h\right)  \xi^{h}$

The vertical space is generated by:

$\varphi_{Hh}^{\prime}\left(  m,h\right)  \xi^{h}=\varphi_{Uu}^{\prime}\left(
m,\Lambda(\varkappa\left(  m\right)  ,h)\right)  \left(  L_{\varkappa\left(
m\right)  h}^{\prime}1\right)  \overrightarrow{\theta}^{h}$

\paragraph{2)}

\subparagraph{a)}

We assume as above that there is a principal connection \textbf{A} on $U_{M}$
with one-form $\widehat{\grave{A}}\left(  p\right)  =\sum_{a\alpha}%
\widehat{\grave{A}}_{\alpha}^{a}\left(  p\right)  dx^{\alpha}\otimes
\overrightarrow{\theta}_{a}$ and potential $\grave{A}\left(  m\right)
=\widehat{\grave{A}}\left(  \varphi_{U}\left(  m,1_{U}\right)  \right)  .$

$\widehat{\grave{A}}\left(  \varphi_{G}\left(  m,u\right)  \right)
=Ad_{u^{-1}}\grave{A}\left(  m\right)  $

$\varphi_{U}^{\ast}\mathbf{A}\left(  p\right)  \left(  \xi^{m},\xi^{u}\right)
=\varphi_{Uu}^{\prime}(m,u)\left(  \xi^{u}+R_{u}^{\prime}\left(  1\right)
\grave{A}\left(  m\right)  \xi^{m}\right)  $

It splits along the two subspaces $h_{0},l_{0}$:

$\xi^{u}=R_{u}^{\prime}1\overrightarrow{\theta}^{l}+L_{u}^{\prime
}1\overrightarrow{\theta}^{h}$

$\grave{A}\left(  m\right)  $

$=\sum_{\alpha}\sum_{a=1}^{r}\grave{A}_{\alpha}^{a}\left(  m\right)
dx^{\alpha}\otimes\overrightarrow{\theta}_{a}+\sum_{\alpha}\sum_{a=r+1}%
^{m}\grave{A}_{\alpha}^{a}\left(  m\right)  dx^{\alpha}\otimes\overrightarrow
{\theta}_{a}$

$=\grave{A}_{h}+\grave{A}_{l}$

$\varphi_{U}^{\ast}\mathbf{A}\left(  p\right)  \left(  \xi^{m},\left(
R_{u}^{\prime}1\right)  \overrightarrow{\theta}^{l}+\left(  L_{u}^{\prime
}1\right)  \overrightarrow{\theta}^{h}\right)  $

$=\varphi_{Uu}^{\prime}(m,u)\left(  \left(  R_{u}^{\prime}1\right)  \left(
\overrightarrow{\theta}^{l}+\grave{A}_{l}\xi^{m}\right)  +\left(
L_{u}^{\prime}1\right)  \left(  \overrightarrow{\theta}^{h}+\grave{A}_{h}%
\xi^{m}\right)  \right)  $

\subparagraph{b)}

\textbf{A}\ \ induces the linear (not equivariant) connection $\mathbf{\Gamma
}$ on the associated fiber bundle $X_{M}:$

$\varphi_{X}^{\ast}\mathbf{\Gamma}\left(  q\right)  \left(  \xi^{m},\xi
^{x}\right)  =\varphi_{Uu}^{\prime}\left(  m,\Lambda\left(  x,1\right)
\right)  \left(  \xi^{x}+\lambda_{u}^{\prime}(1,x)\grave{A}\left(  m\right)
\xi^{m}\right)  $

With $\lambda_{u}^{\prime}(1,x)=\pi_{H}^{\prime}\left(  x\right)
R_{x}^{\prime}1\in L\left(  T_{1}U;T_{x}X\right)  $

$\lambda_{u}^{\prime}(1,x)\grave{A}\left(  m\right)  \xi^{m}$

$=\pi_{H}^{\prime}\left(  x\right)  \left(  R_{x}^{\prime}1\right)  \left(
\sum_{\alpha}\sum_{a=1}^{m}\grave{A}_{\alpha}^{a}\xi^{m,\alpha}\right)  $

$=\left(  R_{x}^{\prime}1\right)  \left(  \sum_{\alpha}\sum_{a=r+1}^{m}%
\grave{A}_{\alpha}^{a}\xi^{m,\alpha}\right)  $

that is : $\Gamma\left(  q\right)  =\left(  R_{x}^{\prime}1\right)
\sum_{a=r+1}^{m}\grave{A}_{\alpha}^{a}\left(  m\right)  \overrightarrow
{\theta}_{a}\otimes dx^{\alpha}$

The covariant derivative of a section $\varkappa=\varphi_{X}\left(  m,x\left(
m\right)  \right)  $ is :

$\nabla^{X}\varkappa=\varkappa^{\ast}\mathbf{\Gamma}\left(  q\right)
=\varphi_{Uu}^{\prime}\left(  m,\Lambda\left(  x,1\right)  \right)  \left(
\frac{dx}{dm}+\lambda_{u}^{\prime}(1,x\left(  m\right)  )\grave{A}\left(
m\right)  \right)  $

\subparagraph{c)}

A connection $\Gamma_{H}$ on $H_{M}$ is a projection on the vertical space
isomorphic to $T_{h}H:$

$q=\varphi_{H}\left(  m,h\right)  =\varphi_{U}\left(  m,\Lambda(x\left(
m\right)  ,h)\right)  $

$\varphi_{H}^{\ast}\mathbf{\Gamma}_{H}\left(  q\right)  \left(  \xi^{m}%
,\xi^{h}\right)  =\varphi_{Hh}^{\prime}\left(  m,h\right)  \left(  \xi
^{h}+\Gamma_{H}\left(  q\right)  \xi^{m}\right)  ,\Gamma_{H}\left(  q\right)
\in\Lambda\left(  T_{m}M;T_{h}H\right)  $

It is principal iff : $\Gamma_{H}\left(  \varphi_{H}\left(  m,h\right)
\right)  =Ad_{h^{-1}}\Gamma_{H}\left(  \varphi_{H}\left(  m,1\right)  \right)
=Ad_{h^{-1}}\Gamma_{H}\left(  \varphi_{U}\left(  m,\varkappa\left(  m\right)
\right)  \right)  $

that is $\mathbf{\Gamma}_{H}\left(  q\right)  \xi_{q}=\varphi_{Hh}^{\prime
}\left(  m,h\right)  \left(  \xi^{h}+\left(  R_{h}^{\prime}1\right)
\Gamma_{H}\left(  \chi\right)  \xi^{m}\right)  $

A connection \textbf{A} induces a principal connection on $H_{M}$\ iff
$\nabla^{X}\varkappa=0$

\begin{proof}
:
\end{proof}

$\varkappa\left(  m\right)  =\Lambda\left(  \varkappa\left(  m\right)
,1\right)  ,u=\Lambda\left(  \varkappa\left(  m\right)  ,h\right)
,p=\varphi_{Ui}\left(  m,u\right)  $

$\xi_{q}=\varphi_{Hm}^{\prime}\left(  m,h\right)  \xi^{m}+\varphi_{Hh}%
^{\prime}\left(  m,h\right)  \xi^{h}$

$=\left(  \varphi_{Um}^{\prime}\left(  m,u\right)  +\varphi_{Uu}^{\prime
}\left(  m,u\right)  \left(  R_{h}^{\prime}\varkappa\right)  \varkappa
^{\prime}\left(  m\right)  \right)  \xi^{m}+\varphi_{Uu}^{\prime}\left(
m,u\right)  \left(  L_{\varkappa\left(  m\right)  }^{\prime}h\right)  \xi^{h}$

$\xi_{q}\in T_{p}U_{M}$ so one can compute :

$\varphi_{U}^{\ast}\mathbf{A}\left(  p\right)  \left(  \xi^{m},\left(
R_{h}^{\prime}\varkappa\right)  \varkappa^{\prime}\xi^{m}+\left(
L_{\varkappa}^{\prime}h\right)  \xi^{h}\right)  $

$=\varphi_{Uu}^{\prime}(m,u)\left(  \left(  R_{h}^{\prime}\varkappa\right)
\varkappa^{\prime}\xi^{m}+\left(  L_{\varkappa}^{\prime}h\right)  \xi
^{h}+\left(  R_{u}^{\prime}1\right)  \grave{A}\left(  m\right)  \xi
^{m}\right)  $

$\left(  R_{u}^{\prime}1\right)  \grave{A}\left(  m\right)  \xi^{m}=\left(
R_{h}^{\prime}\varkappa\right)  \left(  R_{\varkappa}^{\prime}1\right)
\grave{A}\left(  m\right)  \xi^{m}$

$\left(  R_{\varkappa}^{\prime}1\right)  \grave{A}\left(  m\right)  \xi^{m}\in
T_{\varkappa}U$

$\Rightarrow\exists\overrightarrow{\theta}^{h},\overrightarrow{\theta}%
^{l}:\left(  R_{\varkappa}^{\prime}1\right)  \grave{A}\left(  m\right)
\xi^{m}=R_{\varkappa}^{\prime}1\overrightarrow{\theta}^{l}+L_{\varkappa
}^{\prime}1\overrightarrow{\theta}^{h}$

$\pi_{H}^{\prime}\left(  \varkappa\right)  \left(  R_{\varkappa}^{\prime
}1\right)  \grave{A}\left(  m\right)  \xi^{m}=\pi_{H}^{\prime}\left(
\varkappa\right)  R_{\varkappa}^{\prime}1\overrightarrow{\theta}^{l}+\pi
_{H}^{\prime}\left(  \varkappa\right)  L_{\varkappa}^{\prime}1\overrightarrow
{\theta}^{h}=R_{\varkappa}^{\prime}1\overrightarrow{\theta}^{l}$

but $\lambda_{u}^{\prime}\left(  1,\varkappa\right)  =\pi_{H}^{\prime}\left(
\varkappa\right)  R_{x}^{\prime}1$ so one can write:

$R_{\varkappa}^{\prime}1\overrightarrow{\theta}^{l}=\lambda_{g}^{\prime
}\left(  1,\varkappa\right)  \grave{A}\left(  m\right)  \xi^{m}\Rightarrow
L_{\varkappa}^{\prime}1\overrightarrow{\theta}^{h}=\left(  R_{\varkappa
}^{\prime}1\right)  \grave{A}\left(  m\right)  \xi^{m}-\lambda_{g}^{\prime
}\left(  1,\varkappa\right)  \grave{A}\left(  m\right)  \xi^{m}$

$\overrightarrow{\theta}^{h}=\left(  L_{\varkappa^{-1}}^{\prime}%
\varkappa\right)  \left(  \left(  R_{\varkappa}^{\prime}1\right)  -\lambda
_{g}^{\prime}\left(  1,\varkappa\right)  \right)  \grave{A}\left(  m\right)
\xi^{m}$

$\overrightarrow{\theta}^{h}$ is a linear function of $\xi^{m},$\ which does
not depend on h (but on $\varkappa)$.

Let us define :

$\Gamma_{H}\left(  \varphi_{H}\left(  m,1\right)  \right)  :T_{m}M\rightarrow
T_{1}H::\Gamma_{H}\left(  \chi\left(  m\right)  \right)  =\left(
L_{\varkappa^{-1}}^{\prime}\varkappa\right)  \left(  \left(  R_{\varkappa
}^{\prime}1\right)  -\lambda_{g}^{\prime}\left(  1,\varkappa\right)  \right)
\grave{A}\left(  m\right)  $

$\left(  R_{u}^{\prime}1\right)  \grave{A}\left(  m\right)  \xi^{m}=\left(
R_{h}^{\prime}\varkappa\right)  \left(  R_{\varkappa}^{\prime}1\right)
\grave{A}\left(  m\right)  \xi^{m}=\left(  R_{h}^{\prime}\varkappa\right)
\left(  \lambda_{u}^{\prime}\left(  1,\varkappa\right)  \grave{A}\left(
m\right)  +\left(  L_{\varkappa}^{\prime}1\right)  \Gamma_{H}\left(
\chi\right)  \right)  \xi^{m}$

and going back to :

$\varphi_{U}^{\ast}\mathbf{A}\left(  p\right)  \left(  \xi^{m},\left(
R_{h}^{\prime}\varkappa\right)  \varkappa^{\prime}\xi^{m}+\left(
L_{\varkappa}^{\prime}h\right)  \xi^{h}\right)  $

$=\varphi_{Uu}^{\prime}(m,u)\left(  \left(  R_{h}^{\prime}\varkappa\right)
\varkappa^{\prime}\xi^{m}+\left(  L_{\varkappa}^{\prime}h\right)  \xi
^{h}+\left(  R_{h}^{\prime}\varkappa\right)  \left(  \lambda_{u}^{\prime
}\left(  1,\varkappa\right)  \grave{A}\left(  m\right)  +\left(  L_{\varkappa
}^{\prime}1\right)  \Gamma_{H}\left(  \chi\right)  \right)  \xi^{m}\right)  $

$=\varphi_{Uu}^{\prime}(m,u)\left(  \left(  R_{h}^{\prime}\varkappa\right)
\left(  \varkappa^{\prime}+\lambda_{u}^{\prime}\left(  1,\varkappa\right)
\grave{A}\left(  m\right)  \right)  \xi^{m}\right)  $

$+\varphi_{Uu}^{\prime}(m,u)\left(  \left(  L_{\varkappa}^{\prime}h\right)
\xi^{h}+\left(  R_{h}^{\prime}\varkappa\right)  \left(  L_{\varkappa}^{\prime
}1\right)  \Gamma_{H}\left(  \chi\right)  \xi^{m}\right)  $

$\varphi_{Uu}^{\prime}\left(  m,\varkappa\right)  \left(  \varkappa^{\prime
}+\lambda_{u}^{\prime}(1,x\left(  m\right)  )\grave{A}\left(  m\right)
\right)  \xi^{m}=\left(  \nabla^{X}\varkappa\right)  \xi^{m}$

$\left(  \left(  L_{\varkappa}^{\prime}h\right)  \xi^{h}+\left(  R_{h}%
^{\prime}\varkappa\right)  \left(  L_{\varkappa}^{\prime}1\right)  \Gamma
_{H}\left(  \chi\right)  \xi^{m}\right)  =\left(  L_{\varkappa}^{\prime
}h\right)  \left(  \xi^{h}+R_{h}^{\prime}(1)\Gamma_{H}\left(  \chi\right)
\xi^{m}\right)  $

with : $R_{h}^{\prime}\left(  \varkappa\right)  =R_{\varkappa h}^{\prime
}\left(  1\right)  R_{\varkappa^{-1}}^{\prime}\left(  \varkappa\right)
=L_{\varkappa h}^{\prime}\left(  1\right)  Ad_{\left(  \varkappa h\right)
^{-1}}R_{\varkappa^{-1}}^{\prime}\left(  \varkappa\right)  $

$=L_{g}^{\prime}\left(  1\right)  Ad_{h^{-1}}Ad_{\varkappa^{-1}}%
R_{\varkappa^{-1}}^{\prime}\left(  \varkappa\right)  =L_{u}^{\prime}\left(
1\right)  L_{h^{-1}}^{\prime}(h)R_{h}^{\prime}(1)L_{\varkappa^{-1}}^{\prime
}(\varkappa)R_{\varkappa}^{\prime}(1)R_{\varkappa^{-1}}^{\prime}\left(
\varkappa\right)  $

$=L_{\varkappa}^{\prime}\left(  h\right)  R_{h}^{\prime}(1)L_{\varkappa^{-1}%
}^{\prime}(\varkappa)$

$\varphi_{Uu}^{\prime}(m,u)\left(  L_{\varkappa}^{\prime}h\right)  \left(
\xi^{h}+R_{h}^{\prime}(1)\Gamma_{H}\left(  \chi\right)  \xi^{m}\right)
=\varphi_{Hh}^{\prime}\left(  m,h\right)  \left(  \xi^{h}+R_{h}^{\prime
}(1)\Gamma_{H}\left(  \chi\right)  \xi^{m}\right)  $

$\varphi_{U}^{\ast}\mathbf{A}\left(  p\right)  \left(  \xi^{m},\left(
R_{h}^{\prime}\varkappa\right)  \varkappa^{\prime}\xi^{m}+\left(
L_{\varkappa}^{\prime}h\right)  \xi^{h}\right)  $

$=\left(  \nabla^{X}\varkappa\right)  \xi^{m}+\varphi_{Hh}^{\prime}\left(
m,h\right)  \left(  \xi^{h}+R_{h}^{\prime}(1)\Gamma_{H}\left(  \chi\right)
\xi^{m}\right)  $

If $\left(  \nabla^{X}\varkappa\right)  =0$ then we have the principal connection

$\varphi_{Hh}^{\prime}\left(  m,h\right)  \left(  \xi^{h}+R_{h}^{\prime
}(1)\Gamma_{H}\left(  \chi\right)  \xi^{m}\right)  $

This condition is also necessary (Giachetta [5] 5.10.5).$\blacksquare$

The potential of $\mathbf{\Gamma}_{H}$ is :

$\Gamma_{H}\left(  \chi\left(  m\right)  \right)  =\left(  L_{\varkappa^{-1}%
}^{\prime}\varkappa\right)  \left(  \left(  R_{\varkappa}^{\prime}1\right)
-\lambda_{u}^{\prime}\left(  1,\varkappa\right)  \right)  \grave{A}\left(
m\right)  $

$=Ad_{\varkappa^{-1}}\grave{A}\left(  m\right)  -\left(  L_{\varkappa^{-1}%
}^{\prime}\varkappa\right)  R_{x}^{\prime}1\left(  \sum_{\alpha}\sum
_{a=r+1}^{m}\grave{A}_{\alpha}^{a}\overrightarrow{\theta}_{a}\otimes
dx^{\alpha}\right)  $

$\Gamma_{H}\left(  \chi\left(  m\right)  \right)  =Ad_{\varkappa^{-1}}\left(
\sum_{\alpha}\sum_{a=1}^{r}\grave{A}_{\alpha}^{a}\overrightarrow{\theta}%
_{a}\otimes dx^{\alpha}\right)  $

\paragraph{3)}

The condition $\nabla^{X}\varkappa=0$\ reads :

$\frac{dx}{dm}+\lambda_{u}^{\prime}(1,x\left(  m\right)  )\grave{A}\left(
m\right)  =0$

$\Rightarrow\frac{dx}{dm}=-\left(  R_{x\left(  m\right)  }^{\prime}1\right)
\left(  \sum_{\alpha}\sum_{a=r+1}^{m}\grave{A}_{\alpha}^{a}\overrightarrow
{\theta}_{a}\otimes dx^{\alpha}\right)  $

With $\varkappa\left(  m\right)  =\exp l\left(  m\right)  $ the first step to
solve the problem is to find a map $l:M\rightarrow l_{0}$\ such that :

$\left(  \frac{d}{dl}\exp l\right)  \frac{dl}{dm}=-R_{x\left(  m\right)
}^{\prime}1\left(  \sum_{\alpha}\sum_{a=r+1}^{m}\grave{A}_{\alpha}%
^{a}\overrightarrow{\theta}_{a}\otimes dx^{\alpha}\right)  $

Using the derivative of exp (Duistermaat [4] 1.5):

$\left(  \frac{d}{dl}\exp l\right)  \frac{dl}{dm}=\left(  R_{x\left(
m\right)  }^{\prime}1\right)  \left(  \int_{0}^{1}e^{\tau adl\left(  m\right)
}d\zeta\right)  \frac{dl}{dm}$

$=-R_{x\left(  m\right)  }^{\prime}1\left(  \sum_{\alpha}\sum_{a=r+1}%
^{m}\grave{A}_{\alpha}^{a}\overrightarrow{\theta}_{a}\otimes dx^{\alpha
}\right)  $

So the map l(m) is solution of the equation :%

\begin{equation}
\left(  \int_{0}^{1}e^{\tau adl\left(  m\right)  }d\tau\right)  \frac{dl}%
{dm}=-\sum_{\alpha}\sum_{a=r+1}^{m}\grave{A}_{\alpha}^{a}\overrightarrow
{\theta}_{a}\otimes dx^{\alpha}\label{E99}%
\end{equation}

The map $\overrightarrow{\theta}\rightarrow\int_{0}^{1}e^{\tau
ad\overrightarrow{\theta}}d\tau$ is inversible if ad($\overrightarrow{\theta}%
$) is inversible.\ It is inversible and analytic in a neighbourood of 0 and
its inverse is :

$\sum_{k=0}^{\infty}\frac{B_{k}}{k!}\left(  ad\overrightarrow{\theta}\right)
^{k}$ where $B_{k}$\ are the Bernouilli numbers.

$\int_{0}^{1}e^{\tau adl}d\tau=\sum_{k=0}^{\infty}\int_{0}^{1}\frac{1}{k!}%
\tau^{k}\left(  adl\left(  m\right)  \right)  ^{k}d\tau=\sum_{k=0}^{\infty
}\left(  \int_{0}^{1}\tau^{k}d\tau\right)  \frac{1}{k!}\left(  adl\left(
m\right)  \right)  ^{k}=\sum_{k=1}^{\infty}\frac{1}{k!}\left(  adl\left(
m\right)  \right)  ^{k-1}$

$\left(  \int_{0}^{1}e^{\tau adl\left(  m\right)  }d\tau\right)  \frac{dl}%
{dm}=\sum_{k=0}^{\infty}\frac{1}{k!}\frac{1}{k}\frac{d}{dm}\left(  adl\left(
m\right)  \right)  ^{k}=\frac{d}{dm}\left(  \sum_{k=1}^{\infty}\frac{1}%
{k!}\frac{1}{k}\left(  adl\left(  m\right)  \right)  ^{k}\right)  $

Thus : $\frac{d}{d\xi^{\alpha}}_{\alpha}\left(  \sum_{k=1}^{\infty}\frac
{1}{k!}\frac{1}{k}\left(  adl\left(  m\right)  \right)  ^{k}\right)
=-\sum_{a=r+1}^{m}\grave{A}_{\alpha}^{a}\overrightarrow{\theta}_{a}$

\subsubsection{The Higgs mechanism}

\paragraph{1)}

As previously the physical characteristics of particles are modelled in a
representation $\left(  W,\chi\right)  $\ of U and the associated vector
bundle $W_{U}=U_{M}\times_{U}W$ . The restriction $\left(  W,\chi\right)
$\ to H is a representation of H. With a global section $\chi$\ on $X_{M}%
$\ one has a principal fiber bundle $H_{M}$\ and the associated vector bundle
$W_{H}=H_{M}\times_{H}W$\ which is the restriction of $W_{U}$ to H.

\paragraph{2)}

The forces fields are principal connections (\textbf{G,A})\ on $Q_{M}.$
\textbf{A}\ induces a connection $\mathbf{\Gamma}$ on $X_{M}$ and a principal
connection $\Gamma_{H}$\ on $H_{M}$ iff $\nabla^{X}\varkappa=0.$

\paragraph{3)}

The lagrangians are the same.

\paragraph{4)}

We add the variable $\chi\left(  m\right)  =\varphi_{G}\left(  m,\varkappa
\left(  m\right)  \right)  $\ to characterize the vacuum. $\varkappa\left(
m\right)  $ is valued in X=U/H and with Cartan decomposition $\varkappa
(m)=\exp l(m)$ where $l:M\rightarrow l_{0}$ .\ The fields act on $\chi
$\ through the covariant derivative $\nabla^{X}\chi=\frac{dx}{dm}+\lambda
_{g}^{\prime}(1_{U},x)\grave{A}\left(  m\right)  $

\paragraph{5)}

The principle of least action works in two steps :

- at the $X_{M}$\ level (the vacuum) : the force fields (other than
gravitation) interact with the Higgs field and fix a section $\chi$\ such that
$\nabla^{X}\chi=0.$ This fixes the components a=r+1 to m of \`{A}\ by :
$\sum_{a=r+1}^{m}\grave{A}^{a}\overrightarrow{\theta}_{a}=-R_{\varkappa^{-1}%
}^{\prime}\left(  \varkappa\right)  \frac{d\varkappa}{dm}$

- at the $H_{M}$\ level (the system) : $\chi$\ being fixed the gauge group is
reduced, the fields act with the particles through $\Gamma_{H}.$ The Lagrange
equations fix the r first components of \`{A}.

This is equivalent to change the variables, and replace $\left(  \grave{A}%
^{a}\right)  _{a=r+1}^{m}$ by $\frac{d\varkappa}{dm}$\ in the lagrangian. Thus
one introduces m-r "bosons" which, besides the fermions in the Noether
currents can get a mass.

\paragraph{6)}

In the standard model the Higgs mechanism is more complicated, but the scheme
presented above shows the key ingredient of symmetry breakdown : a structure
of the vacuum more complex than expected.

\paragraph{7)}

This mechanism has been brought up for gravitation : the SO(3,1) structure
would come from a more general GL(4) structure (Sardanaschvily [23])

\newpage

\part{\textbf{APPLICATIONS}}

\section{GENERAL\ RELATIVITY}

\label{General relativity}

\paragraph{1)}

The well known Einstein equation can be deduced by the principle of least
action from a very general lagrangian. Let be :

$%
\mathcal{L}%
=\left(  L_{2}\left(  g,z^{i},z_{\alpha}^{i}\right)  +a_{G}\left(
\sum_{\alpha\beta}g^{\alpha\beta}Ric_{\alpha\beta}+\Lambda\right)  \right)
\sqrt{\left\vert \det g\right\vert }$

The key points are :

a) the metric g is in $L_{2}$ but not its derivatives, the other variables
$z^{i}$ are not involved here

b) the gravitation / gravitation interaction is modelled by the Palatini
action (the cosmological constant $\Lambda$\ is not significant here). We have
seen that in a gravitational field theory based on the metric and the
L\'{e}vy-Civita connection this choice is quite mandatory.

The functional derivative with respect to $g^{\alpha\beta}$ gives :

$\frac{\delta S}{\delta g^{\alpha\beta}}=\left(  \frac{\partial L_{2}%
}{\partial g^{\alpha\beta}}+a_{G}\frac{\partial}{\partial g^{\alpha\beta}%
}\left(  g^{\alpha\beta}Ric_{\alpha\beta}+\Lambda\right)  \right)
\sqrt{\left\vert \det g\right\vert }$

$+\left(  L_{2}+a_{G}\left(  g^{\alpha\beta}R_{\alpha\beta}+\Lambda\right)
\right)  \frac{\partial}{\partial g^{\alpha\beta}}\sqrt{\left\vert \det
g\right\vert }$

$\frac{\partial}{\partial g^{\alpha\beta}}\sqrt{\left\vert \det g\right\vert
}=\frac{\partial}{\partial g^{\alpha\beta}}\left(  -\det\left\vert
g^{-1}\right\vert \right)  ^{-1/2}$

$=\frac{1}{2}\left(  -\det g^{-1}\right)  ^{-3/2}\frac{\partial}{\partial
g^{\alpha\beta}}\left(  \det g^{-1}\right)  $

$=\frac{1}{2}\left(  -\det g^{-1}\right)  ^{-3/2}g_{\beta\alpha}\det
g^{-1}=-\frac{1}{2}g_{\beta\alpha}\sqrt{\det\left\vert g\right\vert }$

(notice that the indexes are reversed)

$\frac{\delta S}{\delta g^{\alpha\beta}}=\left(  \left(  \frac{\partial L_{2}%
}{\partial g^{\alpha\beta}}+a_{G}Ric_{\alpha\beta}\right)  -\frac{1}%
{2}g_{\beta\alpha}\left(  L_{2}+a_{G}\left(  g^{\alpha\beta}R_{\alpha\beta
}+\Lambda\right)  \right)  \right)  \sqrt{\left\vert \det g\right\vert }$

And we get the equation :

$\left(  \frac{\partial L_{2}}{\partial g^{\alpha\beta}}+a_{G}Ric_{\alpha
\beta}\right)  =\frac{1}{2}g_{\alpha\beta}\left(  L_{2}+a_{G}\left(
g^{\lambda\mu}R_{\lambda\mu}+\Lambda\right)  \right)  $

$a_{G}\left(  R_{\alpha\beta}-\frac{1}{2}g_{\alpha\beta}\left(  g^{\lambda\mu
}R_{\lambda\mu}+\Lambda\right)  \right)  =-\frac{\partial L_{2}}{\partial
g^{\alpha\beta}}+\frac{1}{2}g_{\alpha\beta}L_{2}$

Or with : $S_{2}=\int_{\Omega}L_{2}\sqrt{\left\vert \det g\right\vert }%
\varpi_{0}$

$\frac{\delta S_{2}}{\delta g^{\alpha\beta}}=\left(  \frac{\partial L_{2}%
}{\partial g^{\alpha\beta}}-\frac{1}{2}g_{\alpha\beta}L_{2}\right)
\sqrt{\left\vert \det g\right\vert }$

we have the Einstein equation :%

\begin{equation}
Ric_{\alpha\beta}-\frac{1}{2}g_{\alpha\beta}\left(  R+\Lambda\right)
=-\frac{1}{a_{G}\sqrt{\left\vert \det g\right\vert }}\frac{\delta S_{2}%
}{\delta g^{\alpha\beta}}\label{E100}%
\end{equation}

The quantity $T_{\alpha\beta}=-\frac{1}{a_{G}\sqrt{\left\vert \det
g\right\vert }}\frac{\delta S_{2}}{\delta g^{ab}}$ is the "stress energy"
tensor.\ It depends on the system and its specification is based on
phenomenological assumptions about the distribution of matter and its
velocity, and the other fields. It should be a symmetric 2-covariant tensor.
The Einstein equation implies $\nabla^{\alpha}T_{\alpha\beta}=0$. Then
particles usually follow geodesics (Wald [29] p73). This equation is local and
in the vacuum $T_{\alpha\beta}=0.$

\paragraph{2)}

Our model does not use the L\'{e}vy-Civita tensor and g is not a variable, but
is actually present, and the gravitational fields interaction is the same.\ So
one can compute the stress energy tensor. However some adjustments are
necessary :

a) replace g where it is actually used, meaning in the Dirac operator and the
scalar product $\left\langle
\mathcal{F}%
_{A},%
\mathcal{F}%
_{A}\right\rangle :$

$D\psi=\sum_{\alpha ij}\left(  \nabla_{\alpha}\psi^{ij}\right)  \sum_{\beta
l}g^{\alpha\beta}O^{\prime}\left(  m\right)  _{\beta}^{l}\left(  \rho
\circ\Upsilon\left(  \varepsilon_{l}\right)  \left(  e_{i}\right)  \right)
\left(  m\right)  \otimes f_{j}\left(  m\right)  $

$=\sum_{\alpha ij}\left(  \nabla_{\alpha}\psi^{ij}\right)  \sum_{\beta
l}g^{\alpha\beta}O^{\prime}\left(  m\right)  _{\beta}^{l}\left(  i\left[
\gamma_{0}\right]  _{i}^{p}+\sum_{l=1}^{3}\left[  \gamma_{l}\right]  _{i}%
^{p}\right)  e_{p}\left(  m\right)  \otimes f_{j}\left(  m\right)  $

We keep : $\gamma^{0}=-i\gamma_{0};r=1,2,3:\gamma^{r}=\gamma_{r}$

$D\psi=\sum_{\alpha ij}\left(  \nabla_{\alpha}\psi^{ij}\right)  \sum_{\beta
l}g^{\alpha\beta}O_{\beta}^{\prime r}\sum_{r=0}^{3}\eta^{rr}\left[  \gamma
^{r}\right]  _{i}^{p}e_{p}\left(  m\right)  \otimes f_{j}\left(  m\right)  $

$\left[  D\psi\right]  =\sum_{\alpha\beta r}g^{\alpha\beta}O_{\beta}^{\prime
r}\eta^{rr}\left[  \gamma^{r}\right]  \left[  \nabla_{\alpha}\psi\right]  $

$a_{F}\left\langle
\mathcal{F}%
_{A},%
\mathcal{F}%
_{A}\right\rangle =a_{F}\frac{1}{2}\sum_{a}\sum_{\alpha\beta\lambda\mu
}g^{\alpha\lambda}g^{\beta\mu}\overline{%
\mathcal{F}%
}_{A\alpha\beta}^{a}%
\mathcal{F}%
_{A\lambda\mu}^{a}$

b) In the model g is computed from O and automatically symmetric. Rather than
using a constraint it is simpler to take $g_{\alpha\beta}$ and $g_{\beta
\alpha}$ as distincts variables and replace $g_{\alpha\beta}$ with $\frac
{1}{2}\left(  g_{\alpha\beta}+g_{\beta\alpha}\right)  $

c) The Ricci tensor is symmetric only if g is symmetric, so one takes :

$Ric_{\alpha\beta}=\sum_{ij\gamma}\left[
\mathcal{F}%
_{B\alpha\gamma}\right]  _{j}^{i}O_{i}^{\gamma}O_{\beta}^{\prime j}%
\rightarrow\frac{1}{2}\sum_{ij\gamma}\left(  \left[
\mathcal{F}%
_{G\alpha\gamma}\right]  _{j}^{i}O_{\beta}^{\prime j}+\left[
\mathcal{F}%
_{G\beta\gamma}\right]  _{j}^{i}O_{\alpha}^{\prime j}\right)  O_{i}^{\gamma}$

$R=\frac{1}{4}\sum_{ij\gamma}\left(  g^{\alpha\beta}+g^{\beta\alpha}\right)
\left(  \left[
\mathcal{F}%
_{G\alpha\gamma}\right]  _{j}^{i}O_{\beta}^{\prime j}+\left[
\mathcal{F}%
_{G\beta\gamma}\right]  _{j}^{i}O_{\alpha}^{\prime j}\right)  O_{i}^{\gamma}$

d) With these adjusments the lagrangian becomes :

$%
\mathcal{L}%
_{M}=Na_{M}\left\langle \psi,\psi\right\rangle +Na_{I}\operatorname{Im}%
\left\langle \psi,\frac{1}{2}\sum_{\lambda\mu r}\left(  g^{\lambda\mu}%
+g^{\mu\lambda}\right)  O_{\mu}^{\prime r}\eta^{rr}\left[  \gamma^{r}\right]
\left[  \nabla_{\lambda}\psi\right]  \right\rangle $

$+Na_{D}\sum_{\lambda}\left\langle \psi,V^{\lambda}\nabla_{\lambda}%
\psi\right\rangle \sqrt{\left\vert \det g\right\vert }$

$%
\mathcal{L}%
_{F}=\{\frac{1}{8}a_{F}\sum_{a}\sum_{\lambda\mu\sigma\theta}\left(
g^{\lambda\sigma}+g^{\sigma\lambda}\right)  \left(  g^{\mu\theta}+g^{\theta
\mu}\right)  \overline{%
\mathcal{F}%
}_{A\lambda\mu}^{a}%
\mathcal{F}%
_{A\sigma\theta}^{a}$

$+\frac{1}{4}a_{G}\sum_{\lambda\mu ij\gamma}\left(  g^{\lambda\mu}%
+g^{\mu\lambda}\right)  \left(  \left[
\mathcal{F}%
_{G\lambda\gamma}\right]  _{j}^{i}O_{\mu}^{\prime j}+\left[
\mathcal{F}%
_{G\mu\gamma}\right]  _{j}^{i}O_{\lambda}^{\prime j}\right)  O_{i}^{\gamma
}\}\sqrt{\left\vert \det g\right\vert }$

and $L_{2}=L_{M}+\frac{1}{8}a_{F}\sum_{a}\sum_{\lambda\mu\sigma\theta}\left(
g^{\lambda\sigma}+g^{\sigma\lambda}\right)  \left(  g^{\mu\theta}+g^{\theta
\mu}\right)  \overline{%
\mathcal{F}%
}_{A\lambda\mu}^{a}%
\mathcal{F}%
_{A\sigma\theta}^{a}$

\paragraph{3)}

The functional derivatives read :

$\frac{\delta S_{2}}{\delta g^{\alpha\beta}}=\frac{\partial L_{2}%
\sqrt{\left\vert \det g\right\vert }}{\partial g^{\alpha\beta}}=\left(
\frac{\partial L_{2}}{\partial g^{\alpha\beta}}-\frac{1}{2}g_{\beta\alpha
}L_{2}\right)  \sqrt{\left\vert \det g\right\vert }$

$\frac{\delta S_{2}}{\delta g^{\beta\alpha}}=\frac{\partial L_{2}%
\sqrt{\left\vert \det g\right\vert }}{\partial g^{\beta\alpha}}=\left(
\frac{\partial L_{2}}{\partial g^{\beta\alpha}}-\frac{1}{2}g_{\alpha\beta
}L_{2}\right)  \sqrt{\left\vert \det g\right\vert }$

$\frac{\partial L_{2}}{\partial g^{\alpha\beta}}=Na_{I}\frac{1}{2}%
\sum_{\lambda\mu r}\left(  \delta_{\lambda}^{\alpha}\delta_{\mu}^{\beta
}+\delta_{\mu}^{\alpha}\delta_{\lambda}^{\beta}\right)  O_{\mu}^{\prime r}%
\eta^{rr}\operatorname{Im}\left\langle \psi,\left[  \gamma^{r}\right]  \left[
\nabla_{\lambda}\psi\right]  \right\rangle $

$+\frac{1}{8}a_{F}\sum_{a}\sum_{\lambda\mu\sigma\theta}\left(  \left(
\delta_{\lambda}^{\alpha}\delta_{\sigma}^{\beta}+\delta_{\sigma}^{\alpha
}\delta_{\lambda}^{\beta}\right)  \left(  g^{\mu\theta}+g^{\theta\mu}\right)
+\left(  g^{\lambda\sigma}+g^{\sigma\lambda}\right)  \left(  \delta_{\mu
}^{\alpha}\delta_{\theta}^{\beta}+\delta_{\theta}^{\alpha}\delta_{\mu}^{\beta
}\right)  \right)  \overline{%
\mathcal{F}%
}_{A\lambda\mu}^{a}%
\mathcal{F}%
_{A\sigma\theta}^{a}$

$\frac{\partial L_{2}}{\partial g^{\alpha\beta}}=Na_{I}\frac{1}{2}\sum_{r}%
\eta^{rr}\left(  O_{\beta}^{\prime r}\operatorname{Im}\left\langle
\psi,\left[  \gamma^{r}\right]  \left[  \nabla_{\alpha}\psi\right]
\right\rangle +O_{\alpha}^{\prime r}\operatorname{Im}\left\langle \psi,\left[
\gamma^{r}\right]  \left[  \nabla_{\beta}\psi\right]  \right\rangle \right)  $

$+\frac{1}{4}a_{F}\sum_{a\lambda\mu}\left(  \overline{%
\mathcal{F}%
}_{A\alpha\lambda}^{a}%
\mathcal{F}%
_{A\beta\mu}^{a}+\overline{%
\mathcal{F}%
}_{A\beta\lambda}^{a}%
\mathcal{F}%
_{A\alpha\mu}^{a}\right)  \left(  g^{\lambda\mu}+g^{\mu\lambda}\right)  $

and

$\frac{\partial L_{2}}{\partial g^{\beta\alpha}}=Na_{I}\frac{1}{2}\sum_{r}%
\eta^{rr}\left(  O_{\alpha}^{\prime r}\operatorname{Im}\left\langle
\psi,\left[  \gamma^{r}\right]  \left[  \nabla_{\beta}\psi\right]
\right\rangle +O_{\beta}^{\prime r}\operatorname{Im}\left\langle \psi,\left[
\gamma^{r}\right]  \left[  \nabla_{\alpha}\psi\right]  \right\rangle \right)
$

$+\frac{1}{4}a_{F}\sum_{a\lambda\mu}\left(  \overline{%
\mathcal{F}%
}_{A\beta\lambda}^{a}%
\mathcal{F}%
_{A\alpha\mu}^{a}+\overline{%
\mathcal{F}%
}_{A\alpha\lambda}^{a}%
\mathcal{F}%
_{A\beta\mu}^{a}\right)  \left(  g^{\lambda\mu}+g^{\mu\lambda}\right)
=\frac{\partial L_{2}}{\partial g^{\alpha\beta}}$

\paragraph{4)}

$\frac{\delta S_{2}}{\delta g^{\alpha\beta}}-\frac{\delta S_{2}}{\delta
g^{\beta\alpha}}=\frac{1}{2}\left(  g_{\alpha\beta}-g_{\beta\alpha}\right)
L_{2}\sqrt{\left\vert \det g\right\vert }$

$S_{2}$ is symmetric with respect to $g_{\alpha\beta},$ $g_{\beta\alpha}$ so
$g_{\alpha\beta}=g_{\beta\alpha}$

$\frac{1}{\sqrt{\left\vert \det g\right\vert }}\frac{\delta S_{2}}{\delta
g^{\alpha\beta}}=Na_{I}\frac{1}{2}\sum_{r}\eta^{rr}\left(  O_{\alpha}^{\prime
r}\operatorname{Im}\left\langle \psi,\left[  \gamma^{r}\right]  \left[
\nabla_{\beta}\psi\right]  \right\rangle +O_{\beta}^{\prime r}%
\operatorname{Im}\left\langle \psi,\left[  \gamma^{r}\right]  \left[
\nabla_{\alpha}\psi\right]  \right\rangle \right)  $

$+\frac{1}{2}a_{F}\sum_{\lambda\mu}\left(  \left(
\mathcal{F}%
_{A\beta\lambda},%
\mathcal{F}%
_{A\alpha\mu}\right)  +\left(
\mathcal{F}%
_{A\alpha\lambda},%
\mathcal{F}%
_{A\beta\mu}\right)  \right)  g^{\lambda\mu}-\frac{1}{2}g_{\alpha\beta}\left(
L_{M}+\frac{1}{2}a_{F}\sum_{\lambda\mu}\left(
\mathcal{F}%
_{A\lambda\mu},%
\mathcal{F}%
_{A}^{\lambda\mu}\right)  \right)  $

$T_{\alpha\beta}=\frac{1}{2}g_{\alpha\beta}\left(  \frac{1}{a_{G}}L-R\right)
-N\frac{a_{I}}{a_{G}}\frac{1}{2}\sum_{r}\eta^{rr}\left(  O_{\alpha}^{\prime
r}\operatorname{Im}\left\langle \psi,\left[  \gamma^{r}\right]  \left[
\nabla_{\beta}\psi\right]  \right\rangle +O_{\beta}^{\prime r}%
\operatorname{Im}\left\langle \psi,\left[  \gamma^{r}\right]  \left[
\nabla_{\alpha}\psi\right]  \right\rangle \right)  $

$-\frac{1}{2}\frac{a_{F}}{a_{G}}\sum_{a\lambda\mu}\left(  \left(
\mathcal{F}%
_{A\beta\lambda},%
\mathcal{F}%
_{A\alpha\mu}\right)  +\left(
\mathcal{F}%
_{A\alpha\lambda},%
\mathcal{F}%
_{A\beta\mu}\right)  \right)  g^{\lambda\mu}$

The tensor is symmetric with respect to $\alpha,\beta$

With $O_{\alpha}^{\prime r}=g_{\alpha\gamma}\eta^{rq}O_{q}^{\gamma},O_{\beta
}^{\prime r}=g_{\beta\gamma}\eta^{rq}O_{q}^{\gamma}$

$\sum_{r}\eta^{rr}\left(  O_{\alpha}^{\prime r}\operatorname{Im}\left\langle
\psi,\left[  \gamma^{r}\right]  \left[  \nabla_{\beta}\psi\right]
\right\rangle +O_{\beta}^{\prime r}\operatorname{Im}\left\langle \psi,\left[
\gamma^{r}\right]  \left[  \nabla_{\alpha}\psi\right]  \right\rangle \right)
$

$=\sum_{rq\gamma}\left(  \eta^{rr}g_{\alpha\gamma}\eta^{rq}O_{q}^{\gamma
}\operatorname{Im}\left\langle \psi,\left[  \gamma^{r}\right]  \left[
\nabla_{\beta}\psi\right]  \right\rangle +\eta^{rr}g_{\beta\gamma}\eta
^{rq}O_{q}^{\gamma}\operatorname{Im}\left\langle \psi,\left[  \gamma
^{r}\right]  \left[  \nabla_{\alpha}\psi\right]  \right\rangle \right)  $

$=\sum_{\gamma q}O_{q}^{\gamma}\left(  g_{\alpha\gamma}\operatorname{Im}%
\left\langle \psi,\left[  \gamma^{q}\right]  \left[  \nabla_{\beta}%
\psi\right]  \right\rangle +g_{\beta\gamma}\operatorname{Im}\left\langle
\psi,\left[  \gamma^{q}\right]  \left[  \nabla_{\alpha}\psi\right]
\right\rangle \right)  $

$g^{\lambda\mu}\left(  \left(
\mathcal{F}%
_{A\beta\lambda},%
\mathcal{F}%
_{A\alpha\mu}\right)  +\left(
\mathcal{F}%
_{A\alpha\lambda},%
\mathcal{F}%
_{A\beta\mu}\right)  \right)  =2g^{\lambda\mu}\operatorname{Re}\left(
\mathcal{F}%
_{A\alpha\mu},%
\mathcal{F}%
_{A\beta\lambda}\right)  $

$T_{\alpha\beta}=\frac{1}{2}g_{\alpha\beta}\left(  \frac{1}{a_{G}}L-R\right)
-\frac{a_{F}}{a_{G}}\sum_{\lambda\mu}g^{\lambda\mu}\operatorname{Re}\left(
\mathcal{F}%
_{A\alpha\lambda},%
\mathcal{F}%
_{A\beta\mu}\right)  $

$-\frac{1}{2}\frac{a_{I}}{a_{G}}V\sum_{\gamma q}O_{q}^{\gamma}\left(
g_{\alpha\gamma}\operatorname{Im}\left\langle \psi,\left[  \gamma^{q}\right]
\left[  \nabla_{\beta}\psi\right]  \right\rangle +g_{\beta\gamma
}\operatorname{Im}\left\langle \psi,\left[  \gamma^{q}\right]  \left[
\nabla_{\alpha}\psi\right]  \right\rangle \right)  $

\paragraph{5)}

With equation \ref{E87}:

$\forall\alpha,\beta:\delta_{\beta}^{\alpha}L=Na_{I}\operatorname{Im}%
\left\langle \psi,\gamma^{\alpha}\nabla_{\beta}\psi\right\rangle +2a_{F}%
\sum_{\lambda}\operatorname{Re}\left(
\mathcal{F}%
_{A\beta\lambda},%
\mathcal{F}%
_{A}^{\alpha\lambda}\right)  $

$+2a_{G}\sum_{a\lambda}%
\mathcal{F}%
_{G\beta\lambda}^{a}\left(  O_{q_{a}}^{\alpha}O_{p_{a}}^{\lambda}-O_{p_{a}%
}^{\alpha}O_{q_{a}}^{\lambda}\right)  $

which is equivalent to :

$\forall q,\beta:LO_{q}^{\prime\beta}=Na_{I}\operatorname{Im}\left\langle
\psi,\gamma^{q}\nabla_{\beta}\psi\right\rangle +2a_{F}\sum_{\lambda\mu
}O_{\lambda}^{\prime q}\operatorname{Re}\left(
\mathcal{F}%
_{A\beta\mu},%
\mathcal{F}%
_{A}^{\lambda\mu}\right)  $

$+2a_{G}\sum_{a\lambda}%
\mathcal{F}%
_{G\beta\lambda}^{a}\left(  \delta_{q_{a}}^{q}O_{p_{a}}^{\lambda}%
-\delta_{p_{a}}^{q}O_{q_{a}}^{\lambda}\right)  $

$Na_{I}\operatorname{Im}\left\langle \psi,\gamma^{q}\nabla_{\beta}%
\psi\right\rangle =LO_{\beta}^{\prime q}-2a_{F}\sum_{\lambda\mu}O_{\lambda
}^{\prime q}\operatorname{Re}\left(
\mathcal{F}%
_{A\beta\mu},%
\mathcal{F}%
_{A}^{\lambda\mu}\right)  $

$-2a_{G}\sum_{a\lambda}%
\mathcal{F}%
_{G\beta\lambda}^{a}\left(  \delta_{q_{a}}^{q}O_{p_{a}}^{\lambda}%
-\delta_{p_{a}}^{q}O_{q_{a}}^{\lambda}\right)  $

So :

$Na_{I}\sum_{\gamma q}O_{q}^{\gamma}\left(  g_{\alpha\gamma}\operatorname{Im}%
\left\langle \psi,\left[  \gamma^{q}\right]  \left[  \nabla_{\beta}%
\psi\right]  \right\rangle +g_{\beta\gamma}\operatorname{Im}\left\langle
\psi,\left[  \gamma^{q}\right]  \left[  \nabla_{\alpha}\psi\right]
\right\rangle \right)  $

$=\sum_{\gamma q}O_{q}^{\gamma}g_{\alpha\gamma}\{LO_{\beta}^{\prime q}%
-2a_{F}\sum_{\lambda\mu}O_{\lambda}^{\prime q}\operatorname{Re}\left(
\mathcal{F}%
_{A\beta\mu},%
\mathcal{F}%
_{A}^{\lambda\mu}\right)  $

$-2a_{G}\sum_{a\lambda}%
\mathcal{F}%
_{G\beta\lambda}^{a}\left(  \delta_{q_{a}}^{q}O_{p_{a}}^{\lambda}%
-\delta_{p_{a}}^{q}O_{q_{a}}^{\lambda}\right)  \}$

$+\sum_{\gamma q}O_{q}^{\gamma}g_{\beta\gamma}\{LO_{\alpha}^{\prime q}%
-2a_{F}\sum_{\lambda\mu}O_{\lambda}^{\prime q}\operatorname{Re}\left(
\mathcal{F}%
_{A\alpha\mu},%
\mathcal{F}%
_{A}^{\lambda\mu}\right)  $

$-2a_{G}\sum_{a\lambda}%
\mathcal{F}%
_{G\alpha\lambda}^{a}\left(  \delta_{q_{a}}^{q}O_{p_{a}}^{\lambda}%
-\delta_{p_{a}}^{q}O_{q_{a}}^{\lambda}\right)  \}$

$=2Lg_{\alpha\beta}-2a_{F}\sum_{\lambda\mu}\left(  g_{\alpha\lambda
}\operatorname{Re}\left(
\mathcal{F}%
_{A\beta\mu},%
\mathcal{F}%
_{A}^{\lambda\mu}\right)  +g_{\beta\lambda}\operatorname{Re}\left(
\mathcal{F}%
_{A\alpha\mu},%
\mathcal{F}%
_{A}^{\lambda\mu}\right)  \right)  $

$-2a_{G}\sum_{a\lambda\mu}\left(  g_{\alpha\mu}%
\mathcal{F}%
_{G\beta\lambda}^{a}+g_{\beta\mu}%
\mathcal{F}%
_{G\alpha\lambda}^{a}\right)  \left(  O_{q_{a}}^{\mu}O_{p_{a}}^{\lambda
}-O_{p_{a}}^{\mu}O_{q_{a}}^{\lambda}\right)  $

And the stress energy tensor reads:

$T_{\alpha\beta}=\frac{1}{2}g_{\alpha\beta}\left(  \frac{1}{a_{G}}L-R\right)
-\frac{a_{F}}{a_{G}}\sum_{\lambda\mu}g^{\lambda\mu}\operatorname{Re}\left(
\mathcal{F}%
_{A\alpha\lambda},%
\mathcal{F}%
_{A\beta\mu}\right)  $

$-\frac{1}{2}\frac{1}{a_{G}}\{2Lg_{\alpha\beta}-2a_{F}\sum_{\lambda\mu}\left(
g_{\alpha\lambda}\operatorname{Re}\left(
\mathcal{F}%
_{A\beta\mu},%
\mathcal{F}%
_{A}^{\lambda\mu}\right)  +g_{\beta\lambda}\operatorname{Re}\left(
\mathcal{F}%
_{A\alpha\mu},%
\mathcal{F}%
_{A}^{\lambda\mu}\right)  \right)  $

$-2a_{G}\sum_{a\lambda\mu}\left(  g_{\alpha\mu}%
\mathcal{F}%
_{G\beta\lambda}^{a}+g_{\beta\mu}%
\mathcal{F}%
_{G\alpha\lambda}^{a}\right)  \left(  O_{q_{a}}^{\mu}O_{p_{a}}^{\lambda
}-O_{p_{a}}^{\mu}O_{q_{a}}^{\lambda}\right)  \}$

$T_{\alpha\beta}=-\frac{1}{2}g_{\alpha\beta}\frac{1}{a_{G}}L$

$+\frac{a_{F}}{a_{G}}\sum_{\lambda\mu}\operatorname{Re}\left(  g_{\alpha
\lambda}\left(
\mathcal{F}%
_{A\beta\mu},%
\mathcal{F}%
_{A}^{\lambda\mu}\right)  +g_{\beta\lambda}\left(
\mathcal{F}%
_{A\alpha\mu},%
\mathcal{F}%
_{A}^{\lambda\mu}\right)  -g^{\lambda\mu}\left(
\mathcal{F}%
_{A\alpha\lambda},%
\mathcal{F}%
_{A\beta\mu}\right)  \right)  $

$+\sum_{a\lambda\mu}\left(  g_{\alpha\mu}%
\mathcal{F}%
_{G\beta\lambda}^{a}+g_{\beta\mu}%
\mathcal{F}%
_{G\alpha\lambda}^{a}+\frac{1}{2}g_{\alpha\beta}%
\mathcal{F}%
_{G\lambda\mu}^{a}\right)  \left(  O_{q_{a}}^{\mu}O_{p_{a}}^{\lambda}%
-O_{p_{a}}^{\mu}O_{q_{a}}^{\lambda}\right)  $

with $R=-\sum_{a,\alpha\beta}%
\mathcal{F}%
_{G\lambda\mu}^{a}\left(  O_{p_{a}}^{\lambda}O_{q_{a}}^{\mu}-O_{q_{a}%
}^{\lambda}O_{p_{a}}^{\mu}\right)  $

\section{ELECTROMAGNETISM}

\label{Electromagnetism}

\paragraph{1)}

The group U is here $U(1)=\left\{  z=\exp i\theta,\theta\in%
\mathbb{R}
\right\}  $,an abelian group (so the bracket is null) with algebra:

$T_{1}U(1)=i%
\mathbb{R}
\Rightarrow\overrightarrow{\theta}_{1}=i.$

The complexified $T_{1}U(1)^{c}=%
\mathbb{C}
$ and the group $U^{c}=%
\mathbb{C}
.$

The only irreducible representations are 1 complex dimensional :

$W=\left\{  \sigma\overrightarrow{f},\sigma\in%
\mathbb{C}
\right\}  ,\chi\left(  \exp i\theta\right)  \overrightarrow{f}=e^{i\theta
}\overrightarrow{f}$

$\chi^{\prime}\left(  1\right)  =i=\left[  \theta_{a}\right]  ;\left[
\theta_{a}\right]  ^{t}=\left[  \theta_{a}\right]  =i=-\left[  \overline
{\theta}_{a}\right]  $

The state tensor is the sum of 2 right and left components : $\psi=\psi
_{R}+\psi_{L}.$\ If each of these components is decomposable : $\psi=\phi
_{R}\sigma_{R}+\phi_{L}\sigma_{L}$ where $\phi_{R},\phi_{L}$ are 2 complex
dimensional vectors and $\sigma_{R},\sigma_{L}$ complex scalar functions.

\paragraph{2)}

The moments are :

a)

$a<4:P_{a}=\operatorname{Im}\left(  \left[  \phi_{R}^{\ast}\right]  \left[
\sigma_{a}\right]  \left[  \phi_{L}\right]  \left(  \sigma_{L}\overline
{\sigma_{R}}\right)  \right)  $

$a>3:P_{a}=-\operatorname{Re}\left(  \left(  \left[  \phi_{R}^{\ast}\right]
\left[  \sigma_{a-3}\right]  \left[  \phi_{L}\right]  \right)  \left(
\sigma_{L}\overline{\sigma_{R}}\right)  \right)  $

b) $J_{r}=-\frac{1}{2}\left(  \eta^{rr}\left(  \left[  \phi_{R}^{\ast}\right]
\left[  \sigma_{a}\right]  \left[  \phi_{R}\right]  \right)  \left\vert
\sigma_{R}\right\vert ^{2}-\left(  \left[  \phi_{L}^{\ast}\right]  \left[
\sigma_{a}\right]  \left[  \phi_{L}\right]  \right)  \left\vert \sigma
_{L}\right\vert ^{2}\right)  $

c) $\left\langle \psi,\psi\right\rangle =-2\operatorname{Im}\left(  \left(
\left[  \phi_{R}\right]  ^{\ast}\left[  \phi_{L}\right]  \right)  \left(
\sigma_{L}\overline{\sigma_{R}}\right)  \right)  $

d) $\rho=-2\operatorname{Im}\left(  \left(  \left[  \phi_{R}^{\ast}\right]
\left[  \phi_{L}\right]  \right)  \left(  \sigma_{L}\overline{\sigma_{R}%
}\right)  \right)  =\left\langle \psi,\psi\right\rangle $

e) $\left[  \mu_{R}-\mu_{L}\right]  ^{r}=\eta^{rr}\left(  \left[  \phi
_{R}^{\ast}\right]  \sigma_{r}\left[  \phi_{R}\right]  \right)  \left\vert
\sigma_{R}\right\vert ^{2}-\left(  \left[  \phi_{L}^{\ast}\right]  \sigma
_{r}\left[  \phi_{L}\right]  \right)  \left\vert \sigma_{L}\right\vert
^{2}=-2J_{r}$

3) The potential \`{A} is a 1-form over M valued in the complexified, so
$\grave{A}=\grave{A}_{\alpha}dx^{\alpha},\grave{A}_{\alpha}\in%
\mathbb{C}
.$

The curvature 2-form $%
\mathcal{F}%
_{A}=d\grave{A}=\sum_{\left\{  \alpha\beta\right\}  }\left(  \partial_{\alpha
}\grave{A}_{\beta}-\partial_{\beta}\grave{A}_{\alpha}\right)  dx^{\alpha
}\wedge dx^{\beta}$ and we have the first Maxwell equation : $d%
\mathcal{F}%
_{A}=0.$

The equation \ref{E82} reads:

$\forall a,\alpha:-N\left(  a_{D}V^{\alpha}\rho+2ia_{I}\sum_{r}O_{r}^{\alpha
}J_{r}\right)  \det O^{\prime}=2a_{F}\sum_{\beta}\partial_{\beta}\left(
\mathcal{F}%
_{A}^{a\alpha\beta}\left(  \det O^{\prime}\right)  \right)  $

Taking the real and imaginary parts :

$\forall a,\alpha:-a_{D}N\left(  V^{\alpha}\rho\right)  \det O^{\prime}%
=2a_{F}\sum_{\beta}\partial_{\beta}\left(  \operatorname{Re}%
\mathcal{F}%
_{A}^{a\alpha\beta}\left(  \det O^{\prime}\right)  \right)  $

$\forall a,\alpha:-a_{I}N\sum_{r}O_{r}^{\alpha}J_{r}\det O^{\prime}=a_{F}%
\sum_{\beta}\partial_{\beta}\left(  \operatorname{Im}%
\mathcal{F}%
_{A}^{a\alpha\beta}\left(  \det O^{\prime}\right)  \right)  $

The 2nd Maxwell equation, without spin, is usually written in the General
Relativity picture :

$\nabla^{\beta}%
\mathcal{F}%
_{A\beta\alpha}=-\mu_{0}J_{\alpha}\Leftrightarrow\mu_{0}J^{\alpha}\sqrt{-\det
g}=\sum_{\beta}\partial_{\beta}\left(
\mathcal{F}%
_{A}^{\alpha\beta}\sqrt{-\det g}\right)  $

with the 4-vector current density $J_{\alpha}$, which is for a particle
$\rho\sum u^{\alpha}\partial_{\alpha}$ with the velocity measured with respect
to the proper time of the particle. So we are lead to identify the real part
of the field with the "usual" electromagnetic field, $\rho$ with the
electromagnetic charge and the constants are such that : $\mu_{0}=\frac{a_{D}%
}{2a_{F}}.$

Here the electromagnetic field has a real and an imaginary part, the latter
acting on the magnetic moment which is parallel to the angular momentum :
$\overrightarrow{\mu}=\sum_{r}\left[  \mu_{A}\right]  ^{r}\partial_{r}%
=-2\sum_{r}J_{r}\partial_{r}.$

Remark : it is common to have an the converse, with the "real" electromagnetic
field purely imaginary.\ The result here is the consequence of the choice of
the signature.

\paragraph{4)}

The Noether current reads here (equation \ref{E79}) :

$Y_{A}^{\alpha}=N\left(  a_{D}\sum_{\alpha}V^{\alpha}\rho^{a}\partial_{\alpha
}-ia_{I}\sum_{r}\left[  \mu_{R}-\mu_{L}\right]  _{a}^{r}\partial_{r}\right)  $

$\qquad=N\left(  a_{D}\sum_{\alpha}V^{\alpha}\rho\partial_{\alpha}-ia_{I}%
\sum_{r}\left[  \mu_{A}\right]  _{a}^{r}\partial_{r}\right)  $

because the bracket is null.\ So we have a global conservation of the flow of
current density and magnetic moment which are its real and imaginary parts.

Furthermore equation \ref{E92d} reads :$\sum_{\alpha}\frac{d}{d\xi^{\alpha}%
}\left(  \left(  N\det O^{\prime}\right)  V^{\alpha}\rho\right)  =0$ so the
flow of the charge current is conserved.

\paragraph{5)}

Within the same picture the moments for symmetric states would be:

a)

$a<4:P_{a}=-\epsilon r_{a}\left(  u^{2}+v^{2}\right)  \operatorname{Im}\left(
\sigma_{L}\overline{\sigma_{R}}\right)  $

$a>3:P_{a}=\epsilon r_{a-3}\left(  u^{2}+v^{2}\right)  \operatorname{Re}%
\left(  \sigma_{L}\overline{\sigma_{R}}\right)  $

b)

$J_{0}=-\frac{1}{2}\left(  u^{2}+v^{2}\right)  \left(  \left\vert \sigma
_{R}\right\vert ^{2}+\left\vert \sigma_{L}\right\vert ^{2}\right)  ;r>0:$
$J_{r}=\frac{1}{2}\left(  u^{2}+v^{2}\right)  \epsilon r_{r}\left(  \left\vert
\sigma_{R}\right\vert ^{2}-\left\vert \sigma_{L}\right\vert ^{2}\right)  $

$J_{r}=-\frac{1}{2}\left(  u^{2}+v^{2}\right)  \epsilon r_{r}\left(  \eta
^{rr}\left\vert \sigma_{R}\right\vert ^{2}-\left\vert \sigma_{L}\right\vert
^{2}\right)  $ with $r_{0}=\epsilon$

c) $\left\langle \psi,\psi\right\rangle =-2\left(  u^{2}+v^{2}\right)
\operatorname{Im}\left(  \sigma_{L}\overline{\sigma_{R}}\right)  $

d) $\rho=2\left(  u^{2}+v^{2}\right)  \operatorname{Re}\left(  \left[
\sigma_{L}\right]  \left[  i\right]  ^{t}\left[  \sigma_{R}\right]  ^{\ast
}\right)  =-2\left(  u^{2}+v^{2}\right)  \operatorname{Im}\left(  \sigma
_{L}\overline{\sigma_{R}}\right)  $

e) $\ \ \ \ \left[  \mu_{R}-\mu_{L}\right]  ^{r}=-2J_{r}$

So the particle has a charge if $\operatorname{Im}\left(  \sigma_{L}%
\overline{\sigma_{R}}\right)  \neq0,$ this emphasizes the need to use of the
complexified of U(1) (with U(1) the charge would be null) and of different
functions for the right and the left side (we know that the electromagnetic
field is part of the larger electroweak field for which chirality is crucial).
The quantities $\left(  r_{1},r_{2},r_{3}\right)  $ give the spatial direction
of both the angular and the magnetic momentum. It is also the orientation of
the would be linear momentum, up to a sign. Equation \ref{E93} reads :

$a_{D}\operatorname{Im}\left\langle \psi,\frac{d\psi}{d\xi^{0}}\right\rangle
=2\operatorname{Im}\left(  \sigma_{L}\overline{\sigma_{R}}\right)  \left(
a_{M}+a_{D}\sum_{\alpha a}V^{\alpha}\operatorname{Re}\grave{A}_{\alpha
}\right)  $

$+a_{D}\sum_{a}V^{\alpha}\epsilon\sum_{a=1}^{3}\left(  r_{a}\left(
\operatorname{Im}\left(  \sigma_{L}\overline{\sigma_{R}}\right)  G_{\alpha
}^{a}-\operatorname{Re}\left(  \sigma_{L}\overline{\sigma_{R}}\right)
G_{\alpha}^{a+3}\right)  \right)  $

$+a_{I}\frac{1}{2}\epsilon\left(  \eta^{jj}\left\vert \sigma_{R}\right\vert
^{2}-\left\vert \sigma_{L}\right\vert ^{2}\right)  \left(  \sum_{\alpha
j}\left(  \frac{dN\left(  \det O^{\prime}\right)  O_{j}^{\alpha}}{N\left(
\det O^{\prime}\right)  d\xi^{\alpha}}+2\left(  \operatorname{Im}\grave{A}%
_{j}\right)  \right)  r_{j}+\left(  \left[  r\right]  \left[  G_{r}\right]
\right)  _{r}\right)  $

with $\left(  u^{2}+v^{2}\right)  =1$

\newpage

\begin{center}
{\LARGE CONCLUSION}

\bigskip

\bigskip
\end{center}

\label{Conclusion}

Let us sum up the main results :

1) It is possible to model a system with individually interacting particles,
with gravitation and other fields,using the modern concepts of physical theory
(Yan-Mills connections, fiber bundle, Clifford algebra), but standing in the
classical picture. The principle of least action can be implemented, and the
constraints on the lagrangian can be met.

2) It is possible to give a sensible description of gravitation in the general
connection and tetrad framework, without involving a metric. This opens the
path to more general solutions than the L\'{e}vy-Civita connection, that would
be required if, as it seems, the connection is not torsion free.\ Moreover the
calculations are manageable, and can give explicit solutions with respect to
natural variables (the structure coefficients).

3) In the simple model we have seen the crucial role of "moments", clearly
identified with respect to the state tensor and clearly related to basic
physical concepts. Noether currents supply the conservation equations useful
to a further study.

4) The framework used to describe particles and fields provides a good basis
to study symmetries, and give hints for a better understanding of some
"paradoxical quantum phenomenon".

The main outcome of this paper is probably pedagogical, as it covers a great
deal of concepts in theoretical physics, using the tools of the trade.\ But
beyond this, several issues would be worth of further studies.

1) Is it possible to implement the machinery in the pure gravitational case ?
So far General Relativity has suffered both from intractable calculations, and
the metric obsession.\ It would be immensely useful to have manageable models,
pertinent for the hottest topics such as the movements of large systems
(galaxies notably) which are, after all, at the core of the "dark matter issue".

2) That is good to be able to model the individual movements of particles, but
of course it is mostly theoretical. So the next step is to introduce some
probabilistic particles distributions, this should be easy using the initial
state represented by the f function. In the thermodynamic picture it would be
of great interest to find a link between the moments \ and the "function of
state" of the whole system.

3) If this construction makes any sense, does it provide us with a better
understanding of the fundations of quantum mechanics ? I think so, and it will
be the topic of a next paper.

Some last words on more technical issues :

1) Introduce the velocity in the lagrangian is possible, even in the General
Relativty picture, and probably mandatory.\ It clearly enhances the
understanding of the interactions, showing the crucial and distinctive role of
the kinematic and dynamic parts. From this point of view the Dirac operator,
as essential as it is, is not enough as it emphasizes the first part. The
solution implemented here can certainly be improved.

2) Complex fields are mandatory, and they deserve the full treatment, even if
it is cumbersome.\ Any shortcut is hazardous.

3) The issue of the signature of the metric is still open...

\newpage

\begin{center}
{\LARGE BIBLIOGRAPHY}
\end{center}

\label{Bibliography}

[1] N.Ashby \textit{Relativity in the Global Positioning System} Living
reviews in relativity 6,(2003) 1

[2] V.A.Bednyakov, N.D.Giokaris, A.V. Bednyakov \textit{On Higgs mass
generation in the standard model }arXiv:hep-ph/0703280v1 27 March 2007

[3] Y.Choquet-Bruhat,N.Noutchegueme \textit{Syst\`{e}me de Yang-Mills Vlasov
en jauge temporelle} Annales de l'IHP section A tome 5 (1991)

[4] J.J.Duistermaat, J.A.Kolk \textit{Lie groups} Springer (1999)

[5] G.Giachetta, L.Mangiarotti, G.Sardanashvily \textit{Advanced classical
field theory} World Scientific (2009)

[6] A.Grigor'yan \textit{Heat kernel on weighted manifolds and applications}
paper (2005)

[7] M.Guidry \textit{Gauge field theories} John Wiley (1991)

[8] H.Halvorson \textit{Algebraic quantum field theory}
arXiv:math-ph/06022036v/1 14 feb 2006

[9] H.Hofer, E.Zehnder \textit{Symplectic invariants and hamiltonian dynamics}
Birkh\"{a}user Advanced Texts (1994)

[10] D.Husemoller \textit{Fibre bundles} (3d edition) Springer-Verlag (1993)

[11] A.W.Knapp \textit{Lie groups : beyond an introduction} 2nd edition
Birkh\"{a}user (2005)

[12] A.W.Knapp \textit{Representation theory of semi simple groups}
\textit{\ }Princeton landmarks (1986)

[13] S.Kobayashi, K.Nomizu \textit{Foundations of differential geometry}
J.Wiley (1996)

[14] I.Kol\'{a}r, P.W.Michor, J.Slov\`{a}k \textit{Natural operations in
differential geometry} Spinger-Verlag (1993)

[15] D.Krupka \textit{Some geometric aspects of variational problems in
fibered manifolds} Universita J.E.Purkyn\v{e} v Brn\v{e} (2001)

[16] S.Lang \textit{Fundamentals of differential geometry} Springer (1999)

[17] A.N.Lasenby, C.J.L.Doran \textit{Geometric algebra, Dirac wave functions
and black holes}

(see also the site : http://www.mrao.cam.ac.uk/\symbol{126}anthony/index.php).

[18] D.Lovelock, H.Rund \textit{Tensors, differential forms and variational
principles} Dover (1989)

[19] P.J.Morrison \textit{Hamiltonian and Action Principle Formulations of
Plasma Physics} Physics of Plasmas 12, 058102-1--13 (2005).

[20] R.Penrose \textit{The road to reality} Vintage books (2005)

[21] E.Poisson \textit{An introduction to the Lorentz-Dirac equation}
arXiv:gr-qc/9912045v1 10 Dec 1999

[22] T.C.Quinn \textit{Axiomatic approach to radiation reaction of scalar
point particles in curved space time} arXiv:gr-qc/0005030v1 10 may 2000

[23] G.Sardanaschvily \textit{Classical gauge theories of gravitation}
Theor.Math.Phys. 132,1163 (2002)

[24] D.E.Soper \textit{Classical field theory} Dover (2008)

[25] G.Svetlichny \textit{Preparation to gauge theories}
arXiv:math-ph/9902.27v3 12 march (1999)

[26] M.E.Taylor \textit{Partial differential equations} Spinger (1996)

[27] A.Trautman \textit{Einstein-Cartan theory} Encyclopedia of Mathematical
Physics Elsevier (2006)

[28] Wu-Ki Tung \textit{Group theory in Physics} World Scientific (1985)

[29] R.M.Wald \textit{General relativity} The University of Chicago Press (1984)

[30] S.Weinberg \textit{The quantum theory of fields} Cambidge University
Press (1995)

\end{document}